\documentclass[12pt,a4paper]{article}
\usepackage[a4paper]{geometry}

\usepackage{amsmath}
\usepackage{amsfonts}
\usepackage{amssymb}
\usepackage{dsfont}
\usepackage{graphicx}
\usepackage{enumitem}
\usepackage[nosort]{cite}
\usepackage[margin=10pt,font=small,labelfont=bf]{caption}
\usepackage[margin=0pt,font=small,labelfont=normalfont,skip=22pt]{subcaption}
\usepackage{extarrows}
\usepackage{multirow}
\usepackage{arydshln}
\usepackage[T1]{fontenc}
\usepackage[new]{old-arrows}
\usepackage{slashed}
\usepackage{hyperref}

\allowdisplaybreaks[2]
\numberwithin{equation}{section}
\graphicspath{{figures/}}

\newcommand{\ov}[1]{\overline{#1}}
\newcommand{\ul}[1]{\underline{#1}}
\newcommand{\op}{\hspace{1pt}}
\newcommand{\lab}{\mathsf }
\newcommand{\ba}{\mathsf }

\newcommand{\tri}{\hspace{-3.5pt}\vartriangle\hspace{-3.5pt}}

\newcommand{\eq}[1]{\begin{equation}\begin{split} #1 \end{split}\end{equation}}

\begin{document}

%%%%%%%%%%%%%%%%%%%%%%%%%%%%%%%%%%%%%%%%%%%%%%%
%%%%%%%%%%%%%%%%%%%%%%%%%%%%%%%%%%%%%%%%%%%%%%%
%%%%%%%%%%%%%%%%%%%%%%%%%%%%%%%%%%%%%%%%%%%%%%%
%%%%%%%%%%%%%%%%%%%%%%%%%%%%%%%%%%%%%%%%%%%%%%%
%%%%%%%%%%%%%%%%%%%%%%%%%%%%%%%%%%%%%%%%%%%%%%%
%%%%%%%%%%%%%%%%%%%%%%%%%%%%%%%%%%%%%%%%%%%%%%%
%%%%%%%%%%%%%%%%%%%%%%%%%%%%%%%%%%%%%%%%%%%%%%%
%%%%%%%%%%%%%%%%%%%%%%%%%%%%%%%%%%%%%%%%%%%%%%%

\thispagestyle{empty}
\vspace*{2cm}

%%%%%%%%%%%%%%%%%%%%%%%%%%%%%%%%%%%%%%%%%%%%%%
%%%%%%%%%%%%%%%%%%%%%%%%%%%%%%%%%%%%%%%%%%%%%%

\begin{center}
{\LARGE
Non-geometric backgrounds in string theory
}
\end{center}

%%%%%%%%%%%%%%%%%%%%%%%%%%%%%%%%%%%%%%%%%%%%%%
%%%%%%%%%%%%%%%%%%%%%%%%%%%%%%%%%%%%%%%%%%%%%%

\vspace{0.5cm}

\begin{center}
  Erik Plauschinn
\end{center}

%%%%%%%%%%%%%%%%%%%%%%%%%%%%%%%%%%%%%%%%%%%%%%
%%%%%%%%%%%%%%%%%%%%%%%%%%%%%%%%%%%%%%%%%%%%%%

\vspace{0.5cm}

\begin{center} 
\it 
Arnold Sommerfeld Center for Theoretical Physics \\[2pt]
Ludwig-Maximilians-Universit\"at M\"unchen \\[2pt]
Theresienstra\ss e 37 \\[2pt]
80333 Munich \\[2pt]
Germany 
\end{center} 

\vspace{2.5cm}

%%%%%%%%%%%%%%%%%%%%%%%%%%%%%%%%%%%%%%%%%%%%%%
%%%%%%%%%%%%%%%%%%%%%%%%%%%%%%%%%%%%%%%%%%%%%%

\begin{abstract}
\noindent
This review provides an introduction to non-geometric backgrounds in 
string theory. Starting from a  discussion of T-duality, geometric and non-geometric 
torus-fibrations are reviewed, generalised geometry and its relation to non-geometric 
backgrounds are explained and compactifications of string theory with geometric and 
non-geometric fluxes are discussed. Furthermore covered are doubled geometry 
as well as non-commutative and non-associative structures in the context of non-geometric
backgrounds.
\end{abstract}

%%%%%%%%%%%%%%%%%%%%%%%%%%%%%%%%%%%%%%%%%%%%%%
%%%%%%%%%%%%%%%%%%%%%%%%%%%%%%%%%%%%%%%%%%%%%%

\clearpage

%%%%%%%%%%%%%%%%%%%%%%%%%%%%%%%%%%%%%%%%%%%%%%%
%%%%%%%%%%%%%%%%%%%%%%%%%%%%%%%%%%%%%%%%%%%%%%%
%%%%%%%%%%%%%%%%%%%%%%%%%%%%%%%%%%%%%%%%%%%%%%%
%%%%%%%%%%%%%%%%%%%%%%%%%%%%%%%%%%%%%%%%%%%%%%%
%%%%%%%%%%%%%%%%%%%%%%%%%%%%%%%%%%%%%%%%%%%%%%%
%%%%%%%%%%%%%%%%%%%%%%%%%%%%%%%%%%%%%%%%%%%%%%%
%%%%%%%%%%%%%%%%%%%%%%%%%%%%%%%%%%%%%%%%%%%%%%%
%%%%%%%%%%%%%%%%%%%%%%%%%%%%%%%%%%%%%%%%%%%%%%%

\makeatletter
\def\tableofcontents{\section*{Table of Contents}\vspace*{4pt}\@starttoc{toc}}
\makeatother

\tableofcontents

\clearpage

%%%%%%%%%%%%%%%%%%%%%%%%%%%%%%%%%%%%%%%%%%%%%%%
%%%%%%%%%%%%%%%%%%%%%%%%%%%%%%%%%%%%%%%%%%%%%%%
%%%%%%%%%%%%%%%%%%%%%%%%%%%%%%%%%%%%%%%%%%%%%%%
%%%%%%%%%%%%%%%%%%%%%%%%%%%%%%%%%%%%%%%%%%%%%%%
%%%%%%%%%%%%%%%%%%%%%%%%%%%%%%%%%%%%%%%%%%%%%%%
%%%%%%%%%%%%%%%%%%%%%%%%%%%%%%%%%%%%%%%%%%%%%%%
%%%%%%%%%%%%%%%%%%%%%%%%%%%%%%%%%%%%%%%%%%%%%%%
%%%%%%%%%%%%%%%%%%%%%%%%%%%%%%%%%%%%%%%%%%%%%%%
%%%%%%%%%%%%%%%%%%%%%%%%%%%%%%%%%%%%%%%%%%%%%%%
%%%%%%%%%%%%%%%%%%%%%%%%%%%%%%%%%%%%%%%%%%%%%%%
%%%%%%%%%%%%%%%%%%%%%%%%%%%%%%%%%%%%%%%%%%%%%%%
%%%%%%%%%%%%%%%%%%%%%%%%%%%%%%%%%%%%%%%%%%%%%%%
%%%%%%%%%%%%%%%%%%%%%%%%%%%%%%%%%%%%%%%%%%%%%%%
%%%%%%%%%%%%%%%%%%%%%%%%%%%%%%%%%%%%%%%%%%%%%%%
%%%%%%%%%%%%%%%%%%%%%%%%%%%%%%%%%%%%%%%%%%%%%%%
%%%%%%%%%%%%%%%%%%%%%%%%%%%%%%%%%%%%%%%%%%%%%%%
%%%%%%%%%%%%%%%%%%%%%%%%%%%%%%%%%%%%%%%%%%%%%%%
%%%%%%%%%%%%%%%%%%%%%%%%%%%%%%%%%%%%%%%%%%%%%%%
%%%%%%%%%%%%%%%%%%%%%%%%%%%%%%%%%%%%%%%%%%%%%%%
%%%%%%%%%%%%%%%%%%%%%%%%%%%%%%%%%%%%%%%%%%%%%%%
%%%%%%%%%%%%%%%%%%%%%%%%%%%%%%%%%%%%%%%%%%%%%%%
%%%%%%%%%%%%%%%%%%%%%%%%%%%%%%%%%%%%%%%%%%%%%%%
%%%%%%%%%%%%%%%%%%%%%%%%%%%%%%%%%%%%%%%%%%%%%%%
%%%%%%%%%%%%%%%%%%%%%%%%%%%%%%%%%%%%%%%%%%%%%%%
%%%%%%%%%%%%%%%%%%%%%%%%%%%%%%%%%%%%%%%%%%%%%%%
%%%%%%%%%%%%%%%%%%%%%%%%%%%%%%%%%%%%%%%%%%%%%%%
%%%%%%%%%%%%%%%%%%%%%%%%%%%%%%%%%%%%%%%%%%%%%%%
%%%%%%%%%%%%%%%%%%%%%%%%%%%%%%%%%%%%%%%%%%%%%%%
%%%%%%%%%%%%%%%%%%%%%%%%%%%%%%%%%%%%%%%%%%%%%%%
%%%%%%%%%%%%%%%%%%%%%%%%%%%%%%%%%%%%%%%%%%%%%%%
%%%%%%%%%%%%%%%%%%%%%%%%%%%%%%%%%%%%%%%%%%%%%%%
%%%%%%%%%%%%%%%%%%%%%%%%%%%%%%%%%%%%%%%%%%%%%%%

\section{Introduction}

This review is concerned with  non-geometric backgrounds in string theory. 
Such spaces  cannot be described in terms of Riemannian geometry and
point-particles cannot be placed into them. 
String theory on the other hand is  a theory of
strings --- one-dimensionally extended objects ---
and can be well-defined on more general configurations, including
non-geometric backgrounds. 
In this introductory section we briefly review some basic aspects of string theory in view of their
application to non-geometric backgrounds. We give a heuristic description of 
the latter, and we summarise the topics discussed in this work.

%%%%%%%%%%%%%%%%%%%%%%%%%%%%%%%%%%%%%%%%%%%%%%%
%%%%%%%%%%%%%%%%%%%%%%%%%%%%%%%%%%%%%%%%%%%%%%%
%%%%%%%%%%%%%%%%%%%%%%%%%%%%%%%%%%%%%%%%%%%%%%%
%%%%%%%%%%%%%%%%%%%%%%%%%%%%%%%%%%%%%%%%%%%%%%%
%%%%%%%%%%%%%%%%%%%%%%%%%%%%%%%%%%%%%%%%%%%%%%%
%%%%%%%%%%%%%%%%%%%%%%%%%%%%%%%%%%%%%%%%%%%%%%%
%%%%%%%%%%%%%%%%%%%%%%%%%%%%%%%%%%%%%%%%%%%%%%%
%%%%%%%%%%%%%%%%%%%%%%%%%%%%%%%%%%%%%%%%%%%%%%%

\subsection{String theory}

String theory is in some way the most simple generalisation of a point-particle theory:
the one-dimensional world-line is replaced by 
a two-dimensional world-sheet. 
At the level of the action this means that the length of the world-line is replaced by
the area of the world-sheet
\eq{
  \label{actions_001}
  \mathcal S= -m \int_{\Gamma} ds
  \hspace{40pt}\longrightarrow
  \hspace{40pt}
  \mathcal S= -T \int_{\Sigma} dA \,,
}
where $ds$ denotes the line-element on the world-line $\Gamma$ and $dA$ denotes the area element
for the world-sheet $\Sigma$. The latter can be an infinite strip corresponding to an open string, or an infinite cylinder corresponding to a closed string.  Furthermore, in \eqref{actions_001} $m$ denotes the mass of the point-particle and 
correspondingly $T$ is the tension of the string. 
The action for the string shown in \eqref{actions_001} is called the Nambu-Goto action. 
However, for quantising string theory one uses the Polyakov action. The Polyakov action is classically equivalent
to the Nambu-Goto action and  will be introduced in equation \eqref{action_01} below.

%%%%%%%%%%%%%%%%%%%%%%%%%%%%%%%%%%%%%%%%%%%%%%%
%%%%%%%%%%%%%%%%%%%%%%%%%%%%%%%%%%%%%%%%%%%%%%%

\subsubsection*{Conformal field theory}

Even though the replacement \eqref{actions_001} appears rather simple, it has far-reaching 
consequences. The  world-sheet theory of the string
has a conformal symmetry and is therefore 
a  conformal field theory (CFT).
Moreover, this CFT is  two-dimensional for which the corresponding symmetry algebra is infinite dimensional. 
(For an introduction to conformal field theory in view of string theory see for instance \cite{Blumenhagen:2009zz}.)  
Such a large symmetry algebra is an important property of  string theory, which is absent for point particles.

It turns out that the  conformal symmetry of the two-dimensional theory 
has an anomaly. 
More concretely, classically the trace of the  energy-momentum tensor $T_{\alpha\beta}$ with $\alpha,\beta=1,2$ vanishes, but in 
the quantised theory its vacuum expectation value is proportional to the central charge $c$ of the CFT
\eq{
  \label{weyl_anom}
  \langle T^{\alpha}{}_{\alpha}\rangle  =  -\frac{c}{12} \, \mathsf R \,.
}  
Here, $\mathsf R$ denotes the Ricci scalar on the world-sheet $\Sigma$ of the string.
There are many configurations which lead to a vanishing total central charge and therefore 
to an anomaly-free theory. The most common ones are the bosonic string 
in $26$-dimensional Minkowski space, the type I and type II superstring theories in ten-dimensional  Minkowski 
space, and two heterotic string theories which are combinations of the bosonic and type II theories.

But  also more involved settings are possible. For instance, 
take the type II superstring in four-dimensional Minkowski space times an abstract CFT with 
central charge $c=9$. The latter CFT does not need to have an interpretation 
in terms of ordinary geometry, in fact, it does not even need to have the notion of 
a dimension. Nevertheless, string theory is well-defined on such spaces.
In a very broad sense, these configurations are 
non-geometric backgrounds.

%%%%%%%%%%%%%%%%%%%%%%%%%%%%%%%%%%%%%%%%%%%%%%%
%%%%%%%%%%%%%%%%%%%%%%%%%%%%%%%%%%%%%%%%%%%%%%%

\subsubsection*{Quantum gravity}

When quantising the two-dimensional world-sheet theory, one finds that the spectrum of the closed string contains 
a massless mode corresponding to a symmetric traceless two-tensor. 
This tensor is subject to equations corresponding to a variant of \eqref{weyl_anom},
and which are displayed in equation \eqref{eom_beta} below. In particular, this 
symmetric traceless tensor has to satisfy Einstein's equation -- and should therefore
be identified with the graviton in a theory of quantum gravity.  
This observation has been corroborated through
computations of scattering amplitudes, which verify that this mode has
the couplings expected from a graviton. 
String theory therefore is a theory of quantum gravity. 
Furthermore, in string theory -- opposed to naive 
quantum-field theories of point particles -- for scattering amplitudes  certain 
divergencies are absent and the theory is expected to be finite.

String theory is not only expected to be a theory of quantum gravity, but it also contains
gauge degrees of freedom. In type I theories these can 
be realised for instance by open strings ending on  D-branes. The latter are  hyper-surfaces 
on which open strings can end, and their world-volume supports a gauge theory typically with 
a $U(N)$ gauge group. 
String theory therefore is a quantum theory in which gauge (open string) and gravitational (closed string) interactions are unified. 
As such, it provides a valuable framework for studying the nature of our universe.

%%%%%%%%%%%%%%%%%%%%%%%%%%%%%%%%%%%%%%%%%%%%%%%
%%%%%%%%%%%%%%%%%%%%%%%%%%%%%%%%%%%%%%%%%%%%%%%

\subsubsection*{Dualities}

Let us return to superstring theory in a flat Minkowski background. As  mentioned, 
this configuration is consistent only in ten space-time dimensions
and five different (supersymmetric) formulations are known: 
these are the type I superstring, the type IIA and type IIB superstring theories, 
and the heterotic string with gauge groups $SO(32)$ and $E_8\times E_8$. 
However, it turns out that these five superstring theories are related to each other through a web of dualities. 
We will give a more precise definition of a 
duality below, but we consider two different theories to be dual to each other if they 
``describe the same physics''.

\begin{figure}[t]
\centering
\fbox{
\begin{picture}(403,160)(0,10)
\put(50,82){\framebox(30,18)[c]{\footnotesize IIA}}
\put(115,82){\framebox(30,18)[c]{\footnotesize IIB}}
\put(50,22){\framebox(30,18)[c]{\footnotesize IIA}}
\put(115,22){\framebox(30,18)[c]{\footnotesize IIB}}
\put(50,82){\framebox(30,18)[c]{\footnotesize IIA}}
\put(184,82){\framebox(30,18)[c]{\footnotesize I}}
\put(247,82){\framebox(60,18)[c]{\footnotesize het $SO(32)$}}
\put(341,82){\framebox(60,18)[c]{\footnotesize het $E_8\times E_8$}}
\put(247,22){\framebox(60,18)[c]{\footnotesize het $SO(32)$}}
\put(341,22){\framebox(60,18)[c]{\footnotesize het $E_8\times E_8$}}
\put(0,148){\scriptsize $D=11$}
\put(0,88){\scriptsize $D=10$}
\put(0,28){\scriptsize $D=9$}
\put(87,30){\vector(1,0){22}}
\put(109,30){\vector(-1,0){22}}
\put(95,34){\scriptsize \bf T}
\put(313,30){\vector(1,0){22}}
\put(335,30){\vector(-1,0){22}}
\put(321,34){\scriptsize \bf T}
\put(220,90){\vector(1,0){22}}
\put(242,90){\vector(-1,0){22}}
\put(228,94){\scriptsize \bf S}
\put(65,74){\vector(0,-1){27}}
\put(53,57){\scriptsize $S^1$}
\put(277,74){\vector(0,-1){27}}
\put(264,57){\scriptsize $S^1$}
\put(130,74){\vector(0,-1){27}}
\put(134,57){\scriptsize $S^1$}
\put(372,74){\vector(0,-1){27}}
\put(376,57){\scriptsize $S^1$}
\put(151,84){\includegraphics[width=15pt,angle=-90,origin=c]{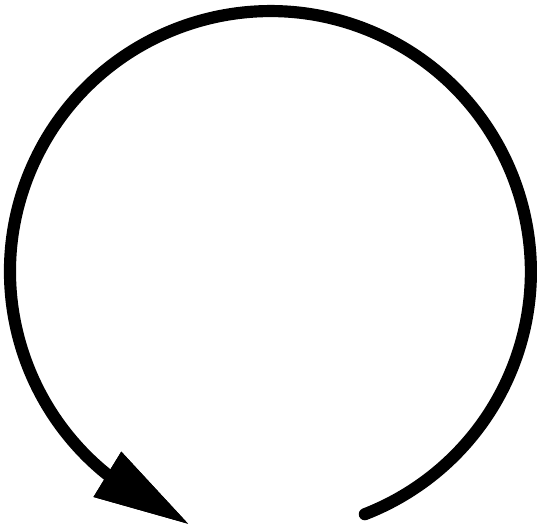}}
\put(155.5,88.5){\scriptsize \bf S}
\put(152.5,95.7){\vector(-2,-3){2}}
\put(190,142){\framebox(50,18)[c]{\footnotesize M-theory}}
\put(180,137){\vector(-4,-1){100}}
\put(250,137){\vector(4,-1){100}}
\put(125,129){\scriptsize $S^1$}
\put(300,129){\scriptsize $S^1/\mathbb Z_2$}
\end{picture}
}
\caption{T-duality and S-duality transformations relating the five superstring theories,  and their relation to M-theory.\label{fig_dual}}
\end{figure}
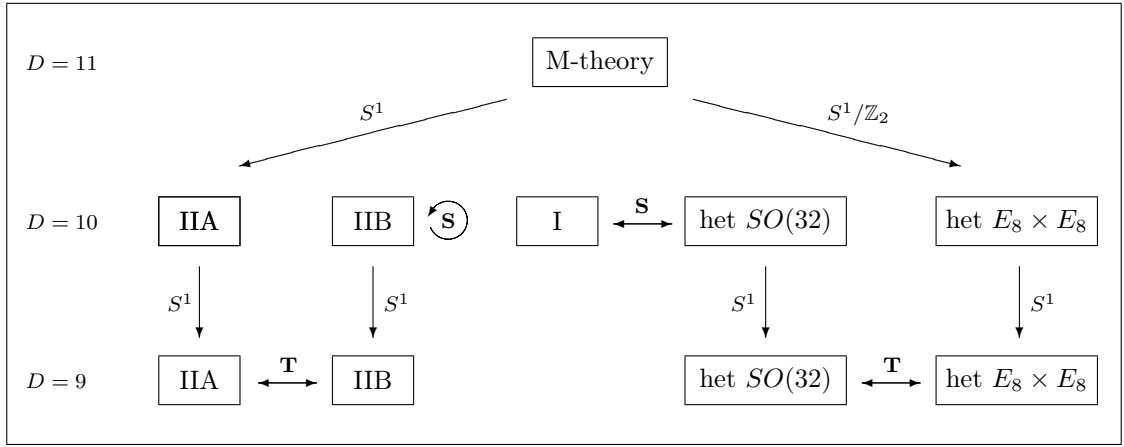

Well-known dualities in string theory
are so-called T-duality and S-duality, and the former plays an important  role in this work. 
\begin{itemize}

\item T-duality is a phenomenon already present for the bosonic string. Namely, 
bosonic string theory compactified on a circle of radius $R$ is 
completely equivalent to a compactification on a circle of radius $1/R$ (in appropriate units).
These two compactifications are two distinct configurations, which however describe the same
physics. 
In the case of the superstring, T-duality interchanges type IIA and type IIB string theory compactified 
on a circle, and the heterotic $SO(32)$ and $E_8\times E_8$ theories.

\item S-duality is a so-called strong-coupling--weak-coupling duality. Examples for S-duality
are type IIB string theory in ten dimensions which is dual to itself, 
and S-duality between the type I superstring and the heterotic $SO(32)$ string theory.

\end{itemize}
Furthermore, type IIA string theory has as a strong-coupling limit an
 eleven-dimensional theory called M-theory. In particular, M-theory compactified on a 
circle has as a low-energy limit the type IIA superstring. 
M-theory is related also to the heterotic $E_8\times E_8$ theory via a compactification  
 on $S^1/\mathbb Z_2$, where the $\mathbb Z_2$ acts on the circle coordinate
$\phi$ as $\mathbb Z_2: \phi \to - \phi$.
In this way all of the five known superstring theories are related to each other \cite{Witten:1995ex}. 
This
web of dualities is summarised in figure~\ref{fig_dual}.
But, many more duality relations can be found:
for instance, mirror symmetry \cite{Greene:1990ud}
relates type IIA and type IIB theories compactified on Calabi-Yau manifolds
to each other, and the AdS/CFT correspondence  \cite{Maldacena:1997re} relates string theory on an AdS space to a conformal field theory 
without gravity on its boundary.

%%%%%%%%%%%%%%%%
%%%%%%%%%%%%%%%%
%%%%%%%%%%%%%%%%
%%%%%%%%%%%%%%%%

The rich structure of dualities is one of the outstanding properties of string theory. 
They make it  possible for instance to relate a difficult-to-solve problem to 
a much easier setting where a solution can be found. 
Also, when applying duality transformations to known configurations 
new ones may be discovered. In fact, this is how the topic of this review has been developed: 
when applying T-duality transformations
to known geometric settings, one can obtain non-geometric backgrounds.

%%%%%%%%%%%%%%%%%%%%%%%%%%%%%%%%%%%%%%%%%%%%%%%
%%%%%%%%%%%%%%%%%%%%%%%%%%%%%%%%%%%%%%%%%%%%%%%

\subsubsection*{Compactifications}

Since string theory is expected to provide a consistent quantum-theory of 
gravity including gauge interactions, it is natural to try to employ it for the description of 
our universe. 
To be able to include ordinary matter we require a supersymmetric world-sheet theory,
and for computational convenience such as stability we require the space-time 
theory to be supersymmetric as well. This leaves us with the five superstring theories
mentioned above, realised in ten dimensions.

However, our universe is four-dimensional. To obtain an effectively
four-di\-men\-sion\-al theory from string theory, one compactifies the latter
on a compact six-dimensional space. The choice of this compactification
space is restricted by various consistency conditions, but a plethora of discrete 
and -- due to dualities -- possibly finite number of choices remains \cite{Vafa:2005ui}. 
This abundance of string-theory solutions is called the string-theory landscape \cite{Susskind:2003kw}.
String-theory compactifications can be characterised in different ways, and here we want to briefly 
mention two approaches:
\begin{itemize}

\item Since string theory is a conformal field theory, a compactification can be  
specified by choosing  a particular CFT. We can for instance split the CFT describing string theory 
into a four-dimensional  Minkowski part and a compact part,
subject to consistency conditions such as \eqref{weyl_anom}. 
Having a  CFT description available allows to determine the spectrum to all orders in string-length 
perturbation theory and to compute for instance scattering amplitudes. 
However, for such compactifications it is difficult to study perturbations of the background
and it is not always possible to obtain a corresponding geometric interpretation. In a broad sense, 
the latter cases can therefore be considered to be non-geometric.

\item In ten dimensions and at lowest order in string-length perturbation theory, 
string theory is described by supergravity.
Compactifications of string theory in this 
effective-field-theory description can be characterised by 
splitting the ten-dimensional Minkowski space into say a four-dimensional part and 
a compact six-dimensional manifold $\mathcal M$ as
\eq{
\mathbb R^{1,9} \;\longrightarrow\; \mathbb R^{1,3} \times \mathcal M \,.  
}
In order to solve the string-theory equations of motion and preserve supersymmetry in four dimensions,
$\mathcal M$ is usually required to be a Calabi-Yau three-fold (with all other background fields trivial). 
However, we can perturb this background by considering non-vanishing vacuum expectation values
for instance for $p$-form field strengths. The latter can be geometric 
as well as non-geometric fluxes, and  we discuss them in detail in this work.

\end{itemize}

%%%%%%%%%%%%%%%%%%%%%%%%%%%%%%%%%%%%%%%%%%%%%%%
%%%%%%%%%%%%%%%%%%%%%%%%%%%%%%%%%%%%%%%%%%%%%%%

\subsubsection*{Applications of non-geometric backgrounds}

Non-geometric backgrounds are the central theme of this review. 
We give a more detailed introduction to this topic  in section~\ref{sec_intro_2}, 
but we want to mention already here some of their applications.
\begin{itemize}

\item Non-geometric fluxes play a role for string-phenomenology,
where they can be used to stabilise moduli. The latter are massless scalar particles
usually arising when compactifying a theory, and these particles are incompatible 
with experimental observations. Non-geometric fluxes can be used to generate a
potential for these fields such that they receive a mass and can be integrated out
from the low-energy theory.

\item In string-cosmology non-geometric fluxes have been used to 
construct potentials for inflation, with the latter being  a period of rapid expansion 
during the early universe.
It has also been argued that non-geometric fluxes can lead to de Sitter vacua with a 
positive cosmological constant.

\item Non-geometric fluxes can furthermore be used to construct 
non-commutative and  non-associative theories of gravity, with the aim to describe
the very early universe shortly after the big bang. 

\item At a more formal level, non-geometric backgrounds can be related to 
gauged supergravity theories. The latter provide effective four- or higher-di\-men\-sio\-nal descriptions
of string-compactifications with fluxes which preserve some supersymmetry.

\end{itemize}

%%%%%%%%%%%%%%%%%%%%%%%%%%%%%%%%%%%%%%%%%%%%%%%
%%%%%%%%%%%%%%%%%%%%%%%%%%%%%%%%%%%%%%%%%%%%%%%
%%%%%%%%%%%%%%%%%%%%%%%%%%%%%%%%%%%%%%%%%%%%%%%
%%%%%%%%%%%%%%%%%%%%%%%%%%%%%%%%%%%%%%%%%%%%%%%
%%%%%%%%%%%%%%%%%%%%%%%%%%%%%%%%%%%%%%%%%%%%%%%
%%%%%%%%%%%%%%%%%%%%%%%%%%%%%%%%%%%%%%%%%%%%%%%
%%%%%%%%%%%%%%%%%%%%%%%%%%%%%%%%%%%%%%%%%%%%%%%
%%%%%%%%%%%%%%%%%%%%%%%%%%%%%%%%%%%%%%%%%%%%%%%

\subsection{Non-geometric backgrounds}
\label{sec_intro_2}

To establish our conventions, let us first note 
that  a transformation which 
leaves an action functional $\mathcal S[\phi]$ invariant (up to boundary terms) will be called a {\em symmetry}. 
On the other hand,
\begin{center}
\vspace*{1mm}
\setlength{\fboxsep}{10pt}
\fbox{
\begin{minipage}[t]{340pt}
The term {\em duality} will be used for  
transformations which ``leave the physics invariant'' -- such as a symmetry of the equations
of motion or a symmetry of the spectrum -- but which is not a symmetry of an action. 
\end{minipage}
}\vspace*{1mm}
\end{center}
We now want to give a more precise definition of non-geometric backgrounds in string 
theory. The term {\it non-geometry} is  used rather broadly in the literature and does
not have a unique meaning. However, let us give the following 
four  characterisations: \label{page_defs_nongeo}
\begin{center}
\vspace*{1mm}
\setlength{\fboxsep}{10pt}
\fbox{
\begin{minipage}[t]{340pt}
A non-geometric background is a string-theory configuration \ldots
\begin{enumerate}[leftmargin=10mm,labelsep=1.75mm,itemindent=0mm,topsep=2.5mm,parsep=1.5mm,itemsep=1mm]

\item which cannot be  described in terms of Riemannian geometry.

\item in which the left- and right-moving sector of a closed string behave differently.

\item in which T-duality transformations are needed to make the background well-defined.

\item in which T-duality transformations are needed to make the background well-defined, but which is {\em not} T-dual to a background described in terms of Riemannian geometry.

\end{enumerate}
\end{minipage}
}
\end{center}
Note that characterisation four is a special case of characterisation three, three is a special case of two, and characterisation two is
a special case of characterisation one. 
Furthermore, 
the term ``string-theory configuration'' has been chosen with some care,  since
not all backgrounds to be considered below are solutions to the equations-of-motion 
of string theory.
In the following we discuss these four characterisations in some more detail.

%%%%%%%%%%%%%%%%%%%%%%%%%%%%%%%%%%%%%%%%%%%%%%%
%%%%%%%%%%%%%%%%%%%%%%%%%%%%%%%%%%%%%%%%%%%%%%%

\subsubsection*{Characterisation 1}

As mentioned  before, string theory is a two-dimensional conformal field theory.
For a consistent quantum theory the Weyl anomaly \eqref{weyl_anom} has to vanish, and one of the 
simplest examples satisfying this
condition is the $26$-fold copy of the free boson CFT
describing the bosonic string moving in $26$-dimensional Minkowski space. 
However,  more complicated CFTs can be used as well 
to obtain a vanishing Weyl anomaly. Examples are  Wess-Zumino-(Novikov-)Witten
models (WZW models) \cite{Wess:1971yu,Novikov:1981,Witten:1983ar} and Gepner models 
\cite{Gepner:1987qi,Gepner:1987vz}.

For the latter two there are regimes in their parameter space where a geometric description  
in terms of a metric and other background fields is not possible, even though these backgrounds
are well-defined in string-theory.
For the $SU(2)$ WZW model at level $k$  this happens for instance at small values of $k$,
and for Gepner models for instance at the Gepner point of the quintic. 
Hence, according to our characterisation 1, these configurations are non-geometric.

Note furthermore that spaces with singularities, such as 
orbifolds with fixed points, are not Riemannian either. String theory is well-defined on such 
backgrounds \cite{Dixon:1985jw,Dixon:1986jc}, however, usually these are not considered to be non-geometric.

%%%%%%%%%%%%%%%%%%%%%%%%%%%%%%%%%%%%%%%%%%%%%%%
%%%%%%%%%%%%%%%%%%%%%%%%%%%%%%%%%%%%%%%%%%%%%%%

\subsubsection*{Characterisation 2}

\begin{figure}[t]
\centering
\vspace*{10pt}
\includegraphics[width=300pt]{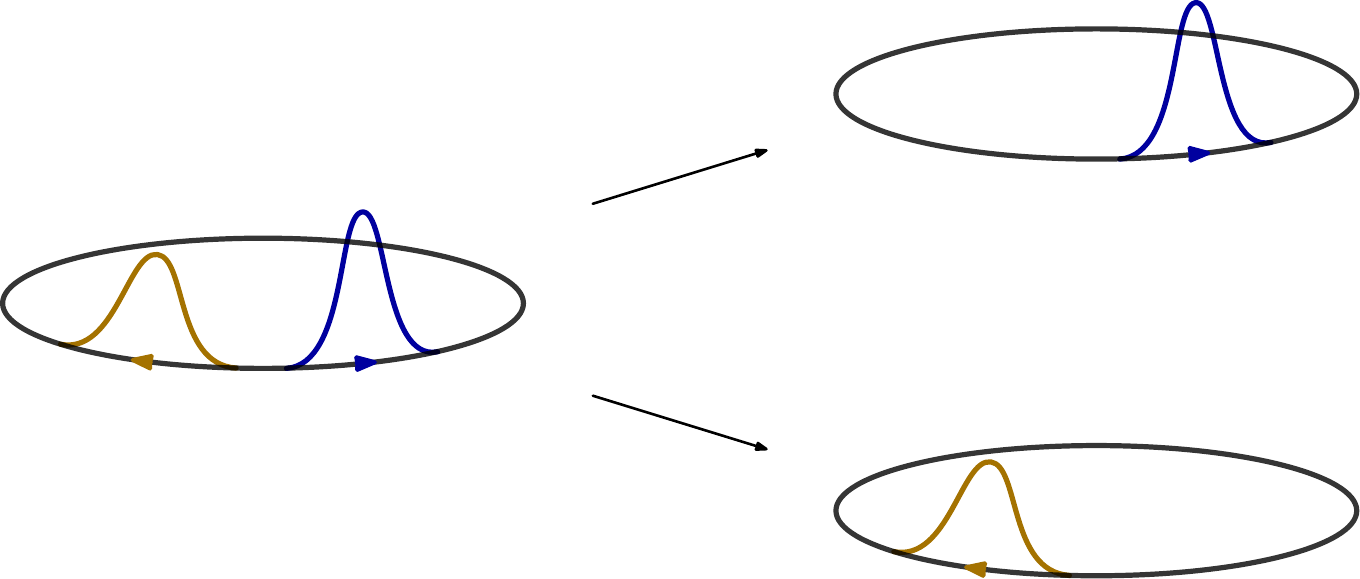}\hspace*{75pt}
\begin{picture}(0,0)
\put(-344,30){\scriptsize Closed string}
\put(-65,114){\scriptsize Left-moving sector}
\put(-52,103){\scriptsize Geometry A}
\put(-68,19){\scriptsize Right-moving sector}
\put(-52,8){\scriptsize Geometry B}
\end{picture}
\vspace*{10pt}
\caption{Illustration of left- and right-moving excitations of the closed string. Each sector 
``sees'' its own geometry A and B, respectively, which do not need to agree. In this case a point-particle 
interpretation of the background is not possible, and the space is called non-geometric according 
to characterisation 2.\label{fig_charac2}} 
\end{figure}
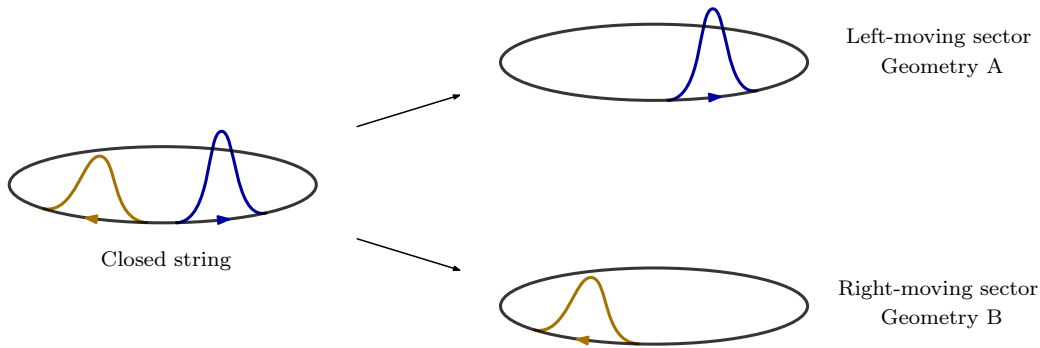

In string theory, the left- and right-moving sector of a closed string  are decoupled at tree-level and
can be treated independently. At one-loop, modular invariance
of the partition function imposes constraints on the coupling between the two sectors,
but this still allows for  non-trivial solutions.  
If the left- and right-moving sector are different from each other, it is in general
not possible to give a geometric interpretation of the background. The
space is therefore called non-geometric. 
Loosely speaking, the left- and right-moving sectors {\it see two different 
geometries} which from a point-particle's point of view 
cannot be combined into a consistent picture (see figure~\ref{fig_charac2}). 
However, for a string such backgrounds are well-defined. 
Examples for such constructions are asymmetric orbifolds \cite{Narain:1986qm,Narain:1990mw},
which we discuss below. 
%%%%%%%%%%%%%%%%
%%%%%%%%%%%%%%%%
%%%%%%%%%%%%%%%%
%%%%%%%%%%%%%%%%

We restrict the difference between the left- and right-moving part of the closed string 
to the sector  describing the metric and Kalb-Ramond $B$-field. 
In particular, the heterotic string (for which one sector is the bosonic string and the other is the superstring)
has in general a geometric interpretation of the target space and 
hence is not considered to be a non-geometric configuration.

%%%%%%%%%%%%%%%%%%%%%%%%%%%%%%%%%%%%%%%%%%%%%%%
%%%%%%%%%%%%%%%%%%%%%%%%%%%%%%%%%%%%%%%%%%%%%%%

\subsubsection*{Characterisation 3}

The third characterisation of non-geometric backgrounds  employed in the literature 
is using T-duality:
non-geometric spaces are string-theory configurations which can be made globally well-defined using T-duality
transformations as transition maps between local charts \cite{Hull:2004in}.
Similarly,  monodromies around defects may contain T-duality transformations, leading to a non-geometric 
background.  

%%%%%%%%%%%%%%%%
%%%%%%%%%%%%%%%%
\begin{figure}[t]
\centering
\begin{subfigure}{0.487\textwidth}
\centering
\vspace*{20pt}
\includegraphics[width=200pt]{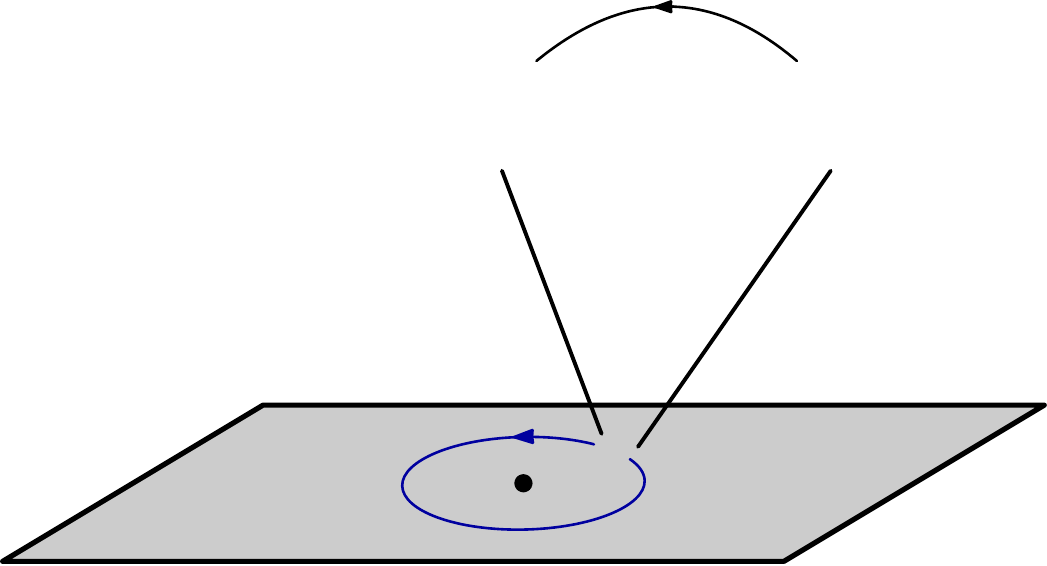}
\begin{picture}(0,0)
\put(-118,83){\scriptsize $\tau(x)$}
\put(-60,83){\scriptsize $\tau(x+1)$}
\put(-92,114){\scriptsize S-duality}
\end{picture}\vspace*{-5pt}
\caption{S-duality\label{fig_dual_a}}
\end{subfigure}
\begin{subfigure}{0.487\textwidth}
\centering
\vspace*{20pt}
\includegraphics[width=200pt]{fig_04a}
\begin{picture}(0,0)
\put(-128,83){\scriptsize $(G,B)(x)$}
\put(-70,83){\scriptsize $(G,B)(x+1)$}
\put(-92,114){\scriptsize T-duality}
\end{picture}\vspace*{-5pt}
\caption{T-duality\label{fig_dual_b}}
\end{subfigure}
\caption{Illustration of how a field configuration can be made well-defined using a duality transformation.
The angular coordinate is denoted by $x$ with identification $x\sim x+1$.
For type IIB string theory shown in figure~\ref{fig_dual_a} the singularity is a $(p,q)$ seven-brane, the field $\tau$ is the axio-dilaton, and the duality transformation is S-duality.
For non-geometric backgrounds illustrated in figure~\ref{fig_dual_b} 
the fields are the metric $G$ and Kalb-Ramond $B$-field and the duality
transformation is T-duality.\label{fig_dual_c}} 
\end{figure}
%%%%%%%%%%%%%%%%
%%%%%%%%%%%%%%%%

Using duality transformations to obtain a global description of a background
is not unique to non-geometric backgrounds. Let us discuss this point in some more detail
using the illustrations in figure~\ref{fig_dual_c}.
\begin{itemize}

\item In type IIB string theory $(p,q)$ seven-branes are defects which have a two-dimensional transversal space
as illustrated in figure~\ref{fig_dual_a}. When encircling this defect with the axio-dilaton field $\tau$, 
the latter undergoes a monodromy transformation which is contained in the 
S-duality group $SL(2,\mathbb Z)$. 
This means that this string-theory configuration is made globally well-defined using a duality 
transformation. 

\item A similar mechanism is at work for non-geometric backgrounds (according to the present
characterisation). When encircling a non-geometric defect with
the metric $G$ and Kalb-Ramond $B$-field, a T-duality transformation is needed in order 
to make the background globally well-defined (see figure~\ref{fig_dual_b}).

\end{itemize}

%%%%%%%%%%%%%%%%%%%%%%%%%%%%%%%%%%%%%%%%%%%%%%%
%%%%%%%%%%%%%%%%%%%%%%%%%%%%%%%%%%%%%%%%%%%%%%%

\subsubsection*{Characterisation 4}

Another definition of non-geometric backgrounds which can be found in the literature 
is as in characterisation 3 above, 
but which furthermore satisfies that it cannot be T-dualised to a 
geometric background. 
More concretely, a subset of T-duality transformation is used to make the 
background well-defined, and there does not exist another T-duality 
transformation which maps the configuration to a geometric setting. 
According to this characterisation, the family of backgrounds arising from a three-dimensional toroidal compactification 
with $H$-flux (discussed in section~\ref{sec_first_steps}) does not contain
a non-geometric background.

%%%%%%%%%%%%%%%%%%%%%%%%%%%%%%%%%%%%%%%%%%%%%%%
%%%%%%%%%%%%%%%%%%%%%%%%%%%%%%%%%%%%%%%%%%%%%%%

\subsubsection*{Remark}

In this work we are not restricting ourselves to one particular definition of non-geometric 
backgrounds in string theory, but discuss examples which fit into different characterisations. 
Nevertheless, the central theme is that of T-duality and therefore our discussion 
is closely related to T-duality in string theory.

%%%%%%%%%%%%%%%%%%%%%%%%%%%%%%%%%%%%%%%%%%%%%%%
%%%%%%%%%%%%%%%%%%%%%%%%%%%%%%%%%%%%%%%%%%%%%%%
%%%%%%%%%%%%%%%%%%%%%%%%%%%%%%%%%%%%%%%%%%%%%%%
%%%%%%%%%%%%%%%%%%%%%%%%%%%%%%%%%%%%%%%%%%%%%%%
%%%%%%%%%%%%%%%%%%%%%%%%%%%%%%%%%%%%%%%%%%%%%%%
%%%%%%%%%%%%%%%%%%%%%%%%%%%%%%%%%%%%%%%%%%%%%%%
%%%%%%%%%%%%%%%%%%%%%%%%%%%%%%%%%%%%%%%%%%%%%%%
%%%%%%%%%%%%%%%%%%%%%%%%%%%%%%%%%%%%%%%%%%%%%%%

\subsection{Structure of this review}

In this review we discuss non-geometric backgrounds  from various points of view.
Since some  of these backgrounds   can be obtained from T-duality transformations
of ordinary geometric configurations,
we begin with a brief review of T-duality:
\begin{itemize}

\item In section~\ref{cha_t_duality} we review T-duality transformations for toroidal  compactifications.
For these spaces a CFT description is available, and hence the dual backgrounds can be obtained to all orders
in string-length perturbation theory. In section~\ref{sec_cft_circ} we first discuss T-duality for the circle, and 
in section~\ref{sec_cft_torus} we study the generalisation to $D$-dimensional toroidal backgrounds. The 
main result is that such T-duality transformations are realised as
$O(D,D,\mathbb Z)$ transformations.

\item In section~\ref{sec_buscher} we turn to T-duality transformations for curved backgrounds.
Here a  CFT description is usually not available, but the dual configurations can be
obtained via the  Buscher rules. The latter are  valid only at leading order in the string length. 
We discuss the Buscher rules from different perspectives, using the ordinary
sigma-model description as well as a description in terms of WZW-models.
We furthermore discuss the equivalence of the original and T-dual backgrounds.

\item In section~\ref{cha_poisson_lie} we give a brief introduction to Poisson-Lie duality, which 
is a framework to study T-duality for backgrounds with non-abelian isometry groups. 
This type of duality will not play a bigger role in this work, however, we include this topic 
for completeness.

\end{itemize}
After having discussed T-duality in string theory, we then turn to non-geometric backgrounds
and mostly follow our characterisation three from section~\ref{sec_intro_2}.
\begin{itemize}

\item In section~\ref{sec_first_steps} we consider the prime example for a non-geometric background
and start from a three-torus with $H$-flux. We show
how applying  successive T-duality transformations first leads to a twisted torus, and 
how a second T-duality leads to a non-geometric T-fold. We also associate corresponding 
geometric and non-geometric fluxes to these
backgrounds. 

\item In section~\ref{cha_torus_fib} we formalise these findings and consider 
torus fibrations. For a $D$-dimensional toroidal fibre the duality group is 
given by $O(D,D,\mathbb Z)$ transformations, which 
can be used as transition functions between local patches. 
In section~\ref{sec_t2_fibr_example} we revisit the three-torus example and 
rephrase it using the language of torus fibrations, and in section~\ref{sec_t2_fibr_general}
we discuss generalisations thereof. We also construct new examples 
of non-geometric torus fibrations. 
In section~\ref{sec_t2_p1} we consider $\mathbb T^2$-fibrations over a two-sphere,
and in section~\ref{sec_tw_r2} we consider
the punctured plane as a base-manifold. The latter setting includes the well-known examples of the
NS5-brane, Kaluza-Klein monopole and non-geometric $5^2_2$-brane.

\end{itemize}
Torus fibrations with T-duality transformations as transition functions
provide explicit examples for non-geometric backgrounds. Using this knowledge, 
we then describe such spaces from a more general point of view.
\begin{itemize}

\item In section~\ref{sec_gen_geo} we review the framework of 
generalised geometry. In this approach one  enlarges the tangent-space of a manifold 
 to a generalised tangent-space. This allows for a
natural action of the group $O(D,D)$ on the geometry, which is related to T-duality transformations. 
After introducing the basic concepts in section~\ref{sec_gg_basics} and 
giving a more mathematical description in section~\ref{sec_lie_courant},
in section~\ref{sec_gg_buscher} we discuss the Buscher rules in this framework.
In section~\ref{sec_gg_fluxes} we give a more precise definition of (non-)geometric 
fluxes using the Courant bracket, and section~\ref{sec_gg_tdual} contains a
treatment of T-duality transformations in generalised geometry. 
Finally, in sections~\ref{sec_gg_frame} and \ref{sec_gg_bianchi} we consider
frame transformations and Bianchi identities.

\item In section~\ref{cha_sugra} we discuss non-geometric backgrounds from an 
effective field-theory point of view. We compactify type II string theory from 
ten to four dimensions on manifolds with $SU(3)\times SU(3)$ structure, 
and include geometric as well as non-geometric fluxes. In section~\ref{sec_sugra_recap} 
we give a review 
of four-dimensional $\mathcal N=2$ and $\mathcal N=1$ supergravity theories, and  
in sections~\ref{sec_cy_flux_echt} and \ref{sec_cy_flux_echt_orient}
we show how fluxes modify the four-dimensional theory by introducing a gauging
of global symmetries. 
We  discuss generalised Scherk-Schwarz reductions in section~\ref{sec_schsch},
and we comment on the validity of 
non-geometric solutions and their applications
to string phenomenology in sections~\ref{sec_fluxes_val} and \ref{sec_fluxes_app}.

\end{itemize}
In the final parts of this review we explain how non-geometric backgrounds
can lead to non-commutative and non-associative structures.
\begin{itemize}

\item First, in section~\ref{cha_dg}, we  review doubled geometry which is a framework 
where not only 
the tangent-space of a manifold is doubled but also the space itself. This allows
for the construction of a world-sheet theory invariant under T-duality transformations
and for a geometric description of non-geometric backgrounds. 
We also briefly discuss double field theory.

\item In section~\ref{cha_nca} we then explain how non-commutative and
non-associative structures can appear in string theory. 
This includes the derivation of a non-associative tri-product via correlation functions 
in section~\ref{sec_nca_closed}, as well as the derivation of a non-associative 
phase-space algebra in sections~\ref{sec_nca_nc} and \ref{sec_nca_phase}.
We also show how the latter are related to asymmetric orbifolds. 
In section~\ref{sec_nca_top_t} we review topological T-duality
and how non-commutativity and non-associativity arises.

\end{itemize}
A summary of the topics discussed in this review as well as of those omitted 
can be found in section~\ref{cha_concl}.

\clearpage
\section{T-duality in conformal field theory}
\label{cha_t_duality}

Non-geometric backgrounds in string theory are closely related to T-duality transformations.
In order to prepare for our subsequent discussion,
in this section we give a brief introduction to T-duality for toroidal compactifications.
For such backgrounds there exists a conformal-field-theory description, which 
makes it possible to obtain results to all orders in string-length perturbation theory. 
The standard review on this topic can be found in \cite{Giveon:1994fu}, and we mention that our 
notation follows in parts \cite{Blumenhagen:2013fgp}.

%%%%%%%%%%%%%%%%%%%%%%%%%%%%%%%%%%%%%%%%%%%%%%%
%%%%%%%%%%%%%%%%%%%%%%%%%%%%%%%%%%%%%%%%%%%%%%%
%%%%%%%%%%%%%%%%%%%%%%%%%%%%%%%%%%%%%%%%%%%%%%%
%%%%%%%%%%%%%%%%%%%%%%%%%%%%%%%%%%%%%%%%%%%%%%%
%%%%%%%%%%%%%%%%%%%%%%%%%%%%%%%%%%%%%%%%%%%%%%%
%%%%%%%%%%%%%%%%%%%%%%%%%%%%%%%%%%%%%%%%%%%%%%%
%%%%%%%%%%%%%%%%%%%%%%%%%%%%%%%%%%%%%%%%%%%%%%%
%%%%%%%%%%%%%%%%%%%%%%%%%%%%%%%%%%%%%%%%%%%%%%%

\subsection{Prerequisites}

We start by fixing our conventions for the world-sheet action of the closed string
and by stating some results needed below. 

%%%%%%%%%%%%%%%%%%%%%%%%%%%%%%%%%%%%%%%%%%%%%%%
%%%%%%%%%%%%%%%%%%%%%%%%%%%%%%%%%%%%%%%%%%%%%%%

\subsubsection*{World-sheet action}

In the following we consider the closed bosonic string, but most of
the results carry over to the superstring. 
The Polyakov action  takes the following general form
\eq{
  \label{action_01}
  \mathcal S = -\frac{1}{4\pi \alpha'} \int_{\Sigma}
  \Bigl[ \,G_{\mu\nu} \, d X^{\mu}\wedge\star d X^{\nu} -  B_{\mu\nu} \,dX^{\mu}\wedge dX^{\nu}
  + \alpha' \op\mathsf R\, \phi \star 1\, \Bigr] 
  \,,
}
where $G_{\mu\nu}=\eta_{\mu\nu}$ with $\mu,\nu=0,\ldots, 25$ is the $26$-dimensional Minkowski space metric, 
$B_{\mu\nu}$ describes a constant $B$-field and $\phi$ denotes the dilaton.
The two-dimensional 
world-sheet (without boundary) is denoted by $\Sigma$, and 
the string length $\ell_{\rm s}$ is related to the dimension-full constant $\alpha'$ via
 $\ell_{\rm s} =2\pi\sqrt{\alpha'}$. 
For later convenience we employed a differential-form notation together with the
Hodge star-operator as follows
\eq{
  \label{ws_ac_conv}
  \arraycolsep1.5pt
  \begin{array}{lcrcrcc}
  \displaystyle d X^{\mu} &  \displaystyle \wedge&  \displaystyle \star d X^{\nu}  &=&   \displaystyle \sqrt{|h|}\: 
  d^2\sigma & h^{\alpha\beta} & \partial_{\alpha}X^{\mu} \,  \partial_{\beta} X^{\nu}  \,, 
  \\[8pt]
  \displaystyle d X^{\mu}&  \displaystyle \wedge &  \displaystyle d X^{\nu}  &=&   \displaystyle d^2\sigma 
  &\epsilon^{\alpha\beta} &\partial_{\alpha}X^{\mu}   \,\partial_{\beta} X^{\nu}  \,,
  \end{array}
  \hspace{40pt}
  \star 1 = \sqrt{|h|}\: d^2\sigma \,,
}
where $\{\sigma^{0},\sigma^1\}$ are the world-sheet time and space coordinates, $h_{\alpha\beta}$ is 
the world-sheet metric, $h$ denotes its determinant, 
and the epsilon-symbol takes values $\epsilon^{\alpha\beta}=\pm1$.
$\mathsf R$ denotes the Ricci scalar corresponding to the world-sheet metric $h_{\alpha\beta}$.

In sections~\ref{sec_cft_circ} and \ref{sec_cft_torus} we will  mostly be interested in cylindrical 
world-sheets $\Sigma$ of the form $\Sigma = \mathbb R \times S^1$, 
with the non-compact direction corresponding to the world-sheet time coordinate $\sigma^0\equiv\tau$
and the circle  corresponding to the world-sheet space coordinate $\sigma^1\equiv\sigma$
defined via the identifications
$\sigma\sim\sigma+\ell_{\rm s}$.
Accordingly, we impose periodicity conditions 
for the fields $X^{\mu}(\tau,\sigma)$ along the 
$\sigma^1$-direction
 as 
$X^{\mu}(\tau,\sigma+\ell_{\rm s}) = X^{\mu}(\tau,\sigma)$.
Using then the reparameterisation and Weyl symmetries of the world-sheet action, we can bring 
\eqref{action_01} 
into conformal gauge 
in which the world-sheet metric takes the form $h_{\alpha\beta} = \eta_{\alpha\beta}$. 
Introducing in addition light-cone coordinates $\sigma^{\pm} = \sigma^0\pm\sigma^1$, the above action 
can  be expressed as
\eq{
  \label{action_02a6}
  \mathcal S = -\frac{1}{2\pi\alpha'} \int_{\Sigma} d^2\sigma\, \partial_+X^{\mu}
  \bigl( G_{\mu\nu} - B_{\mu\nu} \bigr) \op \partial_- X^{\nu} \,,
}
and the equations of motion for $X^{\mu}$ are obtained by varying the action \eqref{action_02a6} with respect to $X^{\mu}$, leading to
\eq{
  \label{eom_closed_9494}
  0 = \partial_+\partial_- X^{\mu} \,.
}

%%%%%%%%%%%%%%%%%%%%%%%%%%%%%%%%%%%%%%%%%%%%%%%
%%%%%%%%%%%%%%%%%%%%%%%%%%%%%%%%%%%%%%%%%%%%%%%
%%%%%%%%%%%%%%%%%%%%%%%%%%%%%%%%%%%%%%%%%%%%%%%
%%%%%%%%%%%%%%%%%%%%%%%%%%%%%%%%%%%%%%%%%%%%%%%
%%%%%%%%%%%%%%%%%%%%%%%%%%%%%%%%%%%%%%%%%%%%%%%
%%%%%%%%%%%%%%%%%%%%%%%%%%%%%%%%%%%%%%%%%%%%%%%
%%%%%%%%%%%%%%%%%%%%%%%%%%%%%%%%%%%%%%%%%%%%%%%
%%%%%%%%%%%%%%%%%%%%%%%%%%%%%%%%%%%%%%%%%%%%%%%

\subsection{Conformal field theory for \texorpdfstring{$S^1$}{S1}}
\label{sec_cft_circ}

Let us now compactify  the closed bosonic string  on a circle $S^1$ 
and study how T-duality transformations act on this background. 
Recall also that we consider the world-sheet $\Sigma$
to be the infinite cylinder $\Sigma = \mathbb R\times S^1$.

%%%%%%%%%%%%%%%%%%%%%%%%%%%%%%%%%%%%%%%%%%%%%%%
%%%%%%%%%%%%%%%%%%%%%%%%%%%%%%%%%%%%%%%%%%%%%%%

\subsubsection*{Compactification}

Compactifying the bosonic string on a circle of radius $R$ means that we identify say
the 25th target-space coordinate as
$X^{25} \sim X^{25} + 2\op\pi R$. For simplicity we also  assume that $B_{\mu25}=0$, which means that the $B$-field has no leg along the circle direction. 
The mode expansion of  $X^{25}(\tau,\sigma)$, solving the equations of motion \eqref{eom_closed_9494} and respecting the periodic identification on the space-time circle,
then becomes
\eq{
  \label{cft_aa849}
  X^{25}(\tau,\sigma) = X^{25}_R(\tau-\sigma) + X^{25}_L(\tau+\sigma) \,,
}
with the right- and left-moving fields
\eq{
  \label{cft_002}
  X^{25}_R(\tau-\sigma) &=  x_R^{25} + \frac{2\pi \alpha'}{\ell_{\rm s}}\op p_R^{25} \op (\tau-\sigma) + 
  i\op \sqrt{\frac{\alpha'}{2}}\, \sum_{n\neq0} \frac{1}{n} \op\alpha^{25}_n \, e^{-\frac{2\pi i}{\ell_{\rm s}}\op n\op (\tau-\sigma)} \,,
  \\
  X^{25}_L(\tau+\sigma) &=  x_L^{25} + \frac{2\pi \alpha'}{\ell_{\rm s}}\op p_L^{25} \op (\tau+\sigma) + 
  i\op \sqrt{\frac{\alpha'}{2}}\, \sum_{n\neq0} \frac{1}{n} \op\ov\alpha^{25}_n \, e^{-\frac{2\pi i}{\ell_{\rm s}}\op n\op (\tau+\sigma)} \,.
}
Here we introduced the centre-of-mass coordinates
\eq{
  x^{25}_R= \frac{x_0^{25} - c}{2}\,,
  \hspace{70pt}
  x^{25}_L= \frac{x_0^{25} + c}{2} \,,
}
with $c$ an arbitrary constant, 
and we have defined the right- and left-moving momenta
\eq{
  \label{cft_001}
  p_R^{25} = \frac{1}{2} \left( \frac{m}{R} -\frac{n\op R}{\alpha'} \right),
  \hspace{70pt}
  p_L^{25} = \frac{1}{2} \left( \frac{m}{R} +\frac{n\op R}{\alpha'} \right),
}
where $n,m\in\mathbb Z$ are the momentum and winding numbers. 
Having a quantised momentum $m$ along a compact direction follows from requiring single-valuedness
of the wave function and is common also for point particles. Having a non-vanishing winding number $n$
is however special to strings. In particular, the closed string can wind $n$ times around the compact direction
which  is not possible for a point particle.
The $n$th winding sector is described by 
the relation $X^{25}(\tau,\sigma+\ell_{\rm s}) = X^{25}(\tau,\sigma) + 2\pi n R$.

Promoting the modes appearing in the  expansions \eqref{cft_002} to operators and replacing Poisson brackets 
by commutators, we obtain the following  non-vanishing commutation relations
\eq{
  \label{comrel}
  \arraycolsep2pt
  \begin{array}{lcl@{\hspace{50pt}}lcl}
  \displaystyle [\op x^{25}_L,p^{25}_L\op] &=& i \,, & 
  \displaystyle [\op\alpha^{25}_m, \alpha^{25}_n\op] &=& m\, \delta_{m+n} \,,
  \\[12pt]
  \displaystyle [\op x^{25}_R,p^{25}_R\op] &=& i \,, & 
  \displaystyle [\op\ov\alpha^{25}_m, \ov\alpha^{25}_n\op] &=& m\, \delta_{m+n} \,.
  \end{array}
}

%%%%%%%%%%%%%%%%%%%%%%%%%%%%%%%%%%%%%%%%%%%%%%%
%%%%%%%%%%%%%%%%%%%%%%%%%%%%%%%%%%%%%%%%%%%%%%%

\subsubsection*{Spectrum}

The spectrum of the closed bosonic string is determined by the mass formula 
together with the level-matching condition. 
These can be written using the following expressions for the left- and right-moving sector
\eq{
  \arraycolsep2pt
  \begin{array}{lclcl}
  \displaystyle \alpha' m_R^2 &=& \displaystyle 2\op\alpha' \bigl(p_R^{25}\bigr)^2 &+& 
    \displaystyle 2\op\bigl( N_R - 1\bigr) \,, \\[4pt]
  \displaystyle \alpha' m_L^2 &=& \displaystyle 2\op\alpha' \bigl(p_L^{25}\bigr)^2 &+& 
    \displaystyle 2\op\bigl( N_L - 1\bigr) \,,
  \end{array}
}
where $p^{25}_{R,L}$ have been defined in \eqref{cft_001} and 
$N_{R,L}=0,1,2,\ldots$ 
denote the number operators counting string excitations in the corresponding sector 
(in light-cone quantisation). They are expressed using the oscillators as 
\eq{
  \label{spec_03}
  N_R = \sum_{\mu=2}^{25}\sum_{n=1}^{\infty}  \op \alpha^{\mu}_{-n} \op\alpha^{\mu}_{+n} \,,
  \hspace{50pt}
  N_L = \sum_{\mu=2}^{25}\sum_{n=1}^{\infty}  \op \ov\alpha^{\mu}_{-n} \op\ov\alpha^{\mu}_{+n} \,.
}  
The combined spectrum is then described by the mass formula
\eq{
  \label{spec_01}
  \alpha' m^2 = \alpha' m_R^2  + \alpha' m_L^2 \,,
}
and is subject to the level-matching condition
\eq{
  \label{spec_02}
  \alpha' m_R^2  = \alpha' m_L^2 \,.
}

%%%%%%%%%%%%%%%%%%%%%%%%%%%%%%%%%%%%%%%%%%%%%%%
%%%%%%%%%%%%%%%%%%%%%%%%%%%%%%%%%%%%%%%%%%%%%%%

\subsubsection*{T-duality}

Next we note that $(p_R^{25})^2$ and $(p_L^{25})^2$ -- both appearing in the mass formula \eqref{spec_01} and in the level-matching condition 
\eqref{spec_02} -- are  invariant 
under the following $\mathbb Z_2$ action
\eq{
  \label{duality_001}
  R \to \frac{\alpha'}{R} \,,
  \hspace{80pt}
  n\leftrightarrow m\,.
}
This symmetry of the spectrum is called T-duality. 
Note that under this action the spectrum is invariant \cite{Kikkawa:1984cp,Sakai:1985cs,Alvarez:1989ad,Nair:1986zn}
but 
the momenta \eqref{cft_001} are mapped as
\eq{
  \label{dual_002}
    \bigl(\,p^{25}_R\,,\,p^{25}_L\,\bigr) \quad \longrightarrow    \quad \bigl(\,-p^{25}_R\,,\,+p^{25}_L\,\bigr) \,.
}
When requiring the physics to be  invariant under the duality transformation, we can deduce from the commutation relations \eqref{comrel} that
also the centre-of-mass positions should be mapped as
$    (\,x^{25}_R\,,\,x^{25}_L\,)  \rightarrow   (\,-x^{25}_R\,,\,+x^{25}_L\,)$.
Since these coordinates and momenta appear in the mode expansions \eqref{cft_002}, 
it is natural to extend the mapping of the zero modes to the full mode expansion 
in the following way
\eq{
  \label{dual_84930}
    \bigl(\,X^{25}_R\,,\,X^{25}_L\,\bigr) \quad \longrightarrow    \quad \bigl(\,-X_R^{25}\,,\,+X^{25}_L\,\bigr) \,.
}
For the  oscillators this implies $(\alpha_n^{25} ,\ov\alpha_n^{25}) \to( - \alpha_n^{25}, + \ov\alpha^{25}_n)$, 
which  leaves the number operators \eqref{spec_03}  as well as 
the commutation relations \eqref{comrel} invariant. 
The mapping \eqref{dual_84930} is therefore indeed a symmetry of the spectrum. 
However, under this $\mathbb Z_2$ transformation the action \eqref{action_02a6}
is not invariant, in particular, we find
\eq{
  \partial_+ X_L^{25} \:\partial_- X_R^{25} \quad \longrightarrow    \quad
  -\op \partial_+ X_L^{25} \:\partial_- X_R^{25} \,,
}
showing that  \eqref{dual_84930} is not a symmetry  but a 
duality transformation.

%%%%%%%%%%%%%%%%%%%%%%%%%%%%%%%%%%%%%%%%%%%%%%%
%%%%%%%%%%%%%%%%%%%%%%%%%%%%%%%%%%%%%%%%%%%%%%%

\subsubsection*{Remarks}

Let us close this section with the following remarks:
\begin{itemize}

\item For the simple example of the closed bosonic string compactified on a circle of radius $R$, we have
seen that the spectrum is invariant under the mapping  $R\to \alpha'/R$
(together with an exchange of momentum and winding numbers).
This means that circle-compactifications with radius $R$ and $\alpha'/R$ are indistinguishable as they 
lead to the same spectrum.

\item In addition to the $\mathbb Z_2$ action shown in \eqref{duality_001}, 
the right- and left-moving momenta-squared $(p_R^{25})^2$ and $(p_L^{25})^2$
are also invariant under 
\eq{
  R \to \frac{\alpha'}{R} \,,
  \hspace{80pt}
  n\leftrightarrow -m\,.
}
The momenta are then mapped as
$(\,p^{25}_R\,,\,p^{25}_L\,) \to(\,+p^{25}_R\,,\,-p^{25}_L\,)$, which correspondingly extends to 
the right- and left-moving fields $X^{25}_R$ and $X^{25}_L$. 
The full T-duality group for a circle compactification is therefore $\mathbb Z_2\times \mathbb Z_2$.

\item The $\mathbb Z_2$ transformation \eqref{duality_001} has a fixed point at $R = \sqrt{\alpha'}$ 
called the self-dual radius. At this point in moduli space additional massless fields
with non-vanishing momentum and winding numbers appear in the spectrum and lead
to a symmetry enhancement. (See for instance \cite{Blumenhagen:2013fgp} for a textbook 
discussion of this mechanism.)

\item The self-dual radius is sometimes interpreted as the minimal length 
scale of the string, since radii $R< \sqrt{\alpha'}$ can be mapped via T-duality to 
$R>\sqrt{\alpha'}$. However, while this is true for the bosonic string, for the superstring 
this reasoning fails as T-duality maps for instance the type IIA superstring to the IIB theory
or the heterotic $E_8\times E_8$ theory to the heterotic $SO(32)$ superstring.

\item \label{page_open_string_cft} In the case of open strings, the two-dimensional world-sheet $\Sigma$ has boundaries and hence 
$\partial\Sigma\neq \varnothing$. 
The simplest example for such a world-sheet is the infinite strip $\Sigma = \mathbb R\times \mathbb I$ where 
$\mathbb I=[0,\ell_{\rm s }]$ is a finite interval, and the fields $X^{\mu}(\tau,\sigma)$ can then have 
either Neumann or Dirichlet boundary conditions
\eq{  
  \arraycolsep2pt
  \begin{array}{l@{\hspace{40pt}}rcl}
  \mbox{Neumann} & \displaystyle 
   \partial_{\sigma} X^{\mu}(\tau,\sigma ) \bigr\rvert_{\partial\Sigma} &=& 0 \,,
  \\[8pt]
  \mbox{Dirichlet} & \displaystyle    X^{\mu}(\tau,\sigma ) \bigr\rvert_{\partial\Sigma} &=& \mathrm{const.} 
  \end{array}
}
The directions with Neumann boundary conditions correspond to the world-volume of a
D$p$-brane, where $p$ denotes the number of spatial dimensions with Neumann boundary
conditions. (The time direction is usually assumed to have Neumann conditions.)
Furthermore, 
in addition to the closed-string background fields $G_{\mu\nu}$, $B_{\mu\nu}$ and $\phi$,
on a D-brane an open-string gauge field $a_{\mu}$ is present. 
On a single circle this gauge field may take a constant non-trivial vacuum expectation value, which 
is called a Wilson loop. 

Now, under a T-duality transformation along a circle the Neumann and Dirichlet boundary conditions 
as well as the momentum and winding numbers are interchanged, and the Wilson loop is interchanged with 
the position of the boundary on the circle. 
For more details we refer for instance to the textbook discussions in \cite{Polchinski:1998rq,Blumenhagen:2013fgp}.

\end{itemize}

%%%%%%%%%%%%%%%%%%%%%%%%%%%%%%%%%%%%%%%%%%%%%%%
%%%%%%%%%%%%%%%%%%%%%%%%%%%%%%%%%%%%%%%%%%%%%%%
%%%%%%%%%%%%%%%%%%%%%%%%%%%%%%%%%%%%%%%%%%%%%%%
%%%%%%%%%%%%%%%%%%%%%%%%%%%%%%%%%%%%%%%%%%%%%%%
%%%%%%%%%%%%%%%%%%%%%%%%%%%%%%%%%%%%%%%%%%%%%%%
%%%%%%%%%%%%%%%%%%%%%%%%%%%%%%%%%%%%%%%%%%%%%%%
%%%%%%%%%%%%%%%%%%%%%%%%%%%%%%%%%%%%%%%%%%%%%%%
%%%%%%%%%%%%%%%%%%%%%%%%%%%%%%%%%%%%%%%%%%%%%%%

\subsection{Conformal field theory for \texorpdfstring{$\mathbb T^D$}{TD}}
\label{sec_cft_torus}

Let us now  generalise the above analysis  from circle  to toroidal compactifications. 
We include a constant Kalb-Ramond field $B$ in our analysis, and we will see that the duality
group is enlarged.

%%%%%%%%%%%%%%%%%%%%%%%%%%%%%%%%%%%%%%%%%%%%%%%
%%%%%%%%%%%%%%%%%%%%%%%%%%%%%%%%%%%%%%%%%%%%%%%

\subsubsection*{Compactification}

A $D$-dimensional toroidal compactification can be specified by the identification of 
$D$ coordinates $X^{\ba I}$ (in a vielbein basis) as follows
\eq{
  \label{cft_t_840934}
  X^{\ba I} \sim X^{\ba I} + 2\pi\op L^{\ba I} \,,
  \hspace{50pt}
  L^{\ba I} = \sum_{i=1}^D e^{\ba I}{}_i\, n^i \,,
  \hspace{50pt}
  n^i\in \mathbb Z \,,
}
where $\ba I= \{25-D, \ldots,25\}$ labels the compactified directions. 
The $D$ vectors $e_i= \{ e^{\ba I}{}_i\}$ are $D$-dimensional and 
are required to be linearly-independent, and  therefore generate a $D$-dimensional 
lattice. Their duals will be denoted by $\ov e^i = \{\ov e^i{}_{\ba I}\}$ which are specified by
$\ov e^i{}_{\ba I}\op e^{\ba I}{}_j  = \delta^i{}_j$.
Turning to the fields $X^{\ba I}(\tau,\sigma)$,
similarly as in \eqref{cft_aa849} the mode expansions can be
split into a left- and right-moving sectors for which we find
\eq{
  \label{cft_003}
  X^{\ba I}_R(\tau-\sigma) &= x_R^{\ba I} + \frac{2\pi\alpha'}{\ell_{\rm s}}\op p_R^{\ba I} \op (\tau-\sigma) + 
  i\op \sqrt{\frac{\alpha'}{2}}\, \sum_{n\neq0} \frac{1}{n} \op\alpha^{\ba I}_n \, e^{-\frac{2\pi i}{\ell_{\rm s}}\op n\op (\tau-\sigma)} \,,
  \\
  X^{\ba I}_L(\tau+\sigma) &=  x_L^{\ba I} + \frac{2\pi\alpha'}{\ell_{\rm s}}\op p_L^{\ba I} \op (\tau+\sigma) + 
  i\op \sqrt{\frac{\alpha'}{2}}\, \sum_{n\neq0} \frac{1}{n} \op\ov\alpha^{\ba I}_n \, e^{-\frac{2\pi i}{\ell_{\rm s}}\op n\op (\tau+\sigma)} \,,
}
where the momenta are expressed in terms of the momentum numbers $m_i\in\mathbb Z$ and winding numbers $n^i$ as
\eq{
  \label{momenta_001}
  p_{R}^{\ba I} &= \frac{1}{2\op\alpha'} \,\delta^{\ba{IJ}}\,\ov e_{\ba J}{}^i \Bigl( \alpha' m_i - g_{ij}\op n^j - b_{ij} \op n^j \Bigr) \,,
  \\[6pt]
  p_{L}^{\ba I} &= \frac{1}{2\op\alpha'} \,\delta^{\ba {IJ}}\,\ov e_{\ba J}{}^i \Bigl( \alpha' m_i + g_{ij}\op n^j - b_{ij} \op n^j \Bigr) \,.
}
The non-trivial information about the metric of the compact space is contained in the 
lattice vectors $e_i$. In the vielbein basis employed in \eqref{cft_t_840934} the corresponding 
metric is trivial, i.e.  $G_{\ba{IJ}} = \delta_{\ba{IJ}}$, whereas in the lattice basis the metric and $B$-field take the form
\eq{
  \label{vielbein_8467}
  g_{ij} = e_i{}^{\ba I}\op \delta_{\ba{IJ}} \op e^{\ba J}{}_j \,, \hspace{50pt}
  b_{ij} = e_i{}^{\ba I}\op B_{\ba{IJ}} \op e^{\ba J}{}_j \,.
}
To make contact with our conventions in section~\ref{sec_cft_circ}, let us choose $e^{25}{}_1 = R$ 
and $\ov e^1{}_{25}=1/R$ from which we find $g_{11}=R^2$. Since furthermore $b_{11}=0$ due to the 
anti-symmetry of the Kalb-Ramond field, we recover the 
expressions \eqref{cft_001} from \eqref{momenta_001}.

Let us finally note that when promoting the modes appearing in the expansion \eqref{cft_003} 
to operators and  replacing Poisson brackets by commutators,
 the corresponding non-trivial commutation relations read
\eq{
  \label{comrel_torus}
  \arraycolsep2pt
  \begin{array}{lcl@{\hspace{50pt}}lcl}
  \displaystyle [\op x^{\ba I}_L,p^{\ba J}_L\op] &=& i \,\delta^{\ba{IJ}}\,, & 
  \displaystyle [\op\alpha^{\ba I}_m, \alpha^{\ba J}_n\op] &=& m\, \delta_{m+n} \,\delta^{\ba{IJ}}\,,
  \\[12pt]
  \displaystyle [\op x^{\ba I}_R,p^{\ba J}_R\op] &=& i\,\delta^{\ba{IJ}} \,, & 
  \displaystyle [\op\ov\alpha^{\ba I}_m, \ov\alpha^{\ba J}_n\op] &=& m\, \delta_{m+n} \,\delta^{\ba{IJ}}\,.
  \end{array}
}

%%%%%%%%%%%%%%%%%%%%%%%%%%%%%%%%%%%%%%%%%%%%%%%
%%%%%%%%%%%%%%%%%%%%%%%%%%%%%%%%%%%%%%%%%%%%%%%

\subsubsection*{Spectrum}

The spectrum of the closed bosonic string compactified on a torus
can again be expressed using 
\eq{
  \arraycolsep2pt
  \begin{array}{lclcl}
  \displaystyle \alpha' m_R^2 &=& \displaystyle 2\op \alpha' \bigl(p_R\bigr)^2 &+& 
    \displaystyle 2\op\bigl( N_R - 1\bigr) \,, \\[4pt]
  \displaystyle \alpha' m_L^2 &=& \displaystyle 2\op \alpha' \bigl(p_L\bigr)^2 &+& 
    \displaystyle 2\op\bigl( N_L - 1\bigr) \,,
  \end{array}
}
where $p_{R,L}^2= p_{R,L}^{\ba I} \op\delta_{\ba{IJ}} \,p_{R,L}^{\ba J}$ and 
$N_{R,L}$ take a similar form as in \eqref{spec_03}.
The right- and left-moving momenta-squared
$p_R^2$ and $p_L^2$ are now expressed in the following way
\eq{
  \label{square_001}
  2\op\alpha'\op p^2_{R,L} &= \frac{\alpha'}{2} \, m^T g^{-1} \op m
  + \frac{1}{2\alpha'} \, n^T \Bigl( g-b\op g^{-1}b\Bigr)\op n
  + n^T b \op g^{-1} m \mp  n^T m
  \\[8pt]
  \arraycolsep2pt
  &= \frac{1}{2} \binom{n}{m}^T
  \left( \begin{array}{cc}
   \frac{1}{\alpha'}\left( g - b\op g^{-1}b\right)& +b\op g^{-1}  \\  - g^{-1} b & \alpha' g^{-1}\end{array} \right)
 \binom{n}{m}
 \\
 &\hspace{180pt}
  \mp \frac{1}{2} \binom{n}{m}^T \left( \begin{array}{cc} 0 & \mathds 1 \\ \mathds 1 & 0 \end{array}\right)\binom{n}{m}
  \,,
}
with the upper sign corresponding to $p^2_R$ and the lower sign to $p_L^2$ and with matrix multiplication 
understood. 
A commonly-used convention is to denote the $2D\times 2D$ dimensional matrices appearing in 
\eqref{square_001} as \cite{Giveon:1988tt}
\eq{
  \label{gen_met_098}
  \arraycolsep4pt
  \mathcal H = \left( \begin{array}{cc}
   \frac{1}{\alpha'}\left( g - b\op g^{-1}b\right)& +b\op g^{-1}  \\[4pt]  - g^{-1} b & \alpha' g^{-1}\end{array} \right)
  ,
  \hspace{50pt}
  \eta =  \left( \begin{array}{cc} 0 & \mathds 1 \\[2pt]  \mathds 1 & 0 \end{array}\right) ,
}
where $\mathcal H$ is also called the generalised metric. 
Note that the index structure of the identity matrix $\mathds 1$ in $\eta$ is 
$\delta_i{}^j$ for the upper-right and $\delta^i{}_j$ for the lower-left part. 
The combined mass formula and the level-matching condition are again  given by
\eq{
  \label{cft_7920}
  \alpha' m^2 = \alpha' m_R^2  + \alpha' m_L^2 \,,
  \hspace{50pt}
  \alpha' m_R^2  = \alpha' m_L^2 \,.
}

%%%%%%%%%%%%%%%%%%%%%%%%%%%%%%%%%%%%%%%%%%%%%%%
%%%%%%%%%%%%%%%%%%%%%%%%%%%%%%%%%%%%%%%%%%%%%%%

\subsubsection*{Invariance of the spectrum}

Next, we want to determine which transformations leave the spectrum determined by \eqref{cft_7920}
invariant. 
This amounts to requiring \eqref{square_001} to be separately invariant for both sign-choices. 
Let us first note that under
\eq{
  \label{odd_001}
  \binom{n}{m} \to \binom{\tilde n}{\tilde m} = \mathcal O \, \binom{n}{m}
  \hspace{50pt}\mbox{with}\qquad
  \mathcal O^T\eta \, \mathcal O = \eta
}
the last term in \eqref{square_001} stays invariant. 
The condition $ \mathcal O^T\eta \, \mathcal O = \eta$ is the defining property of the split orthogonal matrices, and since the transformed $\tilde n^i$ and 
$\tilde m_i$ are again required to be integers, we have in particular
\eq{
  \mathcal O \in O(D,D,\mathbb Z) \,.
}
From \eqref{odd_001} we can furthermore infer that $\mathcal O^{-1} = \eta\, \mathcal O^T  \eta$, 
and  for invariance of the first term in \eqref{square_001} we have to demand
\eq{
  \label{transf_003}
  \mathcal H \to \tilde{\mathcal H} = \mathcal O^{-T} \mathcal H \,\mathcal O^{-1} \,.
}
The  relation \eqref{transf_003} determines  how the background fields $g_{ij}$ and $b_{ij}$
 contained in the matrix $\mathcal H$ 
transform under $O(D,D,\mathbb Z)$.
We thus see that the generalisation of the T-duality group from  circle to toroidal compactifications
is $O(D,D,\mathbb Z)$ \cite{Shapere:1988zv,Giveon:1988tt}. Furthermore, note that $O(1,1,\mathbb Z) = \mathbb Z_2\times \mathbb Z_2$ and 
hence the one-dimensional case is properly included.

%%%%%%%%%%%%%%%%%%%%%%%%%%%%%%%%%%%%%%%%%%%%%%%
%%%%%%%%%%%%%%%%%%%%%%%%%%%%%%%%%%%%%%%%%%%%%%%

\subsubsection*{Duality transformations}

To gain some more insight on how the background fields transform, 
 let us para\-metrise a general $O(D,D,\mathbb Z)$ transformation as
\eq{
  \label{notation_odd_01}
  \mathcal O = \left( \begin{array}{cc} A & B \\ C & D \end{array}\right),
}
where $A,B,C,D$ are $D\times D$ matrices over $\mathbb Z$ (with the appropriate index structure). These matrices are subject to 
the constraint $\mathcal O^T\eta \, \mathcal O = \eta$, which reads
\eq{
  \label{odd_cond_002}
  A^T C + C^T A = 0 \,, 
  \hspace{40pt}
  A^T D + C^T B = \mathds 1\,,
  \hspace{40pt}
  B^T D + D^T B = 0 \,,
}
and similar relations follow from $\mathcal O\,\mathcal O^{-1} = \mathds 1$ as
\eq{
  A \op B^T  + B \op A^T    = 0 \,, 
  \hspace{40pt}
  A \op D^T  + B \op C^T  = \mathds 1\,,
  \hspace{40pt}
  C \op D^T  + D \op C^T  = 0 \,.
}
From equation \eqref{transf_003} we can then determine the transformation behaviour of
the metric $g_{ij}$ and the $B$-field $b_{ij}$. We
find in matrix notation
\label{page_odd_trans_003}
\eq{
  \label{dual_001}
  \tilde g = \Omega_{\pm}^{-T}\op g\: \Omega_{\pm}^{-1} \,, 
  \hspace{50pt}
  \tilde g \pm \tilde  b = \pm\alpha' \op \Bigl[ \op C \pm \tfrac{1}{\alpha'} \op D\op(g\pm b) \op\Bigr]\op\Omega_{\pm}^{-1} \,,
}
where we defined 
\eq{
 \Omega_{\pm}=   A \pm \tfrac{1}{\alpha'} \op B\op(g\pm b) \,.
} 
Note that when introducing $E_{\pm} = g\pm b$, the second relation in \eqref{dual_001}
can also be expressed as
\eq{
  \tilde E_{\pm} = \pm\alpha' \, \frac{\hat A \op E_{\pm} \pm\alpha' \hat B}{\hat C\op E_{\pm} \pm \alpha' \hat D}
  \hspace{40pt}\mbox{for}
  \hspace{40pt}
  \hat{\mathcal O} \equiv \mathcal O^{-1} = \left( \begin{array}{cc} \hat A & \hat B \\ \hat C & \hat D \end{array}\right),
}
with matrix multiplication understood.
For the momenta defined in equation \eqref{momenta_001} we first 
determine the transformation behaviour of $p^i_{R/L} = \ov e^i{}_{\ba I}\op  p^{\ba I}_{R/L}$ (in the lattice basis) under 
$O(D,D,\mathbb Z)$ transformations \eqref{notation_odd_01} as
\eq{
  \tilde p^i_R = (\Omega_-)^i_{\hspace{4pt}j} \, p^j_R \,,
  \hspace{80pt}
  \tilde p^i_L = (\Omega_+)^i_{\hspace{4pt}j} \, p^j_L \,.
}
Similarly as before, we may now extend the duality transformations from the 
momenta to the full mode expansions of $X_R$ and $X_L$. In particular, 
in the lattice basis $X^i_{R,L} = \ov e^i{}_{\ba I} X^{\ba I}_{R,L}$ we have
\eq{
  \label{cft_torus_8494}
  \tilde X^i_R = (\Omega_-)^i_{\hspace{4pt}j} \, X^j_R \,,
  \hspace{80pt}
  \tilde X^i_L = (\Omega_+)^i_{\hspace{4pt}j} \, X^j_L \,.
}
As one can check, this leaves the commutation relations \eqref{comrel_torus} as well as 
the number operators $N_{R,L}$ invariant.
The extended transformation \eqref{cft_torus_8494} is therefore also
a symmetry of the spectrum and of the commutation relations. 
Let us now  determine how the action \eqref{action_02a6} behaves 
under such $O(D,D,\mathbb Z)$ transformations. Using \eqref{dual_001} together with 
\eqref{cft_torus_8494}, we find 
\eq{
  \label{transf_odd_04}
  \partial_+ X_L^T \, \bigl( \op g - b\op\bigr) \,\partial_- X_R
   \quad \longrightarrow    \quad
   \partial_+ X_L^T \: \Omega^T_+ \op \Bigl[ D(g-b) - \alpha' \op C \Bigr] \,\partial_- X_R \,,
}
and thus the action is in general not invariant. The $O(D,D,\mathbb Z)$ transformations
are therefore in general duality transformations.

Next, we  discuss how the fields $X^{\ba I}$ in the vielbein basis
transform under the duality group.
To do so we need the transformation behaviour of the lattice vectors, which
can be  inferred from 
the first relation in \eqref{dual_001}.
We find 
\eq{
  \tilde{e}^{\ba I}{}_i = \mathsf O^{\ba I}{}_{\ba J} \,e^{\ba J}{}_j \, \bigl( \Omega^{-1}_{\pm}\bigr)^j_{\hspace{6pt} i}  \,,
  \hspace{70pt}
  \mathsf O \in O(D)\,,
}
where $\mathsf O^{\ba I}{}_{\ba J}$ is an arbitrary $O(D)$ matrix parametrising changes in the frame bundle
and where we can use both sign choices
of $\Omega_{\pm}$ for the transformed lattice vectors. 
In order to match with the conventions in section~\ref{sec_cft_circ}, 
for convenience we choose the upper sign in $\Omega_{\pm}$ and $\mathsf O = \mathds 1$, which leads to 
\eq{
  \label{transf_odd_03}
  \tilde X^{\ba I}{}_R = \bigl( \op e \,\Omega^{-1}_+\,\Omega_- \op \ov e\, \bigr)^{\ba I}_{\hspace{4pt}\ba J}\, X^{\ba J}_R \,,
  \hspace{60pt}
  \tilde X^{\ba I}{}_L = X^{\ba I}{}_L \,.
}

%%%%%%%%%%%%%%%%%%%%%%%%%%%%%%%%%%%%%%%%%%%%%%%
%%%%%%%%%%%%%%%%%%%%%%%%%%%%%%%%%%%%%%%%%%%%%%%

\subsubsection*{Examples I -- general cases}

To illustrate how T-duality transformations act on the background fields, let us discuss some examples. 
The group $O(D,D,\mathbb Z)$ is generated by elements denoted here as 
$\mathcal O_{\mathsf A}$, $\mathcal O_{\mathsf B}$ and $\mathcal O_{\pm \mathsf i}$ 
\cite{Shapere:1988zv,Giveon:1988tt,Giveon:1989yf,Giveon:1990era} and which we specify below.
In the following we consider the action of these generators 
separately, whereas a general $O(D,D,\mathbb Z)$ is  a combination of these. 
\begin{itemize}

\item We start with transformations  parametrised by a $D\times D$ matrix $\mathsf A\in GL(D,\mathbb Z)$ in the following way
\eq{
  \label{odd_010}
  \arraycolsep5pt
\mathcal O_{\mathsf A} = \left( \begin{array}{cc} \mathsf A^{-1} & 0 \\ 0 & \mathsf A^{T} \end{array}\right)
  \,.
}
Note that $\mathcal O_{\mathsf A}$ has determinant $+1$. Using \eqref{transf_003},  we can work out the 
transformed generalised metric to be 
\eq{
  \tilde{\mathcal H}(\tilde g,\tilde b)  =   \mathcal O_{\mathsf A}^{-T} \:\mathcal H (g,b) \:\mathcal O_{\mathsf A}^{-1}
  = \mathcal H \bigl( \mathsf A^T g\op  \mathsf A,  \mathsf A^T b\op \mathsf A \bigr) \,,
}
and hence $\mathcal O_{\mathsf A}$ describes diffeomorphisms of the 
background geometry. 
From the relations in \eqref{transf_odd_03} we see that the coordinates (in the vielbein basis) are invariant
under such transformations,
\eq{
    \tilde X^{\ba I}{}_R = X^{\ba I}_R \,,
  \hspace{60pt}
  \tilde X^{\ba I}{}_L = X^{\ba I}{}_L \,,
}
and from \eqref{transf_odd_04} we see that also the action is invariant. Therefore, 
transformations of the form \eqref{odd_010} are a symmetry of the 
theory and belong to the so-called geometric group.

\item Next, we consider transformations parametrised by 
an anti-symmetric $D\times D$ matrix $\mathsf B$ with integer entries as
\eq{
  \label{b_transform_94}
  \arraycolsep5pt
  \mathcal O_{\mathsf B} = \left( \begin{array}{cc} \mathds 1 & 0 \\ \mathsf B & \mathds 1 \end{array}\right)
  \,,
}
where the requirement
of anti-symmetry of $\mathsf B$ is due to \eqref{odd_cond_002}. We also note that $\mathcal O_{\mathsf B}$ has determinant $+1$, and we find
\eq{
  \mathcal O_{\mathsf B}^{-T} \:\mathcal H (g,b) \:\mathcal O_{\mathsf B}^{-1}
  = \mathcal H \bigl( g,  b + \alpha' \mathsf B \bigr) \,.
}
The coordinates in the vielbein basis stay invariant under this transformation,
\eq{
    \tilde X^{\ba I}{}_R = X^{\ba I}_R \,,
  \hspace{60pt}
  \tilde X^{\ba I}{}_L = X^{\ba I}{}_L \,,
}
but according to \eqref{transf_odd_04} the action changes as
\eq{
  \partial_+ X_L^T \, \bigl( \op g - b\op\bigr) \,\partial_- X_R
   \quad \longrightarrow    \quad
   \partial_+ X_L^T \, \bigl( \op g - [ \op b + \alpha'\op\mathsf B \op ] \op\bigr) \,\partial_- X_R\,.
}
In general, these shifts of the $B$-field are therefore not a symmetry of the action but a duality 
transformation. However, for the special case of $\mathsf B = d \Lambda$ with
$\Lambda$ a well-defined one-form on the world-sheet $\Sigma$, such shifts are gauge transformations 
which after integration by parts leave the action invariant.

\item There are furthermore transformations parametrised by matrices of the form
$E_{\mathsf i} = \mbox{diag}\,(0,\ldots, 1,\ldots,0)$ with the $1$ at the $\mathsf i$'th position.
Such a transformation is also called a factorised duality and it takes the form
\eq{
  \label{odd_005}
  \mathcal O_{\pm \mathsf i} = \left( \begin{array}{cc} \mathds 1 - E_{\mathsf i}& \pm E_{\mathsf i} \\ \pm E_{\mathsf i} & \mathds 1 -E_{\mathsf i} \end{array}\right)
  .
} 
We note that the determinant of $\mathcal O_{\pm \mathsf i}$ is $-1$.
The transformation of the background fields can be worked out as follows
\begin{equation}
  \label{t-dual_001}
  \arraycolsep2pt
  \begin{array}{lcllcl}
  \tilde g_{\op\mathsf i\op \mathsf i} &=&\displaystyle \frac{\alpha'^2}{g_{\op\mathsf i\op\mathsf i}} \,, \\[5mm]
  \tilde g_{m\op\mathsf i} &=&\displaystyle \pm \alpha' \op\frac{b_{m\op\mathsf i}}{g_{\op\mathsf i\op\mathsf i}} \,, &
  \tilde b_{m\op\mathsf i } &=& \displaystyle \pm\alpha'\op \frac{g_{m\op\mathsf i}}{g_{\op\mathsf i\op \mathsf i}} \,,  \\[5mm]
  \tilde g_{mn} &=& \displaystyle g_{mn} - \frac{g_{m\op\mathsf i} \op g_{n\op \mathsf i} - b_{m\op\mathsf i} \op b_{n\op\mathsf i}}{g_{\op\mathsf i\op\mathsf i}} \,, \hspace{25pt}&
  \tilde b_{mn} &=& \displaystyle b_{mn} - \frac{b_{m\op \mathsf i} g_{n\op \mathsf i} - g_{m\op\mathsf i} b_{n\op\mathsf i}}{g_{\op \mathsf i\op\mathsf i}} \,,   
  \end{array}
\end{equation}
where $m,n\neq \mathsf i$ and where the two sign choices correspond to the two possible signs in \eqref{odd_005}.
Under this transformation, the coordinates in the vielbein basis behave as
\eq{
  \tilde X^{\ba I}{}_R = \bigl( \op e \,\Omega^{-1}_+\,\Omega_- \op \ov e^T\, \bigr)^{\ba I}{}_{\ba J}\, X^{\ba J}_R \,,
  \hspace{60pt}
  \tilde X^{\ba I}{}_L = X^{\ba I}{}_L \,,
}
with $ \Omega_{+}=   \mathds 1 -E_{\mathsf i}\pm \tfrac{1}{\alpha'}\op E_{\mathsf i}\op(g+b)$
and $ \Omega_{-}=   \mathds 1 -E_{\mathsf i}\mp\tfrac{1}{\alpha'}\op E_{\mathsf i}\op(g-b)$, and the action
transforms as
\eq{
  &\partial_+ X_L^T \, \bigl( \op g - b\op\bigr) \,\partial_- X_R
  \\
   &\hspace{60pt} \longrightarrow    \quad
   \partial_+ X_L^T \, \Bigl( \op (g-b)-E_{\mathsf i}\op(g-b)
   - (g-b) E_{\mathsf i} \op\Bigr) \,\partial_- X_R\,.
}
Since the action is not invariant, these transformations are not  symmetry but  duality
transformations.

\end{itemize}
In addition to the generators of the duality group, for later purpose we also consider
so-called $\beta$-transformations which we denote by $\mathcal O_{\beta}$. 
\begin{itemize}

\item Such transformations are parametrised by  an anti-symmetric $D\times D$ matrix $\beta$
and take the form \label{page_beta_trans}
\eq{
  \label{beta_trans_101}
  \mathcal O_{\beta} = \left( \begin{array}{cc} \mathds 1 & \beta \\ 0 & \mathds 1 \end{array}\right)
  ,
}
where the anti-symmetry of $\beta$ is again due to \eqref{odd_cond_002}. Note that $\mathcal O_{\beta}$ has determinant $+1$, and that it can be expressed using $\mathcal O_{\mathsf B}$
and the $O(D,D,\mathbb Z)$ elements
\eq{
  \label{odd_009}
   \mathcal O_{\pm} = \left( \begin{array}{cc} 0 & \pm \delta^{-1} \\  \pm\delta & 0 \end{array}\right),
   \hspace{80pt}
   \mathcal O_{\pm} = \prod_{\mathsf i=1}^D \mathcal O_{\pm \mathsf i} \,,
}
as
\eq{
  \mathcal O_{\beta} = \mathcal O_{\pm} \op \mathcal O_{\mathsf B} \op \mathcal O_{\pm} 
  \hspace{50pt}\mbox{where}\hspace{15pt}
  \beta^{ij} = \delta^{ip}\op \mathsf B_{pq} \op\delta^{qj} \,.
}
These transformations can be interpreted as first performing a factorised duality  
along all directions of the torus, then performing a $\mathsf B$-transformation, and finally 
performing again a factorised duality along all directions.

The coordinates in the vielbein basis  transform under a $\beta$-transformation in the following way
\eq{
  \tilde X^{\ba I}{}_R = \bigl( \op e \,\Omega^{-1}_+\,\Omega_- \op \ov e^T\, \bigr)^{\ba I}{}_{\ba J}\, X^{\ba J}_R \,,
  \hspace{60pt}
  \tilde X^{\ba I}{}_L = X^{\ba I}{}_L \,,
}
with $ \Omega_{\pm}=   \mathds 1 \pm \tfrac{1}{\alpha'} \op \beta \op(g\pm b) $,
and the action transforms as
\eq{
  \label{beta_trans_01}
  &\partial_+ X_L^T \, \bigl( \op g - b\op\bigr) \,\partial_- X_R
  \\
   &\hspace{-4pt} \longrightarrow    \quad
   \partial_+ X_L^T \, \Bigl( \op \bigl[ \op g +\tfrac{1}{\alpha'} (g\op \beta\op b + b\op \beta\op g) \bigr]  
   - \bigl [ \op b + \tfrac{1}{\alpha'}( g\op\beta \op g + b\op \beta\op b) \bigr] \op\Bigr) \,\partial_- X_R\,.
}
Such $\beta$-transformations are therefore in general not a symmetry of the action but are 
duality transformations.

\end{itemize}

%%%%%%%%%%%%%%%%%%%%%%%%%%%%%%%%%%%%%%%%%%%%%%%
%%%%%%%%%%%%%%%%%%%%%%%%%%%%%%%%%%%%%%%%%%%%%%%

\subsubsection*{Examples II -- special cases}

After having discussed the action of the generators of $O(D,D,\mathbb Z)$,
let us now turn to three particular situations. 
For all three examples we consider
a rectangular torus   with vanishing $B$-field of the form
\eq{
  \label{odd_ex_003}
  g_{ij} = \mbox{diag}\,\bigl(\, R_1^2\,,\, R_2^2\,, \,\ldots\,, \,R_D^2 \,\bigr)
  \,,\hspace{50pt} b_{ij} = 0 \,.
}

\begin{itemize}

\item A duality transformation acting on the background \eqref{odd_ex_003} via the matrix 
$\mathcal O_{\pm \mathsf i}$ gives the dual background
\eq{
  \label{odd_t_dual_432}
  \tilde g_{ij} = \mbox{diag}\,\left( R_1^2\,, \,\ldots\,,\, \frac{\alpha'^2}{R_{\mathsf i}^2}\,,\,\ldots \,R_D^2 \right) ,
  \hspace{50pt} \tilde b_{ij} = 0 \,,
}
and hence $\mathcal O_{\pm \mathsf i}$ corresponds to a T-duality transformation \eqref{duality_001}
along the 
direction labelled by $\mathsf i$.

\item As a second example we consider the two $O(D,D,\mathbb Z)$ elements given in \eqref{odd_009}.
For these transformations  the action acquires an overall minus sign
\eq{
  \partial_+ X_L^T \, \bigl( \op g - b\op\bigr) \,\partial_- X_R
   \quad \longrightarrow    \quad
   -\,\partial_+ X_L^T \, \bigl( \op g - b\op\bigr) \,\partial_- X_R\,,
}
implying that $\mathcal O_{\pm}$-transformations  are not a symmetry of the action. 
Moreover, for the example of the rectangular torus with vanishing $B$-field shown in \eqref{odd_ex_003}, the transformations \eqref{odd_009}
result in the dual background \cite{Shapere:1988zv,Giveon:1988tt}
\eq{
  \tilde g_{ij} = \mbox{diag}\,\left( \frac{\alpha'^2}{R_{1}^2}\,, \,\frac{\alpha'^2}{R_{2}^2}\,,\,\ldots \,\frac{\alpha'^2}{R_{D}^2} \right) ,
  \hspace{50pt} \tilde b_{ij} = 0 \,,
}
and therefore correspond to a collective T-duality transformation along all directions of the torus.

\item As a third example, let us consider a $\beta$-transformation acting on the 
background \eqref{odd_ex_003}.
For the anti-symmetric matrix $\beta$ we choose as the only non-vanishing 
entries $\beta_{12}=-\beta_{21}=\beta$, 
and for the transformed background we find
\eq{
  &\tilde g_{ij} = \mbox{diag}\,\left( 
  \frac{\alpha'^2R_1^2}{\alpha'^2+ \beta^2\op R_{1}^2\op R_2^2}\,, \,
  \frac{\alpha'^2R_2^2}{\alpha'^2+ \beta^2\op R_{1}^2\op R_2^2}\,,\,
  R_3^2\,,\,
  \ldots \,R_{D}^2 \right) ,
  \\[10pt]
  &\tilde b_{12} = -\tilde b_{21} =  -\frac{\alpha'\, \beta \op R_1^2\op R_2^2}{\alpha'^2+ \beta^2\op R_{1}^2\op R_2^2}\,,
}
while all other components of $\tilde b_{ij}$ are vanishing.

\end{itemize}

%%%%%%%%%%%%%%%%%%%%%%%%%%%%%%%%%%%%%%%%%%%%%%%
%%%%%%%%%%%%%%%%%%%%%%%%%%%%%%%%%%%%%%%%%%%%%%%

\subsubsection*{Remarks}

We close our discussion with the following three remarks:
\begin{itemize}

\item In this section we have studied $O(D,D,\mathbb Z)$ transformations for torus compactifications
of the closed string. We have seen that a subset of these is symmetries of the action, 
whereas in general $O(D,D,\mathbb Z)$ transformations are  duality transformations which only leave the spectrum 
invariant. 
In the literature $O(D,D,\mathbb Z)$ is often called the T-duality group, although 
not all of these transformations are dualities in the sense of our definition at the beginning of 
section~\ref{sec_intro_2}.

\item The moduli space of toroidal compactifications of the closed string 
is naively of the form $O(D,D,\mathbb R)/[O(D,\mathbb R)\times 
O(D,\mathbb R)]$ \cite{Narain:1985jj}, where the $D^2$ degrees of freedom of the metric and $B$-field 
correspond to $O(D,D,\mathbb R)$. Since the spectrum is invariant under separate $O(D,\mathbb R)$
rotations of the left- and right-moving sector, this part has been divided out.

Furthermore, as we discussed in this section, points in moduli space related by
$O(D,D,\mathbb Z)$ transformations are physically equivalent. 
The true moduli space therefore takes the form
\eq{
\frac{O(D,D,\mathbb R)}{O(D,\mathbb R)\times 
O(D,\mathbb R)} \biggm/ O(D,D,\mathbb Z) \,.
}

%%%%%%%%%%%%%%%%
%%%%%%%%%%%%%%%%
\begin{figure}[t]
\centering
\begin{subfigure}{0.38\textwidth}
\centering
\includegraphics[width=130pt]{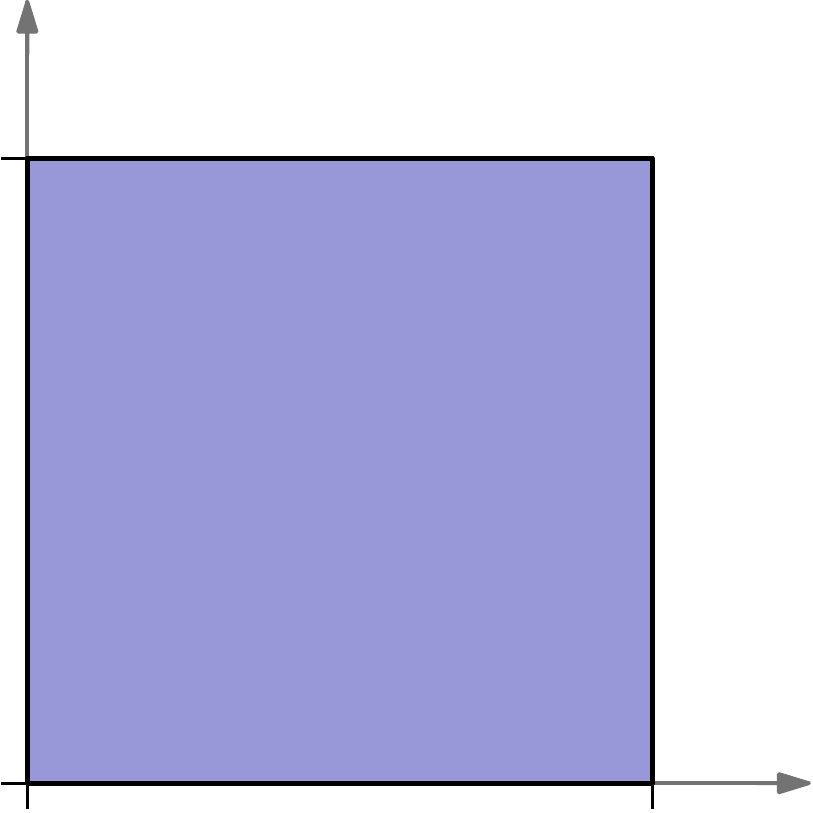}
\begin{picture}(0,0)
\put(-33,-8){\scriptsize $R_1$}
\put(-146,102){\scriptsize $R_2$}
\put(-86,51){\footnotesize $\mathcal F_{12}$}
\end{picture}
\caption{D2-brane on $\mathbb T^2$ with constant gauge flux $\mathcal F_{12}$\label{fig_td2_a}}
\end{subfigure}
\hspace{35pt}
\begin{subfigure}{0.38\textwidth}
\centering
\includegraphics[width=130pt]{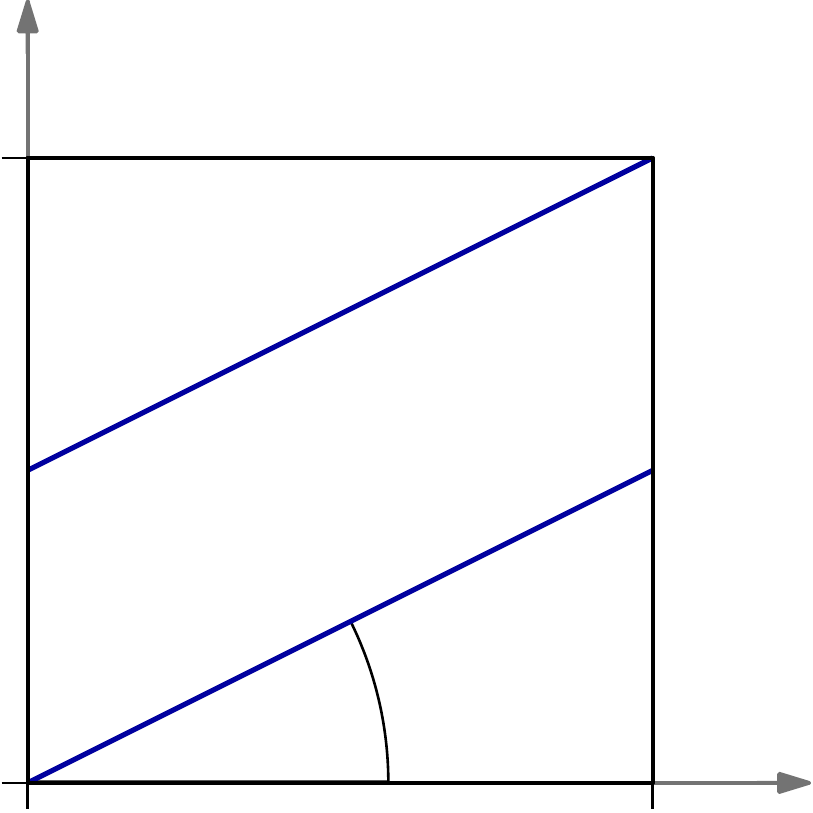}
\begin{picture}(0,0)
\put(-33,-8){\scriptsize $\tilde R_1$}
\put(-146,102){\scriptsize $\tilde R_2$}
\put(-90,12.5){\footnotesize $\varphi$}
\end{picture}
\caption{D1-brane wrapping $\mathbb T^2$ at an angle $\varphi$\label{fig_td2_b}}
\end{subfigure}
\caption{Illustrations of a D2- and a D1-brane wrapping a rectangular two-torus.
In figure~\ref{fig_td2_a} a D2-brane with a constant gauge flux $\mathcal F_{12}$ 
is shown, and in figure~\ref{fig_td2_b} 
a D1-brane wrapping the $\mathbb T^2$ at an angle $\varphi$ is illustrated. 
Under T-duality along one of the circles these two configurations are interchanged.
\label{fig_td2_c}} 
\end{figure}
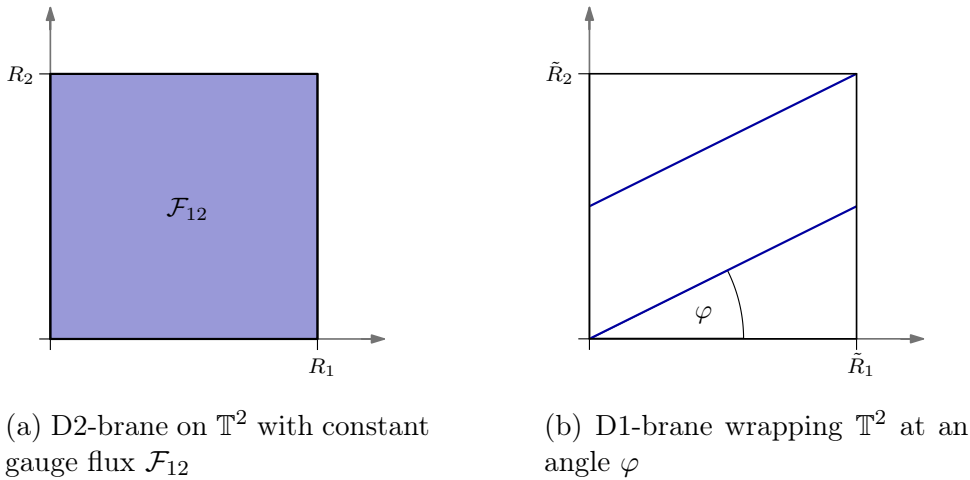
%%%%%%%%%%%%%%%%
%%%%%%%%%%%%%%%%

\item For open strings, a discussion similar to the one on page~\pageref{page_open_string_cft} applies. 
However, for toroidal compactifications with $D\geq2$ the open-string gauge field $a_{\mu}$ 
can be non-constant thus leading to a non-vanishing field-strength $F=da$ on the D-brane. 
This field strength is usually combined with the Kalb-Ramond field into the 
gauge-invariant open-string field strength $\mathcal F$ as \label{page_gauge_flux}
\eq{
  \label{open_string_fs}
  2\pi \alpha' \mathcal F = B + 2\pi \alpha' F \,.
}
Furthermore, D-branes can wrap the torus with non-trivial winding numbers along the torus directions.
These situations have been illustrated for a two-torus in figures~\ref{fig_td2_c}.

Under a T-duality transformation along one of the circles of the $\mathbb T^2$, 
a D2-brane with constant gauge flux $\mathcal F_{12}$ is mapped to a D1-brane 
wrapping the $\mathbb T^2$ at an angle $\varphi$ determined by
\eq{
  \cot \varphi = 2\pi\alpha' \mathcal F_{12} \,,
}  
and vice versa.  
For more details we refer the reader to \cite{Berkooz:1996km,Blumenhagen:2000wh}, 
and for a textbook discussion for instance to \cite{Blumenhagen:2013fgp}.

\end{itemize}

%%%%%%%%%%%%%%%%%%%%%%%%%%%%%%%%%%%%%%%%%%%%%%%
%%%%%%%%%%%%%%%%%%%%%%%%%%%%%%%%%%%%%%%%%%%%%%%
%%%%%%%%%%%%%%%%%%%%%%%%%%%%%%%%%%%%%%%%%%%%%%%
%%%%%%%%%%%%%%%%%%%%%%%%%%%%%%%%%%%%%%%%%%%%%%%
%%%%%%%%%%%%%%%%%%%%%%%%%%%%%%%%%%%%%%%%%%%%%%%
%%%%%%%%%%%%%%%%%%%%%%%%%%%%%%%%%%%%%%%%%%%%%%%
%%%%%%%%%%%%%%%%%%%%%%%%%%%%%%%%%%%%%%%%%%%%%%%
%%%%%%%%%%%%%%%%%%%%%%%%%%%%%%%%%%%%%%%%%%%%%%%
%%%%%%%%%%%%%%%%%%%%%%%%%%%%%%%%%%%%%%%%%%%%%%%
%%%%%%%%%%%%%%%%%%%%%%%%%%%%%%%%%%%%%%%%%%%%%%%
%%%%%%%%%%%%%%%%%%%%%%%%%%%%%%%%%%%%%%%%%%%%%%%
%%%%%%%%%%%%%%%%%%%%%%%%%%%%%%%%%%%%%%%%%%%%%%%
%%%%%%%%%%%%%%%%%%%%%%%%%%%%%%%%%%%%%%%%%%%%%%%
%%%%%%%%%%%%%%%%%%%%%%%%%%%%%%%%%%%%%%%%%%%%%%%
%%%%%%%%%%%%%%%%%%%%%%%%%%%%%%%%%%%%%%%%%%%%%%%
%%%%%%%%%%%%%%%%%%%%%%%%%%%%%%%%%%%%%%%%%%%%%%%
%%%%%%%%%%%%%%%%%%%%%%%%%%%%%%%%%%%%%%%%%%%%%%%
%%%%%%%%%%%%%%%%%%%%%%%%%%%%%%%%%%%%%%%%%%%%%%%
%%%%%%%%%%%%%%%%%%%%%%%%%%%%%%%%%%%%%%%%%%%%%%%
%%%%%%%%%%%%%%%%%%%%%%%%%%%%%%%%%%%%%%%%%%%%%%%
%%%%%%%%%%%%%%%%%%%%%%%%%%%%%%%%%%%%%%%%%%%%%%%
%%%%%%%%%%%%%%%%%%%%%%%%%%%%%%%%%%%%%%%%%%%%%%%
%%%%%%%%%%%%%%%%%%%%%%%%%%%%%%%%%%%%%%%%%%%%%%%
%%%%%%%%%%%%%%%%%%%%%%%%%%%%%%%%%%%%%%%%%%%%%%%
%%%%%%%%%%%%%%%%%%%%%%%%%%%%%%%%%%%%%%%%%%%%%%%
%%%%%%%%%%%%%%%%%%%%%%%%%%%%%%%%%%%%%%%%%%%%%%%
%%%%%%%%%%%%%%%%%%%%%%%%%%%%%%%%%%%%%%%%%%%%%%%
%%%%%%%%%%%%%%%%%%%%%%%%%%%%%%%%%%%%%%%%%%%%%%%
%%%%%%%%%%%%%%%%%%%%%%%%%%%%%%%%%%%%%%%%%%%%%%%
%%%%%%%%%%%%%%%%%%%%%%%%%%%%%%%%%%%%%%%%%%%%%%%
%%%%%%%%%%%%%%%%%%%%%%%%%%%%%%%%%%%%%%%%%%%%%%%
%%%%%%%%%%%%%%%%%%%%%%%%%%%%%%%%%%%%%%%%%%%%%%%

\clearpage
\section{Buscher rules}
\label{sec_buscher}

In this section we  extend the previous discussion of T-duality from toroidal compactifications 
to curved backgrounds. 
For the latter a CFT description is usually not available, which makes it difficult 
to quantise the theory and determine how duality transformations act on the spectrum.
However, a way to derive T-duality transformations for curved 
backgrounds is via Buscher's procedure  
\cite{Buscher:1987sk,Buscher:1987qj}, which gives
the dual background at leading order in string-length perturbation theory.

%%%%%%%%%%%%%%%%%%%%%%%%%%%%%%%%%%%%%%%%%%%%%%%
%%%%%%%%%%%%%%%%%%%%%%%%%%%%%%%%%%%%%%%%%%%%%%%
%%%%%%%%%%%%%%%%%%%%%%%%%%%%%%%%%%%%%%%%%%%%%%%
%%%%%%%%%%%%%%%%%%%%%%%%%%%%%%%%%%%%%%%%%%%%%%%
%%%%%%%%%%%%%%%%%%%%%%%%%%%%%%%%%%%%%%%%%%%%%%%
%%%%%%%%%%%%%%%%%%%%%%%%%%%%%%%%%%%%%%%%%%%%%%%
%%%%%%%%%%%%%%%%%%%%%%%%%%%%%%%%%%%%%%%%%%%%%%%
%%%%%%%%%%%%%%%%%%%%%%%%%%%%%%%%%%%%%%%%%%%%%%%

\subsection{Single T-duality}
\label{section_t_dual_1}

We first describe the general strategy for studying T-duality transformations 
of curved backgrounds. We consider T-duality along a single direction, but 
generalise this discussion to multiple directions in the next section.

%%%%%%%%%%%%%%%%%%%%%%%%%%%%%%%%%%%%%%%%%%%%%%%
%%%%%%%%%%%%%%%%%%%%%%%%%%%%%%%%%%%%%%%%%%%%%%%

\subsubsection*{World-sheet action}

The world-sheet action of the closed bosonic string has been given in equation \eqref{action_01}. 
However, for later convenience let us perform a Wick rotation $\sigma^0\to -i\op \sigma^0$ and go 
to an Euclidean world-sheet.
For the action this implies $\mathcal S \to -i\,\mathcal S_{\rm E}$, where the Euclidean
action is given by
\eq{
  \label{action_02e}
  \mathcal S_{\rm E} =& -\frac{1}{4\pi \alpha'} \int_{\Sigma}
  \Bigl[ \,G_{\mu\nu} \, d X^{\mu}\wedge\star d X^{\nu} - i\op B_{\mu\nu} \,dX^{\mu}\wedge dX^{\nu}
  + \alpha'\op\mathsf R\, \phi \star 1\, \Bigr] 
  \,,
}
but in the following we  drop the subscript $\rm E$.
The world-sheet $\Sigma$ is a two-dimensional (orientable) manifold without boundary,
and the Hodge star-operator has been defined in 
\eqref{ws_ac_conv}.
The fields $X^{\mu}$ can be considered as maps from the world-sheet $\Sigma$ to 
a target space, and the $B$-field appearing in \eqref{action_02e} should be understood 
as the pullback of the target-space quantity $B$ to the world-sheet, i.e. the proper expression 
reads $\int_{\Sigma} X^* B$. 
For notational simplicity we however assume that the distinction between world-sheet and 
target-space quantities is clear from the context. 
Furthermore, we consider 
target space-times of the form 
\eq{
  \label{tdual_split_tst}
  \mathbb R^{1,25-D}\times \mathcal M\,,
}
where 
$\mathcal M$ is a $D$-dimensional compact
manifold parametrised by local  coordinates $X^i$ with $i=1,\ldots,D$.
The non-compact part will not play a role in the discussion in this section.

The equations of motion for the fields $X^i$ are obtained in the usual way from 
the variation of the action with respect to  $X^{i}$. For infinitesimal variations $\delta X^i\ll1$ we find
from \eqref{action_02e}
\eq{
  \label{eom_8494949}
  0 = d\star d X^i + \Gamma^i_{mn} \op dX^m\wedge \star dX^n + \frac{i}{2} \op H^i{}_{mn} 
  dX^m\wedge dX^n 
  -\frac{\alpha'}{2} \op G^{im}\op\partial_m\phi \op R\star 1
  \,,
}
where $\Gamma^i_{jk}$ are the Christoffel symbols computed from  the target-space metric $G_{ij}$, 
$H=dB$ denotes the field strength of the Kalb-Ramond field $B$, and the index of $H_{ijk}$ has been
raised using the inverse of the metric $G_{ij}$.

%%%%%%%%%%%%%%%%%%%%%%%%%%%%%%%%%%%%%%%%%%%%%%%
%%%%%%%%%%%%%%%%%%%%%%%%%%%%%%%%%%%%%%%%%%%%%%%

\subsubsection*{Global symmetry}

In order to apply Buscher's procedure and derive the T-duality transformation rules, we 
require the compact manifold $\mathcal M$ to ``contain a circle''. 
In more precise terms, we
assume that the 
world-sheet action \eqref{action_02e} is invariant under a global symmetry of the form
\eq{
  \label{iso_trafo_01r}
  \delta_{\epsilon} X^{i} = \epsilon\, k^{i} (X) \,,
}
where $\epsilon\ll1$ is constant.
The action \eqref{action_02e} is  invariant under \eqref{iso_trafo_01r} if three conditions
are met: 1) $k= k^i\op \partial_i$ is a Killing vector of the target-space metric $G$, 2) 
there exist a one-form $v$ (globally defined on $\Sigma$) such that $\mathcal L_k B = dv$ 
\cite{Hull:1989jk,Hull:1990ms}, 
and 3) the Lie derivative of the dilaton $\phi$ in the direction $k$ vanishes. 
In terms of equations, these three conditions can be summarised as
\eq{
  \label{gc_009}
  \mathcal L_{k} \op  G = 0\,, \hspace{70pt}
  \mathcal L_{k} \op  B = dv  \,, \hspace{70pt}
  \mathcal L_{k} \op \phi =0\,,
}
where $G = \frac12\op G_{ij} \op dX^i\vee dX^j$ and $B=\frac12\op B_{ij}\op dX^i\wedge dX^j$ 
are interpreted as target-space quantities,\footnote{
The symmetrisation and anti-symmetrisation of the tensor product of differential forms are defined
as $dX^i\vee dX^j = dX^i\otimes dX^j + dX^j\otimes dX^i$ and
$dX^i\wedge dX^j = dX^i\otimes dX^j - dX^j\otimes dX^i$.}
 and 
where the Lie derivative is given by $\mathcal L_k = d\circ\iota_k + \iota_k\circ d$
with $\iota_k$ the contraction operator acting as $\iota_{\partial_i} dX^j = \delta_i{}^j$. 
The requirement of $v$ being globally-defined restricts the allowed $B$-field configurations, 
and we come back to this point on page~\pageref{page_global_v}.

%%%%%%%%%%%%%%%%%%%%%%%%%%%%%%%%%%%%%%%%%%%%%%%
%%%%%%%%%%%%%%%%%%%%%%%%%%%%%%%%%%%%%%%%%%%%%%%

\subsubsection*{Local symmetry}

Following Buscher's procedure, we now gauge the  global symmetry \eqref{iso_trafo_01r}, that is, 
we allow the infinitesimal parameter $\epsilon$  to depend on the world-sheet 
coordinates $\sigma^{\mathsf a}$.
To do so, we introduce a world-sheet gauge field $A$ and replace
$dX^{i}\to dX^i + k^{i} A$ for the term involving the metric. 
For the $B$-field term the gauging  is different due to the one-form $v$.
We furthermore introduce an
additional scalar field $\chi$ whose role will become clear shortly. The resulting gauged action 
(restricted to the compact target-space manifold $\mathcal M$) reads
\eq{
  \label{action_02a}
  &\hat{\mathcal S} =-\frac{1}{2\pi\alpha'} \int_{\Sigma} \;  
  \Bigl[ \op \tfrac{1}{2}\op G_{ij}  (dX^{i} + k^{i} A)\wedge\star(dX^{j} + k^{j} A)  
   \\
  & \hspace{130pt}
   - \tfrac{i}{2}\op B_{ij} \,dX^{i}\wedge dX^{j}
  -   i\op ( v -\iota_k B + d\chi)\wedge A
  \Bigr] \,,
}
 where we ignored the dilaton term which does not get modified in the gauging procedure. 
The corresponding local symmetry transformations are
\eq{
  \label{variantions_01k}
  \hat\delta_{\epsilon} X^{i} = \epsilon\op k^{i} \,,\hspace{60pt}
  \hat\delta_{\epsilon} A = - d\epsilon \,,\hspace{60pt}
  \hat\delta_{\epsilon}\chi = -  \epsilon\, \iota_{k} v 
  \,,
}
and the variation of the action \eqref{action_02a} with respect to \eqref{variantions_01k} gives
\eq{
  \label{variation_35c}
  \hat\delta_{\epsilon} \hat{\mathcal S} =+\frac{i}{2\pi\alpha'} \int_{\Sigma} 
  d \bigl[ \op\epsilon\op(v+ d\chi)\op\bigr] 
  = 0  \,.
}
In the last step we have assumed that the integrand is globally-defined on $\Sigma$
and we have employed Stokes' theorem.

%%%%%%%%%%%%%%%%%%%%%%%%%%%%%%%%%%%%%%%%%%%%%%%
%%%%%%%%%%%%%%%%%%%%%%%%%%%%%%%%%%%%%%%%%%%%%%%

\subsubsection*{Simplifying assumptions}

Before we proceed let us make some simplifying assumptions. More general situations 
are considered below.
\begin{itemize}
\label{page_trivialws}

\item We perform a change of coordinates to so-called  adapted coordinates, 
in which the Killing vector takes the form
\eq{
  \label{adap_01}
  k^{i} = \bigl( \,1\,, \, 0\,, \, \ldots\,, 0\, \bigr)^T \,. 
}
Locally on the target-space manifold this can always be achieved, 
provided that $|k|\neq 0$ at that point. The Killing property
then implies that 
\eq{
  \label{adap_blab_1}
  k^{m} \partial_{m} G_{ij}  = \partial_1 G_{ij} =0 \,,
  \hspace{50pt}
  k^{m} \partial_{m} H_{ijk}  = \partial_1 H_{ijk} =0\,,
}
and hence the components of the metric and of the $H$-flux do not depend on the coordinate $X^1$.

\item We also choose a gauge for the $B$-field 
in which the one-form $v$ vanishes. 
Together with \eqref{adap_blab_1} this implies that the components $B_{ij}$ 
do not depend on the variable 
$X^1$, that is 
\eq{
  \label{adap_blab_2}
  \mathcal L_k B = 0 \hspace{40pt}\longrightarrow \hspace{40pt} \partial_1 B_{ij} = 0\,.
}  
We assume that the above gauge choice  can be achieved via a gauge transformation on $B$, in particular
by $B\to B + d\Lambda$ with $\Lambda$ a globally-defined one-form satisfying $d\iota_k d\Lambda = dv$.

\end{itemize}

%%%%%%%%%%%%%%%%%%%%%%%%%%%%%%%%%%%%%%%%%%%%%%%
%%%%%%%%%%%%%%%%%%%%%%%%%%%%%%%%%%%%%%%%%%%%%%%

\subsubsection*{Back to the ungauged  action}
\label{page_back_single_1}

We now come to the role of the scalar field $\chi$ in the gauged action \eqref{action_02a}. 
In Buscher's procedure we use $\chi$ as
a Lagrange multiplier to ensure that during the gauging procedure no additional 
degrees of freedom are introduced
\cite{Rocek:1991ps,Giveon:1993ai,Alvarez:1993qi}. The latter implies that $A$ has to be pure gauge, i.e. 
$A = d \lambda$ where $\lambda$ is a globally-defined function on $\Sigma$.

In order to discuss this procedure, we need to introduce some notation.
We denote a basis of harmonic one-forms on the two-dimensional world-sheet  $\Sigma$
(oriented, without boundary)
by
\eq{
\omega^{\mathsf m}\in \mathcal H^1(\Sigma,\mathbb R)
\hspace{50pt}\mathsf m=1,\ldots,2g\,, 
}
where $g$ denotes the genus of  $\Sigma$.
The group of harmonic one-forms is isomorphic to the first de Rham 
cohomology group $H^{1}(\Sigma,\mathbb R)$, and 
a basis for the corresponding first homology  group will be denoted by 
$\gamma_{\mathsf m}\in H_{1}(\Sigma,\mathbb Z)$.
The one-cycles and one-forms can be chosen  such that 
$\int_{\gamma_{\mathsf m}}\omega^{\mathsf n} =\delta_{\mathsf m}{}^{\mathsf n}$
and 
$\int_{\Sigma}  \omega^{\mathsf m} \wedge \omega^{\mathsf n}=J^{\mathsf{mn}}$,
where $J^{\mathsf{mn}}$ is a non-degenerate matrix with integer entries 
whose inverse also has integer entries (see for instance
page 250 in \cite{Hori:2003ic}).
Now, using the Hodge decomposition theorem we can express the closed one-form 
$d\chi$ as
\eq{
  \label{hd_9305}
  d\chi = d\chi_{(0)} +  \chi_{(\mathsf m)}\op \omega^{\mathsf m} \,,
}
where $\chi_{(0)}$  is  a globally-defined function  on $\Sigma$
and $\chi_{(\mathsf m)}\in\mathbb R$ are constants, and where a summation over $\mathsf m = 1, \ldots, 2g$ 
is understood.

Let us now determine the equation of motion for the Lagrange multiplier by 
varying the gauged action \eqref{action_02a} with respect to $\chi_{(0)}$ 
\eq{
  \delta_{\chi_{(0)}} \hat{\mathcal S}  =-\frac{i}{2\pi\alpha'} \int_{\Sigma} \;  
  \delta{\chi_{(0)}} \, dA \overset{!}{=}0 
  \hspace{40pt}\longrightarrow \hspace{40pt} 
  F = dA = 0 \,.
}
On a topologically-trivial world-sheet a vanishing field strength means that $A$ has to be pure gauge,
however, this is not true in general. Indeed, for the closed one-form $A$ we can again
perform a Hodge decomposition as 
\eq{
  A = da_{(0)} + a_{(\mathsf m)}\op \omega^{\mathsf m} \,,
}
where $a_{(0)}$ is a globally-defined function on $\Sigma$ and $a_{(m)}\in \mathbb R$ 
correspond to possible Wilson loops $W_{\gamma} = \exp( 2\pi i \oint_{\gamma} A)$ of $A$
around one-cycles $\gamma\in H_1(\Sigma,\mathbb Z)$.
Performing now a variation of the action \eqref{action_02a} with respect to $\chi_{(\mathsf m)}$ we find
\eq{
  \delta_{\chi_{(\mathsf m)}} \hat{\mathcal S}  =\frac{i}{2\pi\alpha'} \op \delta \chi_{(\mathsf m)}
   J^{\mathsf{mn}} a_{(\mathsf n)}
  \overset{!}{=}0 
  \hspace{40pt}\longrightarrow \hspace{40pt} 
  a_{(m)} = 0 \,,
}
and $A$ is therefore pure gauge. Using then the gauge symmetry \eqref{variantions_01k} we can set 
$A=0$  and recover the original action \eqref{action_02e}.

%%%%%%%%%%%%%%%%%%%%%%%%%%%%%%%%%%%%%%%%%%%%%%%
%%%%%%%%%%%%%%%%%%%%%%%%%%%%%%%%%%%%%%%%%%%%%%%

\subsubsection*{Dual action}

In order to obtain the dual action, we integrate out the gauge field $A$. Since
$A$ does not have a kinetic term and is therefore non-dynamical, we can 
solve for $A$ algebraically. 
The variation of the action
\eqref{action_02a} with respect to $A$ takes the form
\eq{
  \hat\delta_{A} \hat{\mathcal S} =+\frac{1}{2\pi\alpha'} \int_{\Sigma} 
  \delta A \wedge \Bigl[ \, G_{11} \star A + \star G_{1i} \, dX^i + i\op \bigl(  d\chi - B_{1i} \, dX^i \bigr)\, \Bigr] 
  \,,
}
leading to the following solution of the equation of motion
\eq{
  \label{sol_94857}
  A = -\frac{1}{G_{11}} \, \Bigl[ \op G_{1i}\, dX^i - i\star\bigl(  d\chi -B_{1i} \, dX^i \bigr)\op \Bigr] \,,
}
where we used that $\star^2 = -1$ acting on a one-form in an Euclidean two-dimensional space.
Defining then 
\eq{
  \label{var_dual_7000}
  d\tilde X^1 = \pm\frac{1}{\alpha'} \op d\chi \,,
}
and using \eqref{sol_94857} in the action  \eqref{action_02a}, we find
\eq{
  \label{action_04}
  \check{\mathcal S} =-\frac{1}{2\pi\alpha'} \int_{\Sigma} \,
  \biggl[ \, \quad&\frac{1}{2}\op \left( G_{mn} - \frac{G_{m1}\op G_{n1}-B_{m1}\op B_{n1}}{G_{11}}\right)  \op  dX^{m} \wedge\star dX^{n}  \\
  &\hspace{40pt}+\frac{1}{2}\op \frac{\alpha'^2}{G_{11}} \op d\tilde X^1 \wedge\star d\tilde X^1
  \pm \alpha'\op \frac{B_{m1}}{G_{11}}\op d\tilde X^1 \wedge\star dX^m
   \\[10pt]
   -\op& \frac{i}{2}\op \left( B_{mn}    - \frac{B_{m1} G_{n1} - G_{m1} B_{n1}}{G_{11}} \right) dX^{m}\wedge dX^{n} \\
  &\hspace{40pt}\mp i \, \alpha'\op\frac{G_{m1}}{G_{11}} \op dX^m\wedge d\tilde X^1
  \mp i\op \alpha' \op dX^1 \wedge d\tilde X^1
  \hspace{25pt}
  \biggr] ,
}
where $m,n = 2,\ldots, D$. 
Due to \eqref{adap_blab_1} and \eqref{adap_blab_2} we note that 
the components $G_{ij}$ and $B_{ij}$ in \eqref{action_04} do not depend on the variable $X^1$. 
From \eqref{action_04} we can now 
read-off the dual background fields. Labelling the $\tilde X^1$-direction again by $1$, we find
\begin{equation}
  \label{dual_back_9494}
  \arraycolsep2pt
  \begin{array}{@{\hspace{-5pt}}lcllcl@{\hspace{-15pt}}}
  \check G_{11} &=&\displaystyle \frac{\alpha'^2}{G_{11}} \,, \\[5mm]
  \check G_{m1} &=&\displaystyle \pm\alpha'\op \frac{B_{m1}}{G_{11}} \,, &
  \check B_{m1} &=& \displaystyle \pm\alpha'\op\frac{G_{m1}}{G_{11}} \,,  \\[5mm]
  \check G_{mn} &=& \displaystyle G_{mn} - \frac{G_{m1} \op G_{n1} - B_{m1} \op B_{n1}}{G_{11}} \,, \hspace{20pt}&
  \check B_{mn} &=& \displaystyle B_{mn} - \frac{B_{m1} G_{n1} - G_{m1} B_{n1}}{G_{11}} \,.   
  \end{array}
\end{equation}
These expressions agree with the ones in \eqref{t-dual_001}, and we 
have therefore shown that through Buscher's procedure we can recover
the known transformation rules of T-duality for a circle.

%%%%%%%%%%%%%%%%%%%%%%%%%%%%%%%%%%%%%%%%%%%%%%%
%%%%%%%%%%%%%%%%%%%%%%%%%%%%%%%%%%%%%%%%%%%%%%%

\subsubsection*{Momentum/winding modes}
\label{page_buscher_global}

Note that we have not yet addressed the last term in the integrated out action \eqref{action_04}.
To do so, we first perform a Hodge decomposition of the closed one-form  $dX^1 \in T^*\Sigma$ as
\eq{
  \label{large_73831}
  dX^1 = dX^1_{(0)} +  X^1_{(\mathsf m)}\op \omega^{\mathsf m} \,,
}
where $X^1_{(0)}$ is again a globally-defined function on $\Sigma$ and  $X^1_{(\mathsf m)}\in\mathbb R$
are constants. 
Comparing this expression to the mode expansion of the closed string shown for instance in \eqref{cft_002}, 
we see that the 
exact part  corresponds to the oscillator terms while the harmonic part corresponds to the 
momentum/winding terms. 
More specifically, let us assume that  the direction $X^1$ is compactified
via the identification
\eq{
  \label{large_93091}
  X^1 \sim X^1 + 2\pi \op n\,, \hspace{50pt} n\in\mathbb Z \,. 
}
For the mode expansion of $X^1$ this implies that when going around a basis one-cycle 
$\gamma_{\mathsf m}\subset\Sigma$ on the world-sheet $\Sigma$, the function $X^1(\sigma^{\mathsf a})$ does not need to be single-valued
but can have integer shifts according to \eqref{large_93091}. 
In formulas this is expressed as
\eq{
  \oint_{\gamma_{\mathsf m}} dX^1 = 2\pi \op n_{(\mathsf m)}\,,
  \hspace{50pt}
  n_{(\mathsf m)} \in \mathbb Z\,.
}
Therefore,  for compactifications \eqref{large_93091} 
the coefficients of the harmonic terms in the expansion \eqref{large_73831}
are quantised as $X^1_{(\mathsf m)} = 2\pi \op n_{(\mathsf m)}$ and $n_{(\mathsf m)}$
are called the momentum/winding numbers.\footnote{
For a two-dimensional world-sheet with Euclidean signature there is 
no preferred time or space direction, and  hence there is no distinction
between momentum and winding numbers.}

Now, when integrating out the gauge field from the action \eqref{action_02a} we 
actually perform the path integral over $A$. Schematically this path integral reads
\eq{
  \hat{\mathcal Z} \sim \int \frac{[\mathsf D X^i] \,[\mathsf D \op\chi]\, [\mathsf DA]}{\mathcal V_{\rm gauge}}
  \: e^{\hat{\mathcal S}[X,\chi,A]} \,,
}
where $\hat{\mathcal S}$ is the gauged action  \eqref{action_02a}  and $\mathcal V_{\rm gauge}$ 
denotes the (infinite) volume of the gauge group. 
Performing the integration over $A$ leads to the following expression
\eq{
  \label{path_93939}
  \check{\mathcal Z} \sim \int \frac{[\mathsf D X^1]\, [\mathsf D X^m] \, [\mathsf D \op\chi]
  }{\mathcal V_{\rm gauge}} \: 
  e^{\check{\mathcal S}[X,\chi]} \,,
}
where $\check{\mathcal S}$ is the integrated out action \eqref{action_04} and $m=2,\ldots, D$.
Let us focus on the $X^1$-dependent terms in \eqref{path_93939}. Taking into account
the decomposition \eqref{large_73831} and that the $X^1_{(\mathsf m)} \in 2\pi \mathbb Z$
are quantised, we compute
\eq{
  \label{path_bremen}
  &\int \frac{[\mathsf DX^1]}{\mathcal V_{\rm gauge}}
   \: \exp\left({\frac{i}{2\pi\alpha'}\int_{\Sigma} dX^1\wedge d\chi} \right)
  \\[6pt]
  =&
   \int \frac{[\mathsf DX^1_{(0)}] }{\mathcal V_{\rm gauge}} \sum_{X^1_{(\mathsf m)}\in 2\pi\mathbb Z}
  \exp\left(\frac{i}{2\pi\alpha'}\int_{\Sigma} 
  X^1_{(\mathsf m)} \op\omega^{\mathsf m} \wedge \chi_{(\mathsf n)} \op\omega^{\mathsf n}
  \right)
  \\
  =&\sum_{k^{(\mathsf m)}\in \mathbb Z} \delta\left( \frac{1}{2\pi\alpha'} \op J^{\mathsf{mn}} \chi_{(\mathsf n)}
  - k^{(\mathsf m)}
  \right).
}
In the first step we performed an integration by parts to eliminate $X^1_{(0)}$ from the action, 
and in the second step the integral over $X^1_{(0)}$ was cancelled by $\mathcal V_{\rm gauge}$. The sum over 
$X^1_{(\mathsf m)}$ produces then a periodic Kronecker-symbol \cite{Rocek:1991ps}.
Recalling that the inverse of $J^{\mathsf{mn}}$ is again a matrix with integer entries and using 
\eqref{path_bremen} in the path integral \eqref{path_93939}, we see that 
the coefficients $\chi_{(\mathsf m)}$ appearing in the Hodge decomposition \eqref{hd_9305}
are quantised as
\eq{
  \chi_{(\mathsf m)} \in 2\pi \alpha' \op \mathbb Z \,.
}
For the dual variable $\tilde X^1$ defined via \eqref{var_dual_7000} this implies that its harmonic
part is quantised in units of $2\pi$, and therefore $\tilde X^1$ describes again a compact direction.
To summarise, if the direction along which a T-duality transformation is performed is compact
with identifications \eqref{large_93091}, then also the dual background has a compact direction
as
\eq{
  \label{dual_back_010101010}
  X^1 \sim X^1 + 2\pi \op n\,, 
  \hspace{30pt}\Longrightarrow\hspace{30pt}
  \tilde X^1 \sim \tilde X^1 + 2\pi \op \tilde n\,, 
  \hspace{50pt} n,\tilde n\in\mathbb Z \,. 
}

%%%%%%%%%%%%%%%%%%%%%%%%%%%%%%%%%%%%%%%%%%%%%%%
%%%%%%%%%%%%%%%%%%%%%%%%%%%%%%%%%%%%%%%%%%%%%%%

\subsubsection*{Dilaton}

The relations in \eqref{dual_back_9494} show how the dual metric and $B$-field  can be expressed 
in terms of the original background fields. The transformation of the dilaton is however not yet included. 
The transformation behaviour of the dilaton can be determined via a one-loop  path-integral computation, 
which we will not review here. We only quote the following result for the dual dilaton from 
\cite{Buscher:1987sk,Buscher:1987qj} as
\eq{
  \label{dual_back_9495}
  \check\phi = \phi - \frac14 \log \frac{\det G}{\det \check G} \,.
}
Note that this transformation leaves  the combination $e^{-2\phi}\op\sqrt{\det G}$ invariant.
At higher loops the relation \eqref{dual_back_9495} is modified, which has been 
discussed in \cite{Tseytlin:1991wr}.

%%%%%%%%%%%%%%%%%%%%%%%%%%%%%%%%%%%%%%%%%%%%%%%
%%%%%%%%%%%%%%%%%%%%%%%%%%%%%%%%%%%%%%%%%%%%%%%

\subsubsection*{Conformal symmetry}

String theory is a two-dimensional conformal field theory, which for the flat backgrounds with constant $B$-field and dilaton 
discussed in section~\ref{cha_t_duality} can  easily be verified. 
However, for curved backgrounds with non-constant $B$-field and dilaton conformality 
imposes restrictions on the background. In particular, the action \eqref{action_02e} is 
conformal (at linear order in $\alpha'$) if the following $\beta$-functionals vanish
\eq{
  \label{eom_beta}
  \arraycolsep2pt
  \hspace*{-15pt}
  \begin{array}{@{}lclcl@{}}
  0 &=& \beta^G_{\mu\nu} &=& \displaystyle \alpha' R_{\mu\nu} + 2\alpha' \nabla_{\mu}\nabla_{\nu} \phi
    -\frac{\alpha'}{4} H_{\mu\rho\sigma} H_{\nu}{}^{\rho\sigma} + \mathcal O(\alpha'^2)\,,
    \\[10pt]
  0 &=& \beta^B_{\mu\nu} &=& \displaystyle -\frac{\alpha'}{2} \nabla^{\rho} H_{\rho \mu\nu}
    +\alpha' \bigl(\nabla^{\rho}\phi\bigr) \op H_{\rho \mu\nu} + \mathcal O(\alpha'^2)\,,
    \\[10pt]
  0 &=& \beta^{\phi} &=& \displaystyle \frac{D-D^{\rm crit.}}{6}-\frac{\alpha'}{2}\op \nabla^{\rho}\nabla_{\rho}
    \phi + \alpha' \bigl( \nabla^{\rho}\phi\bigr)\bigl( \nabla_{\rho}\phi\bigr) -\frac{\alpha'}{24}
    H_{\rho\sigma\lambda}    H^{\rho\sigma\lambda}+ \mathcal O(\alpha'^2)\,,
  \end{array}
  \hspace*{-35pt}
}
where $\nabla_{\mu}$ denotes the covariant derivative with respect to the target-space metric $G_{\mu\nu}$,
and $D^{\rm crit.}$ is the critical dimension of the string ($D^{\rm crit.}=26$ for the bosonic string under consideration). 
Since T-duality should ``leave the physics invariant'',  equations \eqref{eom_beta} have to be 
invariant under the transformations \eqref{dual_back_9494} and \eqref{dual_back_9495}.
This can indeed be checked.

%%%%%%%%%%%%%%%%%%%%%%%%%%%%%%%%%%%%%%%%%%%%%%%
%%%%%%%%%%%%%%%%%%%%%%%%%%%%%%%%%%%%%%%%%%%%%%%

\subsubsection*{Global properties of $B$ and $v$}
\label{page_global_v}

In Buscher's procedure for performing T-duality transformations, we have assumed that
the world-sheet action has a global symmetry. This imposes restrictions on the 
background fields which we summarised in equation \eqref{gc_009}.
In particular, the Kalb-Ramond field has to satisfy 
$\mathcal L_k B = dv$ with $v$ a globally-defined one-form. 
However, since in general the two-form gauge field $B$ is not globally-defined also
$v$ is in general not globally-defined \cite{Hull:2006qs}.

To address this point, let us first note that mathematically the Kalb-Ramond field is a gerbe connection
and let us summarise some properties of  $B$
(see for instance \cite{Belov:2007qj,Howe:2016ggg,Grana:2008yw}). 
We consider open sets $U_{\mathsf a}\subset \mathcal M$ and we let $\{U_{\mathsf a}\}$ be a good open covering
of the compact space $\mathcal M$.\footnote{In 
this paragraph the subscripts $\mathsf a, \mathsf b, \mathsf c,\ldots$ label the open covers $U_{\mathsf a}$.}
On $n$-fold overlaps $U_{\mathsf a_1} \cap U_{\mathsf a_2} \cap \ldots \cap U_{\mathsf a_n}$
the $B$-field has the following properties:
\begin{itemize}

\item The field strength of the Kalb-Ramond field (in an open set $U_{\mathsf a}$) 
is given by $H_{\mathsf a} = d B_{\mathsf a}$. The field strength is closed, that is $dH_{\mathsf a} = 0$.

\item On the two-fold overlap of two covers $U_{\mathsf a}$ and $U_{\mathsf b}$,  with $\Lambda_{\mathsf{ab}}$ a one-form
the Kalb-Ramond field satisfies
\eq{
 \label{bb_001}
  B_{\mathsf a} = B_{\mathsf b} + d \Lambda_{\mathsf {ab}} \,.
}

\item On  three-fold overlaps we use \eqref{bb_001} to derive that
$ B_{\mathsf a} = B_{\mathsf b} + d \Lambda_{\mathsf {ab}} = B_{\mathsf c} + d\Lambda_{\mathsf {bc}} + d\Lambda_{\mathsf {ab}}=B_{\mathsf a} + d\Lambda_{\mathsf {ca}}+ d\Lambda_{\mathsf {bc}} + d\Lambda_{\mathsf {ab}}$.
Since locally every closed form is exact, 
on three-fold overlaps
the one-forms $\Lambda_{\mathsf {ab}}$  satisfy with 
$\lambda_{\mathsf {abc}}$ a zero-form
\eq{
  \Lambda_{\mathsf {ab}} + \Lambda_{\mathsf {bc}} + \Lambda_{\mathsf {ca}} = d \lambda_{\mathsf {abc}} \,.
}

\item Similarly, on four-fold overlaps the functions $\lambda_{\mathsf {abc}}$ are required to satisfy
\eq{
  \lambda_{\mathsf {bcd}} - \lambda_{\mathsf {acd}} + \lambda_{\mathsf {abd}}
  - \lambda_{\mathsf {abc}} = n_{\mathsf {abcd}} \,,
}
where $n_{\mathsf {abcd}}$ are constants. If furthermore $n_{\mathsf {abcd}}\in2\pi\mathbb Z$,
then the field strength $H$ is quantised and $H\in H^3(\mathcal M,\mathbb Z)$.

\end{itemize}

Let us now turn to the one-form $v$ defined via the second relation in \eqref{gc_009}. On two-fold overlaps
we infer from \eqref{bb_001} that 
$dv_{\mathsf a} = dv_{\mathsf b} + d\iota_k d\Lambda_{\mathsf {ab}}$,
which  can be solved by 
$v_{\mathsf a} = v_{\mathsf b} + \iota_k d\Lambda_{\mathsf {ab}} + d \omega_{\mathsf {ab}}$.
Here $\omega_{\mathsf {ab}}$ are functions on two-fold overlaps which
satisfy $\omega_{\mathsf {ab}}+ \omega_{\mathsf {bc}}+ \omega_{\mathsf {ca}}=  \mathrm{const.}$
on three-fold overlaps.
Choosing then for convenience $\omega_{\mathsf {ab}} = \iota_k \Lambda_{\mathsf {ab}}$, we arrive at
\eq{
  v_{\mathsf a} = v_{\mathsf b} + \mathcal L_k \Lambda_{\mathsf {ab}}  \,,
  \hspace{50pt}
  \mathcal L_k (\Lambda_{\mathsf {ab}} + \Lambda_{\mathsf {bc}} + \Lambda_{\mathsf {ca}} )  = 0 \,,
}
on two- and three-fold overlaps, respectively.
Now, if the background admits an open covering such that on two-fold overlaps 
the one-forms $\Lambda_{\mathsf {ab}}$ appearing in \eqref{bb_001}
satisfy
\eq{
  \label{globa_rest_008}
  \mathcal L_k \Lambda_{\mathsf {ab}} = 0 
  \hspace{40pt}\longrightarrow\hspace{40pt}
  v_{\mathsf a} = v_{\mathsf b} \,,
}
then the one-form $v$ can be made globally-defined. 
This is the situation to which we have restricted our analysis in this section. 
A more detailed  discussion of global properties of the $B$-field in relation to 
T-duality can be found in \cite{Hull:2006qs,Belov:2007qj}.

%%%%%%%%%%%%%%%%%%%%%%%%%%%%%%%%%%%%%%%%%%%%%%%
%%%%%%%%%%%%%%%%%%%%%%%%%%%%%%%%%%%%%%%%%%%%%%%

\subsubsection*{Remark}

We close this section with the following remarks.
\begin{itemize}

\item T-duality transformations can also be viewed as canonical transformations on the world-sheet. 
This approach has been 
discussed for instance in the papers \cite{Alvarez:1994wj,Lozano:1995jx}.

\item T-duality for open strings via Buscher's procedure has been 
studied in \cite{Alvarez:1996up,Dorn:1996an,Forste:1996hy} and more 
recently in \cite{Albertsson:2004gr,Cordonier-Tello:2018zdw},
and via canonical transformations in
\cite{Borlaf:1996na,Lozano:1996sc}.
From an effective field theory point this has been investigated for instance 
in \cite{Tseytlin:1996it,Bergshoeff:1996cy,Green:1996bh}.

\item In order to perform a T-duality transformation via Buscher's formalism, 
a (compact) direction of isometry is needed. This includes in particular the case of an angular 
isometry. 
To illustrate this point, let us consider the two-dimensional plane in polar coordinates
with vanishing $B$-field and constant dilaton \cite{Rocek:1991ps}. The background takes the form
\eq{
 \label{radial_001}
 ds^2 = dr^2 + r^2\op d\theta^2 \,,
 \hspace{40pt}
 B = 0\,,
 \hspace{40pt}
 \phi= \phi_0 \,,
}
with $r\in[0,\infty)$ and $\theta\in[0,2\pi)$. The vector $k = \partial_{\theta}$ is a Killing vector of the metric 
along which we can T-dualise, and using the Buscher rules \eqref{dual_back_9494} together with 
\eqref{dual_back_9495} we find for the dual model
\eq{
 \label{radial_002}
 \check{ds}^2 = dr^2 + \frac{\alpha'^2}{r^2}\op d\theta^2 \,,
 \hspace{40pt}
 B = 0\,,
 \hspace{40pt}
 \phi= \phi_0 -\log \frac{r}{\sqrt{\alpha'}} \,.
}
From here we see that at the dual metric is singular at the origin. 
However, since \eqref{radial_002} is related to \eqref{radial_001} by a duality transformation,
the dual background is expected to be well-defined. 

We also mention that T-duality along angular isometries does not necessarily lead to singular 
dual geometries, which has been exemplified in \cite{Plauschinn:2017ism} for the 
NS5-brane solution.

\end{itemize}

%%%%%%%%%%%%%%%%%%%%%%%%%%%%%%%%%%%%%%%%%%%%%%%
%%%%%%%%%%%%%%%%%%%%%%%%%%%%%%%%%%%%%%%%%%%%%%%
%%%%%%%%%%%%%%%%%%%%%%%%%%%%%%%%%%%%%%%%%%%%%%%
%%%%%%%%%%%%%%%%%%%%%%%%%%%%%%%%%%%%%%%%%%%%%%%
%%%%%%%%%%%%%%%%%%%%%%%%%%%%%%%%%%%%%%%%%%%%%%%
%%%%%%%%%%%%%%%%%%%%%%%%%%%%%%%%%%%%%%%%%%%%%%%
%%%%%%%%%%%%%%%%%%%%%%%%%%%%%%%%%%%%%%%%%%%%%%%
%%%%%%%%%%%%%%%%%%%%%%%%%%%%%%%%%%%%%%%%%%%%%%%

\subsection{Collective T-duality}
\label{sec_wzw_action}

We now want to generalise the discussion of the previous section to 
gauging multiple isometries and  performing a collective 
T-duality transformation. 
In this section we employ a Wess-Zumino-(Novikov-)Witten (WZW) formulation \cite{Wess:1971yu,Novikov:1981,Witten:1983ar}
of the world-sheet action, in which the field strength $H$ instead of the Kalb-Ramond field $B$  appears.
The reason is that 
\begin{enumerate}

\item this approach avoids subtleties concerning the gauge choice for the Kalb-Ramond field $B$
and the global restrictions shown in \eqref{globa_rest_008}, and 

\item from the point of view of the
$\beta$-functionals \eqref{eom_beta} the field strength $H$ is the relevant quantity and not
the gauge potential $B$.

\end{enumerate}

%%%%%%%%%%%%%%%%%%%%%%%%%%%%%%%%%%%%%%%%%%%%%%%
%%%%%%%%%%%%%%%%%%%%%%%%%%%%%%%%%%%%%%%%%%%%%%%

\subsubsection*{World-sheet action}

The sigma-model action for the closed string 
is usually defined on a compact two-dimensional manifold without boundaries.
However, in order to incorporate non-trivial field strengths $H\neq 0$ 
for the Kalb-Ramond field, it turns out to be convenient to work with 
a Wess-Zumino term which is defined on a compact  three-dimensional Euclidean world-sheet $\Xi$ 
with two-dimensional boundary $\partial \Xi = \Sigma$.
In this case, the action (restricted to the compact target-space manifold $\mathcal M$ in the splitting \eqref{tdual_split_tst})
takes the form
\eq{
  \label{action_01f}
  \arraycolsep2pt
  \begin{array}{rll}
  \displaystyle \mathcal S = -\frac{1}{2\pi \alpha'} & \displaystyle \int_{\Sigma} 
  \Bigl[  & \displaystyle \tfrac{1}{2}\op G_{ij} \, d X^i\wedge\star d X^j
  + \tfrac{\alpha'}{2}\op \mathsf R\, \phi \star 1 \Bigr] \\[5mm]
  \displaystyle -\frac{i}{2\pi \alpha'} & \displaystyle \int_{\Xi}& 
  \displaystyle \tfrac{1}{3!}\, H_{ijk}\op dX^i\wedge dX^j\wedge dX^k
  \,,
  \end{array}
}
where the Hodge star-operator $\star$ is defined on  $\Sigma$,  
and the differential is understood as $dX^i(\sigma^{\mathsf a}) = \partial_{\mathsf a} X^i d\sigma^{\mathsf a}$ 
with $\sigma^{\mathsf a}$ coordinates on $\Sigma$ or on $\Xi$, depending on the context.
The indices take values $i,j,k= 1,\ldots, D$ with $D$ the dimension of the compact target space $\mathcal M$,
and  $\mathsf R$ denotes the curvature scalar corresponding to the world-sheet metric $h_{\alpha\beta}$
on $\Sigma$.

Note that the choice of three-manifold $\Xi$ for a given boundary $\Sigma=\partial\Xi$ is
not unique. However, if the field strength $H$ is quantised, the path integral only depends on
the data of the two-dimensional theory \cite{Witten:1983tw}. In the above conventions, the quantisation 
condition reads
\eq{
  \label{quantisation}
  \frac{1}{2\pi\alpha'} \int_{\Xi} H \;\in\; 2\pi \op\mathbb Z \,.
}

%%%%%%%%%%%%%%%%%%%%%%%%%%%%%%%%%%%%%%%%%%%%%%%
%%%%%%%%%%%%%%%%%%%%%%%%%%%%%%%%%%%%%%%%%%%%%%%

\subsubsection*{Global symmetry}

As before, we require that the compact target-space manifold
contains at least one circle. More precisely, we assume that the world-sheet action \eqref{action_01f}
is invariant under global transformations of the form
\eq{
  \label{iso_trafo_01}
  \delta_{\epsilon} X^i = \epsilon^{\alpha}\op k_{\alpha}^i(X) 
}
for $\epsilon^{\alpha}$ constant and $\alpha=1,\ldots, N$. This is indeed the case, if the following three conditions are
satisfied \cite{Hull:1989jk,Hull:1990ms}
\eq{
  \label{constraints_35}
  \mathcal L_{k_{\alpha}}  G = 0\,, \hspace{70pt}
  d\bigl( \iota_{k_{\alpha}} H \bigr) = 0 \,, \hspace{70pt}
  \mathcal L_{k_{\alpha}} \phi =0\,,
}
where we used the Bianchi identity $dH=0$.
The  isometry algebra generated by the Killing vectors  is in general  non-abelian
with structure constants $f_{\alpha\beta}{}^{\gamma}$, which is encoded in the Lie bracket
\eq{
   \label{constraints_351}
   \bigl[ k_{\alpha} , k_{\beta} \bigr] = f_{\alpha\beta}{}^{\gamma} \op k_{\gamma} \,.
}

%%%%%%%%%%%%%%%%%%%%%%%%%%%%%%%%%%%%%%%%%%%%%%%
%%%%%%%%%%%%%%%%%%%%%%%%%%%%%%%%%%%%%%%%%%%%%%%

\subsubsection*{Local symmetries}

Let us now promote the global symmetries \eqref{iso_trafo_01} to  local
ones, with $\epsilon^{\alpha}$ depending on the world-sheet coordinates $\sigma^{\mathsf a}$. 
To do so, we introduce world-sheet gauge fields $A^{\alpha}$ and 
as well as Lagrange multipliers $\chi_{\alpha}$, and we solve the second relation 
in \eqref{constraints_35} as
\eq{
  \label{variantions_2084}  
   \iota_{k_{\alpha}} H = d v_{\alpha} \,,
}
where $v_{\alpha}$ are one-forms (defined up to closed terms). Note that  the $v_{\alpha}$ are in general
\emph{not} globally-defined, however, we require the combination 
$v_{\alpha} + d\chi_{\alpha}$ to be a globally-defined one-form in $\Xi$
\cite{Hull:2006qs}.
We come back to this point on page~\pageref{page_global_v2}. 
The resulting gauged action  reads
\eq{
  \label{action_02}
  \hat{\mathcal S} =&-\frac{1}{2\pi\alpha'} \int_{\Sigma} \hspace{8pt}  
  \tfrac{1}{2}\, G_{ij}  (dX^i + k^i_{\alpha} A^{\alpha})\wedge\star(dX^j + k^j_{\beta} A^{\beta})  
  \\[1mm]
  &-\frac{i}{2\pi \alpha'} \int_{\Xi} \hspace{8pt} \tfrac{1}{3!}\, H_{ijk}\op dX^i\wedge dX^j\wedge dX^k
  \\[1mm]
  &-\frac{i}{2\pi \alpha'} \int_{\Sigma} \:\Bigl[ \;
  ( v_{\alpha} + d\chi_{\alpha})\wedge A^{\alpha}
  + \tfrac{1}{2}\op \bigl( \iota_{k_{[\ul \alpha}} v_{\ul \beta]}
  + f_{\alpha\beta}{}^{\gamma} \chi_{\gamma} \bigr)\op A^{\alpha}\wedge A^{\beta}\;
  \Bigr] \,,
}
where we again omitted the dilaton term which does not get modified. The local symmetry 
transformations take the following form
\eq{
  \label{variantions_01}
  \arraycolsep2pt
  \begin{array}{lcl}
  \displaystyle \hat\delta_{\epsilon} X^i &=& \displaystyle \epsilon^{\alpha} k_{\alpha}^i \,,
  \\[6pt]
  \displaystyle \hat\delta_{\epsilon} A^{\alpha} &=& \displaystyle - d\epsilon^{\alpha} - \epsilon^{\beta} A^{\gamma} 
  f_{\beta\gamma}{}^{\alpha} \,,
  \\[6pt]
  \displaystyle \hat\delta_{\epsilon}\chi_{\alpha} &=& \displaystyle 
  - \iota_{k_{(\ov \alpha}} v_{\ov \beta)} \epsilon^{\beta}
  - f_{\alpha\beta}{}^{\gamma}   \epsilon^{\beta} \chi_{\gamma}   \,.
  \end{array}
}
For the abelian case, this realisation appeared in \cite{Alvarez:1993qi,Hull:2006qs},
but here we include the generalisation to the non-abelian case \cite{Plauschinn:2014nha}. 
The action \eqref{action_02} is invariant under \eqref{variantions_01} if the following additional 
restrictions are met
\eq{
\label{variations_45}
\mathcal L_{k_{[\ul \alpha}} v_{\ul \beta]} = f_{\alpha\beta}{}^{\gamma} v_{\gamma} \,,
\hspace{60pt}
\iota_{k_{[\ul \alpha}} \op f_{\ul \beta\ul \gamma ]}{}^{\delta} v_{\delta} = \frac{1}{3} \,
\iota_{k_{\alpha}}\iota_{k_{\beta}}\iota_{k_{\gamma}} H \,.
}
Note that in the literature sometimes the stronger condition $ \iota_{k_{(\ov \alpha}} v_{\ov \beta)}=0$ is required, 
which results in the second relation in \eqref{variations_45} being automatically 
satisfied and, in the case of an abelian symmetry, the Lagrange multipliers not
transforming. 
Turning to the variation of the action \eqref{action_02} with respect to \eqref{variantions_01}, we find
\eq{
  \label{variation_34}
  \hat\delta_{\epsilon} 
  \hat{\mathcal S} =-\frac{i}{2\pi\alpha'} \int_{\Sigma}  d\epsilon^{\alpha}
  \wedge (v_{\alpha}+d\chi_{\alpha})
  -\frac{i}{2\pi \alpha'} \int_{\Xi}  d\epsilon^{\alpha} \wedge dv_{\alpha} \,.
}
Recalling that the combination 
$v_{\alpha}+d\chi_{\alpha}$ is required to be globally-defined on $\Xi$, using 
Stoke's theorem for the first term  we see that the variation \eqref{variation_34} 
vanishes, that is
\raisebox{0pt}[0pt]{$\hat\delta_{\epsilon} \hat{\mathcal S} = 0$},
and hence \eqref{variantions_01} are symmetries of the gauged action \eqref{action_02}.

%%%%%%%%%%%%%%%%%%%%%%%%%%%%%%%%%%%%%%%%%%%%%%%
%%%%%%%%%%%%%%%%%%%%%%%%%%%%%%%%%%%%%%%%%%%%%%%

\subsubsection*{Simplifying assumptions}

In order to proceed, we again make some simplifying assumptions. More concretely:
\begin{itemize}
\label{reqs_7395473974}

\item For most of the formulas in this section we allow for non-abelian isometry algebras with 
non-vanishing structure constants $f_{\alpha\beta}{}^{\gamma}$ (as defined in \eqref{constraints_351}),
however, eventually we have to restrict to an abelian algebra with 
$f_{\alpha\beta}{}^{\gamma}=0$.
An approach to non-abelian T-duality will be discussed in section~\ref{cha_poisson_lie}.

\item We make a choice of coordinates such that $k^m_{\alpha}=0$ for $m=N+1,\ldots, D$.
Since the Killing vectors are required to be linearly independent, this implies that 
the matrix $(k_{\alpha})^{\beta}$ is invertible. 

\item We assume that the symmetric $N\times N$ matrix $ \iota_{(k_{\ov\alpha}} v_{\ov\beta)}$ is constant. 

\end{itemize}

%%%%%%%%%%%%%%%%%%%%%%%%%%%%%%%%%%%%%%%%%%%%%%%
%%%%%%%%%%%%%%%%%%%%%%%%%%%%%%%%%%%%%%%%%%%%%%%

\subsubsection*{Back to the ungauged  action}

In order to recover the original action from the gauged one, we integrate out the
Lagrange multipliers $\chi_{\alpha}$. The  equation of motion for $\chi_{\alpha}$
is obtained by varying the action \eqref{action_02} with respect to $\chi_{\alpha}$, and we find
\eq{
  \delta_{\chi} \hat{\mathcal S} =+\frac{i}{2\pi\alpha'} \int_{\Sigma} 
  \delta\chi_{\alpha} \op \Bigl( dA^{\alpha} - \tfrac{1}{2}\op f_{\beta\gamma}{}^{\alpha} A^{\beta}\wedge A^{\gamma} \Bigr)\,,
}
from which  we can  read off the equations of motion  as 
\eq{
  \label{eom_01}
  0 = d A^{\alpha} -\tfrac{1}{2}\op f_{\beta\gamma}{}^{\alpha} A^{\beta}\wedge A^{\gamma}  \,.
}
Since we have restricted our discussion to abelian isometries, the structure constants $f_{\alpha\beta}{}^{\gamma}$ vanish and we effectively arrive at the situation discussed in the previous section on
page~\pageref{page_back_single_1}. 
In particular, the equations of motion of each Lagrange multiplier $\chi_{\alpha}$ 
restricts each $A^{\alpha}$ to be pure gauge, 
which can then be set to zero using the local symmetries \eqref{variantions_01}.
In this way the ungauged action \eqref{action_01f} is recovered 
from the gauged one \eqref{action_02}.

For the  non-abelian case a vanishing field strength $F^{\alpha}$ means that 
the gauge fields $A^{\alpha}$ are in general not closed. The Hodge decomposition of 
$A^{\alpha}$ then contains coexact terms which makes the analysis more involved. 
We have therefore restricted ourselves to abelian isometries.

%%%%%%%%%%%%%%%%%%%%%%%%%%%%%%%%%%%%%%%%%%%%%%%
%%%%%%%%%%%%%%%%%%%%%%%%%%%%%%%%%%%%%%%%%%%%%%%
%%%%%%%%%%%%%%%%%%%%%%%%%%%%%%%%%%%%%%%%%%%%%%%
%%%%%%%%%%%%%%%%%%%%%%%%%%%%%%%%%%%%%%%%%%%%%%%

\subsubsection*{Dual action}

Turning now to the dual action, we integrate out the gauge fields $A^{\alpha}$ from
the gauged action \eqref{action_02}. Due to the absence of a kinetic term for 
the gauge fields, the equations of motion are algebraic and can be expressed in 
the following way
\eq{
  \label{dual_rick_d_s}
   0 = \mathcal G_{\alpha\beta} \star A^{\beta} + i\op {\mathcal D}_{\alpha\beta} A^{\beta}
   + \star \mathsf k_{\alpha} + i \op  \xi_{\alpha} \,,
}
where we used the definitions
\eq{
  \label{dual_rick_yeah}
  \arraycolsep2pt
  \begin{array}{lclclcl}
  \mathcal G_{\alpha\beta} &=& k_{\alpha}^i G_{ij} k^j_{\beta} \,, &\hspace{60pt} &
  \xi_{\alpha} &=& d\chi_{\alpha} + v_{\alpha} \,,
  \\[4mm]
  \mathcal D_{\alpha\beta} &=&  \iota_{k_{[\ul \alpha}} v_{\ul \beta]} + f_{\alpha\beta}{}^{\gamma} 
  \chi_{\gamma}\,,
  &&
  \mathsf k_{\alpha} & = & k^i_{\alpha} G_{ij} \op dX^j \,.
  \end{array}
}
The relation \eqref{dual_rick_d_s} is sufficient to eliminate the gauge field $A^{\alpha}$ 
from  \eqref{action_02}, and the action with $A^{\alpha}$ integrated out 
 takes the general form
\eq{
  \label{action_05}
  \check{\mathcal S} = -\frac{1}{2\pi \alpha'} \int_{\Sigma} 
  \left( \check G
  + \frac{\alpha'}{2}\op \mathsf R\, \check \phi \star 1 \right)
  -\frac{i}{2\pi \alpha'} \int_{\Xi} \check H
  \,,
}
where we defined world-sheet quantities
\eq{
\label{dual_2344}
\arraycolsep2pt
\begin{array}{lclrrlc}
 \check{G}&=&\displaystyle G-  \frac{1}{2}
   &\displaystyle (\mathsf k+{\xi})^{T}\, \bigl(\mathcal{G}+{\mathcal{D}}\bigr)^{-1}\,\wedge
   &\displaystyle \star (\mathsf k-{\xi}) &&,
 \\[14pt]
 \check{H}&=&\displaystyle H- \frac{1}{2}\, d \op\Bigl[
   &\displaystyle(\mathsf k+{\xi})^{T}\,\bigl(\mathcal{G}+{\mathcal{D}}\bigr)^{-1}\,\wedge
   & (\mathsf k-{\xi}) & \Bigr] &.
\end{array}
}
Here, matrix multiplication in the indices $\alpha,\beta,\ldots$ is understood and we use 
the convention $G = \frac12\op G_{ij} \op  dX^{i}\wedge\star dX^{j}$
and $H = \frac{1}{3!}\op H_{ijk} \op dX^{i}\wedge dX^{j}\wedge dX^k$.

%%%%%%%%%%%%%%%%%%%%%%%%%%%%%%%%%%%%%%%%%%%%%%%
%%%%%%%%%%%%%%%%%%%%%%%%%%%%%%%%%%%%%%%%%%%%%%%

\subsubsection*{Dual background}

From a target-space perspective, the fields in \eqref{dual_2344} depend on
the $D$ original one-forms $dX^i$ as well as on the $N$ one-forms $d\chi_{\alpha}$. 
The components of $\check G$ can hence be
interpreted as a ``metric'' on an  enlarged $(D+N)$-dimensional tangent-space locally spanned 
by $\{ dX^i,d\chi_{\alpha}\}$, and $\check H$ can be interpreted as a corresponding field strength \cite{Plauschinn:2014nha}.
However, the symmetric matrix $\check G$ has $N$ eigenvectors $\check n_{\alpha}$ with vanishing eigenvalue
and similarly the contraction of $\check H$ with $\check n_{\alpha}$ vanishes.
Introducing a basis $dX^I = \{dX^i,d\chi_{\alpha}\}$ with $I=1,\ldots, D+N$, 
we can indeed verify that
\eq{
  \check G_{IJ} \op \check n_{\alpha}^J  =0\,,
  \hspace{60pt}
  \check H_{IJK} \op \check n_{\alpha}^K  =0\,,
}
where the $N$ vectors  are given by
\eq{
  \check n_{\alpha} = k_{\alpha}^i \op\frac{\partial}{\partial X^i} + \bigl( {\mathcal D}_{\alpha\beta}- \iota_{k_{\alpha}}  v_{\beta} \bigr)
  \frac{\partial}{\partial\chi_{\beta}}\,.
}
This means that $\check G$ and $\check H$ are non-vanishing only on a $D$-dimensional subspace of the enlarged
$(D+N)$-dimensional tangent-space. In order to make this explicit, we perform a change of basis in the following way:
\begin{itemize}

\item For the symmetric matrix $\check G_{IJ}$ we define an invertible 
matrix $\mathcal T$ and perform the transformation
\eq{
  \label{coc_01}
  \check{\mathsf G}_{IJ} =  \bigl( \mathcal T^T \check G \,\mathcal T 
  \bigr)_{IJ}\,,
  \hspace{60pt}
  \mathcal T^I{}_J=
  \left( 
  \scalebox{0.8}{$\displaystyle
  \renewcommand{\arraystretch}{1.35}
  \arraycolsep10pt
  \dashlinedash2pt
  \dashlinegap4pt  
  \begin{array}{c@{\hspace{2pt}}c@{\hspace{2pt}}c:c|c}
  \check n^1_1 &\cdots & \check n^1_N&  &  \\ 
  \vdots && \vdots &  0 & 0 \\
  \check n^N_1 &\cdots & \check n^N_N& & \\ \hdashline
  \vdots & & \vdots &  \multirow{2}{*}{$\mathds 1$} & \multirow{2}{*}{$0$}  \\[-2pt]
  \check n_1^{D} &\cdots & \check n_N^{D}&  &  \\ \hline
  \vdots & & \vdots & \multirow{2}{*}{$0$} &\multirow{2}{*}{$\mathds 1$}  \\[-2pt]
  \check n_1^{D+N} &\cdots & \check n_N^{D+N}&  &
  \end{array}
  $}
  \right).
}

In the transformed matrix $\check{\mathsf G}_{IJ}$ 
all entries along the $I,J=1,\ldots, N$ directions vanish, and 
we therefore arrive at
the  expression
\eq{
  \label{coc_02}
  \check{\mathsf G}_{IJ} =
  \renewcommand{\arraystretch}{1.8}
  \arraycolsep6pt
  \dashlinedash2pt
  \dashlinegap4pt  
  \left( \begin{array}{c:c|c}
  0 & 0 & 0 \\ \hdashline
  0 &  \check{\mathsf G}_{mn} & \check {\mathsf G}_{m}{}^{\beta} \\ \hline
  0 & \check {\mathsf G}^{\alpha}{}_{n}  & \check {\mathsf G}^{\alpha\beta}
  \end{array}
  \right) ,
}
where $m,n = N+1, \ldots, D+N$ and $\alpha,\beta = 1,\ldots, N$.
The non-vanishing $D\times D$ block-matrix in \eqref{coc_02}
then corresponds to the dual metric, which takes the explicit form 
\eq{
  \label{dual_g_007}
  &  \arraycolsep2pt\check{\mathsf G}_{mn} = \begin{array}[t]{lcll@{\hspace{2pt}}c@{\hspace{2pt}}ll}
  G_{mn} 
  & - & \mathsf k_{\alpha m} 
  & \displaystyle \bigl[ (\mathcal G + \mathcal D)^{-1}& \displaystyle \mathcal G &\displaystyle (\mathcal G - 
  \mathcal D)^{-1} \bigr]^{\alpha\beta} & \mathsf k_{\beta n}
  \\[2pt]
  & - & \mathsf k_{\alpha m} 
  & \displaystyle \bigl[ (\mathcal G + \mathcal D)^{-1}& \displaystyle \mathcal D &\displaystyle (\mathcal G - 
  \mathcal D)^{-1} \bigr]^{\alpha\beta} &  v_{\beta n}
  \\[2pt]
  & + &  v_{\alpha m} 
  & \displaystyle \bigl[ (\mathcal G + \mathcal D)^{-1}& \displaystyle \mathcal D &\displaystyle (\mathcal G - 
  \mathcal D)^{-1} \bigr]^{\alpha\beta} & \mathsf k_{\beta n}
  \\[2pt]  
  & + &  v_{\alpha m} 
  & \displaystyle \bigl[ (\mathcal G + \mathcal D)^{-1}& \displaystyle \mathcal G &\displaystyle (\mathcal G - 
  \mathcal D)^{-1} \bigr]^{\alpha\beta} &  v_{\beta n}  \,,
  \end{array}
  \\[10pt]
  &  \arraycolsep2pt\check{\mathsf G}^{\alpha}{}_n = \begin{array}[t]{cl@{\hspace{2pt}}c@{\hspace{2pt}}ll}
  +& \displaystyle \bigl[ (\mathcal G + \mathcal D)^{-1}& \displaystyle \mathcal D &\displaystyle (\mathcal G - 
  \mathcal D)^{-1} \bigr]^{\alpha\beta} & \mathsf k_{\beta n}
  \\[2pt]
  +& \displaystyle \bigl[ (\mathcal G + \mathcal D)^{-1}& \displaystyle \mathcal G &\displaystyle (\mathcal G - 
  \mathcal D)^{-1} \bigr]^{\alpha\beta} &  v_{\beta n} \,,
  \end{array}
  \\[10pt]
  &\arraycolsep2pt \check{\mathsf G}^{\alpha\beta} =  \begin{array}[t]{cl@{\hspace{2pt}}c@{\hspace{2pt}}l}
  +&\displaystyle \bigl[ (\mathcal G + \mathcal D)^{-1}& \displaystyle \mathcal G &\displaystyle (\mathcal G - 
  \mathcal D)^{-1} \bigr]^{\alpha\beta} \,.
  \end{array}
}

\item For the field strength a similar analysis applies. Using the matrix $\mathcal T$ defined in 
\eqref{coc_01} we determine 
\eq{
  \label{dual_b_007}
  \check{\mathsf H}_{IJK} = \check H_{LMN} \mathcal T^L{}_I\mathcal T^M{}_J\mathcal T^N{}_K\,,
}
for which we find that the components along the directions $X^{\alpha}$ vanish
\eq{
  \check {\mathsf H}_{\alpha JK}  = 0 \,.
}
Due to the derivative appearing for the dual field strength the explicit expressions 
for the components of $\check H$ are more involved and will  not be presented 
here. However, below we discuss the case of a T-duality along a single direction 
for which we give explicit formulas.

\item The components of the dual metric $\check G$ and the dual field strength $\check H$ 
may still depend on the coordinates $X^{\alpha}$ along which the duality 
transformation has been performed. However, using the local symmetries
\eqref{variantions_01} we can set this dependence to zero (in the abelian case).

\item We also determine the transformed basis on the enlarged tangent-space, which 
is given by $e^I = (\mathcal T^{-1})^I{}_J \, dX^J $. In general, these one-forms take the form
\eq{
  \label{basis_11}
  \arraycolsep2pt
  \begin{array}{lcl}
  \displaystyle e^{\alpha} &=& \displaystyle  \bigl( k^{-1}\bigr)^{\alpha}{}_{\beta} \,dX^{\beta} \,,
  \\[8pt]
  \displaystyle e^m &=& \displaystyle  dX^m - k^m_{\alpha} \op 
  \bigl( k^{-1}\bigr)^{\alpha}{}_{\beta} \,dX^{\beta} \,,
  \\[4pt]
  \displaystyle e_{\alpha} &=& \displaystyle  d\chi_{\alpha} + \bigl( \iota_{(k_{\ov\alpha}} v_{\ov\beta)} + f_{\alpha\beta}{}^{\gamma}
  \chi_{\gamma} \bigr) \op  \bigl( k^{-1}\bigr)^{\beta}{}_{\gamma} \, dX^{\gamma}\,.
  \end{array}
}
The exterior algebra of the dual one-forms $\{e^m,e_{\alpha}\}$ does not close
among itself. 
However, with the assumptions made on page~\pageref{reqs_7395473974} we see that these forms 
are closed, that is
\eq{
  de^{\alpha} = 0 \,, \hspace{50pt}
  de^m = 0 \,, \hspace{50pt} d e_{\alpha} = 0 \,.
}

\end{itemize}
We finally remark that the dual dilaton $\check \phi$ is determined by the same expression
as given already in equation \eqref{dual_back_9495}.

%%%%%%%%%%%%%%%%%%%%%%%%%%%%%%%%%%%%%%%%%%%%%%%
%%%%%%%%%%%%%%%%%%%%%%%%%%%%%%%%%%%%%%%%%%%%%%%

\subsubsection*{Example I -- single T-duality}

We now want to illustrate the above formalism for the case of T-duality 
along a single direction. We assume that we can use adapted coordinates in which 
the Killing vector takes the form 
\eq{
  k = \partial_{1} \,,
}
and we note that due to having only one isometry the conditions \eqref{variations_45} 
are trivially satisfied. 
For $\check G$ given in \eqref{dual_2344} we compute
\eq{
 \check{G}=G-  \frac{1}{2\, G_{11}}\op \Bigl[\, \mathsf k\wedge\star \mathsf k - (d\chi+v)\wedge\star(d\chi+v) \,\Bigr]\,,
}
and performing the change of basis \eqref{coc_01} gives
\eq{
 &\check{G}=
   \frac{1}{2\op G_{11}} \op e_1 \wedge\star e_1
  + \frac{v_n}{G_{11}}  \op e_1 \wedge \star e^n
  \\
 &\hspace{120pt}
 +  \frac{1}{2}  \left( G_{mn} - \frac{G_{1m}G_{1n} - v_m v_n}{G_{11}} \right) 
    e^m  \wedge \star  e^n \,,
}
where $e_1 = d\chi+ v_1 \op dX^1$ and $e^m = dX^m$. Note that $v_1$ is assumed to be constant
and can therefore be absorbed into the definition of $\chi$. 
For the field strength we determine
\eq{
  &\check H = \left[ \frac{1}{6} \op H_{mnk} -\frac{1}{2}\op\frac{G_{1[\ul m} \op H_{1|\ul n\ul k]} }{G_{11}}
  + \partial_{[\ul m} \left( \frac{G_{1|\ul n}}{G_{11}} \right) v_{\ul k]}
  \right] e^m\wedge e^n\wedge e^k
  \\
  &\hspace{120pt}
  + \partial_{[\ul m} \left( \frac{G_{1|\ul n]}}{G_{11}} \right)  e_1 \wedge e^m\wedge e^n \,.
}

%%%%%%%%%%%%%%%%%%%%%%%%%%%%%%%%%%%%%%%%%%%%%%%
%%%%%%%%%%%%%%%%%%%%%%%%%%%%%%%%%%%%%%%%%%%%%%%

\subsubsection*{Global properties of $H$ and $v_{\alpha}$}
\label{page_global_v2}

Similarly as on page~\pageref{page_global_v}, let us also discuss the global properties of 
$H$ and $v_{\alpha}$. Since $H$ is a globally-defined three-form, on the overlap of two open 
covers $U_{\mathsf a}$ and $U_{\mathsf b}$ 
the field strength satisfies $H_{\mathsf a} = H_{\mathsf b}$ with $H_{\mathsf a} \equiv H\rvert_{U_{\mathsf a}}$. 
Assuming the Killing vectors $k_{\alpha}$ to  be globally-defined, we infer from \eqref{variantions_2084} that 
on two-fold overlaps we have
\eq{
  \bigl( \iota_{k_{\alpha}} H \bigr)_{\mathsf a} =  dv_{\alpha|\mathsf a}
 = dv_{\alpha|\mathsf b}
  =   \bigl( \iota_{k_{\alpha}} H \bigr)_{\mathsf b} \,,
}
which can be solved by
\eq{
  \label{global_99928}
  v_{\alpha|\mathsf a}  = v_{\alpha|\mathsf b} + d\omega_{\alpha|\mathsf{ab}} \,.
}
Here, $\omega_{\alpha|\mathsf {ab}}$ are functions on two-fold overlaps which
satisfy on three-fold overlaps $\omega_{\alpha|\mathsf {ab}}+ \omega_{\alpha|\mathsf {bc}}+ \omega_{\alpha|\mathsf {ca}}=  \mathrm{const.}$ Equation \eqref{global_99928} implies that the one-forms $v_{\alpha}$ are
in general \emph{not} globally-defined. 
However, if we require the Lagrange multipliers $\chi_{\alpha}$ introduced in the gauged action 
\eqref{action_02} to satisfy on two-fold overlaps \cite{Hull:2006qs}
\eq{
  \chi_{\alpha|\mathsf a} = \chi_{\alpha|\mathsf b} - \omega_{\alpha|\mathsf {ab}} \,,
}
it follows that $v_{\alpha|\mathsf a} + d\chi_{\alpha|\mathsf a} = v_{\alpha|\mathsf b} + d\chi_{\alpha|\mathsf b} $
and hence the combinations $v_{\alpha}+ d\chi_{\alpha}$ are globally-defined one-forms
as required above.
For more details on the global properties of $H$ and $v_{\alpha}$ we again refer the reader
to \cite{Hull:2006qs,Belov:2007qj}.

%%%%%%%%%%%%%%%%%%%%%%%%%%%%%%%%%%%%%%%%%%%%%%%
%%%%%%%%%%%%%%%%%%%%%%%%%%%%%%%%%%%%%%%%%%%%%%%

\subsubsection*{Example II -- $SU(2)$ WZW model}

As another example for the formalism described above, let us discuss T-duality for 
the $SU(2)$ WZW model. This example has been considered for instance in 
\cite{Kiritsis:1993ju,Alvarez:1993qi,Giveon:1994fu,Bouwknegt:2003vb,Plauschinn:2013wta},
and corresponds to a three-sphere
with $H$-flux. It can be specified as follows
\eq{
\label{tdual_su2_0t_99}
\arraycolsep2pt
\begin{array}{lcl}
G&=& \displaystyle \frac{R^2}{2}\op \Bigl(\op d\eta^2+\sin^2\eta\,  d\zeta_1^2+\cos^2\eta \, d\zeta_2^2 \op\Bigr) \,, 
\\[12pt]
H&=& \displaystyle 2\op R^2 \op \sin\eta\,\cos\eta\, d\eta\wedge d\zeta_1\wedge d\zeta_2\,,
\\[12pt]
\phi &=& \phi_0 \,.
\end{array}
}
Here $\zeta_{1,2}\in[ 0, 2\pi)$ and $\eta\in[0, \pi/2]$, and $R$ denotes the radius of the 
three-sphere. Our conventions for the metric are again $G = \frac{1}{2} \op G_{ij} \op dX^i\vee dX^j$, where
the product $\vee$ is however left implicit.
The dilaton is taken to be constant, and the corresponding $\beta$-functionals \eqref{eom_beta} are vanishing
(up to a constant contribution in $\beta^{\phi}$).
The quantisation condition shown in equation \eqref{quantisation} implies that
\eq{
  \label{wzw_relation}
  R^2 = h\op\alpha' \,,\hspace{50pt} h\in\mathbb Z \,,
}
and the isometry algebra of the three-sphere is $\mathfrak{so}(4)\simeq \mathfrak{su}(2)\times \mathfrak{su}(2)$, 
which contains a $\mathfrak{u}(1)\times\mathfrak{u}(1)$ abelian sub-algebra.

We now want to perform (collective) T-duality transformations along one and two directions
for this background.
\begin{itemize}

\item Let us start with a duality transformation along the direction $k = \partial_{\zeta_1} + \partial_{\zeta_2}$,
which  corresponds to the Hopf-fibre of the three-sphere (see for instance \cite{Bouwknegt:2003vb,Plauschinn:2017ism}). Note that $k$ is globally well-defined and nowhere
vanishing.
Using the above formalism, the dual background can be obtained as  
\eq{
\label{tdual_su2_2t_98}
\arraycolsep2pt
\begin{array}{lcl}
\check G&=& \displaystyle \frac{1}{2} \left[\frac{R^2}{4}\op \Bigl(\op d\tilde\eta^2+\sin^2\tilde\eta\,  d\tilde\zeta^2\op\Bigr)
+ \frac{4}{R^2}\, \xi^2 \right], 
\\[14pt]
\check H&=& \displaystyle \alpha'  \sin\tilde\eta \, d\tilde\zeta \wedge d\tilde\eta\wedge \xi\,,
\\[8pt]
\check \phi &=& \displaystyle \phi_0 + \frac{1}{2} \log \frac{\alpha'}{R^2} \,,
\end{array}
}
where  $\tilde\eta = 2\op \eta$ and $\tilde\zeta=\zeta_1+\zeta_2$, and where the one-form $\xi$ satisfies
\eq{
  d\xi = - \frac{R^2}{4}\op\sin\tilde\eta\, d\tilde\eta \wedge d\tilde\zeta \,.
}
The form of the dual dilaton has been determined using \eqref{dual_back_9495}.
This background corresponds to a circle fibred over a two-sphere \cite{Bouwknegt:2003vb}, which 
in general is a so-called lens space. We come back to this point in section~\ref{sec_nca_top_t}.
Furthermore, the $\beta$-functionals \eqref{eom_beta} corresponding to \eqref{tdual_su2_2t_98} 
are vanishing (up to the same constant contribution to $\beta^{\phi}$) and hence the dual model 
is again a CFT.

We also note that for $h=1$ -- or alternatively $R^2=\alpha'$ -- the original and T-dual 
background are related by a change of coordinates. The $SU(2)$ WZW model is 
therefore self-dual under one T-duality for $R^2 = \alpha'$, i.e. at level $h=1$.

\item We can also perform two T-duality transformations for the three-sphere with $H$-flux. 
As Killing vectors we choose $k_1 = \partial_{\zeta_1}$ and $k_2 = \partial_{\zeta_2}$ --
which vanish at the isolated points $\eta=0$ and $\eta=\pi/2$ --
and for the corresponding one-forms $v_{\alpha}$ we make the choice
\eq{
  v_1 &= +R^2 \bigl( \cos^2\eta +\tfrac{\rho-1}{2} \bigr) \op d\zeta_2 \,, \\[4pt]
  v_2 &= -R^2 \bigl( \cos^2\eta +\tfrac{\rho-1}{2} \bigr) \op d\zeta_1 \,, \\
}
where $\rho\in\mathbb R$ is a parameter related to the gauge freedom in $v_{\alpha}$.
Using then \eqref{dual_g_007} and \eqref{dual_b_007}, for the dual background we obtain \cite{Plauschinn:2017ism}
\eq{
\label{tdual_su2_2t_99}
\arraycolsep2pt
\begin{array}{lcl}
\check G&=& \displaystyle \frac{1}{2} \left[R^2\op d\eta^2+ \frac{4\op \alpha'^2}{R^2} \,\frac{1}{\Delta} \Bigl( 
\sin^2\eta\,  d\tilde\zeta_1^2+\cos^2\eta \, d\tilde\zeta_2^2 \Bigr) \right], 
\\[12pt]
\check H&=& \displaystyle  \frac{8\op\alpha'^2}{R^2} \,\frac{1-\rho^2}{\Delta^2} \op \sin\eta\,\cos\eta\, 
d\eta\wedge d\tilde\zeta_1\wedge d\tilde\zeta_2\,,
\\[12pt]
\check \phi &=& \displaystyle \phi_0 - \frac{1}{2} \log \Delta \,,
\end{array}
}
where $\tilde\zeta_{1,2}$ are the coordinates dual to $\zeta_{1,2}$ defined via
$\alpha' d\tilde \zeta_1 = d\chi_2$ and $\alpha' d\tilde \zeta_2 = d\chi_1$,
and $\Delta$ is given by
\eq{
  \Delta = (1+\rho)^2 \cos^2\eta + (1-\rho)^2 \sin^2\eta \,.
}
For this background the $\beta$-functionals \eqref{eom_beta} are again vanishing.
The parameter $\rho$ only arises when performing a collective T-duality transformation and 
corresponds to a gauge choice for $\iota_{k_{[\ul \alpha}} v_{\ul \beta]}$ in $\mathcal D_{\alpha\beta}$ 
shown in \eqref{dual_rick_yeah}.
On the dual side it cannot be removed by diffeomorphisms, and it is interpreted as parametrising a 
$\beta$-transformations in the context of T-duality for curved backgrounds \cite{Plauschinn:2017ism}.

We also note that for $\rho=0$ and $R^2=2\op\alpha'$, the dual background \eqref{tdual_su2_2t_99} agrees with 
the original background \eqref{tdual_su2_0t_99} and hence the model is self-dual under
two T-dualities for $R^2 = 2\op\alpha'$, that is at level $h=2$.

\end{itemize}

%%%%%%%%%%%%%%%%%%%%%%%%%%%%%%%%%%%%%%%%%%%%%%%
%%%%%%%%%%%%%%%%%%%%%%%%%%%%%%%%%%%%%%%%%%%%%%%

\subsubsection*{Comments on non-abelian T-duality}
\label{page_nonab}

After performing one and two abelian T-dualities for the $SU(2)$ WZW model, it is natural 
to try to perform a collective (non-abelian) T-duality transformations along three directions. 
This has been studied for instance in the papers
\cite{delaOssa:1992vci,Giveon:1993ai,Alvarez:1993qi,Curtright:1994be,Sfetsos:1994vz,Alvarez:1994np,Klimcik:1995ux,Lozano:1995jx,Curtright:1996ig} from different perspectives.

For the present context we note that the Killing vectors $k_{\alpha}$ which -- together with corresponding one-forms $v_{\alpha}$ -- 
satisfy the gauging requirements \eqref{variations_45} belong to 
the so-called vectorial $\mathfrak{su}(2)$ sub-algebra
of the $\mathfrak{su}(2)\times \mathfrak{su}(2)$ isometry algebra.
However, in this case the matrix $\mathcal T$ shown in \eqref{coc_01} is singular, 
and the dual model cannot be obtained. 
Correspondingly, as explained for instance in \cite{Giveon:1993ai}, if the dual background is obtained by a gauge-fixing 
procedure then in the case of the above non-abelian T-duality transformation the gauge symmetry 
does not does not allow to set the original coordinates to zero and hence the dual model is not obtained. 
These observations extend to WZW models based on general Lie groups $G$.
Let us comment on two possibilities to avoid this issue:
\begin{itemize}

\item The non-abelian T-dual of a WZW model can be obtained by starting from a coset construction 
and performing a limiting procedure. This approach has been developed in
\cite{Sfetsos:1994vz}.

\item Another possibility is to consider principle chiral models, for which the Wess-Zumino term of a WZW model 
is set to zero. This implies that the $H$-flux vanishes and that the NS-NS sector of the 
background does not correspond to a CFT.
However, when turning on Ramond-Ramond (R-R) fluxes the $\beta$-functionals \eqref{eom_beta} are modified
and it is possible to obtain vanishing $\beta$-functionals with zero $H$-flux and non-zero R-R fluxes.
In this case the gauging constraints \eqref{variations_45} are trivially satisfied, and for the 
$SU(2)$ model one can choose to 
gauge one of the $\mathfrak{su}(2)$ isometry algebras. 
For more details we refer to the papers \cite{Lozano:2011kb,Sfetsos:2010uq,Itsios:2013wd,Sfetsos:2013wia}.

\end{itemize}
However, in both approaches the full equivalence of the original and dual theory have 
not been established. Moreover, in many cases the isometry algebra of the dual model 
is completely broken, and one cannot invert the duality transformation. 
Nevertheless, non-abelian T-duality is a very useful solution-generating technique
at the level of supergravity \cite{Itsios:2012zv,Itsios:2012dc,Lozano:2012au}. 
We finally mention that using a different approach, non-abelian T-duality transformations can be
described using Poisson-Lie duality which we discuss in section~\ref{cha_poisson_lie}.

%%%%%%%%%%%%%%%%%%%%%%%%%%%%%%%%%%%%%%%%%%%%%%%
%%%%%%%%%%%%%%%%%%%%%%%%%%%%%%%%%%%%%%%%%%%%%%%
%%%%%%%%%%%%%%%%%%%%%%%%%%%%%%%%%%%%%%%%%%%%%%%
%%%%%%%%%%%%%%%%%%%%%%%%%%%%%%%%%%%%%%%%%%%%%%%
%%%%%%%%%%%%%%%%%%%%%%%%%%%%%%%%%%%%%%%%%%%%%%%
%%%%%%%%%%%%%%%%%%%%%%%%%%%%%%%%%%%%%%%%%%%%%%%
%%%%%%%%%%%%%%%%%%%%%%%%%%%%%%%%%%%%%%%%%%%%%%%
%%%%%%%%%%%%%%%%%%%%%%%%%%%%%%%%%%%%%%%%%%%%%%%

\subsection{Equivalence of T-dual theories}

In section~\ref{cha_t_duality} we have discussed T-duality transformations for toroidal string-theory
compactifications with constant Kalb-Ramond and dilaton field. In particular, we
have shown that the spectrum -- encoded in the torus partition function of the closed string -- is invariant under
T-duality. (For higher-genus partition functions the invariance under 
T-duality has been discussed in section~2.3 of \cite{Giveon:1994fu}.)
On the other hand, for curved backgrounds a  CFT description is usually not available and showing that
the spectrum is invariant under T-duality is therefore more difficult. However, we have 
mentioned  that the Buscher rules \eqref{dual_back_9494} leave the $\beta$-functionals \eqref{eom_beta} invariant
at linear order in $\alpha'$. 
These results indicate that the original and  T-dual theory are equivalent, but especially for curved backgrounds
further evidence would be desirable. In this respect we note the following:
\begin{itemize}

\item T-duality transformations can be argued to be the discrete part of a gauge symmetry
\cite{Dine:1989vu,Giveon:1990era,Giveon:1993ph}.
In particular, the original and dual model are related by a gauge transformation, which makes them 
equivalent as conformal field theories. A detailed discussion of this reasoning can be found for instance 
in section~2.6 of \cite{Giveon:1994fu}.

\item Another approach is to interpret T-duality transformations as a discrete 
symmetry of a higher-dimensional theory. 
In particular, in \cite{Kiritsis:1991zt,Dijkgraaf:1991ba,Giveon:1991sy} it was observed that the so-called 
axial- and vector-gaugings of an abelian chiral symmetry of a world-sheet sigma-model 
give two dual versions of that model. 
As shown in \cite{Kiritsis:1993ju}, the coset constructions of these two gaugings (that is, the model obtained 
after integrating out the corresponding gauge field) correspond to the same CFT.
This result has been used in \cite{Rocek:1991ps,Giveon:1991jj}
to show that T-duality is a symmetry of the conformal field theory,
and we discuss this idea in more detail below.

\item We also mention that abelian T-duality for WZW models
leaves the spectrum invariant, as has been shown explicitly in \cite{Gaberdiel:1995mx}. 
The T-dual theory is a certain orbifold of the original WZW model. 

\end{itemize}
In the remainder of this section we explain in some more detail the second approach for showing the equivalence 
of the original and T-dual theories.

%%%%%%%%%%%%%%%%%%%%%%%%%%%%%%%%%%%%%%%%%%%%%%%
%%%%%%%%%%%%%%%%%%%%%%%%%%%%%%%%%%%%%%%%%%%%%%%

\subsubsection*{Setting}

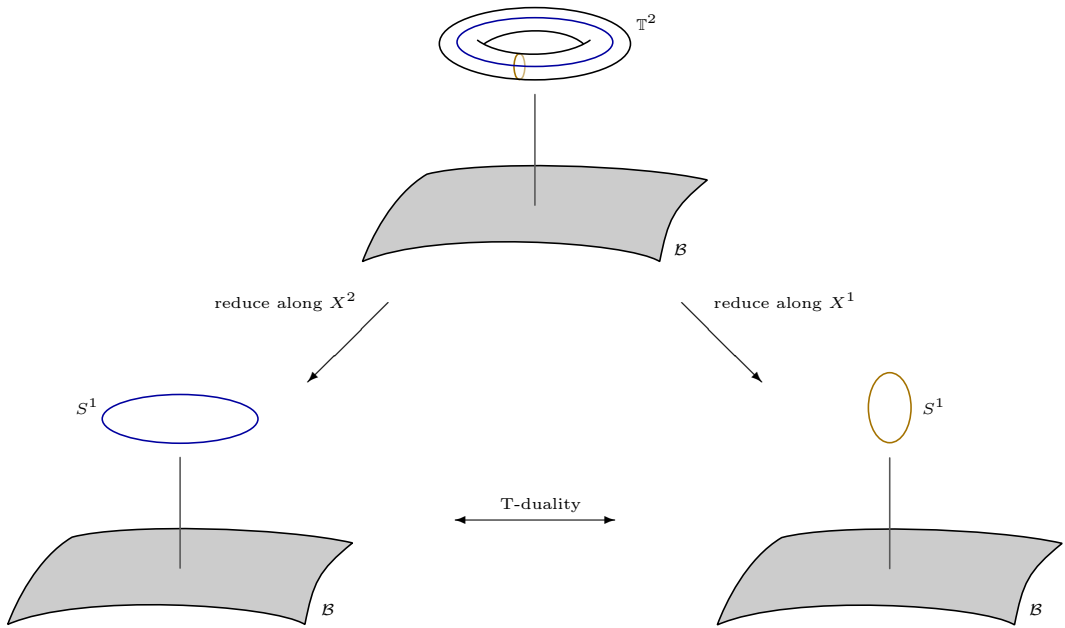
\begin{figure}[t]
\centering
\includegraphics[width=130pt]{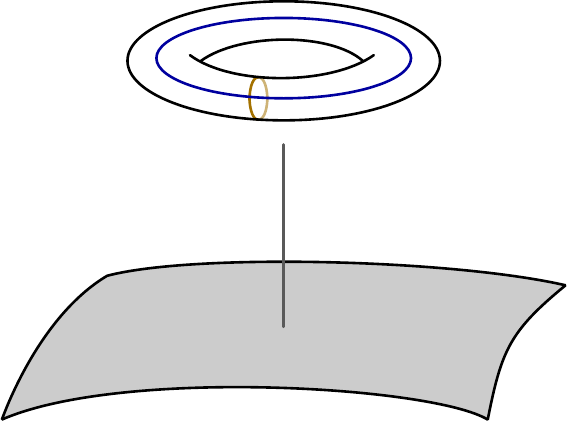}
\\[40pt]
\includegraphics[width=130pt]{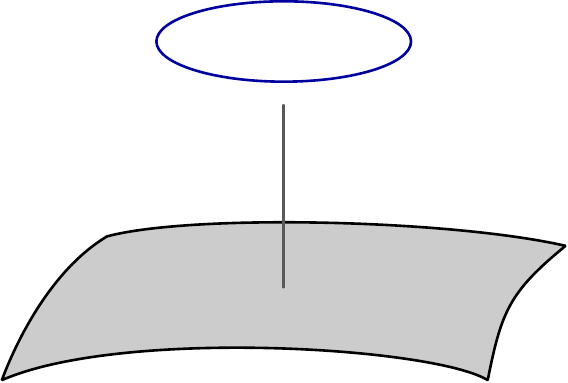}
\hspace{60pt}
\begin{picture}(0,0)
\put(-55,122){\vector(-1,-1){30}}
\put(+55,122){\vector(+1,-1){30}}
\put(-120,120){\tiny reduce along $X^2$}
\put(+67,120){\tiny reduce along $X^1$}
\put(-80,4){\tiny $\mathcal B$}
\put(+185,4){\tiny $\mathcal B$}
\put(52,140){\tiny $\mathcal B$}
\put(38,225){\tiny $\mathbb T^2$}
\put(-172,80){\tiny $S^1$}
\put(+145,80){\tiny $S^1$}
\put(0,40){\vector(+1,0){30}}
\put(0,40){\vector(-1,0){30}}
\put(-13,44){\tiny T-duality}
\end{picture}
\hspace{60pt}
\includegraphics[width=130pt]{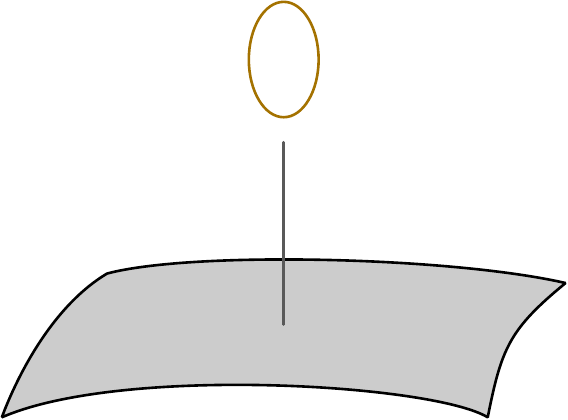}
\vskip1.5em
\caption{Illustration of the main setting in this section. On top, a $\mathbb T^2$-fibration over a
$(D-1)$-dimensional base manifold $\mathcal B$ is shown. If one reduces (gauge the corresponding 
isometry and integrate out the gauge field) along either of the two circles of the two-torus, one obtains two 
circle-fibrations over $\mathcal B$. These two $D$-dimensional backgrounds are related via T-duality. 
\label{fig_rv}}
\end{figure}

Following the work of \cite{Kiritsis:1991zt,Dijkgraaf:1991ba,Giveon:1991sy}, the main idea 
in \cite{Rocek:1991ps} is to consider a two-torus fibration over a $(D-1)$-dimensional 
base manifold. The $(D+1)$-dimensional background is required to have two isometries
along the $\mathbb T^2$-fibre which, similarly as in section~\ref{section_t_dual_1}, correspond 
to symmetries of the world-sheet theory. Depending on which of these two symmetries is 
gauged and integrated out, one obtains two $D$-dimensional backgrounds which are related via
T-duality. This setting is illustrated in figure~\ref{fig_rv}.
%%%%%%%%%%%%%%%%
%%%%%%%%%%%%%%%%
%%%%%%%%%%%%%%%%
%%%%%%%%%%%%%%%%

Let us make this more concrete and consider a two-dimensional non-linear sigma-model
for a $(D+1)$-dimensional target-space. 
Using indices $I,J$ with values $I = 1,\ldots, D+1$ the action reads
\eq{
\label{wsaction9}
\mathcal S =
 -\frac{1}{2\pi \alpha'} \int_{\Sigma}
  \Bigl[ \,\tfrac{1}{2}\op G_{IJ} \, d X^{I}\wedge\star d X^{J} - \tfrac{i}{2}\op B_{IJ} \,dX^{I}\wedge dX^{J}
  + \tfrac{\alpha'}{2}\op\mathsf R\, \phi \star 1\, \Bigr] 
  \,,
}
where $\Sigma$ denotes a two-dimensional world-sheet without boundary. 
The metric and Kalb-Ramond field take the following explicit form
\eq{
  \label{rv_001}
  G_{IJ} = \left( \begin{array}{ccc}
  \frac{\alpha'+B}{2} & 0 & \frac12\op G^+_{b} \\[4pt] 
  0 & \frac{\alpha'-B}{2} & \frac12\op G^-_{b} \\[4pt]
  \frac12 \op G^+_{a} & \frac12\op G^-_{a} & g_{ab}
  \end{array}\right) ,
  \hspace{19pt}
  B_{IJ} = \left( \begin{array}{ccc}
  0 & + \frac12\op B & - \frac12\op G^-_{b} \\[4pt] 
  -\frac12\op B & 0 & -\frac12\op G^+_{b} \\[4pt] 
  +\frac12\op G^-_{a} & +\frac12\op G^+_{a} & b_{ab}
  \end{array}\right) ,
 }
and we assume the 
functions $B$, $G^{\pm}_{a}$ $g_{ab}$ and $b_{ab}$ 
to only depend on $(D-1)$ coordinates $X^a$
with $a = 3,\ldots, D+1$.
Note that $G_{IJ}$ and $B_{IJ}$ do not constitute the most general $(D+1)$-dimensional sigma-model 
background but are of rather restricted form. In particular, they describe 
a $\mathbb T^2$-fibration over a $(D-1)$-dimensional background with fibre coordinates
$X^1$ and $X^2$. 
Given that the metric $G_{IJ}$ in \eqref{rv_001} does not depend on $X^1$ and $X^2$, 
it has at least two abelian isometries generated by the Killing vectors
\eq{
  \label{iso_009}
  \arraycolsep2pt
  k^I_{(1)} = \left( \begin{array}{c} 1 \\ 0 \\  0 \end{array}\right)
  \hspace{50pt}
  {\rm and}
  \hspace{50pt}
  k^I_{(2)} = \left( \begin{array}{c} 0 \\ 1 \\  0 \end{array}\right).
}
We also observe that  the action \eqref{wsaction9} 
with metric and $B$-field \eqref{rv_001}
is invariant under the following ${\mathbb Z}_2$ transformation
\eq{
\label{rv_935u039u}
X^1\; \longleftrightarrow \;X^2\,,
\hspace{40pt}
B \;\longleftrightarrow \;-B\,,
\hspace{40pt}
G^+_{a}\;\longleftrightarrow \;G^-_{a} \,.
}
As we will show below, this symmetry corresponds to T-duality via 
the Buscher rules given in \eqref{dual_back_9494} for the $D$-dimensional theories.

%%%%%%%%%%%%%%%%%%%%%%%%%%%%%%%%%%%%%%%%%%%%%%%
%%%%%%%%%%%%%%%%%%%%%%%%%%%%%%%%%%%%%%%%%%%%%%%

\subsubsection*{Reduced background}

Let us now reduce the above background from $D+1$ to $D$ dimensions 
by gauging one of these isometries \eqref{iso_009}
and integrating out the corresponding gauge field. 
To do so, we first note that 
the action \eqref{wsaction9} is invariant
under global transformations $\delta X^i = \epsilon\op k^i$  since
\eq{
  \mathcal L_{k} \op  G = 0\,, \hspace{70pt}
  \mathcal L_{k} \op  B = dv  \,, \hspace{70pt}
  \mathcal L_{k} \op \phi =0\,,
}
where we included the possibility of a globally-defined one-form $v$. 
The gauged action takes the following form
\eq{
  \label{rv_gauged_action_8}
  &\hat{\mathcal S} =-\frac{1}{2\pi\alpha'} \int_{\Sigma} \;  
  \Bigl[ \op\tfrac{1}{2}\op G_{IJ}  \op(dX^{I} + k^{I} A)\wedge\star(dX^{J} + k^{J} A)  
   \\
  & \hspace{130pt}
   - \tfrac{i}{2}\op B_{IJ} \,dX^{I}\wedge dX^{J}
  -   i\op ( v -\iota_k B )\wedge A \,
  \Bigr] \,,
}
which is invariant under the local symmetry transformations
$\hat\delta_{\epsilon} X^{I} = \epsilon\op k^{I}$ and
$\hat\delta_{\epsilon} A = - d\epsilon$ provided that
$\iota_k v = 0$. 
Note that in contrast to our discussion in 
section~\ref{section_t_dual_1} we have not included a Lagrange multiplier 
in the gauged action. 
We recall that the latter was used to impose that the gauge field $A$ is closed,
which we require here by hand. 
Up to rescalings, this leaves the following two possible choices for $k$ and $v$ \cite{Rocek:1991ps,Bakas:2016nxt}
\eq{
\label{rv_gaugings_992}
\arraycolsep2pt
\begin{array}{l@{\hspace{40pt}}lcl@{\hspace{30pt}}lcl}  
(1) & k &=& \partial_1 \,, & v & = & \displaystyle \tfrac{\alpha'}{2}\op dX^2  \,,
\\[6pt]
(2) & k &=& \partial_2 \,, & v & = & \displaystyle \tfrac{\alpha'}{2}\op dX^1  \,.
\end{array}
}
Integrating out the gauge field, we obtain a world-sheet action describing a  $D$-dimensional 
target space characterised by the following metric and $B$-field
\eq{
\label{red_b_001}
  \check{G}_{ij} &= 
  \left( \begin{array}{@{\hspace{4pt}}c@{\hspace{16pt}}c@{\hspace{4pt}}}
  \displaystyle \alpha' \,\frac{\alpha' \mp  B}{\alpha'\pm B}
  &  \displaystyle  + \frac{\alpha}{\alpha'\pm B}\op G^{\mp}_{b} \\[12pt]
     \displaystyle + \frac{\alpha}{\alpha'\pm B} \op G^{\mp}_{a}
  &   \displaystyle g_{ab} \mp \frac12\, \frac{G_{a}^+G_{b}^+ - G_{a}^-G_{b}^-}{\alpha'\pm B}
  \end{array}
  \right) ,
  \\[12pt]
  \check{B}_{ij} &= 
  \left( \begin{array}{@{\hspace{4pt}}c@{\hspace{16pt}}c@{\hspace{4pt}}}
  0 & \displaystyle - \frac{\alpha'}{\alpha'\pm B} \op G_{b}^{\pm}  \\[12pt]
  \displaystyle + \frac{\alpha'}{\alpha'\pm B} \op G_{a}^{\pm} &
  \displaystyle b_{ab} \mp\frac{1}{2}\, \frac{G_{a}^{-}G_{b}^+ - G_{a}^+G_{b}^-}{\alpha'\pm B}
  \end{array}
  \right) 
  .
}
The first choice in \eqref{rv_gaugings_992} corresponds to the upper sign in \eqref{red_b_001}
for which the indices take values $i = 2,3,\ldots, D+1$, while the second choice corresponds to the lower 
sign with indices  $i =1,3,\ldots, D+1$.

%%%%%%%%%%%%%%%%%%%%%%%%%%%%%%%%%%%%%%%%%%%%%%%
%%%%%%%%%%%%%%%%%%%%%%%%%%%%%%%%%%%%%%%%%%%%%%%

\subsubsection*{T-duality}

We now observe that the reduced metric and $B$-field shown in \eqref{red_b_001} 
for the two choices \eqref{rv_gaugings_992} are related by the Buscher rules
\eqref{dual_back_9494}. For instance, we can easily check that 
\eq{
  \check G_{22} \op\bigr\rvert_{(1)} = \frac{\alpha'^2}{\check G_{11}}\op \biggr\rvert_{(2)}\,,
}
and similarly for the other relations of the Buscher rules. 
We also see that the two backgrounds in \eqref{red_b_001} are related by 
\eqref{rv_935u039u}, which was a symmetry of the $(D+1)$-dimensional theory. 
Hence, T-duality between two $D$-dimensional backgrounds can be 
interpreted as a symmetry of a $(D+1)$-dimensional theory, 
and in this way the two T-dual theories are equivalent.

%%%%%%%%%%%%%%%%%%%%%%%%%%%%%%%%%%%%%%%%%%%%%%%
%%%%%%%%%%%%%%%%%%%%%%%%%%%%%%%%%%%%%%%%%%%%%%%

\subsubsection*{GKO construction}

We can also interpret T-duality from a conformal-field-theory point of view \cite{Giveon:1990era,Kiritsis:1991zt,Rocek:1991ps}. 
Let us first note that the conserved current corresponding to the symmetry
$\delta X^i = \epsilon\op k^i$ of the action \eqref{wsaction9} is given by
\eq{
  J = \star\op \mathsf k + i \op(v-\iota_k B) \,,
}
where $\mathsf k = k^I G_{IJ} dX^J$ is the one-form dual to the vector $k$ and $\star$ denotes the 
Hodge star-operator on the two-dimensional world-sheet $\Sigma$. Assuming the world-sheet theory 
to be conformal, we can choose the world-sheet metric to be flat and introduce a complex coordinate
$z = \sigma^0 + i \sigma^1$.  For the two choices in  \eqref{rv_gaugings_992} 
the current $J$ then takes the form
\eq{
\label{rv_i_saw_the_sign}
\arraycolsep2pt
\begin{array}{l@{\hspace{40pt}}l}  
(1) & \displaystyle J = - J_z \op dz +J_{\ov z} \, d\ov z \,,
\\[6pt]
(2) & \displaystyle J = + J_z \op dz +J_{\ov z} \, d\ov z \,,
\end{array}
}
where we defined the holomorphic and anti-holomorphic currents
\eq{
  \label{rv_242}
  J_z &= \frac{i}{2}\, \Bigl[\, \alpha' \op\bigl( \partial X^1 - \partial X^2 \bigr) + B \op \bigl( \partial X^1 + \partial X^2\bigr) 
  + \bigl( G_a^+ - G_a^-\bigr) \op \partial X^a \,\Bigr] \,,
  \\[8pt]
  J_{\ov z} 
  &= \frac{i}{2}\, \Bigl[\, \alpha' \op\bigl( \ov\partial X^1 + \ov\partial X^2 \bigr) + B \op \bigl( \ov\partial X^1 - 
  \ov \partial X^2\bigr) 
  + \bigl( G_a^+ + G_a^-\bigr) \op \ov\partial X^a \,\Bigr] \,.
}
Here we employed the conventions $\star dz = -i\op dz$ and $\star d\ov z = +i \op d\ov z$ 
as well as $\partial\equiv \partial_z$ and $\ov\partial \equiv \partial_{\ov z}$, and we note that 
the currents $J_z$ and $J_{\ov z}$ are separately conserved, that is $\ov\partial J_z = 0$ and 
$\partial J_{\ov z}=0$. These currents therefore generate at $\mathfrak u(1)\times \mathfrak u(1)$ 
current algebra.

Now, following \cite{Rocek:1991ps}, 
the gauging and integrating out of the two isometries \eqref{iso_009} corresponds to a 
generalisation of the GKO coset construction \cite{Goddard:1984vk,Goddard:1986ee} 
in which one mods out by 
the holomorphic and anti-holomorphic currents \eqref{rv_242}. 
In particular,  one keeps only those fields which are primary with respect to the current algebra.
The only difference between the two possibilities \eqref{rv_gaugings_992}  is the sign of the 
holomorphic charge in \eqref{rv_i_saw_the_sign}, 
and one can go from one model to the other by flipping this sign. 
However, since conformal dimensions and OPEs only depend on quadratic combinations 
of the charges (see for instance \cite{Blumenhagen:2009zz}), the correlation functions are invariant under this operation. 
Hence, T-duality is a symmetry at the level of the conformal field theory.

%%%%%%%%%%%%%%%%%%%%%%%%%%%%%%%%%%%%%%%%%%%%%%%
%%%%%%%%%%%%%%%%%%%%%%%%%%%%%%%%%%%%%%%%%%%%%%%

\subsubsection*{Remarks}

We close this section with the following remarks:
\begin{itemize}

\item A generalisation of the analysis discussed in this section to $\mathbb T^{2n}$-fibrations, which corresponds 
to T-duality transformations along multiple directions, can be found  in
\cite{Bakas:2016nxt}.

\item In this section we have implicitly assumed that the two-dimensional world-sheet 
$\Sigma$ has a trivial topology. In order to extend this analysis to 
non-trivial topologies we have to address the term
\eq{
  \label{rv_extra_term}
  \hat{\mathcal {S}} \supset- \frac{i}{2\pi\alpha'} \int_{\Sigma} \bigl( \pm \alpha'  dX^1\wedge dX^2 \bigr) \,,
}
which arises when integrating out the gauge field from the action 
\eqref{rv_gauged_action_8}. Since $dX^1$ and $dX^2$ are closed, we 
can perform a Hodge decomposition using the notation introduced on 
page~\pageref{page_back_single_1} as
\eq{
   dX^1 = dX^1_{(0)} +  X^1_{(\mathsf m)}\op \omega^{\mathsf m} \,,
   \hspace{40pt}
   dX^2 = dX^2_{(0)} +  X^2_{(\mathsf m)}\op \omega^{\mathsf m}   \,,
}
and the term shown in \eqref{rv_extra_term} then becomes
\eq{
  \hat{\mathcal {S}} \supset  \mp \frac{i}{2\pi} \, X^1_{(\mathsf m)} J^{\mathsf{mn}} X^2_{(\mathsf n)}  \,.
}
If the coordinates $X^1$ and $X^2$ are compactified via the identification $X \sim X + 2\pi \op n$ with 
$n \in \mathbb Z$, in the Hodge decomposition we have $X^{1,2}_{(\mathsf m)} \in 2\pi \mathbb Z$
and the exponential of \eqref{rv_extra_term} gives one. Hence, it does not contribute to the path integral.

\end{itemize}

%%%%%%%%%%%%%%%%%%%%%%%%%%%%%%%%%%%%%%%%%%%%%%%
%%%%%%%%%%%%%%%%%%%%%%%%%%%%%%%%%%%%%%%%%%%%%%%
%%%%%%%%%%%%%%%%%%%%%%%%%%%%%%%%%%%%%%%%%%%%%%%
%%%%%%%%%%%%%%%%%%%%%%%%%%%%%%%%%%%%%%%%%%%%%%%
%%%%%%%%%%%%%%%%%%%%%%%%%%%%%%%%%%%%%%%%%%%%%%%
%%%%%%%%%%%%%%%%%%%%%%%%%%%%%%%%%%%%%%%%%%%%%%%
%%%%%%%%%%%%%%%%%%%%%%%%%%%%%%%%%%%%%%%%%%%%%%%
%%%%%%%%%%%%%%%%%%%%%%%%%%%%%%%%%%%%%%%%%%%%%%%

\subsection{Type II string theories}
\label{sec_buscher_II}

In the above sections we have studied how T-duality transformations act on the metric, 
Kalb-Ramond field and dilaton -- which  comprise the massless sector of the
bosonic string.
In this section we are interested in type II superstring theories, where  the space-time bosons originate from the 
Neveu-Schwarz--Neveu-Schwarz (NS-NS) and Ramond-Ramond (R-R) sectors.
The massless fields of the NS-NS sector are again given by $G$, $B$ and $\phi$,
for which we already discussed the behaviour under T-duality. 
We therefore now turn to the R-R sector and determine the remaining
transformation rules. 
We note that the Ramond-Ramond sector will not play a role in our discussion of 
non-geometric backgrounds,  but we include this topic for completeness.

%%%%%%%%%%%%%%%%%%%%%%%%%%%%%%%%%%%%%%%%%%%%%%%
%%%%%%%%%%%%%%%%%%%%%%%%%%%%%%%%%%%%%%%%%%%%%%%

\subsubsection*{Type II superstring}

Let us start by briefly recalling some aspects of type II superstring theory 
(for a textbook introduction we refer for instance to \cite{Polchinski:1998rr,Blumenhagen:2013fgp}).
In addition to bosonic world-sheet fields
$X^{\mu}(\sigma^{\alpha})$, for the type II string one considers fermionic fields $\psi^{\mu}(\sigma^{\alpha})$ and one
requires the resulting world-sheet theory to be supersymmetric. 
For closed-string world-sheets of the form $\Sigma = \mathbb R \times S^1$ 
these fermions can have periodic (Ramond) or anti-periodic (Neuveu-Schwarz) 
boundary conditions along the $S^1$, which  can be chosen independently 
for the left- and right-moving components $\psi^{\mu}_L$  
and $\psi{}^{\mu}_R$ of the two-component spinor $\psi^{\mu}$. One is therefore left with the 
following four 
combinations of boundary conditions between the left- and right-moving sectors
\eq{
  \label{t_duality_ii_sectors}
  \renewcommand{\arraystretch}{1.1}
  \begin{array}{c@{\hspace{30pt}}c@{\hspace{30pt}}l}
  \mbox{Neveu-Schwarz--Neveu-Schwarz} &
  \mbox{NS-NS} &    \mbox{space-time bosons,} 
  \\
  \mbox{Neveu-Schwarz--Ramond} &
  \mbox{NS-R} &    \mbox{space-time fermions,} 
  \\
  \mbox{Ramond--Neveu-Schwarz} &
  \mbox{R-NS} &    \mbox{space-time fermions,} 
  \\  
  \mbox{Ramond--Ramond} &
  \mbox{R-R} &    \mbox{space-time bosons.} 
  \end{array}
}
After quantising the theory, fields in the NS-NS and R-R sectors correspond to space-time bosons 
whereas fields in the NS-R and R-NS sectors give rise to space-time fermions. 
The critical dimension of the type II superstring is $D=10$.

We furthermore note that for consistency of the theory (e.g. modular invariance)
one has to perform a GSO projection, which is achieved using 
world-sheet fermion-number operators $F_L$ and $F_R$. 
In particular, two interesting theories are obtained by only keeping 
contributions $\lvert\phi_L\rangle\otimes \lvert \phi_R\rangle$
(belonging to the four sectors shown in \eqref{t_duality_ii_sectors})
which satisfy \label{page_type_ii}
\eq{
  \label{gso}
  \renewcommand{\arraystretch}{1.3}
  \arraycolsep2pt
  \begin{array}{@{}l@{\hspace{40pt}}lcl@{\hspace{20pt}}lcl@{}}
  \multirow{2}{*}{type IIA} 
  & (-1)^{F_L} \lvert\phi_L\rangle_{\rm NS} &=& +\lvert\phi_L\rangle_{\rm NS} \,, 
  & (-1)^{ F_R} \lvert\phi_R\rangle_{\rm NS} &=& +\lvert\phi_R\rangle_{\rm NS} \,,
  \\
  & (-1)^{F_L} \lvert\phi_L\rangle_{\rm R} &=& +\lvert\phi_L\rangle_{\rm R} \,, 
  & (-1)^{ F_R} \lvert\phi_R\rangle_{\rm R} &=& -\lvert\phi_R\rangle_{\rm R} \,,
  \\[10pt]
  \multirow{2}{*}{type IIB} 
  & (-1)^{F_L} \lvert\phi_L\rangle_{\rm NS} &=& +\lvert\phi_L\rangle_{\rm NS} \,, 
  & (-1)^{ F_R} \lvert\phi_R\rangle_{\rm NS} &=& +\lvert\phi_R\rangle_{\rm NS} \,,
  \\
  & (-1)^{F_L} \lvert\phi_L\rangle_{\rm R} &=& +\lvert\phi_L\rangle_{\rm R} \,, 
  & (-1)^{ F_R} \lvert\phi_R\rangle_{\rm R} &=& +\lvert\phi_R\rangle_{\rm R} \,,
  \end{array}
}
where we emphasise  the sign difference for $(-1)^{F_R}$ in the Ramond sector. 
Now, the massless spectrum in the NS-NS sector consists 
of the space-time metric $G_{\mu\nu}$, the two-form Kalb-Ramond gauge field $B_{\mu\nu}$ 
and the dilaton scalar field $\phi$, which applies both to the type IIA and  type IIB
theories. In the R-R sector  the massless fields are
a one-form and a three-form gauge potential $C_1$ and $C_3$ for type IIA, 
and a zero-form, a two-form and a self-dual four-form potential $C_0$, $C_2$ and $C_4$ for 
type IIB
\eq{
  \renewcommand{\arraystretch}{1.2}
  \arraycolsep10pt
  \begin{array}{c||c@{\hspace{20pt}}l}
  \mbox{superstring theory} & \multicolumn{2}{c}{\mbox{massless bosonic field content}}
  \\
  \hline\hline
  \mbox{type IIA} & \hspace{10pt}G\,, B\,, \phi\,, & C_1\,, C_3 \,,
  \\
  \mbox{type IIB} & \hspace{10pt}G\,, B\,, \phi\,, & C_0\,, C_2 \,, C_4 \,.
  \end{array}
}  
The effective theory for these fields (and their space-time fermionic superpartners) 
is given by type II supergravity, to which 
we come back in section~\ref{sec_cy_flux}.\footnote{
For the type II superstring one should distinguish two appearances of supersymmetry: 1)
the two-dimensional world-sheet theory has $N=(1,1)$ world-sheet supersymmetry
and $X^{\mu}(\sigma^{\alpha})$ and $\psi^{\mu}(\sigma^{\alpha})$ are superpartners;
2) the ten-dimensional target-space theory has $\mathcal N=2$ space-time supersymmetry,
which is the reason for calling it type II.}

%%%%%%%%%%%%%%%%%%%%%%%%%%%%%%%%%%%%%%%%%%%%%%%
%%%%%%%%%%%%%%%%%%%%%%%%%%%%%%%%%%%%%%%%%%%%%%%

\subsubsection*{T-duality I}

Let us now compactify the type II string on a circle of radius $R$ 
by identifying say $X^9 \sim X^9+2\pi R$ similarly as in section~\ref{cha_t_duality}
(recall that the critical space-time dimension for the superstring is $D=10$).
When performing a T-duality transformation along the circle we noted in equation 
\eqref{dual_84930} that the bosonic world-sheet fields transform as 
$    (\,X^{9}_R\,,\,X^{9}_L\,)  \rightarrow   (\,-X^{9}_R\,,\,+X^{9}_L\,)$.
Now, since in the present case the world-sheet theory is supersymmetric one
can expect that the world-sheet superpartners  of $X^9$ transforms similarly, that is
\eq{
  \label{t_duality_ii_002}
    (\,\psi^{9}_R\,,\,\psi^{9}_L\,)  \rightarrow   (\,-\psi^{9}_R\,,\,+\psi^{9}_L\,) \,.
}
Let us then recall the GSO projection 
shown in \eqref{gso}
and note that $(-1)^{F_R}$ in the Ramond sector (in light-cone quantisation) 
is given by 
$(-1)^{F_R} = 16 \prod_{i=2}^9 b_0^i$ with $b_0^i$ the 
zero modes of $\psi^i_R$ \cite{Blumenhagen:2013fgp}. 
Due to \eqref{t_duality_ii_002} a T-duality transformation changes 
the sign of $b^9_0$ in the right-moving sector
and therefore changes the sign of $(-1)^{F_R}$ acting on  $\lvert\phi\rangle_{\rm R}$.
This means that T-duality maps the type IIA theory to type IIB and vice versa
\eq{
  \label{t-duality_iiab}
  \mbox{type IIA on $S^1$} 
  \hspace{20pt}
  \xleftrightarrow{\hspace{5pt}\mbox{\scriptsize T-duality}\hspace{5pt}} 
  \hspace{20pt}
  \mbox{type IIB on $\tilde S^1$} \,,
}
with $S^1$ and $\tilde S^1$ two T-dual circles.
We note that our argumentation here is somewhat heuristic, however, 
the mapping \eqref{t-duality_iiab} can be checked explicitly.

%%%%%%%%%%%%%%%%%%%%%%%%%%%%%%%%%%%%%%%%%%%%%%%
%%%%%%%%%%%%%%%%%%%%%%%%%%%%%%%%%%%%%%%%%%%%%%%

\subsubsection*{T-duality II}

Let us now turn to the open-string sector for the derivation of the R-R sector transformation 
rules\cite{Bergshoeff:1996cy,Green:1996bh}. We recall from page
\pageref{page_open_string_cft}
that T-duality interchanges Dirichlet and Neumann boundary conditions, in 
particular, performing a T-duality along a direction perpendicular to 
a D$p$-brane results in a D$(p+1)$-brane whereas T-duality 
along a longitudinal direction gives a D$(p-1)$-brane. 
We also note that in an effective theory D-branes can be described by their world-volume
action. We will become more concrete about such actions in 
section~\ref{sec_cy_flux_echt_orient}, but let us state already here that they contain
couplings of the schematic form
\eq{
  \label{t_duality_cs}
  \mathcal{S}_{{\rm D}p} \supset \int_{\Gamma_{p+1}}
  \Bigl(    C_{p+1} + C_{p-1} \wedge B + \tfrac{1}{2} \op C_{p-3} \wedge B\wedge B + \ldots 
  \Bigr)\,,
}
where $\Gamma_{p+1}$ denotes the $(p+1)$-dimensional world-volume of the D$p$-brane
and $C_{p}$ are the  R-R gauge potentials mentioned above. 
Under a T-duality transformation along $\Gamma_{p+1}$ 
the D-brane world-volume is mapped to $\Gamma_p$ and 
-- ignoring for a moment the $B$-field -- 
correspondingly 
the R-R potentials should transform as
$C_{p+1} \to  C_p$. Similarly, a T-duality perpendicular to $\Gamma_{p+1}$ 
results in $ \Gamma_{p+2}$ together with $C_{p+1} \to  C_{p+2}$. 
This mapping agrees with our observation \eqref{t-duality_iiab} 
that T-duality along a single direction interpolates between type IIA and 
type IIB string theory, as all R-R potentials of IIA are odd while the potentials 
in IIB are of even degree.

We now want to make the above argumentation more precise and take into account the 
Kalb-Ramond $B$-field. 
We consider a compactification on a circle along the $X^9$-direction and 
assume that the components $B_{ij}$ are independent of $X^9$. 
This means we can employ our results from section~\ref{section_t_dual_1}, 
in particular, the transformation rules for the NS-NS sector given in 
equation \eqref{dual_back_9494}.
Denoting the components of $C_p$ by $C_{p \op|\op i_1 \ldots }$
and taking $i_n \neq 9$, 
for the transformation of the R-R potentials under T-duality one finds that \cite{Hassan:1999bv} 
\eq{
  \arraycolsep2pt
  \begin{array}{lclcl}
  \displaystyle \check C_{p\op|\op 9 i_2 \ldots i_p} &=& \displaystyle C_{p-1\op|\op i_2 \ldots i_p} &-&
  \displaystyle   (p-1) 
  \frac{ G_{9[\ul{i_2}} \op C_{p-1\op|\op 9 \ul{i_3} \ldots \ul{i_p}]}}{G_{99}} \,,
  \\[12pt]
  \displaystyle \check C_{p\op|\op i_1 \ldots i_p} &=& \displaystyle C_{p+1\op|\op 9i_1 \ldots i_p} &-& 
  \displaystyle  p \,
  B_{9[\ul{\mu_1}} \op \check C_{p\op|\op 9 \ul{i_2} \ldots \ul{i_p}]} \,.
  \end{array}
}
Here we underlined the indices which are part of the anti-symmetrisation, 
we made a particular choice of sign as compared to \cite{Hassan:1999bv},
and we note that the second 
line is defined recursively in terms of the first.

%%%%%%%%%%%%%%%%%%%%%%%%%%%%%%%%%%%%%%%%%%%%%%%
%%%%%%%%%%%%%%%%%%%%%%%%%%%%%%%%%%%%%%%%%%%%%%%

\subsubsection*{Remarks}

We close this section with the following remarks:
\begin{itemize}

\item The transformation rules of the R-R potentials in type II theories 
were first derived from a supergravity point of view 
in \cite{Bergshoeff:1995as,Bergshoeff:1996ui},
and were later generalised in \cite{Meessen:1998qm}.
The D-brane approach described above has been employed
in \cite{Bergshoeff:1996cy,Green:1996bh}, and
in \cite{Hassan:1999bv,Hassan:1999mm} the transformation of the R-R potentials 
has been determined via 
the supersymmetry transformations of fermions.

\item From a world-sheet perspective the transformation rules have been studied in 
\cite{Cvetic:1999zs,Kulik:2000nr} via the Green-Schwarz formalism,
in \cite{Benichou:2008it} via
the pure spinor formalism of \cite{Berkovits:2000fe},
and in \cite{Sfetsos:2010xa} via canonical transformations. 
To our knowledge, however, global properties of the world-sheet and
of R-R fluxes have not been discussed in the same detail as 
for the NS-NS sector.

\end{itemize}

%%%%%%%%%%%%%%%%%%%%%%%%%%%%%%%%%%%%%%%%%%%%%%%
%%%%%%%%%%%%%%%%%%%%%%%%%%%%%%%%%%%%%%%%%%%%%%%
%%%%%%%%%%%%%%%%%%%%%%%%%%%%%%%%%%%%%%%%%%%%%%%
%%%%%%%%%%%%%%%%%%%%%%%%%%%%%%%%%%%%%%%%%%%%%%%
%%%%%%%%%%%%%%%%%%%%%%%%%%%%%%%%%%%%%%%%%%%%%%%
%%%%%%%%%%%%%%%%%%%%%%%%%%%%%%%%%%%%%%%%%%%%%%%
%%%%%%%%%%%%%%%%%%%%%%%%%%%%%%%%%%%%%%%%%%%%%%%
%%%%%%%%%%%%%%%%%%%%%%%%%%%%%%%%%%%%%%%%%%%%%%%
%%%%%%%%%%%%%%%%%%%%%%%%%%%%%%%%%%%%%%%%%%%%%%%
%%%%%%%%%%%%%%%%%%%%%%%%%%%%%%%%%%%%%%%%%%%%%%%
%%%%%%%%%%%%%%%%%%%%%%%%%%%%%%%%%%%%%%%%%%%%%%%
%%%%%%%%%%%%%%%%%%%%%%%%%%%%%%%%%%%%%%%%%%%%%%%
%%%%%%%%%%%%%%%%%%%%%%%%%%%%%%%%%%%%%%%%%%%%%%%
%%%%%%%%%%%%%%%%%%%%%%%%%%%%%%%%%%%%%%%%%%%%%%%
%%%%%%%%%%%%%%%%%%%%%%%%%%%%%%%%%%%%%%%%%%%%%%%
%%%%%%%%%%%%%%%%%%%%%%%%%%%%%%%%%%%%%%%%%%%%%%%

\clearpage
\section{Poisson-Lie duality}
\label{cha_poisson_lie}

In this section we discuss 
Poisson-Lie T-duality, which is a framework  
for describing non-abelian T-duality transformations.
Poisson-Lie T-duality has been developed in the papers
 \cite{Klimcik:1995ux,Klimcik:1995jn,Klimcik:1995dy} and  here
we give an overview of the main idea.
Poisson-Lie duality will not play a role in our discussion of non-geometric backgrounds, 
but we include this topic for completeness.

%%%%%%%%%%%%%%%%%%%%%%%%%%%%%%%%%%%%%%%%%%%%%%%
%%%%%%%%%%%%%%%%%%%%%%%%%%%%%%%%%%%%%%%%%%%%%%%

\subsubsection*{Variation of the action}

We start from the Euclidean world-sheet action of a closed string given in \eqref{action_02e}, which 
we recall for convenience as
\eq{
  \label{plt_11}
  \mathcal S =& -\frac{1}{2\pi \alpha'} \int_{\Sigma}
  \Bigl[ \,\tfrac{1}{2}\op G_{ij} \, d X^{i}\wedge\star d X^{j} - \tfrac{i}{2}\op B_{ij} \,dX^{i}\wedge dX^{j}
  + \tfrac{\alpha'}{2}\op\mathsf R\, \phi \star 1\, \Bigr] 
  \,,
}
with $G_{ij}$ and $B_{ij}$ the components of the target-space metric and Kalb-Ramond field. 
The dilaton is denoted  by $\phi$.
We then perform a variation of the action \eqref{plt_11} with respect to infinitesimal local transformations
\eq{
  \label{plt_20}
  \delta_{\epsilon} X^{i} = \epsilon^{\alpha}\, k_{\alpha}^{i} (X) \,,
}
where $\epsilon^{\alpha}\equiv\epsilon^{\alpha}(\sigma^{\mathsf a})\ll1$ depend on the coordinates $\sigma^{\mathsf a}$ of the two-dimensional world-sheet $\Sigma$ and $\alpha=1,\ldots, N$.
The vector-fields $k_{\alpha}$ are required to satisfy a Lie algebra $\mathfrak g$ with structure constants
$f_{\alpha\beta}{}^{\gamma}$ specified in the following way
\eq{
   \bigl[ k_{\alpha} , k_{\beta} \bigr] = f_{\alpha\beta}{}^{\gamma} \op k_{\gamma} \,.
}
Employing the notation 
$G = \tfrac{1}{2}\op G_{ij} \, d X^{i}\wedge\star d X^{j}$ and 
$B= \tfrac{1}{2}\op B_{ij} \,dX^{i}\wedge dX^{j}$, the variation of the action with respect to \eqref{plt_20}
can be expressed as
\eq{
  \label{plt_01}
  \delta_{\epsilon} \mathcal S = -\frac{1}{2\pi \alpha'} \int_{\Sigma}
  \epsilon^{\alpha} \mathcal L_{k_{\alpha}} \Bigl( G - i\op B + \tfrac{\alpha'}{2}R\, \phi \star 1\Bigr)
  +\frac{1}{2\pi \alpha'} \int_{\Sigma} \epsilon^{\alpha} \, dJ_{\alpha} \,,
}
where $ \mathcal L_{k_{\alpha}} G = \tfrac{1}{2} ( \mathcal L_{k_{\alpha}} G)_{ij}\, d X^{i}\wedge\star d X^{j}$
and $\mathcal L_k$ denotes the target-space Lie derivative along the direction $k$; similar expressions
apply to 
$B$ and $\phi$. The currents $J_{\alpha}$ are defined up to closed terms and read
\eq{
  \label{plt_05}
  J_{\alpha}  = \star \mathsf k_{\alpha} - i\op \iota_{k_{\alpha}} B \,,
}
where $\mathsf k_{\alpha} = k^i_{\alpha} \op G_{ij} dX^j$ are the one-forms dual to the vector-fields
$k_{\alpha}$.
Let us note that using the equations of motion for $X^{i}$ given in \eqref{eom_8494949},
 on-shell the variation \eqref{plt_01} vanishes.

Furthermore, we make the assumption that the dilaton obeys
$\mathcal L_{k_{\alpha}} \phi = 0$.
If this condition is not satisfied, technical subtleties appear which have 
been addressed for instance in 
\cite{Tyurin:1995bu,VonUnge:2002xjf,Hlavaty:2004mr,Hlavaty:2006hu,Jurco:2017gii}.
However, with the Lie derivative of the dilaton vanishing we can then summarise that on-shell
\eq{
  \label{plt_03}
  0 = \mathcal L_{k_{\alpha}}  \bigl( G - i\op B\bigr)  - d J_{\alpha}\op \Bigr\rvert_{\mbox{\scriptsize on-shell}}\,.
}

%%%%%%%%%%%%%%%%%%%%%%%%%%%%%%%%%%%%%%%%%%%%%%%
%%%%%%%%%%%%%%%%%%%%%%%%%%%%%%%%%%%%%%%%%%%%%%%

\subsubsection*{(Non-conserved) Noether currents}

If the Lie derivative acting on $G$, $B$ and $\phi$ vanishes, it follows from \eqref{plt_01}
that the currents $J_{\alpha}$ have to be closed. This is basically the situation we have
studied in section~\ref{sec_buscher} (modulo subtleties regarding the one-forms $v_{\alpha}$).
In particular, the vector-fields $k_{\alpha}$ have to be Killing vector-fields and T-duality 
is along a direction of isometry. 
However, it turns out that imposing a more general condition on the currents leads to an interesting structure.
In particular, let us demand that on-shell
\eq{
  \label{plt_02}
  0 = dJ_{\alpha} - \frac{i}{2} \op \tilde f_{\alpha}{}^{\beta\gamma} \op J_{\beta} \wedge J_{\gamma} \,,
}
where $\tilde f_{\alpha}{}^{\beta\gamma}$ are  structure constants of some Lie algebra $\tilde{\mathfrak g}$ (which 
we come back to below). 
In this case $\mathcal L_{k_{\alpha}} (G-i\op B)$ no longer vanishes on-shell and the 
$k_{\alpha}$ do not correspond to isometries. 
If the currents $J_{\alpha}$ are not closed on-shell but satisfy \eqref{plt_02}, 
the world-sheet theory is said to have a Poisson-Lie symmetry 
\cite{Klimcik:1995ux,Klimcik:1995jn,Klimcik:1995dy} or to have a non-commutative conservation law  \cite{Klimcik:1995ux}.

Let us now use the integrability condition for the Lie derivative given by
$[\mathcal L_{k_{\alpha}}, \mathcal L_{k_{\beta}}] = \mathcal L_{[k_{\alpha},k_{\beta}]}$ together with
\eqref{plt_02} and the equation of motion \eqref{eom_8494949}. 
More concretely, we evaluate using the relation \eqref{plt_02}
\eq{
  [ L_{k_{\alpha}},L_{k_{\beta}} ] \,\bigl( G - i\op B\bigr) = \mathcal L_{[k_{\alpha},k_{\beta}]}
  \,\bigl( G - i\op B\bigr) \,.
}
This leads to a restrictions on the structure constants of the Lie algebras $\mathfrak g$ and $\tilde{\mathfrak g}$ 
of the form
\eq{
  \label{plt_04}
  0 = \tilde f_{\gamma}{}^{\mu\nu}f_{\alpha\beta}{}^{\gamma}  
  &- \tilde f_{\alpha}{}^{\gamma\mu} f_{\beta\gamma}{}^{\nu}
  + \tilde f_{\beta}{}^{\gamma\mu} f_{\alpha\gamma}{}^{\nu} \\
  &+ \tilde f_{\alpha}{}^{\gamma\nu} f_{\beta\gamma}{}^{\mu}
  - \tilde f_{\beta}{}^{\gamma\nu} f_{\alpha\gamma}{}^{\mu} \,,
}
where all indices take values $1,\ldots, N$.
Note that these conditions are invariant under the exchange $f_{\alpha\beta}{}^{\gamma} \leftrightarrow \tilde f_{\gamma}{}^{\alpha\beta}$, which  suggests that there should exist a ``dual'' world-sheet theory 
in which the roles of $\tilde f_{\alpha}{}^{\beta\gamma}$ and $f_{\alpha\beta}{}^{\gamma}$ are interchanged.
In particular, the dual currents $\tilde J^{\alpha}$ in the dual theory should satisfy
\eq{
  0 = d\tilde J^{\alpha} - \frac{i}{2} \op  f_{\beta\gamma}{}^{\alpha} \op \tilde J^{\beta} \wedge \tilde J^{\gamma} \,,
}
and the analogue of \eqref{plt_03} imposes restrictions on  a dual metric and Kalb-Ramond field via
\eq{
    \mathcal L_{\tilde k^{\alpha}}  \bigl( \tilde G - i\op \tilde B\bigr)  = d \tilde J^{\alpha} \,,
}
with dual vector-fields $\tilde k^{\alpha}$ obeying the algebra
\eq{
   \bigl[ \tilde k^{\alpha} , \tilde k^{\beta} \bigr] = \tilde f_{\gamma}{}^{\alpha\beta}\op \tilde k^{\gamma} \,.
}

%%%%%%%%%%%%%%%%%%%%%%%%%%%%%%%%%%%%%%%%%%%%%%%
%%%%%%%%%%%%%%%%%%%%%%%%%%%%%%%%%%%%%%%%%%%%%%%

\subsubsection*{Drinfeld double and Manin triples}

Let us come back to the relations shown in \eqref{plt_04}, where $f_{\alpha\beta}{}^{\gamma}$ 
are the structure constants of a Lie algebra $\mathfrak g$ and $\tilde f_{\alpha}{}^{\beta\gamma}$ 
correspond to a Lie algebra $\tilde{\mathfrak g}$. 
These equations are the standard relations which have to be obeyed by the structure constants
of a Lie bi-algebra $(\mathfrak g,\tilde{\mathfrak g})$ \cite{Drinfeld:1986in,Alekseev:1993qs,Falceto:1992bf}.

A special case of a Lie bi-algebra  is the so-called 
Drinfeld double,  which will become important below.
 A Drinfeld double is any Lie group $D$ such that its Lie algebra
$\mathfrak d$ can be decomposed into a pair of maximally isotropic sub-algebras 
$(\mathfrak g,\tilde{\mathfrak g})$
with respect to a non-degenerate invariant bilinear form $\langle \cdot,\cdot \rangle$ 
on $\mathfrak d$  \cite{Drinfeld:1986in,Klimcik:1995jn}.
Let us explain this terminology:
an isotropic subspace of $\mathfrak d$ is such that the bilinear  pairing of any two 
of its elements vanishes, and maximally isotropic means that the subspace cannot be 
enlarged while preserving the property of isotropy. 
Together with the choice of a canonical pairing between $\mathfrak g$ and $\tilde{\mathfrak g}$ 
this reads in formulas
\eq{
  \label{plt_10a}
  \langle t_{\alpha}, t_{\beta} \rangle =0 \,, \hspace{40pt}
  \langle \tilde t^{\alpha}, \tilde t^{\beta} \rangle =0 \,,  \hspace{40pt}
  \langle t_{\alpha}, \tilde t^{\beta} \rangle = \delta_{\alpha}{}^{\beta} \,,  
}  
where $t_{\alpha}\in\mathfrak g$ and $\tilde t^{\alpha} \in \tilde{\mathfrak g}$ are 
the generators of the two Lie algebras. The structure constants are defined via  the commutators in 
the following way
\eq{
  \arraycolsep2pt
  \begin{array}{lcl}
  \displaystyle [ t_{\alpha},t_{\beta} ] &=&   \displaystyle i\op f_{\alpha\beta}{}^{\gamma}\op t_{\gamma} \,,
  \\[8pt]
    \displaystyle[ \tilde t^{\alpha}, \tilde t^{\beta} ] &=&  \displaystyle
     i\op \tilde f_{\gamma}{}^{\alpha\beta}\op \tilde t^{\gamma} \,,
  \end{array}
  \hspace{50pt}
  [ t_{\alpha}, \tilde t^{\beta} ] = -i\op  f_{\alpha\gamma}{}^{\beta}\op \tilde t^{\gamma} + i\op
  \tilde f_{\alpha}{}^{\beta\gamma} \op t_{\gamma} \,,
}
where we also included the mixed commutator. The latter 
follows from the invariance of the pairing $\langle \cdot,\cdot \rangle$ 
on $\mathfrak d$, but 
can also be determined by comparing
the mixed Jacobi identity in $\mathfrak d$ to \eqref{plt_04}.
Finally, any such decomposition of $\mathfrak d$ into such subspaces is
called a Manin triple, and there are at least two of them:
$\mathfrak d = \mathfrak g + \tilde{\mathfrak g}$ and 
$\mathfrak d =  \tilde{\mathfrak g} + \mathfrak g$.

However, in general a given Drinfeld double Lie algebra $\mathfrak d$ 
can be decomposed into bi-algebras 
in several ways \cite{Klimcik:1995ux},
which leads to the concept of so-called Poisson-Lie {\em plurality} \cite{VonUnge:2002xjf}.
In six dimensions, for instance, the  Drinfeld doubles have been classified in 
\cite{Jafarizadeh:1999xv,Snobl:2002kq}.
Furthermore, the relation between the $O(D,D,\mathbb Z)$ transformations discussed 
in section~\ref{sec_cft_torus} and Poisson-Lie duality has been studied in \cite{Lust:2018jsx}.

%%%%%%%%%%%%%%%%%%%%%%%%%%%%%%%%%%%%%%%%%%%%%%%
%%%%%%%%%%%%%%%%%%%%%%%%%%%%%%%%%%%%%%%%%%%%%%%

\subsubsection*{Duality transformation}

So far we have observed that the consistency condition \eqref{plt_04} is 
invariant under exchanging the algebras $\mathfrak g$ and $\tilde{\mathfrak g}$,
and hence there should exist a dual model with their roles interchanged.
We now want to make this more precise and construct the mapping 
between the two models. 
To do so we restrict our discussion to world-sheet theories with group manifolds as
target-spaces,  though we will present a generalisation below.

We consider a two-dimensional world-sheet theory which has a group $G$ as an
$N$-dimensional target-space. The left- and right-invariant forms 
for the group $G$ can then be written in terms of $g\equiv g(\sigma^{\mathsf a})\in G$ as
\eq{
   \omega_L = g^{-1} (dg) = \omega_L{}^{\alpha} \op t_{\alpha}\,, \hspace{70pt}
  \omega_R = (dg) \op g^{-1}=\omega_R{}^{\alpha}\op t_{\alpha} \,,
}
and they satisfy the Maurer-Cartan equations as follows
\eq{
  \label{wzw_003}
  0 = d\op \omega_L{}^{\alpha} + \frac{i}{2}\op f_{\beta\gamma}{}^{\alpha} \omega_L{}^{\beta}
    \wedge \omega_L{}^{\gamma} \,, 
  \hspace{50pt}
  0 = d\op \omega_R{}^{\alpha} - \frac{i}{2}\op f_{\beta\gamma}{}^{\alpha} \omega_R{}^{\beta}
    \wedge \omega_R{}^{\gamma} \,.
}
Similar expressions apply for the dual Lie group $\tilde G$ with corresponding algebra $\tilde{\mathfrak g}$.
The world-sheet action for the present situation can then be expressed schematically as
\eq{
  \label{plt_06}
  \mathcal S = \frac{1}{2\pi \alpha'} \int_{\Sigma}
  \Bigl[ \,\tfrac{1}{2}\op G_{\alpha\beta} \, \omega_L^{\alpha}\wedge\star \omega_L^{\beta} + \tfrac{i}{2}\op B_{\alpha\beta} \,\omega_L^{\alpha}\wedge \omega_L^{\beta}
\, \Bigr] 
  \,,
}
where $G_{\alpha\beta}$ and $B_{\alpha\beta}$ are the components of the target-space metric and 
Kalb-Ramond field. 
The current \eqref{plt_05} satisfying \eqref{plt_02}  is given in terms of the right-invariant Maurer-Cartan form of the dual group as
\eq{
  \label{plt_07}
  J_{\alpha} = \tilde\omega_{R\op \alpha}\,,
}
which indeed satisfies \eqref{plt_02} as can be seen from the second relation in \eqref{wzw_003}
applied to the dual algebra. 
Now, given a solution $g\equiv g(\sigma^{\mathsf a})\in G$ to the equations of motion of the 
sigma model \eqref{plt_06}, we can lift this solution to the Drinfeld double $D$. More concretely,
we can express $d\in D$ as
\eq{
  d = g\cdot \tilde g \,, \hspace{70pt} g\in G\,,\quad \tilde g\in\tilde G \,,
}
where the multiplication is done in $D$. It is known \cite{Alekseev:1993qs} that for every 
$d\in D$ there are two decompositions applicable, namely 
\eq{
  d = g\cdot \tilde g = \tilde h \cdot h\,.
}
One can then show that $\tilde h\equiv \tilde h(\sigma^{\mathsf a})\in\tilde G$ is a solution 
of the dual sigma-model and $h\equiv h(\sigma^{\mathsf a})\in G$ gives rise to the dual analogue of  
\eqref{plt_07} \cite{Klimcik:1995ux}.

Let us now give some more concrete formulas for the dual sigma model. To do so, we are
going to work with the matrix
\eq{
  \label{plt_071}
  E_{\alpha\beta}(g) = G_{\alpha\beta}(g) + B_{\alpha\beta}(g) \,,
}
which contains the information about the target-space background. The dependence on 
$g\in G$ specifies the position on the group manifold -- and 
since $G$ acts transitively on the target-space, we can consider $E_{\alpha\beta}$ say 
at the origin $g=e$ where $e$ is the identity element, and obtain the 
full dependence via the adjoint action. 
More concretely, for the adjoint action of $G$ on $\mathfrak d = \mathfrak g + \tilde{\mathfrak g}$ 
we define
\eq{
  g^{-1} \binom{t^{\alpha}}{\tilde t_{\alpha}} \op g = 
  \arraycolsep2pt
  \left( \begin{array}{cc} a^{\alpha}{}_{\beta}(g) & b^{\alpha\beta}(g) \\[2pt] 0 & d_{\alpha}{}^{\beta}
  (g) \end{array}\right)
  \binom{t^{\beta}}{\tilde t_{\beta}} \,,
}
where $a(g)$, $b(g)$ and $d(g)$ are $N\times N$-dimensional matrices. 
The dependence of \eqref{plt_071} on $g$ can be expressed using matrix notation as
\cite{Klimcik:1995ux,Klimcik:1995jn}
\eq{
  \label{plt_10}
  E(g) = \bigl[ a^T(g)+ E(e)\op b^T(g) \bigr]^{-1} E(e) \op d^T(g) \,,
}
and for the dual background the corresponding matrix $\tilde E^{\alpha\beta}(\tilde g)$ can
be expressed in a similar way, namely
\eq{
  \label{plt_09}
  \tilde E(\tilde g) = \bigl[ \tilde a^T(\tilde g)+ \tilde E(\tilde e)\op \tilde b^T(\tilde g) \bigr]^{-1} \tilde E(\tilde e) \op \tilde d^T(\tilde g)\,,
}
where the roles of $t_{\alpha}$ and $\tilde t^{\alpha}$ have been interchanged. 
The final point is  that at the origin $e\in D$ of the Drinfeld double $D$ the matrix $E_{\alpha\beta}(e)$
can be regarded as a map $E(e):\mathfrak g \to \tilde{\mathfrak g}$, and the matrix $\tilde E^{\alpha\beta}(\tilde e)$ is a map $\tilde E(\tilde e):\tilde{\mathfrak g}\to\mathfrak g$, where $e\in D$ is also the unit $e\in G$ and 
$\tilde e \in \tilde G$. 
By comparing for instance with the abelian situation discussed around equation \eqref{odd_009},
we see that $\tilde E$ should be the inverse of $E$
\eq{
  \label{plt_09c}
  E(e)\op \tilde E(\tilde e) = \tilde E(\tilde e) \op E(e)  = \mathds 1 \,.
}
This relation has been derived via the equations of motion in \cite{Klimcik:1995ux}
and via a doubled sigma-model (cf. section~\ref{cha_dg}) in \cite{Klimcik:1995dy,Driezen:2016tnz},
and in this way we see that using \eqref{plt_09c} we can be express \eqref{plt_09} in terms of \eqref{plt_10}.
In particular, the dual background can be written as (we simply replace $\tilde E(\tilde e)$ by
$E^{-1}(e)$ in \eqref{plt_09})
\eq{
  \label{plt_12}
  \tilde E(\tilde g) = \bigl[ \tilde a^T(\tilde g)+ E^{-1}(e)\op \tilde b^T(\tilde g) \bigr]^{-1} E^{-1}(e) \op \tilde d^T(\tilde g)\,.
}

%%%%%%%%%%%%%%%%%%%%%%%%%%%%%%%%%%%%%%%%%%%%%%%
%%%%%%%%%%%%%%%%%%%%%%%%%%%%%%%%%%%%%%%%%%%%%%%

\subsubsection*{Example}

As an example for a Poisson-Lie duality, let us take $\mathfrak g$ to be a
non-abelian Lie algebra and $\tilde{\mathfrak g}$ to be an abelian one. 
In this case, the constraint \eqref{plt_04} is trivially satisfied.
The standard
example 
is given by a so-called principal chiral model on a simple group $G$ \cite{delaOssa:1992vci,Fridling:1983ha,Fradkin:1984ai}, for which 
the matrix \eqref{plt_07} takes the form
\eq{
  E_{\alpha\beta} =  k\op \delta_{\alpha\beta} \,,
}
where $k\in \mathbb Z^+$ denotes the level of the group $G$.
Going then through the dualisation procedure discussed above, one finds for 
the Poisson-Lie dual the expression
\eq{
  \label{plt_13}
  (\tilde E^{-1})_{\alpha\beta} = \delta_{\alpha\beta} + f_{\alpha\beta}{}^{\gamma}\chi_{\gamma} \,,
}
where $f_{\alpha\beta}{}^{\gamma}$ are the structure constants of $\mathfrak g$ and 
$\chi_{\alpha}$ are $N$ coordinates for the dual circles. 
The dual metric and $B$-field correspond to the symmetric and anti-symmetric part in \eqref{plt_13}, 
which agree with the 
expressions found in \cite{delaOssa:1992vci,Fridling:1983ha,Fradkin:1984ai}.
In particular, for $G=SU(2)$ with $f_{\alpha\beta}{}^{\gamma} = \epsilon_{\alpha\beta}{}^{\gamma}$ 
one finds
\eq{
  \tilde G^{\alpha\beta} = \frac{1}{k\op (k^2+\chi^2)} \bigl( \op k^2 \op \delta^{\alpha\beta} + \chi^{\alpha}\chi^{\beta}\bigr)\,,
  \hspace{30pt}
  \tilde B^{\alpha\beta} = - \frac{1}{k^2 + \chi^2}\, \epsilon^{\alpha\beta}{}_{\gamma} \chi^{\gamma} \,,
}
where $\chi^{\alpha} = \delta^{\alpha\beta}\chi_{\beta}$ and $\chi^2 = \chi_1^2 + \chi_2^2+\chi_3^2$.

%%%%%%%%%%%%%%%%%%%%%%%%%%%%%%%%%%%%%%%%%%%%%%%
%%%%%%%%%%%%%%%%%%%%%%%%%%%%%%%%%%%%%%%%%%%%%%%

\subsubsection*{Generalisation}

Instead of considering an $N$-dimensional Lie group $G$ as a target-space, we can also study manifolds 
where $G$ is fibered over a base-manifold $\mathcal B$. Let us denote local 
coordinates in the base by $y^m$ with $m=1,\ldots, d_{\mathcal B}$. 
The generalisation of \eqref{plt_07} then takes the form
\eq{
  E_{IJ}(g,y) = \left( \arraycolsep2pt \begin{array}{cc}
  E_{\alpha\beta}(g,y) & E_{\alpha n}(g,y) \\[4pt]
  E_{m \beta}(g,y) & E_{mn} (y) \end{array}\right),
}
where we used the combined index $I = \{\alpha,m\}$ and the corresponding basis 
of one-forms reads $\{ \omega_L^{\alpha}, dy^m\}$. 
For the analogue of \eqref{plt_10} we then define
\eq{
  A(g) =  \left( \arraycolsep2pt \begin{array}{cc}
  a(g) & 0 \\[4pt]
  0 & \mathds 1 \end{array}\right),
  \hspace{30pt}
  B(g) =  \left( \arraycolsep2pt \begin{array}{cc}
  b(g) & 0  \\[4pt]
  0 & 0 \end{array}\right),
  \hspace{30pt}
  D(g) =  \left( \arraycolsep2pt \begin{array}{cc}
  d(g) & 0 \\[4pt]
  0 & \mathds 1 \end{array}\right),
}
and we find
\eq{
  E(g,y) = \bigl[ A^T(g)+ E(e,y)\op B^T(g) \bigr]^{-1} E(e,y) \op D^T(g) \,.
}
The background after dualisation along the fibre is encoded in the generalisation 
of \eqref{plt_12}, namely
\eq{
  \tilde E(\tilde g,y) = \bigl[ \tilde A^T(\tilde g,y)+ E^{-1}(e,y)\op \tilde B^T(\tilde g) \bigr]^{-1} E^{-1}(e,y) 
  \op \tilde D^T(\tilde g,y)\,.
}

%%%%%%%%%%%%%%%%%%%%%%%%%%%%%%%%%%%%%%%%%%%%%%%
%%%%%%%%%%%%%%%%%%%%%%%%%%%%%%%%%%%%%%%%%%%%%%%

\subsubsection*{Remarks}

Many  aspects of Poisson-Lie duality and generalisations thereof have been discussed in the 
literature. Here we want to briefly mention some of them:
\begin{itemize}

\item As one can see from the last relation in \eqref{plt_10a}, the Lie algebras $\mathfrak g$ and 
$\tilde{\mathfrak g}$ are dual to each other. The latter can therefore be identified with 
the dual space $\mathfrak g^*$, and one is lead to the framework of Courant algebroids
(to be discussed in section~\ref{sec_lie_courant}). 
This has been investigated for instance in \cite{Severa:2015hta}.

\item Poisson-Lie duality can also be understood at the classical level as a 
canonical transformation. 
This has been discussed in \cite{Sfetsos:1997pi}, where the explicit form of the generating
functional is given, and for Poisson-Lie plurality this has been addressed in \cite{Hlavaty:2006yz}.

\item In order to establish Poisson-Lie duality not only at the classical but also at the 
quantum level, a path-integral derivation of the duality is needed. 
This has been investigated  
in \cite{Alekseev:1995ym,Tyurin:1995bu}.

\item Poisson-Lie duality in the context of open strings and D-branes has been discussed in 
\cite{Klimcik:1995np,Klimcik:1996hp},
and in \cite{Hassler:2017yza} the transformation of the
Ramond-Ramond sector of 
type II string theory under Poisson-Lie duality has been studied
(using the framework of double field theory, cf. section~\ref{sec_dft}).

\end{itemize}

%%%%%%%%%%%%%%%%%%%%%%%%%%%%%%%%%%%%%%%%%%%%%%%
%%%%%%%%%%%%%%%%%%%%%%%%%%%%%%%%%%%%%%%%%%%%%%%
%%%%%%%%%%%%%%%%%%%%%%%%%%%%%%%%%%%%%%%%%%%%%%%
%%%%%%%%%%%%%%%%%%%%%%%%%%%%%%%%%%%%%%%%%%%%%%%
%%%%%%%%%%%%%%%%%%%%%%%%%%%%%%%%%%%%%%%%%%%%%%%
%%%%%%%%%%%%%%%%%%%%%%%%%%%%%%%%%%%%%%%%%%%%%%%
%%%%%%%%%%%%%%%%%%%%%%%%%%%%%%%%%%%%%%%%%%%%%%%
%%%%%%%%%%%%%%%%%%%%%%%%%%%%%%%%%%%%%%%%%%%%%%%
%%%%%%%%%%%%%%%%%%%%%%%%%%%%%%%%%%%%%%%%%%%%%%%
%%%%%%%%%%%%%%%%%%%%%%%%%%%%%%%%%%%%%%%%%%%%%%%
%%%%%%%%%%%%%%%%%%%%%%%%%%%%%%%%%%%%%%%%%%%%%%%
%%%%%%%%%%%%%%%%%%%%%%%%%%%%%%%%%%%%%%%%%%%%%%%
%%%%%%%%%%%%%%%%%%%%%%%%%%%%%%%%%%%%%%%%%%%%%%%
%%%%%%%%%%%%%%%%%%%%%%%%%%%%%%%%%%%%%%%%%%%%%%%
%%%%%%%%%%%%%%%%%%%%%%%%%%%%%%%%%%%%%%%%%%%%%%%
%%%%%%%%%%%%%%%%%%%%%%%%%%%%%%%%%%%%%%%%%%%%%%%
%%%%%%%%%%%%%%%%%%%%%%%%%%%%%%%%%%%%%%%%%%%%%%%
%%%%%%%%%%%%%%%%%%%%%%%%%%%%%%%%%%%%%%%%%%%%%%%
%%%%%%%%%%%%%%%%%%%%%%%%%%%%%%%%%%%%%%%%%%%%%%%
%%%%%%%%%%%%%%%%%%%%%%%%%%%%%%%%%%%%%%%%%%%%%%%
%%%%%%%%%%%%%%%%%%%%%%%%%%%%%%%%%%%%%%%%%%%%%%%
%%%%%%%%%%%%%%%%%%%%%%%%%%%%%%%%%%%%%%%%%%%%%%%
%%%%%%%%%%%%%%%%%%%%%%%%%%%%%%%%%%%%%%%%%%%%%%%
%%%%%%%%%%%%%%%%%%%%%%%%%%%%%%%%%%%%%%%%%%%%%%%
%%%%%%%%%%%%%%%%%%%%%%%%%%%%%%%%%%%%%%%%%%%%%%%
%%%%%%%%%%%%%%%%%%%%%%%%%%%%%%%%%%%%%%%%%%%%%%%
%%%%%%%%%%%%%%%%%%%%%%%%%%%%%%%%%%%%%%%%%%%%%%%
%%%%%%%%%%%%%%%%%%%%%%%%%%%%%%%%%%%%%%%%%%%%%%%
%%%%%%%%%%%%%%%%%%%%%%%%%%%%%%%%%%%%%%%%%%%%%%%
%%%%%%%%%%%%%%%%%%%%%%%%%%%%%%%%%%%%%%%%%%%%%%%
%%%%%%%%%%%%%%%%%%%%%%%%%%%%%%%%%%%%%%%%%%%%%%%
%%%%%%%%%%%%%%%%%%%%%%%%%%%%%%%%%%%%%%%%%%%%%%%

\clearpage  
\section{Non-geometry}
\label{sec_first_steps}

After having studied T-duality transformations from different perspectives, we now
turn to non-geometric backgrounds. In this section we discuss the 
standard example for a non-geometric background, namely the three-torus with 
$H$-flux \cite{Dasgupta:1999ss,Kachru:2002sk,Hull:2004in,Shelton:2005cf} with its T-dual configurations, 
and in later sections  generalise this example. 

%%%%%%%%%%%%%%%%%%%%%%%%%%%%%%%%%%%%%%%%%%%%%%%
%%%%%%%%%%%%%%%%%%%%%%%%%%%%%%%%%%%%%%%%%%%%%%%
%%%%%%%%%%%%%%%%%%%%%%%%%%%%%%%%%%%%%%%%%%%%%%%
%%%%%%%%%%%%%%%%%%%%%%%%%%%%%%%%%%%%%%%%%%%%%%%

\subsection{Three-torus with \texorpdfstring{$H$}{H}-flux}
\label{sec_ex1_hhh}

Let us consider a flat three-torus $\mathbb T^3$
with non-trivial field strength $H=dB$ for the Kalb-Ramond $B$-field.
The components of the metric  in the coordinate basis of
one-forms $\{dX^1,dX^2,dX^3\}$ are taken to be of the form
\eq{
  \label{metric_01}
  G_{ij} = \left( \begin{array}{ccc} 
  R_1^2 & 0 & 0 \\ 
  0 & R_2^2 & 0 \\
  0 & 0 & R_3^2
  \end{array} \right),
}  
and the topology is characterised by the identifications $X^i\simeq X^i+2\pi$ for $i=1,2,3$.
The components of the field strength $H$  are chosen to be constant,
which, keeping in mind the quantisation condition \eqref{quantisation}, leads to
\eq{
  \label{ex1_metric_98}
  H = \frac{\alpha'}{2\pi}\, h \, dX^1 \wedge dX^2\wedge dX^3 \,,
  \hspace{60pt}
  h\in  \mathbb Z \,.
  \hspace{-30pt}
}
The dilaton for this background is also taken to be constant, that is $\phi = \phi_0 = {\rm const.}$
We  note that in the basis $\{\partial_1,\partial_2,\partial_3\}$ dual to the one-forms $dX^i$, 
the Killing vectors respecting the periodic identification of the torus   
can be chosen as
\eq{
  \label{ex1_killing_02}
  \arraycolsep2pt
  k_{1}^i = \left( \begin{array}{c} 1 \\ 0 \\ 0 \end{array} \right)  ,\hspace{50pt}
  k_{2}^i = \left( \begin{array}{c} 0 \\ 1 \\ 0 \end{array} \right)  ,\hspace{50pt}
  k_{3}^i = \left( \begin{array}{c} 0 \\ 0 \\ 1 \end{array} \right)  ,
}  
which satisfy an abelian algebra and hence $[k_{\alpha},k_{\beta}] = 0$.
Note that the Killing vectors $k_{\alpha}$ can be rescaled by 
non-vanishing constants, which however does not change the results discussed below. 
Finally, for later reference let us also introduce a vielbein basis $e^a$ with $a=1,2,3$ for 
the metric \eqref{metric_01}
as
\eq{
  e^1 = R_1\op dX^1 \,, \hspace{60pt}
  e^2 = R_2\op dX^2 \,, \hspace{60pt}
  e^3 = R_3\op dX^3 \,, 
}
which  satisfies 
$G = \tfrac{1}{2}\op G_{ij} \op dX^i\vee dX^j = \tfrac{1}{2}\op \delta_{ab} \op e^a \vee e^b$. 
The $H$-flux \eqref{ex1_metric_98} in this basis is expressed as
\eq{
  \label{ex1_2050}
  H = \frac{\alpha'}{2\pi}\,\frac{h}{R_1\op R_2 \op R_3} \, e^1\wedge e^2\wedge e^3 \,.
}

%%%%%%%%%%%%%%%%%%%%%%%%%%%%%%%%%%%%%%%%%%%%%%%
%%%%%%%%%%%%%%%%%%%%%%%%%%%%%%%%%%%%%%%%%%%%%%%
%%%%%%%%%%%%%%%%%%%%%%%%%%%%%%%%%%%%%%%%%%%%%%%
%%%%%%%%%%%%%%%%%%%%%%%%%%%%%%%%%%%%%%%%%%%%%%%

\subsection{Twisted torus}
\label{sec_twisted_torus_ex}

We now  perform a T-duality transformation along
one of the Killing vectors shown in \eqref{ex1_killing_02} 
and discuss the resulting background.

%%%%%%%%%%%%%%%%%%%%%%%%%%%%%%%%%%%%%%%%%%%%%%%
%%%%%%%%%%%%%%%%%%%%%%%%%%%%%%%%%%%%%%%%%%%%%%%

\subsubsection*{The background}

Let us consider say the Killing vector $k_1=\partial_1$ in \eqref{ex1_killing_02} 
and perform a T-duality transformation 
along this direction. 
Using either the Buscher rules \eqref{dual_back_9494} or following the approach of section~\ref{sec_wzw_action}, we find that the dual background is 
given by
\eq{
  \label{ex1_76}
  \arraycolsep2pt
  \begin{array}{lcl}
  \displaystyle \check G 
  &  = & \displaystyle  \frac{\alpha'^2}{R_1^2}\, \xi\otimes \xi + 
  R_2^2\, dX^2\otimes dX^2 + R_3^2\, dX^3\otimes dX^3  \,, 
  \\[11pt]
  \displaystyle \check H  & = & \displaystyle 0 \,, \\[4pt]
  \displaystyle \check \phi &=& \displaystyle \phi_0 - \log\left[\frac{R_1}{\sqrt{\alpha'}}\right] .
  \end{array}
}
Note that the one-form $\xi$ employed in \eqref{ex1_76} satisfies
\eq{
  \label{ex1_78}
  d\xi = \frac{h}{2\pi}\, dX^2\wedge dX^3 \,,
}
which means that $\xi$ depends on the $X^2$- or $X^3$-direction. 
In fact, \eqref{ex1_76} together with \eqref{ex1_78} describes 
a  principal $U(1)$-bundle over a two-dimensional base, which is also known 
as a twisted three-torus \cite{Dasgupta:1999ss,Kachru:2002sk}.

To be more concrete, let us choose the following parametrisation of $\xi$ (more details
on the choice of parametrisation can be found for instance in \cite{Plauschinn:2013wta})
\eq{
  \label{ex1_94892}
  \xi  = d\tilde X^1 - \frac{h}{2\pi}\op X^3\op dX^2 \,,
}
where $\tilde X^1$ denotes the dual coordinate.
From here we can infer the global structure of the twisted three-torus by demanding 
the metric $\check G$ in \eqref{ex1_76} to be well-defined. In particular, we find
\eq{
  \label{ex1_2051}
  \arraycolsep2pt
  \renewcommand{\arraystretch}{1.3}
  \begin{array}{c@{\hspace{15pt}}lcl@{\hspace{40pt}}lcl}
  1) & \tilde X^1 &\rightarrow&\tilde X^1 +2\pi \,, \\
  2) & X^2 &\rightarrow&X^2 +2\pi \,, \\
  3) & X^3 &\rightarrow&X^3 +2\pi \,,  & \tilde X^1 &\rightarrow&\tilde X^1 + h\op X^2 \,,
  \end{array}
}
which describes a two-torus along the direction $\tilde X^1$ and $X^2$ which is twisted when
going around the circle in the $X^3$-direction. This $\mathbb T^2$-fibration over $S^1$ is  called a twisted torus.

%%%%%%%%%%%%%%%%%%%%%%%%%%%%%%%%%%%%%%%%%%%%%%%
%%%%%%%%%%%%%%%%%%%%%%%%%%%%%%%%%%%%%%%%%%%%%%%

\subsubsection*{Geometric flux}

The background we started from carries a non-vanishing $H$-flux, and one can ask whether a similar quantity 
can be defined for the twisted torus. 
Clearly, for the dual background we have a vanishing $H$-flux $\check H=0$, but
we also observed  a non-trivial twisting \eqref{ex1_78} of the geometry. 
To investigate this point, we introduce a 
vielbein-basis $\check e^a$ for the dual metric $\check G$ in \eqref{ex1_76} as
follows
\eq{
  \check e^1 = \frac{\alpha'}{R_1}\op \xi\,, \hspace{60pt}
  \check e^2 = R_2\op dX^2\,, \hspace{60pt}
  \check e^3 = R_3\op dX^3\,.
}
The exterior algebra of these vielbeins is given by 
\eq{
  \label{ex1_2049}
  d\check e^1 =  \frac{\alpha'}{2\pi}\,\frac{h}{R_1\op R_2 \op R_3} \, e^2\wedge e^3 \,,
  \hspace{50pt}
  d\check e^2 = 0 \,,
  \hspace{60pt}
  d\check e^3 = 0 \,,  
}
from which we can read-off the (torsion-free) spin connection. 
Using the convention $de^a= \frac12 \op f_{bc}{}^a e^b\wedge e^c$, 
from \eqref{ex1_2049} we can find the non-vanishing structure constants as
\eq{
  \label{ex1_2052}
  f_{23}{}^1 = \frac{\alpha'}{2\pi}\, \frac{h}{R_1\op R_2 \op R_3} \,.
}
Comparing now with \eqref{ex1_2050} we see that the flux $h$ of the original model 
is encoded in the 
structure constants, and for this reason
$f_{ab}{}^c$ is also called a geometric flux. 
We come back to this point below.

%%%%%%%%%%%%%%%%%%%%%%%%%%%%%%%%%%%%%%%%%%%%%%%
%%%%%%%%%%%%%%%%%%%%%%%%%%%%%%%%%%%%%%%%%%%%%%%

\subsubsection*{Global properties}

Let us remark on the global properties of the twisted torus. First we note
that in general the Buscher rules discussed in section~\ref{sec_buscher} 
give the dual metric and $B$-field only locally. That means, the identifications
which describe the global structure of the dual background are not always
known.
\begin{itemize}

\item In particular, let us  recall from
page~\pageref{page_buscher_global} that if the circle along which one T-dualises 
allows for the standard quantisation of the corresponding 
coordinate $X^i$ with momentum and winding modes, then also the dual direction will be compact
with an appropriate quantisation \eqref{dual_back_010101010}. 
In the present situation of a three-torus with $H$-flux as the starting point, 
we see that the corresponding equation of motion \eqref{eom_8494949} is not 
a wave equation but reads for say the $X^1$-coordinate
\eq{
  0 = d\star dX^1 + i\, \frac{\alpha'}{2\pi }\op \frac{h}{R_1^2}\op dX^2\wedge dX^3 \,.
}
In general it is not known how to quantise this theory and deduce the momentum and winding 
modes. Strictly speaking, we therefore cannot conclude that 
the dual coordinate is compact with identification 
$\tilde X^1\to \tilde X^1 + 2\pi$ --- although this is very strongly expected.

\item Let us also recall our discussion from page~\pageref{page_global_v2} and
compare \eqref{ex1_94892} with \eqref{dual_rick_yeah}  to deduce the one-form
\eq{
  v = -\frac{\alpha'}{2\pi} \op h\, X^3 \op dX^2 
}
for the $H$-flux background. 
This one-form is not globally-defined, however, in the dualisation 
procedure only the combination $v+d\chi$ is required to be globally-defined. 
Noting that we have relabelled the dual coordinate as
$\tilde X^1 \equiv \chi/\alpha'$, we can therefore conclude that under $X^3\to X^3 + 2\pi$  we have
to demand that
\eq{
  \tilde X^1 \rightarrow\tilde X^1 + h\op X^2 \,,
}
which agrees with the identifications of the twisted torus shown in the third line of 
equation \eqref{ex1_2051}. 

\end{itemize}

%%%%%%%%%%%%%%%%%%%%%%%%%%%%%%%%%%%%%%%%%%%%%%%
%%%%%%%%%%%%%%%%%%%%%%%%%%%%%%%%%%%%%%%%%%%%%%%
%%%%%%%%%%%%%%%%%%%%%%%%%%%%%%%%%%%%%%%%%%%%%%%
%%%%%%%%%%%%%%%%%%%%%%%%%%%%%%%%%%%%%%%%%%%%%%%

\subsection{T-fold}

After a first T-duality transformation on the three-torus with $H$-flux which resulted in
the twisted torus, we now perform a second T-duality transformation
along the direction $k_2 = \partial_2$.

%%%%%%%%%%%%%%%%%%%%%%%%%%%%%%%%%%%%%%%%%%%%%%%
%%%%%%%%%%%%%%%%%%%%%%%%%%%%%%%%%%%%%%%%%%%%%%%

\subsubsection*{The background}

The T-dual background can be obtained
either by applying the Buscher rules \eqref{dual_back_9494} to the 
twisted torus \eqref{ex1_76} \cite{Kachru:2002sk,Lowe:2003qy},  or by performing a
collective T-duality transformation  along two directions for the three-torus with  $H$-flux given below \eqref{metric_01} \cite{Plauschinn:2014nha}.
The result of both approaches is the same. If we define the quantity
\eq{
  \label{ex1_37}
    \rho = \frac{R_1^2 R_2^2}{\alpha'^2} + \left[\frac{h}{2\pi}\op X^3\right]^2 \,,
}
we find for the dual background
\eq{
  \label{ex1_83}
  \arraycolsep2pt
  \begin{array}{lcl}
  \displaystyle \check{\mathsf G} &=& \displaystyle  \frac{1}{\rho} \,\Bigl[ R_2^2 \, d\tilde X^1\otimes d\tilde X^1
  + R_1^2 \, d\tilde X^2\otimes d\tilde X^2 \Bigr]
  + R_3^2 \, dX^3\otimes dX^3 \,, \\[12pt]
  \displaystyle \check{\mathsf H} &=& \displaystyle  
  -\frac{\alpha'}{2\pi}\, \frac{h}{\rho^2}\left(\frac{R_1^2R_2^2}{\alpha'^2} - \left[\frac{h}{2\pi}\op X^3\right]^2\right) \, d\tilde X^1\wedge d\tilde X^2\wedge dX^3
  \,,
  \\[14pt]
  \check{\boldsymbol{\phi}} & = & \displaystyle  \phi_0 -\frac12 \log\rho \,,
  \end{array}
}
where $\tilde X^1$ and $\tilde X^2$ denote the dual coordinates. 
The metric and field strength shown in \eqref{ex1_83}
describe the so-called T-fold \cite{Hull:2004in}.

Note that the background \eqref{ex1_83} is peculiar: in order to make it globally well-defined,
we would need a diffeomorphism which relates $\check{\mathsf G}$ at $X^3=2\pi$ to 
$X^3=0$. However, no such diffeomorphism exists -- which can be seen
from the Ricci scalar corresponding to $\check{\mathsf G}$ given by
\eq{
  \label{ex1_ricciscal}
  \check{\mathsf R} = \frac{h^2}{\pi^2 R_3^2}\, \frac{1}{\rho^2} 
  \left(\frac{R_1^2R_2^2}{\alpha'^2} - \frac52\left[\frac{h}{2\pi}\op X^3\right]^2\right)
  \,.
}
More concretely, since the Ricci scalar should be 
invariant under diffeomorphisms, the expression \eqref{ex1_ricciscal}
should be invariant under $X^3\to X^3+2\pi$ if the background admits a description in terms of Riemannian geometry. 
Since this is not the case, we can conclude that the above background does not allow
for a geometric description and is therefore  non-geometric.

%%%%%%%%%%%%%%%%%%%%%%%%%%%%%%%%%%%%%%%%%%%%%%%
%%%%%%%%%%%%%%%%%%%%%%%%%%%%%%%%%%%%%%%%%%%%%%%

\subsubsection*{Duality transformations}

The geometric symmetries of a string-theory background include
diffeomorphisms as well as gauge transformations of the Kalb-Ramond $B$-field. 
However, if we enlarge these symmetry transformation and include duality transformations, 
we can obtain a well-defined interpretation of the T-fold. 

To illustrate this point, let us choose a particular gauge for the 
$B$-field and write the components of the dual metric and Kalb-Ramond field as follows
\eq{
  \label{t-fold_001}
  \check{\mathsf G}_{ij} & = \frac{1}{\rho}\left( \begin{array}{ccc}
  R_2^2 & 0 & 0 \\
  0 & R_1^2 & \makebox[15pt]{0} \\
  \makebox[45pt]{0} & \makebox[45pt]{0} & \rho\op R_3^2
  \end{array}\right) ,
  \\
  \check{\mathsf B}_{ij} &= \frac{1}{\rho}\left( \begin{array}{ccc}
  0 & - \frac{\alpha'}{2\pi}\op h\op X^3 & \makebox[20.5pt]{0}\\
  + \frac{\alpha'}{2\pi}\op h\op X^3 & 0 & 0\\
  0 & 0 & 0
  \end{array}\right) .
}
Next, we recall that part of the $O(D,D,\mathbb Z)$ duality transformations discussed in 
section~\ref{sec_cft_torus} are  $\beta$-transformations.
They act on the generalised metric $\mathcal H$ -- which encodes the metric and $B$-field as shown in 
\eqref{gen_met_098} -- via the matrix given in equation  \eqref{beta_trans_101}. 
Using \eqref{dual_001}, we can then check that
\eq{
  \label{t-fold_007}
  \begin{array}{c}
  \displaystyle\check{\mathsf G}_{ij}(X^3+2\pi) = \mbox{$\beta$-transform}\Bigl[ \check{\mathsf G}(X^3) \Bigr]_{ij} \,,
  \\[10pt]
  \displaystyle\check{\mathsf B}_{ij}(X^3+2\pi) = \mbox{$\beta$-transform}\Bigl[ \check{\mathsf B}(X^3) \Bigr]_{ij} \,,
  \end{array}
  \hspace{30pt}
  \beta =  \left( \begin{array}{ccc}
  0 & - h & 0 \\ +h & 0 & 0 \\ 0 & 0 & 0 \end{array} \right).
}
This means, we can make the space globally-defined by using a $\beta$-transformation
as a transition function. Because these transformations are not part of the geometric transformations,
 the space is  non-geometric. 
However, since locally we do have a description in terms of a metric and only globally 
Riemannian geometry fails, the space is also called {\em globally non-geometric}.

%%%%%%%%%%%%%%%%%%%%%%%%%%%%%%%%%%%%%%%%%%%%%%%
%%%%%%%%%%%%%%%%%%%%%%%%%%%%%%%%%%%%%%%%%%%%%%%

\subsubsection*{Non-geometric flux}

As in the previous cases, we want to identify a flux for this background. Since 
the $H$-flux is not well-defined under $X^3\to X^3+2\pi$, and since similarly
the vielbein one-forms are not well-defined, we should look for a different quantity. 
The required formalism will be discussed in section~\ref{sec_gg_frame}, but let us already now
define a metric $\mathsf g^{ij}$ and an anti-symmetric  bivector-field $\beta^{ij}$ via
\eq{
  \bigl[ \check{\mathsf G} - \check{\mathsf B} \bigr]^{-1} = \mathsf g - \beta \,.
}
For the T-fold background \eqref{t-fold_001} this leads to
\eq{
  \renewcommand{\arraystretch}{1.2}
  \arraycolsep4pt
  \label{ex1_2053}
    \mathsf g^{ij}  = \frac{1}{\alpha'^2}\left( \begin{array}{ccc}
  R_1^2 & 0 & 0 \\
  0 & R_2^2 & 0 \\
  0 & 0 & \frac{\alpha'^2}{R_3^2}
  \end{array}\right) ,
  \hspace{35pt}
  \beta^{ij} = \frac{1}{\alpha'^2}\left( \begin{array}{ccc}
  0 &+\frac{\alpha'}{2\pi}\op h\op X^3 & 0\\
  -\frac{\alpha'}{2\pi}\op h\op X^3 & 0 & 0\\
  0 & 0 & 0
  \end{array}\right) ,
}
which allows us to define the so-called $Q$-flux (in the coordinate basis) as follows
\eq{
  Q_i{}^{jk} = \partial_i \op \beta^{jk} \,.
}
However, in order to compare $Q_i{}^{jk}$ with the expressions in the vielbein basis \eqref{ex1_2050} 
and \eqref{ex1_2052}, let us use the vielbein basis
\eq{
  \label{ex1_33712728}
  \mathsf e^i{}_{a } = \frac{1}{\alpha'}\left( \begin{array}{ccc}
  R_1 & 0 & 0 \\
  0 & R_2 & 0 \\
  0 & 0 & \frac{\alpha'}{R_3}
  \end{array}\right) ,
}
corresponding to the metric  in \eqref{ex1_2053} as $\mathsf g^{ij} = \mathsf e^i{}_a\op\delta^{ab} \op e_b{}^j$.
In this basis, the $Q$-flux reads 
$Q_a{}^{bc} = Q_i{}^{jk} \mathsf e^i{}_a \op\mathsf e^b{}_j\op\mathsf e^c{}_k$ and we find for the example
of the T-fold in the vielbein basis
\eq{
  \label{ex1_2052222}
  Q_3{}^{12} =  \frac{\alpha'}{2\pi}\,  \frac{h}{R_1\op R_2 \op R_3}\,.
}
Thus, for the T-fold background the flux-quantum $h$ is now encoded in a so-called non-geometric 
$Q$-flux. 
We also note that because the vielbein matrices \eqref{ex1_33712728} are constant, 
a corresponding geometric flux $f^{ab}{}_c$ vanishes.

%%%%%%%%%%%%%%%%%%%%%%%%%%%%%%%%%%%%%%%%%%%%%%%
%%%%%%%%%%%%%%%%%%%%%%%%%%%%%%%%%%%%%%%%%%%%%%%

\subsubsection*{Remarks}

Let us close this section about the T-fold with the following  two remarks.
\begin{itemize}

\item The dilaton for the T-fold background is shown in \eqref{ex1_83}, which depends 
 on the coordinate $X^3$. 
It is somewhat involved to derive the transformation rules of the dilaton 
under $\beta$-transformations from first principles, however, 
the transformation  under T-duality was given in \eqref{dual_back_9495}. The latter implies
 that the combination
\eq{
  \label{dila_inva}
   e^{-2\phi} \sqrt{\det G}
}
is invariant under T-duality.  Since T-duality  and $\beta$-transformations
(for toroidal backgrounds with constant $B$-field) are both part of the duality group $O(D,D,\mathbb Z)$,
it is natural to require \eqref{dila_inva} to be invariant also under $\beta$-transformations. 
Using this requirement, for the T-fold we then determine
\eq{
   e^{-2\check{\boldsymbol{\phi}} }\sqrt{\det \check{\mathsf G}} = R_1R_2R_3 \op e^{-2\phi_0} \,,
}
which does not depend on $X^3$. From here we can derive 
the transformation of the dilaton using \eqref{t-fold_007}, for which we can conclude that 
\eq{
   \check{\boldsymbol{\phi}}(X^3+2\pi) = \mbox{$\beta$-transform}\bigl[ \check{\boldsymbol{\phi}}(X^3) \bigr]
   \,.
}

\item Let us also apply our discussion on page~\pageref{page_global_v2}
to the T-fold background.
If we consider a collective T-duality transformation acting on the three-torus with $H$-flux 
along the directions $k_1$ and $k_2$ (defined in \eqref{ex1_killing_02}), then the two 
one-forms $v_{\alpha}$ can be chosen as
\eq{
   v_1 = -\frac{\alpha'}{2\pi} \op h\, X^3 \op dX^2 \,,
   \hspace{50pt}
   v_2 = +\frac{\alpha'}{2\pi} \op h\, X^3 \op dX^1 \,.
}
These are not globally-defined
on the $H$-flux background, but only 
the combinations $d\chi_{\alpha} + v_{\alpha}$ are required to be globally-defined. This leads to the 
identifications 
\eq{
  X^3\to X^3 + 2\pi \,, \hspace{40pt}
  \left\{ \begin{array}{l}
  \tilde X^1 \to \tilde X^1 + h\op X^2 \,,
  \\[2pt]
  \tilde X^2 \to \tilde X^2 - h\op X^1 \,,
  \end{array}
  \right.
}
where $\tilde X^{1} = \chi_{1}/\alpha'$ and $\tilde X^{2} = \chi_{2}/\alpha'$ are the dual coordinates and 
$X^{1}$ and $X^2$ are the original ones. 
We therefore see that for the T-fold background the original and dual coordinates 
are mixed when going around the circle in the $X^3$-direction, which 
suggests that the background should be thought of as a twisted torus involving the 
original as well as the dual coordinates. 
The corresponding framework to describe such configurations 
is called doubled geometry \cite{Hull:2004in}, which we discuss in 
section~\ref{cha_dg}.

\end{itemize}

%%%%%%%%%%%%%%%%%%%%%%%%%%%%%%%%%%%%%%%%%%%%%%%
%%%%%%%%%%%%%%%%%%%%%%%%%%%%%%%%%%%%%%%%%%%%%%%
%%%%%%%%%%%%%%%%%%%%%%%%%%%%%%%%%%%%%%%%%%%%%%%
%%%%%%%%%%%%%%%%%%%%%%%%%%%%%%%%%%%%%%%%%%%%%%%

\subsection{\texorpdfstring{$R$}{R}-space}

In order to arrive at the T-fold background discussed in the previous section, we have performed 
two T-duality transformations on the three-torus with $H$-flux. 
Given that the original space is three-dimensional, it is natural to  try to
perform a third  duality transformations and arrive at a so-called non-geometric $R$-space.

%%%%%%%%%%%%%%%%%%%%%%%%%%%%%%%%%%%%%%%%%%%%%%%
%%%%%%%%%%%%%%%%%%%%%%%%%%%%%%%%%%%%%%%%%%%%%%%

\subsubsection*{Duality transformations}

To arrive at the $R$-space we would either collectively T-dualise along all directions of the 
three-torus, or alternatively perform a T-duality transformation along the $X^3$-direction 
of the T-fold.  However, both of these approaches violate consistency requirements:
\begin{itemize}

\item For the three-torus with $H$-flux defined in \eqref{metric_01} and 
\eqref{ex1_metric_98}, the constraint for gauging \eqref{variations_45}
is not satisfied. 
Indeed, as the Killing vectors \eqref{ex1_killing_02} are abelian, we have
\eq{
3\,\iota_{k_{[\ul \alpha}} \op f_{\ul \beta\ul \gamma ]}{}^{\delta} v_{\delta} = 
\iota_{k_{\alpha}}\iota_{k_{\beta}}\iota_{k_{\gamma}} H 
\qquad\longrightarrow\qquad
0=h  \,,
}
which can only be solved for vanishing $H$-flux. A collective T-duality transformation 
along three directions for the $\mathbb T^3$ with $H$-flux is therefore not allowed.

\item Correspondingly, 
when trying to perform a single T-duality transformation 
along the $X^3$-direction of the T-fold background \eqref{t-fold_001}, we
see that the direction along $\partial_3$ is not an isometry. Indeed, we determine
\eq{
  \mathcal L_{\partial_3} \check {\mathsf G}_{\mbox{\op\scriptsize T-fold}} = 
  - \frac{h^2 \op X^3}{2\pi^2\rho^2} \:
  \Bigl[ R_2^2 \, d\tilde X^1\otimes d\tilde X^1
  + R_1^2 \, d\tilde X^2\otimes d\tilde X^2 \Bigr] \neq 0 \,.
}

\end{itemize}
On the other hand, the structure which we have observed so far is rather suggestive. 
In particular, let us summarise the family of backgrounds discussed in this section
as follows
\eq{
  \renewcommand{\arraystretch}{1.2}
 \boxed{ \begin{array}{p{2.75cm}}
 \centering three-torus~$\mathbb T^3$ \\
 \centering flux:~$H_{123}$
 \end{array}
 }
 &
  \renewcommand{\arraystretch}{1.2}
 \qquad
 \xrightarrow{\hspace{10pt}\mbox{\scriptsize T-duality~}T_1\hspace{13pt}}
 \qquad
 \boxed{ \begin{array}{p{2.75cm}}
 \centering twisted torus \\
 \centering  flux:~$f_{23}{}^1$
 \end{array}
 }
 \\[8pt]
 &
  \renewcommand{\arraystretch}{1.2} 
 \qquad
 \xrightarrow{\hspace{10pt}\mbox{\scriptsize T-duality~}T_{12}\hspace{10pt}}
 \qquad
 \boxed{ \begin{array}{p{2.75cm}}
 \centering T-fold \\
 \centering flux:~$Q_3{}^{12}$
 \end{array}
 }
 \\[18pt]
 &
 \qquad
 \xrightarrow{\hspace{8pt}\mbox{\scriptsize T-duality~}T_{123}\hspace{8pt}}
 \qquad
 \hspace{1.4cm}\ldots
 \\[14pt]
}
Thus, continuing this line, we can conjecture that three duality transformations on 
the three-torus with $H$-flux will lead to a space  characterised by
an object with three anti-symmetric upper indices. 
Taking the letter following $Q$, this object is usually
called the $R$-flux $R^{ijk}$ \cite{Shelton:2005cf,Shelton:2006fd}. Comparing with \eqref{ex1_2050},
\eqref{ex1_2052} and \eqref{ex1_2052222},
in the present case the flux (in an appropriate vielbein basis) is then expected to take the form
\eq{
  \label{ex1_flux_773}
  R^{123} =  \frac{\alpha'}{2\pi}\, \frac{h}{R_1\op R_2 \op R_3}\,.
}

%%%%%%%%%%%%%%%%%%%%%%%%%%%%%%%%%%%%%%%%%%%%%%%
%%%%%%%%%%%%%%%%%%%%%%%%%%%%%%%%%%%%%%%%%%%%%%%

\subsubsection*{Non-geometric properties}

Since we cannot reach the $R$-space through application of the Buscher rules, its properties 
are not well-understood. However, we can make
the following observation \cite{Wecht:2007wu}:
\begin{itemize}

\item Let us start from the $H$-flux background introduced in section~\ref{sec_ex1_hhh} 
and consider a D3-brane 
wrapping the three-torus $\mathbb T^3$ and extending along the external time direction. 
Due to the Freed-Witten anomaly \cite{Freed:1999vc}, such a configuration is not allowed.

\item Nevertheless, when performing three T-duality transformations along the three-torus the 
D3-brane is expected to become a D0-brane. The latter is point-like in the three-dimensional $R$-space.

\item Since the original configuration is forbidden, also a point-like D0-brane on the 
$R$-space has to be forbidden. This suggests that a description of the $R$-space in terms of 
ordinary Riemann geometry of point particles is not possible, and hence 
we are led to the notion of a {\em locally non-geometric space}.

\end{itemize}

%%%%%%%%%%%%%%%%%%%%%%%%%%%%%%%%%%%%%%%%%%%%%%%
%%%%%%%%%%%%%%%%%%%%%%%%%%%%%%%%%%%%%%%%%%%%%%%
%%%%%%%%%%%%%%%%%%%%%%%%%%%%%%%%%%%%%%%%%%%%%%%
%%%%%%%%%%%%%%%%%%%%%%%%%%%%%%%%%%%%%%%%%%%%%%%

\subsection{Summary}

Let us summarise the discussion of this section: starting from the three-torus with $H$-flux we have 
performed  T-duality transformations leading to the twisted torus, 
the T-fold and the $R$-space. 
To each of these backgrounds we can associate a flux which was
given in equations \eqref{ex1_2050},
\eqref{ex1_2052}, \eqref{ex1_2052222} and \eqref{ex1_flux_773}.
Generalising this example to higher-dimensional tori with $H$-flux, 
schematically the chain of T-duality transformations can be 
expressed as \cite{Kachru:2002sk,Shelton:2005cf}
\eq{
  \label{ex1_chain}
  H_{ijk} \qquad\xrightarrow{\hspace{10pt}T_i\hspace{10pt}}\qquad
  f_{jk}{}^i \qquad\xrightarrow{\hspace{10pt}T_j\hspace{10pt}}\qquad
  Q_{k}{}^{ij} \qquad\xrightarrow{\hspace{10pt}T_k\hspace{10pt}}\qquad
  R^{ijk} \,.
}

The example discussed in this section should be understood as a toy example, 
which illustrates the main properties of a non-geometric background. 
Up to this point it is however not a rigorous construction: 
\begin{itemize}

\item The  $H$-flux background shown in \eqref{metric_01} and 
\eqref{ex1_metric_98} does not solve the string-theoretical equations of motion
\eqref{eom_beta}. It is therefore not 
a proper string-theory background and applying 
T-duality transformations is a priori not justified. 
Furthermore, when performing a similar analysis for the three-sphere with $H$-flux --
which does solve the equations of motion \eqref{eom_beta}  --
then no non-geometric features arise (see equation \eqref{tdual_su2_2t_99}). In fact, for the three-sphere the question whether 
a T-fold analogue appears can be traced to the question whether the string-equations of motion 
are satisfied \cite{Plauschinn:2014nha}.

\item The vanishing of the $\beta$-functionals \eqref{eom_beta} can be 
interpreted as the equations of motion of an effective theory. 
For superstring theory this is for instance type IIA or type IIB supergravity. 
In these theories it is possible to turn on also  Ramond-Ramond 
fluxes, which do allow for configurations  solving the equations of motion. 
However, even though such backgrounds can be solutions to the supergravity equations of motion,
a string-theoretical CFT description is usually difficult.

\end{itemize}
On the other hand, there is evidence from a variety of examples 
that non-geometric spaces are indeed relevant backgrounds
for string theory. 
We mention some of them here, and discuss these in more detail in the 
subsequent sections. 
\begin{itemize}

\item The example of the three-torus with $H$-flux is a particular example of a parabolic $\mathbb T^2$-fibration over
a circle, which contains the T-fold as one of its T-dual backgrounds. However, 
also so-called  elliptic fibrations exist which do solve the string-theory equations of motion. 
We discuss these backgrounds in section~\ref{sec_tw_r2}.

\item From a supergravity point of view, upon compactification the fluxes 
appearing in \eqref{ex1_chain} give rise to gauged supergravity theories.
In particular, the charges corresponding to the local gauge symmetries 
are related to the various fluxes. 
Moreover, in order to reproduce all gaugings which are possible from a  supergravity point of view,
non-geometric fluxes are needed. We discuss this point in section~\ref{cha_sugra}.

\item T-duality transformations can be described  using the framework of generalised geometry.
In this approach, non-geometric fluxes appear naturally and can be given a
microscopic description.  We explain this point in section~\ref{sec_gen_geo}.

\end{itemize}

%%%%%%%%%%%%%%%%%%%%%%%%%%%%%%%%%%%%%%%%%%%%%%%
%%%%%%%%%%%%%%%%%%%%%%%%%%%%%%%%%%%%%%%%%%%%%%%
%%%%%%%%%%%%%%%%%%%%%%%%%%%%%%%%%%%%%%%%%%%%%%%
%%%%%%%%%%%%%%%%%%%%%%%%%%%%%%%%%%%%%%%%%%%%%%%
%%%%%%%%%%%%%%%%%%%%%%%%%%%%%%%%%%%%%%%%%%%%%%%
%%%%%%%%%%%%%%%%%%%%%%%%%%%%%%%%%%%%%%%%%%%%%%%
%%%%%%%%%%%%%%%%%%%%%%%%%%%%%%%%%%%%%%%%%%%%%%%
%%%%%%%%%%%%%%%%%%%%%%%%%%%%%%%%%%%%%%%%%%%%%%%
%%%%%%%%%%%%%%%%%%%%%%%%%%%%%%%%%%%%%%%%%%%%%%%
%%%%%%%%%%%%%%%%%%%%%%%%%%%%%%%%%%%%%%%%%%%%%%%
%%%%%%%%%%%%%%%%%%%%%%%%%%%%%%%%%%%%%%%%%%%%%%%
%%%%%%%%%%%%%%%%%%%%%%%%%%%%%%%%%%%%%%%%%%%%%%%
%%%%%%%%%%%%%%%%%%%%%%%%%%%%%%%%%%%%%%%%%%%%%%%
%%%%%%%%%%%%%%%%%%%%%%%%%%%%%%%%%%%%%%%%%%%%%%%
%%%%%%%%%%%%%%%%%%%%%%%%%%%%%%%%%%%%%%%%%%%%%%%
%%%%%%%%%%%%%%%%%%%%%%%%%%%%%%%%%%%%%%%%%%%%%%%

\clearpage
\section{Torus fibrations}
\label{cha_torus_fib}

After having studied the three-torus with $H$-flux and its T-dual configurations, in this section 
we discuss non-geometric backgrounds in a more systematic way.
We consider $n$-dimensional torus fibrations over various base manifolds,
and first revisit in section~\ref{sec_t2_fibr_example}
the three-torus example in this framework. In the following sections we 
study more general $\mathbb T^2$-fibrations over the circle, over $\mathbb P^1$ 
and over the punctured plane, and we connect some of these fibrations 
to the compactified NS5-brane, Kaluza-Klein monopole and non-geometric $5^2_2$-brane.

%%%%%%%%%%%%%%%%%%%%%%%%%%%%%%%%%%%%%%%%%%%%%%%
%%%%%%%%%%%%%%%%%%%%%%%%%%%%%%%%%%%%%%%%%%%%%%%

\subsubsection*{Setting and notation}

\begin{figure}[t]
\centering
\vspace*{20pt}
\includegraphics[width=200pt]{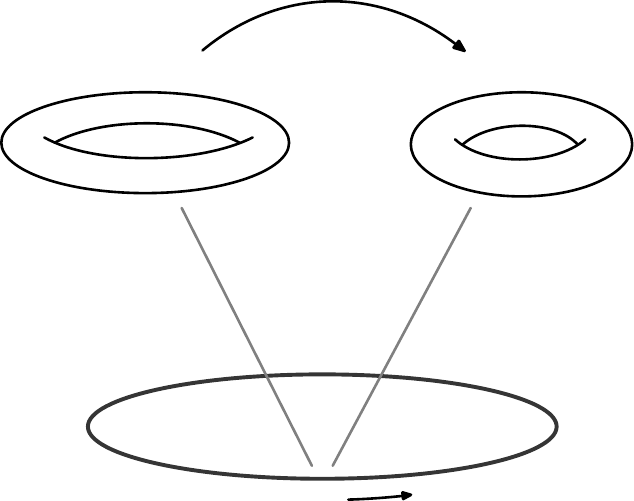}
\begin{picture}(0,0)
\put(-4,125){\footnotesize $(G_{\mathsf{ab}}, B_{\mathsf{ab}})(x)$}
\put(-280,125){\footnotesize $(G_{\mathsf{ab}}, B_{\mathsf{ab}})(x+2\pi)$}
\put(-118,165){\footnotesize $O(n,n,\mathbb Z)$}
\put(-86,-8){\footnotesize $x$}
\end{picture}
\vspace*{14pt}
\caption{Illustration of how a torus fibration over a circle can be patched in order to obtain a globally well-defined
space. For a geometric background the $O(n,n,\mathbb Z)$ transformations 
can be diffeomorphisms and gauge transformations, whereas for a non-geometric background for instance also $\beta$-transformations are allowed.\label{fig_forus_fibration}} 
\end{figure}
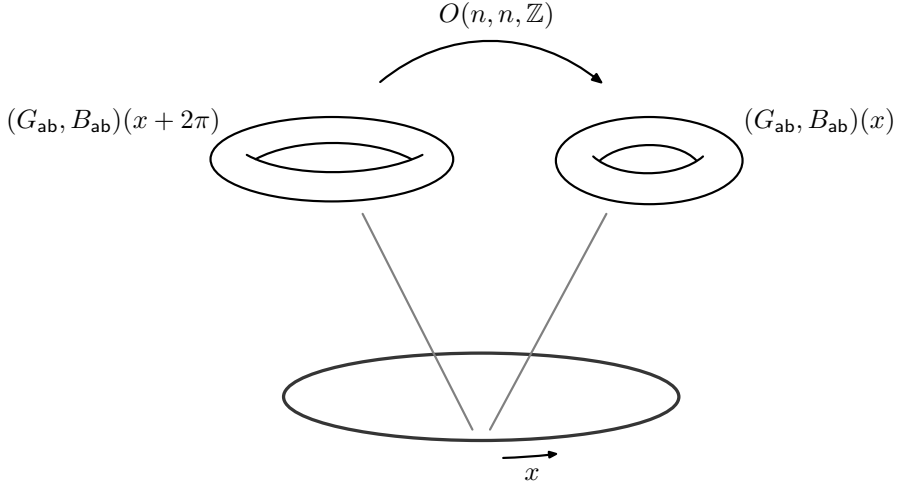

The setting we are interested in is that of $n$-dimensional torus fibrations 
$\mathcal M$ over a $(D-n)$-dimensional base-manifold
$\mathcal B$
\eq{
  \mathbb T^n \; \varlonghookrightarrow\; \begin{array}[t]{@{\op}c}\mathcal M \\[4pt]
  \big\downarrow \\[8pt]
  \mathcal B
  \end{array}
}
Local coordinates on the base-manifold will be denoted by $x^m$ with $m=1,\ldots, D-n$ and 
coordinates in the fibre are $y^{\mathsf a}$ with $\mathsf a=1,\ldots, n$.
The metric and $B$-field are assumed to only depend on the coordinates $x^m$ of the base.
We furthermore assume that the base-manifold $\mathcal B$ has at least one 
non-contractible one-cycle $\gamma$, and examples for such manifolds are 
$\mathcal B= S^1$ and $\mathcal B = \mathbb R^2\setminus \{0\}$. 
The non-triviality of the fibration is encoded in the monodromy of the metric and $B$-field 
along the cycle $\gamma$. In particular, if we parametrise going around $\gamma$ as 
$x\to x+2\pi$ we can ask  how 
$(G_{\mathsf{ab}}, B_{\mathsf{ab}})(x+2\pi)$ is related to $(G_{\mathsf{ab}},B_{\mathsf{ab}})(x)$:
\begin{itemize}

\item In an ordinary geometric background, we can use the symmetries of the theory
to relate the torus fibre at  $x+2\pi$ to $x$ \cite{Scherk:1978ta,Scherk:1979zr,Hull:1998vy,Dabholkar:2002sy,Hull:2005hk}. These symmetries are diffeomorphisms and 
gauge transformations, which are sometimes referred to as geometric transformations.

\item However,  for non-geometric backgrounds also proper T-duality transformations can be used to 
relate the fibre at $x+2\pi$ to $x$ \cite{Hull:2004in,Hull:2007jy}, which mix the metric and the $B$-field.
These are also called non-geometric transformations.

\end{itemize}
Note that both, the symmetry and duality  transformations are part of the duality group of the $n$-dimensional
torus fibre, which in the present case is $O(n,n,\mathbb Z)$.
The monodromy along the cycle $\gamma\subset \mathcal B$ and the patching of the fibre 
are illustrated in figure~\ref{fig_forus_fibration}.
%%%%%%%%%%%%%%%%
%%%%%%%%%%%%%%%%
%%%%%%%%%%%%%%%%
%%%%%%%%%%%%%%%%

%%%%%%%%%%%%%%%%%%%%%%%%%%%%%%%%%%%%%%%%%%%%%%%
%%%%%%%%%%%%%%%%%%%%%%%%%%%%%%%%%%%%%%%%%%%%%%%
%%%%%%%%%%%%%%%%%%%%%%%%%%%%%%%%%%%%%%%%%%%%%%%
%%%%%%%%%%%%%%%%%%%%%%%%%%%%%%%%%%%%%%%%%%%%%%%

\subsection{\texorpdfstring{$\mathbb T^2$}{T2}-fibrations over the circle I}
\label{sec_t2_fibr_example}

To start, let us revisit the example of the three-torus with $H$-flux from the previous section. 
In the present notation the local coordinates are $\{ y^1, y^2, x\}$, which are normalised 
as $y^{\mathsf a} \sim y^{\mathsf a}+2\pi$ and $x\sim x +2\pi$. The metric and $B$-field can be brought into 
the following general form
\eq{
  \label{back_ex_939393}
  \arraycolsep2pt
    G_{ij} = \left( \begin{array}{cc} 
    G_{\mathsf{ab}}(x) & 0 \\ 0 & R_3^2 \end{array}\right),
    \hspace{60pt}
    B_{ij} = \left( \begin{array}{cc} 
    B_{\mathsf{ab}}(x) & 0 \\ 0 & 0 \end{array}\right).
}

%%%%%%%%%%%%%%%%%%%%%%%%%%%%%%%%%%%%%%%%%%%%%%%
%%%%%%%%%%%%%%%%%%%%%%%%%%%%%%%%%%%%%%%%%%%%%%%

\subsubsection*{$O(2,2,\mathbb Z)$ transformations}

Let us now focus on the two-dimensional toroidal part and the corresponding
$O(2,2,\mathbb Z)$ duality group, 
and investigate 
how the fibre at the point $x+2\pi$ on the base is related to the point $x$.
\begin{itemize}

\item We begin with the three-torus with non-vanishing $H$-flux. The $\mathbb T^2$-fibre of this
 background can be described by
\eq{
  \label{fibr_1_h}
  \arraycolsep4pt
  \renewcommand{\arraystretch}{1.2}
  \mbox{$H$-flux:} \hspace{40pt}
    G_{\mathsf{ab}} = \left( \begin{array}{cc} 
    R_1^2 & 0 \\ 0 & R_2^2 \end{array}\right),
    \hspace{20pt}
    B_{\mathsf{ab}} = \left( \begin{array}{cc} 
    0 & +\frac{\alpha'}{2\pi}\op h\op x \\ - \frac{\alpha'}{2\pi}\op h \op x & 0 \end{array}\right),
}
with $h\in\mathbb Z$.
Let us then employ the framework from section~\ref{sec_cft_torus}
and consider an $O(2,2,\mathbb Z)$ transformation. 
Defining  
\eq{
  \label{monodro_01}
  \arraycolsep5pt
  \mathcal O_{\mathsf B} = \left( \begin{array}{cc} \mathds 1 & 0 \\ \mathsf B & \mathds 1 \end{array}\right)
  \,,
  \hspace{50pt}
  \mathsf B = \left( \begin{array}{cc} 0 & +h \\ -h & 0 \end{array}\right),
}
with $\mathcal O_{\mathsf B}\in O(2,2,\mathbb Z)$, 
we can check that the generalised metric $\mathcal H_H$ corresponding to 
\eqref{fibr_1_h} transforms as
\eq{
   \mathcal H_H \bigl( x+2\pi \bigr)= \mathcal O_{\mathsf B}^{-T} \:\mathcal H_H (x) \:\mathcal O_{\mathsf B}^{-1}
   \,.
}
This transformation describes a gauge transformation of the $B$-field, 
which is needed  to relate the $\mathbb T^2$-fibre at $x+2\pi$ to $x$.   
This is a geometric transformation.

\item Next, we turn to the twisted torus. The metric 
and $B$-field of the torus fibre can be determined from \eqref{ex1_76} as follows
\eq{
  \label{fibr_1_f}
  \arraycolsep4pt
  \renewcommand{\arraystretch}{1.2}
  \mbox{$f$-flux:} \hspace{40pt}
    G_{\mathsf{ab}} = \left( \begin{array}{cc} 
    \frac{\alpha'^2}{R_1^2} & -\frac{\alpha'^2}{R_1^2}\op \frac{h}{2\pi}\op x \\ -\frac{\alpha'^2}{R_1^2}\op \frac{h}{2\pi}\op x
    & R_2^2 +\frac{\alpha'^2}{R_1^2}\bigl[\frac{h}{2\pi x}\bigr]^2 \end{array}\right),
    \hspace{30pt}
    B_{\mathsf{ab}} = 0\,.
}
The $O(2,2,\mathbb Z)$ transformation of interest is 
given by
\eq{
  \label{monodro_02}
  \arraycolsep5pt
\mathcal O_{\mathsf A} = \left( \begin{array}{cc} \mathsf A^{-1} & 0 \\ 0 & \mathsf A^{T} \end{array}\right)
  \,,
  \hspace{50pt}
  \mathsf A =  \left( \begin{array}{cc} 1 & -h \\ 0 & 1 \end{array}\right),
}
and for the generalised metric $\mathcal H_f$ corresponding to \eqref{fibr_1_f}
we can compute
\eq{
   \mathcal H_f \bigl( x+2\pi \bigr)= \mathcal O_{\mathsf A}^{-T} \:\mathcal H_f (x) \:\mathcal O_{\mathsf A}^{-1}
   \,.
}
This transformation describes a diffeomorphism, which is used to make the twisted torus
a globally-defined space. Again, this is a geometric transformation.

\item We finally turn to T-fold background specified in \eqref{ex1_83}. The corresponding
metric and $B$-field of the $\mathbb T^2$-fibre along the $\tilde X^1$- and $\tilde X^2$-direction
can be determined from \eqref{t-fold_001} as
\eq{
  \arraycolsep4pt
  \renewcommand{\arraystretch}{1.2}
  \mbox{$Q$-flux:} \hspace{20pt}
    G_{\mathsf{ab}} = \frac{1}{\rho} \left( \begin{array}{cc} 
    R_2^2 & 0 \\ 0  & R_1^2  \end{array}\right),
    \hspace{20pt}
    B_{\mathsf{ab}} = \frac{1}{\rho} \left( \begin{array}{cc} 
    0 & - \frac{\alpha'}{2\pi}\op h\op x \\ + \frac{\alpha'}{2\pi}\op h\op x  & 0 \end{array}\right),
}
where $\rho$ was defined in equation \eqref{ex1_37}.
We have already discussed that in the case of the T-fold a $\beta$-transformation is 
needed to make the background globally-defined. In particular, for 
\eq{
  \label{monodro_03}
 \mathcal O_{\beta} = \left( \begin{array}{cc} \mathds 1 & \beta \\ 0 & \mathds 1 \end{array}\right) ,
 \hspace{50pt}
  \beta =  \left( \begin{array}{cc} 0 & +h \\ -h & 0 \end{array}\right),
}
we can check that
\eq{
   \mathcal H_Q \bigl( x+2\pi \bigr)= \mathcal O_{\beta}^{-T} \:\mathcal H_Q (x) \:\mathcal O_{\beta}^{-1}
   \,.
}
As we discussed on page~\pageref{page_beta_trans}, such a transformation is 
not a symmetry but a duality transformation. The T-fold background is therefore
a non-geometric background.

\end{itemize}

%%%%%%%%%%%%%%%%%%%%%%%%%%%%%%%%%%%%%%%%%%%%%%%
%%%%%%%%%%%%%%%%%%%%%%%%%%%%%%%%%%%%%%%%%%%%%%%

\subsubsection*{Chain of duality transformations}

Let us also revisit the chain of T-duality transformations shown in \eqref{ex1_chain}.
Applying a first T-duality transformation to the $H$-flux background
along the $X^1$-direction gives a twisted torus with geometric $f$-flux, and 
performing a second T-duality transformation along the $X^2$-direction results in a T-fold with $Q$-flux. 
In terms of $O(2,2,\mathbb Z)$ transformations acting on the generalised metric this reads in formulas
\eq{
  \mathcal H_{f} = \mathcal O_{+1}^{-T} \:\mathcal H_{H} \:\mathcal O_{+1}^{-1} \,,
  \hspace{50pt}
  \mathcal H_{Q} = \mathcal O_{+2}^{-T} \:\mathcal H_{f} \:\mathcal O_{+2}^{-1} \,,
}
where subscript of $\mathcal H$ indicates the corresponding background and 
where the  matrices $\mathcal O_{\pm\mathsf i}$ have been defined in \eqref{odd_005}.\footnote{
Using $\mathcal O_{-1}$ and  $\mathcal O_{-2}$ reverses the sign of the flux quantum number, which corresponds to the parity symmetry $\Omega$ of the world-sheet theory.}
We therefore arrive at the following picture
\eq{
  \label{chain_777}
  \arraycolsep10pt
  \begin{array}{@{}ccccc@{}}
  \\[-14pt]
  \mathcal O_{\mathsf B} && \mathcal O_{\mathsf A} && \mathcal O_{\beta} 
  \\[4pt]
  \includegraphics[width=25pt]{fig_01} &&
  \includegraphics[width=25pt]{fig_01} &&
  \includegraphics[width=25pt]{fig_01}
  \\[6pt]
  \boxed{\begin{array}{c}\mbox{$H$-flux}\end{array}} 
  & \xleftrightarrow{\quad\:\mathcal O_{\mathsf E}\quad} &
  \boxed{\begin{array}{c}\mbox{$f$-flux}\end{array}} 
  & \xleftrightarrow{\quad\:\mathcal O_{\mathsf E}\quad} &
  \boxed{\begin{array}{c}\mbox{$Q$-flux}\end{array}} 
  \end{array}
}
with $\mathcal O_{\mathsf B}$, $\mathcal O_{\mathsf A}$ and $\mathcal O_{\beta}$ denoting the 
patching-transformations. 
From \eqref{chain_777} it becomes also clear how the latter  are related. For instance,
the transformation $\mathcal O_{\mathsf A}$ for the $f$-flux background or
$\mathcal O_{\beta}$ for the $Q$-flux background is determined by conjugation as
\eq{
  \label{chain_779}
  \mathcal O^{(f)}_{\mathsf A} = \mathcal O_{+1}^{-1} \, \mathcal O^{(H)}_{\mathsf B} \op \mathcal O_{+1}
  \,,
  \hspace{50pt}
  \mathcal O^{(Q)}_{\beta} = \mathcal O_{+2}^{-1} \, \mathcal O^{(f)}_{\mathsf A} \op \mathcal O_{+2}
  \,.
}
Note that in \eqref{chain_777} all generators of $O(2,2,\mathbb Z)$ play a role, 
and that we have not included the $R$-space background. 
In order to obtain the latter a T-duality transformation along the base-manifold --
that is the $x$-direction in \eqref{back_ex_939393} -- has to be performed, which 
cannot be described within the present framework.

%%%%%%%%%%%%%%%%%%%%%%%%%%%%%%%%%%%%%%%%%%%%%%%
%%%%%%%%%%%%%%%%%%%%%%%%%%%%%%%%%%%%%%%%%%%%%%%

\subsubsection*{Three-torus with $H$-, $f$- and $Q$-flux}

We also want to briefly discuss a generalisation of three-torus example,
where we consider $H$-, $f$- and $Q$-flux simultaneously.
The metric and Kalb-Ramond field for this example are specified by
\eq{
  \label{super_fluxes}
    G_{\mathsf{ab}} & 
  \arraycolsep4pt
  \renewcommand{\arraystretch}{1.3}    
    = \frac{1}{ 1 + \left[\frac{R_1 R_2}{\alpha'}\op\frac{q\op x}{2\pi} \right]^2} 
    \left( \begin{array}{cc} 
    R_1^2 & R_1^2\op\frac{f\op x}{2\pi} \\ R_1^2\op\frac{f\op x}{2\pi}  & 
    R_2^2+R_1^2  \left[\frac{f\op x}{2\pi} \right]^2 \end{array}\right),
    \\[10pt]
    B_{\mathsf{ab}} &
  \arraycolsep2pt
  \renewcommand{\arraystretch}{1.3}      
    = \alpha'\left( \begin{array}{cc} 
    0 & +\frac{h\op x}{2\pi} \\ - \frac{h\op x}{2\pi} & 0 \end{array}\right)
    + \frac{\alpha'}{ \frac{\alpha'^2}{R_1^2 R_2^2} + \left[\frac{q\op x}{2\pi} \right]^2} 
    \left( \begin{array}{cc} 
    0 & -\frac{q\op x}{2\pi} \\ + \frac{q\op x}{2\pi} & 0 \end{array}\right),
}
where $h,f,q\in\mathbb Z$ denote the corresponding flux quanta. Note that setting two of these fluxes
to zero reproduces the  above backgrounds (up to inversion of the radii and changing the sign of the geometric flux). 
The $O(2,2,\mathbb Z)$ transformation connecting the torus fibre at $x+2\pi$ to $x$
is naively expected to be a combination of the transformations discussed above. However, only for
\eq{
  \widetilde{ \mathcal O} = \mathcal O_{\mathsf B(h)} \, \mathcal O_{\mathsf A(f)}\, \mathcal O_{\beta(q)} 
  \hspace{40pt}\mbox{with}\hspace{10pt} h\op q = 0
}
this is possible. The situation with $h\op q\neq0$ requires a more involved construction,
and we address this question in the next section. Let us however
summarise that 
\begin{itemize}

\item the background \eqref{super_fluxes} with $q=0$ and $h,f\neq0$ describes a twisted three-torus with $H$-flux,

\item the background \eqref{super_fluxes} with $h=0$ and $f,q\neq0$ describes a T-fold with geometric flux. 

\end{itemize}

%%%%%%%%%%%%%%%%%%%%%%%%%%%%%%%%%%%%%%%%%%%%%%%
%%%%%%%%%%%%%%%%%%%%%%%%%%%%%%%%%%%%%%%%%%%%%%%

\subsubsection*{Remarks}

\begin{itemize}

\item As we have mentioned earlier, the non-triviality of the fibration is encoded in 
the various geometric and non-geometric fluxes. 
In particular, from the monodromies $\mathcal O_{\mathsf B}$, $\mathcal O_{\mathsf A}$
and $\mathcal O_{\beta}$ shown in \eqref{monodro_01}, \eqref{monodro_02} and 
\eqref{monodro_03} we can read off the $H$-, $f$- and $Q$-flux of the corresponding 
background. 
For more general monodromies it might be more difficult to identify the fluxes, 
and we come back to this question below.

\item The transformations $\mathcal O_{\mathsf A}$, $\mathcal O_{\mathsf B}$
and $\mathcal O_{\beta}$ belong to the $\det \mathcal O=+1$ part of $O(2,2,\mathbb Z)$
and are connected to the identity. The fibration can therefore be constructed as a continuous path 
starting from the identity, which we explain in more detail in the next section.
However, T-duality transformations 
$\mathcal O_{\mathsf E}$ have determinant $\det \mathcal O_{\mathsf E}=-1$ and thus
belong to the disconnected part of the group.
This means that fibrations with monodromies of the form $\mathcal O_{\mathsf E}$ cannot be constructed 
in the same way.

\item Since the geometric and non-geometric backgrounds discussed in this section are related 
by duality transformations, according to our definition 4 on page~\pageref{page_defs_nongeo}
this family of backgrounds is called geometric. However, according to our definition 3
the T-fold is a non-geometric background.

\end{itemize}

%%%%%%%%%%%%%%%%%%%%%%%%%%%%%%%%%%%%%%%%%%%%%%%
%%%%%%%%%%%%%%%%%%%%%%%%%%%%%%%%%%%%%%%%%%%%%%%
%%%%%%%%%%%%%%%%%%%%%%%%%%%%%%%%%%%%%%%%%%%%%%%
%%%%%%%%%%%%%%%%%%%%%%%%%%%%%%%%%%%%%%%%%%%%%%%
%%%%%%%%%%%%%%%%%%%%%%%%%%%%%%%%%%%%%%%%%%%%%%%
%%%%%%%%%%%%%%%%%%%%%%%%%%%%%%%%%%%%%%%%%%%%%%%
%%%%%%%%%%%%%%%%%%%%%%%%%%%%%%%%%%%%%%%%%%%%%%%
%%%%%%%%%%%%%%%%%%%%%%%%%%%%%%%%%%%%%%%%%%%%%%%

\subsection{\texorpdfstring{$\mathbb T^2$}{T2}-fibrations over the circle II}
\label{sec_t2_fibr_general}

In section~\ref{sec_t2_fibr_example} we have formulated the three-torus 
with $H$-flux and its T-dual configurations as $\mathbb T^2$-fibrations 
over a circle. This example falls into a particular category 
of fibrations -- so-called parabolic fibrations --  but more general constructions are possible. 
In this section we now want to take a different approach and first specify
the monodromy of the $\mathbb T^2$-fibre along the base-circle, and 
then construct a corresponding metric and Kalb-Ramond field. 
Our notation follows \cite{Lust:2015yia}, 
which in parts is based on  \cite{Hull:2005hk}.

%%%%%%%%%%%%%%%%%%%%%%%%%%%%%%%%%%%%%%%%%%%%%%%
%%%%%%%%%%%%%%%%%%%%%%%%%%%%%%%%%%%%%%%%%%%%%%%

\subsubsection*{K\"ahler and complex-structure moduli}

Let us start by characterising the two-torus background in terms of its complex-structure modulus $\tau$ and the complexified K\"ahler modulus $\rho$. These moduli are defined in terms of the metric and Kalb-Ramond field 
in the following way
\eq{
  \label{moduli_99}
  \tau = \frac{G_{12}}{G_{22}}+i\,\frac{\sqrt{\det G}}{G_{22}} \,,
  \hspace{50pt}
  \rho = \frac{1}{\alpha'} \left( B_{12} + i\, \sqrt{\det G} \right)\,.
}
Roughly speaking, $\tau$ encodes the shape of the $\mathbb T^2$ and $\rho$ determines 
its volume plus the $B$-field. 
Next, we note that the duality group $O(2,2,\mathbb Z)$ of the two-torus 
is isomorphic to 
\eq{
  \label{isos_849490}
  O(2,2,\mathbb Z) \simeq
  SL(2,\mathbb Z) \times SL(2,\mathbb Z) \times \mathbb Z_2\times \mathbb Z_2 \,,
}  
and for a more detailed discussion we refer for instance to \cite{Giveon:1994fu}.
The duality group acts on the two moduli 
in the following way:
\begin{itemize}

\item The two $SL(2,\mathbb Z)$ factors are M\"obius transformations on the 
complex-struc\-ture and complexified K\"ahler modulus 
\eq{
  \label{moebius_003}
  \arraycolsep3pt
  \begin{array}{lcl@{\hspace{50pt}}l@{\hspace{2pt}}c@{\hspace{2pt}}lcl}
  \tau &\to& \displaystyle \frac{a\op\tau + b}{c\op\tau + d} \,, & M_{\tau} &=&
  \displaystyle  \begin{pmatrix} a & b \\ c & d \end{pmatrix} &\in& SL(2,\mathbb Z) \,,
  \\[18pt]
  \rho &\to& \displaystyle \frac{\tilde a\op\rho + \tilde b}{\tilde c\op\rho + d} \,, &M_{\rho}&=&
  \displaystyle   \begin{pmatrix} \tilde a & \tilde b \\ \tilde c & \tilde d \end{pmatrix} &\in& SL(2,\mathbb Z) \,.
  \end{array}
}
Note that $SL(2,\mathbb Z)$ is generated by $T$- and $S$-transformations, which for 
$\tau$ and $\rho$ means
\eq{
  \label{moebius_005}
  \arraycolsep2pt
  \begin{array}{c@{\hspace{20pt}}lcl@{\hspace{80pt}}c@{\hspace{20pt}}lcl}
  T: & \tau &\to& \displaystyle \tau+1  \,,
  &
  \tilde T: & \rho &\to& \displaystyle \rho + 1 \,,  
  \\[10pt]
  S: & \tau &\to& \displaystyle - \frac{1}{\tau} \,,
  &
  \tilde S: & \rho &\to& \displaystyle - \frac{1}{\rho} \,.
  \end{array}
}
Let us emphasise that $SL(2,\mathbb Z)$ transformations of $\tau$ 
are large diffeomorphisms of the two-torus and therefore correspond 
to  geometric symmetries. 
A $\tilde T$-transformation acting on $\rho$ can be interpreted as a 
gauge transformation which is again geometric, but a $\tilde S$-transformation 
acting on $\rho$ 
will in general invert the volume of the $\mathbb T^2$ and is not a 
geometric transformation.

\item One of the $\mathbb Z_2$ factors in \eqref{isos_849490} corresponds to mirror symmetry 
$(\tau,\rho)\to (\rho,\tau)$. This is a T-duality transformation, which can be seen for a particular case by 
setting $G_{12}$ and 
$B_{12}$ to zero in \eqref{moduli_99}.

\item The remaining $\mathbb Z_2$ factor in \eqref{isos_849490} corresponds to a 
reflection of the form $(\tau,\rho)\to (-\ov\tau,-\ov \rho)$.

\end{itemize}
Including in addition the world-sheet parity transformation $\Omega$ 
which acts on the Kalb-Ramond field as $B_{12}\to -B_{12}$ and leaves the 
metric invariant,  the minimal set of generators of the  duality group \eqref{isos_849490}
turns out to be \cite{Giveon:1994fu}
\eq{
  \arraycolsep2pt
  \begin{array}{r@{\hspace{40pt}}lcl@{\hspace{30pt}}lcl}
  1) & \tau &\to& \displaystyle -\frac{1}{\tau} \,, & \rho & \to & \rho \,, \\[10pt]
  2) & \tau &\to& \displaystyle \tau+ 1 \,, & \rho & \to & \rho \,, \\[10pt]
  3) & \tau &\to& \displaystyle \rho \,, & \rho & \to & \tau \,,  \\[10pt]
  4) & \tau &\to& \displaystyle \tau \,, & \rho & \to & - \ov{\rho} \,.
  \end{array}
}

%%%%%%%%%%%%%%%%%%%%%%%%%%%%%%%%%%%%%%%%%%%%%%%
%%%%%%%%%%%%%%%%%%%%%%%%%%%%%%%%%%%%%%%%%%%%%%%

\subsubsection*{$O(2,2,\mathbb Z)$ versus $SL(2,\mathbb Z) \times SL(2,\mathbb Z) \times \mathbb Z_2\times \mathbb Z_2$}

In order to compare our present discussion to the results in section~\ref{sec_t2_fibr_example},
it is useful to translate $O(2,2,\mathbb Z)$ transformations 
into  transformations acting on 
$\tau$ and $\rho$.
To do so, we express the metric and Kalb-Ramond field of the two-torus
in terms of the complex-structure and complexified K\"ahler modulus
$\tau = \tau_1+ i\op\tau_2$ and $\rho = \rho_1 + i\op \rho_2$ as
\eq{
  \label{monodro_392}
  \arraycolsep3pt
  G_{\mathsf{ab}} =  \alpha' \,\frac{\rho_2}{\tau_2} \left( \begin{array}{c@{\hspace{8pt}}c} 
  \tau_1^2 + \tau_2^2 & \tau_1 \\[4pt] \tau_1 & 1 \end{array}\right) ,
  \hspace{50pt}
  B_{\mathsf{ab}} =   \alpha' \left( \begin{array}{c@{\hspace{8pt}}c} 
  0 & + \rho_1 \\[4pt] -\rho_1 & 0 \end{array}\right) .
}
Using these relations, we can write the transformations we encountered in section~\ref{sec_t2_fibr_example}
as follows
\eq{
  \label{monodro_849}
  \arraycolsep2pt
  \begin{array}{l@{\hspace{10pt}}lc@{\hspace{20pt}}lcl@{\hspace{20pt}}lcl}
  \mbox{transformation \eqref{monodro_01} } & \mathcal O_{\mathsf B} &:& \tau & \to & \tau \,,
  & \rho & \to & \rho + h \,,
  \\[16pt]
  \mbox{transformation \eqref{monodro_02} } & \mathcal O_{\mathsf A} &:& \tau & \to & \displaystyle 
   \frac{\tau}{-h\op\tau+1} \,,  & \rho & \to & \rho  \,,
  \\[16pt]
  \mbox{transformation \eqref{monodro_03} } & \mathcal O_{\beta} &:& \tau & \to & 
   \tau \,,  & \rho & \to & \displaystyle  \frac{\rho}{-h\op \rho+1}  \,,
  \\[14pt]
  \mbox{transformation } & \mathcal O_{+1} &:& \tau & \to & 
   \displaystyle -\frac{1}{\rho} \,,  & \rho & \to & \displaystyle  -\frac{1}{\tau} \,,
  \\[16pt]
  \mbox{transformation } & \mathcal O_{+2} &:& \tau & \to & 
   \displaystyle \rho \,,  & \rho & \to & \displaystyle  \tau \,.
  \end{array}
}
The last two lines show how a T-duality transformation along the two directions
$y^1$ and $y^2$ of the two-torus act on the moduli.

%%%%%%%%%%%%%%%%%%%%%%%%%%%%%%%%%%%%%%%%%%%%%%%
%%%%%%%%%%%%%%%%%%%%%%%%%%%%%%%%%%%%%%%%%%%%%%%

\subsubsection*{Constructing the background from the monodromy}

We now want to construct a metric and Kalb-Ramond field for the two-torus
from a given monodromy transformation acting on the complex-structure modulus $\tau$ 
and the complexified K\"ahler modulus $\rho$. 
Let us consider say $M_{\tau}\in SL(2,\mathbb Z)$ acting on $\tau$ via 
M\"obius transformations \eqref{moebius_003}. We let $\tau\equiv \tau(x)$ vary over the 
base-circle and impose that \cite{Dabholkar:2002sy}
\eq{
  \label{mondro_5724}
  \tau(0)=\tau_0\,, \hspace{50pt} 
  \tau(2\pi) = M_{\tau} [ \tau_0 ] \,.
}
The coordinate dependence of $\tau(x)$ is contained in $M_{\tau}(x)$ such that 
$M_{\tau}(2\pi) = M_{\tau}$. 
In order to construct $M_{\tau}(x)$ we consider an element 
$\mathfrak m= \log M_{\tau}$ in the Lie algebra $\mathfrak{sl}(2,\mathbb R)$
and exponentiate, that is
$M_{\tau}(x) =  \exp ( \mathfrak{m} \op x/2\pi)$. 
The $x$-dependence of $\tau(x)$ is then given by
\eq{
  \label{monodro_444}
  \tau(x) = M_{\tau}(x) [ \tau_0 ] \,,
}
which indeed satisfies \eqref{mondro_5724}.
For the complexified K\"ahler modulus $\rho$ a similar discussion with a corresponding 
monodromy matrix $M_{\rho}$ applies.
However, for the two $\mathbb Z_2$ factors in \eqref{isos_849490}  -- containing a T-duality 
transformation -- 
such a construction is not possible since these are not connected to the identity. 
Using then $\tau(x)$ and $\rho(x)$ in \eqref{monodro_392} together with \eqref{back_ex_939393} determines the metric and $B$-field of the full background.

Let us now classify $SL(2,\mathbb Z)$ transformations. As we have seen, such transformations  
can be described by two-by-two matrices $M\in SL(2,\mathbb Z)$ according to
\eqref{moebius_003}, which fall into 
three classes:\label{page_monodro_class}
\begin{flushleft}
\begin{tabular}{@{\hspace{15pt}}l@{\hspace{6pt}}l@{\hspace{4pt}}l}
1. & {\em Elliptic type},  & for which the trace of $M$ satisfies $|{\rm tr}\op M|<2$.
\\[8pt]
2. & {\em Parabolic type}, & for which the trace of $M$ satisfies $|{\rm tr}\op M|=2$.
\\[8pt]
3. & {\em Hyperbolic type}, & for which the trace of $M$ satisfies $|{\rm tr}\op M|>2$.
\end{tabular}
\end{flushleft}
Elliptic transformations are of finite order, in particular of order six, four or three, 
and there are six conjugacy classes. \label{page_sl2_classes}
Parabolic transformations are of infinite order, and there is an infinite number of conjugacy 
classes. For hyperbolic transformations there is a conjugacy class for each value of 
the trace plus additional sporadic classes. More details can be found for instance 
in \cite{DeWolfe:1998pr,Dabholkar:2002sy,Lust:2015yia}.

Following the procedure outlined above, one can now construct 
the background for a given monodromy in $\tau$ and $\rho$. Depending 
on the type of the $SL(2,\mathbb Z)$ transformation, the resulting 
expressions for $\tau(x)$ and $\rho(x)$ differ significantly. 
We do not want to give general discussion of such solutions but  
refer for instance to \cite{Hull:2005hk,Lust:2015yia}.
However, below we illustrate some features of more general 
$\mathbb T^2$-fibrations  through the example of the three-torus 
with simultaneous $H$-, geometric and $Q$-flux.

%%%%%%%%%%%%%%%%%%%%%%%%%%%%%%%%%%%%%%%%%%%%%%%
%%%%%%%%%%%%%%%%%%%%%%%%%%%%%%%%%%%%%%%%%%%%%%%

\subsubsection*{Three-torus with $H$-, $f$- and $Q$-flux -- revisited}

Let us recall \eqref{monodro_849} and note that the $O(2,2,\mathbb Z)$ transformations
$\mathcal O_{\mathsf B}$, $\mathcal O_{\mathsf A}$ 
and $\mathcal O_{\beta}$ individually are all of parabolic type both in $\tau$ and $\rho$.
For these we have discussed the corresponding (non-)geometric backgrounds in 
section~\ref{sec_t2_fibr_example}.
In equation \eqref{super_fluxes} we have also shown a three-dimensional background
with either $H$- and geometric flux or geometric and $Q$-flux. As one can see from
\eqref{monodro_849}, the corresponding transformations 
$  \mathcal O_{\mathsf B(h)} \, \mathcal O_{\mathsf A(f)}$
and   $\mathcal O_{\mathsf A(f)}\, \mathcal O_{\beta(q)}$ are again of 
parabolic type.

However, when all three types of fluxes are present simultaneously we expect to have 
a monodromy transformation of the form
\eq{
  \label{allflux_73748}
  \widetilde{\mathcal O}=\mathcal O_{\mathsf B(h)} \, \mathcal O_{\mathsf A(f)}\, \mathcal O_{\beta(q)} 
  \hspace{5pt}: \hspace{20pt}\tau \to \frac{\tau}{f\,\tau + 1} \,,
  \hspace{20pt}
  \rho \to \frac{(1-h\op q) \op\rho + h}{-q\op \rho + 1}\,,
} 
which are of parabolic type for $\tau$ but which are of varying type for $\rho$ depending on 
the value of $h\op q$.  This can be seen from the corresponding $SL(2,\mathbb Z)$ matrices
\eq{
  M_{\tau} = \left( \begin{array}{cc} 1 & 0 \\ f & 1 \end{array}\right) ,
  \hspace{50pt}
  M_{\rho} = \left( \begin{array}{cc} 1-h\op q & h \\ -q & 1 \end{array}\right) .
}
Depending on the type of monodromy transformation for $\rho$, the functional form 
of $\rho(x)$ is different for each of the three cases mentioned above. In the following we do not 
determine the background for general choices of fluxes $(h,f,q)$, but only want to 
illustrate the main features through some examples.
\begin{itemize}

\item \underline{$(h,f,q) = (+1, f,+1)$:} For this choice of fluxes the $\tau$-transformation
is pa\-ra\-bo\-lic and the $\rho$-transformation
is of elliptic type of order six. The corresponding metric and $B$-field of the torus 
fibre (recall \eqref{back_ex_939393} for our notation) read
\eq{
  \scalebox{0.95}{
  $\begin{split}  
  \arraycolsep4pt
  \renewcommand{\arraystretch}{1.3}    
 G^{(+1,+1)}_{\mathsf{ab}} &= \frac{3\op\alpha'^2}{\Omega^{(+1,+1)}}
    \left( \begin{array}{cc} 
    R_1^2 & R_1^2\op\frac{f\op x}{2\pi} \\ R_1^2\op\frac{f\op x}{2\pi}  & 
    R_2^2+R_1^2  \left[\frac{f\op x}{2\pi} \right]^2 \end{array}\right),
    \\
    B_{12}^{(+1,+1)} &= \frac{2\op\alpha'}{\Omega^{(+1,+1)}}\,\sin\bigl[\tfrac{x}{6}\bigr]
    \biggl( \bigl( \alpha'^2+ R_1^2\op R_2^2 \bigr) \sin\bigl[\tfrac{x}{6}\bigr]
    + \sqrt{3} \op\bigl( \alpha'^2- R_1^2\op R_2^2 \bigr) \cos\bigl[\tfrac{x}{6}\bigr]
    \biggr),
    \\[14pt]
    \Omega^{(+1,+1)}&=2\op R_1^2\op R_2^2 \op\Bigl( 1- \cos\bigl[\tfrac{x}{3}\bigr] \Bigr)
    + \alpha'^2 \Bigl( 2 + \cos\bigl[\tfrac{x}{3}\bigr] + \sqrt{3}\op \sin\bigl[\tfrac{x}{3}\bigr] \Bigr)\,.
    \end{split}$
    }
}

\item \underline{$(h,f,q) = (+2, f,+1)$:} For this choice  the $\tau$-transformation
is again pa\-ra\-bo\-lic and the $\rho$-transformation
is  elliptic of order four. The  metric and $B$-field of the torus 
fibre read
\eq{
  \scalebox{0.95}{
  $\begin{split}  
  \arraycolsep4pt
  \renewcommand{\arraystretch}{1.3}    
 G^{(+2,+1)}_{\mathsf{ab}} &= \frac{2\op\alpha'^2}{\Omega^{(+2,+1)}}
    \left( \begin{array}{cc} 
    R_1^2 & R_1^2\op\frac{f\op x}{2\pi} \\ R_1^2\op\frac{f\op x}{2\pi}  & 
    R_2^2+R_1^2  \left[\frac{f\op x}{2\pi} \right]^2 \end{array}\right),
    \\
    B_{12}^{(+2,+1)} &= \frac{2\op\alpha'}{\Omega^{(+2,+1)}}\,\sin\bigl[\tfrac{x}{4}\bigr]
    \biggl( \bigl( 2\op\alpha'^2+ R_1^2\op R_2^2 \bigr) \sin\bigl[\tfrac{x}{4}\bigr]
    + \bigl( 2\op\alpha'^2- R_1^2\op R_2^2 \bigr) \cos\bigl[\tfrac{x}{4}\bigr]
    \biggr),
    \\[14pt]
    \Omega^{(+2,+1)}&= R_1^2\op R_2^2 \op\Bigl( 1- \cos\bigl[\tfrac{x}{2}\bigr] \Bigr)
    + 2\op\alpha'^2 \Bigl( 1 + \sin\bigl[\tfrac{x}{2}\bigr] \Bigr)\,.
    \end{split}$
    }
}

\item \underline{$(h,f,q) = (+2, f,+2)$:} Here the $\tau$-transformation
is again parabolic  but also the $\rho$-transformation is
parabolic. The  metric and $B$-field  take the form
\eq{
  \scalebox{0.95}{
  $\begin{split}  
  \arraycolsep4pt
  \renewcommand{\arraystretch}{1.3}    
 G^{(+2,+2)}_{\mathsf{ab}} &= \frac{\alpha'^2}{\Omega^{(+2,+2)}}
    \left( \begin{array}{cc} 
    R_1^2 & R_1^2\op\frac{f\op x}{2\pi} \\ R_1^2\op\frac{f\op x}{2\pi}  & 
    R_2^2+R_1^2  \left[\frac{f\op x}{2\pi} \right]^2 \end{array}\right),
    \\
    B_{12}^{(+2,+2)} &= \frac{\alpha'}{\Omega^{(+2,+2)}}\,\frac{x}{\pi}
    \biggl( \bigl( \alpha'^2+ R_1^2\op R_2^2 \bigr) \op\frac{x}{\pi}
    - \bigl( \alpha'^2- R_1^2\op R_2^2 \bigr) 
    \biggr),
    \\[14pt]
    \Omega^{(+2,+2)}&= R_1^2\op R_2^2 \left( \frac{x}{\pi}\right)^2 + \alpha'^2 \left( 1- \frac{x}{\pi}\right)^2\,.
    \end{split}$
    }
}

\item \underline{$(h,f,q) = (+1, f,-1)$:} Finally, this is an example where the $\tau$-transformation
is parabolic and where the $\rho$-transformation is
hyperbolic. The  metric and $B$-field  read
\eq{
  \scalebox{0.95}{
  $\begin{split}  
  \arraycolsep4pt
  \renewcommand{\arraystretch}{1.3}    
 G^{(+1,-1)}_{\mathsf{ab}} &= \frac{\bigl(35+15\sqrt{5}\bigr) \bigl( 6+2 \sqrt{5}\bigr)^{\frac{x}{\pi}}\alpha'^2}{\Omega^{(+1,-1)}}
    \left( \begin{array}{cc} 
    R_1^2 & R_1^2\op\frac{f\op x}{2\pi} \\ R_1^2\op\frac{f\op x}{2\pi}  & 
    R_2^2+R_1^2  \left[\frac{f\op x}{2\pi} \right]^2 \end{array}\right),
    \\
    B_{12}^{(+1,-1)} &= \frac{\alpha'}{\Omega^{(+1,-1)}}
    \Bigl( \bigl(3+\sqrt{5}\bigr)^{\frac{x}{\pi}}- 2^{\frac{x}{\pi}} \Bigr)
    \\
    &\hspace{40pt} \biggl[ \Bigl( \bigl( 4+2\sqrt{5} \bigr)\op 2^{\frac{x}{\pi}}    
    + \bigl(11+5\sqrt{5}\bigr)\bigl( 3 + \sqrt{5}\bigr)^{\frac{x}{\pi}}\Bigr) R_1^2\op R_2^2
    \\
    &\hspace{80pt}+\Bigl( \bigl( 4+2\sqrt{5} \bigr)\bigl( 3+\sqrt{5}\bigr)^{\frac{x}{\pi}}+ \bigl( 11+5\sqrt{5} \bigr)\op 2^{\frac{x}{\pi}} \Bigr) \alpha'^2 \biggr] \,,
    \\[14pt]
    \Omega^{(+1,-1)}&= \bigl( 7+3\sqrt{5} \bigr) \Bigl( 2^{\frac{x}{\pi}} - 
    \bigl( 3+\sqrt{5}\bigr)^{\frac{x}{\pi}}\Bigr)^2 R_1^2\op R_2^2 
    \\
    &\hspace{80pt}+
    \Bigl(\bigl( 14+6\sqrt{5} \bigr)\bigl( 6+2\sqrt{5}\bigr)^{\frac{x}{\pi}}
    +\bigl( 18+8\sqrt{5} \bigr)\op 4^{\frac{x}{\pi}} \Bigr)\alpha'^2\,.
    \end{split}$
    }
}

\end{itemize}
As one can see from these examples, the  form of the background
can be rather complicated and depends on the type of $SL(2,\mathbb Z)$ 
transformation.
These are explicit examples of toroidal backgrounds with simultaneously 
$H$-, geometric and $Q$-flux present.

%%%%%%%%%%%%%%%%%%%%%%%%%%%%%%%%%%%%%%%%%%%%%%%
%%%%%%%%%%%%%%%%%%%%%%%%%%%%%%%%%%%%%%%%%%%%%%%

\subsubsection*{T-duality transformations}

In contrast to the situation discussed in section~\ref{sec_t2_fibr_example}, in general 
T-duality transformations  do  not simply exchange 
the various fluxes.
Let us illustrate this observation again with the three-torus with 
$H$-, geometric and $Q$-flux simultaneously present. 
A T-duality transformation along a direction of two-torus 
fibre changes the monodromy transformation through conjugation, 
similarly as in \eqref{chain_779}. 
Starting from \eqref{allflux_73748} with
$\widetilde {\mathcal O} =  \mathcal O_{\mathsf B(h)} \, \mathcal O_{\mathsf A(f)}\, \mathcal O_{\beta(q)} $, we have
\eq{
  \arraycolsep2pt
  \begin{array}{@{}cc@{\hspace{20pt}}lcl@{\hspace{18pt}}lcl@{}}
  \displaystyle \widetilde {\mathcal O}
  &:& \tau &\to&\displaystyle  \frac{\tau}{f\,\tau + 1} \,,
  &
  \rho &\to&\displaystyle  \frac{(1-h\op q) \op\rho + h}{-q\op \rho + 1}\,,
  \\[16pt]
  \displaystyle \mathcal O^{-1}_{+1}\: \widetilde {\mathcal O} \: \mathcal O_{+1}
  &:& \tau &\to&\displaystyle  \frac{\tau+q}{-h\,\tau + (1-h\op q)} \,,
  &
  \rho &\to&\displaystyle  \rho- f\,,
  \\[16pt]
  \displaystyle \mathcal O^{-1}_{+2}\: \widetilde {\mathcal O} \: \mathcal O_{+2}
  &:& \tau &\to&\displaystyle  \frac{(1-h\op q)\op\tau+h}{-q\,\tau + 1} \,,
  &
  \rho &\to&\displaystyle  \frac{\rho}{f\op\rho-+1}\,,
  \\[16pt]
  \displaystyle \mathcal O^{-1}_{+2} \mathcal O^{-1}_{+1}\: \widetilde {\mathcal O} \:  \mathcal O_{+1}\mathcal O_{+2}
  &:& \tau &\to&\displaystyle  \tau -f \,,
  &
  \rho &\to&\displaystyle  \frac{\rho+q}{-h\op \rho+(1-h\op q)}\,.
  \end{array}
} 
Let us note that after a T-duality along the $y^1$-direction of the $\mathbb T^2$
the resulting monodromy acts on the complexified K\"ahler modulus as
$\rho\to \rho-f$, which corresponds to a gauge transformation of the 
$B$-field. Therefore, this background is geometric -- although with a 
rather complicated transformation behaviour of the complex-structure modulus.
According to our definition 4 on page~\pageref{page_defs_nongeo}, the family of backgrounds 
would therefore be called geometric.

%%%%%%%%%%%%%%%%%%%%%%%%%%%%%%%%%%%%%%%%%%%%%%%
%%%%%%%%%%%%%%%%%%%%%%%%%%%%%%%%%%%%%%%%%%%%%%%
%%%%%%%%%%%%%%%%%%%%%%%%%%%%%%%%%%%%%%%%%%%%%%%
%%%%%%%%%%%%%%%%%%%%%%%%%%%%%%%%%%%%%%%%%%%%%%%
%%%%%%%%%%%%%%%%%%%%%%%%%%%%%%%%%%%%%%%%%%%%%%%
%%%%%%%%%%%%%%%%%%%%%%%%%%%%%%%%%%%%%%%%%%%%%%%
%%%%%%%%%%%%%%%%%%%%%%%%%%%%%%%%%%%%%%%%%%%%%%%
%%%%%%%%%%%%%%%%%%%%%%%%%%%%%%%%%%%%%%%%%%%%%%%

\subsection{\texorpdfstring{$\mathbb T^2$}{T2}-fibrations over \texorpdfstring{$\mathbb P^1$}{P1}}
\label{sec_t2_p1}

In this section we consider $\mathbb T^2$-fibrations over 
the two-sphere $\mathbb P^1$ instead over the circle. We follow the 
papers \cite{Hellerman:2002ax} and \cite{Lust:2015yia}, 
which are based on work  in \cite{Greene:1989ya}.

%%%%%%%%%%%%%%%%%%%%%%%%%%%%%%%%%%%%%%%%%%%%%%%
%%%%%%%%%%%%%%%%%%%%%%%%%%%%%%%%%%%%%%%%%%%%%%%

\subsubsection*{Setting}

\begin{figure}[t]
\centering
\includegraphics[width=200pt]{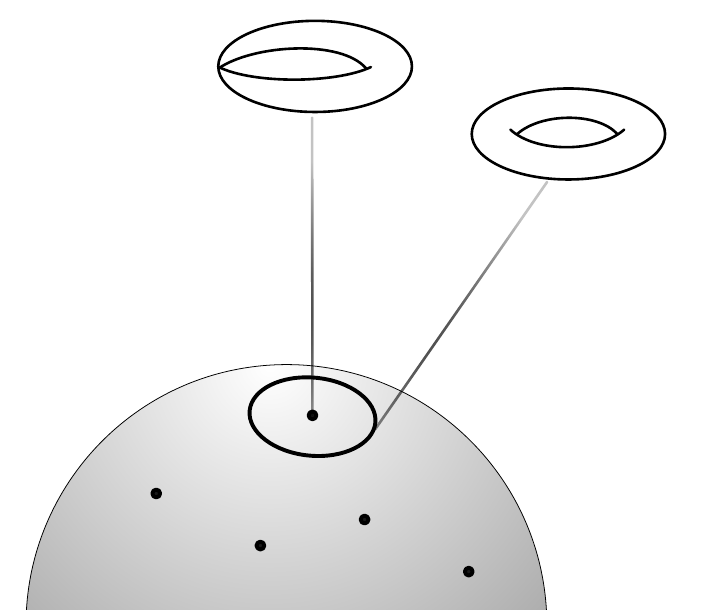}
\caption{Illustration of a $\mathbb T^2$-fibration over a two-sphere $\mathbb P^1$, 
with a number of points where the $\mathbb T^2$-fibre 
degenerates. Along a path around the point of degeneration the $\mathbb T^2$ undergoes
a monodromy transformation.\label{fig_p1t2}}
\end{figure}

More concretely,  we consider $\mathbb T^2$-fibrations over a punctured sphere 
$\mathbb P^1 \setminus \Delta$, where $\Delta=(x_1, \ldots, x_{\mathsf n})$ 
is a set of $\mathsf n$ points in $\mathbb P^1$ where the fibre is allowed to degenerate. 
The degeneration of the fibre will be characterised by the monodromy of 
the complex-structure and K\"ahler moduli of the two-torus 
along a path surrounding the corresponding point in $\mathbb P^1$ (see figure~\ref{fig_p1t2}).
%%%%%%%%%%%%%%%%
%%%%%%%%%%%%%%%%
%%%%%%%%%%%%%%%%
%%%%%%%%%%%%%%%%

Focusing on a single point of degeneration, locally we can choose 
a complex coordinate $z\in \mathbb C$ such that the degeneration is located at the 
origin $z=0$. The geometric data of the $\mathbb T^2$-fibre is encoded in 
the complex-structure and (complexified) K\"ahler moduli, 
and for a non-trivial fibration we let $\tau$ and $\rho$ depend on $z$. 
Here, we require this dependence to be holomorphic, which is 
related to solving the string equations of motion 
and to preserving supersymmetry \cite{Hellerman:2002ax,Bergshoeff:2006jj,deBoer:2012ma}.

%%%%%%%%%%%%%%%%%%%%%%%%%%%%%%%%%%%%%%%%%%%%%%%
%%%%%%%%%%%%%%%%%%%%%%%%%%%%%%%%%%%%%%%%%%%%%%%

\subsubsection*{Degenerations}

Let us now outline the general strategy for determining $\tau(z)$ and $\rho(z)$ from a 
given monodromy. We focus on say the complex-structure modulus, and 
encircling the degeneration corresponds to 
$z\to e^{2\pi \op i}z$. For a given monodromy $M_{\tau}$ we  require that 
the complex structure at $\tau(z+2\pi)$ is related to $\tau(z)$ by a $SL(2,\mathbb Z)$ transformation
\eq{
  \tau \left( e^{2\pi \op i}z \right) = M_{\tau}\bigl[ \tau(z) \bigr] \equiv
  \frac{a\op\tau(z) + b}{c\op \tau(z) + d} \,.
}
In section~\ref{sec_t2_fibr_general} we have encountered a very similar situation. 
Choosing polar coordinates $z=r\op e^{i\op\theta}$, we have already explained 
how to obtain $\tau(\theta)$ for a fixed value of $r=r_0>0$. In particular, 
recalling \eqref{monodro_444} and denoting by $\mathfrak m$ the Lie-algebra element 
corresponding to $M_{\tau}$, we have seen that 
\eq{
  \label{monodro_19046}
  \tau(r_0,\theta) = M_{\tau}(\theta)\bigl[\tau_0(r_0)\bigr] \,,
  \hspace{40pt}
  M_{\tau}(x) =  \exp ( \mathfrak{m} \op \theta/2\pi) \,.
}
Now, in order to determine $\tau(z)$ we allow for arbitrary values of the radius and replace $r_0\to r$
in \eqref{monodro_19046}.
Requiring then that $\tau$ depends holomorphically on $z$ leads to the Cauchy-Riemann
equations
\eq{
  \frac{\partial\tau(r,\theta)}{\partial r} + \frac{i}{r}\, \frac{\partial \tau(r,\theta)}{\partial \theta} =0
  \,.
}
For all three classes of $SL(2,\mathbb Z)$ transformations a solution to these equations always exists, which 
then determines $\tau(z)$ for a given monodromy $M_{\tau}$ (up to integration constants).
Of course, a similar analysis can be performed also for the complexified K\"ahler modulus $\rho$
and corresponding monodromies $M_{\rho}\in SL(2,\mathbb Z)$.

%%%%%%%%%%%%%%%%%%%%%%%%%%%%%%%%%%%%%%%%%%%%%%%
%%%%%%%%%%%%%%%%%%%%%%%%%%%%%%%%%%%%%%%%%%%%%%%

\subsubsection*{Simple examples}

Let us now discuss some examples of $\mathbb T^2$-fibrations with varying complex structure.
The simplest solution for $\tau(z)$ with a non-trivial monodromy is given by \cite{Greene:1989ya}
\eq{
  \tau  = \frac{1}{2\pi \op i} \log z\,,
}
which, when going around the origin via $z\to e^{2\pi \op i} z$, behaves as
\eq{
  \tau \to \tau +1 \,.
}
Other examples for non-trivial monodromies of the complex-structure modulus together 
with their solutions $\tau(z)$ are the following \cite{Lust:2015yia}
\eq{
  \arraycolsep2pt
  \begin{array}{l@{\hspace{30pt}}lcl@{\hspace{30pt}}lcl}
  \mbox{parabolic} & \tau & \to &\tau + b\,, & \tau &=& \displaystyle \frac{b}{2\pi \op i} \log z \,,
  \\[12pt]
  \mbox{elliptic order 6} & \tau & \to &\displaystyle \frac{\tau + 1}{-\tau}\,, 
  & \tau &=& \displaystyle \frac{1-z^{1/3}}{e^{2\pi \op i/3}- e^{4\pi \op i/3}z^{1/3}} \,,  
  \\[12pt]
  \mbox{elliptic order 4} & \tau & \to &\displaystyle -\frac{1}{\tau}\,, 
  & \tau &=& \displaystyle \frac{1-\sqrt{z}}{i+ i \sqrt{z}} \,,  
  \\[12pt]
  \mbox{elliptic order 3} & \tau & \to &\displaystyle -\frac{1}{\tau+1}\,, 
  & \tau &=& \displaystyle \frac{1-z^{2/3}}{e^{2\pi \op i/3}- e^{4\pi \op i/3}z^{2/3}} \,.  
  \end{array}
}

%%%%%%%%%%%%%%%%%%%%%%%%%%%%%%%%%%%%%%%%%%%%%%%
%%%%%%%%%%%%%%%%%%%%%%%%%%%%%%%%%%%%%%%%%%%%%%%

\subsubsection*{Classification}

A convenient way to encode the monodromy of the complex structure $\tau$
for a certain class of fibrations 
is by describing the $\mathbb T^2$-fibre  as an elliptic curve satisfying the 
Weierstrass equation
\eq{
  \label{mondro_weier_1}
  y^2 = x^3 + f(z) \op x + g(z) \,,
}
with $z$ the local coordinate on the base-manifold. The discriminant locus where
the fibre degenerates is given by $\Delta: 0 = 4f^3+27\op g^2$. 
Furthermore, the functional form of $\tau(z)$ is specified implicitly by Klein's $j$-invariant
which is expressed in terms of $f(z)$ and $g(z)$ as
\eq{
  j(\tau) = \frac{(12f)^2}{4f^3+27g^2} \,.
}
Inverting this relation then gives the form of $f(z)$ and $g(z)$, which determine $\tau(z)$. 
Degenerations of the complex structure of an elliptic fibration were classified by
Kodaira \cite{kodaira1,kodaira2,kodaira3}. Depending on the 
functions $f(z)$ and $g(z)$ appearing in \eqref{mondro_weier_1} one
can identify a corresponding monodromy around the degeneration point,
which we summarise in table~\ref{tab_kodaira}.
%%%%%%%%%%%%%%%
%%%%%%%%%%%%%%%
\begin{table}[t]
\centering
\arraycolsep4pt
\begin{tabular}{c@{\hspace{30pt}}c@{\hspace{30pt}}c@{\hspace{30pt}}c}
order of singularity & singularity & Kodaira type & monodromy
\\
\hline\hline
&&&\\[-6pt]
$0$ & smooth & none & 
\scalebox{0.75}{$\left( \begin{array}{cc} 1 & 0 \\ 0 & 1 \end{array}\right)$}
\\[12pt]
$n$ & $A_{n-1}$ & $I_n$ & 
\scalebox{0.75}{$\left( \begin{array}{cc} 1 & n \\ 0 & 1 \end{array}\right)$}
\\[12pt]
$2$ & cusp & $II$ & 
\scalebox{0.75}{$\left( \begin{array}{cc} 1 & 1 \\ -1 & 0 \end{array}\right)$}
\\[12pt]
$3$ & $A_{1}$ & $III$ & 
\scalebox{0.75}{$\left( \begin{array}{cc} 0 & 1 \\ -1 & 0 \end{array}\right)$}
\\[12pt]
$4$ & $A_{2}$ & $IV$ & 
\scalebox{0.75}{$\left( \begin{array}{cc} 0 & 1 \\ -1 & -1 \end{array}\right)$}
\\[12pt]
$6+n$ & $D_{4+n}$ & $I^*_n$ & 
\scalebox{0.75}{$\left( \begin{array}{cc} -1 & -n \\ 0 & -1 \end{array}\right)$}
\\[12pt]
$8$ & $E_6$ & $IV^*$ & 
\scalebox{0.75}{$\left( \begin{array}{cc} -1 & -1 \\ 1 & 0 \end{array}\right)$}
\\[12pt]
$9$ & $E_7$ & $III^*$ & 
\scalebox{0.75}{$\left( \begin{array}{cc} 0 & -1 \\ 1 & 0 \end{array}\right)$}
\\[12pt]
$10$ & $E_8$ & $II^*$ & 
\scalebox{0.75}{$\left( \begin{array}{cc} 0 & -1 \\ 1 & 1 \end{array}\right)$}
\end{tabular}
\caption{Kodaira classification of singularities. The first column
lists the vanishing order of the discriminant locus $\Delta$,
the second column gives the type of singularity in the total space,
the third column lists Kodaira's name for the fibre type, 
and the last column gives the corresponding monodromy matrix.
(For more details see for instance table~1 in 
\cite{Hellerman:2002ax} or table~2 in \cite{Lust:2015yia}.)
\label{tab_kodaira}
}
\end{table}
%%%%%%%%%%%%%%%
%%%%%%%%%%%%%%%
Note however, that not all monodromies can be obtained
via elliptic fibrations, for instance, hyperbolic monodromies 
do not appear in Kodaira's classification.

%%%%%%%%%%%%%%%%%%%%%%%%%%%%%%%%%%%%%%%%%%%%%%%
%%%%%%%%%%%%%%%%%%%%%%%%%%%%%%%%%%%%%%%%%%%%%%%

\subsubsection*{Global model}

Let us finally return to the global setting with $\mathbb P^1$ as the base-manifold. 
There are two conditions we have to impose in order to have a consistent 
background \cite{Hellerman:2002ax}:
\begin{enumerate}

\item The metric on the base-manifold should be that of a two-sphere. By
choosing local complex coordinates $z$ and $\ov z$ the metric on $\mathbb P^1$
can always be brought into the form
\eq{
  ds^2  = e^{\varphi(z,\ov z)}\op dz\op d\ov z \,,
}
where the function $\varphi$ encodes the monodromy.
In the limit $z\to\infty$ and hence far away from the points of degenerations,
the metric should behave as
\eq{
  ds^2 \sim \left\lvert \frac{dz}{z^2} \right\rvert^2 \,,
}
so that in terms of the variable $u=1/z$ the point $z=\infty$ is a smooth point $u=0$ on $\mathbb P^1$. 
Since a codimension two singularity has a deficit angle $\pi/6$ \cite{Greene:1989ya}, a two-sphere is obtained
if the orders of the singularities add up to $24$.

\item The second condition which has to be imposed is that the 
monodromy encircling all singularities should be trivial.
More concretely, as illustrated in figure~\ref{fig_deform_p1}, a monodromy surrounding all singularities
on a $\mathbb P^1$
can always be deformed such that it surrounds a non-degenerate point and hence should be trivial. 
%%%%%%%%%%%%%%%%
%%%%%%%%%%%%%%%%
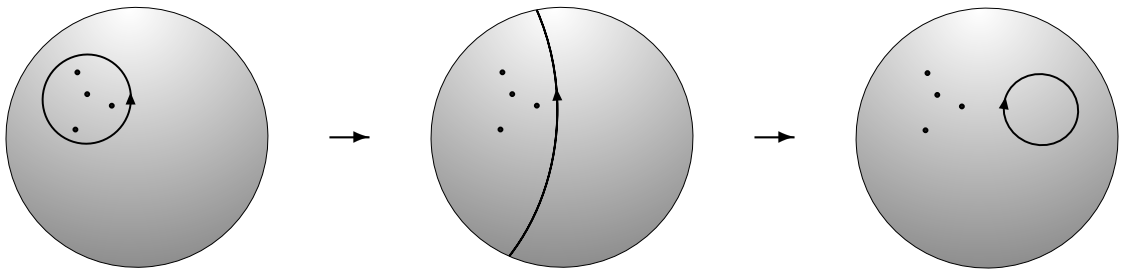
\begin{figure}[t]
\centering
\includegraphics[width=100pt]{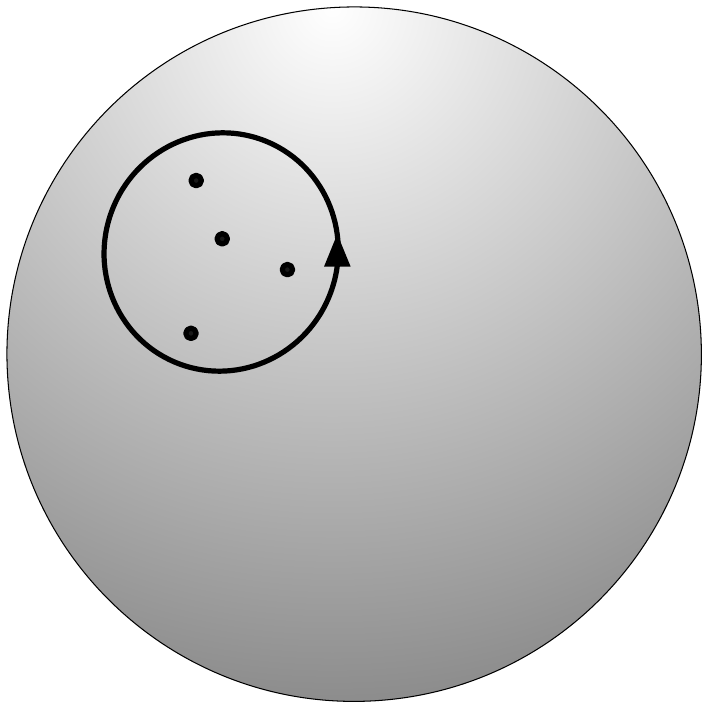}
\hspace{15pt}
\begin{picture}(0,0)
\thicklines
\put(0,50){\vector(1,0){15}}
\end{picture}
\hspace{30pt}
\includegraphics[width=100pt]{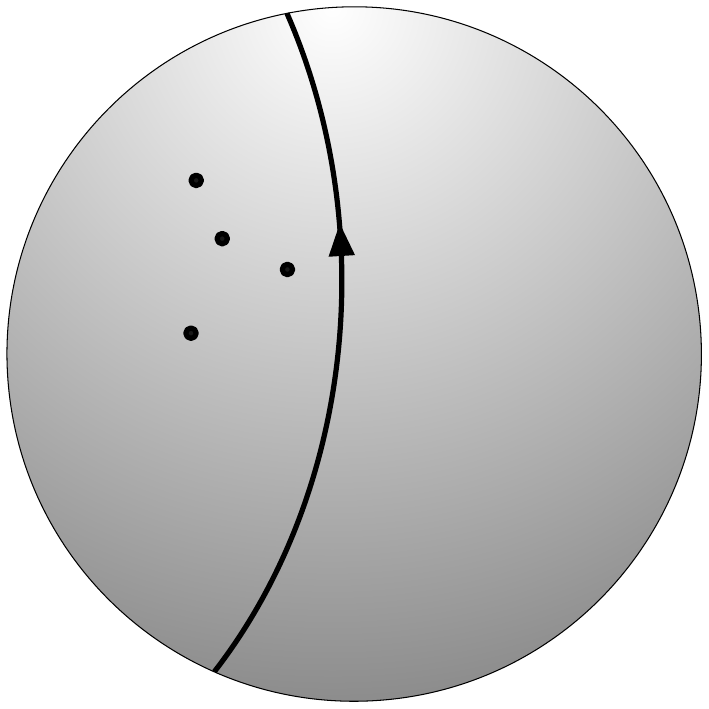}
\hspace{15pt}
\begin{picture}(0,0)
\thicklines
\put(0,50){\vector(1,0){15}}
\end{picture}
\hspace{30pt}
\includegraphics[width=100pt]{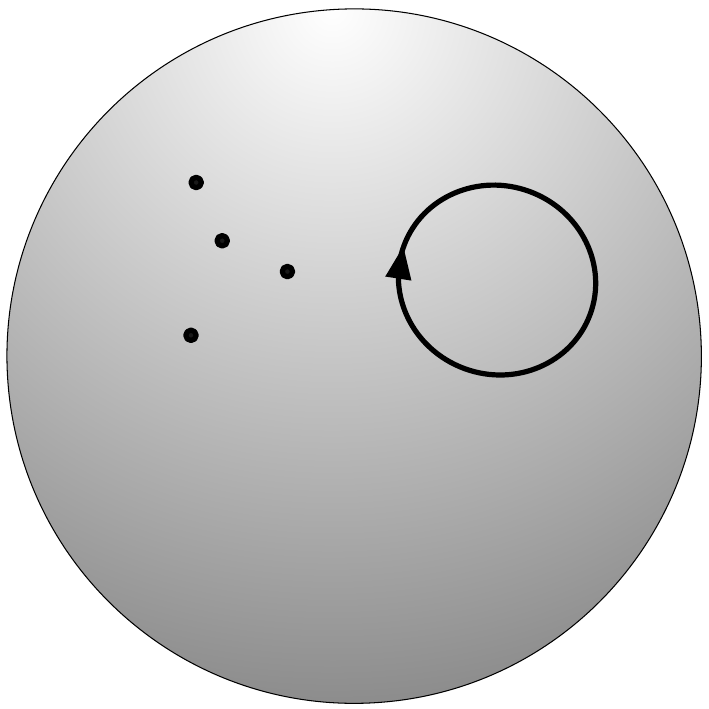}

\caption{Illustration of how a contour on a two-sphere encircling a set of  marked points can be deformed 
to a contour encircling none of these points.
\label{fig_deform_p1}}
\end{figure}
%%%%%%%%%%%%%%%%
%%%%%%%%%%%%%%%%

\end{enumerate}

%%%%%%%%%%%%%%%%%%%%%%%%%%%%%%%%%%%%%%%%%%%%%%%
%%%%%%%%%%%%%%%%%%%%%%%%%%%%%%%%%%%%%%%%%%%%%%%

\subsubsection*{Remark}

Let us recall from \eqref{moebius_005} that the $SL(2,\mathbb Z)$ transformations acting 
on $\tau$ and $\rho$ are generated by $T$- and $S$-transformations. Comparing 
then these transformations with the example of the three-torus (cf. \eqref{monodro_849}), we find\,\footnote{A 
monodromy in $\tau$ of the form $T:\tau\to \tau+1$ leads to a twisted torus with one unit of geometric flux, 
in which the coordinates and radii are interchanged as $y^1\leftrightarrow y^2$, $R_1\leftrightarrow R_2$
as compared to the twisted torus given in \eqref{fibr_1_f}.}
\eq{
  \arraycolsep2pt
  \begin{array}{lcc@{\hspace{30pt}}l@{\hspace{6pt}:\hspace{10pt}}lcl}
  \mbox{three-torus} &\mbox{with} & h=1 & \tilde T & \rho &\to & \rho+1 \,,
  \\[10pt]
  \mbox{twisted torus}\, &\mbox{with} & f=1 & U = S\, T\op S & \tau &\to 
     & \displaystyle \frac{\tau}{-\tau + 1} \,,
  \\[10pt]
  \mbox{T-fold} &\mbox{with} & q=1 & \tilde U = \tilde S\, \tilde T\op \tilde S & \rho &\to 
     & \displaystyle \frac{\rho}{-\rho + 1} \,.
  \end{array}
}  
It is now tempting to identify a monodromy generated by $\tilde T$ with one unit of 
$H$-flux, a monodromy generated by $U$ with one unit of geometric flux and so on. 
However, it turns out that a general monodromy in $SL(2,\mathbb Z)$ can be
expressed in multiple ways in terms of $T$- and $S$-transformations. Hence, 
there is no unique assignment of a flux to a given monodromy in this way \cite{Lust:2015yia}.

%%%%%%%%%%%%%%%%%%%%%%%%%%%%%%%%%%%%%%%%%%%%%%%
%%%%%%%%%%%%%%%%%%%%%%%%%%%%%%%%%%%%%%%%%%%%%%%
%%%%%%%%%%%%%%%%%%%%%%%%%%%%%%%%%%%%%%%%%%%%%%%
%%%%%%%%%%%%%%%%%%%%%%%%%%%%%%%%%%%%%%%%%%%%%%%
%%%%%%%%%%%%%%%%%%%%%%%%%%%%%%%%%%%%%%%%%%%%%%%
%%%%%%%%%%%%%%%%%%%%%%%%%%%%%%%%%%%%%%%%%%%%%%%
%%%%%%%%%%%%%%%%%%%%%%%%%%%%%%%%%%%%%%%%%%%%%%%
%%%%%%%%%%%%%%%%%%%%%%%%%%%%%%%%%%%%%%%%%%%%%%%

\subsection{\texorpdfstring{$\mathbb T^2$}{T2}-fibrations over \texorpdfstring{$\mathbb R^2$}{R2}}
\label{sec_tw_r2}

We also want to discuss $\mathbb T^2$-fibrations over $\mathbb R^2\setminus \Delta$, 
where $\Delta$ is in general  a set of $\mathsf n$ points in $\mathbb R^2$ at which 
the $\mathbb T^2$-fibre is allowed to degenerate. 
However, in this section we focus again on a single point of degeneration located at the 
origin. Furthermore, we embed these four-dimensional fibrations into
ten dimensions in the following way
\eq{
\mathbb R^{1,5}\times \bigl( \mathcal B\ltimes \mathbb T^2 \bigr) \,,
\hspace{40pt} \mathcal B = \mathbb C\setminus \{0\} \,,
}
and require them to be consistent supergravity backgrounds.

%%%%%%%%%%%%%%%%%%%%%%%%%%%%%%%%%%%%%%%%%%%%%%%
%%%%%%%%%%%%%%%%%%%%%%%%%%%%%%%%%%%%%%%%%%%%%%%

\subsubsection*{Setting}

More concretely, we consider ten-dimensional backgrounds with a in general non-trivial 
metric, Kalb-Ramond $B$-field and dilaton $\phi$ which are required to solve the 
string-theoretical equations of motion given in \eqref{eom_beta}.
We make the following ansatz
\eq{
  \label{back_48494}
  \arraycolsep2pt
  \begin{array}{lcl}
  ds^2 & = & \displaystyle \eta_{\mu\nu} \op dx^{\mu}dx^{\nu} + e^{2\varphi_1} \tau_2 \op\rho_2 \,\alpha' dz \op d\ov z
  + G_{\mathsf{ab}}(z)\op dy^{\mathsf a} dy^{\mathsf b} \,,
  \\[6pt]
  B &=& \alpha'\op \rho_1\op dy^1\wedge dy^2 \,,
  \\[6pt]
  e^{2\phi} & = & \rho_2 \,,
  \end{array}
}  
where $x^{\mu}$ with $\mu = 0,\ldots,5$ denote local coordinates in six-dimensional Minkowski
space, $z\in\mathbb C$ is a complex coordinate in the base-manifold $\mathcal B$, 
and $y^{\mathsf a}$ with $\mathsf a=1,2$ are local coordinates on the $\mathbb T^2$-fibre. 
As before, $\tau = \tau_1 + i\op \tau_2$ denotes the 
complex-structure modulus and $\rho=\rho_1+ i\op \rho_2$ denotes the 
complexified K\"ahler modulus of the fibre, and $\varphi = \varphi_1 + i\op \varphi_2$ 
is a meromorphic function on $\mathbb C$.\footnote{In  \eqref{back_48494} 
the real part of $\varphi$ appears as a warp factor. The imaginary part is 
related to supersymmetry transformations \cite{deBoer:2012ma} which
we are not discussing here.}
Requiring the metric on $\mathcal B$ to be single-valued, one can show that
$\varphi(z)$ has to transform as
\eq{
  e^{\varphi(z)} \to   e^{\varphi(z)}\bigl( c\op \tau(z) + d \bigr) \,,
}
under the  monodromy $M$ when encircling the defect \cite{Hellerman:2002ax,Flournoy:2004vn,Bergshoeff:2006jj,deBoer:2012ma}. 
In particular, this ensures that -- depending on the monodromy -- $e^{2\varphi_1} \tau_2$
or $e^{2\varphi_1} \rho_2$ is single-valued. (We do not consider monodromies in 
$\tau$ and $\rho$ simultaneously, for which we refer for instance to \cite{Lust:2015yia}.)

%%%%%%%%%%%%%%%%%%%%%%%%%%%%%%%%%%%%%%%%%%%%%%%
%%%%%%%%%%%%%%%%%%%%%%%%%%%%%%%%%%%%%%%%%%%%%%%

\subsubsection*{Examples for $\tau$-monodromies}

Let us now discuss examples for the three different cases of elliptic, parabolic and hyperbolic 
$\tau$-monodromies mentioned on page~\pageref{page_monodro_class}. Additional examples can be found in 
\cite{Schulgin:2008fv}.
\begin{itemize}

\item Let us start with a parabolic monodromy $\tau\to \tau+1$ along a path around the origin $z=0$. 
We have already discussed the main structure of the solution above, and we find
\eq{
  \label{monodro_9289238928c}
  \tau(z) = \frac{1}{2\pi \op i} \log\left( \frac{z}{\mu} \right) ,
  \hspace{40pt}
  e^{\varphi(z)} = \frac{\alpha'}{R_1^2} \,,
  \hspace{40pt}
  \rho = i\op \frac{R_1\op R_2}{\alpha'} \,,
}
where we included an integration constant $\mu$, $R_{1,2}$ are the radii of the two-torus
and where a possible constant $B$-field has been set to zero. 
Using polar coordinates for $z$ as $z = r\op e^{i\theta}$ and \eqref{back_48494}, we can express the 
corresponding background as follows
\eq{
  \label{monodro_harm_934203}
  \arraycolsep2pt
  \begin{array}{lcl}
  \displaystyle
  ds^2 &=& \displaystyle \eta_{\mu\nu} \op dx^{\mu}dx^{\nu} 
  +  h(r) \Bigl[ 
  \alpha'\bigl(dr^2 + r^2\op d\theta^2\bigr) + R_1^2 \bigl(dy^1\bigr)^2 \Bigr]
  \\[8pt]
  &&\displaystyle \hspace{53pt}{}
  + \frac{\alpha'}{h(r)}\left( dy^2 + \frac{\theta}{2\pi}\op 
  dy^1 \right)^2 ,
  \\[10pt]
  B &=& 0 \,,
  \\[10pt]
  e^{2\phi} &=& \displaystyle \frac{R_1\op R_2}{\alpha'} \,.
  \end{array}
}
The function $h(r)$  depends only on the radial direction and is given by
\eq{
  \label{monodro_harm_934204}
  h(r) =  \frac{R_2}{R_1}\, \frac{\log\left[ \frac{\mu}{r}\right]}{2\pi} \,.
}
This configuration is a solution to the string equations of motion \eqref{eom_beta},
and it is known as the  semi-flat limit of the compactified Kaluza-Klein monopole.
We come back to this point below.

\item Next, we consider an elliptic monodromy of order four, which acts on the 
complex-structure $\tau$ as $\tau \to -1/\tau$ when encircling the degeneration. 
The solution for the holomorphic functions appearing in 
\eqref{back_48494} is given by (see for instance \cite{Lust:2015yia})
\eq{
  \label{monodro_970221}
  \tau(z) = \frac{1-e^{\frac{i\op\kappa}{2}}\sqrt{z}}{i+i\op e^{\frac{i\op\kappa}{2}}\sqrt{z}}\,,
  \hspace{40pt}
  e^{\varphi(z)} = \frac{z^{1/4}}{1-e^{\frac{i\op\kappa}{2}}\sqrt{z}}\,,
  \hspace{40pt}
  \rho = i\op \frac{R_1\op R_2}{\alpha'} \,,
}
where $\kappa$ is an integration constant. The singularity is of 
Kodaira type III, and the  explicit form of the background can  be obtained
using  \eqref{monodro_970221} in \eqref{back_48494} and \eqref{monodro_392}.
Without further discussing this solution, let us simply state the explicit form
\eq{
  \label{monodro_900392}
  \arraycolsep2pt
  \begin{array}{lcl}
  \displaystyle
  ds^2 &=& \displaystyle \eta_{\mu\nu} \op dx^{\mu}dx^{\nu} 
  + \frac{R_1R_2\op \bigl( r^{3/2} - r^{1/2} \bigr)}{1-2\cos\left[ \theta+\kappa\right] r + r^2 }
   \bigl(dr^2 + r^2\op d\theta^2\bigr)
  \\[10pt]
  &&\displaystyle \hspace{53pt}{}
  + \frac{R_1 R_2}{r-1} \biggl[ \hspace{15pt} 
  \bigl( 1 - 2\cos\left[ \tfrac{\theta+\kappa}{2}\right]\sqrt{r} + r \bigr)\op \bigl( dy^1\bigr)^2
  \\
  &&\displaystyle\hspace{107pt}{}-4\sin\left[ \tfrac{\theta+\kappa}{2}\right]\sqrt{r} \op dy^1 dy^2
  \\
  &&\displaystyle\hspace{107pt}{}+
  \bigl( 1 - 2\cos\left[ \tfrac{\theta+\kappa}{2}\right]\sqrt{r} + r \bigr)\op \bigl( dy^2\bigr)^2
  \hspace{10pt}
  \biggr]
  ,
  \\[0pt]
  B &=& 0 \,,
  \\[10pt]
  e^{2\phi} &=& \displaystyle \frac{R_1\op R_2}{\alpha'} \,.
  \end{array}
}

\item Let us also give an example for a hyperbolic monodromy in $\tau$. 
Taking a transformation of the form $\tau\to -N- 1/\tau$ for $N\geq3$, the solution for the holomorphic functions
is given by (see e.g. \cite{Lust:2015yia})
\eq{
  \label{monodro_8610383}
  \arraycolsep2pt
  \begin{array}{lcl}
  \displaystyle \tau(z) &=& \displaystyle \frac{1}{\lambda}\left( \frac{(\kappa_1)^{\lambda^2}(\lambda^2-1)}
  {\kappa_1\op e^{i\op\tilde z}- (\kappa_1)^{\lambda^2}}-1 \right), 
  \\[16pt]
  \displaystyle e^{\varphi(z)} &=& \displaystyle  \kappa_2\, \lambda\, e^{-\frac{i}{2}\op \tilde z}
  \left( \kappa_1 \op e^{i\op \tilde z} - (\kappa_1)^{\lambda^2}  \right),
  \end{array}
  \hspace{40pt}
  \rho = i\op \frac{R_1\op R_2}{\alpha'} \,,
}
where $\kappa_{1,2,3}$ are again integration constants and
\eq{
  & \tilde z = \kappa_3\op (1-\lambda^2)+\frac{1}{\pi}\log(\lambda)
      \log\bigl(\pi \op z(\lambda^2-1)\bigr) \,,
\\
&\lambda = \frac{1}{2} \left( N + \sqrt{N^2-4}\right). 
}
It is not clear how to interpret such solutions near the point of degeneration, since the imaginary part of 
$\tau$ is highly oscillating near $z=0$. 
This is consistent with the fact that diffeomorphisms of hyperbolic type cannot be obtained 
as monodromies of a degenerating elliptic curve and hence cannot be associated with a 
degeneration point of the fibre. 
The resulting background has a rather involved form and will not be presented here.

\end{itemize}

%%%%%%%%%%%%%%%%%%%%%%%%%%%%%%%%%%%%%%%%%%%%%%%
%%%%%%%%%%%%%%%%%%%%%%%%%%%%%%%%%%%%%%%%%%%%%%%

\subsubsection*{Examples for $\rho$-monodromies}

We now turn to the discussion of non-trivial monodromies for the complexified K\"ahler modulus
around the point $z=0$ at which the fibre degenerates. 
For single-valuedness of the metric now $\rho$ determines the transformation of the warp factor 
$\varphi(z)$.
\begin{itemize}

\item We again start with a parabolic monodromy of the form $\rho\to \rho+1$. 
By comparing with the definition of $\rho$ given in \eqref{moduli_99}, we see that 
this transformation corresponds to a gauge transformation of the $B$-field. 
Going through the procedure explained above and fixing the complex structure to a particular 
form, one obtains the following 
solution for the holomorphic functions
\eq{
  \label{monodro_9289238928}
  \tau(z) = i\op \frac{R_1}{R_2} ,
  \hspace{40pt}
  e^{\varphi} = \frac{\alpha'}{R_1^2} \,,
  \hspace{40pt}
  \rho = \frac{1}{2\pi \op i} \log\left( \frac{z}{\mu} \right)   \,.
}
Using these expressions in \eqref{back_48494} and changing to polar coordinates via 
$z = r\op e^{i\op\theta}$, we find the following background
\eq{
  \label{monodro_ns5_flat}
  \arraycolsep2pt
  \begin{array}{lcl}
  ds^2 & = & \displaystyle \eta_{\mu\nu} \op dx^{\mu}dx^{\nu} + h(r) \Bigl[ 
  \alpha'\bigl(dr^2 + r^2\op d\theta^2\bigr) + R_1^2 \bigl(dy^1\bigr)^2 + R_2^2 \bigl(dy^2\bigr)^2 \Bigr]\,,
  \\[10pt]
  B &=& \displaystyle \alpha'\op\frac{\theta}{2\pi} \op dy^1\wedge dy^2 \,,
  \\[10pt]
  e^{2\phi} & = & \displaystyle \frac{R_1\op R_2}{\alpha'} \, h(r) \,,
  \end{array}
}
with
\eq{
  h(r) =  \frac{\alpha'}{R_1\op R_2}\, \frac{\log\left[ \frac{\mu}{r}\right]}{2\pi} \,.
}
As we will explain further below, this is the semi-flat limit of the 
NS5-brane solution compactified on a two-torus.

\item As a second example, we consider a parabolic monodromy of the form $\rho\to \rho/(1-\rho)$.
The holomorphic functions are now specified by
\eq{
  \label{monodro_9289238928b}
  \tau(z) = i\op \frac{R_1}{R_2} ,
  \hspace{40pt}
  e^{\varphi} = \frac{R_2}{2\pi\sqrt{\alpha'}} \log\left(\frac{z}{\mu}\right)\,,
  \hspace{40pt}
  \rho = - \frac{2\pi\op i}{ \log\bigl( \frac{z}{\mu}\bigr)}   \,.
}
This leads to the background
\eq{
  \label{monodro_93473455}
  \arraycolsep2pt
  \begin{array}{lcl}
  ds^2 & = & \displaystyle \eta_{\mu\nu} \op dx^{\mu}dx^{\nu} +  h(r) \op
  \alpha'\op \Bigl[dr^2 + r^2\op d\theta^2\Bigr] 
  \\[8pt]
  &&\displaystyle \hspace{56pt}
  +{}\frac{h(r)}{h(r)^2 + \left[ \frac{R_1\op R_2}{\alpha'} \op \frac{\theta}{2\pi}\right]^2}
  \Bigl[ R_1^2\bigl( dy^1\bigr)^2 + R_2^2\bigl( dy^2\bigr)^2 \Bigr]\,,
  \\[12pt]
  B &=& \displaystyle -\frac{R_1^2\op R_2^2}{2\pi\alpha'} \op\frac{\theta}{h(r)^2 + \left[ \frac{R_1\op R_2}{\alpha'} \op \frac{\theta}{2\pi}\right]^2}  \,,
  \\[16pt]
  e^{2\phi} & = & \displaystyle \frac{R_1\op R_2}{\alpha'}\, \frac{h(r)}
  {h(r)^2 + \left[ \frac{R_1\op R_2}{\alpha'} \op \frac{\theta}{2\pi}\right]^2}  \,,
  \end{array}
}
with
\eq{
  h(r) =  \frac{R_1\op R_2}{\alpha'}\, \frac{\log\left[ \frac{\mu}{r}\right]}{2\pi} \,.
}
This is a non-geometric background, which is also known as the $5^2_2$-brane \cite{deBoer:2010ud,deBoer:2012ma,Hassler:2013wsa}.

\item An elliptic  monodromy in $\rho$ of the form $\rho\to -1/\rho$ can be obtained by 
applying a T-duality transformation along the $y^2$-direction
to the background \eqref{monodro_970221}. 
According to \eqref{monodro_849}, such a duality transformation 
interchanges $\tau$ and $\rho$ leading to 
\eq{
  \tau = i\op \frac{R_1}{R_2} \,,
  \hspace{40pt}
  e^{\varphi(z)} = \frac{z^{1/4}}{1-e^{\frac{i\op\kappa}{2}}\sqrt{z}}\,,
  \hspace{40pt}
  \rho(z) = \frac{1-e^{\frac{i\op\kappa}{2}}\sqrt{z}}{i+i\op e^{\frac{i\op\kappa}{2}}\sqrt{z}}\,,
}
where $\kappa$ is again an integration constant.
The corresponding ten-dimensional solution is obtained by using these expressions
in \eqref{back_48494} for which one finds
\eq{
  \arraycolsep2pt
  \begin{array}{lcl}
  ds^2 & = & \displaystyle \eta_{\mu\nu} \op dx^{\mu}dx^{\nu} 
  + \frac{R_1}{R_2}\op \frac{\alpha' \bigl( r^{3/2} - r^{1/2} \bigr)}{1-2\cos\left[ \theta+\kappa\right] r + r^2 }
   \,\bigl(dr^2 + r^2\op d\theta^2\bigr)
  \\[8pt]
  &&\displaystyle \hspace{56pt}
  +\frac{\alpha'(r-1)}{1+2\cos\left[ \frac12( \theta+\kappa)\right] r^{1/2} + r }\left[ \frac{R_1}{R_2}\op 
  \bigl( dy^1\bigr)^2 + \frac{R_2}{R_1}\op 
  \bigl( dy^2\bigr)^2 \right]
  \\[16pt]
  B &=& \displaystyle -\frac{2\op\alpha'\sin\left[ \frac12( \theta+\kappa)\right] r^{1/2} }{1+2\cos\left[ \frac12( \theta+\kappa)\right] r^{1/2} + r } dy^1\wedge dy^2 \,,
  \\[16pt]
  e^{2\phi} & = & \displaystyle \frac{r-1}{1+2\cos\left[ \frac12( \theta+\kappa)\right] r^{1/2} + r }\,.
  \end{array}
}

\item Finally, an example for a hyperbolic $\rho$-monodromy can be obtained by 
applying again a T-duality transformation to the existing solution \eqref{monodro_8610383}
interchanging $\rho$ and $\tau$. However, we do not discuss this solution further.

\end{itemize}

%%%%%%%%%%%%%%%%%%%%%%%%%%%%%%%%%%%%%%%%%%%%%%%
%%%%%%%%%%%%%%%%%%%%%%%%%%%%%%%%%%%%%%%%%%%%%%%

\subsubsection*{Remark}

In this section we have embedded $\mathbb T^2$-fibrations over $\mathbb R^2\setminus\{0\}$
with non-trivial monodromies into ten dimensions. 
These solutions satisfy the leading-order string equations of motion summarised in
\eqref{eom_beta}. 
However,  a proper string-theory background
will in general have higher-order $\alpha'$-corrections which are not captured via 
\eqref{eom_beta}.
This can be seen also in some of the solutions constructed in this section.
For instance, the background \eqref{monodro_900392} with an 
elliptic monodromy in $\tau$ has a singular behaviour near the core of the defect at $r=1$.
This signals that the effective supergravity description breaks down 
and $\alpha'$-corrections have to be taken into account. 
A similar behaviour can be observed for the example of a hyperbolic monodromy in
$\tau$ given in \eqref{monodro_8610383}.

%%%%%%%%%%%%%%%%%%%%%%%%%%%%%%%%%%%%%%%%%%%%%%%
%%%%%%%%%%%%%%%%%%%%%%%%%%%%%%%%%%%%%%%%%%%%%%%
%%%%%%%%%%%%%%%%%%%%%%%%%%%%%%%%%%%%%%%%%%%%%%%
%%%%%%%%%%%%%%%%%%%%%%%%%%%%%%%%%%%%%%%%%%%%%%%
%%%%%%%%%%%%%%%%%%%%%%%%%%%%%%%%%%%%%%%%%%%%%%%
%%%%%%%%%%%%%%%%%%%%%%%%%%%%%%%%%%%%%%%%%%%%%%%
%%%%%%%%%%%%%%%%%%%%%%%%%%%%%%%%%%%%%%%%%%%%%%%
%%%%%%%%%%%%%%%%%%%%%%%%%%%%%%%%%%%%%%%%%%%%%%%

\subsection{The NS5-brane, KK-monopole and \texorpdfstring{$5^2_2$}{522}-brane}

Comparing the form of the complex-structure and complexified K\"ahler moduli
for the solutions \eqref{monodro_9289238928}, \eqref{monodro_9289238928c}
and \eqref{monodro_9289238928b}, we see that these are related by 
T-duality transformations along the fibre-directions. 
In particular, their monodromies are those of the three-torus with $H$-flux, 
the twisted torus and the T-fold, respectively. 
In this section we now want to study the higher-dimensional 
origin of these backgrounds.

%%%%%%%%%%%%%%%%%%%%%%%%%%%%%%%%%%%%%%%%%%%%%%%
%%%%%%%%%%%%%%%%%%%%%%%%%%%%%%%%%%%%%%%%%%%%%%%

\subsubsection*{The NS5-brane}

The NS5-brane solution of string theory is a well-known solitonic solution to the ten-di\-men\-sio\-nal 
equations of motion \eqref{eom_beta}. It is extended along $(5+1)$ space-time directions, and 
has a four-dimensional Euclidean transverse space. The corresponding ten-dimensional  background fields
take the form
\eq{
  \arraycolsep2pt
  \begin{array}{lcl}
  ds^2 & = & \displaystyle \eta_{\mu\nu} \op dx^{\mu}dx^{\nu} + \mathsf H(\vec x)\op d\vec x^2\,,
  \\[10pt]
  H &=& \displaystyle \star_4 d\mathsf H(\vec x)\,,
  \\[10pt]
  e^{2\phi} & = & \displaystyle e^{2\phi_0} \mathsf H(\vec x) \,,
  \end{array}
  \hspace{50pt}
  \mathsf H(\vec x) = 1+ \frac{1}{|\vec x|^2} \,,
}
where $\mu,\nu=0,\ldots,5$, $\vec x=(x^6,x^7,x^8,x^9)^T$ denote the transversal coordinates 
and $\star_4$ is the Hodge star-operator in the transverse space. Furthermore, 
$e^{\phi_0}= g_s$ is the string-coupling at spatial infinity.

%%%%%%%%%%%%%%%%
%%%%%%%%%%%%%%%%
\begin{figure}[t]
\centering
\begin{subfigure}{360pt}
\centering
\includegraphics[width=350pt]{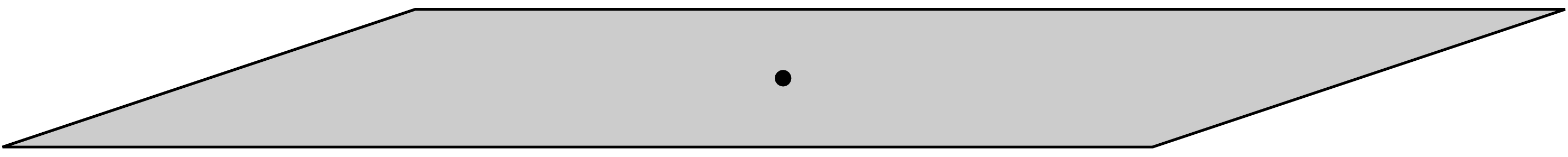}\vspace{-5pt}
\caption{Single NS5-brane.\label{fig_ns5_1}}
\end{subfigure}
\\[28pt]
\begin{subfigure}{360pt}
\centering
\includegraphics[width=350pt]{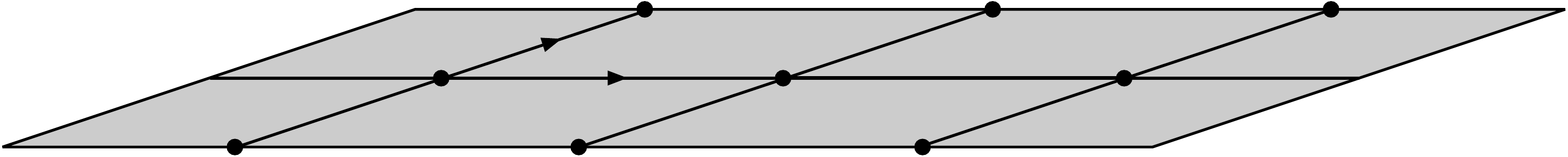}\begin{picture}(0,0)
\put(-224,9){\scriptsize $y^1$}
\put(-242,25){\scriptsize $y^2$}
\end{picture}\vspace{-5pt}
\caption{Infinite array of NS5-branes.\label{fig_ns5_2}}
\end{subfigure}\vspace{5pt}
\caption{Illustration of the space transversal to the NS5-brane solution
for $(x^6,x^7)=(0,0)$. 
In figure~\ref{fig_ns5_1} a single NS5-brane is shown, and in figure~\ref{fig_ns5_2} an infinite 
array of NS5-branes is illustrated.} 
\end{figure}
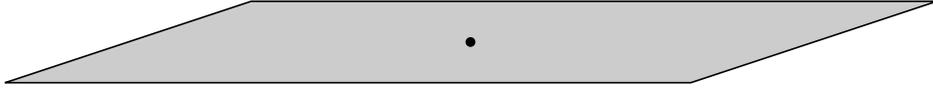
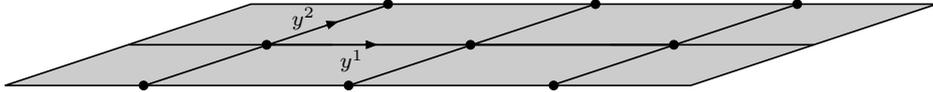
%%%%%%%%%%%%%%%%
%%%%%%%%%%%%%%%%

The NS5-brane solution can be compactified on a two-torus by placing it on an infinite array with 
length $2\pi R_1$ and $2\pi R_2$, as illustrated in figure~\ref{fig_ns5_2}. 
This results in an infinite sum of harmonic functions $\mathsf H(\vec x)$, 
which can be regularised as \cite{Becker:2009df}
\eq{
  \label{monodro_728299a}
  \begin{array}{lcl}
  h(r) &=& \displaystyle  1+\sum_{\vec{n}\in\mathbb{Z}^2}\frac{1}{r^2+\bigl(y^1-2\pi\op\frac{ R_1}{\sqrt{\alpha'}}\op  n_1\bigr)^2+\bigl(y^2-2\pi\op\frac{ R_2}{\sqrt{\alpha'}}\op  n_1\bigr)^2}
  \\[18pt]
  &\rightarrow & \displaystyle  \frac{\alpha'}{R_1\op R_2}\, \frac{\log\left[ \frac{\mu}{r}\right]}{2\pi} 
  + \mathcal O\bigl(e^{-r} \bigr) \,,
  \end{array}
}
where we re-labelled $(x^8,x^9)\to (y^1,y^2)$ and where $r^2 = (x^6)^2+(x^7)^2$ denotes the radial distance in the uncompactified two-dimensional transversal 
space. 
The constant $\mu$ controls the regularisation of the sum.
At leading order in $r$ the compactified solution matches with the background given 
in \eqref{monodro_ns5_flat} and thus provides a ten-dimensional origin for this $\mathbb T^2$-fibration.
The limit shown in \eqref{monodro_728299a}
is also called the semi-flat limit \cite{Strominger:1996it,Vegh:2008jn}.

%%%%%%%%%%%%%%%%%%%%%%%%%%%%%%%%%%%%%%%%%%%%%%%
%%%%%%%%%%%%%%%%%%%%%%%%%%%%%%%%%%%%%%%%%%%%%%%

\subsubsection*{The KK-monopole}

\begin{figure}[t]
\centering
\begin{subfigure}{360pt}
\centering
\includegraphics[height=80pt]{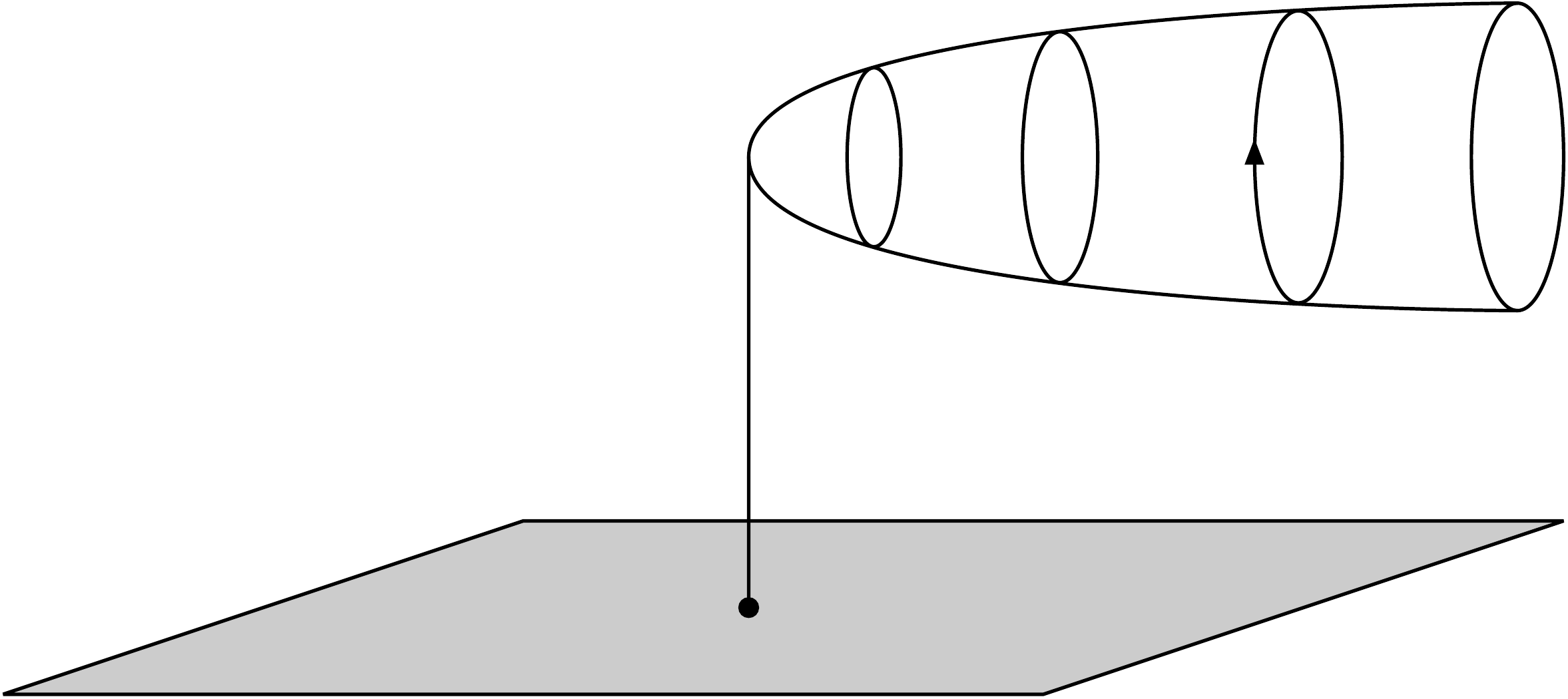}\begin{picture}(0,0)
\put(-46.5,60.5){\scriptsize $y^2$}
\end{picture}\vspace{-5pt}
\caption{Single KK-monopole.\label{fig_kk_1}}
\end{subfigure}
\\[28pt]
\begin{subfigure}{360pt}
\centering
\includegraphics[height=85pt]{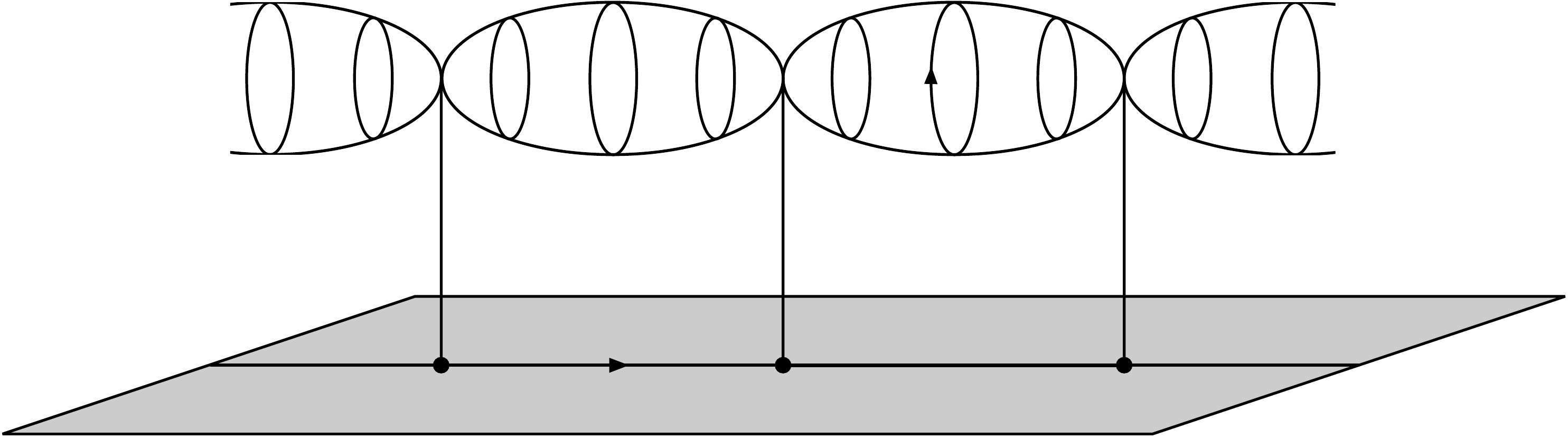}\begin{picture}(0,0)
\put(-191,4.5){\scriptsize $y^1$}
\put(-134,68.5){\scriptsize $y^2$}
\end{picture}\vspace{-5pt}
\caption{Infinite array of KK-monopoles.\label{fig_kk_2}}
\end{subfigure}\vspace{5pt}
\caption{Illustration of the space transversal to the KK-monopole solution
for $x^6=0$. 
In figure~\ref{fig_kk_1} a single KK-monopole is shown, and in figure~\ref{fig_kk_2} an infinite 
array of KK-monopoles is illustrated.} 
\end{figure}
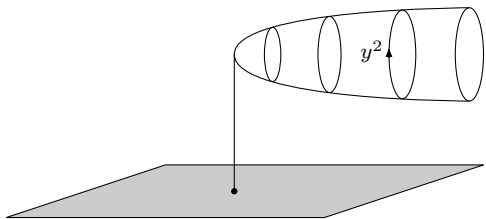
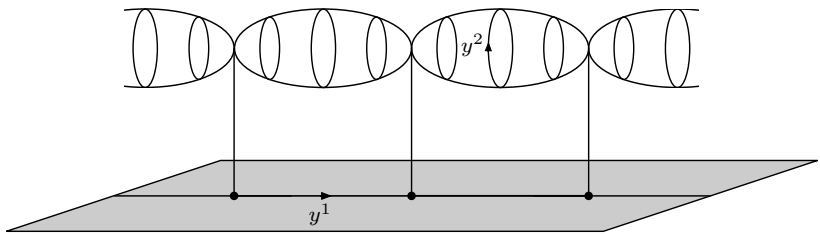

Let us now turn to Kaluza-Klein monopole with compact circle direction $y^2$.
This background is specified by the following field configuration \cite{Sorkin:1983ns,Gross:1983hb}
\eq{
  \arraycolsep2pt
  \begin{array}{lcl}
  ds^2 & = & \displaystyle \eta_{\mu\nu} \op dx^{\mu}dx^{\nu} + \mathsf H(\vec x)\op d\vec x^2
  + \frac{1}{\mathsf H(\vec x)}\op \bigl( dy^2 + A \bigr)^2
  \,,
  \\[10pt]
  H &=& \displaystyle 0\,,
  \\[10pt]
  e^{2\phi} & = & \displaystyle e^{2\phi_0} \,,
  \end{array}
  \hspace{17pt}
  dA = \star_3 d\op \mathsf H(\vec x)\,,
}
where again $\mu,\nu=0,\ldots,5$ but where now $\vec x= (x^6,x^7,x^8)$, and where 
$\star_3$ denotes the Hodge star-operator along the non-compact transversal part.
The non-triviality of the circle-fibration is encoded in the connection one-form $A$, and 
the harmonic function is given by
\eq{
  \mathsf H(\vec x) = 1+ \frac{R_2}{2\sqrt{\alpha'}\,|\vec x|} \,,
}
with $R_2$ denoting the radius of the $y^2$-direction. 
The $H$-flux vanishes, and the dilaton $\phi=\phi_0$ is constant.
An illustration of the space transversal to the KK-monopole can be found 
in figure~\ref{fig_kk_1}.

%%%%%%%%%%%%%%%%
%%%%%%%%%%%%%%%%
%%%%%%%%%%%%%%%%
%%%%%%%%%%%%%%%%

This solution can now be compactified on a on a circle by placing it on an infinite array
with length $2\pi R_1$ (see figure~\ref{fig_kk_2}), which results in an infinite sum of harmonic functions. 
This sum can be regularised as \cite{Ooguri:1996me}
\eq{
  \label{monodro_728299b}
  \begin{array}{lcl}
  h(r) &=& \displaystyle  1+\sum_{n\in\mathbb{Z}}\:\frac{R_2}
  {2\sqrt{\alpha'}\sqrt{r^2+\bigl(y^1-2\pi\op\frac{ R_1}{\sqrt{\alpha'}}\op  n\bigr)^2}}
  \\[22pt]
  &\rightarrow & \displaystyle  \frac{R_2}{R_1}\, \frac{\log\left[ \frac{\mu}{r}\right]}{2\pi} 
  + \mathcal O\bigl(e^{-r} \bigr) \,,
  \end{array}
}
where we re-labelled  $x^8\to y^1$ and where $r^2 = (x^6)^2+(x^7)^2$.
The constant $\mu$ is again related to the regulator, and 
at leading order in $r$  the harmonic function agrees with \eqref{monodro_harm_934204},
and the background is the same as shown in \eqref{monodro_harm_934203}.
The limit \eqref{monodro_728299b} is again called the semi-flat limit.

%%%%%%%%%%%%%%%%%%%%%%%%%%%%%%%%%%%%%%%%%%%%%%%
%%%%%%%%%%%%%%%%%%%%%%%%%%%%%%%%%%%%%%%%%%%%%%%

\subsubsection*{The $5^2_2$-brane}

Unfortunately, for the background \eqref{monodro_93473455} with the non-geometric 
parabolic monodromy $\rho\to \rho/(1-\rho)$ a ten-dimensional origin 
similarly to the two examples discussed above is not known. 
We therefore cannot give a corresponding higher-dimensional solution (within string theory).

%%%%%%%%%%%%%%%%%%%%%%%%%%%%%%%%%%%%%%%%%%%%%%%
%%%%%%%%%%%%%%%%%%%%%%%%%%%%%%%%%%%%%%%%%%%%%%%

\subsubsection*{Family of solutions}

In analogy to the chain of backgrounds connected by T-duality transformations illustrated in \eqref{chain_777}, 
we can now summarise the above solutions in the following way 
\eq{
  \label{chain_778}
  \arraycolsep6pt
  \begin{array}{@{}ccccc@{}}
  \\[-6pt]
  \boxed{\begin{array}{c}\mbox{NS5-brane} \\ \mbox{on $\mathbb T^2$}\end{array}} 
  &&
  \boxed{\begin{array}{c}\mbox{KK-monopole} \\ \mbox{on $S^1$}\end{array}} 
  \\[16pt]
  \rotatebox{90}{$\xleftarrow{\hspace{30pt}}$}\hspace{1pt}
  &&
  \rotatebox{90}{$\xleftarrow{\hspace{30pt}}$}\hspace{1pt}
  \\[2pt]
  \boxed{\begin{array}{c}\mbox{NS5-brane} \\ \mbox{semi-flat}\end{array}} 
  & \xleftrightarrow{\hspace{8pt}\tau\leftrightarrow\rho\hspace{8pt}} &
  \boxed{\begin{array}{c}\mbox{KK-monopole} \\ \mbox{semi-flat}\end{array}} 
  & \xleftrightarrow{\hspace{8pt}\tau\leftrightarrow -\frac{1}{\rho}\hspace{8pt}} &
  \boxed{\begin{array}{c}\mbox{$5^2_2$-brane}\end{array}} 
  \\[12pt]
  \includegraphics[width=25pt,angle=180]{fig_01} &&
  \includegraphics[width=25pt,angle=180]{fig_01} &&
  \includegraphics[width=25pt,angle=180]{fig_01}
  \\[22pt]
  \scriptstyle\rho\to\rho+1 && \scriptstyle\tau\to\tau+1 && \scriptstyle\rho\to \frac{\rho}{-\rho+1}
  \end{array}
}
where on top we have shown the full NS5-brane and KK-monopole solution compactified
on $\mathbb T^2$ and $S^1$, respectively. 
Performing then the semi-flat limit and ignoring higher-order terms in the transversal radius (see 
equations \eqref{monodro_728299a} and \eqref{monodro_728299b}),
we arrive at a family of $\mathbb T^2$-fibrations over $\mathbb C\setminus\{0\}$
which are related by T-duality transformations. We furthermore indicated the 
patching transformations for each of these fibrations.

%%%%%%%%%%%%%%%%%%%%%%%%%%%%%%%%%%%%%%%%%%%%%%%
%%%%%%%%%%%%%%%%%%%%%%%%%%%%%%%%%%%%%%%%%%%%%%%

\subsubsection*{Remarks}

Let us close this section with the following remarks:
\begin{itemize}

\item Around equations \eqref{monodro_728299a} and \eqref{monodro_728299b} 
we have illustrated how -- at leading-order in $r$ -- compactifications 
of the NS5-brane solution and the KK-monopole correspond to the 
backgrounds \eqref{monodro_ns5_flat} and  \eqref{monodro_harm_934203}, respectively.
However, taking into account higher-order corrections in $r$ (and thus capturing 
higher-order $\alpha'$-corrections) modifies this picture \cite{Gregory:1997te}.
On the other hand, it has been shown in \cite{Tong:2002rq,Harvey:2005ab} that corrections to the compactified 
NS5-brane correspond to instanton corrections. 
This analysis has been extended in \cite{Harvey:2005ab,Jensen:2011jna,Kimura:2013fda,Kimura:2013zva,Lust:2017jox} 
to include $\mathbb T^2$-compactifications.

\item We also remark that a discussion of the T-duality chain \eqref{chain_778} in the context of the heterotic string 
can be found in \cite{Sasaki:2016hpp}, and an effective world-volume action for the $5^2_2$-brane
has been proposed in \cite{Chatzistavrakidis:2013jqa}.

\item In this section we have studied the NS5-brane and its T-dual backgrounds. 
However, string-theory also features S-duality which for instance maps 
a NS5-brane to a D5-brane. The latter is then related to other D$p$-branes via T-duality. 
This gives rise to a plethora of localised sources, which can have more general non-geometric
properties. 

The combination of T-duality and S-duality is called U-duality,
and an appropriate framework to study corresponding sources is exceptional field theory (EFT).
Since it is beyond the scope of this work to discuss EFT,  we want to refer the reader to the review
\cite{Hohm:2013vpa,Hohm:2013uia,Hohm:2014fxa}. 
We note however that brane solutions obtained from U-duality transformations have been studied 
for instance in 
\cite{
Meessen:1998qm,
Eyras:1998hn,
Obers:1998fb,
LozanoTellechea:2000mc,
Bergshoeff:2011ee,
Kleinschmidt:2011vu,
deBoer:2012ma,
Chatzistavrakidis:2014sua,
Bergshoeff:2015cba,
Bakhmatov:2017les,
Berman:2018okd
}.

\end{itemize}

%%%%%%%%%%%%%%%%%%%%%%%%%%%%%%%%%%%%%%%%%%%%%%%
%%%%%%%%%%%%%%%%%%%%%%%%%%%%%%%%%%%%%%%%%%%%%%%
%%%%%%%%%%%%%%%%%%%%%%%%%%%%%%%%%%%%%%%%%%%%%%%
%%%%%%%%%%%%%%%%%%%%%%%%%%%%%%%%%%%%%%%%%%%%%%%
%%%%%%%%%%%%%%%%%%%%%%%%%%%%%%%%%%%%%%%%%%%%%%%
%%%%%%%%%%%%%%%%%%%%%%%%%%%%%%%%%%%%%%%%%%%%%%%
%%%%%%%%%%%%%%%%%%%%%%%%%%%%%%%%%%%%%%%%%%%%%%%
%%%%%%%%%%%%%%%%%%%%%%%%%%%%%%%%%%%%%%%%%%%%%%%

\subsection{Remarks}
\label{sec_t2_remarks}

In this last section we want to give an overview of further aspects and developments
in the context of torus fibrations which we did not cover above.
\begin{itemize}

\item A dimensional reduction of the backgrounds
studied in section~\ref{sec_t2_fibr_example} and \ref{sec_t2_fibr_general}
along the compact directions corresponds to a generalised Scherk-Schwarz
reduction \cite{Scherk:1978ta,Scherk:1979zr}. In the reduced lower-dimensional theory
a potential is generated, which can be specified in terms of the 
monodromy matrix of the fibration \cite{Dabholkar:2002sy}.
Monodromies of elliptic $SL(2,\mathbb Z)$-type have finite order and are conjugate
to rotations. This implies that the corresponding potential will always 
have a stable minimum at the fixed-point of the monodromy. 
Parabolic monodromies have fixed points which correspond to 
decompactification limits, and hyperbolic monodromies do not have 
critical points on the upper half-plane. 
We discuss this point in detail in section~\ref{sec_schsch}.

\item In the above sections we have focused on $\mathbb T^2$-fibrations over various
base-manifolds. Of course, one can also consider  $\mathbb T^n$-fibrations for $n\geq 3$, however, in this 
case the structure of the monodromy group becomes more involved \cite{Achmed-Zade:2018rfc}. 
Nevertheless, in \cite{Vegh:2008jn} a number of models in different dimensions are constructed.

More generally, one can also fibre $K3$-manifolds non-trivially over a base and 
consider monodromies in the duality group of $K3$. For mirror symmetry such constructions have been 
called mirror-folds and have been 
discussed in 
\cite{Kawai:2007nb,Israel:2013wwa,Hull:2017llx}.

\item Non-geometric backgrounds can also be studied for the heterotic string. 
A difference to the type II constructions presented in this section
is that on the heterotic side also Wilson-moduli have to be taken into account, 
which enlarge the duality group. 
For some explicit constructions see for instance 
\cite{Flournoy:2004vn}.
Via the heterotic--F-theory duality, non-geometric heterotic models can be 
mapped to the F-theory side
\cite{McOrist:2010jw,Malmendier:2014uka,Gu:2014ova,Lust:2015yia,Font:2016odl,Kimura:2018oze}.
Here one finds that some of the heterotic non-geometric constructions 
correspond to geometric F-theory models. This means that 
non-geometric backgrounds are a natural  part of string theory.

\item A useful way to understand geometric as well as non-geometric compactifications
is by geometrising the duality group. 
This is inspired by F-theory \cite{Vafa:1996xn}, where the (in general varying) axio-dilaton is interpreted as 
the complex-structure modulus of a two-torus $\mathbb T^2$ fibred over 
the space-time. 
The S-duality group $SL(2,\mathbb Z)$ then acts on $\mathbb T^2$ in a geometric 
way as large diffeomorphisms, similarly as we discussed above. 
For a recent review on F-theory see for instance \cite{Weigand:2018cod}.

For the  $\mathbb T^2$-fibrations considered in the above sections we 
compactify along the $\mathbb T^2$-directions such that 
we are left with a lower-dimensional theory with a complex-structure modulus
$\tau(x)$ and a complexified K\"ahler modulus $\rho(x)$. 
The corresponding T-duality group is 
$SL(2,\mathbb Z) \times SL(2,\mathbb Z) \times \mathbb Z_2\times \mathbb Z_2$.
Since $SL(2,\mathbb Z)$ is the modular group of a two-torus,
one can associate to the full duality group a degenerate genus two-surface of the form
\eq{
\includegraphics[width=170pt]{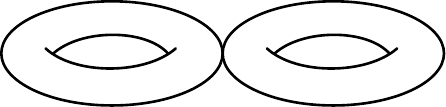}
}
On this surface the T-duality group acts in a geometric way, namely as the 
mapping-class group. 
For larger duality groups a corresponding discussion can be found in
\cite{Candelas:2014jma,Candelas:2014kma}.

\item Instead of restricting oneself to the T-duality group, one can also construct 
string-theory backgrounds with patching-transformations contained 
in larger duality groups. For certain type II compactifications this
is for instance the $U$-duality group (a combination of T- and S-duality) \cite{Hull:1994ys,Witten:1995ex}, 
and corresponding $U$-folds have been discussed for instance in 
\cite{Kumar:1996zx,Liu:1997mb,Martucci:2012jk,Braun:2013yla}.

\item Another way to describe non-geometric $\mathbb T^n$-fibrations is through Hull's doubled 
formalism \cite{Hull:2004in} to be discussed in section~\ref{cha_dg}.
Here, one doubles the dimension of the torus fibre to $\mathbb T^{2n}$ where -- roughly speaking --
one considers the left- and right-moving modes shown for instance in \eqref{cft_003}
as independent coordinates. 
The physical torus-fibre is obtained by choosing a $n$-dimensional subspace
of $\mathbb T^{2n}$, which is also called a polarisation.
Duality transformations $O(n,n,\mathbb Z)$ then change this polarisation within the doubled space, 
and hence lead to dual configurations. 
The doubled formalism is particularly useful to describe non-geometric backgrounds 
and we come back to this question in section~\ref{cha_dg}.

\end{itemize}

%%%%%%%%%%%%%%%%%%%%%%%%%%%%%%%%%%%%%%%%%%%%%%%
%%%%%%%%%%%%%%%%%%%%%%%%%%%%%%%%%%%%%%%%%%%%%%%
%%%%%%%%%%%%%%%%%%%%%%%%%%%%%%%%%%%%%%%%%%%%%%%
%%%%%%%%%%%%%%%%%%%%%%%%%%%%%%%%%%%%%%%%%%%%%%%
%%%%%%%%%%%%%%%%%%%%%%%%%%%%%%%%%%%%%%%%%%%%%%%
%%%%%%%%%%%%%%%%%%%%%%%%%%%%%%%%%%%%%%%%%%%%%%%
%%%%%%%%%%%%%%%%%%%%%%%%%%%%%%%%%%%%%%%%%%%%%%%
%%%%%%%%%%%%%%%%%%%%%%%%%%%%%%%%%%%%%%%%%%%%%%%
%%%%%%%%%%%%%%%%%%%%%%%%%%%%%%%%%%%%%%%%%%%%%%%
%%%%%%%%%%%%%%%%%%%%%%%%%%%%%%%%%%%%%%%%%%%%%%%
%%%%%%%%%%%%%%%%%%%%%%%%%%%%%%%%%%%%%%%%%%%%%%%
%%%%%%%%%%%%%%%%%%%%%%%%%%%%%%%%%%%%%%%%%%%%%%%
%%%%%%%%%%%%%%%%%%%%%%%%%%%%%%%%%%%%%%%%%%%%%%%
%%%%%%%%%%%%%%%%%%%%%%%%%%%%%%%%%%%%%%%%%%%%%%%
%%%%%%%%%%%%%%%%%%%%%%%%%%%%%%%%%%%%%%%%%%%%%%%
%%%%%%%%%%%%%%%%%%%%%%%%%%%%%%%%%%%%%%%%%%%%%%%

\clearpage
\section{Generalised geometry}
\label{sec_gen_geo}

In this section we take a different perspective on T-duality, 
non-geometric backgrounds and non-geometric fluxes. The framework we are going to discuss 
is that of generalised geometry, which
has been developed in \cite{Hitchin:2004ut,Gualtieri:2003dx}.
For a treatment in the physics literature we refer for instance to
\cite{Grana:2006kf,Zabzine:2006uz,Ellwood:2006ya,Grana:2008yw,Berman:2010is}, which we  follow in parts in this section. 

%%%%%%%%%%%%%%%%%%%%%%%%%%%%%%%%%%%%%%%%%%%%%%%
%%%%%%%%%%%%%%%%%%%%%%%%%%%%%%%%%%%%%%%%%%%%%%%
%%%%%%%%%%%%%%%%%%%%%%%%%%%%%%%%%%%%%%%%%%%%%%%
%%%%%%%%%%%%%%%%%%%%%%%%%%%%%%%%%%%%%%%%%%%%%%%

\subsection{Basic concepts}
\label{sec_gg_basics}

The main motivation for generalised geometry was to combine complex 
and symplectic manifolds -- admitting a complex structure and a symplectic structure, respectively --
into a common framework \cite{Hitchin:2004ut,Gualtieri:2003dx}.
It turns out that this allows to describe diffeomorphisms and gauge transformations
of the Kalb-Ramond $B$-field in a combined way.
In this section we introduce the basic concepts of generalised geometry, and in later 
sections use them for the description of non-geometric backgrounds.

%%%%%%%%%%%%%%%%%%%%%%%%%%%%%%%%%%%%%%%%%%%%%%%
%%%%%%%%%%%%%%%%%%%%%%%%%%%%%%%%%%%%%%%%%%%%%%%

\subsubsection*{Generalised tangent-bundle}

In generalised geometry vector-fields and differential one-forms are combined into 
a unified framework. The main idea is to consider 
a so-called generalised tangent-bundle $E$ over a $D$-dimensional manifold $M$, which  can be introduced 
via the sequence
\eq{
  \label{gg_sequ}
  0 \longrightarrow T^*M \longrightarrow E \longrightarrow TM \longrightarrow 0 \,.
}
Locally $E$ is the direct sum $TM\oplus T^*M$ of the tangent-bundle $TM$ and the cotangent-bundle $T^*M$ of a manifold $M$,
and the sections of $E$ are called generalised vectors 
which contain a vector-part $x$ and a one-form part $\xi$. 
Again locally, these can be expressed as
\eq{
  X = x + \xi \,, \hspace{60pt} x\in \Gamma( TM)\,, \quad \xi\in \Gamma(T^*M) \,.
}
The non-triviality of the generalised tangent-bundle $E$ is encoded in transition functions 
between local patches $U_{\mathsf a}\subset M$. 
When going from one patch $U_{\mathsf a}$ to another patch $U_{\mathsf b}$, 
diffeomorphisms are used to relate vectors and one-forms --  
describing the non-triviality of $TM$. But, additional transformations of the one-forms 
encode how $T^*M$ is fibred over $TM$. 
In formulas, this reads\,\footnote{In this section we will employ a coordinate-free notation for vector-fields and differential forms for most of the time. However, sometimes we also use $\{\partial_i\} \in \Gamma(TM)$ 
and $\{dx^i\}\in \Gamma(T^*M)$ as  local bases for the tangent- and cotangent-space.}
\eq{
  \label{transf_e01}
  x_{\mathsf a} + \xi_{\mathsf a} = \mathsf A^{-1}{}_{\hspace{-10pt}\mathsf {ab}} \, x_{\mathsf b} + \Bigl[\op
  \mathsf A^{T}{}_{\hspace{-5pt}\mathsf {ab}}\op \xi_{\mathsf b} - \iota_{\mathsf A^{-1}{}_{\hspace{-8pt}\mathsf {ab}}\, x_{\mathsf b}} 
  \mathsf  B_{\mathsf {ab}}
  \op\Bigr] \,,
}
where $\mathsf A_{\op\mathsf {ab}}\in GL(D, \mathbb R)$ is an invertible matrix describing diffeomorphisms, 
$\mathsf B_{\op\mathsf {ab}}$ is  a two-form\footnote{In the following we use the notation
$\mathsf B$ both for a two-form $\mathsf B= \frac12 \mathsf B_{ij} dx^i\wedge dx^j$
as well as for its components $\mathsf B_{ij}$. 
The distinction between a two-form and an anti-symmetric matrix should be clear from the context.
\label{foot_notation}}
 and $\iota_x$ denotes the contraction with a  vector-field $x$.
The notation $\mathsf {ab}$ indicates that we are working on the overlap
$U_{\mathsf a}\cap U_{\mathsf b}$ of  two local patches $U_{\mathsf a}$ and $U_{\mathsf b}$.
Using a two-component notation for the generalised vector and recalling 
the $2D\times 2D$ matrices \eqref{odd_010} and \eqref{b_transform_94},
we can express
\eqref{transf_e01} as
\eq{
  \label{gen_vec_02}
  \arraycolsep5pt
  X_{\mathsf a} = \binom{x_{\mathsf a}}{\xi_{\mathsf a}}
  =  \left( \begin{array}{cc} \mathds 1 & 0 \\ \mathsf B_{\op\mathsf {ab}} & \mathds 1 \end{array}\right)
  \left( \begin{array}{cc} \mathsf A_{\op\mathsf {ab}}^{-1} & 0 \\ 0 & \mathsf A_{\op\mathsf {ab}}^{T} \end{array}\right)
  \binom{x_{\mathsf b}}{\xi_{\mathsf b}}
  = \mathcal O_{\mathsf B_{\op\mathsf {ab}}} \mathcal O_{\mathsf A_{\op\mathsf {ab}}}
  X_{\mathsf b}\,.
}
Furthermore, one usually restricts the two-form as $\mathsf B_{\op\mathsf {ab}} = d\Lambda_{\mathsf {ab}}$,
where 
on the triple overlap $U_{\mathsf a}\cap U_{\mathsf b} \cap U_{\mathsf c}$
the one-forms $\Lambda_{\mathsf {ab}}$ have to satisfy (see also our discussion on page~\pageref{page_global_v})
\eq{
  \Lambda_{\mathsf {ab}} + \Lambda_{\mathsf {bc}} + \Lambda_{\mathsf {ca}}  = g^{-1}_{\mathsf {abc}}
  \bigl( d g_{\mathsf {abc}} \bigr) \,.
}
The function 
$g_{\mathsf {abc}}$ is an element of $U(1)$ and is given by
$g_{\mathsf {abc}} = e^{i\op \lambda_{\mathsf {abc}}}$, 
which describes the structure of a gerbe. 
The generalised tangent-bundle therefore geometrises diffeomorphisms and $B$-field 
gauge transformations.

%%%%%%%%%%%%%%%%%%%%%%%%%%%%%%%%%%%%%%%%%%%%%%%
%%%%%%%%%%%%%%%%%%%%%%%%%%%%%%%%%%%%%%%%%%%%%%%

\subsubsection*{Bi-linear form and $O(D,D,\mathbb R)$}

Given the generalised tangent-bundle $E$, there is a natural bilinear form of signature $(D,D)$.
Denoting by  $\iota_x$ again the contraction with a vector-field $x$, we have
\eq{
  \label{bilin_04}
  \bigl\langle X , Y \bigr\rangle = 
  \bigl\langle x+ \xi , y + \chi \bigr\rangle = \frac{1}{2} \bigl( \iota_x \chi + \iota_y \xi \bigr) \,,
}
where $x,y\in \Gamma(TM)$ and $\xi,\chi\in \Gamma(T^*M)$.
Employing the $2D$-component vector-notation shown in \eqref{gen_vec_02},
 we can express \eqref{bilin_04} as
\eq{
  \label{bilin_0444493}
   \bigl\langle X , Y \bigr\rangle = 
   \frac{1}{2} \op X^T \eta \, Y\,,
  \hspace{80pt}
   \eta =  \left( \begin{array}{cc} 0 & \mathds 1 \\[2pt] \mathds 1 & 0 \end{array}\right) ,
}
where we adopted the same notation as in equation \eqref{gen_met_098}.
The transformations which leave this inner product invariant are 
$O(D,D,\mathbb R)$ transformations $\mathcal O$ characterised by $\mathcal O^T \eta\, \mathcal O = \eta$. 
Note that in contrast to our discussion in section~\ref{sec_cft_torus}, there are no restrictions
on the vector-fields being integers and hence the transformations take values in $\mathbb R$.

Let us now  study the basic transformations belonging to the structure group $O(D,D,\mathbb R)$. 
This discussion is  similar to the one  in section~\ref{sec_cft_torus},
which we recall here from a slightly different perspective:\label{page_odd_action_83}
\begin{itemize}

\item As we already mentioned, diffeomorphisms are described by matrices of the form \eqref{odd_010}, which are  expressed in terms of 
$\mathsf A\in GL(D,\mathbb R)$. On the 
generalised vectors these act as
\eq{
  \label{gg_a_transform_07687}
  X' = \mathcal O_{\mathsf A} \op X 
  \hspace{40pt} \arraycolsep5pt
\mathcal O_{\mathsf A} = \left( \begin{array}{cc} \mathsf A^{-1} & 0 \\ 0 & \mathsf A^{T} \end{array}\right)
  .
}

\item The so-called $\mathsf B$-transforms with $O(D,D,\mathbb R)$ a matrix of the form \eqref{b_transform_94} is
expressed in terms of an anti-symmetric $D\times D$ matrix $\mathsf B$.
Here we have
\eq{
  \label{gg_b-transform_048}
  &X' = \mathcal O_{\mathsf B} \op X \,,
  \hspace{40pt}
    \arraycolsep5pt
  \mathcal O_{\mathsf B} = \left( \begin{array}{cc} \mathds 1 & 0 \\ \mathsf B & \mathds 1 \end{array}\right)
  \,,
  \\[10pt]
  &X=x+\xi \;\mapsto\; X'= x + \bigl( \xi - \iota_x\op  \mathsf B\bigr) \,,
}
where in the second line $\mathsf B$ is interpreted as a two-form with components given by the 
anti-symmetric matrix $\mathsf B_{ij}$ (cf. footnote \ref{foot_notation}). 

\item The so-called  $\beta$-transforms are described by matrices of the form
\eqref{beta_trans_101}, which are expressed in terms of an anti-symmetric $D\times D$ matrix $\beta$. For such
transformations we have
\eq{
  \label{gg_beta_transform_120}
  &X' = \mathcal O_{\beta} \op X  \,,
  \hspace{40pt}
    \arraycolsep5pt
  \mathcal O_{\beta} = \left( \begin{array}{cc} \mathds 1 & \beta \\ 0 & \mathds 1 \end{array}\right)
  \,,
  \\[10pt]
  &X=x+\xi \;\mapsto\; X'= \bigl( x + \beta \op\llcorner \xi \bigr) + \xi \,,
}
where the action of the bivector-field $\beta= \frac12 \beta^{ij}\partial_i\wedge \partial_j$ on forms is 
defined via the contraction. In particular, for a one-form $\xi$ one has $ \beta \op\llcorner \xi = - \xi_i \op\beta^{ij} \op\partial_j$, where $\{\partial_i\}$ is a local basis on the  tangent-space $TM$.

\item Finally, transformations expressed in terms of $O(D,D,\mathbb R)$ matrices 
of the form \eqref{odd_005} interchange vector-field and one-form 
components. With $E_{\mathsf i}$ a $D\times D$ matrix whose only non-vanishing component is $(E_{\mathsf i})_{\mathsf i\mathsf i}=1$, we have the transformations
\eq{
  \label{gg_transf_344244}
&X' = \mathcal O_{\pm\mathsf i} \op X \,,
  \hspace{40pt}
  \mathcal O_{\pm \mathsf i} = \left( \begin{array}{cc} \mathds 1 - E_{\mathsf i}& \pm E_{\mathsf i} \\ \pm E_{\mathsf i} & \mathds 1 -E_{\mathsf i} \end{array}\right)
  .
}

\end{itemize}

%%%%%%%%%%%%%%%%%%%%%%%%%%%%%%%%%%%%%%%%%%%%%%%
%%%%%%%%%%%%%%%%%%%%%%%%%%%%%%%%%%%%%%%%%%%%%%%

\subsubsection*{Courant bracket}

Similarly to having  a Lie bracket for vector-fields on the tangent-space, one can define a corresponding bracket for the 
generalised tangent-space $E$. 
In the present case this is the   Courant bracket which  is given by \cite{courant}
\eq{
  \label{courant_brack_620}
  \bigl[ X, Y \bigr]_{\rm C} = 
  \bigl[ x+ \xi , y + \chi \bigr]_{\rm C} = 
  [\op x,y\op]_{\rm L} + \mathcal L_x \chi - \mathcal L_y \xi - \frac12 \op d \bigl( \iota_x \chi - \iota_y \xi \bigr)
  \,,
}
where $[\op\cdot\op,\cdot\op]_{\rm L}$ denotes the usual Lie bracket of vector-fields and
$\iota_x$ denotes the contraction with the vector-field $x$. 
Note that the Lie derivative $\mathcal L$ and the contraction $\iota$ satisfy the 
following relations
\eq{
  \label{lie_rel_9846}
  \mathcal L_x = d\, \iota_x + \iota_x \, d\,,
  \hspace{40pt}
  \mathcal L_{[x,y]_{\rm L}} = [\mathcal L_x, \mathcal L_y] \,,
  \hspace{40pt}
  \iota_{[x,y]_{\rm L}} = [\mathcal L_x, \iota_y] \,.
}
The Courant bracket is anti-symmetric and maps 
two generalised vectors to another generalised vector. 
In general, however, the Courant bracket is not a Lie bracket as it fails to satisfy the 
Jacobi identity. The latter can be represented by the Jacobiator 
\eq{
  \label{jacobiator_39479357}
  \mbox{Jac}\bigl( X,Y,Z \bigr)_{\rm C} = 
  \bigl[ [ X,Y]_{\rm C}, Z\bigr]_{\rm C} +
      \bigl[ [ Z,X]_{\rm C}, Y\bigr]_{\rm C} +
    \bigl[ [ Y,Z]_{\rm C}, X\bigr]_{\rm C} \,,
}
with $X,Y,Z\in \Gamma(E)$ three generalised vectors.
Let us also define the so-called Nijenhuis operator
\eq{
  \label{nijenhuis}
  \mbox{Nij}\bigl( X,Y,Z \bigr)_{\rm C} =  \frac{1}{3} \Bigl( 
  \bigl\langle [X,Y]_{\rm C} , Z \bigr\rangle  
  + \bigl\langle [Z,X]_{\rm C} , Y \bigr\rangle  
  + \bigl\langle [Y,Z]_{\rm C} , X \bigr\rangle  
  \Bigr) \,,
}
where the inner product was defined in equation \eqref{bilin_04}. For 
the above Courant bracket, one can then show that \cite{Gualtieri:2003dx}
\eq{
  \label{jacobi_courant_7373}
    \mbox{Jac}\bigl( X,Y,Z \bigr)_{\rm C} = d\, \mbox{Nij}\bigl( X,Y,Z \bigr)_{\rm C} \,.
}
The right-hand side is in general non-vanishing, and imposes a non-trivial constraint.
If the Nijenhuis operator vanishes, the generalised structure is said to be integrable.

Transformations which preserve the Courant bracket are diffeomorphisms and 
$B$-transforms. Indeed, as one can check we have
\eq{
  \label{auto_courant}
  \mathcal O_{\mathsf A} \left( \bigl[ X, Y \bigr]_{\rm C} \right)
  = \bigl[\mathcal O_{\mathsf A}  X,\mathcal O_{\mathsf A}  Y \bigr]_{\rm C} \,,
  \hspace{40pt}
  \mathcal O_{\mathsf B} \left( \bigl[ X, Y \bigr]_{\rm C} \right)
  = \bigl[\mathcal O_{\mathsf B}  X,\mathcal O_{\mathsf B}  Y \bigr]_{\rm C} \,,
}
for $d \mathsf B = 0$ \cite{Gualtieri:2003dx}.
This explains the restriction on $\mathsf B$ mentioned below \eqref{gen_vec_02}.
These transformations form the so-called  geometric group, which is also the group used in the patching 
\eqref{transf_e01}. 
Elements not belonging to the geometric group -- such as $\beta$-transformations -- 
change the differentiable structure. 
Let us however also note that, as we will exemplify on page~\pageref{page_auto_more},
if the generalised vectors satisfy certain restrictions then the Courant bracket can be preserved 
also by additional $O(D,D,\mathbb R)$ transformations.

%%%%%%%%%%%%%%%%%%%%%%%%%%%%%%%%%%%%%%%%%%%%%%%
%%%%%%%%%%%%%%%%%%%%%%%%%%%%%%%%%%%%%%%%%%%%%%%

\subsubsection*{Dirac structure}

\label{page_dirac}
A structure which will be useful later is the so-called Dirac structure. 
Its definition is that of a subspace $L\subset TM\oplus T^*M$ which is 
\begin{itemize}

\item maximal (dimension of $L$ is $D$) 
\item isotropic ($\eta(V,V)=0$ for all $V\in L$) and which is 
\item involutive (closed under the Courant bracket). 

\end{itemize}
It turns out that the Nijenhuis operator defined in \eqref{nijenhuis} vanishes for 
elements in $L$, and hence the Jacobiator on $L$ vanishes too. 
This means that on the subspace $L\subset TM\oplus T^*M$
the Courant bracket is a Lie bracket.

%%%%%%%%%%%%%%%%%%%%%%%%%%%%%%%%%%%%%%%%%%%%%%%
%%%%%%%%%%%%%%%%%%%%%%%%%%%%%%%%%%%%%%%%%%%%%%%

\subsubsection*{Dorfman bracket}

Another bracket which is useful in the context of generalised geometry is
the Dorfman bracket  \cite{KosmannSchwarzbach:2003en,Gualtieri:2003dx}. It is defined as
\eq{
  X\circ Y =  [x,y]_{\rm L} + \mathcal L_x \chi - \iota_y d \xi \,,
}
for generalised vectors $X,Y\in\Gamma(E)$.
This bracket is not skew-symmetric, but its anti-symmetrisation gives the Courant bracket
\eqref{courant_brack_620}.
The Dorfman bracket satisfies a Leibniz rule of the form
\eq{
  X \circ \bigl( Y \circ Z \bigr) = \bigl( X \circ Y \bigr) \circ Z + Y \circ \bigl( X\circ Z \bigr) \,,
}
and can therefore be used to define a so-called generalised Lie 
derivative as
\eq{
  \label{gen_lie_846}
  \mathcal L_X Y = X\circ Y \,.
}
On functions $f$ the generalised Lie derivative acts as $\mathcal L_X f = \iota_x df$.

%%%%%%%%%%%%%%%%%%%%%%%%%%%%%%%%%%%%%%%%%%%%%%%
%%%%%%%%%%%%%%%%%%%%%%%%%%%%%%%%%%%%%%%%%%%%%%%

\subsubsection*{Generalised metric}

We now want to define a positive-definite metric for the generalised tangent-bundle $E$. 
Since the inner product \eqref{bilin_04} has split signature, it is not suitable candidate. 
However, let us choose a $D$-dimensional sub-bundle $C_+\subset E$ of the generalised tangent-bundle
which is positive definite with respect to the inner product \eqref{bilin_04}.
The orthogonal complement, which is negative-definite, is denoted by
$C_-$ and we have $E = C_+ \oplus C_-$.  The generalised metric can then be defined as 
\eq{
  \label{gen_met_split}
  \mathcal H = \eta\,\big|_{C_+} - \,\eta\,\big|_{C_-} \,.
}
This decomposition defines a reduction of the structure group from $O(D,D,\mathbb R)$ to $O(D,\mathbb R)\times O(D,\mathbb R)$. 
Note that $\mathcal H$ is symmetric and non-degenerate. In the basis chosen above, let us then 
specify\,\footnote{This choice is well-motivated in the context of generalised K\"ahler geometry. 
For more details see for instance chapter 6 of \cite{Gualtieri:2003dx}.}
\eq{
  \mathcal H   = \left( \begin{array}{cc} \frac{1}{\alpha'}\, g & 0 \\[2pt] 0 & \alpha' \op g^{-1} \end{array}\right),
}
where $g$ is the ordinary metric on $M$. 
Indeed, this choice of $\mathcal H$ is positive definite. 
However, if we now perform a $\mathsf B$-transform \eqref{gg_b-transform_048} with 
an anti-symmetric matrix $\mathsf B_{ij}$ given by the components of the 
the Kalb-Ramond field $b$ as $\mathsf B_{ij} = \tfrac{1}{\alpha'}\, b_{ij}$, we find
\eq{
  \label{gen_met_again}
  \arraycolsep4pt
  \mathcal H 
  \to \mathcal H' = \mathcal O_{\mathsf B}^{-T}\, \mathcal H \:\mathcal O_{\mathsf B}^{-1}
  = \left( \begin{array}{cc}
   \frac{1}{\alpha'}\left( g - b\op g^{-1}b\right)& +b\op g^{-1}  \\[4pt]  - g^{-1} b & \alpha' g^{-1}\end{array} \right),
}
which is the generalised metric we  defined already in \eqref{gen_met_098}.
Let us mention that an arbitrary  $\mathsf B$-transform is in general not an automorphism
of the Courant bracket, and therefore changes the differentiable structure. 
Only when $db = 0$, the transformed and the original 
background are equivalent. 
However, for our purpose here of motivating the generalised metric locally this distinction is not important.

%%%%%%%%%%%%%%%%%%%%%%%%%%%%%%%%%%%%%%%%%%%%%%%
%%%%%%%%%%%%%%%%%%%%%%%%%%%%%%%%%%%%%%%%%%%%%%%

\subsubsection*{Generalised vielbein}

We finally want to introduce a generalised vielbein basis for the generalised metric. 
To do so, we use a similar notation as in \eqref{vielbein_8467} and write for the components of
the ordinary metric in a local basis $\{dx^i\}\in\Gamma(TM)$
\eq{
  \label{vilebein_93479375}
  g_{ij} = (e^T)_i{}^a\op \delta_{ab} \, e^b{}_j \,, 
}
and the inverse of $e^a{}_i$ is denoted again as $\ov e^i{}_a$.
For the $B$-field we employ a similar notation, that is $  b_{ij} = (e^T)_i{}^a\op b_{ab} \, e^b{}_j $.
Using then $2D$-dimensional indices $I$ and $A$, the generalised 
vielbein $\mathcal E_I = \{\mathcal E^A{}_I\}$ 
with index structure 
\eq{
  \label{gen_viel_947}
  \mathcal E = \{\mathcal E^A{}_I \} = 
  \left( \begin{array}{cc} \mathcal E^a{}_i & \mathcal E^{a\op i} \\[2pt] \mathcal E_{a\op i} & 
  \mathcal E_a{}^i \end{array}\right) 
}
can be defined via the relations
\eq{
  \label{vielbein_again_649}
  \eta = \mathcal E^T  \left( \begin{array}{cc} 0 & \mathds 1 \\[2pt] \mathds 1 & 0 \end{array}\right)
  \mathcal E \,,
  \hspace{50pt}
  \mathcal H = \mathcal E^T  \left( \begin{array}{cc} \,\delta\, & 0 \\[2pt] 0 & \delta^{-1} \end{array}\right)
  \mathcal E \,,
}
where matrix multiplication is understood. Note that the
index-structure of the generalised metric is $\mathcal H_{IJ}$,
for the Kronecker $\delta$-symbol we use $\delta_{ab}$ and $\delta^{ab}\equiv(\delta^{-1})^{ab}$,
and that the inverse of $\mathcal E$ will be denoted
by $\ov{\mathcal E}$.
We can now solve \eqref{vielbein_again_649} for the generalised vielbeins by introducing two sets of ordinary vielbein matrices $(e_{\pm})^a{}_i$ as $ g_{ij} = (e^T_{\pm})_i{}^a\op \delta_{ab} \, (e_{\pm})^b{}_j $. The subscripts $\pm$ indicate 
to which of the two subspaces shown in \eqref{gen_met_split} the vielbeins correspond to,
and $e_+$ and $e_-$ are related to each other by an $O(D,\mathbb R)$ transformation acting on 
the flat index $a$ which leaves
the right-hand side of \eqref{vilebein_93479375} invariant.
The generalised vielbein can then be expressed as \cite{Grana:2008yw}
\eq{
  \label{vielbein_242242}
  \arraycolsep1pt
  \mathcal E = \frac{1}{2\sqrt{\alpha'}}
  \left( \begin{array}{@{\hspace{4pt}}clcrl@{\hspace{14pt}}cl@{\hspace{4pt}}} 
  &\bigl(e_+ + e_-\bigr)^a{}_i &-&\delta^{ab}& \bigl(\ov e^T_+- \ov e_-^T\bigr)_b^{\hspace{4pt}m}\op b_{mi}  & 
  \delta^{ab}& \alpha'\op\bigl( \ov e^T_+ - \ov e^T_-\bigr)_b{}^i 
  \\[5pt] 
  \delta_{ab} &\bigl(e_+ - e_-\bigr)^b{}_i &-&&\bigl(\ov e^T_++ \ov e_-^T\bigr)_a^{\hspace{4pt}m}\op b_{mi}
  && 
  \alpha'\op\bigl( \ov e^T_+ + \ov e^T_-\bigr)_a{}^i  \end{array}\right) .
}

However, as for the usual frame-bundle the ordinary vielbeins take values in, also here
there are transformations acting on $\mathcal E_A$ which do not change the defining 
equations \eqref{vielbein_again_649}. Indeed, 
we see that \label{page_odod_939}
\eq{
\label{gg_7382929}
\mathcal E^A \to \mathcal K^A{}_B\op \mathcal E^B 
}
with $\mathcal K\in O(D,D,\mathbb R)$
leaves the first relation in \eqref{vielbein_again_649} invariant. 
Though, the second relation is left invariant only by transformations 
$\mathcal K \in O(D,\mathbb R)\times O(D,\mathbb R) \subset O(D,D,\mathbb R)$. 
In analogy to the usual frame bundle which defines an $O(D,\mathbb R)$ structure, in the generalised-geometry situation we therefore have a  $O(D,\mathbb R)\times O(D,\mathbb R)$ structure. 
In particular, for the vielbein \eqref{vielbein_242242} the corresponding transformations are given by
\cite{Grana:2008yw}
\eq{
  \label{odod}
  \arraycolsep0.5pt
  \mathcal K =  \frac12 \left( \begin{array}{@{\hspace{4pt}}cccc@{\hspace{8pt}}ccccc@{\hspace{4pt}}} &&O_+ + O_- && &\bigl( &O_+-O_-&\bigr) &\delta^{-1} \\[4pt] 
  \delta &\bigl(  & O_+-O_- &\bigr) & \delta & \bigl(&  O_++O_-& \bigr) &\delta^{-1} \end{array}\right),
  \hspace{40pt}
  \bigl( O_{\pm} \bigr)^a{}_b\in O(D,\mathbb R) \,.
}
Using these transformations we can set for instance $e_+=e_-=e$, which simplifies the generalised vielbein 
\eqref{vielbein_242242} to a convention often used in the literature
\eq{
  \label{vielbein_again_648}
  \mathcal E = 
  \frac{1}{\sqrt{\alpha'}}\left( \begin{array}{cc} e^a{}_i & 0 \\[2pt] - \ov e_a{}^m\op b_{mi} & 
  \alpha'\,\ov e_a{}^i \end{array}\right) .
}

%%%%%%%%%%%%%%%%%%%%%%%%%%%%%%%%%%%%%%%%%%%%%%%
%%%%%%%%%%%%%%%%%%%%%%%%%%%%%%%%%%%%%%%%%%%%%%%
%%%%%%%%%%%%%%%%%%%%%%%%%%%%%%%%%%%%%%%%%%%%%%%
%%%%%%%%%%%%%%%%%%%%%%%%%%%%%%%%%%%%%%%%%%%%%%%

\subsection{Lie and Courant algebroids}
\label{sec_lie_courant}

The introduction of generalised geometry in the last section was done with an
application to T-duality and non-geometric backgrounds in mind.
In this section we want to explain the underlying mathematical structures in some
more detail. For further discussions we refer to \cite{Gualtieri:2003dx}, and 
for a discussion in the physics literature for instance to \cite{Blumenhagen:2013aia,Deser:2013pra}.

%%%%%%%%%%%%%%%%%%%%%%%%%%%%%%%%%%%%%%%%%%%%%%%
%%%%%%%%%%%%%%%%%%%%%%%%%%%%%%%%%%%%%%%%%%%%%%%

\subsubsection*{Lie algebroid}

Let us start by introducing the concept of a Lie algebroid \cite{mackenzie_1987}.
To specify a Lie algebroid one needs three pieces of information:
\begin{itemize}

\item a vector bundle $E$ over a manifold $M$,

\item a bracket $[\op\cdot\,,\cdot\op ]_E : E \times E \rightarrow E$, and

\item a homomorphism $\rho : E \rightarrow TM$ called the anchor map. \label{def_anchor}

\end{itemize}
Similar to the usual Lie bracket, we require the bracket $[\op\cdot\,,\cdot\op]_E$ to satisfy a Leibniz rule.
Denoting functions by $f\in {\cal C}^{\infty}(M)$ and sections  of $E$ by $s_i\in\Gamma(E)$, this reads
\eq{
\label{Leibniz}
[ s_1, f s_2]_E = f \hspace{1pt}[ s_1,s_2]_E + \rho(s_1)(f) \op s_2  \,,
} 
where $\rho(s_1)$ is a vector-field which acts on functions $f$ as a derivation.
If  in addition the bracket $[ \cdot\,,\cdot]_E$ satisfies a Jacobi identity
\eq{
\label{jacobii}
\bigl[  s_1, [ s_2, s_3 ]_E \bigr]_E = \bigl[ [  s_1, s_2 ]_E , s_3 \bigr]_E +  \bigl[ s_2, [ s_1, s_3 ]_E\bigr]_E\, ,
}
then $(E,[\op\cdot\,,\cdot\op]_E, \rho)$ is  called
a Lie algebroid. (If the Jacobi identity is not satisfied, the resulting structure is called a quasi-Lie algebroid.)
Therefore, roughly speaking, when replacing vector-fields and their Lie bracket $[\op\cdot\op,\cdot\op]_L$ 
by sections of $E$ and the corresponding bracket $[\op\cdot\op,\cdot\op]_E$ one obtains a Lie algebroid. 
The relation between the brackets is established by the anchor $\rho$. Indeed, the requirement that $\rho$ is a homomorphism implies that
\eq{
\label{anhom}
	\rho\bigl( [ s_1,s_2]_E\bigr) = \bigl[\rho(s_1),\rho(s_2)\bigr]_L \,,
	\hspace{50pt} s_{1,2} \in \Gamma(E)\,.
}

%%%%%%%%%%%%%%%%%%%%%%%%%%%%%%%%%%%%%%%%%%%%%%%
%%%%%%%%%%%%%%%%%%%%%%%%%%%%%%%%%%%%%%%%%%%%%%%

\subsubsection*{Examples}

Let us illustrate this construction with two examples.
\begin{itemize}

\item We start with considering the tangent-bundle $E=TM$ with the usual Lie bracket $[\op \cdot\,,\cdot\op]_E=[\op\cdot\op,\cdot\op]_L$. The anchor is chosen to be the identity map, i.e.\ $\rho={\rm id}$.
Then, the conditions \eqref{Leibniz} and \eqref{jacobii} reduce to the well-known properties of the Lie bracket, and \eqref{anhom} is trivially satisfied. 
Therefore, $ E=(TM,[\op\cdot\op,\cdot\op]_{L},\rho= \textrm{id})$ is indeed a Lie algebroid.
\label{LA_exp1}

\item As a second example, we consider a Poisson manifold $(M,\beta)$ with 
Poisson tensor $\beta =  \frac{1}{2}\,\beta^{ij} \op \partial_i \wedge \partial_j$, where $\{\partial_i\} \in \Gamma(TM)$ denotes again a local basis of vector-fields.
A Lie algebroid is  given by
$E=(T^* M,[\op\cdot\,,\cdot\op ]_{\rm KS},\rho =  \beta^\sharp)$, in which 
the anchor $\beta^{\sharp}$ is defined as
\eq{ 
\beta^{\sharp} (dx^i) = \beta^{ij}\op\partial_j \,,
}
with $\{dx^i\} \in\Gamma(T^*M)$ the standard basis of one-forms dual to the vector-fields.
The bracket $[\op\cdot\,,\cdot\op ]_{\rm KS}$ on $T^*M$ is the Koszul-Schouten bracket, which for one-forms $\xi$ and $\eta$ is defined as
\eq{
\label{koszul}
	[\op\xi,\eta\op]_{\rm KS} = \mathcal L_{\beta^\sharp(\xi)}\op \eta
 -\iota_{\beta^\sharp(\eta)}\,d\xi \, .
}
 The conditions \eqref{Leibniz}, \eqref{jacobii} and \eqref{anhom} are satisfied, provided that $\beta$ is a Poisson tensor, 
that is
\eq{
  \label{gg_lie_alg_938649386}
 \beta^{[\ul i|m}  \partial_m \beta^{|\ul j\ul k]}=0 \,. 
}

\end{itemize}

%%%%%%%%%%%%%%%%%%%%%%%%%%%%%%%%%%%%%%%%%%%%%%%
%%%%%%%%%%%%%%%%%%%%%%%%%%%%%%%%%%%%%%%%%%%%%%%

\subsubsection*{Differential geometry}
\label{page_dg_lie_alg}

Using the bracket $[\op\cdot\op,\cdot\op]_E$ of a Lie algebroid $E$, one can define a corresponding 
differential $d_E:\Gamma(\wedge^k E^*)\to \Gamma(\wedge^{k+1}E^*)$ through the relation
\eq{
\label{gg_ll_244}
 \bigl( d_E \,\omega\bigr)(s_0,\ldots,s_k)=\quad &\sum_{i=0}^k (-1)^i\rho(s_i)\,  \omega(s_0,\ldots,\hat s_i ,\ldots, s_k)
 \\
 +&\sum_{i<j}(-1)^{i+j}\, \omega \bigl( [s_i,s_j]_E, s_0, \ldots, \hat s_i , \ldots, \hat s_j , \ldots, s_k\bigr)\,,
}
where $\omega\in \Gamma(\wedge^k E^*)$ with $E^*$ the vector-space dual to $E$, $s_i\in\Gamma(E)$
and the hat stands for deleting the corresponding entry.
One can show that this differential is nil-potent for a Lie algebroid.
Furthermore, using this differential we can define a Lie derivative acting in the following way
\eq{
  \mathcal L_{s_1} s_2 = [s_1,s_2]_E \,,
  \hspace{40pt}
  \mathcal L_s \,\omega = \iota_s \circ d_E \,\omega + d_E \circ \iota_s \, \omega\,,
}
where $s,s_i\in \Gamma(E)$ and $\omega\in\Gamma(E^*)$. This Lie derivative satisfies the standard 
properties.
Finally, a covariant derivative for a Lie algebroid $E$ 
is a bilinear map $\nabla:\Gamma(E)\times \Gamma(E)\to \Gamma(E)$
which has the properties
\eq{
  \nabla_{f\op s_1} s_2 = f\op \nabla_{s_1} s_2 \,,
  \hspace{40pt}
  \nabla_{s_1} ( f\op s_2) = \rho(s_1)(f) \, s_2 + f \nabla_{s_1} s_2 \,,
}
for $s_1,s_2\in\Gamma(E)$ and $f\in\mathcal C^{\infty}(M)$. Using this covariant derivative
and the Lie-algebroid bracket,
one can now construct curvature and torsion tensors similarly to ordinary differential geometry as
\eq{
  \begin{array}{l@{\hspace{2.5pt}}l}
  \displaystyle R(s_1,s_2)\, s_3 &\displaystyle = \bigl[\nabla_{s_1},\nabla_{s_2}\bigr]\, s_3 
  -  \nabla_{[ s_1,s_2]_E}\, s_3 \,,
  \\[8pt]
  \displaystyle T(s_1,s_2) &\displaystyle = \nabla_{s_1} s_2 - \nabla_{s_2} s_1-[ s_1,s_2]_E \,,
  \end{array}
}
for $s_i\in \Gamma(E)$. 
For more details on these constructions we refer the reader for instance to 
\cite{Blumenhagen:2013aia,Deser:2013pra} and references
therein.

%%%%%%%%%%%%%%%%%%%%%%%%%%%%%%%%%%%%%%%%%%%%%%%
%%%%%%%%%%%%%%%%%%%%%%%%%%%%%%%%%%%%%%%%%%%%%%%

\subsubsection*{Courant algebroid}

Let us now turn to the mathematical structure relevant for generalised geometry, 
which is that of a Courant algebroid. It is a combination of a Lie algebroid 
with its dual into a Lie bi-algebroid \cite{Liu:1995lsa,1999royten}.
To be more precise, 
\begin{flushright}
\begin{minipage}{0.96\textwidth}
Let $M$ be a manifold and $E\to M$ a vector-bundle over $M$
together with a non-degenerate symmetric bilinear form $\langle \cdot\,,\cdot\rangle$,
a skew-symmetric bracket $[\op \cdot\,,\cdot\op ]_E$ on its sections $\Gamma(E)$ and a 
bundle map $\rho:E\to TM$.
Then $(E,[ \op\cdot\,,\cdot\op ]_E,\langle \cdot\,,\cdot\rangle,\rho)$ is called a Courant 
algebroid if the following properties hold: 
\begin{itemize}

\item For $s_1,s_2\in \Gamma(E)$ one has $\rho([s_1,s_2]_E) = [\rho(s_1),\rho(s_2)]_E$.

\item For $s_1,s_2\in \Gamma(E)$ and $f\in\mathcal C^{\infty}(M)$ one has
\eq{
\bigl[ s_1, f s_2\bigr]_E = f \hspace{1pt}\bigl[ s_1,s_2\bigr]_E + \rho(s_1)(f)\op s_2  - \langle s_1,s_2\rangle \,
\mathfrak D f\,.
} 

\item For $f_1,f_2\in \mathcal C^{\infty}(M)$ one has $\langle \mathfrak Df_1,\mathfrak D f_2 \rangle=0$,
which means that $\rho\circ \mathfrak D=0$.

\item For $s_0,s_1,s_2\in \Gamma(E)$ one has
\eq{
  &\rho(s_0) \langle s_1,s_2\rangle  =
  \left\langle [s_0,s_1]_E+\mathfrak D\langle s_0,s_1\rangle,s_2\right\rangle
  \\
  &\hspace{120pt}+
  \left\langle s_1, [s_0,s_2]_E+\mathfrak D\langle s_0,s_2\rangle\right\rangle \,.
}

\item For $s_1,s_2,s_3\in\Gamma(E)$ one has $\mbox{Jac}\op(s_1,s_2,s_3)_E = \mathfrak D\op  \mbox{Nij}(s_1,s_2,s_3)_E$.

\end{itemize}

\end{minipage}
\end{flushright}
Here we used a definition of the Jacobiator $\mbox{Jac}\op(\cdot\,,\cdot\,,\cdot)_E$
and Nijenhuis tensor $\mbox{Nij}\op(\cdot\,,\cdot\,,\cdot)_E$ for a bracket $[\op\cdot\op,\cdot\op]_E$
similar to the ones in \eqref{jacobiator_39479357} and \eqref{nijenhuis}. In particular, we
have defined
\eq{
  \mbox{Jac}\bigl( s_1,s_2,s_3 \bigr)_{E} &= 
  \bigl[ [ s_1,s_2]_{E}, s_3\bigr]_{E} +
      \bigl[ [ s_3,s_1]_{E}, s_2\bigr]_{E} +
    \bigl[ [ s_2,s_3]_{E}, s_1\bigr]_{E} \,,
    \\[8pt]
  \mbox{Nij}\bigl( s_1,s_2,s_3 \bigr)_{E} &=  \frac{1}{3} \Bigl( 
  \bigl\langle [s_1,s_2]_{E} , s_3 \bigr\rangle  
  + \bigl\langle [s_3,s_1]_{E} , s_2 \bigr\rangle  
  + \bigl\langle [s_2,s_3]_{E} , s_1 \bigr\rangle  
  \Bigr) \,,
}
and we defined
a map $\mathfrak D: \mathcal C^{\infty}(M) \to \Gamma(E)$ through
\eq{
  \langle \mathfrak Df,s\rangle = \frac{1}{2}\, \rho(s) \op f \,.
}

If we now compare this definition to our discussion in section~\ref{sec_gg_basics}, we
see that the generalised tangent-bundle \eqref{gg_sequ}
together with the bilinear pairing \eqref{bilin_04}
and the Courant bracket \eqref{courant_brack_620}
is a Courant algebroid, in which the anchor map $\rho$ is the projection from $E$ to $TM$.
It then follows that the operator $\mathfrak D$ is the exterior derivative $d$.

%%%%%%%%%%%%%%%%%%%%%%%%%%%%%%%%%%%%%%%%%%%%%%%
%%%%%%%%%%%%%%%%%%%%%%%%%%%%%%%%%%%%%%%%%%%%%%%
%%%%%%%%%%%%%%%%%%%%%%%%%%%%%%%%%%%%%%%%%%%%%%%
%%%%%%%%%%%%%%%%%%%%%%%%%%%%%%%%%%%%%%%%%%%%%%%

\subsection{Buscher rules}
\label{sec_gg_buscher}

In this section we discuss an application of the generalised-geometry framework
to the Buscher rules considered in section~\ref{sec_buscher}.
More concretely, we want to express the restrictions for performing T-duality in terms 
of generalised vectors on the generalised tangent-space.

%%%%%%%%%%%%%%%%%%%%%%%%%%%%%%%%%%%%%%%%%%%%%%%
%%%%%%%%%%%%%%%%%%%%%%%%%%%%%%%%%%%%%%%%%%%%%%%

\subsubsection*{Restrictions}

Recall that the Buscher rules were derived by identifying a global symmetry of the 
world-sheet theory, gauging this symmetry, and integrating out the gauge field. 
In order for this to be possible, certain constraints on the background have to be satisfied. 
These are constraints for the existence of a global and a local symmetry, and we summarise the 
corresponding equations \eqref{constraints_35}, \eqref{constraints_351} and \eqref{variations_45} here as follows 
(we ignore the dilaton
constraint)
\begin{subequations}\label{const_global}
\begin{align}
\mbox{\begin{minipage}[t][1.3em][c]{150pt}global symmetry\end{minipage}} 
&  
\mbox{\begin{minipage}{140pt}$\displaystyle   \mathcal L_{k_{\alpha}}  G = 0\,,$\end{minipage}} 
\\[-0.5em]
\label{const_global_b}
&\mbox{\begin{minipage}{140pt}$\displaystyle   \iota_{k_{\alpha}} H = dv_{\alpha} \,,$\end{minipage}} 
\end{align}
\end{subequations}
\vspace*{-0.5em}
\begin{align}\label{const_iso}
\mbox{\begin{minipage}{150pt}\hspace*{2pt}isometry algebra\end{minipage}} 
&  
\mbox{\begin{minipage}{150pt}$\displaystyle    \bigl[ k_{\alpha} , k_{\beta} \bigr]_{\rm L} = f_{\alpha\beta}{}^{\gamma} \op k_{\gamma}\,,$\end{minipage}}
\end{align}
\vspace*{-0.5em}
\begin{subequations}
\begin{align}
\label{const_local_a}
\mbox{\begin{minipage}[t][1.3em][c]{150pt}local symmetry\end{minipage}} 
&  
\mbox{\begin{minipage}{140pt}$\displaystyle   \mathcal L_{k_{[\ul \alpha}} v_{\ul \beta]} = f_{\alpha\beta}{}^{\gamma} v_{\gamma} \,,$\end{minipage}} 
\\[-0.5em]
\label{const_local_b}
&\mbox{\begin{minipage}{140pt}$\displaystyle 3\op\iota_{k_{[\ul \alpha}} \op f_{\ul \beta\ul \gamma ]}{}^{\delta} v_{\delta} = 
\iota_{k_{\alpha}}\iota_{k_{\beta}}\iota_{k_{\gamma}} H \,.$\end{minipage}} 
\end{align}
\end{subequations}
%%
%%
%%
%%

%%%%%%%%%%%%%%%%%%%%%%%%%%%%%%%%%%%%%%%%%%%%%%%
%%%%%%%%%%%%%%%%%%%%%%%%%%%%%%%%%%%%%%%%%%%%%%%

\subsubsection*{Reformulation using the generalised Lie derivative}

In order to describe the global-symmetry requirements \eqref{const_global}, let us 
determine the generalised Lie derivative \eqref{gen_lie_846} of the 
generalised metric shown in \eqref{gen_met_again}.
For a local basis on $TM\oplus T^*M$ of the form $\{d\op\mathsf X^I\} = \{dx^i,\partial_i\}$,
this Lie derivative is determined by computing
\eq{
  \label{mieke}
  \mathcal L_X   \mathcal H = 
  \mathcal L_X   \Bigl( \mathcal H_{IJ}\, d\op\mathsf X^I \vee d\op\mathsf X^J \Bigr)\,.
}
If we choose for the generalised vector $X = x + \xi$
 one finds that the components of \eqref{mieke} read \cite{Grana:2008yw}
\eq{
  \bigl(\mathcal L_X \mathcal H \bigr)_{IJ} = 
  \left( \begin{array}{cc}
  \scalebox{0.8}{$
  \begin{array}{lll}
  \tfrac{1}{\alpha'} \Bigl[(g+b)\bigl(\mathcal L_x \op g^{-1}\bigr)(g-b) \\[4pt]
  \qquad+\bigl[d\xi + \mathcal L_x(g+b) \bigr] \,g^{-1}(g-b)\\[4pt]
  \qquad-(g+b)\,g^{-1} \bigl[ d\xi - \mathcal L_x(g-b) \bigr] \Bigr]
  \end{array}
  $}
  & +b \op\bigl(\mathcal L_x \op g^{-1}\bigr)+ \bigl[\op d\xi +  \mathcal L_x\op b \bigr] g^{-1}
  \\[26pt]
  - \bigl(  \mathcal L_x \op g^{-1} \bigr) \op b - g^{-1} \bigl[\op  d\xi +  \mathcal L_x \op b \bigr]
  &  \alpha'\op \mathcal L_{k_{\alpha}} g^{-1}
  \end{array}
  \right).
}
In a similar way, one shows that $\mathcal L_X \eta=0$ for all choices of generalised vectors $X$.
Rewriting \eqref{const_global_b} as $\mathcal L_{k_{\alpha}} B = d(v_{\alpha} + \iota_{k_{\alpha}}B)$
and defining 
\eq{
\label{gen_kv_840}
K_{\alpha} = k_{\alpha} - \bigl( v_{\alpha} + \iota_{k_{\alpha}}B\bigr)\,,
}
we see that the global symmetry requirements can be 
stated using the generalised metric as
\eq{
  \mathcal L_{K_{\alpha}} \mathcal H = 0\,.
}
The generalised vectors $K_{\alpha}$ are then called generalised Killing vectors.
Using \eqref{lie_rel_9846}, one can furthermore show that the isometry-algebra relation 
\eqref{const_iso} and the local relations \eqref{const_local_a}
are encoded in the closure of the so-called $H$-twisted Courant bracket 
\eq{
  \label{cour_twi}
  \bigl[ X, Y \bigr]^H_{\rm C} = 
  [\op x,y\op]_{\rm L} + \mathcal L_x \chi - \mathcal L_y \xi - \frac12 \op d \bigl( \iota_x \chi - \iota_y \xi \bigr)
  +\iota_x\iota_y H
}
as
\eq{
  \label{gen_geo_closure_749}
  \bigl [ K_{\alpha} , K_{\beta} \bigr]_{\rm C}^{H} = f_{\alpha\beta}{}^{\gamma} \op K_{\gamma} \,.
}
In addition, the remaining local symmetry relation \eqref{const_local_b} is encoded in the vanishing
of the Nijenhuis tensor \eqref{nijenhuis} with respect to the twisted Courant bracket \eqref{cour_twi}, that is
\cite{Plauschinn:2014nha}
\eq{
    \mbox{Nij}\,\bigl( K_{\alpha},K_{\beta} ,K_{\gamma} \bigr)^H_{\rm C} = 0 \,.
}
To summarise, the constraints on performing a T-duality transformation using 
Buscher's approach can be expressed in the framework of generalised geometry
as \eqref{gen_kv_840} being a generalised Killing vector with respect to the generalised metric, which is closed
with respect to the $H$-twisted Courant bracket \eqref{cour_twi} and whose Nijenhuis tensor vanishes.

%%%%%%%%%%%%%%%%%%%%%%%%%%%%%%%%%%%%%%%%%%%%%%%
%%%%%%%%%%%%%%%%%%%%%%%%%%%%%%%%%%%%%%%%%%%%%%%

\subsubsection*{Remark}

We remark that  a particular solution to the constraint
\eqref{const_local_b} is given by requiring $\iota_{k{(\ov \alpha}} v_{\ov \beta)}=0$.
This is what is often done in the literature, however, in general other solutions 
to \eqref{const_local_b} may exist. Using the inner product \eqref{bilin_04}, we can 
express $\iota_{k{(\ov \alpha}} v_{\ov \beta)}=0$ as
\eq{
  \bigl\langle K_{\alpha} , K_{\beta} \bigr\rangle = 0\,,
}
which is an isotropy condition. Together with \eqref{gen_geo_closure_749}, the 
generalised Killing vectors then define an isotropic and involutive sub-bundle
of the generalised tangent-bundle. If this sub-bundle is in addition maximal, 
it defines a Dirac structure (cf. our discussion on page~\pageref{page_dirac}).

%%%%%%%%%%%%%%%%%%%%%%%%%%%%%%%%%%%%%%%%%%%%%%%
%%%%%%%%%%%%%%%%%%%%%%%%%%%%%%%%%%%%%%%%%%%%%%%
%%%%%%%%%%%%%%%%%%%%%%%%%%%%%%%%%%%%%%%%%%%%%%%
%%%%%%%%%%%%%%%%%%%%%%%%%%%%%%%%%%%%%%%%%%%%%%%

\subsection{Fluxes}
\label{sec_gg_fluxes}

In this section we present a description of geometric as well as non-geometric fluxes
within the framework of generalised geometry. We show
how different flux-backgrounds can be generated by acting with $O(D,D)$ transformations
on the generalised vielbeins; if these transformations are \textit{not} automorphisms of the 
Courant bracket a different flux-background is obtained.

%%%%%%%%%%%%%%%%%%%%%%%%%%%%%%%%%%%%%%%%%%%%%%%
%%%%%%%%%%%%%%%%%%%%%%%%%%%%%%%%%%%%%%%%%%%%%%%

\subsubsection*{Generalised vielbein}

In section~\ref{sec_twisted_torus_ex} we have seen that the  geometric flux can be defined 
via the exterior derivative of vielbein one-forms. 
Let us formalise this approach 
and introduce the connection one-form $\omega^a{}_b$ which satisfies
Cartan's structure equations. 
(For a textbook introduction to this topic see for instance section 7.8 in \cite{Nakahara:2003nw}.)
In particular, choosing the torsion to be vanishing
the connection one-form is specified by the relation
\eq{
  \label{spin_conn_510}
  d e^a + \omega^a{}_b \wedge e^b = 0\,,
  \hspace{50pt}
  \omega^a{}_b = \Gamma^a{}_{cb} \,e^c \,,
}
where $e^a = e^a{}_i \,dx^i$ is a basis of one-forms. As before, the $e^a{}_i$ are determined via
$g_{ij} = (e^T)_i{}^a\op \delta_{ab} \, e^b{}_j $
and $\Gamma^a{}_{cb}$ are the Christoffel symbols for the vielbein-basis.
The algebra of the vielbein vector-fields $\ov e_a = (\ov e^T)_a{}^i \partial_i$ is specified by  structure constant $f_{ab}{}^c$ 
as 
\eq{
  \label{vielbeins_7856}
  [ \op \ov e_a, \ov e_b\op ]_{\rm L} = f_{ab}{}^c\,  \ov e_c \,.
}
In case of vanishing torsion, we can express the structure constants as 
$f_{ab}{}^c = \Gamma^c{}_{ab} - \Gamma^c{}_{ba}$, and 
equation \eqref{spin_conn_510} can therefore be written as
\eq{
  d e^a = - \frac{1}{2} \, f_{bc}{}^a \op e^b \wedge e^c \,.
}
With respect to the example of the geometric flux, we therefore see that 
this flux is alternatively encoded in the structure constants of the vielbein vector-field algebra \eqref{vielbeins_7856}.

We now apply a similar reasoning to the generalised vectors:
we replace the ordinary vielbein vector-fields $\ov e_a=\{\ov e_a{}^i\}$ by the generalised 
vielbein vector-fields 
$\ov{\mathcal E}_A=\{\ov{\mathcal E}_A{}^I\}$ introduced in equation \eqref{gen_viel_947},
and we replace the Lie bracket by the
Courant bracket \eqref{courant_brack_620}.
In analogy to \eqref{vielbeins_7856} we then define generalised structure constants $F_{AB}{}^C$ through
\eq{
  \label{gen_vielbein_284}
  \bigl[ \ov{\mathcal E}_A, \ov{\mathcal E}_B \bigr]_{\rm C} = F_{AB}{}^C \,\ov{\mathcal E}_C \,.
}

%%%%%%%%%%%%%%%%%%%%%%%%%%%%%%%%%%%%%%%%%%%%%%%
%%%%%%%%%%%%%%%%%%%%%%%%%%%%%%%%%%%%%%%%%%%%%%%

\subsubsection*{Fluxes via $O(D,D)$ transformations}

Let us now construct explicit examples for backgrounds with  geometric and non-geometric 
fluxes. We start from a trivial generalised vielbein and act on them with non-constant  $O(D,D)$ 
transformations. Such transformation are in general not symmetries of the Courant bracket and
therefore change the corresponding background. However, this approach provides
us with a technique to generate new flux backgrounds.

Our starting point is a locally-flat metric $g_{ij} = \alpha' \op \delta_{ij}$ and a vanishing 
Kalb-Ramond $B$-field. The generalised metric $\mathcal H$ then takes a diagonal form, and the 
generalised vector-fields $\ov{\mathcal E}_A$ -- which are the inverse-transpose of \eqref{gen_viel_947} -- can be expressed as
\eq{
  \label{gg_vielbein_flat_09087}
  \bigl(\ov{\mathcal E}_{(0)}\bigr)_A\mathop{\vphantom{\bigr)}}\nolimits^I = 
  \left( \begin{array}{cc} \delta_a{}^i & 0 \\[2pt] 0 & \delta^a{}_i \end{array}\right) .
}
Employing a local basis $\{\partial_I\} = \{ \partial_i, dx^i\}$ of the generalised tangent-space 
and defining $\ov{\mathcal E}_A = \ov{\mathcal E}_A{}^I\op  \partial_I$,  for the Courant bracket \eqref{courant_brack_620} of  generalised vielbeins
we find
\eq{
  \label{gv_algebra_pp1}
  \arraycolsep1.5pt
  \begin{array}{cccrcr}
  \bigl[\, \ov{\mathcal E}_{(0)\,a} \, ,\, \ov{\mathcal E}_{(0)\,b} \,\bigr]_{\rm C} &=&
  0\,,
  \\[8pt]
  \bigl[\, \ov{\mathcal E}_{(0)\,a} \, ,\, \ov{\mathcal E}_{(0)}{}^{\hspace{1pt}b} \,\bigr]_{\rm C} &=&
  0 \,,
  \\[8pt]
  \bigl[\, \ov{\mathcal E}_{(0)}{}^{\hspace{1pt}a} \, ,\, \ov{\mathcal E}_{(0)}{}^{\hspace{1pt}b} \,\bigr]_{\rm C} &=&
  0\,.
  \end{array}
}
Next, we generate new backgrounds by applying $O(D,D)$ transformations $\mathcal O$ to these generalised vectors, for which we have in matrix notation 
\eq{
  \label{gv_77729}
  \ov{\mathcal E}_{(\mathcal O)} = \ov{\mathcal E}_{(0)} \op \mathcal O^T \,,
  \hspace{60pt}
  \mathcal O = \left( \begin{array}{cc} A & B \\ C & D \end{array}\right).
}
Let us emphasise that these are in general non-constant transformations
which are also not automorphisms of the Courant bracket. We discuss now four examples of 
$O(D,D)$ transformations acting on the generalised vector-field \eqref{gg_vielbein_flat_09087}:
\begin{itemize}

\item We start with an $\mathsf A$-transformation which was mentioned in equation \eqref{gg_a_transform_07687}.
Choosing $\mathsf A^i{}_j = \delta^i{}_b\op \hat e^b{}_j$ with $\hat e^a{}_i=  e^a{}_i/\sqrt{\alpha'}$ non-trivial 
and dimension-less
vielbein matrices, we obtain\,\footnote{We use dimension-less quantities in order to make the generalised vielbeins dimension-less, in accordance with our conventions in \eqref{vielbein_again_649}.}
\eq{
  \label{gv_geom_flux_vb}
  \bigl(\ov{\mathcal E}_{(\mathsf A)}\bigr)_A\mathop{\vphantom{\bigr)}}\nolimits^I = 
  \left( \begin{array}{cc} \hat{\ov e}_a{}^i & 0 \\[2pt] 0 & \hat e^a{}_i \end{array}\right) .
}
The Courant brackets between the generalised vectors then read
\eq{
  \label{gv_algebra_pp2}
  \arraycolsep1.5pt
  \begin{array}{cccrcr}
  \bigl[\, \ov{\mathcal E}_{(\mathsf A)\,a} \, ,\, \ov{\mathcal E}_{(\mathsf A)\,b} \,\bigr]_{\rm C} &=&
  +& \hat f_{ab}{}^c\,\ov{\mathcal E}_{(\mathsf A)\,c}  \,,
  \\[8pt]
  \bigl[\, \ov{\mathcal E}_{(\mathsf A)\,a} \, ,\, \ov{\mathcal E}_{(\mathsf A)}{}^{\hspace{1pt}b} \,\bigr]_{\rm C} &=&
  - & \hat f_{ac}{}^b \, \ov{\mathcal E}_{(\mathsf A)}{}^{\hspace{1pt}c} \,,
  \\[8pt]
  \bigl[\, \ov{\mathcal E}_{(\mathsf A)}{}^{\hspace{1pt}a} \, ,\, \ov{\mathcal E}_{(\mathsf A)}{}^{\hspace{1pt}b} \,\bigr]_{\rm C} &=&
  &0\,,
  \end{array}
}
where $\hat f_{ab}{}^c$ are the structure constants of the dimension-less
vielbein vector-fields $\hat {\ov e}_a$ defined  as $ [ \op \hat{\ov e}_a, \hat{\ov e}_b\op ]_{\rm L} = f_{ab}{}^c\,  \hat{\ov e}_c$.
We therefore see that an $O(D,D)$ transformation with non-trivial matrices $\mathsf A$ 
gives rise to a geometric flux.

\item The next case we consider is a $\mathsf B$-transformation \eqref{gg_b-transform_048}
acting on the trivial generalised vielbeins \eqref{gg_vielbein_flat_09087}. 
For $\mathsf B_{ij} = \hat b_{ij}$ the (dimension-less) components of a non-trivial Kalb-Ramond field $b=\frac{\alpha'}{2} \op \hat b_{ij}\op dx^i\wedge dx^j$, we find
\eq{
  \label{gv_h_flux_case_vb}
  \bigl(\ov{\mathcal E}_{(\mathsf B)}\bigr)_A\mathop{\vphantom{\bigr)}}\nolimits^I = 
  \left( \begin{array}{cc} \delta_a{}^i & -\delta_a{}^m \hat b_{mi} \\[2pt] 0 & \delta^a{}_i \end{array}\right) .
}
The Courant bracket for the corresponding generalised vectors takes the form
\eq{
  \label{gv_algebra_pp3}
  \arraycolsep1.5pt
  \begin{array}{cccrcr}
  \bigl[\, \ov{\mathcal E}_{(\mathsf B)\,a} \, ,\, \ov{\mathcal E}_{(\mathsf B)\,b} \,\bigr]_{\rm C} &=&
  -& \hat H_{abc}\,\ov{\mathcal E}_{(\mathsf B)}{}^{\hspace{1pt}c}  \,,
  \\[8pt]
  \bigl[\, \ov{\mathcal E}_{(\mathsf B)\,a} \, ,\, \ov{\mathcal E}_{(\mathsf B)}{}^{\hspace{1pt}b} \,\bigr]_{\rm C} &=&
  & 0 \,,
  \\[8pt]
  \bigl[\, \ov{\mathcal E}_{(\mathsf B)}{}^{\hspace{1pt}a} \, ,\, \ov{\mathcal E}_{(\mathsf B)}{}^{\hspace{1pt}b} \,\bigr]_{\rm C} &=&
  &0\,,
  \end{array}
}
where $\hat H_{abc} = \delta_a{}^i\,\delta_b{}^j\,\delta_c{}^k\, \hat H_{ijk}$ and $\hat H_{ijk} = 3\op \partial_{[\ul i} \hat b_{\ul j\ul k]}$
are the (dimension-less) components of the $H$-flux $H=db$. 
Here we see that  a non-trivial $\mathsf B$-transform can give rise to a non-trivial $H$-flux.

\item Let us also discuss a $\beta$-transformation \eqref{gg_beta_transform_120} acting again on the 
trivial generalised vielbeins \eqref{gg_vielbein_flat_09087}. We find
\eq{
  \bigl(\ov{\mathcal E}_{(\beta)}\bigr)_A\mathop{\vphantom{\bigr)}}\nolimits^I = 
  \left( \begin{array}{cc} \delta_a{}^i &0 \\[2pt] -\delta^a{}_m \op \beta^{mi} & \delta^a{}_i \end{array}\right) ,
}
where $\beta^{ij}$ is an anti-symmetric matrix,
and for the Courant brackets we obtain
\eq{
  \label{gv_algebra_pp4}
  \arraycolsep1.5pt
  \begin{array}{cccrcr}
  \bigl[\, \ov{\mathcal E}_{(\beta)\,a} \, ,\, \ov{\mathcal E}_{(\beta)\,b} \,\bigr]_{\rm C} &=&
  &&& 0 \,,
  \\[8pt]
  \bigl[\, \ov{\mathcal E}_{(\beta)\,a} \, ,\, \ov{\mathcal E}_{(\beta)}{}^{\hspace{1pt}b} \,\bigr]_{\rm C} &=&
  &&
   -& Q_a{}^{bc} \, \ov{\mathcal E}_{(\beta)\,c}\,,
  \\[8pt]
  \bigl[\, \ov{\mathcal E}_{(\beta)}{}^{\hspace{1pt}a} \, ,\, \ov{\mathcal E}_{(\beta)}{}^{\hspace{1pt}b} \,\bigr]_{\rm C} &=&
  -& Q_c{}^{ab}\, \ov{\mathcal E}_{(\beta)}{}^{\hspace{1pt}c}
  &+& R^{abc}\,\ov{\mathcal E}_{(\beta)\,c}  \,.
  \end{array}
}
Here we defined the non-geometric $Q$- and $R$-flux as
$Q_a{}^{bc} = \delta_a{}^i\,\delta^b{}_j\,\delta^c{}_k\, Q_i{}^{jk}$ and 
$R^{abc} = \delta^a{}_i\,\delta^b{}_j\,\delta^c{}_k\, R^{ijk}$ with
\eq{
  \label{fluxes_qr9386493856}
  Q_i{}^{jk} = \partial_i \op \beta^{jk} \,,
  \hspace{70pt}
  R^{ijk} = 3\, \beta^{[\ul i|m}\partial_m\op\beta^{\ul j\ul k]} \,.
}
Let us point out that the expression for the $R$-flux is similar to equation
\eqref{gg_lie_alg_938649386} in the context of a Lie algebroid, and that 
a non-vanishing $R$-flux gives rise to a quasi-Lie algebroid.

\item Finally, $O(D,D)$-transformations $\mathcal O_{\pm \mathsf i}$ defined in \eqref{gg_transf_344244}
act on the trivial generalised vielbeins \eqref{gg_vielbein_flat_09087} 
by interchanging vector-field and one-form components. 
Since the matrices $\mathcal O_{\pm \mathsf i}$ are constant the resulting
Courant brackets vanish, however, when $\mathcal O_{\pm \mathsf i}$ acts on non-trivial 
vielbein matrices this changes. 
We come back to this point in section~\ref{sec_gg_tdual}.

\end{itemize}
To summarise, when acting with non-trivial $O(D,D)$ transformations on the generalised 
vector-fields the corresponding Courant brackets are modified. This is expected since
such transformations are in general not automorphisms of the Courant bracket, however,
in this way we can generate backgrounds with non-vanishing geometric and non-geometric
fluxes. In particular, 
$\mathsf A$-transformations lead to a geometric flux, 
$\mathsf B$-transformations give an $H$-flux, and $\beta$-transformations
can generate $Q$- and $R$-fluxes. 
Using the notation \eqref{gen_vielbein_284}, we see that the 
geometric and  non-geometric fluxes are contained in $F_{AB}{}^C$,
where the indices $A,B,C$ can be upper or lower ones:
\eq{
  \label{gen_vielbein_284ccc}
  \arraycolsep2pt
  \begin{array}{lccl@{\hspace{50pt}}l}
  F_{abc} & = & -& \hat H_{abc} & \mbox{$H$-flux}\,, \\[4pt]
  F_{ab}{}^c & = & +& \hat f_{ab}{}^c & \mbox{geometric flux}\,, \\[4pt]
  F_{a}{}^{bc} & = & - &Q_a{}^{bc} & \mbox{non-geometric $Q$-flux}\,, \\[4pt]
  F^{abc} & = & +&R^{abc} & \mbox{non-geometric $R$-flux}\,.
  \end{array}
}

%%%%%%%%%%%%%%%%%%%%%%%%%%%%%%%%%%%%%%%%%%%%%%%
%%%%%%%%%%%%%%%%%%%%%%%%%%%%%%%%%%%%%%%%%%%%%%%

\subsubsection*{General form of fluxes}

For completeness, let us also give a general expression for the fluxes. 
We start again from the trivial generalised vector-field \eqref{gg_vielbein_flat_09087}
and apply to it a general $O(D,D)$ transformation of the form \eqref{gv_77729}.
The corresponding Courant brackets take the form \eqref{gen_vielbein_284}.
Expressing then the structure constants $F_{AB}{}^C$ using $\delta_A{}^I$ as
$F_{ab}{}^c = \delta_a{}^i\,\delta_b{}^j \, F_{ij}{}^k \,\delta_k{}^c$,
$F_{abc} = \delta_a{}^i\,\delta_b{}^j \, \delta_c{}^k \, F_{ijk} $ and so on, we find
{\allowdisplaybreaks
\begin{subequations}
\label{wim}
\begin{align}
  \arraycolsep2pt
  &\begin{array}{lc@{\hspace{10pt}}crll@{\hspace{10pt}}cc}
  F_{ijk} &=&& \displaystyle \bigl(A^T\bigr)_i^{\hspace{5pt}m} 
     & \partial_m \bigl(A^T\bigr)_j^{\hspace{5pt}n} & C_{nk} &-& { i\leftrightarrow j}
     \\[4pt]
  &&+& \displaystyle \bigl(A^T\bigr)_i^{\hspace{5pt}m} 
     & \partial_m \bigl(C^T\bigr)_{jn} & A^n_{\hspace{5pt}k} &-& { i\leftrightarrow j}
     \\[4pt]
  &&+& \displaystyle \bigl(A^T\bigr)_k^{\hspace{5pt}m} 
     & \partial_m \bigl(A^T\bigr)_i^{\hspace{5pt}n} & C_{nj} &
     \\[4pt]
  &&+& \displaystyle \bigl(A^T\bigr)_k^{\hspace{5pt}m} 
     & \partial_m \bigl(C^T\bigr)_{in} & A^n_{\hspace{5pt}j} \,,
  \end{array}
\\[12pt]
  &\begin{array}{lc@{\hspace{10pt}}crll@{\hspace{10pt}}cc}
  F_{ij}{}^k &=&& \displaystyle \bigl(A^T\bigr)_i^{\hspace{5pt}m} 
     & \partial_m \bigl(A^T\bigr)_j^{\hspace{5pt}n} & D_n^{\hspace{5pt}k} &-& { i\leftrightarrow j}
     \\[4pt]
  &&+& \displaystyle \bigl(A^T\bigr)_i^{\hspace{5pt}m} 
     & \partial_m \bigl(C^T\bigr)_{jn} & B^{nk} &-& { i\leftrightarrow j}
     \\[2pt]
  &&+& \displaystyle \bigl(B^T\bigr)^{km} 
     & \partial_m \bigl(A^T\bigr)_i^{\hspace{4pt}n} & C_{nj} &
     \\[2pt]
  &&+& \displaystyle \bigl(B^T\bigr)^{km} 
     & \partial_m \bigl(C^T\bigr)_{in} & A^n_{\hspace{5pt}j} \,,
  \end{array}
\\[12pt]
  &\begin{array}{lc@{\hspace{10pt}}crll@{\hspace{10pt}}cc}  
  F_i{}^{jk} &=&& \displaystyle \bigl(B^T\bigr)^{km} 
     & \partial_m \bigl(A^T\bigr)_i^{\hspace{5pt}n} & D_n^{\hspace{5pt}j} &-& { j\leftrightarrow k}
     \\[2pt]
  &&+& \displaystyle \bigl(B^T\bigr)^{km} 
     & \partial_m \bigl(C^T\bigr)_{in} & B^{nj} &-& { j\leftrightarrow k}
     \\[2pt]
  &&+& \displaystyle \bigl(A^T\bigr)_i^{\hspace{5pt}m} 
     & \partial_m \bigl(B^T\bigr)^{jn} & D_{n}^{\hspace{5pt}k} &
     \\[2pt]
  &&+& \displaystyle \bigl(A^T\bigr)_i^{\hspace{5pt}m} 
     & \partial_m \bigl(D^T\bigr)^j_{\hspace{5pt}n} & B^{nk} \,,
  \end{array}
\\[12pt]
  &\begin{array}{lc@{\hspace{10pt}}crll@{\hspace{10pt}}cc}
  F^{ijk} &=&& \displaystyle \bigl(B^T\bigr)^{im} 
     & \partial_m \bigl(B^T\bigr)^{jn} & D_n^{\hspace{5pt}k} &-& { i\leftrightarrow j}
     \\[2pt]
  &&+& \displaystyle \bigl(B^T\bigr)^{im} 
     & \partial_m \bigl(D^T\bigr)^j_{\hspace{5pt}n} & B^{nk} &-& { i\leftrightarrow j}
     \\[2pt]
  &&+& \displaystyle \bigl(B^T\bigr)^{km} 
     & \partial_m \bigl(B^T\bigr)^{in} & D_{n}^{\hspace{5pt}j} &
     \\[2pt]
  &&+& \displaystyle \bigl(B^T\bigr)^{km} 
     & \partial_m \bigl(D^T\bigr)^i_{\hspace{5pt}n} & B^{nj} \,.
  \end{array}
\end{align}
\end{subequations}
}

%%%%%%%%%%%%%%%%%%%%%%%%%%%%%%%%%%%%%%%%%%%%%%%
%%%%%%%%%%%%%%%%%%%%%%%%%%%%%%%%%%%%%%%%%%%%%%%

\subsubsection*{Example}

Let us  finally mention a particular example which is often discussed in the literature. We consider
the generalised vielbein \eqref{vielbein_again_648}, which we can 
express as
\eq{
  \label{gen_viel_6541}
  \ov{\mathcal E}_A\mathop{\vphantom{\bigr)}}\nolimits^I = 
  \left( \begin{array}{cc} \hat{\ov e}_a{}^i & - \hat{\ov e}_a{}^m \hat b_{mi} \\[2pt] 0 & \hat e^a{}_i \end{array}\right) 
  = \Bigl[\,  \ov{\mathcal E}_{(0)} \,\mathcal O_{\mathsf A}^T \, \mathcal O_{\mathsf B}^T \,\Bigr]_A
  \mathop{\vphantom{\Bigr]}}\nolimits^I \,,
}
where $\ov{\mathcal E}_{(0)} $ has been defined in \eqref{gg_vielbein_flat_09087} 
and $\mathsf A^i{}_j = \delta^i{}_b\op \hat e^b{}_j$ and $\mathsf B_{ij} = \hat b_{ij}$
are given in terms of the dimension-less vielbein matrices and the Kalb-Ramond field.
For the generalised vectors we then have
\eq{
  \label{gen_viel_6540}
  \ov{\mathcal E}_a = \hat{\ov e}_a - \iota_{\hat{\ov e}_a} \hat b \,,  \hspace{80pt}
  \ov{\mathcal E}^a = \hat e^a \,,
}
and the Courant bracket \eqref{gen_vielbein_284} reads
\eq{
  \label{gv_algebra_6492}
  \arraycolsep1.5pt
  \begin{array}{cccrcr}
  \bigl[\op \mathcal E_a , \mathcal E_b \op\bigr]_{\rm C} &=&
  +& \hat f_{ab}{}^c \,\mathcal E_c & - &\hat H_{abc}\, \mathcal E^c \,,
  \\[8pt]
  \bigl[\op \mathcal E_a , \mathcal E^b \op\bigr]_{\rm C} &=&
  & & - &\hat f_{ac}{}^{b}\, \mathcal E^c \,,
  \\[8pt]
  \bigl[\op \mathcal E^a , \mathcal E^b \op\bigr]_{\rm C} &=& 
  & &  & 0\,.
  \end{array}
}
The $H$-flux in the vielbein basis is given by $\hat H_{abc} = \hat{\ov e}_a{}^i\, \hat{\ov e}_b{}^j\,\hat{\ov e}_c{}^k\, \hat H_{ijk}$.
We therefore see that the generalised vielbein \eqref{gen_viel_6540} encodes
the $H$-flux as well as the geometric flux, as one would expect from 
\eqref{gen_viel_6541}.

%%%%%%%%%%%%%%%%%%%%%%%%%%%%%%%%%%%%%%%%%%%%%%%
%%%%%%%%%%%%%%%%%%%%%%%%%%%%%%%%%%%%%%%%%%%%%%%
%%%%%%%%%%%%%%%%%%%%%%%%%%%%%%%%%%%%%%%%%%%%%%%
%%%%%%%%%%%%%%%%%%%%%%%%%%%%%%%%%%%%%%%%%%%%%%%

\subsection{T-duality}
\label{sec_gg_tdual}

Let us now discuss T-duality transformations in the context of generalised geometry.
These can be realised via $O(D)\times O(D)$ transformations of the form \eqref{gg_7382929} acting on the 
generalised vielbein vector-fields.

%%%%%%%%%%%%%%%%%%%%%%%%%%%%%%%%%%%%%%%%%%%%%%%
%%%%%%%%%%%%%%%%%%%%%%%%%%%%%%%%%%%%%%%%%%%%%%%

\subsubsection*{$O(D)\times O(D)$ transformations}

In section~\ref{sec_gg_fluxes} we have considered $O(D,D)$ transformations acting on the generalised vector
$\ov{\mathcal E}_A{}^I$ from the right, that is contracting the transformation matrix $\mathcal O\in O(D,D)$ with the index $I$.  
However, as noted in \eqref{gg_7382929}  we have an $O(D)\times O(D)$ structure, whose 
transformations act on the generalised vector from the left. 
Recalling the form of the transformations given in \eqref{odod}, 
\begin{itemize}

\item we see that $O_+=O_-$ corresponds to $O(D)$ transformations which rotate the 
generalised-vector components 
$\ov{\mathcal E}_a$ and  $\ov{\mathcal E}^a$ in a similar way. 
The transformation of the 
corresponding Courant brackets is then worked out along similar lines as above. 

\item On the other hand, for $O_+\neq O_-$ 
the $O(D)\times O(D)$ transformations mix the components $\ov{\mathcal E}_a$ and  $\ov{\mathcal E}^a$.
The corresponding Courant brackets are  modified, in particular, the type of fluxes appearing on the 
right-hand side will in general change. 

\end{itemize}
As a specific example for the second situation, let us consider two $D\times D$ matrices of the form
\eq{
  \label{gengeo_93472}
  \arraycolsep2pt
  O_+ =  \left( \begin{array}{ccccc} +1&  \\
  &&+1 \\
  &&& \ddots \\
  &&&&+1  \end{array}\right),
  \hspace{40pt}
  O_- =  \left( \begin{array}{ccccc} -1&  \\
  &&+1 \\
  &&& \ddots \\
  &&&&+1  \end{array}\right),
}
which via \eqref{odod} lead to a matrix $\mathcal K_{+1}$. The latter is an example of $\mathcal K_{\pm\mathsf i}$, which we define as
\eq{
  \label{gg_transf_080802}
  \mathcal K_{\pm \mathsf i} = 
  \left( \begin{array}{cc} \mathds 1 - E_{\mathsf i}& \pm E_{\mathsf i} \\ \pm E_{\mathsf i} & \mathds 1 -E_{\mathsf i} \end{array}\right)
  \equiv \mathcal O_{\pm \mathsf i} \,,  
}  
where $E_{\mathsf i}$ and $\mathcal O_{\pm \mathsf i}$ were given around equation \eqref{gg_transf_344244}.
Applying the transformation $\mathcal K_{+1}$ induced by \eqref{gengeo_93472} 
to the generalised vector from the left, we interchange 
the first vector-field and one-form components
\eq{
  \label{gg_transf_080803}
  \ov{\mathcal E}_A \;\to\;  \ov{\mathcal E}'_A = \bigl( \mathcal K_{+ 1}\bigr){}_A^{\hspace{6pt}B }\,
    \ov{\mathcal E}_B  
  = \left( \begin{array}{c}
  \delta_{11} \ov{\mathcal E}^1 \\[2pt]
  \ov{\mathcal E}_{\hat a} \\[2pt]
  \delta^{11} \ov{\mathcal E}_1 \\[2pt]
  \ov{\mathcal E}^{\hat a}
  \end{array}
  \right),
  \hspace{50pt}
  \hat a = 2,\ldots D\,.
}
Computing now the Courant bracket of the transformed generalised vector-fields,  a similar interchange can be found.
In particular, for \eqref{gg_transf_080803} the index $1$ of the various fluxes is raised or lowered according to
the following mapping
\eq{
  \label{gg_flux_88492}
  \arraycolsep2pt
  \begin{array}{@{}lcl@{\hspace{21pt}}lcl@{\hspace{21pt}}lcl@{\hspace{21pt}}lcl@{}}
  \hat H_{\hat a\hat b1} &\to& -\hat f_{\hat a\hat b}{}^1 \,,
  &
  \hat f_{\hat a\hat b}{}^1 & \to & - \hat H_{\hat a\hat b1}\,,
  &
  \\[4pt]
  &&& \hat f_{\hat a1}{}^{\hat b} &\to&  -Q_{\hat a}{}^{1\hat b} \,,
  &
  Q_{\hat a}{}^{1\hat b} &\to& - \hat f_{\hat a1}{}^{\hat b}\,,
  \\[4pt]
  &&&&&&
  Q_1{}^{\hat a\hat b} &\to& -R^{1\hat a\hat b} \,,
  &
  R^{1\hat a\hat b} &\to& - Q_1{}^{\hat a\hat b}  \,,
  \end{array}
}
where $\hat a,\hat b= 2,\ldots, D$ and where the other flux-components do not change. 
For transformations which are a combination 
of $\mathcal K_{\pm\mathsf i}$ a similar exchange 
between upper- and lower-indices can be found. 
The behaviour of the fluxes shown in \eqref{gg_flux_88492} is in agreement with our results from 
section~\ref{sec_first_steps}, as well as with 
\cite{Shelton:2005cf}, where the same observation has been made in the context of supergravity.
To summarise, 
\begin{quote}
A T-duality transformation along the direction $x^{\mathsf i}$ can be realised by 
acting with $\mathcal K_{\pm\mathsf i}$ on the generalised vector-field  $\ov{\mathcal E}_A$ from the left. 
For the fluxes, a corresponding lower index $\mathsf i$ is raised 
and a corresponding upper index $\mathsf i$ is lowered. 
\end{quote}

 %%%%%%%%%%%%%%%%%%%%%%%%%%%%%%%%%%%%%%%%%%%%%%%
%%%%%%%%%%%%%%%%%%%%%%%%%%%%%%%%%%%%%%%%%%%%%%%

\subsubsection*{$O(D,D)$ transformations}

Let us now also relate the above results to $O(D,D)$ transformations acting on the generalised vectors 
from the right. Recall from \eqref{gv_77729} that a general generalised vector can be 
expressed as $\ov{\mathcal E}_{(\mathcal O)} = \ov{\mathcal E}_{(0)} \op \mathcal O^T$. 
Acting with a transformation $\mathcal O_{\pm\mathsf i}$ from the right can be written as
\eq{
  \label{gg_dual_auto_7391}
  \ov{\mathcal E}_{(\mathcal O)}\;\rightarrow\; \ov{\mathcal E}'_{(\mathcal O')}
  &=
  \ov{\mathcal E}_{(\mathcal O)}\op \mathcal O^T_{\pm \mathsf i}
  \\[2pt]
  &=
  \ov{\mathcal E}_{(0)}\op \mathcal O^T\op\mathcal O^T_{\pm \mathsf i}
  \\
  &= \mathcal K_{\pm\mathsf i}\, \ov{\mathcal E}_{(0)} \bigl( \mathcal O^{-1}_{\pm \mathsf i} \op
  \mathcal O \op \mathcal O_{\pm \mathsf i} \bigr)^T \,,
}
where the expression in parenthesis encodes the duality transformation of the metric and Kalb-Ramond 
field. We thus see that a $\mathcal O_{\pm\mathsf i}$ transformation from the right 
leads to a transformation of the background quantities, and raises or lowers the position 
of the corresponding structure index.

Let us  now note that transformations of the form 
$\ov{\mathcal E}\to \ov{\mathcal E}'= \ov{\mathcal E}\, \mathcal O^T_{\pm \mathsf i}$ are in general 
not automorphisms of the Courant bracket. However, for backgrounds satisfying
additional conditions -- such as the T-duality requirements \eqref{const_global} -- this can change
\cite{Ellwood:2006ya}.\label{page_auto_more}
Let us illustrate this observation with the example of the three-torus with $H$-flux 
discussed in section~\ref{sec_first_steps}. Using \eqref{gen_viel_6541}, from the metric \eqref{metric_01} and 
the $H$-flux \eqref{ex1_metric_98} we determine a generalised vector-field as 
\eq{
  \label{gg_flux_ex_739}
   \ov{\mathcal E}_A\mathop{\vphantom{\bigr)}}\nolimits^I
   = \left( 
   \scalebox{0.9}{$\displaystyle
   \renewcommand{\arraystretch}{1.4}
   \begin{array}{ccc|ccc}
   \frac{\sqrt{\alpha'}}{R_1} & 0 & 0 & 0 & -\frac{h}{2\pi}\,\frac{x^3}{R_1} & 0 \\
   0 & \frac{\sqrt{\alpha'}}{R_2} & 0 & +\frac{h}{2\pi}\,\frac{x^3}{R_2} & 0 & 0 \\
   0 & 0 &  \frac{\sqrt{\alpha'}}{R_3} & 0 & 0 & 0\\
   \hline
   0 & 0 & 0 & \frac{R_1}{\sqrt{\alpha'}} & 0 & 0 \\
   0 & 0 & 0 & 0 & \frac{R_2}{\sqrt{\alpha'}} & 0 \\
   0 & 0 & 0 & 0 & 0 & \frac{R_3}{\sqrt{\alpha'}} \\   
   \end{array}
   $}
   \right),
}
from which we can determine the only non-vanishing flux via the Courant brackets 
as $\hat H_{123} = \frac{h}{2\pi} \,\frac{1}{R_1\op R_2\op R_3}$
in accordance with \eqref{ex1_2050}.
We now consider two types of $O(D,D)$ transformations:
\begin{itemize}

\item Let us first consider a matrix $\mathcal O_{+1}$ acting on \eqref{gg_flux_ex_739} from the right. 
This transformation leaves the Courant bracket between the generalised vectors \eqref{gg_flux_ex_739} 
invariant, and hence the flux does not change. 
This can be understood from \eqref{gg_dual_auto_7391} by observing  that $\mathcal O_{+1}$
generates a T-duality transformation of the background along the direction $x^1$, which is however
un-done by  $\mathcal K_{+1}$ acting from the left.

\item As a second type we consider a matrix $\mathcal O_{+3}$ acting on \eqref{gg_flux_ex_739} from the right.
Note that since \eqref{gg_flux_ex_739} depends explicitly on $x^3$, this transformation is not 
an automorphism of the corresponding Courant bracket \cite{Ellwood:2006ya}.
Indeed, the transformed flux reads 
$\hat f_{12}{}^{3} = -\frac{h}{2\pi} \,\frac{R_3}{R_1\op R_2}$.
When comparing with \eqref{gg_dual_auto_7391}, 
we see that $\mathcal O_{+3}$ generates a T-duality transformation along the direction $x^3$ which 
maps $R_3\to \alpha'/R_3$, and $K_{+3}$ maps the $H$-flux to a geometric flux.

\end{itemize}
To summarise, for the example of the three-torus with $H$-flux we have illustrated 
that T-duality transformations acting on the generalised vectors as $O(D,D)$-transformations 
from the right leave the background invariant, provided that the conditions discussed
in section~\ref{sec_gg_buscher} are satisfied. If the latter are not satisfied, 
the background changes.

%%%%%%%%%%%%%%%%%%%%%%%%%%%%%%%%%%%%%%%%%%%%%%%
%%%%%%%%%%%%%%%%%%%%%%%%%%%%%%%%%%%%%%%%%%%%%%%
%%%%%%%%%%%%%%%%%%%%%%%%%%%%%%%%%%%%%%%%%%%%%%%
%%%%%%%%%%%%%%%%%%%%%%%%%%%%%%%%%%%%%%%%%%%%%%%
%%%%%%%%%%%%%%%%%%%%%%%%%%%%%%%%%%%%%%%%%%%%%%%
%%%%%%%%%%%%%%%%%%%%%%%%%%%%%%%%%%%%%%%%%%%%%%%
%%%%%%%%%%%%%%%%%%%%%%%%%%%%%%%%%%%%%%%%%%%%%%%
%%%%%%%%%%%%%%%%%%%%%%%%%%%%%%%%%%%%%%%%%%%%%%%

\subsection{Frame transformations}
\label{sec_gg_frame}

On page~\pageref{page_odod_939} we discussed the $O(D)\times O(D)$
structure of the generalised tangent-bundle. Let us now first consider 
a particular $O(D)\times O(D)$ transformation \eqref{odod}
which replaces the Kalb-Ramond $B$-field by a bivector-field $\beta$, 
and then construct an effective action for the transformed fields.

%%%%%%%%%%%%%%%%%%%%%%%%%%%%%%%%%%%%%%%%%%%%%%%
%%%%%%%%%%%%%%%%%%%%%%%%%%%%%%%%%%%%%%%%%%%%%%%

\subsubsection*{Change of frame}

We start by recalling the generalised vielbein of a background with geometric and $H$-flux shown in equation \eqref{vielbein_again_648} as 
\eq{
  \label{gg_frame_985693856}
  \mathcal E = 
  \frac{1}{\sqrt{\alpha'}}\left( \begin{array}{cc} e^a{}_i & 0 \\[2pt] - \ov e_a{}^m\op b_{mi} & 
  \alpha'\,\ov e_a{}^i \end{array}\right) ,
}
and then we perform an $O(D)\times O(D)$ transformation \eqref{odod} specified by the following two $O(D)$ 
matrices \cite{Grana:2008yw}
\eq{
  \label{gg_frame_0482}
  O_+= \mathds 1 \,,\hspace{50pt}
  O_- =  \bigl( \op e - \delta^{-1}\op \ov e^T \op b\op\bigr)
  \bigl( \op e + \delta^{-1}\op \ov e^T \op b\op\bigr)^{-1}\,.
}
The transformed generalised vielbein then takes the following form
\eq{
  \mathcal E' = 
  \frac{1}{\sqrt{\alpha'}}\left( \begin{array}{cc} e'{}^a{}_i & -e'{}^a{}_m\op\beta^{mi} \\[2pt] 0 & 
  \alpha'\,\ov e'{}_a{}^i \end{array}\right) ,
}
which is expressed in terms of a transformed vielbein $e'{}^a{}_i$ (with corresponding transformed metric $g'_{ij}$)
and a bivector-field $\beta^{ij}$ of the form
\eq{
  \label{gg_frame_83645}
  e' =e\op g^{-1} \op (g-b)  \,,\hspace{40pt}
  \arraycolsep1pt
  \begin{array}[t]{l@{\hspace{3pt}}c@{\hspace{3pt}}clcl}
  g' &=&  & \displaystyle  (g+b) & \displaystyle g^{-1}&\displaystyle (g-b) \,, 
  \\[4pt]
  \beta &=&  - &\displaystyle  (g-b)^{-1}&\displaystyle  b&\displaystyle  (g+b)^{-1} \,.
  \end{array}
}
Let us note that the expressions for the transformed metric and the bivector-field can also be 
encoded via the relation
\eq{
  ( g-b)^{-1} = g'^{-1} - \beta \,.
}
Since the background can be specified in terms of a generalised vielbein, we see that locally 
the information is contained either in 
a metric and Kalb-Ramond field $g$ and $b$ -- or equivalently in a different metric and a bivector-field 
$g'$ and $\beta$.

%%%%%%%%%%%%%%%%%%%%%%%%%%%%%%%%%%%%%%%%%%%%%%%
%%%%%%%%%%%%%%%%%%%%%%%%%%%%%%%%%%%%%%%%%%%%%%%

\subsubsection*{Example}

Even though locally one can always perform a change of frames from $(g,b)$ to $(g',\beta)$, globally 
the transformation \eqref{gg_frame_0482} may not be well-defined. Let us illustrate this situation with 
the example of the T-fold. 
Using our conventions from section~\ref{sec_t2_fibr_example}, we recall the metric and $B$-field of the 
three-dimensional T-fold background as follows
\eq{
  g_{ij} = \left( \begin{array}{ccc} 
  \frac{R_2^2}{\rho} & 0 & 0 \\ 
  0 & \frac{R_1^2}{\rho} & 0 \\
  0 & 0 & R_3^2
  \end{array} \right),
  \hspace{40pt}
   b_{ij} = \frac{1}{\rho} \left( \begin{array}{ccc} 
    0 & - \frac{\alpha'}{2\pi}\op h\op x^3  & 0 \\[4pt] + \frac{\alpha'}{2\pi}\op h\op x^3  & 0 & 0 \\[4pt]
    0 & 0 & 0
    \end{array}\right),
}  
where $  \rho = \frac{R_1^2 R_2^2}{\alpha'^2} + \left[\frac{h}{2\pi}\op x^3\right]^2$
and $h\in\mathbb Z$.  The transformation \eqref{gg_frame_0482} for this example takes the 
following explicit form
\eq{
  \label{gg_frame_34235}
  O_+= \mathds 1 \,,\hspace{50pt}
  O_- =   \frac{1}{\rho}\left( 
\begin{array}{ccc}  \frac{R_1^2 R_2^2}{\alpha'^2} - \left[\frac{h}{2\pi} x^3\right]^2 & +\frac{R_1 R_2}{\alpha'}
  \op\frac{h}{\pi} \op x^3 & 0 \\
  -\frac{R_1 R_2}{\alpha'}  \op\frac{h}{\pi} \op x^3 & \frac{R_1^2 R_2^2}{\alpha'^2} - \left[\frac{h}{2\pi} x^3\right]^2
  & 0 \\ 0 & 0 & \rho
  \end{array}
  \right),
}
and the resulting metric and bivector-field is determined from \eqref{gg_frame_83645} as
\eq{
  \label{gg_frame_934}
  g'_{ij} = \left( \begin{array}{ccc} 
  \frac{\alpha'}{R_1^2} & 0 & 0 \\ 
  0 & \frac{\alpha'}{R_2^2} & 0 \\
  0 & 0 & R_3^2
  \end{array} \right),
  \hspace{40pt}
   \beta^{ij} = \left( \begin{array}{ccc} 
    0 & +  \frac{h\op x^3}{2\pi\op\alpha'}  & 0 \\[4pt] -\frac{h\op x^3}{2\pi\op\alpha'}   & 0 & 0 \\[4pt]
    0 & 0 & 0
    \end{array}\right).
}  
Note that these expressions are very similar to the metric and $B$-field of the three-torus with 
$H$-flux shown in equation \eqref{fibr_1_h}.
However, even though locally \eqref{gg_frame_934} takes a rather simple form, the T-fold background
is nevertheless non-geometric. This can be seen by noting that the change of frame
\eqref{gg_frame_34235} is not well-defined under the identification $x^3\to x^3 + 2\pi$,
and hence the frame \eqref{gg_frame_934} is globally not well-defined \cite{Grana:2008yw}.
This shows the globally non-geometric nature of the T-fold.

%%%%%%%%%%%%%%%%%%%%%%%%%%%%%%%%%%%%%%%%%%%%%%%
%%%%%%%%%%%%%%%%%%%%%%%%%%%%%%%%%%%%%%%%%%%%%%%

\subsubsection*{Effective action I -- example}

We now want to discuss how frame transformations change the effective description of 
the theory. 
In particular, let us recall the ten-dimensional string-theory action for  the NS-NS sector 
in type II theories as
\eq{
  \label{action_nsns_77394}
  \mathcal S = \frac{1}{2\op\kappa^2} \int e^{-2\phi} \left[
  R \star 1 - \frac{1}{2} \op H\wedge \star H + 4\op d \phi \wedge \star d\phi \op \right] ,
}
where $R$ denotes the Ricci scalar for the metric $g$, $H=db$ is the field strength of 
the Kalb-Ramond $B$-field, $\phi$ denotes the dilaton and $\star$ is the ten-dimensional 
Hodge star-operator. 
We also note that the generalised vielbein shown in \eqref{gg_frame_985693856} 
contains information about the metric and $B$-field, which appear in the action 
\eqref{action_nsns_77394}.
However, when performing the change of frames shown in \eqref{gg_frame_0482}
the degrees of freedom of $g$ and $b$ are re-packaged into a new metric $g'$ and 
a  bivector-field $\beta$. 
In view of \eqref{action_nsns_77394}, a natural question to ask is how
an action for $g'$ and $\beta$ can be constructed.
This program has been followed in the papers 
\cite{Andriot:2011uh,Andriot:2012wx,Andriot:2012an,Blumenhagen:2012nk,Blumenhagen:2012nt,Blumenhagen:2013aia,
Andriot:2013xca}
and the resulting formulation has been called symplectic gravity or $\beta$-supergravity.

In order to address this question we first 
recall that \eqref{action_nsns_77394} is invariant 
under diffeomorphisms $x^i\to x^i+ \xi^i(x)$ and gauge transformations of the Kalb-Ramond field. In particular, 
with $\xi$ a vector-field and $\Lambda$ a one-form, the metric and $B$-field
transform as
\eq{
  \arraycolsep2pt
  \begin{array}{lcl@{\hspace{80pt}}lcl}
  \delta_{\xi} \op g &=& \mathcal L_{\xi} \op g \,,
  &
  \delta_{\Lambda}\op g &=&  0\,,
  \\[4pt]
  \delta_{\xi} \op b &=& \mathcal L_{\xi} \op b \,,
  &
  \delta_{\Lambda} \op b &=&  d\Lambda\,,
  \end{array}
}
where $\mathcal L_{\xi}$ denotes the usual Lie-derivative along the direction $\xi$. 
Now, when performing the change of frames \eqref{gg_frame_83645}, 
diffeomorphisms and gauge transformations of $g$ and $b$ become intertwined.
Clearly, since $g$ and $b$ behave as ordinary tensors also $g'$ and $\beta$ 
will transform as expected under diffeomorphisms. 
However, under gauge transformations of $b$ now both $g'$ and $\beta$ will transform,
and these transformations have been called momentum- \cite{Andriot:2011uh} or $\beta$-diffeomorphisms \cite{Blumenhagen:2012nk}
in the literature.

The main task is now to construct an action for $g'$ and $\beta$ (and the dilaton $\phi$), which
is invariant under ordinary diffeomorphisms as well as $\beta$-diffeomorphisms. 
This has been investigated in the papers
\cite{Andriot:2011uh,Andriot:2012wx,Andriot:2012an}, where the 
explicit form of the action can be found. 
The latter is motivated from double field theory (see section~\ref{sec_dft} for 
a brief introduction to double field theory) and its explicit form is somewhat involved. 
We therefore do not recall it here but refer to the above-mentioned literature.
However, we can comment on the appearance of non-geometric fluxes in this action.
In particular, the analogue of the field strength for the Kalb-Ramond field, transforming covariantly under $\beta$-diffeomorphisms, 
is given by 
\eq{
  \Theta_{ijk} = -\op g'_{im}\op g'_{jn}\op g'_{kl} \bigl( 3\op \beta^{[\ul m|p} \partial_p \beta^{\ul n\ul l]}\bigr)
  + \mathcal O( \partial g') + \mathcal O(\partial \beta) \,,
}
where the $ \mathcal O( \partial g')$ and $ \mathcal O( \partial \beta)$ terms can be made explicit.
This expression contains the $R$-flux 
$R^{mnl} = 3 \op \beta^{[\ul m|p} \partial_p \beta^{\ul n\ul l]}$
defined in equation \eqref{fluxes_qr9386493856},
which shows that the transformed action is suitable for describing non-geometric flux-backgrounds. 
We also mention that questions concerning the equations of motion of 
the $\beta$-supergravity, its dimensional reduction and specific examples have been discussed in 
detail in \cite{Andriot:2013xca}.

%%%%%%%%%%%%%%%%%%%%%%%%%%%%%%%%%%%%%%%%%%%%%%%
%%%%%%%%%%%%%%%%%%%%%%%%%%%%%%%%%%%%%%%%%%%%%%%

\subsubsection*{Effective action II -- Lie algebroid}

The frame transformation \eqref{gg_frame_0482} is only 
on particular example of an $O(D)\times O(D)$ transformation acting on the 
generalised vielbein \eqref{gg_frame_985693856}. 
A framework which incorporates more general changes of frame 
\cite{Blumenhagen:2012nk,Blumenhagen:2012nt,Blumenhagen:2013aia}
are  Lie algebroids which we introduced in section~\ref{sec_lie_courant}.
To explain this construction, we start from a general Lie algebroid $(E,[\cdot,\cdot]_E,\rho)$ 
and require the anchor $\rho:E\to TM$ 
to be invertible. We then consider the following additional structure:
\begin{itemize}

\item We equip the Lie algebroid $E$ with a metric 
$\mathsf  g\in\Gamma(E^*\otimes_{\mathrm{sym}}E^*)$, which 
is related to a Riemannian metric $ g$ on $TM$ (appearing in  \eqref{gg_frame_985693856}) through the 
dual anchor $\rho^*:E^*\to T^*M$ as
\eq{
\label{frames_8263}
g = \big(\!\otimes^2\!\rho^*\big)\op\mathsf g \,.
}
As  
briefly discussed on page~\pageref{page_dg_lie_alg}, and more
detailedly explained in \cite{Blumenhagen:2013aia},
for the metric on $E$ we can construct a corresponding differential 
geometry. The corresponding Ricci tensor 
$\mathsf{Ric}\in\Gamma(E^*\!\otimes_{\mathrm{sym}}\!E^*)$ on 
the Lie algebroid can be constructed explicitly and is
related to the ordinary Ricci tensor $Ric\in\Gamma(T^*M\otimes_{\mathrm{sym}}T^*M)$ via
\eq{
Ric = \big(\!\otimes^2\!\rho^*\big)\op\mathsf { Ric} \,.
}

\item Next, we turn to a Kalb-Ramond field 
$\mathsf{b}\in\Gamma(\Lambda^2 E^*)$ on the Lie algebroid. It is related 
to the usual Kalb-Ramond field $b$ via the dual anchor as
\eq{
b=\big(\Lambda^2\rho^*\big)(\mathsf{b})   \,.
}
Using the differential $d_E$ on the Lie algebroid given in \eqref{gg_ll_244}, 
we can define a corresponding field strength as
\eq{
\label{frames_29499}
	\Theta = d_E\op \mathsf{b}\quad\in\Gamma(\Lambda^3E^*) \,,
}
and due to $d_E$ being nilpotent we see that $\Theta$ is 
invariant under Lie-algebroid gauge transformations
\eq{
  \mathsf b \to \mathsf b + d_E\op \mathsf a\,,
  \hspace{50pt}
  \mathsf a\in\Gamma(E^*) \,.
}
In \cite{Blumenhagen:2013aia} these transformations have been called 
\textit{$\rho$-gauge transformations}, and they are the analogue of the 
usual Kalb-Ramond $B$-field gauge transformations.

\end{itemize}
After having defined a metric and Kalb-Ramond field on the Lie algebroid, 
we can construct an action for $\mathsf g$ and $\mathsf b$ which
is invariant under diffeomorphisms and $\rho$-gauge transformations.
Including the dilaton, such an action is given by \cite{Blumenhagen:2013aia}
\eq{
  \label{frames_dual_action}
 \mathcal S = \frac{1}{2\op\kappa^2} \int e^{-2{\phi} }  \left|\rho^*\right|  \left[\op
  \mathsf R \star 1 - \frac{1}{2} \op \Theta\wedge \star \Theta + 4\op d_E {\phi} \wedge \star d_E{\phi} \op \right] ,
}
where $\mathsf R$ is the Ricci scalar constructed from the Ricci tensor $\mathsf {Ric}$ 
and the metric $\mathsf g$ on the Lie algebroid, the Hodge star-operator is 
defined with respect to the metric $\mathsf g$, and $\Theta$ has been 
given in \eqref{frames_29499} in terms of the Lie algebroid differential
$d_E$ defined in \eqref{gg_ll_244}.
The determinant of the dual anchor $\rho^*$ is denoted as $ \left|\rho^*\right| $.
This action is by construction invariant under diffeomorphisms and 
two-form gauge transformations.
Furthermore, using the anchor $\rho$ we can for instance relate 
the Lie algebroid metric $\mathsf g$ to the ordinary metric $g$ 
as shown in \eqref{frames_8263}, and more generally show that 
\eqref{frames_dual_action} is equivalent to \eqref{action_nsns_77394}.

Let us finally connect our discussion here to the change of frames 
considered above. 
To do so, we  have to specify the anchor and first note 
that the index structure of $\rho$  is 
$\rho^i{}_{\mathsf a}$, where $i$ is an index on the tangent-space and 
$\mathsf a$ denotes a general index on the Lie algebroid (which can be upper or lower depending on 
the Lie algebroid $E$). The dual anchor 
is given by $\rho^* = \rho^{-T}$. 
For a given change of frames \eqref{odod}, 
specified in terms of two $O(D)$-transformations $O_+$ and 
$O_-$, the anchor reads
\eq{
  \rho = \frac{1}{2} \, \ov{ e} \left[ \op O_+ \op e \op g^{-1} ( g -  b)
  + O_- \op  e\op  g^{-1} ( g +  b)
  \right] \mathbf 1\,,
}
where $e$ is the vielbein matrix corresponding to the metric $g$
and the identity matrix $\mathbf 1$ with index structure $\delta^i{}_{\mathsf a}$
has been included to properly match the indices on $E$ and $TM$.
Now, for the example \eqref{gg_frame_0482} the anchor takes the explicit form
\eq{
  \rho = \mathds 1 - g^{-1} b = \mathds 1 + \beta\op g'\,,
}
but for instance also $\rho= \beta$ has been analysed 
\cite{Blumenhagen:2012nk,Blumenhagen:2012nt}.
To summarise, given a $O(D)\times O(D)$ frame transformation of the generalised 
vielbein which mixes the metric and $B$-field, we can use the framework 
of Lie algebroids to construct a corresponding transformed effective action 
shown in \eqref{frames_dual_action}.
Further aspects of this action, such as its equations of motion, its extension 
to the Ramond-Ramond sector of type II string theory and its relation 
to double field theory, can be found in
\cite{Blumenhagen:2012nk,Blumenhagen:2012nt,Blumenhagen:2013aia}.

%%%%%%%%%%%%%%%%%%%%%%%%%%%%%%%%%%%%%%%%%%%%%%%
%%%%%%%%%%%%%%%%%%%%%%%%%%%%%%%%%%%%%%%%%%%%%%%
%%%%%%%%%%%%%%%%%%%%%%%%%%%%%%%%%%%%%%%%%%%%%%%
%%%%%%%%%%%%%%%%%%%%%%%%%%%%%%%%%%%%%%%%%%%%%%%

\subsection{Bianchi identities}
\label{sec_gg_bianchi}

In this section we  discuss Bianchi identities for the geometric and non-geometric fluxes
introduced above. 
In the case of only $H$-flux, the Bianchi identity takes the well-known form $dH=0$, however,
if other fluxes are present this condition is modified.

%%%%%%%%%%%%%%%%%%%%%%%%%%%%%%%%%%%%%%%%%%%%%%%
%%%%%%%%%%%%%%%%%%%%%%%%%%%%%%%%%%%%%%%%%%%%%%%

\subsubsection*{Roytenberg algebra}

Let us first summarise the Courant brackets between generalised vielbein vector-fields
determined in \eqref{gv_algebra_pp2}, \eqref{gv_algebra_pp3} and 
\eqref{gv_algebra_pp4} as follows
\eq{
  \label{royten_29990}
  \arraycolsep1.5pt
  \begin{array}{cccrcr}
  \bigl[\, \ov{\mathcal E}_{a} \, ,\, \ov{\mathcal E}_{b} \,\bigr]_{\rm C} &=&
  +& \hat f_{ab}{}^c\,\ov{\mathcal E}_{c}&-& \hat H_{abc}\,\ov{\mathcal E}^{\hspace{1pt}c} \,,
  \\[8pt]
  \bigl[\, \ov{\mathcal E}_{a} \, ,\, \ov{\mathcal E}^{\hspace{1pt}b} \,\bigr]_{\rm C} &=&
  - & \hat f_{ac}{}^b \, \ov{\mathcal E}^{\hspace{1pt}c}&
   -& Q_a{}^{bc} \, \ov{\mathcal E}_{c}\,,
  \\[8pt]
  \bigl[\, \ov{\mathcal E}^{\hspace{1pt}a} \, ,\, \ov{\mathcal E}^{\hspace{1pt}b} \,\bigr]_{\rm C} &=&
  -& Q_c{}^{ab}\, \ov{\mathcal E}^{\hspace{1pt}c}
  &+& R^{abc}\,\ov{\mathcal E}_{c}  \,,
  \end{array}
}
where $\hat f_{ab}{}^c$ and $\hat H_{abc}$ denote the dimension-less geometric and $H$-flux, 
and $Q_a{}^{bc}$ and $R^{abc}$ denote the non-geometric $Q$- and $R$-flux. 
In the mathematical literature this algebra is also known as the Roytenberg algebra 
\cite{1999royten,Roytenberg:2001am}.\footnote{
In the context of gauged supergravities a similar structure has appeared \cite{Kaloper:1999yr},
and for a discussion of this algebra from a world-sheet point of view see \cite{Halmagyi:2008dr,Halmagyi:2009te}.
}
Using the conventions \eqref{gen_vielbein_284ccc}, the algebra \eqref{royten_29990} can be 
collectively written as in 
\eqref{gen_vielbein_284}
\eq{
  \bigl[ \ov{\mathcal E}_A, \ov{\mathcal E}_B \bigr]_{\rm C} = F_{AB}{}^C \,\ov{\mathcal E}_C \,.
}
Since the bracket for this algebra is the Courant bracket \eqref{courant_brack_620},
we can demand that the Jacobi identity \eqref{jacobi_courant_7373} is satisfied. 
In general, this will impose non-trivial restrictions on the various fluxes.
In particular, starting from
\eq{
   0 =  \mbox{Jac}\bigl( \,\ov{\mathcal E}_A,\ov{\mathcal E}_B,\ov{\mathcal E}_C\op \bigr)_{\rm C} - 
   d\, \mbox{Nij}\bigl( \,\ov{\mathcal E}_A,\ov{\mathcal E}_B,\ov{\mathcal E}_C\op \bigr)_{\rm C} \,,
}
and assuming $F_{ABC}=F_{AB}{}^D\eta_{DC}$ to be completely anti-symmetric, we obtain
\eq{
  \label{bianchi_83999190}
  0 =D_{[\ul A} F_{\ul B \ul C\ul D]} - \frac{3}{4} \, F_{[\ul A \ul B}{}^{M} F_{M|\ul C\ul D]} \,.
}
The derivative $D_A = \rho(\ov{\mathcal E}_A)$ is the anchor-projection of the generalised vielbein vector-field
$\ov{\mathcal E}_A$ (cf. section~\ref{sec_lie_courant}), in particular, for a generalised vector-field
\eq{
  \ov{\mathcal E}_A = \delta_A{}^I \op \bigl( \mathcal O^T\bigr){}_I^{\hspace{5pt} J} \op \partial_J \,,
  \hspace{40pt}\mbox{with}\hspace{20pt}
  \mathcal O = \left( \begin{array}{cc} A & B \\ C & D \end{array}\right),
}
the derivative $D_A$ is given by
\eq{
  \label{gg_bianchi_der_38756}
  D_a = \delta_a{}^i (A^T)_i{}^j \,\partial_j \,,\hspace{50pt}
  D^a = \delta^a{}_i (B^T)^{ij}\,\partial_j \,.
}

%%%%%%%%%%%%%%%%%%%%%%%%%%%%%%%%%%%%%%%%%%%%%%%
%%%%%%%%%%%%%%%%%%%%%%%%%%%%%%%%%%%%%%%%%%%%%%%

\subsubsection*{Bianchi identities}

We can now work out the explicit form of the Bianchi identities \eqref{bianchi_83999190}
for the Roytenberg algebra \eqref{royten_29990}.
Using the conventions \eqref{gen_vielbein_284ccc} for the fluxes and recalling that 
$H_{abc} = \delta_a{}^i\op \delta_b{}^j\op \delta_c{}^k\op H_{ijk}$, and so on, we have
\eq{
\label{gg_bianchi_83948016}
\arraycolsep1.15pt
\begin{array}{c@{\hspace{4pt}}lclccllccllccllccll}
0 = &2& D_{[\ul a} & \hat H_{\ul b \ul c \ul d]} &&&&&- & 3 & \hat f_{\,[\ul a \ul b}{}^m & \hat H_{m|\ul c\ul d]} 
\,,\hspace{-5pt}
\\[8pt]
0 = &3& D_{[\ul a} & \hat f_{\,\ul b \ul c]}{}^d & +&  & D^d & \hat H_{abc} &-&3& \hat f_{\,[\ul a\ul b}{}^m & 
\hat f_{\,m|\ul c]}{}^d &+&3&\hat H_{[\ul a \ul b|m} & Q_{\ul c]}{}^{md}
\,,\hspace{-5pt}
\\[8pt]
0= &2& D_{[\ul a} & Q_{\ul b]}{}^{cd} & - &2& D^{[\ul c} & \hat f_{\, ab}{}^{\ul d]}
&-& & \hat f_{\,ab}{}^m & Q_m{}^{cd} &- & 4& \hat f_{\, m[\ul a}{}^{[\ul c} & Q_{\ul b]}{}^{m|\ul d]}
&- && \hat H_{abm} & R^{mcd}\,,\hspace{-5pt}
\\[8pt]
0=& &D_a & R^{abc} &+&3 & D^{[\ul b} & Q_a{}^{\ul c\ul d]} &+&3 &Q_a{}^{m[\ul b} & Q_m{}^{\ul c\ul d]}
&+&3& \hat f_{\,a m}{}^{[\ul b} & R^{m|\ul c\ul d ]}\,,\hspace{-5pt}
\\[10pt]
0=&&&&&2&D^{[\ul a} & R^{\ul b\ul c\ul d]} &+&3& Q_m{}^{[\ul a\ul b} & R^{m|\ul c\ul d]} \,,\hspace{-5pt}
\end{array}
}
where for an easier distinction we underscored the indices which are anti-sym\-me\-trised. 
Let us make three remarks concerning these Bianchi identities:
\begin{itemize}

\item The relations \eqref{gg_bianchi_83948016} have appeared in the literature in various forms:
for constant fluxes the derivatives $D_a$ and $D^a$ vanish and one finds 
the Bianchi identities of \cite{Shelton:2005cf,Ihl:2007ah,Aldazabal:2008zza}.
Using Lie algebroids similar expressions have appeared in 
\cite{Blumenhagen:2012pc},
and in the context of double field theory the above Bianchi identities  can be found for instance 
in the review \cite{Geissbuhler:2013uka}.
The Bianchi identities can also be derived requiring a twisted differential to be nil-potent
\cite{Shelton:2005cf}, which we discuss in the next section.

\item It is also worth pointing out that 
when contracting all indices of the third relation in \eqref{gg_bianchi_83948016}, one finds that 
$\hat H_{abc}R^{abc}=0$. 
Applying this observation to a three-torus, we see that on $\mathbb T^3$ the $H$- and $R$-flux 
cannot be present simultaneously. 

\item For a discussion of Bianchi identities for geometric and non-geometric fluxes 
in the presence of NS-NS sources such as the NS5-brane, Kaluza-Klein mono\-pole or
$5^2_2$-brane see for instance \cite{Villadoro:2007tb,Andriot:2014uda}.

\end{itemize}

%%%%%%%%%%%%%%%%%%%%%%%%%%%%%%%%%%%%%%%%%%%%%%%
%%%%%%%%%%%%%%%%%%%%%%%%%%%%%%%%%%%%%%%%%%%%%%%
%%%%%%%%%%%%%%%%%%%%%%%%%%%%%%%%%%%%%%%%%%%%%%%
%%%%%%%%%%%%%%%%%%%%%%%%%%%%%%%%%%%%%%%%%%%%%%%
%%%%%%%%%%%%%%%%%%%%%%%%%%%%%%%%%%%%%%%%%%%%%%%
%%%%%%%%%%%%%%%%%%%%%%%%%%%%%%%%%%%%%%%%%%%%%%%
%%%%%%%%%%%%%%%%%%%%%%%%%%%%%%%%%%%%%%%%%%%%%%%
%%%%%%%%%%%%%%%%%%%%%%%%%%%%%%%%%%%%%%%%%%%%%%%
%%%%%%%%%%%%%%%%%%%%%%%%%%%%%%%%%%%%%%%%%%%%%%%
%%%%%%%%%%%%%%%%%%%%%%%%%%%%%%%%%%%%%%%%%%%%%%%
%%%%%%%%%%%%%%%%%%%%%%%%%%%%%%%%%%%%%%%%%%%%%%%
%%%%%%%%%%%%%%%%%%%%%%%%%%%%%%%%%%%%%%%%%%%%%%%
%%%%%%%%%%%%%%%%%%%%%%%%%%%%%%%%%%%%%%%%%%%%%%%
%%%%%%%%%%%%%%%%%%%%%%%%%%%%%%%%%%%%%%%%%%%%%%%
%%%%%%%%%%%%%%%%%%%%%%%%%%%%%%%%%%%%%%%%%%%%%%%
%%%%%%%%%%%%%%%%%%%%%%%%%%%%%%%%%%%%%%%%%%%%%%%

\clearpage
\section{Flux compactifications}
\label{cha_sugra}

In this section we discuss non-geometric backgrounds
from an effective field-theory point of view. We  
compactify type II superstring theory from ten to four dimensions on manifolds with 
$SU(3)\times SU(3)$ structure, include geometric as well as non-geometric fluxes, 
and investigate the resulting effective theory.
We are particularly interested in how fluxes modify the effective four-dimensional theory.
We also mention that reviews on (non-geometric) flux-compactifications can be found in  
\cite{Grana:2005jc,Douglas:2006es,Blumenhagen:2006ci,Wecht:2007wu}.

%%%%%%%%%%%%%%%%%%%%%%%%%%%%%%%%%%%%%%%%%%%%%%%
%%%%%%%%%%%%%%%%%%%%%%%%%%%%%%%%%%%%%%%%%%%%%%%
%%%%%%%%%%%%%%%%%%%%%%%%%%%%%%%%%%%%%%%%%%%%%%%
%%%%%%%%%%%%%%%%%%%%%%%%%%%%%%%%%%%%%%%%%%%%%%%

\subsection{\texorpdfstring{$SU(3)\times SU(3)$}{SU(3)xSU(3)} structures}
\label{sec_su3su3}

A suitable framework for discussing compactifications with non-geometric fluxes
is that of $SU(3)\times SU(3)$ structures. In this section we
give a brief review of the main concepts and ideas, but refer for more details
to the original literature
\cite{
Grana:2005ny,
Grana:2006kf,
Grana:2006hr,
Cassani:2007pq,
Grana:2008yw
}
or for instance to the lecture notes \cite{Koerber:2010bx}.

%%%%%%%%%%%%%%%%%%%%%%%%%%%%%%%%%%%%%%%%%%%%%%%
%%%%%%%%%%%%%%%%%%%%%%%%%%%%%%%%%%%%%%%%%%%%%%%

\subsubsection*{Pair of $SU(3)$ structures}

We are interested in compactifications of type II string theory from ten to four dimensions.
For the ten-dimensional space-time we make the following ansatz
\eq{
   \mathbb M^{3,1} \times \mathcal M\,,
}
where $\mathbb M^{3,1}$ is a four-dimensional  space-time with Lorentz signature and 
$\mathcal M$ is a
compact six-dimensional Euclidean space. 
If we  demand  that $\mathcal{N}=2$ supersymmetry is preserved in four dimensions,
then  
the two ten-dimensional Majorana-Weyl spinors $\epsilon^A$ which parametrise the supersymmetry transformations 
in ten dimensions
have to decompose as
\eq{
  \label{decomp_gg_938}
  \mbox{type IIA:}\hspace{40pt}
  &
  \arraycolsep2pt  
  \begin{array}{lcl}
  \epsilon^1 &=& \epsilon^1_+\otimes \eta^1_++ \epsilon^1_-\otimes \eta^1_- \,, \\[6pt]
  \epsilon^2 &=& \epsilon^2_+\otimes \eta^2_-+ \epsilon^2_-\otimes \eta^2_+ \,, 
  \end{array}
  \\[16pt]
  \mbox{type IIB:}\hspace{40pt}
  &
  \arraycolsep2pt
  \begin{array}{lcl}
  \epsilon^A &=& \epsilon^A_+\otimes \eta^A_-+ \epsilon^A_-\otimes \eta^A_+ \,, 
  \end{array}
}
with $A=1,2$. Here, $\epsilon^A_{\pm}$ are positive/negative chirality spinors in four dimensions 
 and $\eta_{\pm}^A$ are globally-defined 
and nowhere-vanishing spinors on the internal  manifold $\mathcal M$.
Due to the Majorana condition on $\epsilon^A$ in ten dimensions, $\epsilon^A_-$ and $\eta^A_-$ are the charge conjugates of
$\epsilon^A_+$ and $\eta^A_+$, respectively. 
The requirement of having two globally-defined and nowhere-vanishing spinors on $\mathcal M$ 
implies that the 
structure group of the internal manifold has to be reduced to $SU(3)$
for each spinor, and hence the internal manifold should admit a pair of $SU(3)$ structures. 
In the special case of a Calabi-Yau three-fold the two spinors are parallel everywhere, that is 
$\eta^1_+=\eta^2_+$, and we have   a single $SU(3)$ structure.

The two spinors $\eta^A_{\pm}$ shown in \eqref{decomp_gg_938}
can be used to introduce two globally-defined real two-forms 
$J^A_{ab}$ and two complex three-forms $\Omega^A_{abc}$ on the compact space.
Using the usual anti-symmetrised product of $\gamma$-matrices along the compact six dimensions,
we have
\eq{
  \label{structures_734830}
  ( J^A)_{ab} = i\, \eta_+^{A\,\dagger} \op\gamma_{ab} \,\eta_+^A \,,
  \hspace{50pt}
  ( \Omega^A)_{abc} = -i\, \eta_-^{A\,\dagger} \op\gamma_{abc} \,\eta_+^A \,.
}
These expressions provide an alternative definition of a pair of $SU(3)$ structures, 
and they illustrate the general  correspondence between 
almost complex structures $J^A$ and Weyl spinors $\eta^A_+$.

%%%%%%%%%%%%%%%%%%%%%%%%%%%%%%%%%%%%%%%%%%%%%%%
%%%%%%%%%%%%%%%%%%%%%%%%%%%%%%%%%%%%%%%%%%%%%%%

\subsubsection*{$SU(3)\times SU(3)$ structure}

It turns out that the two spinors $\eta^A_+$ along the compact directions can be described conveniently using 
generalised geometry \cite{Hitchin:2004ut,Gualtieri:2003dx,Jeschek:2004wy}.
In particular,
$(\eta^1_+,\eta^2_+) $ 
transforms as a $Spin(D,D)$ spinor of the generalised tangent-bundle
$E$ introduced in section~\ref{sec_gg_basics}.
For our case of interest the dimension is $D=6$, and the basic 
$Spin(6,6)$ spinor representations are Majorana-Weyl.
The pair of $SU(3)$ structures can then be viewed as an 
$SU(3)\times SU(3)$ structure on $E$ (provided  the compatibility condition \eqref{sg_compa_87456}
discussed below is satisfied).

It  turns out that for the generalised tangent-bundle the spinor bundle $S$ is isomorphic
to the bundle of forms $\wedge^* (T^*\mathcal M)$ \cite{Hitchin:2004ut}. More concretely, the positive-helicity spin bundle 
$S^+$ is isomorphic to poly forms of even degree and the negative-helicity spin bundle $S^-$ 
is isomorphic to odd forms
\eq{
  \label{iso_834839}
   S^+ \simeq \bigwedge_{\rm even} T^*\mathcal M \,,
   \hspace{50pt}
   S^- \simeq \bigwedge_{\rm odd} T^*\mathcal M \,.
}
Let us make that more precise and consider the following two globally-defined spinors \cite{Grana:2005sn}
\eq{
  \Phi^+_{(0)} = \eta^1_+ \otimes\ov\eta^2_+ \,, \hspace{50pt}
  \Phi^-_{(0)} =  \eta^1_+ \otimes\ov\eta^2_- \,,
}
where $\Phi^+_{(0)}\in\Gamma(S^+)$ and $\Phi^-_{(0)}\in\Gamma(S^-)$. The product of the spinors $\eta^A_{\pm}$ is defined using $n_D\times n_D$ anti-symmetrised $\gamma$-matrices as
\eq{
\arraycolsep2pt
\begin{array}{ccccccc}
\displaystyle \eta^1_+ \otimes\eta^2_+ &=& \displaystyle  \frac{1}{n_D} & \displaystyle 
\sum_{p\in2\mathbb Z} & \displaystyle \frac{1}{p!} & \displaystyle   \bigl( \op\ov\eta^2_+\op\gamma_{a_1 \ldots a_p}\op \eta^1_+ \op\bigr) & \displaystyle  \gamma^{a_p\ldots a_1} \,,
  \\
\displaystyle  \eta^1_+ \otimes\eta^2_- &=& \displaystyle  \frac{1}{n_D} &\displaystyle \sum_{p\in2\mathbb Z+1} 
& \displaystyle \frac{1}{p!} & \displaystyle 
  \bigl( \op\ov\eta^2_-\op\gamma_{a_1 \ldots a_p}\op \eta^1_+ \op\bigr) & \displaystyle  \gamma^{a_p\ldots a_1} \,.
  \end{array}
}
Using now the Clifford map we can relate these expressions to elements of $\wedge^* (T^*\mathcal M)$ in the following way
\eq{
  \slashed \omega = \sum_k \frac{1}{k!}\op \omega^{(k)}_{m_1 \ldots m_k}\op  \gamma^{m_1 \ldots m_k} 
  \hspace{12pt}\longleftrightarrow\hspace{12pt}
  \omega = \sum_k \frac{1}{k!}\op \omega^{(k)}_{m_1 \ldots m_k}\op  dx^{m_1} \wedge \ldots \wedge   dx^{m_k}\,.
}
This shows that $\Phi^+$ is an even multiform and $\Phi^-$ is an odd multiform, in agreement with 
\eqref{iso_834839}. 
The degrees of freedom of the Kalb-Ramond $B$-field  can be included in the above spinors  as follows,
\eq{
  \label{sg_pure_24}
  \Phi^+ =  e^{B} \,\Phi^+_{(0)} \,, \hspace{50pt}
  \Phi^- = e^{B} \, \Phi^-_{(0)} \,,
}
where the exponential is understood as a series expansion and where the wedge product is left implicit.

%%%%%%%%%%%%%%%%%%%%%%%%%%%%%%%%%%%%%%%%%%%%%%%
%%%%%%%%%%%%%%%%%%%%%%%%%%%%%%%%%%%%%%%%%%%%%%%

\subsubsection*{Generalised spinors}

Let us now briefly summarise some of the main formulas and concepts relevant for generalised spinors, 
which will be useful  for our subsequent discussion. 
\begin{itemize}

\item We denote a generalised spinor by $\Phi$,  a generalised vector will be written again as $X=(x,\xi)$ 
and $O(D,D)$ $\gamma$-matrices are denoted by
$\Gamma^M$. The Clifford action of $X$ on a generalised spinor is then given by \label{page_clifford}
\eq{
  \label{gs_3928765}
  X\cdot \Phi= X_M \op \Gamma^M \, \Phi = \bigl( x^m\op\Gamma_m + \xi_m \op\Gamma^m \bigr) \Phi\,.
}
Alternatively, as pointed out above, $\Phi$ can be interpreted as a multi-form (for which we use the same symbol). The corresponding 
Clifford action is then realised by $\Gamma^m = dx^m\wedge$ and $\Gamma_m= \iota_{\partial_m}$
as follows
\eq{
  X \cdot \Phi = \bigl( \iota_x + \xi \wedge\bigr) \,\Phi \,.
}
Note that $\Gamma^m$ and $\Gamma_m$ are satisfy a Clifford algebra, since the operators $dx^i\wedge$ and $\iota_{\partial_i}$ satisfy 
\eq{
  \bigl\{ dx^i \wedge, dx^j\wedge \bigr\} = 0 \,, \hspace{40pt}
  \bigl\{ \iota_{\partial_i}, \iota_{\partial_j} \bigr\} = 0 \,, \hspace{40pt}
  \bigl\{ dx^i \wedge, \iota_{\partial_j} \bigr\} = \delta^i{}_j \,.
}
These relations can be combined into the Clifford algebra $\{\Gamma^M,\Gamma^N\} =   \eta^{MN}$, where the metric $\eta_{MN}$ has been defined 
in \eqref{bilin_0444493} which we recall for convenience as
\eq{
   \eta_{IJ} =  \left( \begin{array}{cc} 0 & \delta_i{}^j \\[2pt] \delta^i{}_j & 0 \end{array}\right) .
}

\item Let us also note  that $O(D,D)$ transformations act on 
generalised spinors. For $\mathsf B$- and $\beta$-transformations one finds in particular
\eq{
  \label{gg_b_transform_83920}
  \arraycolsep2pt
  \begin{array}{l@{\hspace{8pt}}c@{\hspace{8pt}}lcll}
  \displaystyle\Phi & \displaystyle \xrightarrow{\hspace{10pt}\mbox{\scriptsize $\mathsf B$-transform}\hspace{10pt}}
  & \displaystyle  \Phi' &=& \displaystyle e^{\mathsf B \wedge} & \Phi \,,
  \\[4pt]
  \displaystyle\Phi & \displaystyle \xrightarrow{\hspace{10pt}\mbox{\scriptsize $\beta$-transform}\hspace{10pt}}
  & \displaystyle  \Phi' &=& \displaystyle e^{\beta  \op\llcorner} & \Phi \,,
  \end{array}
 }
where the exponential is understood again as an expansion
and the action of $\beta$ has been given below \eqref{gg_beta_transform_120}.
The action of an $\mathsf A$-transform is somewhat more involved, and can be found for instance in 
section 2.3 of 	\cite{Gualtieri:2003dx}.

\item The pairing for generalised spinors can be expressed using the 
 Mukai paring. The latter is defined as\,\footnote{
In the literature one can also find a convention
where instead of $ (-1)^{\left[ \frac{p+1}{2}\right]}$ one uses $ (-1)^{\left[ \frac{p}{2}\right]}$ in the Mukai pairing. 
This leads to some sign-differences in subsequent formulas.}
\eq{
  \label{gg_mukai}
  \bigl\langle \Phi^{(1)} , \Phi^{(2)} \bigr\rangle = 
  \sum_{p=0}^D (-1)^{\left[ \frac{p+1}{2}\right]} \, \Phi^{(1)}_p \wedge \Phi^{(2)}_{D-p}\,,
}
where $[a]$ denotes the integer part of some number $a$ and where $\Phi_p$ denotes the 
$p$-form part of the multiform $\Phi$. 
Note furthermore that the Mukai pairing is invariant under the action of $O(D,D)$.

\item Let us also introduce the annihilator space of a spinor as $L_{\Phi} = \{ X \in \Gamma(E): X\cdot \Phi = 0\}$,
where $E$ is the generalised tangent bundle locally expressed as $E=T\mathcal M \oplus T^*\mathcal M$
(see our discussion around equation \eqref{gg_sequ}). 
It is isotropic, 
and if $L_{\Phi}$ is of maximal dimension $D$
the corresponding spinor is called pure. Alternatively, a pure spinor is annihilated by half of the $\Gamma$-matrices. 
Note that the spinors defined in \eqref{sg_pure_24} are pure spinors,
and that any pure spinor can be represented as a wedge product of an exponentiated complex two-form with
a complex $k$-form  \cite{Gualtieri:2003dx}.
 
Pure spinors are in one-to-one correspondence with a generalised almost  complex structure on the 
generalised tangent-space.

 \item Each of the pure spinors \eqref{sg_pure_24} defines an $SU(3)$ structure on the generalised tangent-space
 $E$. If these spinors are compatible, together they form an $SU(3)\times SU(3)$ structure. 
 The requirements for compatibility are that $\mbox{dim}\op ( L_{\Phi^+}\cap L_{\Phi^-})=3$ and that 
 $\Phi^+$ and $\Phi^-$ have the same normalisation \cite{Gualtieri:2003dx}. Using the Mukai pairing, these conditions can be 
 written as \cite{Grana:2005ny,Cassani:2007pq}
 \eq{
   \label{sg_compa_87456}
   &\bigl\langle \Phi^+ , X\cdot \Phi^- \bigr\rangle = 
   \bigl\langle \ov \Phi{}^+ , X\cdot \Phi^- \bigr\rangle = 0\,,
   \hspace{40pt}\forall X\in\Gamma(E)\,,
   \\[4pt]
  &
   \bigl\langle \Phi{}^+ ,  \ov \Phi{}^+ \bigr\rangle = \bigl\langle  \Phi{}^- ,  \ov \Phi{}^- \bigr\rangle \,.
}

\end{itemize}

%%%%%%%%%%%%%%%%%%%%%%%%%%%%%%%%%%%%%%%%%%%%%%%
%%%%%%%%%%%%%%%%%%%%%%%%%%%%%%%%%%%%%%%%%%%%%%%

\subsubsection*{Example: Calabi-Yau three-fold}

Finally, let us  come back to the example of a Calabi-Yau three-fold. In this case the 
spinors are parallel to each other everywhere, that is $\eta^1_+=\eta^2_+=\eta_+$.
The pair of $SU(3)$ structures then reduces to a single $SU(3)$ structure, for
which we have from \eqref{structures_734830}
\eq{
   J_{ab} = i\, \eta_+^{\dagger} \op\gamma_{ab} \,\eta_+ \,,
  \hspace{50pt}
   \Omega_{abc} = -i\, \eta_-^{\dagger} \op\gamma_{abc} \,\eta_+ \,.
}
The corresponding spinors, written as differential forms, then read
\eq{
  \label{sg_39561}
  \Phi^+ = 
  e^{B-i\op J} \,,
  \hspace{50pt}
  \Phi^- = 
  \Omega \,,
}
which are familiar expressions from Calabi-Yau compactifications. 
One can furthermore check that the compatibility conditions \eqref{sg_compa_87456} are 
satisfied, using that $\Omega\wedge J=0$ as well as the normalisation 
$\frac{i}{8}\op\Omega\wedge\ov\Omega = \frac{1}{3!} J^3$.
We discuss the case of Calabi-Yau three-folds in more detail in section~\ref{sec_sg_cy}.

%%%%%%%%%%%%%%%%%%%%%%%%%%%%%%%%%%%%%%%%%%%%%%%
%%%%%%%%%%%%%%%%%%%%%%%%%%%%%%%%%%%%%%%%%%%%%%%
%%%%%%%%%%%%%%%%%%%%%%%%%%%%%%%%%%%%%%%%%%%%%%%
%%%%%%%%%%%%%%%%%%%%%%%%%%%%%%%%%%%%%%%%%%%%%%%

\subsection{Four-dimensional supergravity}
\label{sec_sugra_recap}

We are interested in the effective theory resulting from 
compactifications of type II string theory to four dimensions preserving some supersymmetry.
In this section we therefore review some aspects of four-dimensional supergravity theories. 
We focus on $\mathcal N=2$ or $\mathcal{N}=1$ local supersymmetry, 
and for a reviews on this topic see for instance \cite{Andrianopoli:1996cm,Trigiante:2016mnt} and for a  textbook
treatment see \cite{Freedman:2012zz}.

%%%%%%%%%%%%%%%%%%%%%%%%%%%%%%%%%%%%%%%%%%%%%%%
%%%%%%%%%%%%%%%%%%%%%%%%%%%%%%%%%%%%%%%%%%%%%%%

\subsubsection*{$\mathcal{N}=2$ supergravity in $D=4$}

Let us start with $\mathcal{N}=2$ supergravity in four dimensions. 
 The relevant supergravity multiplets
are the gravitational multiplet, vector-multiplet and hyper-multiplet, which are summarised in 
table~\ref{sg_n2_83756}.
%%%%%%%%%%%%%%
%%%%%%%%%%%%%%
\begin{table}[t]
\eq{
  \nonumber
  \arraycolsep15pt
  \renewcommand{\arraystretch}{1.5}
  \begin{array}{c||cc}
  \multicolumn{1}{c||}{\mbox{multiplet}} & \mbox{bosonic fields} & \mbox{fermionic fields}
  \\ \hline\hline
  \mbox{gravity} & \mbox{metric $g_{\mu\nu}$, gravi-photon $A^0_{\mu}$} & \mbox{$2$ gravitini}
  \\
  \mbox{vector} & \mbox{vector $A^i_{\mu}$, complex scalar $z^i$} & \mbox{$2$ gaugini}
  \\
  \mbox{hyper} & \mbox{$4$ real scalars $q^u$} & \mbox{$4$ fermions}
  \end{array}
}
\caption{
Multiplets relevant for four-dimensional $\mathcal N=2$ supergravity theories. The index 
$i=1,\ldots, n_V$ labels the vector-multiplets
and $u=1,\ldots, 4\op n_H$ labels the real scalars of the hyper-multiplets. 
The total number of vector- and hyper-multiplets is denoted by $n_V$ and $n_H$, respectively.
\label{sg_n2_83756}
}
\end{table}
%%%%%%%%%%%%%%
%%%%%%%%%%%%%%
The scalar fields of these  multiplets parametrise the 
vector- and hyper-multiplet moduli spaces, and due to supersymmetry the vector moduli space
has to be a special K\"ahler manifold and the hyper-multiplet scalars 
span a  quaternionic-K\"ahler manifold. For $\mathcal{N}=2$ supersymmetry, they form a direct product
\eq{
  \label{sg_n2_837563}
  \mathcal M^{\rm scalar} = \mathcal M_{\rm SK}^{\rm vector} \times
  \mathcal M^{\rm hyper}_{\rm quaternionic} \,.
}
Next, we turn to the supergravity action for the multiplets shown 
in table~\ref{sg_n2_83756}. The kinetic terms of the bosonic fields for the $\mathcal N=2$ 
multiplets are determined as follows:
\begin{itemize}

\item The moduli space of the vector-multiplet scalars $z^i$ is a K\"ahler manifold, and can therefore
be described using a K\"ahler potential $\mathcal K(z,\ov z)$. The corresponding K\"ahler metric
is given by
\begin{equation}
  \label{eff_genK_839290}
  \mathcal G_{i\ov j} = \partial_i\op\partial_{\ov j}\, \mathcal K
  \,,
  \hspace{70pt}\partial_i\equiv \frac{\partial}{\partial z^i}\,.
\end{equation}
The metric for the hyper-multiplet scalars will be denoted by $h_{uv}$ 
with $u,v=1,\ldots, 4\op n_H$. It describes a  quaternionic-K\"ahler manifold,
which is related to a triplet of almost complex structures $J^x$ with $x=1,2,3$ 
satisfying  a quaternionic algebra. However, 
note that a quaternionic-K\"ahler manifold is not K\"ahler. 
We refer to  \cite{Freedman:2012zz} for more details on the geometry of 
such spaces.
The K\"ahler metric \eqref{eff_genK_839290} and the quaternionic-K\"ahler metric $h_{uv}$
then determine the kinetic terms of the scalar fields  in the action.

\item For the kinetic terms of the vector-fields, we first note that the moduli space of the
vector-multiplet scalars 
 is special K\"ahler and therefore is equipped  with an additional structure. In particular, 
the corresponding K\"ahler potential $\mathcal K$ can be expressed using a holomorphic pre-potential 
$\mathcal F$.\footnote{The notion of special K\"ahler geometry is more general than discussed here, and a pre-potential 
does not need to exist. However, in our subsequent discussion a pre-potential always exists.}
Introducing projective coordinates $Z^I=(Z^0,Z^i)$ for the vector-multiplet scalars 
via $z^i = Z^i/Z^0$, the K\"ahler potential is given by
\eq{
  \mathcal K = - \log \left[\, i\op \bigl( \op\ov Z{}^I \mathcal F_I - Z^I \op \ov{\mathcal F}_I \bigr) \right]
  ,
  \hspace{50pt}
  \mathcal F_I = \frac{\partial \mathcal F}{\partial Z^I}\,,
}
with $I=0,\ldots, n_V$. Since the $Z^I$ are projective coordinates, $Z^0$ cancels out in all physically-relevant quantities. 
For convenience, one therefore often chooses $Z^0=1$. 
Furthermore, in order to preserve $\mathcal N=2$ supersymmetry the holomorphic pre-potential 
$\mathcal F(Z)$ has to be a homogeneous function of degree two. 
Using the pre-potential, one can construct a period matrix of the following form
\begin{equation}
\label{sg_938569910}
\mathcal N_{IJ}=\overline{\mathcal F}_{IJ}+2\op i \, \frac{
{\rm Im}(\mathcal F_{IM}) Z^M \, {\rm Im}(\mathcal F_{JN}) Z^N}{
           Z^M \op{\rm Im}(\mathcal F_{MN}) Z^N}  \,,
             \hspace{50pt}
  \mathcal F_{IJ} = \frac{\partial \mathcal F_I}{\partial Z^J}\,.
\end{equation}
This matrix encodes the kinetic term for the combined gravi-photon and vector-multiplet vector-fields
$A^I = (A^0,A^i)$.

\item Finally, the kinetic term of the four-dimensional metric is given by the usual Einstein-Hilbert term in the action.

\end{itemize}
The bosonic part of the (ungauged) $\mathcal N=2$ supergravity action in four dimensions can now be expressed 
using the above quantities. It takes the following general form\,\footnote{
Here we assume here that the pre-potential $\mathcal F(Z)$ is 
invariant under gauge transformations of the vector-fields $A^I$. 
If this is not the case, a Chern-Simons term has to be added to 
the action \eqref{eff_genK_839291}. See for instance \cite{Freedman:2012zz} for more details.
}
\eq{
  \label{eff_genK_839291}
  \mathcal S = \frac{1}{2\op\kappa_4^2}\int \,\Bigl[ \;
  R\star 1 
  &+  \mbox{Im}\op (\mathcal N_{IJ}) \op F^I\wedge \star F^J
  -\mbox{Re}\op (\mathcal N_{IJ}) \op F^I\wedge  F^J    
  \\
  &- 2\op  \mathcal G_{i\ov j} \, dz^{i} \wedge\star d\ov z{}^{j}
  - h_{uv} \op dq^u \wedge \star dq^v
  \hspace{45pt}  
  \Bigr] \,,
}
where $\star$ is the usual Hodge star-operator in four dimensions,
$\kappa_4^2$ is the four-dimensional gravitational coupling constant and 
$F^I$ denote the field strengths of the vector-fields $A^I$.

%%%%%%%%%%%%%%%%%%%%%%%%%%%%%%%%%%%%%%%%%%%%%%%
%%%%%%%%%%%%%%%%%%%%%%%%%%%%%%%%%%%%%%%%%%%%%%%

\subsubsection*{$\mathcal{N}=2$ gauged supergravity in $D=4$}

It is possible to deform the $\mathcal N=2$ supergravity theory  to a gauged supergravity
(for  reviews see \cite{Samtleben:2008pe,Trigiante:2016mnt}). 
More concretely, the scalar manifolds \eqref{sg_n2_837563}
described by the metrics \eqref{eff_genK_839290} and $h_{uv}$
can have isometries. These isometries  usually extend to global symmetries
of the supergravity action  \eqref{eff_genK_839291} and can be gauged.

The gauging procedure promotes the global symmetries to local ones using 
the vector-fields $A^I$. 
Let us assume that  for infinitesimal transformation parameters $\epsilon^I\ll 1$ 
the K\"ahler metrics are invariant under 
\eq{
  \delta z^i = \epsilon ^I  k_I^{i} \,,
  \hspace{70pt}
  \delta q^u =\epsilon ^I  k_I^{u} \,,
}
which implies that the vector-field components  $k_I^i$ and $k_I^u$ correspond to  Killing vectors
$k_I = k_I^i \partial_{z^i} + k_I^u \partial_{q^u}$.
These can in general be non-abelian with structure constants $f_{IJ}{}^K$ determined through
$  [k_I, k_J]_{\rm L} = f_{IJ}{}^K k_K$.
Furthermore, since the Killing vectors $k_I^i$ for the vector-multiplet scalars are holomorphic they can be expressed 
as
\eq{
  k_I^i = -i \, \mathcal G^{i\ov j} \op\partial_{\ov j} \op \mathcal P^0_I \,,
}
where $\mathcal G^{i\ov j} $ is the inverse of the K\"ahler metric \eqref{eff_genK_839290}
and 
where the real functions $\mathcal P^0_I$ are called moment maps. 
For the Killing vectors of the hyper-multiplet scalars a similar result can be obtained: 
using an appropriate covariant derivative $\nabla_u$ one finds that
\eq{
  \nabla_u \mathcal P^x_I = J^x_{uv} \op k^v_I \,,
}
where $J^x$ with $x=1,2,3$ is the triplet of complex structures mentioned above. 
There are further conditions and restrictions on the moment maps $\vec {\mathcal P}_I = \{\mathcal P^x_I\}$
which we do not discuss here, but which can be found for instance in \cite{Freedman:2012zz}.

Finally, when gauging the $\mathcal N=2$ supergravity theory one  essentially has to apply the following changes to 
the ungauged action shown in \eqref{eff_genK_839291}:
\begin{itemize}

\item One replaces the ordinary derivative for the scalar fields $z^i$ and $q^u$  by 
covariant derivatives
\eq{
  \label{sg_gauging_878929}
  dz^{i} \;\rightarrow \; Dz^{i} = dz^{i} -  k_I^{i}\op A^I  \,,
  \hspace{50pt}
  dq^{u} \;\rightarrow \; Dq^{u} = dq^{u} -  k_I^{u}\op A^I  \,.
}

\item The field strength of the gauge fields $A^I$ is replaced in the following way
\eq{
  F^I  \;\rightarrow \;  \mathbf F^I = F^I - \frac12 \op f_{MN}{}^I A^M\wedge A^N\,.
}

\item The gauging generates a scalar potential, which can be expressed using 
the K\"ahler potential $\mathcal K$, the moment maps $\vec{\mathcal P}_I$ and the K\"ahler covariant derivative
$D_i \equiv \partial_{i} + \partial_{i} \mathcal K$ as
\eq{
  \label{sg_pot_n2_38745}
  V = \quad&e^{\mathcal K} \Bigl(\op \mathcal G_{i\ov j}\op k^i_{I} \op k^{\ov j}_J + 4\op h_{uv} \op k^u_I \op k^v_J \op \Bigr) 
  \op \ov Z{}^I Z^J\\
   +\,& e^{\mathcal K} \Bigl(\op \mathcal G^{i\ov j}\op
   D_{i\vphantom{\ov{ j}}} Z^I  D_{\ov j} \ov Z{}^J - 3\op Z^I \ov Z{}^J \Bigr) \op \vec{\mathcal P}_I \cdot \vec{\mathcal P}_J \,.
}

\end{itemize}

%%%%%%%%%%%%%%%%%%%%%%%%%%%%%%%%%%%%%%%%%%%%%%%
%%%%%%%%%%%%%%%%%%%%%%%%%%%%%%%%%%%%%%%%%%%%%%%

\subsubsection*{$\mathcal{N}=1$ supergravity in $D=4$}

We now turn to the case of $\mathcal{N}=1$ supergravity in four dimensions. The 
multiplets relevant for our subsequent discussion are the gravity multiplet,  vector-multiplet and chiral multiplet, 
and their field content is summarised in table~\ref{sg_29849354}.
%%%%%%%%%%%%%%
%%%%%%%%%%%%%%
\begin{table}[t]
\eq{
  \nonumber
  \arraycolsep15pt
  \renewcommand{\arraystretch}{1.5}
  \begin{array}{c||cc}
  \multicolumn{1}{c||}{\mbox{multiplet}} & \mbox{bosonic fields} & \mbox{fermionic fields}
  \\ \hline\hline
  \mbox{gravity} & \mbox{metric $g_{\mu\nu}$} & \mbox{$1$ gravitino}
  \\
  \mbox{vector} & \mbox{vector $A^i_{\mu}$} & \mbox{$1$ gaugino}
  \\
  \mbox{chiral} & \mbox{complex scalar $\phi^{\alpha}$} & \mbox{$2$ fermions}
  \end{array}
}
\caption{Multiplets relevant for four-dimensional $\mathcal N=1$ supergravity theories. The index    
$i=1,\ldots, n_V$ labels the vector-multiplets and 
$\alpha = 1, \ldots, n_C$ labels chiral multiplets. 
The total numbers of vector- and chiral multiplets are denoted by $n_V$ and  $n_C$, respectively.
\label{sg_29849354}
}
\end{table}
%%%%%%%%%%%%%%
%%%%%%%%%%%%%%
For $\mathcal N=1$ supergravity the moduli space parametrised by the complex scalar fields $\phi^{\alpha}$ 
 of the chiral multiplets 
is a  K\"ahler manifold.

Let us now discuss the kinetic terms for the bosonic fields shown in 
table~\ref{sg_29849354}. They are characterised in the following way:
\begin{itemize}

\item Since the manifold of the complex scalar fields in the chiral multiplets is K\"ahler, 
its metric can be expressed using a K\"ahler potential $\mathcal K$ similarly as in 
\eqref{eff_genK_839290}
\eq{
  \label{eff_genK_839290b}
  \mathcal G_{\alpha \ov \beta} = \partial_{\alpha}\op\partial_{\ov \beta}\, \mathcal K
  \,,
  \hspace{70pt}\partial_{\alpha}\equiv \frac{\partial}{\partial \phi^{\alpha}}\,.
}
This K\"ahler metric in turn determines the kinetic terms of the scalar fields in the action.

\item The kinetic term of the vector-fields can be expressed in terms of 
 a holomorphic function $f_{ij}(\phi)$. The real part of this function is symmetric in its indices
 and is required to be invertible. 
 Note that this function can depend on the scalars $\phi^{\alpha}$.

\item The kinetic term for the four-dimensional metric is again the Einstein-Hilbert term. 

\end{itemize}
Next, we turn to the interactions. There are the following two sources of interaction terms in 
$\mathcal N=1$ supergravity theories:
\begin{itemize}

\item Interactions of the ungauged theory are encoded in a superpotential $W(\phi)$, which is an
arbitrary holomorphic function of the complex scalars.

\item In a  gauged theory,  isometries of the scalar manifold
have been promoted to local symmetries.
Say that the moduli-space metric stays invariant under 
$\delta \phi^{\alpha} = \epsilon^i \op k_i^{\alpha}$ for $\epsilon^i\ll 1$, then 
$k^{\alpha}_i$ are holomorphic Killing vectors.
The latter are determined again from real moment maps $\mathcal P_i$ as
$k_i^{\alpha} = -i \, \mathcal G^{\alpha\ov \beta} \op\partial_{\ov \beta} \op \mathcal P_i$,
where $\mathcal G_{\alpha\ov\beta}$ is the K\"ahler metric \eqref{eff_genK_839290b}. 
Integrating this relation determines the moment maps as
\eq{
  \mathcal P_i = i\op \bigl( k_i^{\alpha} \partial_{\alpha} \mathcal K - \xi_i \bigr) \,,
}
where the constants $\xi_i$ are called Fayet-Iliopoulos parameters. For an $U(1)$ symmetry 
they can take arbitrary real values, whereas for non-abelian symmetries they are required to vanish
\cite{Freedman:2012zz}. 
We furthermore note that gauge invariance of the superpotential leads to the requirement
that
$k^{\alpha}_i D_{\alpha} W + i \op \mathcal P_i \op W = 0$, where 
$D_{\alpha} \equiv \partial_{{\alpha}} + \partial_{{\alpha}} \mathcal K$.

\end{itemize}
We can now write down the bosonic part of the $\mathcal N=1$ supergravity action in four dimensions. 
It is given by
\eq{
  \mathcal S = \frac{1}{2\op\kappa_4^2}\int \,\Bigl[ \;
  R\star 1 
  &-  \mbox{Re}\op (f_{ij}) \op F^i\wedge \star F^j
  +\mbox{Im}\op (f_{ij}) \op F^I\wedge  F^J    
  \\[-4pt]
  &- 2\,  \mathcal G_{\alpha\ov \beta} \, d\phi^{\alpha} \wedge\star d\ov \phi{}^{\ov\beta}
  \\[2pt]
  & - \bigl( V_F + V_D \bigr)\star 1
  \hspace{120pt}  
  \Bigr] \,,
}
where the scalar potential can be split into an F- and D-term 
contribution.
The F-term scalar potential is expressed in terms of the superpotential and K\"ahler potential in the following way
\eq{
  \label{f-term_pot}
  V_F= e^{\mathcal{K}} \,\biggl(\, \mathcal G^{\alpha\ov \beta} D_{\alpha} W\,
  D_{\ov \beta}\ov W 
  - 3 \bigl|W \bigr|^2 \,\biggr) \,,
}
where $\mathcal G^{\alpha\ov \beta}$ is the inverse of the K\"ahler metric of the scalar fields shown in \eqref{eff_genK_839290b}. 
The K\"ahler-covariant derivative $D_{\alpha} W$ is given as above by
\eq{
  \label{cov_deriv}
  D_{\alpha} W = 
  \partial_{\alpha} W +  \bigl( \partial_{\alpha}\mathcal K \bigr)\op W \,,
}
and the D-term potential is expressed using the gauge kinetic function $f_{ij}$ and the moment maps $\mathcal P_i$ 
as follows
\eq{
  V_D = \frac12 \left[(\mbox{Re}\op f)^{-1}\right]^{ij}\op \mathcal P_i\op \mathcal P_j \,.
}

%%%%%%%%%%%%%%%%%%%%%%%%%%%%%%%%%%%%%%%%%%%%%%%
%%%%%%%%%%%%%%%%%%%%%%%%%%%%%%%%%%%%%%%%%%%%%%%

\subsubsection*{Summary}

To summarise the discussion in this section, we recall that
four-dimensional $\mathcal N=2$ and $\mathcal N=1$ supergravity theories
are characterised by only a few quantities:
\begin{itemize}

\item For $\mathcal N=2$ supergravity, a holomorphic pre-potential $\mathcal F$ (if it exists) 
describes the vector-multiplet sector,  and a quaternionic-K\"ahler metric $h_{uv}$ describes the 
hyper-multiplets. In the gauged theory one additionally specifies local isometries by their 
Killing vectors, which in turn determine moment maps $\mathcal P^0_I$ and $\vec{ \mathcal P}_I$.

\item For the $\mathcal N=1$ theory, a K\"ahler potential $\mathcal K$ encodes the dynamics of  the chiral multiplets, 
a holomorphic gauge-kinetic function $f_{ij}$ describes the vector-multiplets, and 
a holomorphic superpotential $W$ gives rise to an F-term potential.
If the theory is  gauged, Killing vectors specify the local symmetries which 
determine moment maps $\mathcal P_i$. The latter generate a D-term potential.

\end{itemize}
In the following sections we determine these quantities for Calabi-Yau compactifications
with geometric and non-geometric fluxes.

%%%%%%%%%%%%%%%%%%%%%%%%%%%%%%%%%%%%%%%%%%%%%%%
%%%%%%%%%%%%%%%%%%%%%%%%%%%%%%%%%%%%%%%%%%%%%%%
%%%%%%%%%%%%%%%%%%%%%%%%%%%%%%%%%%%%%%%%%%%%%%%
%%%%%%%%%%%%%%%%%%%%%%%%%%%%%%%%%%%%%%%%%%%%%%%

\subsection{Calabi-Yau manifolds}
\label{sec_sg_cy}

As a starting point for compactifications of  type II string theory 
from ten to four dimensions, we consider  Calabi-Yau three-folds
and therefore want to briefly establish our notation for the latter. 
A Calabi-Yau $n$-fold is a compact K\"ahler manifold of complex dimension $n$ with vanishing first Chern class.
It comes with a holomorphic $n$-form $\Omega$ and a K\"ahler form $J$,
which satisfy 
the following relations
\eq{
  d\Omega = 0 \,, \hspace{40pt}
  dJ = 0 \,, \hspace{40pt}
  \Omega\wedge J = 0  \,.
}
We furthermore include a common normalisation condition for Calabi-Yau three-folds, 
which takes the form 
\eq{
  \frac{i}{8}\,\Omega\wedge\ov\Omega = \frac{1}{3!} \op J^3 \,.
}  
The cohomology of a Calabi-Yau three-fold can be characterised in terms of the 
Hodge numbers $h^{p,q}= \mbox{dim} \op H^{p,q}(\mathcal M)$, which are the dimensions 
of the corresponding Dolbeault cohomology groups.
The only non-vanishing Hodge numbers are 
\eq{
  \arraycolsep2pt
  \begin{array}{lclcl@{\hspace{60pt}}lcl}
  h^{0,0} &= & h^{3,3} &=& 1\,, 
  &
  h^{1,1} &=& h^{2,2}\,,
  \\[4pt]
  h^{3,0} &=& h^{0,3} & =& 1\,, 
  &
 h^{2,1} &= & h^{1,2}\,. 
 \end{array}
}

%%%%%%%%%%%%%%%%%%%%%%%%%%%%%%%%%%%%%%%%%%%%%%%
%%%%%%%%%%%%%%%%%%%%%%%%%%%%%%%%%%%%%%%%%%%%%%%

\subsubsection*{Odd cohomology}

Let us first discuss the third cohomology group $H^3(\mathcal M)$ of a Calabi-Yau three-fold.
We  note that its dimension is $2 \op h^{2,1}+2$, and a symplectic basis for this group will be denoted by
\eq{
  \label{basis_001}
  \{\alpha_{I},\beta^{I}\} \in H^3(\mathcal M) \,, \hspace{60pt}
  I=0,\ldots, h^{2,1} \,.
}
This basis can be chosen such that the only non-vanishing pairings satisfy
\eq{
  \label{symp_01}
  \int_{\mathcal M} \alpha_{I}\wedge \beta^{J} = \delta_{I}{}^{J} \,.
}
The holomorphic three-form $\Omega$ of the Calabi-Yau three-fold can then be expanded in the basis \eqref{basis_001} as
\cite{Candelas:1990pi}
\eq{
\label{hol_three}
\Omega=Z^I\, \alpha_I - \mathcal F_I\, \beta^I\,,
} 
where the periods $Z_I$ and $\mathcal F_I$ are functions of the complex-structure moduli 
$z^i$ with $i=1,\ldots , h^{2,1}$. Furthermore, the $\mathcal F_{I}$ can be 
determined from a holomorphic pre-potential $\mathcal F$ as 
$\mathcal F_{I} = \mathcal \partial \mathcal F/\partial Z^{I}$.
Using the corresponding period matrix $\mathcal N_{IJ}$ defined in 
equation \eqref{sg_938569910},
one  furthermore finds
\eq{
\label{hodgeperiod}
\int_{\mathcal M}  \alpha_{I} \wedge \star \op \alpha_{J}
   &=- \bigl({\rm Im}\,{\cal N}\op\bigr)_{IJ}
   -\left[ \bigl({\rm Re}\,{\cal N}\op\bigr)\bigl({\rm Im}\,{\cal N}\op\bigr)^{-1} 
   \bigl({\rm Re}\,{\cal N}\op\bigr)\right]_{IJ} \,, \\[8pt]
\int_{\mathcal M}   \alpha_{I} \wedge \star\op \beta^{J}
   &=- \left[ \bigl({\rm Re}\,{\cal N}\op\bigr)\bigl({\rm Im}\,{\cal N}\op\bigr)^{-1}\right]
   _{I}^{\hspace{6pt}J}\,, \\[8pt]
\int_{\mathcal M}  \beta^{I} \wedge\star\op  \beta^{J}
  &=- \left[ \bigl({\rm Im}\,{\cal N}\op\bigr)^{-1} \right]^{IJ}\,,
}
with $\star$ denoting the six-dimensional Hodge star-operator on $\mathcal M$.
For later purpose let us also define a $(2h^{2,1}+2)\times (2 h^{2,1}+2)$ matrix as
\eq{
\label{res_016}
{\cal M}^-
&=\left(\begin{matrix} 
   \mathds 1 & {\rm Re}\,{\cal N} \\
   0 & \mathds 1 \end{matrix}\right)
\left(\begin{matrix} -{\rm Im}\,{\cal N} & 0 \\
 0 &   -{\rm Im}\,{\cal N}^{-1} \end{matrix}\right)
\left(\begin{matrix} \mathds 1 & 0 \\
{\rm Re}\,{\cal N} & \mathds 1 \end{matrix}\right) \\[4pt]
&=\int_{\mathcal M} \left( \begin{matrix} 
\alpha_{\Lambda} \wedge\star\op\alpha_{\Sigma} & 
\alpha_{\Lambda} \wedge\star\op\beta^{\Sigma} \\[4pt] 
\beta^{\Lambda} \wedge\star\op\alpha_{\Sigma} & 
\beta^{\Lambda} \wedge\star\op\beta^{\Sigma} 
\end{matrix}
\right).
}

%%%%%%%%%%%%%%%%%%%%%%%%%%%%%%%%%%%%%%%%%%%%%%%
%%%%%%%%%%%%%%%%%%%%%%%%%%%%%%%%%%%%%%%%%%%%%%%

\subsubsection*{Even cohomology}

Turning now to the even cohomology, for the $(1,1)$- and $(2,2)$-part of a Calabi-Yau three-fold $\mathcal M$  we introduce bases of 
the form
\eq{
  \label{res_033a}
  \arraycolsep2pt
  \begin{array}{ccl}
  \{ \omega_{\lab A} \}  &\in&  H^{1,1}(\mathcal M) \,, \\[5pt]
  \{ \sigma^{\lab A} \} & \in & H^{2,2}(\mathcal M) \,,
  \end{array}
  \hspace{50pt}\lab A = 1,\ldots, h^{1,1} \,.
}
We can group these two- and four-forms together with the zero- and six-form on $\mathcal M$ in the following way
\eq{
  \label{res_033}
  \arraycolsep2pt
  \begin{array}{ccl}
  \{ \omega_{ A} \}  &=&  \displaystyle \bigl\{1,\, \omega_{\lab A} \bigr\}\,, \\[6pt]
  \{ \sigma^{ A} \} &=& \bigl\{  \tfrac{\sqrt{g}}{{\cal V}} \op dx^6,\,\sigma^{\lab A}\bigr\} \,,
  \end{array}
  \hspace{50pt}A = 0,\ldots, h^{1,1} \,,
}
where $\mathcal V = \frac{1}{6} \int_{\mathcal M} J^3$ is the volume of 
 $\mathcal M$. 
These two bases can be chosen such that
\eq{
  \label{symp_01a}
  \int_{\mathcal M} \omega_{A}\wedge \sigma^{ B} = \delta_{ A}{}^{ B} \,,
}
and the triple intersection numbers corresponding to the $\omega_{\mathsf A}$ in \eqref{res_033a} are given by
\eq{
  \label{res_025}
  \kappa_{\lab{ABC}} = \int_{\mathcal M} \omega_{\lab A}\wedge \omega_{\lab B}
  \wedge \omega_{\lab C} \,.
}
The K\"ahler form $J$
and the Kalb-Ramond field $B$ can be expanded in the basis $\{\omega_{\lab A}\}$ in the following way
\eq{
\label{dim_red_79}
J=t^{\lab A} \op \omega_{\lab A} \,, \hspace{70pt}
B = b^{\lab A}\op\omega_{\lab A}\,,
}
which are then combined into a so-called complexified K\"ahler form $\mathcal J$ as
\eq{
  \label{sg_214343535}
  \mathcal J = B - i J = \bigl(b^{\lab A} - i \op t^{\lab A} \bigr)\op \omega_{\lab A}
  = \mathcal J^{\lab A} \omega_{\lab A}\,.
}
We also note that similarly to the complex-structure moduli space, also for the K\"ahler moduli
of a Calabi-Yau three-fold
one finds a special K\"ahler structure \cite{Candelas:1990pi}. The corresponding pre-potential is given by
\eq{
  \mathcal F = \frac{1}{6} \op \frac{\kappa_{\mathsf{ABC}} \op \mathfrak J^{\mathsf A} \mathfrak J^{\mathsf B}
  \mathfrak J^{\mathsf C}}
  {\mathfrak J^0} \,,
}
where the triple intersection numbers have been defined in equation \eqref{res_025} and where
we introduced projective coordinates $\mathfrak J^A$ through $\mathcal J^{\mathsf A}= \mathfrak J^A/\mathfrak J^0$.
Using this pre-potential we can determine a matrix similar to \eqref{res_016}. 
We will discuss this point in the following paragraph.

%%%%%%%%%%%%%%%%%%%%%%%%%%%%%%%%%%%%%%%%%%%%%%%
%%%%%%%%%%%%%%%%%%%%%%%%%%%%%%%%%%%%%%%%%%%%%%%

\subsubsection*{$B$-twisted Hodge star-operator}

Let us define a so-called $B$-twisted Hodge star-operator, which is needed in order to 
describe the dynamics of $B$-twisted pure spinors of the form \eqref{sg_pure_24}. 
Following \cite{Jeschek:2004wy,Benmachiche:2006df,Cassani:2007pq}, we write
\eq{
  \star_B= e^{-B}  \wedge \star\,\lambda \, e^{+B} \,,
}
where the projection operator $\lambda$ acts on $2n$-forms  as  
$\lambda(\rho^{(2n)}) = (-1)^n \rho^{(2n)}$
and on $(2n-1)$-forms as
$\lambda(\rho^{(2n-1)}) = (-1)^n \rho^{(2n-1)}$.
For the Mukai pairings of the basis elements we then find for instance that
\eq{
  \arraycolsep1pt
  \begin{array}{clclcccccl}
  \bigl\langle & \alpha_{I}  &, \star_B & \alpha_{J} &\bigr\rangle 
  &=&&\bigl(\alpha_{I}\wedge e^{+B} \bigr) &\wedge \op\star & \bigl( \alpha_{J} \wedge e^{+B} \bigr)  
  \,,
 \\[6pt]
  \bigl\langle & \omega_A &, \star_B & \omega_B &\bigr\rangle &
  =&-&\bigl(\omega_A \wedge e^{+B} \bigr) &\wedge\op \star& \bigl( \omega_B \wedge e^{+B} \bigr) \,,
  \end{array}
}
and similarly for the other combinations. We can then 
modify the matrix \eqref{res_016} in the following way
\eq{
\label{res_0162}
{\cal M}^-
&=+\int_{\mathcal M} \left( \begin{matrix} 
\langle \alpha_{\Lambda} ,\star_B\op\alpha_{\Sigma} \rangle& 
\langle \alpha_{\Lambda} ,\star_B\op\beta^{\Sigma}\rangle \\[4pt] 
\langle \beta^{\Lambda} ,\star_B\op\alpha_{\Sigma} \rangle& 
\langle \beta^{\Lambda} ,\star_B\op\beta^{\Sigma} \rangle
\end{matrix}
\right).
}
The analogue of $\mathcal M^-$  in \eqref{res_0162} for the even cohomology takes a similar form. In particular,
we have
\eq{
\label{res_019}
{\cal M}^+
= -\int_{\mathcal M} \left( \begin{matrix} 
\langle \omega_A ,\star_B\op\omega_B\rangle & 
\langle \omega_A ,\star_B\op\sigma^B \rangle\\[4pt] 
\langle \sigma^A ,\star_B\op\omega_B\rangle & 
\langle \sigma^A ,\star_B\op\sigma^B \rangle 
\end{matrix}
\right).
}
Note that both ${\cal M}^+$ and ${\cal M}^-$ are positive-definite matrices.

%%%%%%%%%%%%%%%%%%%%%%%%%%%%%%%%%%%%%%%%%%%%%%%
%%%%%%%%%%%%%%%%%%%%%%%%%%%%%%%%%%%%%%%%%%%%%%%
%%%%%%%%%%%%%%%%%%%%%%%%%%%%%%%%%%%%%%%%%%%%%%%
%%%%%%%%%%%%%%%%%%%%%%%%%%%%%%%%%%%%%%%%%%%%%%%

\subsection{Calabi-Yau compactifications}
\label{sec_cy_flux}

In this section we  briefly summarise  the main features of 
type II string-theory compactifications on Calabi-Yau three-folds.
This will serve as a starting point for compactifications 
with fluxes
\cite{
Ferrara:1988ff,
Bodner:1989cg,
Ferrara:1989ik,
Bodner:1990zm,
Michelson:1996pn,
Bohm:1999uk,
DallAgata:2001brr,
Louis:2002ny
}, 
which we consider in section~\ref{sec_cy_flux_echt}. 
For a textbook treatment of Calabi-Yau compactifications 
see for instance \cite{Blumenhagen:2013fgp}, and in this section we focus on type IIB string theory
for simplicity. Similar results are obtained for type IIA.

%%%%%%%%%%%%%%%%%%%%%%%%%%%%%%%%%%%%%%%%%%%%%%%
%%%%%%%%%%%%%%%%%%%%%%%%%%%%%%%%%%%%%%%%%%%%%%%

\subsubsection*{Type II string theory in ten dimensions}

The massless field content of type IIB string theory is described by type IIB supergravity 
in ten dimensions, whose bosonic field content in the Neveu-Schwarz--Neveu-Schwarz (NS-NS) and 
Ramond--Ramond (R-R) sector is given by (see also our discussion in section~\ref{sec_buscher_II})
\eq{
  \arraycolsep4pt
  \begin{array}{lcl@{\hspace{50pt}}lcl}
  g & \ldots& \mbox{metric} \,,
  \\[6pt]
  B & \ldots& \mbox{Kalb-Ramond field} \,,
  &
  C_p & \ldots & \mbox{R-R $p$-form potential} \,,
  \\[6pt]
  \phi & \ldots& \mbox{dilaton} \,.
  \end{array}
}
In type IIB string theory  R-R $p$-form potentials of degree $p=0,2,4$ are present,
however, for our purposes the so-called 
democratic formulation \cite{Bergshoeff:2001pv} turns out to be more useful in which 
$p$-form potentials of degree $p=0,2,4,6,8$ are considered. 
The corresponding field strengths are defined as
\eq{
  \label{sg_fs_82375693846}
  \widetilde F_p = d\,C_{p-1}+H\wedge C_{p-3} \,,
}
where $H=dB$ is the field strength of the Kalb-Ramond field. 
In order to obtain the field content of type IIB supergravity, the following duality 
relations are imposed on the equations of motion
\eq{
  \label{duality}
  \widetilde F_p = (-1)^{\frac{p+3}{2}}\star 
  \widetilde F_{10-p} \,.
} 
For later convenience, we also note that the field strengths \eqref{sg_fs_82375693846} can be encoded in  a multiform $\widetilde F$
using a multiform potential $\mathcal C$ in the following way
\eq{
  \label{sg_fs_8346}
  \widetilde F = \bigl( d + H \wedge \bigr)\, \mathcal C\,,
  \hspace{60pt}
  \mathcal C = \sum_{p=0,2,4,6,8} C_p \,.
}
With $(2\kappa_{10}^2)^{-1}\!=2\pi\,\ell_{\rm s}^{-8}$ and $\ell_{\rm s}$ the string length, 
the  bosonic part of the democratic type IIB (pseudo-)action reads \cite{Bergshoeff:2001pv}
\eq{
  \label{action_iib_pre}
  \mathcal{S}_{\rm IIB} = \frac{1}{2\kappa_{10}^2} 
  \int \biggl[\, e^{-2\phi}\biggl( R\star1 
  + 4\,d\phi\wedge\star d\phi 
  &-\frac{1}{2}\, H\wedge\star H\biggr)  \\
  &
  -\frac{1}{4} \sum_{p=1,3,5,7,9} \widetilde F_p\wedge \star 
  \widetilde F_p \,\biggr] \;.
}

%%%%%%%%%%%%%%%%%%%%%%%%%%%%%%%%%%%%%%%%%%%%%%%
%%%%%%%%%%%%%%%%%%%%%%%%%%%%%%%%%%%%%%%%%%%%%%%

\subsubsection*{Compactification}

We now compactify the ten-dimensional type IIB theory on a Calabi-Yau three-fold and
determine the effective theory of the massless modes in four dimensions. 
This is done by expanding the ten-dimensional fields in the cohomology of 
the Calabi-Yau manifold introduced above and integrating over the compact space.

The massless degrees of freedom of the
internal metric are encoded in the K\"ahler form $J$ and the holomorphic 
three-form $\Omega$. 
These are the K\"ahler moduli $t^{\lab A}$ and the complex structure moduli $z^i$. 
Furthermore, the Kalb-Ramond two-form can be expanded as in \eqref{dim_red_79} for the internal
part, and we note that a massless two-form in four dimensions is dual to a scalar. 
The dilaton provides an additional  scalar degree of freedom. 
Thus, the massless field-content in four dimensions originating from the type IIB NS-NS sector 
is summarised as
\eq{
  \mbox{NS-NS sector:}\hspace{70pt}
  \begin{array}{l@{\hspace{20pt}}c@{\hspace{20pt}}lll}
  g_{MN} & \rightarrow& g_{\mu\nu}\,,&t^{\lab A}\,,& z^i \,,
  \\[6pt]
  B_{MN} & \rightarrow& B_{\mu\nu}\,,& b^{\lab A} \,,
  \\[6pt]
  \phi &\rightarrow& \phi \,,
  \end{array}
}
where $\mu,\nu=0,\ldots,3$ label the four-dimensional non-compact directions. 
For the Ramond-Ramond sector, the physical degrees of freedom are the ten-dimensional zero-, two-, and self-dual 
four-form. Expanding again in the cohomology basis of the Calabi-Yau three-fold, the massless
degrees in four-dimensions are determined as
\eq{
  \mbox{R-R sector:}\hspace{70pt}
  \begin{array}{l@{\hspace{20pt}}c@{\hspace{20pt}}lll}
  C_0 & \rightarrow& C_0 \,,
  \\[6pt]
  C_2 & \rightarrow& (C_2)_{\mu\nu}\,,& (C_2)^{\lab A} \,,
  \\[6pt]
  C_4 &\rightarrow& (C_4)_{\mu\nu}^{\lab A} \,, & (C_4)_{\mu}^I \,,
  \end{array}
}
where a two-form in four dimensions is again dual to a scalar. 
Due to $C_4$ being self-dual, we only keep half of the degrees of freedom in $C_4$. 
The massless field content can now be combined into massless $\mathcal{N}=2$ supergravity multiplets,  which is summarised in table~\ref{table_n2_fields}.

%%%%%%%%%%%%%%
%%%%%%%%%%%%%%
\begin{table}[t]
\eq{
  \nonumber
  \arraycolsep15pt
  \renewcommand{\arraystretch}{1.5}
  \begin{array}{c||c||c}
  \mbox{multiplet} & \mbox{multiplicity} & \mbox{bosonic field content}
  \\
  \hline\hline
  \mbox{gravity} & 1 & g_{\mu\nu}\,,\:  (C_4)_{\mu}^0 
  \\
  \mbox{vector} & h^{2,1} & (C_4)_{\mu}^i\,,\: z^i
  \\
  \mbox{hyper} & 1 & \phi\,,\: C_0 \,,\: B_{\mu\nu} \,,\: (C_2)_{\mu\nu}
  \\
  \mbox{hyper} & h^{1,1} & (C_4)_{\mu\nu}^{\lab A} \,,\: (C_2)^{\lab A}\,, \: t^{\lab A}\,,\: B^{\lab A}
  \end{array}
}
\caption{Massless bosonic field content of type IIB string theory compactified on a Calabi-Yau three-fold. We have 
indicated how these fields are combined into massless multiplets of $\mathcal N=2$ supergravity in four dimensions. 
\label{table_n2_fields}
}
\end{table}
%%%%%%%%%%%%%%
%%%%%%%%%%%%%%

%%%%%%%%%%%%%%%%%%%%%%%%%%%%%%%%%%%%%%%%%%%%%%%
%%%%%%%%%%%%%%%%%%%%%%%%%%%%%%%%%%%%%%%%%%%%%%%

\subsubsection*{Supergravity data and  generalised spinors}

The four-dimensional  theory preserves $\mathcal N=2$ supersymmetry. 
As we have seen in section~\ref{sec_sugra_recap}, the 
corresponding
supergravity data is encoded in a K\"ahler potential for the vector-multiplet scalar fields, a pre-potential $\mathcal F$ 
for the vector-multiplet vectors, 
a quaternionic-K\"ahler metric for the hyper-multiplet scalars,
and Killing pre-potentials $\mathcal P^x$ describing possible gaugings. 
Let us now express (some of) these quantities using the pure spinors \eqref{sg_39561}.
\begin{itemize}

\item The scalar fields of the vector-multiplets span a special-K\"ahler manifold. 
The geometry of this manifold is therefore described by a K\"ahler potential $\mathcal K^-$, 
which can be expressed using the pure spinor $\Phi^-$ given in   \eqref{sg_39561}
as \cite{Grana:2005ny,Coimbra:2011nw} 
\eq{
  \label{sg_kaeher_01}
  \Phi^- =  \Omega \,,
  \hspace{40pt}
  e^{-\mathcal K^-} = -i \int_{\mathcal M} \bigl\langle \Phi{}^-, \ov\Phi{}^-\bigr\rangle
   = - i \int_{\mathcal M} \Omega\wedge\ov\Omega \,,
}
where $\langle \cdot,\cdot \rangle$ denotes the Mukai pairing \eqref{gg_mukai}.
From this K\"ahler potential the corresponding K\"ahler metric for the 
complex-structure moduli can be determined, similarly as in \eqref{eff_genK_839290}.

\item For  the vector-fields the kinetic terms are given by the period matrix 
\eqref{sg_938569910}, which is determined from $Z^{I}$ and 
$\mathcal F_{I}$ appearing in \eqref{hol_three}.

\item For the hyper-multiplets we first recall that the scalar fields parametrise a 
quaternionic-K\"ahler manifold. However, this manifold contains a
special K\"ahler manifold as a sub-manifold, which is parametrised by the 
NS-NS sector scalars. 
Using the pure spinor $\Phi^+$ given in   \eqref{sg_39561},
the corresponding K\"ahler potential 
can be written
as \cite{Grana:2005ny}
\eq{
  \label{sg_kaeher_02}
  \Phi^+ = e^{B-i\op J} \,,
  \hspace{40pt}
  e^{-\mathcal K^+} = + i \int_{\mathcal M} \bigl\langle \Phi{}^+, \ov\Phi{}^+\bigr\rangle
  = \frac{4}{3} \int_{\mathcal M} J^3 \,,
}
where the pairing between the pure spinors is again the Mukai pairing.

\end{itemize}
The scalar potential of the $\mathcal N=2$ theory can be determined from the 
moment maps $\mathcal P^x = \ov Z {}^I \mathcal P_I^x$.\footnote{
Gaugings of the vector-multiplet scalars $z^i$ are not important for our discussion, and hence 
the Killing vectors $k^i_I$ appearing in the potential \eqref{sg_pot_n2_38745} are set to zero.}
For type IIB compactifications they can be written using the pure spinors 
and the Mukai pairing as 
\cite{Grana:2005ny}
\begin{subequations}
  \label{sg_coupl_pp}
\begin{align}
  \label{sg_coupl_ppa}
  \mathcal P^1- i \op\mathcal P^2 &\sim
  \int_{\mathcal M}
  \bigl\langle \Phi^-, d\Phi^+\bigr\rangle \,,
  \\[6pt]
  \label{sg_coupl_ppb}
  \mathcal P^3 &\sim
  \int_{\mathcal M}
  \bigl\langle \Phi^-, \widetilde F\bigr\rangle \,,
\end{align}
\end{subequations}
where $\widetilde F$ is the combined field strength given in \eqref{sg_fs_8346}. 
The precise normalisation of the Killing pre-potentials is not important here but can be found for instance in 
\cite{Grana:2006hr}.
However, 
for Calabi-Yau compactifications without fluxes it follows that $B$ in $\Phi^+$ is closed 
and hence $d\Phi^+=0$, and that  the R-R fluxes $\widetilde F$ are set to zero.
The moment maps \eqref{sg_coupl_pp} therefore vanish, and the four-dimensional 
theory is an ungauged $\mathcal N=2$ supergravity.

%%%%%%%%%%%%%%%%%%%%%%%%%%%%%%%%%%%%%%%%%%%%%%%
%%%%%%%%%%%%%%%%%%%%%%%%%%%%%%%%%%%%%%%%%%%%%%%
%%%%%%%%%%%%%%%%%%%%%%%%%%%%%%%%%%%%%%%%%%%%%%%
%%%%%%%%%%%%%%%%%%%%%%%%%%%%%%%%%%%%%%%%%%%%%%%

\subsection{Calabi-Yau compactifications with fluxes}
\label{sec_cy_flux_echt}

In this section we discuss Calabi-Yau compactifications with 
geometric as well as non-geometric fluxes
\cite{Michelson:1996pn,
Bohm:1999uk,
DallAgata:2001brr,
Louis:2002ny
}. In the NS-NS sector, 
we generate these fluxes by performing non-trivial 
$O(D,D)$ transformations of the background, similarly as in section~\ref{sec_gg_fluxes}.
In particular, we consider $O(D,D)$ transformations of the generalised 
spinors which in general will lead to non-vanishing moment maps \eqref{sg_coupl_pp}.
The resulting theory is then a gauged $\mathcal N=2$ supergravity theory in four 
dimensions. 
In this section we  illustrate the role played by non-geometric 
fluxes in the effective theory, and refer  to the original literature
\cite{
Grana:2005ny,
Grana:2006hr,
Cassani:2007pq
}
for a more detailed analysis.

%%%%%%%%%%%%%%%%%%%%%%%%%%%%%%%%%%%%%%%%%%%%%%%
%%%%%%%%%%%%%%%%%%%%%%%%%%%%%%%%%%%%%%%%%%%%%%%

\subsubsection*{NS-NS sector -- examples}

We implement an $O(D,D)$ transformation of the background 
via $O(D,D)$ transformations of the pure spinors $\Phi^+$ and $\Phi^-$. 
As we have mentioned before, $O(D,D)$ transformations leave the Mukai pairing 
\eqref{gg_mukai} invariant, and hence 
the K\"ahler potentials \eqref{sg_kaeher_01} and 
\eqref{sg_kaeher_02} stay invariant. 
On the other hand, the interactions described via the Killing pre-potentials \eqref{sg_coupl_pp}
will be modified. 
Let us discuss this point  for two examples:
\begin{itemize}

\item First, we consider a non-trivial $\mathsf B$-transform of the pure spinors. 
According to \eqref{gg_b_transform_83920} we take
$\Phi^{\pm} \to e^{\mathsf B} \op\Phi^{\pm}$ with $\mathsf B$ a two-form with field strength
$H=d\op\mathsf B$.  Using the invariance of the Mukai pairing, for  \eqref{sg_coupl_ppa} 
this implies
\eq{
  \label{sg_def_398756a}
  \bigl\langle \Phi^-, d\Phi^+\bigr\rangle \rightarrow
  \hspace{20pt} &
  \bigl\langle e^{+\mathsf B}\op\Phi^-, d\bigl( e^{+\mathsf B}\op\Phi^+\bigr)\bigr\rangle 
  \\[4pt]
  =\,&\bigl\langle \Phi^-, e^{-\mathsf B}\op d\bigl( e^{+\mathsf B}\op\Phi^+\bigr)\bigr\rangle 
  \\[4pt]
  =\,&\bigl\langle \Phi^-, \bigl( d + H \bigr)\op\Phi^+\bigr\rangle  \,,
}
where the exponentials are again understood as a series expansion and wedge products between differential
forms are left implicit. 
Using that $\Phi^+$ is closed, we therefore see that a non-trivial $\mathsf B$-transform generates an interaction term 
of the form
\eq{
   \mathcal P^1- i \op\mathcal P^2 \sim
  \int_{\mathcal M} \Omega\wedge H \,.
}

\item As a second example we consider a $\beta$-transform  of the pure spinors. 
Following \eqref{gg_b_transform_83920} we take 
$\Phi^{\pm} \to e^{\beta} \op\Phi^{\pm}$, where the contraction of the 
bivector-field is again left implicit (cf. below equation \eqref{gg_beta_transform_120}).
We then find that 
\eq{
  \label{sg_def_398756b}
  \bigl\langle \Phi^-, d\Phi^+\bigr\rangle \rightarrow
  \hspace{20pt} &
  \bigl\langle e^{+\beta}\op\Phi^-, d\bigl( e^{+\beta}\op\Phi^+\bigr)\bigr\rangle 
  \\[4pt]
  =\,&\bigl\langle \Phi^-, e^{-\beta}\op d\bigl( e^{+\beta}\op\Phi^+\bigr)\bigr\rangle 
  \\[4pt]
  =\,&\bigl\langle \Phi^-, \bigl( \op D + Q - R\op \bigr)\op\Phi^+\bigr\rangle  \,,
}
where we use the following notation 
\eq{
  \arraycolsep2pt
  \begin{array}{lcl@{\hspace{50pt}}lcl}
  D &=& dx^m \op\delta_m{}^n \op\partial_n - \iota_m \hspace{1pt} \beta^{mn} \op \partial_n\,,
  \\[6pt]
  Q &=& \tfrac{1}{2} \op Q_i{}^{jk} \op dx^i \wedge \iota_j\wedge \iota_k\,,
  &Q_i{}^{jk} &=& \partial_i \beta^{jk} \,,
  \\[6pt]
  R&=& \tfrac{1}{6} \op R^{ijk} \op\iota_i\wedge \iota_j\wedge \iota_k \,,
  & R^{ijk} &=& 3 \op \beta^{[\ul i| m} \partial_m \beta^{\ul j\ul k]}\,.
  \end{array}
}  
Details of this computation can be found in the appendix of \cite{Andriot:2014qla}.
The expressions for $Q_i{}^{jk}$ and $R^{ijk}$ have appeared already in 
\eqref{fluxes_qr9386493856}, where we identified them with the non-geometric 
$Q$- and $R$-flux, and the derivative $D$ is related to 
the expressions shown in \eqref{gg_bianchi_der_38756}. 
A non-trivial $\beta$-transformation therefore generates a potential due to non-vanishing $Q$- and $R$-fluxes.
Using the two-form $\mathcal J$ defined in \eqref{sg_214343535} which encodes 
the complexified K\"ahler moduli, 
we find in particular
\eq{
   \mathcal P^1- i \op\mathcal P^2 \sim
    \int_{\mathcal M} \Omega\wedge \Bigl[\,  \tfrac{1}{2}\op Q \op \mathcal J^2 
    - \tfrac{1}{6} \op R\op \mathcal J^3\,
    \Bigr]\,.
}
Note that since $\mathcal J$ is a $(1,1)$-form, $D\op e^{\mathcal J}$ in \eqref{sg_def_398756b}
cannot become a $(0,3)$-form
and therefore $\Phi^-\wedge D\op \Phi^+=0$ on a Calabi-Yau three-fold.

\end{itemize}

%%%%%%%%%%%%%%%%%%%%%%%%%%%%%%%%%%%%%%%%%%%%%%%
%%%%%%%%%%%%%%%%%%%%%%%%%%%%%%%%%%%%%%%%%%%%%%%

\subsubsection*{NS-NS sector -- general expression}

We can now combine these two examples into a  general expression including
 all types of fluxes. Using the Clifford action discussed on page~\pageref{page_clifford}, 
under a general $O(D,D)$ transformation we have
\eq{
    \bigl\langle \Phi^-, d\Phi^+\bigr\rangle \rightarrow
    \bigl\langle \Phi^-, \mathcal D \op\Phi^+\bigr\rangle \,,
}
where the generalised Dirac operator for the generalised spinor 
is given by
\eq{
  \label{sg_dirac_8275}
   \mathcal D = \slashed\nabla = \nabla_A \Gamma^A\,,
  \hspace{60pt}
  \nabla_A =  \rho(\ov{\mathcal E}_A) - \frac{1}{3!} \op F_{ABC}\op \Gamma^B\op \Gamma^C \,.
}
Here we use the same conventions as in section~\ref{sec_gg_fluxes}
and express the $O(D,D)$ $\gamma$-matrices $\Gamma^M=(dx^m\wedge, \iota_m)$ as
$\Gamma^A = \delta^A{}_M \Gamma^M$,
and the anchor-projection $ \rho(\ov{\mathcal E}_A)=D_A$ of the generalised vielbein vector-field was discussed below
\eqref{bianchi_83999190}.
The spin-connection is expressed in terms of the fluxes $F_{ABC} = F_{AB}{}^D\eta_{DC}$ 
introduced in section~\ref{sec_gg_fluxes}.\footnote{We assume, similarly as in section~\ref{sec_gg_bianchi},
that
the flux components $F_{ABC}$ are completely anti-symmetric in their indices.
This excludes terms of the form $F_{im}{}^m$ and $Q_m{}^{mi}$.
}
Using \eqref{gg_bianchi_der_38756} as well as the conventions \eqref{gen_vielbein_284ccc}, 
one reproduces \eqref{sg_def_398756a} and \eqref{sg_def_398756b}.
In components, we can write \eqref{sg_dirac_8275} in the following way \cite{Shelton:2005cf}
\eq{
  \label{sg_twisted_d}
    \mathcal D = D + H -F +Q - R \,,
}
where 
\eq{
  \arraycolsep2pt
  D = dx^i\op(A^T)_i{}^j \op\partial_j  + \iota_i\op (B^T)^{ij}\op\partial_j \,,
  \hspace{50pt}
  \begin{array}{lcclccccc}
  H&=& \tfrac{1}{6} & H_{ijk} &dx^i & \wedge & dx^j & \wedge & dx^k \,,
  \\[6pt]
    F &=& \tfrac{1}{2} & F_{ij}{}^k & dx^i &\wedge &dx^j &\wedge &\iota_k\,,
  \\[6pt]
    Q &=& \tfrac{1}{2} & Q_i{}^{jk} & dx^i &\wedge &\iota_j &\wedge &\iota_k\,,
  \\[6pt]
  R&=& \tfrac{1}{6} & R^{ijk} &\iota_i & \wedge & \iota_j & \wedge & \iota_k \,.
  \end{array}
}
The flux-components $H_{ijk}$, $F_{ij}{}^k$, $Q_i{}^{jk}$ and $R^{ijk}$ 
in a local basis
have been defined in \eqref{wim}, which  in turn are given by a 
choice of generalised vielbein. The generalised vielbein can be expressed 
in terms of an $O(D,D)$ matrix as in \eqref{gv_77729}, which then determines $D$.
We can now give the general form 
for \eqref{sg_coupl_ppa} using \eqref{sg_twisted_d} as
\eq{
   \mathcal P^1- i \op\mathcal P^2 \sim
    \int_{\mathcal M} \Omega\wedge \Bigl[\,  H - F\op \mathcal J + \tfrac{1}{2}\op Q \op \mathcal J^2 
    - \tfrac{1}{6} \op R\op \mathcal J^3\,
    \Bigr]\,.
}
To conclude, we see that deforming the background geometry by a
non-trivial $O(D,D)$ transformation generates 
non-vanishing moment maps \eqref{sg_coupl_ppa}. This implies that 
the four-dimensional theory becomes a gauged $\mathcal N=2$ supergravity theory, 
in which the gaugings are determined by the geometric and non-geometric fluxes. 

%%%%%%%%%%%%%%%%%%%%%%%%%%%%%%%%%%%%%%%%%%%%%%%
%%%%%%%%%%%%%%%%%%%%%%%%%%%%%%%%%%%%%%%%%%%%%%%

\subsubsection*{R-R sector}

Let us next turn to the moment map $\mathcal P^3$ shown in \eqref{sg_coupl_ppb}.
Non-vanishing field strengths for R-R potentials $C_p$ cannot be generated by 
an $O(D,D)$ transformation, so here we have to choose them by hand.
We start again from the situation of vanishing
NS-NS fluxes for which we have $\widetilde F = d\op\mathcal C$.
Performing  a non-trivial $\mathsf B$-transform of the pure spinors as well as of $\mathcal C$, 
we find 
\eq{
   \bigl\langle \Phi^-, \widetilde F\bigr\rangle
   = \bigl\langle \Phi^-, d \op\mathcal C\bigr\rangle
\rightarrow
  \hspace{20pt} &
  \bigl\langle e^{+\mathsf B}\op\Phi^-, d\bigl( e^{+\mathsf B}\op\mathcal C\bigr)\bigr\rangle 
  \\[4pt]
  =\,&\bigl\langle \Phi^-, e^{-\mathsf B}\op d\bigl( e^{+\mathsf B}\op\mathcal C\bigr)\bigr\rangle 
  \\[4pt]
  =\,&\bigl\langle \Phi^-, \bigl( d +H \bigr)\op\mathcal C\bigr\rangle  \,,
}
where $H=d\op\mathsf  B$.
This suggests that  $\widetilde F = (d+H)\op\mathcal C$ should be identified as the R-R field strength
in the case of non-vanishing $H$-flux,
which agrees  with the definition given already in \eqref{sg_fs_82375693846}.
Using this example together with  our results from the NS-NS sector, for 
a general $O(D,D)$ transformation of the background we are therefore led to 
the field strength
\eq{
  \label{sg_fs_rr}
  \widetilde F  = \mathcal D\op \mathcal C \,.
}
In the four-dimensional theory, this gives rise to a modification of the kinetic terms of the
R-R sector scalars in the hyper-multiplets, which is expected from the 
general expression given in \eqref{sg_gauging_878929}.
We therefore see again that a non-trivial $O(D,D)$ transformation
leads to geometric and non-geometric fluxes, which generates gaugings in the four-dimensional theory.

%%%%%%%%%%%%%%%%%%%%%%%%%%%%%%%%%%%%%%%%%%%%%%%
%%%%%%%%%%%%%%%%%%%%%%%%%%%%%%%%%%%%%%%%%%%%%%%

\subsubsection*{Cohomology}

Let us also consider the action of  \eqref{sg_twisted_d} on the cohomology 
of the Calabi-Yau three-fold. 
The operator $\mathcal D$ is often also called
a twisted differential \cite{Shelton:2005cf}, and its action on the 
cohomology bases \eqref{basis_001} and \eqref{res_033}
can be parametrised as\op\footnote{
For the action of $\mathcal D$ on the cohomology a local basis on the tangent- and 
cotangent-space as  in \eqref{sg_twisted_d} is not suitable. Furthermore, the flux components 
are required to be constants in order to have elements in $H^{p,q}(\mathcal M,\mathbb R)$.}
 \cite{Grana:2006hr}
\eq{
\label{deffluxes}
\arraycolsep1.5pt
\begin{array}{lccllcll@{\hspace{50pt}}lccllcll}
{\mathcal D}\op \alpha_I &\sim& &q_{I}{}^{ A} & \omega_{ A}
&+&  f_{I \op A}& \sigma^{ A}\,,
&
{\mathcal D}\op \beta^I &\sim& &\tilde q^{I \op A} &\omega_{ A}
&+ &\tilde f^{I}{}_ { A} &\sigma^{ A}\,, 
\\[8pt]
{\mathcal  D}\op\omega_{ A}&\sim&- &\tilde f^{I}{}_{ A} &\alpha_I &+ &
f_{I\op  A}& \beta^I\,,
&
{\mathcal D}\op\sigma^{ A} &\sim& &\tilde q^{I\op  A} &\alpha_I &-  &
q_{I}{}^{ A}& \beta^I\,.
\end{array}
}
Here $\sim$ denotes equality up to terms which vanish under the Mukai pairing 
\eqref{gg_mukai} with any other basis element. 
Furthermore, $f_{I \op \lab A}$ and $\tilde f^{I}{}_ {\lab A}$ correspond to the geometric
$F$-fluxes, while $q_{I}{}^{\lab A}$ and $\tilde q^{I\op \lab A}$ are the
components of the 
non-geometric $Q$-fluxes.
For the $H$- and $R$-flux we use the conventions
\eq{
\label{fluxzerocomp}
\arraycolsep2pt
\begin{array}{lcl@{\hspace{70pt}}lcl}
f_{I \op0}&=&h_I\,, & \tilde f^{I}{}_0&=&\tilde h^I\,,\\[5pt]
q_{I}{}^0&=&r_I\,,& \tilde q^{I\op 0}&=&\tilde r^I\, .
\end{array}
}
Let us furthermore define a $(2\op h^{2,1}+2)\times (2\op h^{1,1}+2)$ matrix
as follows
\eq{
  \label{res_015}
{\mathcal Q}=\left(\begin{array}{@{}rr@{\hspace{2pt}}}
  -\tilde f^{I}{}_ { A} & \tilde q^{I\op  A} \\
  f_{I \op A} & -q_{I}{}^{ A}
\end{array}\right),
}
as well as  two symplectic structures  $\mathcal S_{\pm}$ as
\eq{
  \label{res_018}
    \mathcal S_{\pm}=\left(\begin{matrix} 0 & +\mathds 1 \\ -\mathds 1 & 0\end{matrix}\right),
}
where $\mathcal S_+$ is of dimensions 
$(2\op h^{1,1}+2)\times(2\op h^{1,1}+2)$ and $\mathcal S_-$ is a matrix with dimensions  $(2\op h^{2,1}+2)\times(2\op h^{2,1}+2)$. Defining finally
$  \tilde{\mathcal Q} = \mathcal S_- \mathcal Q \,\mathcal S^T_+ $, we can write \eqref{deffluxes} more compactly as \cite{Grana:2006hr}
\eq{
  \label{res_017}
  \mathcal D \,\binom{\omega_A}{\sigma^A} \sim \mathcal Q^T \binom{\alpha_{I}}{\beta^{I}} \,,
  \hspace{50pt}
  \mathcal D \, \binom{\alpha_{I}}{\beta^{I}}\sim  -\tilde{\mathcal Q} \,
  \binom{\omega_A}{\sigma^A} 
  \,.
}

%%%%%%%%%%%%%%%%%%%%%%%%%%%%%%%%%%%%%%%%%%%%%%%
%%%%%%%%%%%%%%%%%%%%%%%%%%%%%%%%%%%%%%%%%%%%%%%

\subsubsection*{Bianchi identities}

Let us now come back to the twisted differential $\mathcal D$ shown in \eqref{sg_dirac_8275},
and determine its square. We find the following expression
\eq{
  \label{sg_dft_bianchi}
  \arraycolsep2pt
  \begin{array}{lccl}
  \mathcal D^2 = \slashed\nabla^2 =
  &-&\displaystyle\frac{1}{24} &
  \displaystyle \Bigl[ 
  D_{[\ul M} F_{\ul N \ul O\ul P]} - \frac{3}{4} \, F_{[\ul M \ul N}{}^{R} F_{R|\ul O\ul P]}
  \Bigr] \op \Gamma^{MNOP}
  \\[10pt]
  &-&\displaystyle\frac{1}{8} &
  \displaystyle D^M  F_{MPQ} \,
  \op \Gamma^{PQ}
  \\[10pt]
  &-&\displaystyle\frac{1}{48} & F^{MNO} F_{MNO} \,.
  \end{array}
}
Note that the first line is proportional to the Bianchi identity \eqref{bianchi_83999190},
whereas the third line can be obtained by suitable index contractions.
These expressions have appeared in the literature before, see for instance
\cite{Geissbuhler:2013uka} for the context of double field theory.
Requiring then the extended Bianchi identities, that is including  $D^M F_{MPQ}=0$,  to be satisfied gives $\mathcal D^2 =0$.

We can also turn this reasoning around and require the twisted differential  to 
be nil-potent, that is $\mathcal D^2=0$, leading to the Bianchi identities for
the fluxes. 
For the action of the twisted differential on the cohomology shown in \eqref{res_017},
this implies that  \cite{Grana:2006hr}\footnote{In the presence of NS-NS sources such as 
the NS5-brane, KK-monopole or $5^2_2$-brane the condition \eqref{bianchi_838884} 
is modified. See  \cite{Villadoro:2007tb,Andriot:2014uda} for details.}
\eq{
  \label{bianchi_838884}
  \mathcal D^2 = 0 \hspace{30pt}\longrightarrow\hspace{30pt}
  \arraycolsep1.5pt
  \begin{array}{lclc}
  \mathcal Q^T &\cdot \hspace{4pt}\mathcal S_- \op\cdot &\mathcal Q &= 0 \,,
  \\[4pt]
  \mathcal Q &\cdot \hspace{4pt}\mathcal S_+ \op\cdot  &\mathcal Q^T &= 0 \,.
  \end{array}
}
Furthermore, we can derive the Bianchi identity (in the absence of sources, see section~\ref{sec_cy_flux_echt_orient} for their inclusion) for the R-R field strength shown in equation \eqref{sg_fs_rr}
as 
\eq{
  \label{bianchi_023030948}
  \mathcal D \op\widetilde F = 0\,.
}

However, there appears to be a slight mismatch between the Bianchi identities \eqref{bianchi_838884}
computed from the action of $\mathcal D$ on the cohomology and 
the Bianchi identities in a coordinate basis shown in \eqref{gg_bianchi_83948016}. This issue has been 
studied in \cite{Shukla:2016xdy,Gao:2018ayp} for toroidal examples,
but to our knowledge  is currently still under investigation.

%%%%%%%%%%%%%%%%%%%%%%%%%%%%%%%%%%%%%%%%%%%%%%%
%%%%%%%%%%%%%%%%%%%%%%%%%%%%%%%%%%%%%%%%%%%%%%%

\subsubsection*{Scalar potential}

Given the action \eqref{res_017} of the twisted differential on the cohomology of the 
Calabi-Yau three-fold, we can now derive  explicit expressions for the moment maps \eqref{sg_coupl_pp}
and determine
the resulting scalar potential. To do so, we first expand the pure spinors $\Phi^+$ and $\Phi^-$
of the Calabi-Yau three-fold in the bases  \eqref{res_033} and \eqref{basis_001} as
\eq{
  &\Phi^+ =  \bigl( \omega_0 \; \omega_{\mathsf A} \; \sigma^0 \; \sigma^{\mathsf A}\bigr) \cdot
  \left( \begin{array}{c}
  1  \\
  \mathcal J^{\lab A} \\
  \tfrac{1}{6}\op \kappa_{\lab{ABC}} \mathcal J^{\lab A} \mathcal J^{\lab B} J^{\lab C} \\
   \tfrac{1}{2}\op \kappa_{\lab{ABC}} \mathcal J^{\lab B} \mathcal J^{\lab C} 
  \end{array}
  \right) ,
  \\[10pt]
  &
  \Phi^- = \bigl( \alpha_{I}\;\beta^{I}\bigr) \cdot 
  \left(\begin{array}{@{}r@{}}X^{I} \\ -\mathcal F_{I} \end{array}\right)  .
}
These expansions define a $(2\op h^{2,1}+2)$-dimensional vector $V^+$ and
a $(2\op h^{2,1}+2)$-dimensional vector $V^-$ as follows
\eq{
  V^+ = 
  \left( \begin{array}{c}
  1  \\
  \mathcal J^{\lab A} \\
  \tfrac{1}{6}\op \kappa_{\lab{ABC}} \mathcal J^{\lab A} \mathcal J^{\lab B} J^{\lab C} \\
   \tfrac{1}{2}\op \kappa_{\lab{ABC}} \mathcal J^{\lab B} \mathcal J^{\lab C} 
  \end{array}
  \right) ,
  \hspace{40pt}
  V^- =\left(\begin{array}{@{}r@{}}X^{I} \\ -\mathcal F_{I} \end{array}\right).
}
For the Ramond-Ramond sector we expand the $\tilde F$-flux 
and the potentials $\mathcal C$ along the Calabi-Yau manifold $\mathcal M$
in a similar way
\eq{
  \mathcal C\bigr\rvert_{\mathcal M} =  \;&
  \renewcommand{\arraystretch}{1.2}
  \arraycolsep2pt
  \bigl( \omega_0 \; \omega_{\mathsf A} \; \sigma^0 \; \sigma^{\mathsf A}\bigr) \cdot
   \left(\begin{array}{@{}l@{}}
  (C_0) \\
  (C_2)^{\lab A} \\
  \hspace{4pt}C_6 \\
  (C_4)_{\lab A} 
  \end{array}\right)
  ,
  \\[10pt]
  \widetilde F\bigr\rvert_{\mathcal M}  =\;
  &
  \renewcommand{\arraystretch}{1.2}
  \arraycolsep1pt
 \bigl( \alpha_{I}\;\beta^{I}\bigr) \cdot 
  \left(\begin{array}{cl} & (F_3)^{I} \\ -& (F_3)_I \end{array}
  \right),
}
which defines two vectors as
\eq{
  \renewcommand{\arraystretch}{1.2}
  \arraycolsep1pt
  \mathsf C = \left(\begin{array}{@{}l@{}}
  (C_0) \\
  (C_2)^{\lab A} \\
  \hspace{4pt}C_6 \\
  (C_4)_{\lab A} 
  \end{array}\right),
  \hspace{70pt}
  \mathsf F = \left(\begin{array}{cl} & (F_3)^{I} \\ -& (F_3)_I \end{array}
  \right).
}
Here, $C_6$ denotes the component of the R-R six-form potential along the 
compact manifold. 
Using these definitions, we can evaluate the moment maps $\mathcal P^x$ 
discussed above, and use the general expression \eqref{sg_pot_n2_38745} 
to evaluate the scalar potential.
Using the definition of the matrices $\mathcal M^+$ and $\mathcal M^-$ given in 
\eqref{res_019} and \eqref{res_0162}
we find 
\eq{
  \label{res_022}
  V = &\hspace{14pt}
  \frac{1}{2} \op \bigl( \mathsf F^T + \mathsf C^T \cdot \mathcal Q^T \bigr) \cdot \mathcal M^- \cdot
  \bigl( \mathsf F + \mathcal Q \cdot \mathsf C \bigr) \\
  & + \frac{e^{-2\phi}}{2} \op V^{+\op T}\cdot {\cal Q}^T \cdot {\cal M}^-\cdot{\cal Q}\cdot \ov V{}^+\\
  & + \frac{e^{-2\phi}}{2} \op   V^{-\op T} \cdot \tilde{\mathcal Q} \cdot \mathcal M^+ \cdot \tilde{\mathcal Q}^T \cdot \ov V^-
  \\
  &-\frac{e^{-2\phi}}{4\op \mathcal V} \, 
  V^{-\op T} \cdot \mathcal S_- \cdot \mathcal Q\cdot \Bigl(  V^+ \times \ov V{}^{+\op T} + \ov V{}^+ \times  V^{+\op T}  \Bigr)
   \cdot \mathcal Q^T\cdot \mathcal S_-^T \cdot \ov V{}^- \,,
}
where $\mathcal V$ is the overall volume of the Calabi-Yau three-fold
and where $\cdot$ denotes matrix multiplication while $\times$ stands for the 
ordinary product of scalars. 
This 
expression agrees with the scalar potential of $\mathcal N=2$ gauged supergravity found in \cite{DAuria:2007axr}
(after going to Einstein frame). In the context of $SU(3)\times SU(3)$ structure compactifications
this potential  has appeared for instance in \cite{Cassani:2008rb}, 
and in the context of double field theory it has been derived in \cite{Blumenhagen:2015lta}.

%%%%%%%%%%%%%%%%%%%%%%%%%%%%%%%%%%%%%%%%%%%%%%%
%%%%%%%%%%%%%%%%%%%%%%%%%%%%%%%%%%%%%%%%%%%%%%%

\subsubsection*{Mirror symmetry}

With the help of the scalar potential \eqref{res_022}, we can also illustrate 
mirror symmetry \cite{Greene:1990ud}.
This is a  well-established duality between 
 compactifications of type IIA and type IIB string theory on Calabi-Yau three-folds,
which essentially interchanges the even and odd cohomology groups. 
In the scalar potential this is realised by interchanging the $+$ and $-$ labels of the vectors
$V^+$ and $V^-$, and by exchanging the fluxes as
\eq{
  \mathcal Q \;\longrightarrow\; \tilde{\mathcal Q}^T \,.
}
For the R-R sector one finds that the fluxes between type IIB and type IIA are interchanged. 
Working out how the components of $\mathcal Q$ transform under mirror symmetry, we see that 
some of the geometric and non-geometric fluxes are mapped as follows \cite{Grana:2006hr}
\eq{
  \arraycolsep3pt
  \begin{array}{lcl@{\hspace{70pt}}lcl}
  f  & \longrightarrow &- f^T  \,, & \tilde f & \longrightarrow & q^T \,, 
  \\[4pt]
  \tilde q  & \longrightarrow &- \tilde q^T  \,, & q & \longrightarrow & \tilde f^T \,.
  \end{array}
}
Therefore, when compactifying string theory on a Calabi-Yau three-fold for instance with geometric $H$-flux, 
the mirror dual will in general be a Calabi-Yau compactification with non-geometric $R$-flux. 
This shows that non-geometric fluxes are a natural part of string-theory compactifications,
which are connected via mirror symmetry to geometric fluxes.

%%%%%%%%%%%%%%%%%%%%%%%%%%%%%%%%%%%%%%%%%%%%%%%
%%%%%%%%%%%%%%%%%%%%%%%%%%%%%%%%%%%%%%%%%%%%%%%

\subsubsection*{Remark}

We also briefly mention partial supersymmetry breaking from 
$\mathcal N=2$ to $\mathcal N=1$ in four-dimensions, for which non-geometric fluxes
play an important role. 
In the series of papers \cite{Cassani:2009na,Louis:2009xd,Louis:2010ui,Hansen:2013dda}
it was studied how partial supersymmetry breaking can be realised in string-theory compactifications. 
It turns out that for partially-broken Minkowski vacua one needs at least two gauged
isometries, which can be generated by at least two entries in the flux matrix \eqref{res_015} -- out of which 
one has to be a non-geometric flux.\footnote{
More concretely, one needs both electric and magnetic gaugings
with one of them corresponding to a non-geometric flux. We did not introduce a distinction between 
electric and magnetic gaugings in this section, but schematically an electric gauging is done with 
a gauge field whose field strength appears in the action explicitly. A magnetic gauging on the other hand 
is done with a gauge field corresponding to the dual field strength and which does not explicitly appear 
in the action. A formalism to deal with such gaugings in a systematic way has been developed in
\cite{deWit:2005ub}.
}
This observation has been employed in \cite{Blumenhagen:2016axv,Blumenhagen:2016rof}
to determine the back-reaction of non-geometric fluxes on Calabi-Yau three-fold compactifications.
The back-reacted solutions are asymmetric Gepner models, which are related 
to non-geometric constructions previously studied in \cite{Israel:2013wwa,Israel:2015efa}.

%%%%%%%%%%%%%%%%%%%%%%%%%%%%%%%%%%%%%%%%%%%%%%%
%%%%%%%%%%%%%%%%%%%%%%%%%%%%%%%%%%%%%%%%%%%%%%%
%%%%%%%%%%%%%%%%%%%%%%%%%%%%%%%%%%%%%%%%%%%%%%%
%%%%%%%%%%%%%%%%%%%%%%%%%%%%%%%%%%%%%%%%%%%%%%%

\subsection{Calabi-Yau orientifolds with fluxes}
\label{sec_cy_flux_echt_orient}

In string theory one is often interested in $\mathcal N=1$ supergravity theories in 
four dimensions, which  can be obtained from $\mathcal N=2$ compactifications  by 
performing an orientifold projection \cite{Grimm:2004uq,Grimm:2004ua}. 
In this section we want to determine 
the scalar potential  of the four-dimensional 
$\mathcal N=1$ theory corresponding to the orientifold-projected version of \eqref{res_022}, and for a full analysis we refer to \cite{Benmachiche:2006df,Micu:2007rd}.

%%%%%%%%%%%%%%%%%%%%%%%%%%%%%%%%%%%%%%%%%%%%%%%
%%%%%%%%%%%%%%%%%%%%%%%%%%%%%%%%%%%%%%%%%%%%%%%

\subsubsection*{Orientifold projection}

We focus again on type IIB string theory and perform an orientifold projection of the form $\Omega_{\rm P}(-1)^{F_{L}}\sigma$.
Here, $\Omega_{\rm P}$ denotes the world-sheet parity operator, $F_L$ is the left-moving fermion number
(cf. page~\pageref{page_type_ii}), and 
$\sigma$ is a holomorphic involution on the compact space $\mathcal M$. 
We choose the action of $\sigma^*$ on the K\"ahler and holomorphic three-form as
\eq{
  \label{orient_choice}
  \sigma^*J=+J\,,\hspace{60pt}
  \sigma^*\op\Omega=-\Omega\,.
}
The fixed loci of this involution on $\mathcal M$ are zero- and four-dimensional which  -- taking into 
account that $\sigma$ leaves the four-dimensional space invariant -- gives rise to orientifold 
three- and seven-planes.\footnote{If $\sigma^*$ leaves $\Omega$ invariant, one obtains O$5$- and 
O$9$-planes. The analysis done in this section for this situation can be found in \cite{Benmachiche:2006df}.}
The orientifold projection gives rise to a splitting of the cohomology groups 
into even and odd eigenspaces of $\sigma^*$ as follows
\eq{
  \label{cohom_split}
  H^{p,q}(\mathcal{M}) = 
  H^{p,q}_+(\mathcal{M})\oplus
  H_-^{p,q}(\mathcal{M}) \,,
}
and the corresponding Hodge numbers will be denoted by 
$h^{p,q}_{\pm} = \mbox{dim}\op  H^{p,q}_{\pm}(\mathcal{M})$. According to 
\eqref{cohom_split} they satisfy $h^{p,q} = h^{p,q}_+ + h^{p,q}_-$. 
One can furthermore determine the following relations
\cite{Grimm:2004uq}
\eq{
  \arraycolsep2pt
  \renewcommand{\arraystretch}{1.25}
  \begin{array}{lcllclcllcl}
  h^{1,1}_{\pm} & = & h^{2,2}_{\pm} \;, \qquad\qquad\;\;
    h^{3,0}_{+} & = & h^{0,3}_{+} & = & 0 \;, \qquad\qquad\;\;
      h^{0,0}_{+} & = & h^{3,3}_{+} & = & 1 \;, 
      \\[4pt]
  h^{2,1}_{\pm} & = & h^{1,2}_{\pm} \;, \qquad\qquad\;\;
    h^{3,0}_{-} & = & h^{0,3}_{-} & = & 1 \;, \qquad\qquad\;\;
      h^{0,0}_{-} & = & h^{3,3}_{-} & = & 0 \;.      
  \end{array}
}  
Turning to the world-sheet parity operator and the left-moving fermion number, one finds that 
their action on the ten-dimensional bosonic fields is given by
\eq{
  \label{orient_signs}
  \arraycolsep2pt
  \begin{array}{lclclcl}
  \arraycolsep1.5pt
  \displaystyle \Omega_{\rm P}\, (-1)^{F_L}\: g &=& + \:g \;,  & \hspace{60pt} &
  \displaystyle \Omega_{\rm P}\, (-1)^{F_L}\: B &=& - B \;, \\[6pt]
  \displaystyle \Omega_{\rm P}\, (-1)^{F_L}\: \phi &=& + \:\phi \;,  & \qquad &
  \displaystyle \Omega_{\rm P}\, (-1)^{F_L}\: C_p &=& (-1)^{\frac{p}{2}} 
    \:C_p \;.
  \end{array}
}
For the $H$-flux and the $\tilde F_3$-form flux one can infer their transformation behaviour from \eqref{orient_signs}, and 
for the geometric and non-geometric fluxes one chooses  \cite{Shelton:2005cf,Blumenhagen:2015lta}
\eq{
  \label{sg_orient_82946}
  \arraycolsep2pt
  \begin{array}{lccl}
  \Omega_{\rm P}(-1)^{F_{L}} \,H &=& -& H \,, \\[4pt]
  \Omega_{\rm P}(-1)^{F_{ L}} \,F &=& +& F \,, \\[4pt]
  \Omega_{\rm P}(-1)^{F_{ L}} \,Q &=& - &Q \,, \\[4pt]
  \Omega_{\rm P}(-1)^{F_{ L}} \,R &=& + &R \,.
  \end{array}
  \hspace{50pt}
  \Omega_{\rm P}(-1)^{F_{ L}} \,\widetilde F_3 = - \widetilde F_3 \,, 
}

%%%%%%%%%%%%%%%%%%%%%%%%%%%%%%%%%%%%%%%%%%%%%%%
%%%%%%%%%%%%%%%%%%%%%%%%%%%%%%%%%%%%%%%%%%%%%%%

\subsubsection*{Massless spectrum }

The massless spectrum of the theory compactified on a Calabi-Yau orientifold can 
be determined as before by expanding the ten-dimensional fields into 
elements of the appropriate cohomology groups and integrating over the compact space. 
Only fields invariant under the orientifold projection are kept, which means for instance
that we consider
\eq{
  J \in H^{1,1}_{+}(\mathcal{X}) \,,
  \hspace{40pt}
  \Omega \in H^{3}_{-}(\mathcal{X}) \,,  
  \hspace{40pt}
  B\bigr\rvert_{\mathcal M} \in H^{1,1}_{-}(\mathcal{X}) \,,
}
where $B\bigr\rvert_{\mathcal M}$ denotes the restriction of the Kalb-Ramond field to the compact space.
For the R-R potentials similar results apply, depending on the degree of $C_p$. 
%%%%%%%%%%%%%%
%%%%%%%%%%%%%%
\begin{table}[t]
\eq{
  \nonumber
  \arraycolsep15pt
  \renewcommand{\arraystretch}{1.3}
  \begin{array}{c||c||c}
  \mbox{multiplet} & \mbox{multiplicity} & \mbox{bosonic field content}
  \\
  \hline\hline
  \mbox{gravity} & 1 & g_{\mu\nu} 
  \\
  \mbox{vector} & h^{2,1}_+ & (C_4)_{\mu}^i\
  \\
  \mbox{chiral} & h^{2,1}_- & z^{\hat i}\
  \\
  \mbox{chiral} & 1 & \tau
  \\
  \mbox{chiral} & h^{1,1}_+ & T_{\lab A}
  \\
  \mbox{chiral} & h^{1,1}_- & G^{\hat{\mathsf A}}
  \\
  \end{array}
}
\caption{Massless bosonic field content of type IIB string theory compactified on a Calabi-Yau orientifold with 
O$3$- and O$7$-planes in four dimensions.
\label{sg_ori_87829}
}
\end{table}
%%%%%%%%%%%%%%
%%%%%%%%%%%%%%
The resulting spectrum is summarised in table~\ref{sg_ori_87829}, 
where we indicated the index running over the $\sigma^*$-odd cohomology by a hat, that is
\eq{
  \arraycolsep2pt
  \begin{array}{lcl@{\hspace{80pt}}lcl}
  i &=& 1,\ldots, h^{2,1}_+ \,, & 
  \mathsf A &=& 1,\ldots, h^{1,1}_+ \,, 
  \\[4pt]
  \hat i &=& 1,\ldots, h^{2,1}_- \,, & 
  \hat{\mathsf A} &=& 1,\ldots, h^{1,1}_- \,.
  \end{array}
}
The complex-structure moduli $z^{\hat i}$ in table~\ref{sg_ori_87829} originate from the projection of the 
$\mathcal N=2$ moduli \eqref{hol_three} by keeping only the odd cohomology expansion coefficients. 
The remaining scalar fields $\tau$, $T_{\mathsf A}$ and $G{}^{\hat{\mathsf A}}$ are a combination of the 
R-R potentials $C_p$, the Kalb-Ramond field $B$, the dilaton $\phi$ and the
K\"ahler moduli $t^{\mathsf A}$. 
We specify their precise form below.

%%%%%%%%%%%%%%%%%%%%%%%%%%%%%%%%%%%%%%%%%%%%%%%
%%%%%%%%%%%%%%%%%%%%%%%%%%%%%%%%%%%%%%%%%%%%%%%

\subsubsection*{Generalised spinors and $\mathcal N=1$ supergravity data}

Similar to the $\mathcal N=2$ situation, one can encode the properties of the $\mathcal N=1$ 
theory in terms of generalised spinors. To do so, let us define the following 
quantities  \cite{Benmachiche:2006df}
\eq{
  \label{sg_n1_89919466}
  \arraycolsep2pt
   \begin{array}{lcl}
   \Phi^+ &=& \displaystyle  e^{-\phi}\, e^{B- i\op J} \,,\\[4pt]
   \Phi^- &=& \Omega \,,
   \end{array}
   \hspace{60pt}
   \Phi^+_{\rm c} = e^{B}\op \mathcal C_{\rm mod} + i \,\mbox{Re}\op \Phi^+ \,,
}
where the  sum over all R-R potential $C_p$ 
defined in \eqref{sg_fs_8346} has been separated into a flux contribution and a moduli contribution 
as
$\mathcal C=\mathcal C_{\rm flux} + \mathcal C_{\rm mod}$.
The scalar fields $\tau$, $T_{\mathsf A}$ and $G^{\hat{\mathsf A}}$ shown in 
table~\ref{sg_ori_87829} can be determined by expanding $\Phi^+_{\rm c}$ as
follows
\eq{
  \Phi^+_{\rm c} = \tau + G^{\hat{\mathsf A}} \omega_{\hat{\mathsf A}} + 
  T_{\mathsf A} \sigma^{\mathsf A} \,,
}
where we employed the basis of $(1,1)$- and $(2,2)$-forms introduced in 
\eqref{res_033a}.
Let us now discuss the $\mathcal N=1$ supergravity data of these theories:
\begin{itemize}

\item Since these scalar fields are part of chiral multiplets of a $\mathcal N=1$ supergravity theory in 
four dimensions, the  metrics of their scalar manifolds are K\"ahler. 
The corresponding K\"ahler potentials
can be expressed using \eqref{sg_n1_89919466} 
in the following way \cite{Benmachiche:2006df,Cassani:2007pq}
\eq{
  \arraycolsep2pt
  \begin{array}{lcrcl}
  \displaystyle\mathcal K^+ &=& -2 \displaystyle\log \left[ + i \int_{\mathcal M} \bigl\langle \Phi{}^+, \ov\Phi{}^+\bigr\rangle \right] 
  &=& \displaystyle -\log\bigl[ -\tfrac{i}{2} \op (\tau- \ov \tau) \bigr]
  -2 \log\bigl[ \op 8 \op\hat{\mathcal V}\op \bigr]\,,
  \\[15pt]
  \displaystyle\mathcal K^- &=& \displaystyle -\log\left[ -i \int_{\mathcal M} \bigl\langle \Phi{}^-, \ov\Phi{}^-\bigr\rangle \right]
   &=& \displaystyle -\log\left[- i \int_{\mathcal M} \Omega\wedge\ov\Omega \,\right],
  \end{array}
}
where $\hat{\mathcal V} = e^{-\frac{3}{2}\phi} \tfrac{1}{6}\int J^3 $ denotes the volume of $\mathcal M$ in 
Einstein frame. 
Note that $\mathcal K^-$ is the projection of the $\mathcal N=2$ result given in \eqref{sg_kaeher_01},
whereas $\mathcal K^+$ differs from   \eqref{sg_kaeher_02} by a factor of two and the dilaton dependence.
Furthermore, these expressions agree with the standard results for the K\"ahler potential of 
Calabi-Yau orientifold compactifications \cite{Grimm:2004uq}.

\item Turning now to the interactions, the superpotential $W$ of the $\mathcal N=1$ supergravity theory
is given by a suitable combination of the moment maps \eqref{sg_coupl_pp} of the 
$\mathcal N=2$ theory, subject to the orientifold projection. 
In the absence of $H$-flux, one finds  \cite{Benmachiche:2006df,Cassani:2007pq}
\eq{
  \label{sg_w_sup}
  W =   \int_{\mathcal M} \bigl\langle \Phi^-, \widetilde F + d\op\Phi^+_{\rm c} \bigr\rangle \,.
}
where the field strength of the R-R potentials $\widetilde F= d\op\mathcal C_{\rm flux}$
contains only terms invariant under the orientifold projection
and where we employed the Mukai pairing \eqref{gg_mukai}.

\item Note that in $W$ only the real part of the generalised spinor $\Phi^+$ appears. The imaginary part contributes to 
the $\mathcal N=1$ D-term potential, which takes the form (see for instance \cite{Cassani:2007pq,Blumenhagen:2015lta})
\eq{
  \label{sg_d_termp}
  V_D \sim \int_{\mathcal M} \bigl\langle 
  \op d \op\mbox{Im}\op \Phi^+ , \star  \op d\op \mbox{Im}\op \Phi^+ \op\bigr\rangle \,,
}
where $\star$ denotes the Hodge star-operator on $\mathcal M$
and where we employed again the Mukai pairing \eqref{gg_mukai}.
Note that $d\op \mbox{Im}\op \Phi^+$ is part of the $\sigma^*$-even cohomology.

However, recall that the D-term potential arises from gaugings of the chiral-multiplet scalars 
using the vector-fields. From the spectrum of the  $\mathcal N=1$ theory shown in table~\ref{sg_ori_87829}
we see that only vector-fields $(C_4)_{\mu}^i$ with index
\raisebox{0pt}[0pt]{$i=1,\ldots, h^{2,1}_+$} are present. 
Other vector-field components of $C_4$ are either pro\-jec\-ted out through the orientifold projection, 
or are removed due to the self-duality of $C_4$. 
In the scalar potential \eqref{sg_d_termp} we therefore have to restrict $d\op \mbox{Im}\op \Phi^+$
to those parts proportional to $\{\alpha_{i}\}\in H^{2,1}_+(\mathcal M)$. 
This situation changes when considering magnetic gaugings of the isometries, which 
we do not discuss here.

\end{itemize}
For the case of a Calabi-Yau orientifold without fluxes, we note that $\Phi^+$ is closed 
and that $\widetilde F$ is vanishing. This implies that the scalar F- and D-term potentials 
vanish, and that the four-dimensional theory is an ungauged $\mathcal N=1$ supergravity.

%%%%%%%%%%%%%%%%%%%%%%%%%%%%%%%%%%%%%%%%%%%%%%%
%%%%%%%%%%%%%%%%%%%%%%%%%%%%%%%%%%%%%%%%%%%%%%%

\subsubsection*{Deformations I -- F-term potential}

Similarly as in section~\ref{sec_cy_flux_echt}, we now want to deform the Calabi-Yau orientifold background 
by performing non-trivial $O(D,D)$ transformations of the generalised spinors 
$\Phi^{\pm}$ and of the R-R potentials $\mathcal C$. 
Let us start by considering a non-trivial $\mathsf B$-transform acting 
as
$\Phi^{\pm} \to e^{\mathsf B} \Phi^{\pm}$ and 
as $\mathcal C \to e^{\mathsf B} \op\mathcal C$.
For the superpotential \eqref{sg_w_sup} this means that 
\eq{
  \bigl\langle \Phi^-, d\mathcal C_{\rm flux} + d\op\Phi^+_{\rm c} \bigr\rangle
   \rightarrow
  \hspace{20pt} &
  \bigl\langle e^{+\mathsf B}\Phi^-, d \bigl( e^{+\mathsf B} \mathcal C_{\rm flux}\bigr)  
  + d \bigl( e^{\mathsf B}\op\Phi^+_{\rm c}\bigr) \bigr\rangle  
  \\[4pt]
  =\,&\bigl\langle \Phi^-, e^{-\mathsf B}d \bigl( e^{+\mathsf B} \mathcal C_{\rm flux}\bigr)  + 
  e^{-\mathsf B}d \bigl( e^{\mathsf B}\op\Phi^+_{\rm c}\bigr) \bigr\rangle    
  \\[4pt]
  =\,&\bigl\langle \Phi^-, \bigl( d +H \bigr) \op \mathcal C_{\rm flux} + 
  \bigl( d +H \bigr)\op\Phi^+_{\rm c} \bigr\rangle     \,,
}
where $H=d\op\mathsf B$. This suggests that in the case of non-vanishing $H$-flux we should identify the 
R-R field strength as $\widetilde F = (d+H ) \op\mathcal C_{\rm flux}$, which agrees again with 
our definition already given in  \eqref{sg_fs_82375693846}.
Using this example as well as our results from  section~\ref{sec_cy_flux_echt}, 
for a background originating from a general $O(D,D)$ transformation the superpotential should be 
given by \cite{Shelton:2005cf,Villadoro:2006ia,Shelton:2006fd}
\eq{
  \label{sg_n1_or_22947}
   W =   \int_{\mathcal M} \bigl\langle \Phi^-, \widetilde F + \mathcal D\op\Phi^+_{\rm c} \bigr\rangle\,,
}
where $\widetilde F = \mathcal D\op\mathcal C_{\rm flux}$ and where $\mathcal D$ has been given  in 
\eqref{sg_twisted_d}. Due to $\Phi^- = \Omega$ being part of the orientifold-odd cohomology, 
in the superpotential only the orientifold-odd part of $ \widetilde F + \mathcal D\op\Phi^+_{\rm c} $ contributes. 
Explicitly, the superpotential reads
\eq{
  \label{sg_824827444}
  W = \int_{\mathcal M}\Omega \wedge \left( \widetilde F_3 +\tau \op H 
  - F\op \omega_{\hat{\mathsf A}} \op G^{\hat{\mathsf A}}
  + Q\op \sigma^{\mathsf A} \op T_{\mathsf A}
  \right).
}
Note that the first two terms are the familiar Gukov-Vafa-Witten superpotential \cite{Gukov:1999ya}, 
whereas the remaining terms provide the generalisation for all types of fluxes. 
Furthermore, we point out that the $R$-flux does not contribute to the superpotential. 

%%%%%%%%%%%%%%%%%%%%%%%%%%%%%%%%%%%%%%%%%%%%%%%
%%%%%%%%%%%%%%%%%%%%%%%%%%%%%%%%%%%%%%%%%%%%%%%

\subsubsection*{Deformations II -- D-term potential}

A similar analysis can be performed for the D-term potential \eqref{sg_d_termp}.
We replace the exterior derivative in $d\op \mbox{Im}\op \Phi^+$ by the twisted differential $\mathcal D$, and
we use the matrix $\mathcal M^-$ defined in \eqref{res_019} to write
\eq{
  V_D 
  = \frac{1}{2}\,
   \Bigl[ ({\rm Im}\, \mathcal N)^{-1} \Bigr]^{ij} \op
   \bigl( \mathcal D \op\mbox{Im}\op \Phi^+ \bigr)_i\,
   \bigl( \mathcal D \op\mbox{Im}\op \Phi^+ \bigr)_j \,.
}
Here, only the even part of the third cohomology contributes which is again related to 
the self-duality condition to be imposed on the R-R four-form potential. 
We can then determine explicitly \cite{Blumenhagen:2015lta}
\eq{
  \label{sg_n1_or_22947b}
  \bigl( \mathcal D \op\mbox{Im}\op \Phi^+ \bigr)_i = 
  e^{-\phi} \left[ - \frac{1}{6}\, R \op J^3 + \frac12 R\op \bigl(B^2 \op J\bigr)
  -  Q\op \bigl( B\op J\bigr)
  + F \op J
  \right]_i \,,
}
where the index $i$ labels the corresponding component of the orientifold-even third cohomology. 
We also note that the flux-components appearing in the superpotential \eqref{sg_n1_or_22947}
do not appear in \eqref{sg_n1_or_22947b} and vice versa.
In particular, the $H$-flux does not contribute to the D-term potential.

%%%%%%%%%%%%%%%%%%%%%%%%%%%%%%%%%%%%%%%%%%%%%%%
%%%%%%%%%%%%%%%%%%%%%%%%%%%%%%%%%%%%%%%%%%%%%%%

\subsubsection*{Localised sources I -- tadpole conditions}

When performing the orientifold projection described above, in string theory new 
localised sources are introduced. These are orientifold planes, and 
the choice \eqref{orient_choice} gives rise to orientifold three- and seven-planes
which fill out four-dimensional space-time and wrap zero- and four-dimensional 
sub-manifolds in the compactification space $\mathcal M$.  
These O-planes are non-dynamical objects, however, they couple to the Ramond-Ramond 
potentials $C_p$. 
In the democratic formulation of type II supergravity \cite{Bergshoeff:2001pv}
the resulting equations of motion/Bianchi identities \eqref{bianchi_023030948} of the 
$\mathcal N=2$ theory are therefore modified to 
\eq{
    \mathcal D \op\widetilde F = \mbox{sources} \,.
}
In order to solve such Bianchi identities one typically has to introduce 
D-branes as additional local sources, and the integrated Bianchi identities are known as tadpole 
cancellation conditions. 
In the context of non-geometric fluxes this question has been discussed 
in \cite{Micu:2007rd,Guarino:2008ik,Aldazabal:2011yz,Blumenhagen:2015kja}.

Let us  make this more precise: 
D$p$-branes and orientifold O$p$-planes are hyper-surfaces in 
ten-dimensional space-time. 
The corresponding world-volume actions
contain Chern-Simons couplings to the R-R potentials $C_q$ which read
(for reviews see e.g. \cite{Blumenhagen:2006ci,Plauschinn:2009izh})
\eq{
  \label{actions_cs_do}
  \mathcal{S}_{{\rm D}p} &\supset -\hphantom{Q_p \op}\mu_p \int_{\Gamma} 
    \mbox{ch}\left( \mathcal F\right) \wedge
    \sqrt{ \frac{\hat{\mathcal A}(\mathcal R_T)}{\hat{\mathcal A}(\mathcal R_N)}} \wedge
    \bigoplus_q \varphi^*C_q \,, \\[2mm]
  \mathcal{S}_{{\rm O}p} &\supset -Q_p \op\mu_p
  \int_{\Gamma} 
    \sqrt{ \frac{\mathcal{L}(\mathcal R_T/4)}{\mathcal{L}(\mathcal R_N/4)}} \wedge
    \bigoplus_q \varphi^* C_q \,.
}
These expressions require some further clarification and explanation:
\begin{itemize}

\item The D-branes and O-planes wrap sub-manifolds $\Gamma$ of the ten-dimensional
space-time and are therefore localised. 
In \eqref{actions_cs_do}  $\varphi^*$ denotes  the pull-back from ten dimensions to $\Gamma$.

\item The expressions $\mathcal R_T$ and $\mathcal R_N$ stand for the restrictions of the curvature two-form $\mathcal R$ to the 
tangent and normal bundle of $\Gamma$. 
We have furthermore employed the Chern character $\mbox{ch}(\mathcal F)$
of the open-string field strength $\mathcal F$ (cf.~our discussion on page~\pageref{page_gauge_flux})
as well as
the $\hat{\mathcal A}$-genus and the Hirzebruch polynomial $\mathcal{L}$.  The  definitions of these 
quantities can be found for instance in \cite{Nakahara:2003nw}, 
and we note that the square-roots in \eqref{actions_cs_do} can be expanded as $(1+ \mbox{4-form} + \mbox{8-form} + \ldots )$.

\item The tension of the D-banes and O-planes $\mu_p$ and the charge of the O-planes $Q_p$ are given by
$\mu_p=2\pi/l_{\rm s}^{p+1}$ and $Q_p=-2^{p-4}$.
In particular, (mutually supersymmetric) D-branes and O-planes have opposite 
charges.

\item We are using the democratic formulation of type IIA/B supergravity 
in which all odd/even R-R potentials $C_q$ appear in the action. In order to connect to 
ordinary type II supergravity, one imposes self-duality constraints for the R-R field strengths $\widetilde F_q$
of the form $\widetilde F_q \sim \star \widetilde F_{10-q}$ on the
equations of motion.
This means in particular that 
equations of motion for the potentials of the form $d\star \widetilde F + \ldots = 0$
can equivalently be expressed as
Bianchi identities $d\widetilde F + \ldots = 0$
(see \cite{Bergshoeff:2001pv} for more details).

\end{itemize}
After having introduced the D-brane and O-plane actions involving the R-R potentials $C_q$, we can now determine the Bianchi
identities for the R-R field strengths. We do this by first computing the equations of 
motion for the R-R potentials $C_q$, and then using the above-mentioned duality to 
obtain the Bianchi identities. The action from which we want to determine the equations of motion reads schematically
\eq{
  \mathcal S = \mathcal S_{\mbox{\scriptsize type II}} + \sum_{{\rm D}p} \mathcal S_{{\rm D}p}
  + \sum_{{\rm O}p}  \mathcal S_{{\rm O}p}\,,
}
where the sums are over all D-branes and O-planes present in the background. 
In particular, the D-brane sum includes the orientifold images. 
In the absence of NS-NS fluxes,
we determine from the equations of motion  the following  Bianchi identities 
(see e.g. \cite{Plauschinn:2008yd} for an explicit computation)
\eq{
  \label{bianchi_82828}
  d\widetilde F_q =   \sum_{{\rm D}p} \mathcal Q_{{\rm D}p}
  + \sum_{{\rm O}p}  \mathcal Q_{{\rm O}p} \;\biggr\rvert_{q+1}\,.
}
The charges $\mathcal Q$ are multi-forms and we have restricted them to 
their $(q+1)$-form part.
With $[\op\Gamma\op]$ denoting the Poincar\'e dual to the cycle $\Gamma$ wrapped by the 
D-brane or O-plane,\footnote{For instance, when considering space-times of the form 
$\mathbb R^{3,1}\times \mathcal M$ with $\mathcal M$ a compact six-dimensional space
and D$p$-branes filling $\mathbb R^{3,1}$, 
$\Gamma_{{\rm D}3}$ is point-like in $\mathcal M$ and $[\Gamma_{{\rm D}3}]\in H^6(\mathcal M)$ is a six-form in $\mathcal M$,
$\Gamma_{{\rm D}5}$ wraps a two-cycle in $\mathcal M$ and $[\Gamma_{{\rm D}5}]\in H^4(\mathcal M)$ is a four-form in $\mathcal M$ 
and $\Gamma_{{\rm D}7}$ wraps a four-cycle in $\mathcal M$ and
$[\Gamma_{{\rm D}7}]\in H^2(\mathcal M)$ is a two-form in $\mathcal M$.}
the $\mathcal Q$ are
defined  as \cite{Minasian:1997mm,Aspinwall:2004jr}
\eq{
  \label{bianchi_928367762}
  \mathcal Q_{{\rm D}p} =  \mbox{ch}\left( \mathcal F\right) \wedge
    \sqrt{ \frac{\hat{\mathcal A}(\mathcal R_T)}{\hat{\mathcal A}(\mathcal R_N)}} \wedge
    [\Gamma_{{\rm D}p}] \,,
    \hspace{30pt}
  \mathcal Q_{{\rm O}p} = Q_p \sqrt{ \frac{\mathcal{L}(\mathcal R_T/4)}{\mathcal{L}(\mathcal R_N/4)}} \wedge 
   [\Gamma_{{\rm O}p}] \,.
}
So far we have  assumed that the NS-NS fluxes are vanishing. 
However, for non-vanishing $H$-flux we find from 
the type II supergravity action \eqref{action_iib_pre}
that the left-hand side of 
\eqref{bianchi_82828} should be replaced by  $(d+H\wedge )\widetilde F$. 
Furthermore, for non-vanishing geometric fluxes $F$ and non-geometric $Q$- and $R$-fluxes
we have argued in section~\ref{sec_cy_flux_echt} that the Bianchi identity 
should involve the twisted differential $\mathcal D$ (see equation \eqref{bianchi_023030948}). 
In the presence of NS-NS fluxes and localised sources, the Bianchi identity \eqref{bianchi_82828} 
therefore becomes
\eq{
  \mathcal D\widetilde F =   \sum_{{\rm D}p} \mathcal Q_{{\rm D}p}
  + \sum_{{\rm O}p}  \mathcal Q_{{\rm O}p} \,.
}
Finally, when integrating these Bianchi identities over the transversal space 
one finds the tadpole-cancellation 
conditions.

%%%%%%%%%%%%%%%%%%%%%%%%%%%%%%%%%%%%%%%%%%%%%%%
%%%%%%%%%%%%%%%%%%%%%%%%%%%%%%%%%%%%%%%%%%%%%%%

\subsubsection*{Localised sources II -- Freed-Witten anomaly}

D-branes in flux-backgrounds furthermore have to satisfy the Freed-Witten a\-nom\-a\-ly 
cancellation condition \cite{Freed:1999vc}. In the case of only $H$-flux this 
means that $H$ restricted to the D-brane has to be exact, which reads in formulas
\eq{
  [H] \bigr\rvert_{\mbox{\scriptsize D-brane}} = 0 \,.
}
Here $[H]\in H^3$ denotes the cohomology class of the $H$-flux. 
For general fluxes it has been argued that the Freed-Witten condition can be expressed as
\cite{Camara:2005dc,Villadoro:2006ia,Aldazabal:2011yz,Camara:2005dc,
LoaizaBrito:2006se,Font:2008vd,Aldazabal:2008zza,Blumenhagen:2015kja}
\eq{
  \label{freed_284727}
  \mathcal D \op [\Gamma_{{\rm D}p}] = 0\,,
}
where $[\Gamma_{{\rm D}p}]$ denotes again the Poincar\'e dual of the cycle $\Gamma_{{\rm D}p}$ wrapped 
by the D-brane.
In \eqref{freed_284727} we assumed that the open-string gauge flux $\mathcal F$ vanishes,
but the natural generalisation to non-trivial $\mathcal F$ 
can be expressed using the charges \eqref{bianchi_928367762} as
\cite{Blumenhagen:2015kja}
\eq{
  \mathcal D\op \mathcal Q_{{\rm D}p} = 0 \,.
}

%%%%%%%%%%%%%%%%%%%%%%%%%%%%%%%%%%%%%%%%%%%%%%%
%%%%%%%%%%%%%%%%%%%%%%%%%%%%%%%%%%%%%%%%%%%%%%%
%%%%%%%%%%%%%%%%%%%%%%%%%%%%%%%%%%%%%%%%%%%%%%%
%%%%%%%%%%%%%%%%%%%%%%%%%%%%%%%%%%%%%%%%%%%%%%%

\subsection{Scherk-Schwarz reductions}
\label{sec_schsch}

In our discussion above we have studied how geometric and
non-geometric fluxes of a higher-dimensional theory affect the 
lower-dimensional one. In particular, we have interpreted
 fluxes as operators acting on the cohomology of the 
compactification space -- as shown in equation \eqref{deffluxes} -- 
which gave rise to a gauged supergravity theory.
However, as we have argued in section~\ref{cha_torus_fib}, fluxes
can also be considered  as encoding the non-triviality of torus fibrations. 
We now want to describe compactifications of type II string theory
on such backgrounds, which  fall into the 
class of (generalised) Scherk-Schwarz reductions \cite{Scherk:1978ta,Scherk:1979zr}.
In regard to non-geometric backgrounds, these
have been investigated for instance in the papers
\cite{Dabholkar:2002sy,Flournoy:2004vn,Hull:2005hk,Hull:2006tp,Hull:2007jy,DallAgata:2007egc,Hull:2009sg}.

%%%%%%%%%%%%%%%%%%%%%%%%%%%%%%%%%%%%%%%%%%%%%%%
%%%%%%%%%%%%%%%%%%%%%%%%%%%%%%%%%%%%%%%%%%%%%%%

\subsubsection*{General idea I}

Let us start by reviewing the main idea of Scherk-Schwarz compactifications in the 
present context, following in parts \cite{Dabholkar:2002sy,Hull:2005hk}. We consider a
gravity theory in $D+1$ dimensions together with a number of scalar fields $\Phi(\hat x)$.
The latter are assumed to take values in a coset space $G/K$, where 
$G$ is typically a non-compact group with maximal compact subgroup $K$. 
In this case the scalar fields can be combined into a vielbein matrix $\mathcal V(\hat x)$, 
transforming under global transformations $g\in G$ and local transformations
$k(\hat x)\in K$ as $\mathcal V\to k(\hat x) \op\mathcal V \op g$.
Restricting  to real matrices $\mathcal V$, the action for the 
combined system (invariant under the above transformations) reads
\eq{
  \label{schsch_001}
  \hat{\mathcal S} = \frac{1}{2\op\hat\kappa^2} \int  \,\Bigl[ \,
  \hat R\star 1 + \tfrac{1}{2}\op\mbox{Tr} \op\bigl( d\mathcal V^{-1}  \wedge\star d\mathcal V \bigr) \Bigr]\,,
}
where $\hat R$ is the Ricci scalar in $D+1$ dimensions and $\star$ denotes the corresponding 
Hodge star-operator. Note that the scalar potential for this theory vanishes and hence
the scalar fields $\Phi(\hat x)$ contained in $\mathcal V(\hat x)$ are massless.

We now want to reduce this theory on a circle, over which $\mathcal V$ is non-trivially fibred. 
We therefore split the $(D+1)$-dimensional coordinates as $\hat x^I \to (x^{\mu},y)$ with $\mu=0,\ldots, D-1$ and 
identify $y\sim y+2\pi$ which gives rise to a circle.
For $\mathcal V(\hat x)$ we consider the ansatz
\eq{
  \label{schsch_002}
  \mathcal V(\hat x) = \mathsf V(x) \, \exp\bigl[ \tfrac{\mathfrak m \op y}{2\pi} \bigr] \,,
}
where $\mathfrak m=\log M$ is the Lie-algebra element corresponding to a monodromy 
matrix 
$M\in G$. This  ensures that the theory is well-defined under
$y\to y+2\pi$, since $\mathcal V(x,y+2\pi) =\mathcal V(x,y) \op M$ is a symmetry of the $(D+1)$-dimensional 
action.
We can then perform a dimensional  reduction of \eqref{schsch_001} on 
the circle: the $(y,y)$-component of the $(D+1)$-dimensional metric becomes 
a scalar field $\phi(x)$ in the $D$-dimensional theory, and the off-diagonal 
$(x,y)$-components of the metric give rise to a $D$-dimensional gauge field $A$.
After going to Einstein frame, we arrive 
at the following $D$-dimensional theory
\eq{
  \label{schsch_005}
  \mathcal S =  \frac{1}{2\op\kappa^2} \int  \,\Bigl[ \,
  R\star 1 
  &-\tfrac{1}{2} \op d\phi\wedge\star d\phi
  -e^{-\gamma\phi} \op F\wedge\star F 
  \\
  &+ \tfrac{1}{2}\op\mbox{Tr} \op\bigl( D \mathsf V^{-1}  \wedge\star D \mathsf V \bigr) 
  - V\star 1\,
  \Bigr]\,,
}
where $\kappa^2 = \hat \kappa^2/2\pi$ is the $D$-dimensional coupling constant, 
$F=dA$ denotes the field strength of the gauge field $A$, $\gamma$ is a positive constant satisfying 
$\gamma^2 = 2\op\frac{D-1}{D-2}$
and the gauge-covariant derivative of $\mathsf V$ reads $D\mathsf V = d\mathsf V + A \op\mathsf V \frac{\mathfrak m}{2\pi}$. 
The scalar potential is non-zero and takes the form
\eq{
  V = \frac{1}{8\pi^2} \op e^{\gamma \phi} \, \mbox{Tr} \op\bigl( \mathfrak m^2 \bigr) \,.
}
We thus see that in $D$ dimensions a potential $V$ is generated by the Lie-algebra element $\mathfrak m$ corresponding 
to the monodromy $M$.
The ansatz \eqref{schsch_002} therefore gives rise to 
a non-trivial potential \cite{Scherk:1979zr}.

%%%%%%%%%%%%%%%%%%%%%%%%%%%%%%%%%%%%%%%%%%%%%%%
%%%%%%%%%%%%%%%%%%%%%%%%%%%%%%%%%%%%%%%%%%%%%%%

\subsubsection*{General idea II}

For our subsequent discussion we are interested in a slightly different setting. In particular, 
we want to consider a kinetic term in the $(D+1)$-dimensional theory given by
\eq{
  \label{schsch_003}
 \hat{\mathcal S} \supset \frac{1}{2\op\hat\kappa^2} \int  \,
  \tfrac{1}{2}\op\mbox{Tr} \op\bigl( d\mathcal H^{-1}  \wedge\star d\mathcal H \bigr) \,,
  \hspace{50pt}
  \mathcal H = \mathcal V^T \eta \op \mathcal V \,,
}
where $\mathcal H$ is a non-degenerate real symmetric matrix.
In this ansatz the matrix $\eta$ is a constant metric invariant under the local $K$-transformations specified above. 
Under global transformations $g\in G$ the field $\mathcal H$ transforms
as $\mathcal H \to g^T \mathcal H \op g$, under which \eqref{schsch_003} is invariant. 
Upon dimensional reduction with the ansatz \eqref{schsch_002} for $\mathcal V$, the 
$D$-dimensional action contains in addition to the first line in \eqref{schsch_005} the terms
\eq{
  \label{schsch_011}
  \mathcal S \supset \frac{1}{2\op\kappa^2} \int  \,\Bigl[ \,
   \mbox{Tr} \op\Bigl( D \mathsf V^{-1}  \wedge\star D \mathsf V 
   - \bigl(\mathsf V^{T} \eta \op \mathsf V \bigr)^{-1} D\mathsf V^T \eta \wedge\star D \mathsf V
   \Bigr) 
  - V\star 1\, \Bigr] \,,
}
where again $D\mathsf V = d\mathsf V + A \op\mathsf V \frac{\mathfrak m}{2\pi}$ and where the  scalar potential $V$ is given by \cite{Dabholkar:2002sy}
\eq{
  \label{schsch_012}
  V = \frac{1}{4\pi^2} \op e^{\gamma \phi} \, \mbox{Tr} \op\Bigl[\op
  \mathfrak m^2 
  +  \bigl(\mathsf V^{T} \eta \op \mathsf V \bigr)^{-1} \mathfrak m^T \op \mathsf V^T \op\eta \,\mathsf V \op \mathfrak m
  \op
  \Bigr] \,.
}
Note that this potential depends on the scalar fields $\Phi(x)$ implicitly via  the reduced vielbein matrix 
$\mathsf V(x)$, and that $V$ will in general generate a mass-term for these scalars.

Let us now study the properties of this potential in some more detail. First, we note
that we can rewrite the scalar potential in the following way 
\eq{
  \label{schsch_008}
  V = \frac{1}{8\pi^2} \op e^{\gamma \phi} \, \mbox{Tr} \op\Bigl[\op
   \bigl( \tilde{\mathfrak m} + \eta^{-1} \op \tilde{\mathfrak m}^T \eta\op \bigr)^2 
  \op
  \Bigr] \,,
  \hspace{50pt}
  \tilde{\mathfrak m} = \mathsf V \op \mathfrak m \op \mathsf V^{-1} \,,
}
in which $\tilde{\mathfrak m}$ depends on the moduli fields $\Phi(x)$. 
Since above $\mathcal V(\hat x)$ was assumed to be real, 
the potential has an absolute minimum $V=0$ which is reached either for $\phi\to - \infty$
or for \cite{Dabholkar:2002sy}
\eq{
  \label{schsch_010}
 \eta \, \tilde{\mathfrak m} +\tilde{\mathfrak m}^T  \eta =0 \,.
} 
The situation described by \eqref{schsch_010} is more interesting as it fixes (some of) the moduli fields $\Phi(x)$. 
From the minimum-condition \eqref{schsch_010} we see that $r = \eta\, \tilde{\mathfrak m}_0$ is a generator of rotations
since it satisfies $r^T = - r$, where the subscript indicates that we are at the minimum 
of the potential. 
We can then determine from $\mathfrak m = \mathsf V_0^{-1} \eta^{-1} r\, \mathsf V_0$
the monodromy matrix $M = \mathsf V_0^{-1} \exp(\eta^{-1}\op r )\op \mathsf V_0$,
and we can show that in the minimum the matrix $\mathcal H$  in 
\eqref{schsch_003} is given by $\mathcal H_0 = \mathsf V_0^T \eta\op \mathsf V_0$. 
This implies now that at the minimum of the scalar potential, $\mathcal H_0$ is invariant under the 
monodromy group generated by $\mathfrak m$, that is
\eq{
  M^T \op \mathcal H_0 \op M = \mathcal H_0 \,,
}
and hence such a critical point of the potential is a fixed point under the monodromy group
\cite{Dabholkar:2002sy}.  
We finally remark the following:
\begin{itemize}

\item The minimum-condition \eqref{schsch_010} can also be expressed in 
terms of the full matrix $\mathcal H$ and the un-dressed Lie-algebra element 
$\mathfrak m$ as
\eq{
  \label{schsch_010b}
  \mathcal H \, {\mathfrak m} +{\mathfrak m}^T  \mathcal H  =0 \,.
}
In fact, the kinetic term \eqref{schsch_011} as well as the scalar potential \eqref{schsch_012}
can be formulated entirely in terms 
of $\mathcal H(\hat x) =  \exp\bigl[ \mathfrak m^T  y/2\pi \bigr] H(x)  \exp\bigl[ \mathfrak m \op y/2\pi \bigr]$.

\item For complex Lie-algebra elements $\mathfrak m$, one has to use $\mathcal H = \mathcal V^{\dagger} \eta \mathcal V$
instead of the expression given in \eqref{schsch_003}. The minimum condition
\eqref{schsch_010b} then for instance becomes $\mathcal H \, {\mathfrak m} +{\mathfrak m}^{\dagger}  \mathcal H  =0$.

\end{itemize}

%%%%%%%%%%%%%%%%%%%%%%%%%%%%%%%%%%%%%%%%%%%%%%%
%%%%%%%%%%%%%%%%%%%%%%%%%%%%%%%%%%%%%%%%%%%%%%%

\subsubsection*{$\mathbb T^2$-fibrations}
\label{page_schsch_t2}

Let us now connect our present analysis to our discussion in section~\ref{cha_torus_fib}. 
In particular, we want to describe the  effective theory obtained by compactifying 
string theory on $\mathbb T^2$-fibrations over a circle. 
We do this in two steps: 
\begin{enumerate}

\item First, we perform a Kaluza-Klein reduction of string theory on a two-torus from $D+3$ to $D+1$ dimensions.

\item In a second step we perform a Scherk-Schwarz reduction of the $(D+1)$-dimensional
theory on a circle from $D+1$  to $D$ dimensions.

\end{enumerate}
Let us therefore recall from equation \eqref{monodro_392} that the metric 
of a two-torus and the Kalb-Ramond $B$-field can be parametrised in terms of two complex scalar fields in 
the following way
\eq{
  \label{schsch_006}
  \arraycolsep3pt
  G_{\mathsf{ab}} =  \alpha' \,\frac{\rho_2}{\tau_2} \left( \begin{array}{c@{\hspace{8pt}}c} 
  \tau_1^2 + \tau_2^2 & \tau_1 \\[4pt] \tau_1 & 1 \end{array}\right) ,
  \hspace{50pt}
  B_{\mathsf{ab}} =   \alpha' \left( \begin{array}{c@{\hspace{8pt}}c} 
  0 & + \rho_1 \\[4pt] -\rho_1 & 0 \end{array}\right) ,
}
where $\mathsf a,\mathsf b = 1,2$, $\tau = \tau_1 + i\op\tau_2$ is the complex-structure modulus and 
$\rho= \rho_1 + i\op \rho_2$ denotes the complexified K\"ahler modulus. 
Now, when compactifying a $(D+3)$-dimensional theory on a two-torus with metric and $B$-field
\eqref{schsch_006},
we obtain a $(D+1)$-dimensional theory which contains the kinetic terms
of the moduli as
\eq{
  \label{schsch_007}
  \hat{\mathcal S}_{\mathbb T^2} \supset
   \frac{1}{2\op\hat\kappa^2} \int  \left[\,
   \frac{2}{\tau_2^2} \, d\tau\wedge\star d\ov\tau + \frac{2}{\rho_2^2 }\, d\rho\wedge \star d\ov\rho \,\right] .
}
Using then the generalised metric $\mathcal H$ defined in \eqref{gen_met_098}, we can bring 
\eqref{schsch_007} into the form
\eq{
  \label{schsch_009}
  \hat{\mathcal S}_{\mathbb T^2} \supset\frac{1}{2\op\hat\kappa^2} \int  \,
  \tfrac{1}{2}\op\mbox{Tr} \op\bigl( d\mathcal H^{-1}  \wedge\star d\mathcal H \bigr) \,,
}
which agrees precisely with the general expression shown in equation \eqref{schsch_003}.
Next, we perform a Scherk-Schwarz compactification of this $(D+1)$-dimensional theory on a circle to $D$ dimensions.
To do so, we note that \eqref{schsch_009} is in general invariant under 
global $GL(2D,\mathbb R)$ transformations acting on $\mathcal H$. However, 
as we discussed in section~\ref{sec_cft_torus}, in string theory this is broken 
to $O(D,D,\mathbb Z)$. 
Following the Scherk-Schwarz procedure discussed above, we now choose a vielbein $\mathcal V$
as in \eqref{schsch_002} with for instance
\eq{
   \mathsf V =\sqrt{\frac{\rho_2}{\tau_2}} 
  \renewcommand{\arraystretch}{1.3}
   \left( \begin{array}{cc|cc}
   \tau_2 & 0 & 0 & 0 \\
   \tau_1 & 1 & 0 & 0 \\ \hline
   -\frac{\rho_1}{\rho_2} \op\tau_1 & - \frac{\rho_1}{\rho_2} & \frac{1}{\rho_2} & - \frac{\tau_1}{\rho_2} \\
   +\frac{\rho_1}{\rho_2} \op \tau_2 & 0 & 0 & +\frac{\tau_2}{\rho_2}
   \end{array}
   \right),
   \hspace{50pt} \mathfrak m \in \mathfrak{so}(D,D) \,,
}
and $\eta$ is taken as the four-by-four identity matrix.
The potential in the $D$-dimensional theory has been determined in \eqref{schsch_008}, and it 
has a global minimum if the twisting $\mathfrak m$ satisfies \eqref{schsch_010} -- or 
in terms of $\mathcal H$ -- \eqref{schsch_010b}.

%%%%%%%%%%%%%%%%%%%%%%%%%%%%%%%%%%%%%%%%%%%%%%%
%%%%%%%%%%%%%%%%%%%%%%%%%%%%%%%%%%%%%%%%%%%%%%%

\subsubsection*{Examples}

Let us now discuss some examples for Scherk-Schwarz reductions of $\mathbb T^2$-fibrations
over the circle. We recall from equation \eqref{isos_849490} that the corresponding
$O(2,2,\mathbb Z)$ duality group splits into two $SL(2,\mathbb Z)$ factors acting on 
$\tau$ and $\rho$ and two $\mathbb Z_2$ factors. Since the latter are not connected
to the identity, we are not able to describe them in the present approach.
However, for $SL(2,\mathbb Z)$ this is possible, and the corresponding monodromies $M$ can be 
of  parabolic, elliptic or hyperbolic type 
\cite{Dabholkar:2002sy,Hull:2005hk}
(see also our discussion on page~\pageref{page_monodro_class}). We discuss these situations
 in turn:
\begin{itemize}

\item First, for parabolic monodromies in $\tau$ or in $\rho$ the condition \eqref{schsch_010}
cannot be satisfied and hence the potential has no minimum  \cite{Dabholkar:2002sy}.
Let us illustrate this result with the examples of the
three-torus with $H$-flux, the twisted three-torus and the three-torus T-fold discussed in section 
\eqref{sec_t2_fibr_example}.
We note that the Lie algebra elements $\mathfrak m$ corresponding to 
\eqref{monodro_01}, \eqref{monodro_02} and \eqref{monodro_03} are given by
\eq{
  &\mathfrak m_{\mathsf B} =
  \scalebox{0.8}{$
  \renewcommand{\arraystretch}{1.3}
   \left( \begin{array}{p{16pt}p{16pt}|p{16pt}p{16pt}@{\hspace{5pt}}p{0pt}@{}}
   \centering0 &  \centering0 &  \centering0 & \centering0 &\\
   \centering0 &  \centering0 &  \centering0 & \centering0 & \\ \hline
   \centering0 &  \centering$+h$ &  \centering0 & \centering0 & \\
   \centering$-h$ &  \centering0 &  \centering0 & \centering0 & 
   \end{array}
   \right)
   $},
   \hspace{50pt}
  \mathfrak m_{\mathsf A} =
  \scalebox{0.8}{$
  \renewcommand{\arraystretch}{1.3}
   \left( \begin{array}{p{16pt}p{16pt}|p{16pt}p{16pt}@{\hspace{5pt}}p{0pt}@{}}
   \centering0 &  \centering$+h$ &  \centering0 & \centering0 &\\
   \centering0 &  \centering0 &  \centering0 & \centering0 & \\ \hline
   \centering0 &  \centering0 &  \centering0 & \centering0 & \\
   \centering0 &  \centering0 &  \centering$-h$ & \centering0 & 
   \end{array}
   \right)
   $},
\\[10pt]
  &
  \mathfrak m_{\beta} =
  \scalebox{0.8}{$
  \renewcommand{\arraystretch}{1.3}
   \left( \begin{array}{p{16pt}p{16pt}|p{16pt}p{16pt}@{\hspace{5pt}}p{0pt}@{}}
   \centering0 &  \centering0 &  \centering0 & \centering$+h$ &\\
   \centering0 &  \centering0 &  \centering$-h$ & \centering0 & \\ \hline
   \centering0 &  \centering0 &  \centering0 & \centering0 & \\
   \centering0 &  \centering0 &  \centering0 & \centering0 & 
   \end{array}
   \right)
   $},
 }
respectively.
Each of these elements satisfies the minimum condition \eqref{schsch_010b} (with $\mathcal H$ computed from \eqref{schsch_006}
according to \eqref{gen_met_098}) only in some degeneration limit where for instance 
$\rho_2\to \infty$ or $|\tau|^2\to 0$. For finite values of the moduli the Scherk-Schwarz potential 
\eqref{schsch_008}  has no minimum. 
This shows that the family of three-tori with $H$-flux, geometric flux and $Q$-flux provides only 
toy-models.

\item For elliptic monodromies we mentioned on page~\pageref{page_sl2_classes} that these are of finite 
orders six, four and three. 
In all of these cases the minimum-condition \eqref{schsch_010b} has a finite solution, and 
hence the scalar potential has a minimum \cite{Dabholkar:2002sy}.
An overview on the Lie-algebra generators $\mathfrak m$ together with stabilised values of $\tau$ and $\rho$ 
for some combinations of monodromies can be found in table~\ref{table_schsch1}.
%%%%%%%%%%%%%%%%%%%%
%%%%%%%%%%%%%%%%%%%%
\begin{table}[p]
\centering
\scalebox{0.73}{
\tabcolsep8pt
\begin{tabular}{@{\hspace{10pt}}cc||l@{\hspace{2pt}}r||ll@{\hspace{10pt}}}
$\tau$-monodromy   & 
$\rho$-monodromy  & 
\multicolumn{2}{c||}{generator $\mathfrak m$ }&
\multicolumn{2}{c}{minimum of potential}
\\ \hline\hline
&&&&& \\[-8pt]
elliptic $4$ & $\varnothing$ &
$\mathfrak m = $ & $ \displaystyle \frac{\pi}{2}\,
\scalebox{0.9}{$
  \renewcommand{\arraystretch}{1.3}
   \left( \begin{array}{p{23pt}p{23pt}|p{23pt}p{23pt}@{\hspace{5pt}}p{0pt}@{}}
   \centering0 &  \centering$+1$ &  \centering0 & \centering0 &\\
   \centering$-1$ &  \centering0 &  \centering0 & \centering0 & \\ \hline
   \centering0 &  \centering0 &  \centering0 & \centering$+1$ & \\
   \centering0 &  \centering0 &  \centering$-1$ & \centering0 & 
   \end{array}
   \right)
   $}
$   
&
$\tau=+i $ &    
\\[32pt]
elliptic $6$ & $\varnothing$ &
$\mathfrak m = $ & $\displaystyle \frac{\pi}{3\sqrt{3}}
\scalebox{0.9}{$
  \renewcommand{\arraystretch}{1.3}
   \left( \begin{array}{p{23pt}p{23pt}|p{23pt}p{23pt}@{\hspace{5pt}}p{0pt}@{}}
   \centering$+1$ &  \centering$-2$ &  \centering0 & \centering0 &\\
   \centering$+2$ &  \centering$-1$ &  \centering0 & \centering0 & \\ \hline
   \centering0 &  \centering0 &  \centering$-1$ & \centering$-2$ & \\
   \centering0 &  \centering0 &  \centering$+2$ & \centering$+1$ & 
   \end{array}
   \right)
   $}
$   
&
$\displaystyle \tau=\frac{-1+i\sqrt{3}}{2} $ &    
\\[32pt]
$\varnothing$ & elliptic $4$ & 
$\mathfrak m =$ & $ \displaystyle \frac{\pi}{2}\,
\scalebox{0.9}{$
  \renewcommand{\arraystretch}{1.3}
   \left( \begin{array}{p{23pt}p{23pt}|p{23pt}p{23pt}@{\hspace{5pt}}p{0pt}@{}}
   \centering0 &  \centering0 &  \centering0 & \centering$+1$ &\\
   \centering0 &  \centering0 &  \centering$-1$ & \centering0 & \\ \hline
   \centering0 &  \centering$+1$ &  \centering0 & \centering0 & \\
   \centering$-1$ &  \centering0 &  \centering0 & \centering0 & 
   \end{array}
   \right)
   $}
$  
&&
$\rho=+i $ 
\\[32pt]
 $\varnothing$ & elliptic $6$ &
$\mathfrak m = $ & $\displaystyle \frac{2\pi}{3\sqrt{3}}
\scalebox{0.9}{$
  \renewcommand{\arraystretch}{1.3}
   \left( \begin{array}{p{23pt}p{23pt}|p{23pt}p{23pt}@{\hspace{5pt}}p{0pt}@{}}
   \centering$-1$ &  \centering0 &  \centering0 & \centering$+2$ &\\
   \centering0 &  \centering$-1$ &  \centering$-2$ & \centering0 & \\ \hline
   \centering0 &  \centering$+2$ &  \centering$+1$ & \centering0 & \\
   \centering$-2$ &  \centering0 &  \centering0 & \centering$+1$ & 
   \end{array}
   \right)
   $}
$   
&&
$\displaystyle \rho=\frac{-1+i\sqrt{3}}{2} $ 
\\[32pt]
elliptic $4$ & elliptic $4$ &
$\mathfrak m = $ & $ \displaystyle \frac{i\op\pi}{2}\,
\scalebox{0.9}{$
  \renewcommand{\arraystretch}{1.3}
   \left( \begin{array}{p{23pt}p{23pt}|p{23pt}p{23pt}@{\hspace{5pt}}p{0pt}@{}}
   \centering$+1$ &  \centering0 &  \centering$+1$ & \centering0 &\\
   \centering0 &  \centering$+1$ &  \centering0 & \centering$+1$ & \\ \hline
   \centering$+1$ &  \centering0 &  \centering$+1$ & \centering0 &\\
   \centering0 &  \centering$+1$ &  \centering0 & \centering$+1$ & 
   \end{array}
   \right)
   $}
$   
&
$\tau=+i $ & $\rho=+i$
\\[32pt]
elliptic $4$ & elliptic $6$ & 
$\mathfrak m =$ & $ \displaystyle \frac{\pi}{6}\,
\scalebox{0.9}{$
  \renewcommand{\arraystretch}{1.3}
   \left( \begin{array}{p{23pt}p{23pt}|p{23pt}p{23pt}@{\hspace{5pt}}p{0pt}@{}}
   \centering$+\frac{2}{\sqrt{3}}$ &  \centering$-3$ &  \centering0 & \centering$-\frac{4}{\sqrt{3}}$ &\\
   \centering$3$ &  \centering$+\frac{2}{\sqrt{3}}$ &  \centering$+\frac{4}{\sqrt{3}}$ & \centering0 & \\ \hline
   \centering0 &  \centering$-\frac{4}{\sqrt{3}}$ &  \centering$-\frac{2}{\sqrt{3}}$ & \centering$-3$ & \\
   \centering$+\frac{4}{\sqrt{3}}$ &  \centering0 &  \centering$3$ & \centering$-\frac{2}{\sqrt{3}}$ & 
   \end{array}
   \right)
   $}
$  
&$\tau=+i$ &
$\displaystyle \rho=\frac{-1+i\sqrt{3}}{2} $ 
\\[32pt]
elliptic $6$ & elliptic $4$ & 
$\mathfrak m =$ & $ \displaystyle \frac{\pi}{6}\,
\scalebox{0.9}{$
  \renewcommand{\arraystretch}{1.3}
   \left( \begin{array}{p{23pt}p{23pt}|p{23pt}p{23pt}@{\hspace{5pt}}p{0pt}@{}}
   \centering$+\frac{2}{\sqrt{3}}$ &  \centering$-\frac{4}{\sqrt{3}}$ &  \centering0 & \centering$+3$ &\\
   \centering$+\frac{4}{\sqrt{3}}$ &  \centering$-\frac{2}{\sqrt{3}}$ &  \centering$-3$ & \centering0 & \\ \hline
   \centering0 &  \centering$3$ &  \centering$-\frac{2}{\sqrt{3}}$ & \centering$-\frac{4}{\sqrt{3}}$ & \\
   \centering$-3$ &  \centering0 &  \centering$+\frac{4}{\sqrt{3}}$ & \centering$+\frac{2}{\sqrt{3}}$ & 
   \end{array}
   \right)
   $}
$  
&
$\displaystyle \tau=\frac{-1+i\sqrt{3}}{2} $ 
&$\rho=+i$ 
\end{tabular}
}
\caption{Overview of Lie-algebra generators $\mathfrak m$ and values of $\tau$ and $\rho$ satisfying the 
corresponding minimum condition \eqref{schsch_010b} for some combinations of elliptic monodromies. 
Note that in the case of elliptic monodromies of orders four in $\tau$ and $\sigma$ the 
Lie-algebra element $\mathfrak m$ is complex valued, and hence the minimum condition
reads $\mathcal H \, {\mathfrak m} +{\mathfrak m}^{\dagger}  \mathcal H  =0$.
\label{table_schsch1}}
\end{table}
%%%%%%%%%%%%%%%%%%%%
%%%%%%%%%%%%%%%%%%%%

\item For hyperbolic monodromies the analysis is rather involved. 
However, one can show that the  scalar potential \eqref{schsch_008} does not have a minimum 
specified by \eqref{schsch_010b} \cite{Dabholkar:2002sy}.
Simple examples for hyperbolic monodromies in $\tau$ are of the form (with $N\in\mathbb Z$ and $|N|>2$)
\eq{
  \tau \to N - \frac{1}{\tau} \,,
  \hspace{60pt}
  M = 
    \renewcommand{\arraystretch}{1.3}
   \left( \begin{array}{cc|cc}
   N & +1 & 0 & 0 \\
   -1 & 0 & 0 & 0 \\ \hline
   0 & 0 & 0 & +1 \\
   0 & 0 & -1 & N
   \end{array}
   \right).
}

\end{itemize}

%%%%%%%%%%%%%%%%%%%%%%%%%%%%%%%%%%%%%%%%%%%%%%%
%%%%%%%%%%%%%%%%%%%%%%%%%%%%%%%%%%%%%%%%%%%%%%%

\subsubsection*{Asymmetric orbifolds}

As we have argued above,  if the scalar potential \eqref{schsch_008} 
has a minimum characterised by \eqref{schsch_010b}, then at the minimum 
the monodromy $M$ leaves the generalised metric invariant, that is
\eq{
  M^{-T} \op \mathcal H_0 \op M^{-1} = \mathcal H_0 \,.
}
In other words, at this fixed point the background specified by a metric and $B$-field
is invariant under $M$. However, this does not imply that the action of the monodromy group on 
the fibre-coordinates $X^{\mathsf a}$ is trivial. More concretely, as explained in more detail in 
\cite{Dabholkar:2002sy}, if a discrete symmetry is gauged -- which is the case in our situation -- 
one should think of the resulting space as an orbifold construction.\footnote{
Orbifolds are manifolds subject to identifications under a discrete symmetry group. For instance, the circle $S^1$
can be seen as the freely-acting orbifold $S^1=\mathbb R/\mathbb Z$ where $\mathbb Z$ denotes the identification
of points $x\sim x+2\pi$. However, more general orbifold groups are possible, which do not need to 
be freely acting. For more information on orbifold constructions in string theory see for instance the 
original papers \cite{Dixon:1985jw,Dixon:1986jc} or for instance \cite{Kiritsis:2007zza}.\label{foot_orbifold}
}
To illustrate this point let us recall from section~\ref{sec_cft_torus} that
under transformations of the form \eqref{transf_003} the left- and right-moving 
coordinates in the lattice basis transform as \eqref{cft_torus_8494}.
In particular, we have for
\eq{
  M = \left( \begin{array}{cc} A & B \\ C & D \end{array}\right)
  \,,
  \hspace{50pt}
   \Omega_{\pm}=   A \pm \tfrac{1}{\alpha'} \op B\op(g\pm b) \,,
}  
that the coordinates transform as (cf. equation \eqref{cft_torus_8494})
\eq{
  X^i_R \to (\Omega_-)^i_{\hspace{4pt}j} \, X^j_R \,,
  \hspace{80pt}
  X^i_L \to  (\Omega_+)^i_{\hspace{4pt}j} \, X^j_L \,.
}
Now, for the sub-block $B$ in $M$ non-zero, we see that the left- and
right-moving coordinates transform in general differently. 
The corresponding orbifold is therefore an asymmetric orbifold, 
in which the group acts differently in the left- and right-moving sectors. 

As an example, let us consider the situation of elliptic monodromies of 
order four and six in $\tau$ and $\rho$, respectively (cf. table~\ref{table_schsch1}).
The corresponding monodromy element $M\in G$ takes the form
\eq{
  M = 
    \renewcommand{\arraystretch}{1.3}
   \left( \begin{array}{cc|cc}
   0 & +1 & +1 & 0 \\
   -1 & 0 & 0 & +1 \\ \hline
   +1 & 0 & 0 & 0 \\
   0 & +1 & 0 & 0
   \end{array}
   \right),
}
and at the minimum $(\tau_0,\rho_0) = (+i\op,\frac{-1+i\sqrt{3}}{2})$ the left- and right-moving 
coordinates (in the lattice basis) transform under $M$ as
\eq{
  \arraycolsep2pt
  \begin{array}{lcl@{\hspace{50pt}}lcl}
  X^{\mathsf a}_L &\to & \displaystyle
  \arraycolsep2pt
  \left( \begin{array}{rr} 
  \cos \phi_L & +\sin \phi_L \\ -\sin\phi_L & \cos\phi_L
  \end{array}\right)^{\mathsf a}_{\hspace{7pt}\mathsf b} X^{\mathsf b}_L \,, 
  & \phi_L &=& \displaystyle \frac{\pi}{6} \,,
  \\[18pt]
  X^{\mathsf a}_R &\to & \displaystyle
  \arraycolsep2pt
  \left( \begin{array}{rr} 
  \cos \phi_R & +\sin \phi_R \\ -\sin\phi_R & \cos\phi_R
  \end{array}\right)^{\mathsf a}_{\hspace{7pt}\mathsf b} X^{\mathsf b}_R \,, 
  & \phi_R &=& \displaystyle \frac{5\pi}{6} \,,
  \end{array}
}
with $\mathsf a,\mathsf b = 1,2$ labelling the coordinates of the $\mathbb T^2$.
Note that this action is of order twelve, and that the action on the left- and right-moving 
sector is indeed different.

%%%%%%%%%%%%%%%%%%%%%%%%%%%%%%%%%%%%%%%%%%%%%%%
%%%%%%%%%%%%%%%%%%%%%%%%%%%%%%%%%%%%%%%%%%%%%%%

\subsubsection*{Remarks}

We close this section on generalised Scherk-Schwarz reductions with the following remarks:
\begin{itemize}

\item Our discussion was focused on the bosonic sector of the theory. When 
considering superstring theory,  space-time fermions have to be 
included and in this case Scherk-Schwarz reductions generically break supersymmetry
\cite{Scherk:1978ta}. For a discussion of this result in the present context see for instance
\cite{Dabholkar:2002sy,Gray:2005ea}.

\item The Scherk-Schwarz compactifications discussed above lead to 
theories in which some of the symmetries are gauged. This can be seen for instance
from comparing the kinetic terms of the scalars in \eqref{schsch_011}  with
gauge-covariant derivative
\eq{
D\mathsf V = d\mathsf V + A \op\mathsf V \frac{\mathfrak m}{2\pi} \,,
}
with the general expression \eqref{sg_gauging_878929} of gauged supergravity theories. 
The gauging parameter is related to the matrix $\mathfrak m$,
and in regard to our discussion in 
sections~\ref{sec_cy_flux_echt} and \ref{sec_cy_flux_echt_orient} we 
note that $\mathfrak m$ encodes the geometric and non-geometric fluxes of the 
higher-dimensional theory \cite{Andrianopoli:2005jv}.

For Scherk-Schwarz reductions on $n$-dimensional tori $\mathbb T^n$ from $D+n$ to 
$D$ dimensions, the $(D+n)$-dimensional metric gives rise to $n$ gauge fields in 
$D$ dimensions. 
In this case there are $n$ monodromy generators $\mathfrak m$, which in general 
are non-commuting.

\item A discussion of the relation between asymmetric orbifolds at the fixed point of a 
monodromy
and non-geometric backgrounds
from a world-sheet perspective can be found in \cite{Flournoy:2005xe,Tan:2015nja} 
as well as in \cite{Lust:2010iy,Condeescu:2012sp,Condeescu:2013yma},
and we discuss this point in some more detail in section~\ref{sec_nca_nc}.

\end{itemize}

%%%%%%%%%%%%%%%%%%%%%%%%%%%%%%%%%%%%%%%%%%%%%%%
%%%%%%%%%%%%%%%%%%%%%%%%%%%%%%%%%%%%%%%%%%%%%%%
%%%%%%%%%%%%%%%%%%%%%%%%%%%%%%%%%%%%%%%%%%%%%%%
%%%%%%%%%%%%%%%%%%%%%%%%%%%%%%%%%%%%%%%%%%%%%%%

\subsection{Validity of solutions}
\label{sec_fluxes_val}

In the sections  above we have argued that non-geometric fluxes are a natural part of 
string theory. In particular, at the level of the \textit{theory} we have seen that 
\begin{enumerate}

\item non-geometric fluxes naturally combine  into a twisted differential and 
can be described using the framework of $SU(3)\times SU(3)$ structures.

\item We have also illustrated how non-geometric backgrounds can be incorporated into 
generalised Scherk-Schwarz reductions, in which  monodromies around compact directions
can contain T-duality transformations. 

\end{enumerate}
In this section we now want to discuss \textit{solutions} to theories
with non-geometric features, in particular their validity from an effective-field-theory point of 
view.\footnote{We consider a \textit{theory} to give rise 
to a set of equations of motion, and \textit{solutions} to a theory are 
solutions to the equations of motion.}

%%%%%%%%%%%%%%%%%%%%%%%%%%%%%%%%%%%%%%%%%%%%%%%
%%%%%%%%%%%%%%%%%%%%%%%%%%%%%%%%%%%%%%%%%%%%%%%

\subsubsection*{Flux compactifications I -- perturbing the background}

One way to approach compactifications of string theory on Calabi-Yau manifolds with fluxes, is
to start from a Calabi-Yau background without fluxes. Such 
configurations solve the string-equations of motion \eqref{eom_beta}
and have a vanishing scalar potential. In a second step one  perturbs these 
backgrounds by including fluxes, which in turn generates a potential. 
If these perturbations are small one can expect that minima of the potential 
correspond to small deformations of the Calabi-Yau background, which include the back-reaction 
of the fluxes on the geometry.

The quantity which encodes the perturbation generated by the fluxes is the flux density.
Indeed, for the case of $H$-flux we see that the $\beta$-functionals \eqref{eom_beta}
contain for instance a term $H_{ijk} G^{ii'} G^{jj'}G^{kk'}H_{i'j'k'}$ which is the flux-density 
squared. 
Let us illustrate this point for the 
example of the three-torus introduced in section~\ref{sec_first_steps}. 
For a rectangular three-torus with radii $R_1,R_2,R_3$ the densities 
of the $H$-flux $H_{123}$, geometric flux $f_{23}{}^1$, $Q$-flux $Q_3{}^{12}$ and $R$-flux $R^{123}$ read, respectively,
\eq{
  \label{fluxes_val_001}
  \arraycolsep1pt
  \begin{array}{l@{}l@{\hspace{50pt}}cc}
  H&\mbox{-flux density:} & 
  \displaystyle \frac{h}{2\pi}  & \displaystyle \frac{\alpha'}{R_1R_2R_3}  \,,
  \\[14pt]
  f&\mbox{-flux density:} & 
  \displaystyle \frac{f}{2\pi} & \displaystyle \frac{R_1}{R_2R_3}  \,,
  \\[14pt]
  Q&\mbox{-flux density:} & 
  \displaystyle \frac{q}{2\pi}  & \displaystyle \frac{R_1R_2}{\alpha'R_3}  \,,
  \\[14pt]
  R&\mbox{-flux density:} & 
  \displaystyle \frac{r}{2\pi}  & \displaystyle \frac{R_1R_2R_3}{\alpha'^2}  \,,
  \end{array}
}
where $h,f,q,r\in\mathbb Z$. From \eqref{fluxes_val_001} we see 
that in certain parameter regimes of the radii the flux-densities can be made small, 
however, not all densities can be made small 
at the same time. Let us consider two cases:
\begin{itemize}

\item In the scaling limit 
$R_1\sim \sqrt{\alpha'} L$, $R_2\sim \sqrt{\alpha'} L$, $R_3\sim\sqrt{\alpha'} L^3$ with $L\gg 1$,
the densities of the $H$-, $f$- and $Q$-flux  become small whereas the $R$-flux density becomes large.
The requirement of small perturbations can therefore be realised only for 
the first three fluxes but not for the $R$-flux. 
On the other hand, we note that due to the Bianchi identities \eqref{gg_bianchi_83948016} --
in particular $H_{ijk}R^{ijk}=0$ --
a simultaneous presence of $H$- and $R$-flux on the three-torus is excluded. 
  
\item As a second case let us also consider the scaling limit
$R_1\sim \sqrt{\alpha'} /L^3$, $R_2\sim \sqrt{\alpha'} L$, $R_3\sim\sqrt{\alpha'} L$ with $L\gg 1$.
Here the $f$-, $Q$- and $R$-flux densities become small whereas the $H$-flux density becomes
large. However, again the Bianchi identities for the fluxes do not allow for an $H$- and $R$-flux 
to be simultaneously present on a three-torus.

\end{itemize}
It is expected (but to our knowledge not investigated in detail) 
that this observation is a general feature: the Bianchi identities for the fluxes ensure that a scaling 
limit can be found in which all of the allowed flux-densities become small.

%%%%%%%%%%%%%%%%%%%%%%%%%%%%%%%%%%%%%%%%%%%%%%%
%%%%%%%%%%%%%%%%%%%%%%%%%%%%%%%%%%%%%%%%%%%%%%%

\subsubsection*{Flux compactifications II -- validity of approximation}

However, for backgrounds with non-geometric fluxes one generically faces the following issues:
\begin{itemize}

\item In string theory the length-scales of the compactification space (such as the radii of 
the three-torus) cannot be chosen by hand.  They correspond to moduli fields which have to be stabilised 
dynamically, and one has to ensure that in a specific model the stabilised values of these fields 
indeed lead to small flux-densities.

\item Furthermore, even though the flux-densities may be made small, in the presence of 
non-geometric fluxes typically some of the length-scales of the compactification 
space become small. For instance, in the second example above we have 
$R_1 \ll \sqrt{\alpha'}$. 
This implies that in general the supergravity approximation breaks down and 
that string-theoretical effects have to be taken into account. Hence,
naively such solutions are not trustworthy which
is a generic problem of non-geometric backgrounds.

\end{itemize}
These issues have to be addressed in order for a particular solution to be reliable.
We want to point out however that these are questions concerning the solutions of the 
theory -- at the level of the theory we have argued that non-geometric fluxes
are a natural part of string theory.

%%%%%%%%%%%%%%%%%%%%%%%%%%%%%%%%%%%%%%%%%%%%%%%
%%%%%%%%%%%%%%%%%%%%%%%%%%%%%%%%%%%%%%%%%%%%%%%

\subsubsection*{Scherk-Schwarz reductions}

For the Scherk-Schwarz reductions discussed in section~\ref{sec_schsch} the situation
is slightly different. 
To illustrate this point, let us first recall that for ordinary Kaluza-Klein reductions 
of a scalar field $\phi(x,y)$ on a circle of radius $R$,
one expands $\phi(x,y)$ in eigenfunctions of the Laplace operator on the circle as
\eq{
  \phi(x,y) = \sum_{n\in\mathbb Z} \phi_n(x) \op e^{i\frac{n y}{R}} \,.
}
Here $x^{\mu}$ denotes $D$-dimensional coordinates and $y\sim y+2\pi R$ parametrises the 
circle. The mass of the Kaluza-Klein modes $\phi_n(x)$ in $D$ dimensions can 
be determined from the Klein-Gordon equation 
as $m_n^2 = (n/R)^2$ (see for instance \cite{Blumenhagen:2013fgp} for a textbook
treatment). 
In string theory, we are now interested in the following limits:
\begin{enumerate}

\item We want to decouple the massive Kaluza-Klein modes $\phi_n(x)$ with $n\neq0$ and 
make them heavier than some observable energy-scale $m_{\rm obs}\sim 1/L_{\rm obs}$,
where $L_{\rm obs}$ is some minimal observable length-scale.
This implies that $R\ll L_{\rm obs}$ and hence the compact dimension is not
observed in experiments.

\item However, in the supergravity approximation employed for Kaluza-Klein reductions
of string theory, we also want to decouple higher string-excitations. We therefore 
require in addition that $\sqrt{\alpha'}\ll R$. 

\end{enumerate}

Let us now come to our discussion of string-theory compactifications 
on $\mathbb T^2$-fibrations over a circle
on page~\pageref{page_schsch_t2}. We have split this procedure 
into two steps: 1)  we performed an ordinary Kaluza-Klein compactification 
on $\mathbb T^2$ and kept only the massless modes, and 2) we performed a
Scherk-Schwarz reduction of these massless modes on a circle.
When determining the minima of the resulting scalar potential (for elliptic monodromies), we
observed that they correspond to fixed points of the monodromy group.
More concretely,
\begin{itemize}

\item from for instance table~\ref{table_schsch1} we see that typical values
for the stabilised K\"ahler and complex-structure moduli of the $\mathbb T^2$
satisfy $|\tau| \sim \mathcal O(1)$ and $|\rho| \sim \mathcal O(1)$. This means that the length-scales
$R_{\mathbb T^2}$
of the $\mathbb T^2$ are of order of the string length, i.e. $R_{\mathbb T^2} \sim \sqrt{\alpha'}$. 
Such values  however violate our second requirement from above $\sqrt{\alpha'}\ll R_{\mathbb T^2}$,
and therefore  string-effects have to be taken into account in order 
for these solutions to be reliable.

\item On the other hand, at the minimum of the potential the theory has an orbifold description.
For such orbifolds of $\mathbb T^2$ a CFT description exists \cite{Dabholkar:2002sy}, which 
includes all higher string-modes (at the perturbative level).
At the minimum we therefore have a string-theoretical description and  can trust this solution.

\end{itemize}
These examples of non-geometric Scherk-Schwarz reductions indicate, that finding 
reliable non-geometric solutions is rather a technical problem of controlling 
string-theory corrections and not a conceptual problem.

%%%%%%%%%%%%%%%%%%%%%%%%%%%%%%%%%%%%%%%%%%%%%%%
%%%%%%%%%%%%%%%%%%%%%%%%%%%%%%%%%%%%%%%%%%%%%%%
%%%%%%%%%%%%%%%%%%%%%%%%%%%%%%%%%%%%%%%%%%%%%%%
%%%%%%%%%%%%%%%%%%%%%%%%%%%%%%%%%%%%%%%%%%%%%%%

\subsection{Applications}
\label{sec_fluxes_app}

We have seen that when compactifying string theory on backgrounds with 
(non-geometric) fluxes, a potential 
for the moduli fields 
is generated in the lower-di\-men\-sio\-nal theory. This potential 
provides mechanisms to give masses to the moduli, 
to realise inflation
and to construct solutions with a positive cosmological constant.

%%%%%%%%%%%%%%%%%%%%%%%%%%%%%%%%%%%%%%%%%%%%%%%
%%%%%%%%%%%%%%%%%%%%%%%%%%%%%%%%%%%%%%%%%%%%%%%

\subsubsection*{Moduli stabilisation}

For compactifications of  string theory from ten to say four dimensions, 
the degrees of freedom of the metric, dilaton and 
$p$-form potentials along the compact directions appear as scalar 
and vector fields in the four-dimensional theory. 
The massive excitations can be ignored below a certain cut-off scale 
(typically the $m_{\rm obs}$ mentioned above), 
however, the massless excitations contribute to the lower-dimensional spectrum. 
These massless scalar fields are called moduli and parametrise 
deformations of the compactification background. 
For the example of Calabi-Yau compactifications discussed in section~\ref{sec_cy_flux},
the massless  fields have been summarised in table~\ref{table_n2_fields}.

From a phenomenological point of view, the presence of massless particles 
during the early universe can modify the abundances 
of hydrogen and helium and thereby destroy the 
very successful predictions of big bang nucleosynthesis.
Massless scalar fields (apart from the Higgs field) furthermore give rise to fifth forces, which 
are highly constrained by experiment.
Moduli fields should therefore acquire a mass, which is known as moduli stabilisation
(for reviews see for instance \cite{Grana:2005jc,Blumenhagen:2006ci}).
Moduli can be stabilised by generating a scalar potential for them and
-- as we have seen in the above sections -- fluxes achieve this requirement. 
Let us briefly discuss moduli stabilisation for type IIB and type IIA orientifold
compactifications:
\begin{itemize}

\item As we explained in section~\ref{sec_cy_flux_echt_orient}, for type IIB 
orientifolds with O3-/O7-planes the superpotential \eqref{sg_824827444} contains 
couplings between the  $H$-flux and the
axio-dilaton $\tau$, between the geometric $F$-flux and the moduli $G^{\hat{\mathsf A}}$ and 
between the non-geometric $Q$-flux and the K\"ahler moduli $T_{\mathsf A}$. 
If all fluxes compatible with the Bianchi identities are non-vanishing, generically all moduli 
fields appear in the scalar potential and
will receive a mass in the four-dimensional theory.

Let us emphasise that this is an interesting result: 
for vanishing $Q$-flux the K\"ahler moduli do not appear 
in the superpotential $W$ and therefore remain massless at the perturbative level. 
On the other hand, non-perturbative corrections to $W$ coming from D-brane instantons or gaugino 
condensates can generate a dependence of the superpotential on 
the K\"ahler moduli, which leads to the  KKLT \cite{Kachru:2003aw}
or large-volume  \cite{Balasubramanian:2005zx} scenarios.
In these approaches the K\"ahler moduli are stabilised through non-perturbative effects,
whereas the $Q$-flux allows to stabilise K\"ahler moduli perturbatively.

\item For type IIB orientifolds with O5-/O7-planes the situation is slightly different
\cite{Benmachiche:2006df}. The four-dimensional complex scalar fields 
take a different form as compared to type IIB with O3-/O7-planes, but in the superpotential 
the K\"ahler moduli couple to the geometric $F$-flux, 
the analogues of the \raisebox{0pt}[0pt]{$G^{\hat{\mathsf A}}$}-moduli couple to the 
non-geometric $Q$-flux and the axio-dilaton couples to the non-geometric 
$R$-flux \cite{Villadoro:2006ia}.

\item For type IIA orientifolds the fixed loci of the orientifold projection are O6-planes. In this setting the 
NS-NS fluxes couple to the complex-structure moduli, as expected from mirror symmetry. 
For more details we refer the reader 
to \cite{Benmachiche:2006df}.

\end{itemize}
Let us now give an overview of string-theory constructions with non-geometric fluxes,
where moduli have been stabilised.\label{page_lit_sol}
\begin{itemize}

\item On toroidal compactification backgrounds moduli stabilisation 
using non-geometric fluxes has been studied in 
\cite{Shelton:2005cf, 
Aldazabal:2006up,
Villadoro:2006ia,
Shelton:2006fd}.
In particular, systematic studies of allowed flux combinations and resulting 
solutions  can be found in
\cite{Font:2008vd,
Guarino:2008ik,
deCarlos:2009fq,
deCarlos:2009qm,
Aldazabal:2011yz, 
Dibitetto:2011gm,
Damian:2013dwa}.

\item On more general Calabi-Yau or $SU(3)$-structure backgrounds  moduli stabilisation using non-geometric fluxes  
has been studied
for instance in \cite{Micu:2007rd,Palti:2007pm,Blumenhagen:2015kja}. In \cite{Micu:2007rd} it was argued that non-geometric flux-vacua in a par\-a\-met\-rically-controlled 
regime can be constructed. 

\end{itemize}

%%%%%%%%%%%%%%%%%%%%%%%%%%%%%%%%%%%%%%%%%%%%%%%
%%%%%%%%%%%%%%%%%%%%%%%%%%%%%%%%%%%%%%%%%%%%%%%

\subsubsection*{Inflation}

There are strong experimental indications that our universe underwent a period of inflation in which 
it expanded rapidly (see for instance \cite{Baumann:2014nda} for a review).
In order to realise inflation a non-trivial potential for a scalar field has to be generated, 
which satisfies certain slow-roll conditions. We
do not want to go into further details here, but only mention that 
background fluxes can give rise to such potentials.
Using non-geometric fluxes, this has been studied for instance in 
\cite{Damian:2013dq,Hassler:2014mla,Blumenhagen:2015qda,Blumenhagen:2015xpa}.

%%%%%%%%%%%%%%%%%%%%%%%%%%%%%%%%%%%%%%%%%%%%%%%
%%%%%%%%%%%%%%%%%%%%%%%%%%%%%%%%%%%%%%%%%%%%%%%

\subsubsection*{De Sitter vacua}

In \cite{Maldacena:2000mw} a no-go theorem has been formulated
which -- under certain conditions -- forbids vacua with a positive 
cosmological constant. In particular, string theory with usual 
geometric fluxes does not allow for de Sitter vacua. 
However, non-geometric fluxes violate the assumptions of 
\cite{Maldacena:2000mw}  and are therefore believed to 
circumvent the no-go theorem.

Explicit constructions of de Sitter vacua using non-geometric fluxes 
can be found for instance in the papers 
\cite{Hertzberg:2007wc,
deCarlos:2009fq,
deCarlos:2009qm,
Dibitetto:2010rg,
Dibitetto:2011gm,
Danielsson:2012by,
Blaback:2013ht,
Blaback:2013qza,
Hassler:2014mla,
Blaback:2015zra,
Blumenhagen:2015xpa,
Junghans:2016abx
}.
However, as we have mentioned in section~\ref{sec_fluxes_val}, typically 
the validity of such solutions is difficult to show. 
This may be in accordance with the very recent de Sitter conjectures  
\cite{Obied:2018sgi,Ooguri:2018wrx}, which exclude meta-stable de Sitter 
vacua in any theory of quantum gravity.

%%%%%%%%%%%%%%%%%%%%%%%%%%%%%%%%%%%%%%%%%%%%%%%
%%%%%%%%%%%%%%%%%%%%%%%%%%%%%%%%%%%%%%%%%%%%%%%
%%%%%%%%%%%%%%%%%%%%%%%%%%%%%%%%%%%%%%%%%%%%%%%
%%%%%%%%%%%%%%%%%%%%%%%%%%%%%%%%%%%%%%%%%%%%%%%

\subsection{Summary}

Let us close our discussion of type II string-theory compactifications with a brief summary 
of the main points discussed in this section:
\begin{itemize}

\item In section~\ref{sec_sugra_recap} we have reviewed some basic results of
$\mathcal N=2$ and $\mathcal N=1$ supergravity theories in four dimensions. 
We found that these theories are characterised by only a few quantities.

For $\mathcal N=2$ theories these are a K\"ahler potential $\mathcal K$ (often with a corresponding 
pre-potential $\mathcal F$) which describes vector-multiplets, and a quaternionic-K\"ahler metric 
$h_{uv}$ 
describing hyper-multiplets. For Calabi-Yau compactifications the latter contains
a K\"ahler sub-manifold, which is described again by a K\"ahler potential. 
A scalar potential is generated by moment maps $\mathcal P$, which correspond to
gaugings of vector- and hyper-multiplet isometries. 

For the $\mathcal N=1$ theory the vector-multiplets are characterised in terms of a
holomorphic gauge kinetic function $f$,  and the chiral multiplets are described by a
K\"ahler potential $\mathcal K$. A scalar potential is generated by a holomorphic superpotential $W$
as well as by gaugings of isometries.

\item In section~\ref{sec_cy_flux} we then compactified type IIB string theory on a Calabi-Yau 
three-fold (without fluxes). The resulting theory is a $\mathcal N=2$ supergravity in 
four dimensions. The relevant supergravity quantities of this theory, such as 
the K\"ahler potential and the moment maps, can be expressed 
using the framework of generalised geometry. In particular, two generalised 
spinors $\Phi^{\pm}$ determine two K\"ahler potentials $\mathcal K^{\pm}$ 
and the moment maps $\mathcal P$.

\item In section~\ref{sec_cy_flux_echt} we then discussed how non-trivial $O(D,D)$ 
transformations can generate geometric as well as geometric fluxes. The effect of these
transformations is encoded in a so-called twisted differential $\mathcal D$, which 
contains the various fluxes. These fluxes give rise to gaugings in the 
$\mathcal N=2$ theory.

We furthermore mentioned mirror symmetry, which is a symmetry between type IIB and type IIA compactifications
that exchanges 
geometric and non-geometric flux-components. This shows that 
non-geometric fluxes are an integral part of flux compactifications.

\item In section~\ref{sec_cy_flux_echt_orient} we discussed the orientifold projection of 
type IIB compactions from $\mathcal N=2$ to $\mathcal N=1$ theories. Here, 
the gaugings of the $\mathcal N=2$ theory are split into an F-term contribution contained in
the generalisation of the Gukov-Vafa-Witten superpotential, and a D-term potential.

\item In section~\ref{sec_schsch} we  studied generalised Scherk-Schwarz reductions.
This analysis connects to our discussion of torus fibrations in section~\ref{cha_torus_fib},
and we have described how the scalar potential of such compactifications can be
obtained. The properties of the minimum have been analysed and we observed 
that, if the minimum exists, it is in general  described by asymmetric orbifold constructions.

\item In section~\ref{sec_fluxes_val} we have discussed the validity of non-geometric 
solutions. We have recalled that non-geometric fluxes are a 
natural part of string theory, however, non-geometric solutions (i.e. minima of the potential)
typically violate the supergravity approximation. They are therefore naively not reliable. 
On the other hand, for certain non-geometric Scherk-Schwarz reductions
the minimum of the potential has a CFT description and such solutions can therefore
be trusted.

\item In section~\ref{sec_fluxes_app} we gave an overview on applications of 
non-geometric fluxes to moduli stabilisation, realising inflation and constructing 
de Sitter vacua in string theory.

\end{itemize}

%%%%%%%%%%%%%%%%%%%%%%%%%%%%%%%%%%%%%%%%%%%%%%%
%%%%%%%%%%%%%%%%%%%%%%%%%%%%%%%%%%%%%%%%%%%%%%%
%%%%%%%%%%%%%%%%%%%%%%%%%%%%%%%%%%%%%%%%%%%%%%%
%%%%%%%%%%%%%%%%%%%%%%%%%%%%%%%%%%%%%%%%%%%%%%%
%%%%%%%%%%%%%%%%%%%%%%%%%%%%%%%%%%%%%%%%%%%%%%%
%%%%%%%%%%%%%%%%%%%%%%%%%%%%%%%%%%%%%%%%%%%%%%%
%%%%%%%%%%%%%%%%%%%%%%%%%%%%%%%%%%%%%%%%%%%%%%%
%%%%%%%%%%%%%%%%%%%%%%%%%%%%%%%%%%%%%%%%%%%%%%%
%%%%%%%%%%%%%%%%%%%%%%%%%%%%%%%%%%%%%%%%%%%%%%%
%%%%%%%%%%%%%%%%%%%%%%%%%%%%%%%%%%%%%%%%%%%%%%%
%%%%%%%%%%%%%%%%%%%%%%%%%%%%%%%%%%%%%%%%%%%%%%%
%%%%%%%%%%%%%%%%%%%%%%%%%%%%%%%%%%%%%%%%%%%%%%%
%%%%%%%%%%%%%%%%%%%%%%%%%%%%%%%%%%%%%%%%%%%%%%%
%%%%%%%%%%%%%%%%%%%%%%%%%%%%%%%%%%%%%%%%%%%%%%%
%%%%%%%%%%%%%%%%%%%%%%%%%%%%%%%%%%%%%%%%%%%%%%%
%%%%%%%%%%%%%%%%%%%%%%%%%%%%%%%%%%%%%%%%%%%%%%%

\clearpage
\section{Doubled geometry}
\label{cha_dg}

In this section we give an introduction to doubled geometry \cite{Hull:2004in,Dabholkar:2005ve,Hull:2006va}. In this  approach to T-duality
and non-geometric backgrounds,
a world-sheet action invariant under the duality group $O(D,D,\mathbb Z)$ is 
constructed. In section~\ref{sec_dg_tor} we motivate such two-dimensional theories
from the example of toroidal compactifications, and in section~\ref{sec_dg_tf}
we give a more general derivation following \cite{Hull:2006va}.
In section~\ref{sec_dft} we comment on a target-space approach to a doubled geometry, 
called double field theory  \cite{Hull:2009mi,Hohm:2010pp,Hohm:2010jy}.

%%%%%%%%%%%%%%%%%%%%%%%%%%%%%%%%%%%%%%%%%%%%%%%
%%%%%%%%%%%%%%%%%%%%%%%%%%%%%%%%%%%%%%%%%%%%%%%
%%%%%%%%%%%%%%%%%%%%%%%%%%%%%%%%%%%%%%%%%%%%%%%
%%%%%%%%%%%%%%%%%%%%%%%%%%%%%%%%%%%%%%%%%%%%%%%

\subsection{Toroidal compactifications}
\label{sec_dg_tor}

We start this section by introducing doubled geometry through the example of toroidal compactifications 
with constant Kalb-Ramond field.
We work with a world-sheet theory with flat world-sheet metric of Lorentzian signature,
and we follow the original papers \cite{Hull:2004in,Hull:2006va} where further details can be found.

%%%%%%%%%%%%%%%%%%%%%%%%%%%%%%%%%%%%%%%%%%%%%%%
%%%%%%%%%%%%%%%%%%%%%%%%%%%%%%%%%%%%%%%%%%%%%%%

\subsubsection*{Doubled coordinates}

Let us recall from page~\pageref{page_odd_trans_003} that on a $D$-dimensional 
torus with constant Kalb-Ramond
field the T-duality group acts as $O(D,D,\mathbb Z)$ transformations.
Using the formulas \eqref{cft_torus_8494} and \eqref{dual_001}, for the left- and right-moving 
target-space coordinates $X^i_{L}$ and $X^i_R$ 
this means that they transform as
\eq{ 
  \label{db_993}
  \arraycolsep2pt
  \left(\begin{array}{c}
  \tilde X_L \\
  +\frac{1}{\alpha'} (\tilde g + \tilde b) \tilde X_L 
  \end{array}
  \right)
  &=
  \arraycolsep2pt
   \left( \begin{array}{cc} A & B \\ C & D \end{array}\right)
  \left(\begin{array}{c}
  X_L \\
  +\frac{1}{\alpha'} (g + b)  X_L 
  \end{array}
  \right) ,
\\[10pt]
  \arraycolsep2pt
  \left(\begin{array}{c}
  \tilde X_R \\
  -\frac{1}{\alpha'} (\tilde g - \tilde b) \tilde X_R 
  \end{array}
  \right)
  &=
  \arraycolsep2pt
   \left( \begin{array}{cc} A & B \\ C & D \end{array}\right)
  \left(\begin{array}{c}
  X_R \\
  -\frac{1}{\alpha'} (g - b)  X_R 
  \end{array}
  \right) ,
}
where the $O(D,D,\mathbb Z)$ transformation is parametrised in terms 
of $D\times D$ matrices $A$, $B$, $C$, $D$ as in \eqref{notation_odd_01}. Here and in the following 
matrix multiplication is  understood. 
Next, as we explained around equation \eqref{odd_009}, a T-duality transformation 
along all $D$ directions of the torus corresponds to a transformation parametrised by the $O(D,D,\mathbb Z)$ matrix 
\eq{
   \mathcal O_+ = \left( \begin{array}{@{\hspace{8pt}}cc} 0 &  \delta^{-1} \\  \delta & 0 \end{array}\right),
}
where for notational convenience we chose the positive sign.
From \eqref{db_993} we can then infer that the fully-dualised coordinates  take the form
\eq{
  \tilde X^i = \tilde X^i_L + \tilde X^i_R = \Bigl[ \tfrac{1}{\alpha'}\op \delta^{-1}\op (g+b) X_L
  - \tfrac{1}{\alpha'}\op \delta^{-1} \op (g-b) X_R \Bigr]^i \,,
}
with $i=1,\ldots, D$.
Motivated by this result, we define an additional set of coordinates as
\eq{
  \label{dg_872848}
  \hat X_i =  \tfrac{1}{\alpha'}\op  (g+b)_{ij} X^j_L
  - \tfrac{1}{\alpha'} \op (g-b)_{ij} X^j_R \,,
}
which together with the original coordinates $X^i = X^i_L + X^i_R$ transform under 
$O(D,D,\mathbb Z)$ in the following way (cf. equation \eqref{db_993})
\eq{
  \arraycolsep2pt
  \left(\begin{array}{c}
  \tilde X^i \\
  \tilde {\hat X}_i
  \end{array}
  \right)
  =
   \left( \begin{array}{cc} A^i{}_j & B^{ij} \\ C_{ij} & D_i{}^j \end{array}\right)
  \left(\begin{array}{c}
  X^j \\
  {\hat X}_j
  \end{array}
  \right).
}
It is then convenient to introduce a set of doubled coordinates $\mathbb X^I$ with $I=1,\ldots, 2D$ 
and their transformation 
under $\mathcal O \in O(D,D,\mathbb Z)$ as
\begin{align}
  \label{dg_297a}
  &\arraycolsep2pt
  \mathbb X^I =   \left(\begin{array}{c}
  X^i \\
  {\hat X}_i
  \end{array}
  \right) 
  =
    \left(\begin{array}{c}
  X^i_L + X^i_R \\[2pt]
  \tfrac{1}{\alpha'}\op  (g+b)_{ij} X^j_L
  - \tfrac{1}{\alpha'} \op (g-b)_{ij} X^j_R
  \end{array}
  \right) 
  \\[10pt]
  \label{dg_297b}
  &
  \tilde{\mathbb X}^I = \mathcal O^I{}_J \mathbb X^J \,.
\end{align}
The definition of the coordinates $\mathbb X^I$ can also be motivated by considering the 
tachyon vertex operator for the closed string compactified on a torus.
(For an introduction to vertex operators in two-dimensional CFTs see for instance \cite{Blumenhagen:2009zz}.)
Denoting normal ordering by $:\!\ldots\! :$ 
and recalling the form of the left- and right-moving momenta from \eqref{momenta_001}, we have
\eq{
  {\cal V} &= \, :\!\exp \,\Bigl( \op i\op p_L\cdot   X_L +
  i\op p_R \cdot X_R \op \Bigr) \!: 
  \\[4pt]
  &=\, :\!\exp \left(\,\frac{1}{2} \op m_i \op [X_L + X_R]^i 
  + \frac12 \op n^i \op\left[  \tfrac{1}{\alpha'}\op  (g+b) X_L
  - \tfrac{1}{\alpha'} \op (g-b) X_R \right]_i\op
  \right)\!:
  \\[4pt]
  &=\, :\!\exp \left(\,\frac{1}{2} \op m_i\op \mathbb X^i  + \frac{1}{2} \op n^i\op \mathbb X_i \right) \!: \,,
}
where $m_i,n^i\in\mathbb Z$ are the momentum and winding numbers. 
Note that this expression is invariant under $O(D,D,\mathbb Z)$ transformations
of the form \eqref{odd_001} provided that $\mathbb X^I$ transforms as in \eqref{dg_297b}. 
We also mention that $X^i$ are the coordinates conjugate to the momentum numbers $m_i$, while
the dual coordinates $\hat X_i$ are dual to the winding numbers $n^i$. The $\hat X_i$  are
therefore also called winding coordinates.

%%%%%%%%%%%%%%%%%%%%%%%%%%%%%%%%%%%%%%%%%%%%%%%
%%%%%%%%%%%%%%%%%%%%%%%%%%%%%%%%%%%%%%%%%%%%%%%

\subsubsection*{Constraint}

From the definition of the doubled coordinates $\mathbb X^I$ shown in \eqref{dg_297a}
we see that $X^i$ and $\hat X_i$ contain similar information. We
can make this more precise by 
separating the left- and right-moving sectors through the following relations
\eq{
  \label{dg_9927}
  &\eta^{-1} \mathcal H \op\partial_+ \mathbb X = \hphantom{+}\partial_+ \mathbb X \,,
  \\[2pt]
  &\eta^{-1} \mathcal H \op\partial_- \mathbb X = -\partial_- \mathbb X \,,  
}
where matrix multiplication is understood and where $\eta$ and  the generalised 
metric $\mathcal H$ were given in \eqref{gen_met_098}. We furthermore employed   
that the metric and $B$-field are constant as well as that $X_L^i \equiv X_L^i(\sigma^+)$
and  $X_R^i \equiv X_R^i(\sigma^-)$ with $\sigma^{\pm} = \tau \pm \sigma$. 
Using the Hodge star-operator on a two-dimensional flat world-sheet with 
Lorentzian signature we have $\star d\sigma^{\pm} = \pm d\sigma^{\pm}$,
which allows us to express \eqref{dg_9927} as
\eq{
  \label{dg_9927b}
   d\mathbb X = \eta^{-1}\mathcal H \star d\mathbb X \,.
}
Consistency of this constraint  requires that 
$(\eta^{-1} \mathcal H)^2 = \mathds 1$, which is indeed satisfied for the generalised metric 
$\mathcal H$. 
To conclude, we see that the doubled coordinates $\mathbb X^I$ have to satisfy a
self-duality relation.

%%%%%%%%%%%%%%%%%%%%%%%%%%%%%%%%%%%%%%%%%%%%%%%
%%%%%%%%%%%%%%%%%%%%%%%%%%%%%%%%%%%%%%%%%%%%%%%

\subsubsection*{World-sheet action}

We have seen above that the doubled coordinates $\mathbb X^I$ transform covariantly under 
the duality group
$O(D,D,\mathbb Z)$, and we now want to construct a world-sheet action
which is invariant under such duality transformations. 
A natural guess is the following expression \cite{Hull:2004in}
\eq{
  \label{dg_9927c}
  \mathcal S = -\frac{1}{4\pi} \int_{\Sigma}\op \Bigl[  \op
   \tfrac{1}{2}\op \mathcal H_{IJ} \,d\mathbb X^I \wedge\star d\mathbb X^J 
   + \Omega_{IJ} \op d\mathbb X^I \wedge d\mathbb X^J
   \op\Bigr]
   \,,
}
where $\mathcal H$ denotes again the generalised metric \eqref{gen_met_098}. 
The topological term corresponding to an anti-symmetric 
matrix $\Omega_{IJ}$  has been introduced for later use and does not affect the dynamics.
The $2D$ components of $\mathbb X^I$ appearing in \eqref{dg_9927c}
are considered to be independent fields, which are 
however subject to the duality condition \eqref{dg_9927b}
to be imposed on the equations of motion. 
The action \eqref{dg_9927c} is then 
invariant under the following $O(D,D,\mathbb Z)$ transformations
\eq{
    \mathbb X \to \mathcal O\op  \mathbb X  \,,
    \hspace{40pt}
     \mathcal H \to  \mathcal O^{-T} \mathcal H \,\mathcal O^{-1}  \,,
    \hspace{40pt}
     \Omega \to  \mathcal O^{-T} \Omega \,\mathcal O^{-1}  \,,
}
which agrees with the transformation behaviour 
of the coordinates $\mathbb X^I$ shown in \eqref{dg_297b} and with that
of the generalised metric given in \eqref{transf_003}.

Turning now to the equations of motion for the doubled action \eqref{dg_9927c}, we see that they take the form
\eq{
  d\bigl( \mathcal H_{IJ} \star d \mathbb X^J \bigr) = 0 \,.
}
When imposing the constraint \eqref{dg_9927b} they
are automatically satisfied, which means that the dynamics is governed by the constraint.

%%%%%%%%%%%%%%%%%%%%%%%%%%%%%%%%%%%%%%%%%%%%%%%
%%%%%%%%%%%%%%%%%%%%%%%%%%%%%%%%%%%%%%%%%%%%%%%

\subsubsection*{Equivalence with the standard formulation}

We now want to show that the doubled world-sheet action \eqref{dg_9927c} is 
equivalent to the standard formulation \eqref{action_01} \cite{Hull:2006va}. 
We follow an approach similar to our discussion of the Buscher rules in 
section~\ref{sec_buscher},
and first expand  \eqref{dg_9927c}  as follows
\eq{
  \mathcal S = -\frac{1}{4\pi} \int_{\Sigma}\op \Bigl[  \hspace{17pt}
  &   \tfrac{1}{2} \op \tfrac{1}{\alpha'} g_{ij}\op  d X^i \wedge\star d X^j 
  \\
   +\,& \tfrac{1}{2} \op\alpha'  g^{ij} \bigl( d\hat X - \tfrac{1}{\alpha'}\op b \op dX \bigr)_i
   \wedge \star \bigl( d\hat X - \tfrac{1}{\alpha'}\op b \op dX \bigr)_j
   \\[2pt]
   +\,&  d\hat X_i  \wedge d X^i
   \hspace{165pt}\Bigr]
   \,,
}
where we made a specific choice for the matrix $\Omega_{IJ}$.   
This action has a number of global symmetries, for instance it is invariant under
$\hat X_i \to \hat X_i + \epsilon_i$ for $\epsilon_i = {\rm const.}$
In a second step we make this global symmetry local by introducing 
one-form world-sheet gauge fields $C_i$. This leads to the gauged action\label{page_dg_gauging}
\eq{
  \label{dg_9927d}
  \hat{\mathcal S} = -\frac{1}{4\pi} \int_{\Sigma}\op \Bigl[  \hspace{17pt}
  &   \tfrac{1}{2} \op \tfrac{1}{\alpha'} \op g_{ij}\,  d X^i \wedge\star d X^j 
  \\
   +\,& \tfrac{1}{2} \op\alpha'  g^{ij} \bigl( d\hat X + C- \tfrac{1}{\alpha'}\op b \, dX \bigr)_i
   \wedge \star \bigl( d\hat X + C- \tfrac{1}{\alpha'}\op b \, dX \bigr)_j
   \\[2pt]
   +\,&  \bigl( d\hat X + C\bigr)_i  \wedge d X^i
   \hspace{180pt}\Bigr]
   \,,
}
which, since $g_{ij}$ and $b_{ij}$ are assumed to be constant, is invariant under the local transformations
\eq{
  \hat X_i \to \hat X_i + \epsilon_i\,,
  \hspace{60pt}
  C_i \to C_i - d\epsilon_i \,,
}
for  $\epsilon_i$ depending on the world-sheet coordinates. 
The last step for showing the equivalence with the standard formulation
is  to integrate out the gauge fields $C_i$. 
The solutions to the equations of motion for the latter are determined as
\eq{
  C_i = - d\hat X_i - \frac{1}{\alpha'} \op g_{ij} \star dX^j + \frac{1}{\alpha'} \op b_{ij} \op dX^j \,,
}
and using them in  \eqref{dg_9927d} gives the standard expression \eqref{action_01}
(up to the dilaton contribution to be discussed below)
\eq{
  \check{\mathcal S} = -\frac{1}{4\pi \alpha'} \int_{\Sigma}
  \Bigl[ \,g_{ij} \op d X^{i}\wedge\star d X^{j} - b_{ij} \op dX^{i}\wedge dX^{j}
  \, \Bigr] 
  \,.
}
This shows that the doubled world-sheet action \eqref{dg_9927c} is equivalent to the 
usual string-theory action.

%%%%%%%%%%%%%%%%%%%%%%%%%%%%%%%%%%%%%%%%%%%%%%%
%%%%%%%%%%%%%%%%%%%%%%%%%%%%%%%%%%%%%%%%%%%%%%%

\subsubsection*{Polarisation and T-duality}

In the doubled world-sheet action \eqref{dg_9927c} the $2D$ coordinates $\mathbb X^I$ are 
considered to be independent of each other. From the identification 
\eqref{cft_t_840934} we 
know already that the ordinary coordinates $X^i$ parametrise a $D$-dimensional torus $\mathbb T^D$, 
but using the mode expansion \eqref{cft_003} we can similarly 
determine an identification for the dual coordinates $\hat X_i$. Together they read
\eq{
  \label{dg_240091}
  \arraycolsep2pt
  \begin{array}{lcl@{\hspace{50pt}}lcl}
  X^i &\sim& X^i + 2\pi \op n^i \,, & n^i & \in& \mathbb Z \,,
  \\[4pt]
  \hat X_i &\sim& \hat X_i + 2\pi \op m_i \,, & m_i & \in& \mathbb Z \,,
  \end{array}
}
where $n^i$ and $m_i$ denote the momentum and winding numbers of the closed string. 
We therefore see that the doubled coordinates $\mathbb X^I$ parametrise a
$2D$-dimensional doubled torus $\mathbb T^{2D} = \mathbb T^D \times \hat{\mathbb T}^D$.

\begin{figure}[t]
\centering
\begin{subfigure}{100pt}
\centering
\hspace*{5pt}
\includegraphics[width=90pt]{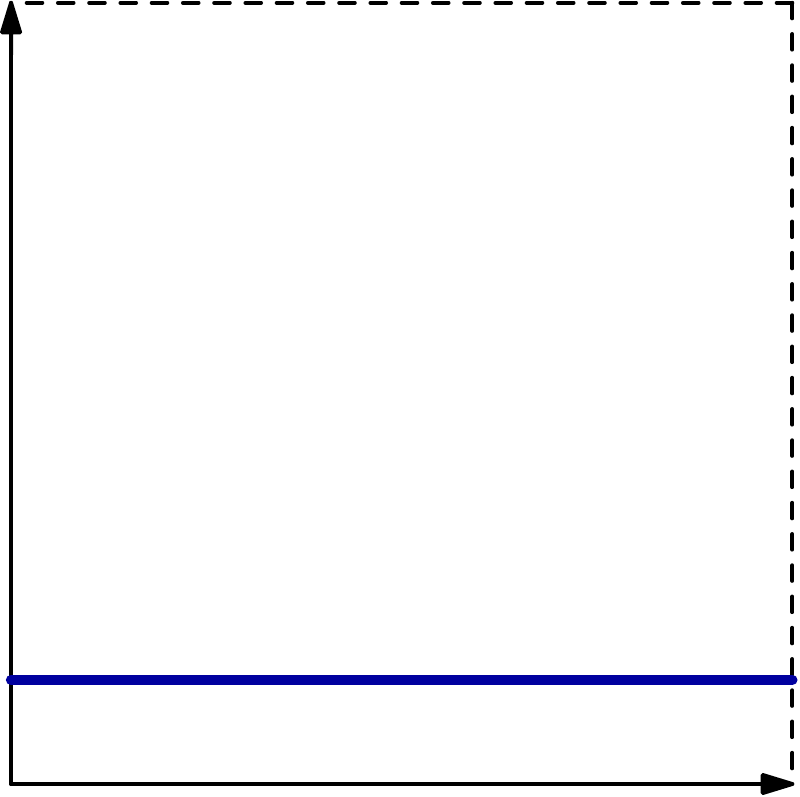}
\caption{Subspace 1}
\begin{picture}(0,0)
\put(39,36){\scriptsize$\mathbb T^D$}
\put(-52,128){\scriptsize$\hat{\mathbb T}^D$}
\end{picture}
\label{fig_doub_1}
\end{subfigure}
\hspace{30pt}
\begin{subfigure}{100pt}
\centering
\hspace*{5pt}
\includegraphics[width=90pt]{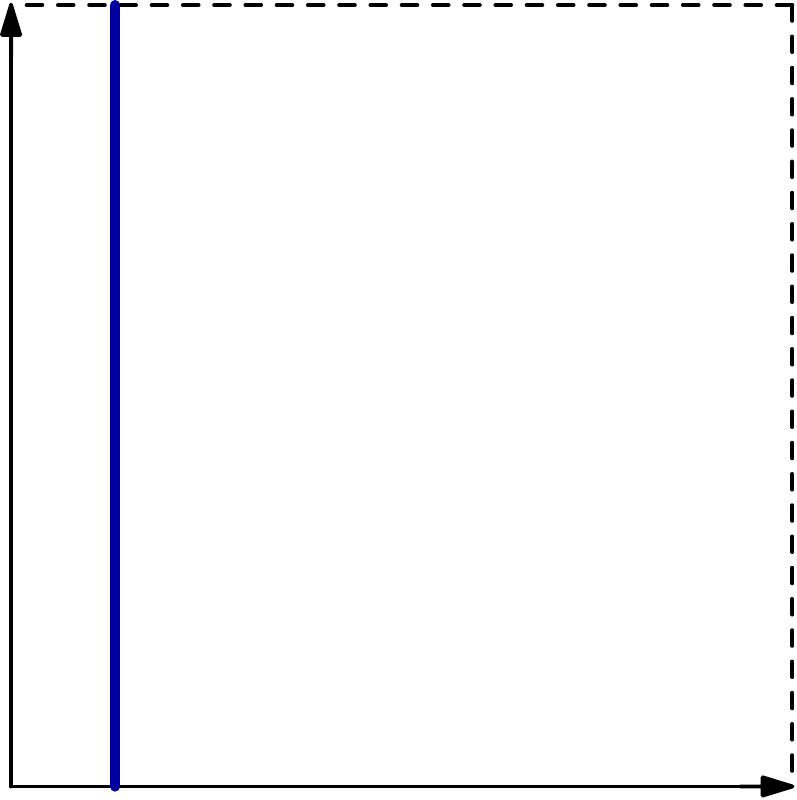}
\caption{Subspace 2}
\begin{picture}(0,0)
\put(39,36){\scriptsize$\mathbb T^D$}
\put(-52,128){\scriptsize$\hat{\mathbb T}^D$}
\end{picture}
\label{fig_doub_2}
\end{subfigure}
\hspace{30pt}
\begin{subfigure}{100pt}
\centering
\hspace*{5pt}
\includegraphics[width=90pt]{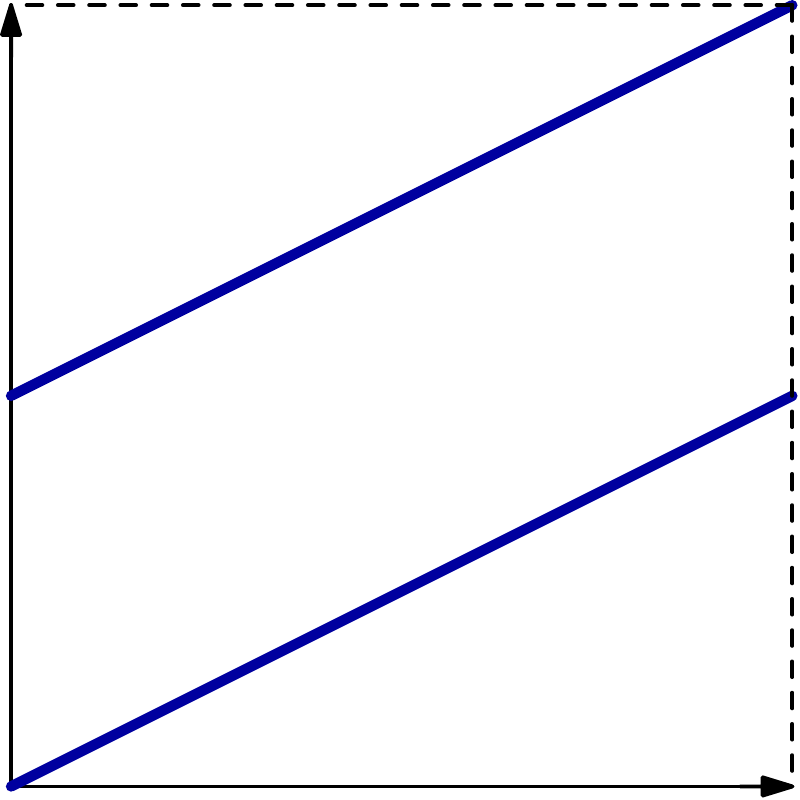}
\caption{Subspace 3}
\begin{picture}(0,0)
\put(39,36){\scriptsize$\mathbb T^D$}
\put(-52,128){\scriptsize$\hat{\mathbb T}^D$}
\end{picture}
\label{fig_doub_3}
\end{subfigure}
\caption{Illustration of how a $D$-dimensional physical subspace  inside
the $2D$-dimensional doubled space $\mathbb T^D\times \hat{\mathbb T}^D$ can be identified.
In figure~\ref{fig_doub_1} the projector $\Pi^I{}_J$ selects the ordinary space $\mathbb T^D$, 
in figure~\ref{fig_doub_2} the fully T-dual space $\hat{\mathbb T}^D$ is chosen, 
and in figure~\ref{fig_doub_3} a linear combination of both is chosen. 
\label{fig_doub}
}
\end{figure}
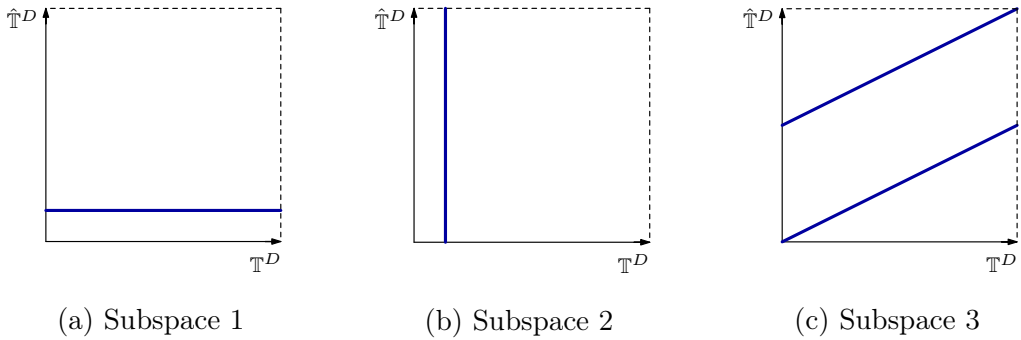

From the doubled-geometry perspective the coordinates $\mathbb X^i$ and $\mathbb X_i$ are
on equal footing. The identification of $\mathbb X^i = X^i$ with the physical coordinates
and $\mathbb X_i=\hat X_i$ with the dual ones introduced in  \eqref{dg_297a}
is arbitrary, and 
we can equally-well take another $D$-dimensional subset of $\mathbb X^I$ 
to represent the physical space. 
This freedom of choice is related to $O(D,D,\mathbb Z)$ transformations, which 
we can make more precise by considering a  projector $\Pi^I{}_J$
separating the physical from the dual coordinates as
\eq{
  X^I =\binom{X^i}{\hat X_i} = 
  \binom{ \Pi^i{}_J \mathbb X^J }{
  \hat{\Pi}_{i\op J} \mathbb X^J }
  = \Pi^I{}_J \op\mathbb X^J \,.
}
In our conventions \eqref{dg_297a} the projector has been chosen as the identity. 
Let us also note that the projector has to be consistent with the boundary conditions \eqref{dg_240091}, 
which means the entries of the matrix $\Pi^I{}_J$ have to be integers. 
A T-duality transformation can now be interpreted in two ways: 
\begin{itemize}

\item It is either an active $O(D,D,\mathbb Z)$ transformation acting on the doubled coordinates $\mathbb X^I$ and the generalised metric $\mathcal H_{IJ}$. This changes the background encoded in $\mathcal H_{IJ}$
while keeping the projector $\Pi^I{}_J$ fixed.

\item Alternatively, a duality transformation can be seen as a passive transformation which only acts on 
the projector $\Pi^I{}_J$. This changes the identification of the physical subspace 
inside the doubled torus $\mathbb T^{2D}$.
This point  is illustrated in figures~\ref{fig_doub}.

\end{itemize}
%%%%%%%%%%%%%%%%
%%%%%%%%%%%%%%%%
%%%%%%%%%%%%%%%%
%%%%%%%%%%%%%%%%

%%%%%%%%%%%%%%%%%%%%%%%%%%%%%%%%%%%%%%%%%%%%%%%
%%%%%%%%%%%%%%%%%%%%%%%%%%%%%%%%%%%%%%%%%%%%%%%

\subsubsection*{Remarks}

Let us close this section with the following remarks on doubled geometry.
\begin{itemize}

\item We note that when imposing the constraint \eqref{dg_9927b} directly on
\eqref{dg_9927c} the doubled action vanishes. 
This is familiar from theories with odd self-dual forms, such as the five-form field strength of type IIB 
string theory. The constraint  has to be imposed on the equations of motion, 
and the action \eqref{dg_9927c} therefore is only a pseudo-action.

\item Determining a general $D$-dimensional background from the doubled geometry can also be 
achieved via a gauging procedure, similar as on page~\pageref{page_dg_gauging}.
One considers a
global symmetry
\eq{
  \mathbb X^I \to \mathbb X^I + \epsilon^{\alpha} k_{\alpha}^I \,,
}
with $\alpha = 1, \ldots, D$, where the $D$ vectors $k_{\alpha}^I$ 
are required to be linearly-independent and
parametrise which global symmetries are gauged.
After constructing the gauged action by introducing gauge fields and integrating the gauge fields out, 
one obtains the action in which the directions corresponding to $k_{\alpha}$ have
been removed. 
Note that this procedure is similar to our discussion of the Buscher rules in 
section~\ref{sec_buscher}, except that no Lagrange multipliers are introduced.

\end{itemize}

%%%%%%%%%%%%%%%%%%%%%%%%%%%%%%%%%%%%%%%%%%%%%%%
%%%%%%%%%%%%%%%%%%%%%%%%%%%%%%%%%%%%%%%%%%%%%%%
%%%%%%%%%%%%%%%%%%%%%%%%%%%%%%%%%%%%%%%%%%%%%%%
%%%%%%%%%%%%%%%%%%%%%%%%%%%%%%%%%%%%%%%%%%%%%%%

\subsection{Torus fibrations}
\label{sec_dg_tf}

We now extend the  formalism of the previous section 
from toroidal compactifications to non-trivial torus fibrations
over some base-manifold. We mainly follow the original papers \cite{Hull:2004in,Hull:2006va},
to which we refer for further details.

%%%%%%%%%%%%%%%%%%%%%%%%%%%%%%%%%%%%%%%%%%%%%%%
%%%%%%%%%%%%%%%%%%%%%%%%%%%%%%%%%%%%%%%%%%%%%%%

\subsubsection*{The setting}

Let us start by specifying the setting we are working in. 
Similarly as in section~\ref{cha_torus_fib}, we
consider $D$-dimensional torus fibrations 
with $n$-dimensional fibres $\mathbb T^n$
over a $(D-n)$-dimensional base-manifold $\mathcal B$. 
However, in order to be compatible with the notation in \cite{Hull:2006va} we change 
our convention  to the following 
\eq{
  \arraycolsep2pt
  \begin{array}{l@{\hspace{40pt}}l@{\hspace{20pt}}l@{\hspace{20pt}}lcl}
  \mbox{coordinates on $\mathbb T^n$} 
  &
  X^i &\mbox{with} & i&=&1,\ldots, n \,,
  \\[6pt]
  \mbox{coordinates on $\mathcal B$}
  &
  Y^m &\mbox{with} & m&=&1,\ldots, D-n \,.
 \end{array}
}
Choosing suitable coordinates on the torus fibre, the metric for such torus fibrations 
can be brought into the following form
\eq{
  \label{dg_tf_24a}
  G = \tfrac{1}{2} \op g_{mn} \, dY^m\vee dY^n + \tfrac{1}{2}\op g_{ij}\, P^i \vee P^j   \,,
}
where $P^i$ are globally-defined one-forms 
\eq{
  P^i = dX^i + A^i \,.
}  
The $A^i = A^i{}_m(Y) \op dY^m$ can be interpreted as connection one-forms, whose field strength 
will be denoted by $F^i = dA^i$. This data encodes the non-triviality of the fibration of the metric. 
Furthermore, the components $g_{mn}$ and $g_{ij}$ are independent of $X^i$. 
For the Kalb-Ramond field we use the same basis $\{ dY^m,P^i\}$ of the cotangent-space to 
express $B$ as
\eq{
  \label{dg_tf_24b}
  B =  
  \frac12\,  b_{mn} \op dY^m\wedge dY^n
  - \alpha' \op P^i \wedge \hat{A}_i 
  + \frac12 \,  b_{ij}\op  P^i \wedge  P^j 
   \,,
}
where $\hat A_i = \hat A_{im}(Y)\op dY^m$ are one-forms on the base-manifold $\mathcal B$.
The components $b_{mn}$ and $b_{ij}$ are again required to be independent of $X^i$.

Let us then note that the Kalb-Ramond $B$-field does not need to be globally-defined but can 
have a non-vanishing field strength $H=dB\neq0$. 
This implies in particular that the $\hat A_i$ are in general not globally-defined, similarly as the 
$A^i$. In fact, the $\hat A_i$  can be interpreted as connection one-forms on a dual bundle \cite{Hull:2004in,Hull:2006va}
for which we denote the corresponding field strength by $\hat F_i = d \hat A_i$. 
This data encodes the non-triviality of the fibration related to the $B$-field. 
By introducing  coordinates $\hat X_i$ on a dual torus fibre $\hat{\mathbb T}^n$, we can then define
globally-defined 
one-forms for a dual torus fibration as
\eq{
  \hat P_i = d\hat X_i + \hat A_i \,.
}

%%%%%%%%%%%%%%%%%%%%%%%%%%%%%%%%%%%%%%%%%%%%%%%
%%%%%%%%%%%%%%%%%%%%%%%%%%%%%%%%%%%%%%%%%%%%%%%

\subsubsection*{The doubled action }

Motivated by these observations, let us consider general 
doubled coordinates $\mathbb X^I$  on 
a doubled torus-fibre $\mathbb T^{2n}$ together with corresponding doubled connections 
$\mathcal A^I$. 
We then introduce globally-defined one-forms as
\eq{
  \label{dg_tf_2948}
  \mathcal P^I = d\mathbb X^I + \mathcal A^I \,,
  \hspace{60pt}
  I = 1,\ldots, 2n\,.
}  
A metric on the doubled torus fibre will be denoted by $\mathcal H_{IJ}$, 
and we required it to be independent of $\mathbb X$ as well as  to 
satisfy $(  \eta^{-1} \mathcal H )^2 = \mathds 1$
with $\eta_{IJ}$  the $O(n,n,\mathbb Z)$ invariant metric.
The corresponding doubled world-sheet action takes the form \cite{Hull:2006va}
\eq{
  \label{dg_tf_001}
  \mathcal S = -\frac{1}{4\pi} \int_{\Sigma}\op \Bigl[  \op
   \tfrac{1}{2}\op \mathcal H_{IJ} \,\mathcal P^I \wedge\star \mathcal P^J 
   &+ \eta_{IJ}\op \mathcal P^I \wedge \mathcal A^J
   \\
   &+ \Omega_{IJ} \op d\mathbb X^I \wedge d\mathbb X^J
   + \tfrac{1}{\alpha'} \op \mathcal L(Y) 
   \op\Bigr]
   \,,
}
where 
$\mathcal P^I$, $\mathcal A^I$ and $d\mathbb X^I$ denote the pull-backs of the 
corresponding target-space quantities to the world-sheet $\Sigma$. We use the
same symbols for the world-sheet and target-space quantities, but the distinction 
should be clear from the context. 
The term in \eqref{dg_tf_001} containing $\Omega_{IJ}$ is topological and is needed for showing the equivalence 
with the standard formulation. 
The Lagrangian for the base-manifold is given by
\eq{
  \mathcal L(Y) =  g_{mn} \op dY^m \wedge\star dY^n - b_{mn} \op dY^m\wedge dY^n +\alpha'\op 
  \Omega_{IJ}\op  \mathcal A^I \wedge   {\mathcal A}^J \,.
}
The theory is furthermore subject to the constraint 
\eq{
  \label{dg_tf_829479247}
  \mathcal P = \eta^{-1} \op\mathcal H \star \mathcal P \,,
}
and with $\mathcal F^I = d\mathcal A^I$ the field strength of $\mathcal A^I$ 
the equations of motion are obtained as
\eq{
  d\star\bigl( \mathcal H_{IJ} \mathcal P^J \bigr) = \eta_{IJ}  \mathcal F^J\,.
}
Note that the latter can be expressed also as 
$d\star ( \eta^{-1} \mathcal H \mathcal P - \star \mathcal P)=0$, and hence the constraint implies
the equations of motion.

The action \eqref{dg_tf_001} is invariant under the following $GL(2n,\mathbb R)$ transformations acting on the 
coordinates and metrics 
\eq{
  \arraycolsep2pt
  \begin{array}{lccl@{\hspace{50pt}}lclcl}
  \mathcal P &\to&  \mathcal O &  \mathcal P  \,,
  &
   \mathcal H &\to & \mathcal O^{-T} &\mathcal H  &\mathcal O^{-1}  \,,
 \\[4pt]
  \mathcal A &\to&  \mathcal O &  \mathcal A  \,,
  &
  \Omega& \to & \mathcal O^{-T} &\Omega  &\mathcal O^{-1}  \,,
  \end{array}
}
which is however broken to $GL(2n,\mathbb Z)$ by the boundary conditions of the doubled torus $\mathbb T^{2n}$ imposed on $\mathbb X^I$. 
Furthermore, the constraint \eqref{dg_tf_829479247} breaks $GL(2n,\mathbb Z)$
to $O(n,n,\mathbb Z)$ which shows that the action \eqref{dg_tf_001}
is invariant under T-duality transformations.

%%%%%%%%%%%%%%%%%%%%%%%%%%%%%%%%%%%%%%%%%%%%%%%
%%%%%%%%%%%%%%%%%%%%%%%%%%%%%%%%%%%%%%%%%%%%%%%

\subsubsection*{Equivalence with the standard formulation}

The equivalence with the standard formulation is again shown using a gauging procedure, 
similarly as in section~\ref{sec_dg_tor}.
To do so, we first have to choose a polarisation $\Pi^I{}_J$ specifying which of the 
coordinates $\mathbb X^I$ correspond to the physical space. 
In particular, we define
\eq{
  \arraycolsep2pt
  \begin{array}{lclclcl}
  \displaystyle X^I &=& \displaystyle  \binom{X^i}{\hat X_i} 
  &=& \displaystyle  \binom{\Pi^i{}_J \mathbb X^J}{
  \Pi_{iJ} \mathbb X^J} 
  &=& \displaystyle  \Pi^I{}_J\op \mathbb X^J \,,
  \\[14pt]
  \displaystyle P^I &=& \displaystyle  \binom{P^i}{\hat P_i} 
  &=& \displaystyle  \binom{\Pi^i{}_J \mathcal P^J}{
  \Pi_{iJ} \mathcal P^J} 
  &=& \displaystyle  \Pi^I{}_J\op \mathcal P^J \,,
  \\[14pt]
  \displaystyle A^I &=& \displaystyle  \binom{A^i}{\hat A_i} 
  &=& \displaystyle  \binom{\Pi^i{}_J \mathcal A^J}{
  \Pi_{iJ} \mathcal A^J} 
  &=& \displaystyle  \Pi^I{}_J\op \mathcal A^J \,.
  \end{array}
}
Using this polarisation, we can identify $\mathcal H$ as the generalised metric \eqref{gen_met_098},
and we take $\Omega_{IJ} =\frac12 \left( \begin{smallmatrix} 0 & -\mathds 1 \\ + \mathds 1 & 0 \end{smallmatrix}\right)$.
The doubled action \eqref{dg_tf_001} can then be written as
\eq{
  \label{df_tf_2444}
  \mathcal S = -\frac{1}{4\pi} \int_{\Sigma}\op \Bigl[  \hspace{17pt}
  &   \tfrac{1}{2} \op \tfrac{1}{\alpha'} g_{ij}\op  P^i \wedge\star P^j 
   + \tfrac{1}{2} \op\alpha'  g^{ij} \bigl( \hat P - \tfrac{1}{\alpha'}\op b \op P \bigr)_i
   \wedge \star \bigl( \hat P - \tfrac{1}{\alpha'}\op b \op P \bigr)_j
   \\[2pt]
   +\,& \bigl( P^i \wedge \hat A_i + \hat P_i \wedge A^i \bigr) + d\hat X_i  \wedge d X^i
   \\[4pt]
   +\,& \tfrac{1}{\alpha'}\op  \mathcal L(Y)
   \hspace{241pt}\Bigr]
   .
}
Since $g_{ij}$ and $b_{ij}$ do not depend on $\mathbb X$, this action is 
invariant under transformations of the form 
$\hat X_i \to \hat X_i + \epsilon_i$ for $\epsilon_i = {\rm const.}$
Similarly as on page~\pageref{page_dg_gauging}, 
this global symmetry can be made local by introducing gauge fields 
$C_i$ and replacing $\hat P_i \to \hat P_i + C_i $ 
as well as $d \hat X_i \to d\hat X_i +  C_i $ in the action \eqref{df_tf_2444}.
The equations of motion for the gauge fields $C_i$ are solved by
\eq{
  C_i = - \hat P_i - \frac{1}{\alpha'} \op g_{ij} \star P^j + \frac{1}{\alpha'} \op b_{ij} \op P^j \,,
}
which when inserted into the gauged action gives
\eq{
  \check{\mathcal S} = -\frac{1}{4\pi \alpha'} \int_{\Sigma}
  \Bigl[ \hspace{16pt} 
  &g_{mn} \, dY^m\wedge dY^n + g_{ij}\, P^i \wedge P^j 
  \\
  -\,& b_{mn} \op dY^m\wedge dY^n
  +2 \alpha' \op P^i \wedge \hat{A}_i 
  -  b_{ij}\op  P^i \wedge  P^j 
  \, \Bigr] 
  \,.
}
Comparing now with the metric and $B$-field shown in \eqref{dg_tf_24a} and 
\eqref{dg_tf_24b}, we see that this is the ordinary string-theory action 
for the $D$-dimensional torus fibration considered in this section. 
We have therefore established the equivalence of the doubled action 
with the standard formulation \cite{Hull:2006va}.

%%%%%%%%%%%%%%%%%%%%%%%%%%%%%%%%%%%%%%%%%%%%%%%
%%%%%%%%%%%%%%%%%%%%%%%%%%%%%%%%%%%%%%%%%%%%%%%

\subsubsection*{Non-geometric backgrounds -- T-folds}

The doubled formalism provides a suitable framework to describe non-geometric backgrounds.
Let us recall that 
according to our characterisation 3 on 
page~\pageref{page_defs_nongeo},
for non-geometric backgrounds 
the transition functions between local patches are duality transformations. These spaces are also 
called T-folds \cite{Hull:2004in}.
More concretely, as discussed in section~\ref{cha_torus_fib}, for  torus fibrations 
with fibre $\mathbb T^n$ the monodromy group along non-contractible loops in the base-manifold 
is contained in $O(n,n,\mathbb Z)$. For geometric backgrounds the monodromy belongs to 
the geometric subgroup, while for non-geometric backgrounds the monodromy is a proper 
duality transformation.

Let us now compare the description of non-geometric backgrounds
in  section~\ref{cha_torus_fib} to our present discussion:
\begin{itemize}

\item In the approach of  section~\ref{cha_torus_fib}, we have constructed non-geometric backgrounds 
whose transition functions are duality transformations. In this case the action is not invariant when changing 
from one local patch to another. 

\item In the doubled formalism on the other hand, the world-sheet action is invariant 
under $O(n,n,\mathbb Z)\subset GL(2n,\mathbb Z)$ transformations and duality transformations
are diffeomorphisms for the doubled space. 
Thus, non-geometric backgrounds in doubled geometry
actually have a geometric description, and 
the action is invariant when changing between local patches.

The non-geometric nature of this background arises because it is not possible to 
find a globally-consistent choice of a physical subspace $\mathbb T^n$ inside the doubled torus $\mathbb T^{2n}$.

\end{itemize}

%%%%%%%%%%%%%%%%%%%%%%%%%%%%%%%%%%%%%%%%%%%%%%%
%%%%%%%%%%%%%%%%%%%%%%%%%%%%%%%%%%%%%%%%%%%%%%%

\subsubsection*{Dilaton}

The dilaton has not been included in the above discussion. However, following \cite{Hull:2006va},
we can add a term of the form 
\eq{
  \label{dg_tf_dil_22}
   -\frac{1}{4\pi} \int_{\Sigma}
  \mathsf R\, \Phi \star 1
}
to the doubled action \eqref{dg_tf_001}, where $\mathsf R$ is the Ricci scalar of the world-sheet metric
and $\Phi=\Phi(Y)$ denotes the doubled dilaton. Note that since $\Phi$ is independent of $\mathbb X$,
 \eqref{dg_tf_dil_22}  is invariant under $O(n,n,\mathbb Z)$ transformations.

Now, when relating the doubled action to the standard formulation by in\-tegra\-ting out a
 local symmetry, there is a one-loop effect when performing the path-integral \cite{Tseytlin:1990nb,Tseytlin:1990va}. 
This implies that the doubled dilaton $\Phi$ in \eqref{dg_tf_dil_22} gets an additional contribution,
so that the physical dilaton $\phi$ takes the form
\eq{
  \phi = \Phi - \frac{1}{4}\log \bigl( \det G_{ij}\bigr) \,,
}
where $G_{ij}$ denotes the metric on the physical space shown in \eqref{dg_tf_24a}.
Under duality transformations, $\phi$ then transforms in the standard way \eqref{dual_back_9495}.

%%%%%%%%%%%%%%%%%%%%%%%%%%%%%%%%%%%%%%%%%%%%%%%
%%%%%%%%%%%%%%%%%%%%%%%%%%%%%%%%%%%%%%%%%%%%%%%
%%%%%%%%%%%%%%%%%%%%%%%%%%%%%%%%%%%%%%%%%%%%%%%
%%%%%%%%%%%%%%%%%%%%%%%%%%%%%%%%%%%%%%%%%%%%%%%

\subsection{Double field theory}
\label{sec_dft}

In this section we briefly discuss double field theory (DFT) \cite{Hull:2009mi,Hohm:2010pp,Hohm:2010jy},
which is a target-space
formulation manifestly invariant under $O(D,D,\mathbb Z)$ transformations. 
A detailed discussion of DFT is beyond the scope of this work, 
for which we want to refer the reader to the reviews
 \cite{Aldazabal:2013sca,Berman:2013eva,Hohm:2013bwa}. Here
we only point out relations with doubled geometry 
and the relevance of DFT for non-geometric backgrounds.

%%%%%%%%%%%%%%%%%%%%%%%%%%%%%%%%%%%%%%%%%%%%%%%
%%%%%%%%%%%%%%%%%%%%%%%%%%%%%%%%%%%%%%%%%%%%%%%

\subsubsection*{Doubled world-sheet actions}

In section~\ref{sec_dg_tor} we have illustrated that a world-sheet theory with 
doubled coordinates is subject to a self-duality constraint. 
Theories with self-dual odd forms are difficult to deal with, since 
when the constraint is imposed at the level of the action the latter vanishes. 
Examples for such theories are type IIB string theory with a self-dual five-form 
field strength, the world-volume theory of M5-branes with a self-dual three-form field strength and 
the above-mentioned doubled world-sheet theory with a self-dual one-form. 
In all these cases it turns out to be difficult to quantise the theory.

Let us now be somewhat more precise concerning the doubled world-sheet formalism. 
The approach of Hull \cite{Hull:2004in} discussed in this section,
involving both coordinates $X^i$ and $\hat X_i$, 
has previously appeared in a similar form in
\cite{Cremmer:1997ct} and is related to the approaches in 
\cite{Duff:1989tf,Maharana:1992my}.
The important question is, however, how the self-duality constraint \eqref{dg_tf_829479247} 
is imposed:
\begin{itemize}

\item If the  constraint is implemented using the non-covariant formalism of \cite{Floreanini:1987as}, 
then the doubled-geometry formulation of \cite{Hull:2004in} is closely related to  that of Tseytlin
\cite{Tseytlin:1990nb,Tseytlin:1990va}. 
In this formulation  manifest Lorentz invariance of the two-dimensional world-sheet theory
is lost, which makes it difficult to compute for instance the $\beta$-functionals.

\item One can impose additional constraints to restore Lorentz invariance for the action by  Tseytlin
\cite{Sfetsos:2009vt,Avramis:2009xi}, which however complicates the computation of 
the Weyl anomaly. 
Other approaches are that of twisted double-tori which can be found in \cite{DallAgata:2008ycz},
or by following the Pasti-Sorokin-Tonin procedure  \cite{Pasti:1996vs} which has been discussed 
in \cite{Driezen:2016tnz}.

\item When imposing an additional gauge symmetry for the doubled action, one
can obtain the Tseytlin model as a particular gauge choice \cite{Nibbelink:2012jb,Nibbelink:2013zda}.
This approach is based on \cite{Rocek:1997hi}.

\item When following the approach to self-dual forms by Siegel \cite{Siegel:1983es},
one is led to a formulation similar to that of \cite{Hull:1988si}.

\item In Hull's formalism \cite{Hull:2004in,Hull:2006va} the constraint is imposed through a gauging procedure 
as discussed above. Here a particular polarisation separating physical from 
dual coordinates has to be chosen by hand, but it is argued that this method 
is suitable for quantising the theory.

\end{itemize}
More proposals for such actions can be found in the literature, which 
are all classically-equivalent to the ordinary string-theory action. 
However, it is usually difficult to quantise these actions.

%%%%%%%%%%%%%%%%%%%%%%%%%%%%%%%%%%%%%%%%%%%%%%%
%%%%%%%%%%%%%%%%%%%%%%%%%%%%%%%%%%%%%%%%%%%%%%%

\subsubsection*{Double field theory}

In string theory, the vanishing of the $\beta$-functionals \eqref{eom_beta} is interpreted
as the equations of motion of an effective target-space theory. 
For the doubled world-sheet theories one can try to follow a similar reasoning, 
and the corresponding one-loop $\beta$-functionals   have been computed for instance
in \cite{Berman:2007xn,Berman:2007yf} and \cite{Nibbelink:2013zda}.
However, these have technical difficulties concerning the validity of a perturbative 
$\alpha'$-expansion or they do not reproduce the equations of motion 
expected from a doubled target-space theory, respectively.

On the other hand,  in \cite{Hull:2009mi} a target-space theory has been constructed
using closed-string field theory
which is manifestly invariant under $O(D,D)$ transformations. This formulation is expected to be a
target-space description of the doubled world-sheet action (or a variant thereof), for which a number of indications
have been collected. The current status in the literature is, however, that this question has not yet been completely
settled.

%%%%%%%%%%%%%%%%%%%%%%%%%%%%%%%%%%%%%%%%%%%%%%%
%%%%%%%%%%%%%%%%%%%%%%%%%%%%%%%%%%%%%%%%%%%%%%%

\subsubsection*{Basics of DFT}

We now want to give a brief introduction to 
double field theory, which  is defined on a space with a doubled number of target-space dimensions. Similarly 
to what we have seen for the world-sheet action, in DFT the ordinary coordinates $X^i$ 
are supplemented by dual coordinates $\hat X_i$ (also called winding coordinates). 
These are combined into double coordinates $X^I = (X^i,\hat X_i)$, 
with   $i=1,\ldots, D$ and $I=1,\ldots, 2D$.
Relevant quantities for double field theory are the $O(D,D)$ invariant metric $\eta_{IJ}$
and the generalised metric $\mathcal H_{IJ}$ shown in 
\eqref{gen_met_098}. We recall them here as
\eq{
  \label{gen_met_098aaa}
  \arraycolsep4pt
  \mathcal H = \left( \begin{array}{cc}
   \frac{1}{\alpha'}\left( g - b\op g^{-1}b\right)& +b\op g^{-1}  \\[4pt]  - g^{-1} b & \alpha' g^{-1}\end{array} \right)
  ,
  \hspace{50pt}
  \eta =  \left( \begin{array}{cc} 0 & \mathds 1 \\[2pt]  \mathds 1 & 0 \end{array}\right) .
}
For these expressions we can introduce vielbein matrices $\mathcal E^A{}_I$ as
in \eqref{vielbein_again_649}. In particular, we have 
\eq{
  \label{vielbein_again_649aaa}
  \eta = \mathcal E^T  \eta \,
  \mathcal E \,,
  \hspace{70pt}
  \mathcal H = \mathcal E^T \mathcal S\, 
  \mathcal E \,,
}
where the $2D\times 2D$ matrix $\mathcal S$ is given by the following expression in the generalised vielbein
basis 
\eq{
  \mathcal S_{AB} =  \left( \begin{array}{cc} \,\delta\, & 0 \\[2pt] 0 & \delta^{-1} \end{array}\right).
}
Here we restrict the doubling of the coordinates to say a compact part of the space-time such 
that the time direction is not doubled. This can however also be extended to the full space-time. 
An action for double field theory can be  determined by invoking the following symmetries:
\begin{itemize}

\item First, one requires the action  to be invariant under local diffeomorphisms of the
doubled coordinates $X^I$, that is  $(X^i,\hat X_i)\to (X^i + \xi^i(X),\hat X_i+\hat \xi_i(X))$.

\item Second, the action should be  invariant under a global  $O(D,D)$ symmetry.
It has been realised that for manifest $O(D,D)$ invariance and for closure of the algebra 
of infinitesimal diffeomorphisms, one has to impose the so-called strong constraint
\eq{
  \label{strong_c}
  \partial_i A\, \hat\partial^i B +  \hat\partial^i A\, \partial_i B=0\, ,
}
where $\hat \partial^i$ denotes the derivative with respect to the winding coordinate $\hat X_i$.
The quantities $A$ and $B$ in \eqref{strong_c} can be any function or matrix. 

\end{itemize}
We note that there exist two formulations of a DFT action, 
which differ by terms that  are either total derivatives or are
vanishing due to the strong constraint \eqref{strong_c}.

%%%%%%%%%%%%%%%%%%%%%%%%%%%%%%%%%%%%%%%%%%%%%%%
%%%%%%%%%%%%%%%%%%%%%%%%%%%%%%%%%%%%%%%%%%%%%%%

\subsubsection*{Flux-formulation of DFT}

For our purposes it is convenient to use the  flux formulation of double field theory,
which has been developed in \cite{Aldazabal:2011nj,Geissbuhler:2011mx,Grana:2012rr}
and is, as has been shown in \cite{Hohm:2010xe}, 
related to earlier work of Siegel \cite{Siegel:1993xq,Siegel:1993th}.
In a frame with flat indices, the action is given by 
\begin{align}
  \nonumber
  &\mathcal S_{\rm DFT}=\frac{1}{2\op \kappa_{10}^2}\int d^{2D} X \: e^{-2d}\, \bigg[ \\[4pt]
  \nonumber
  &\hspace{32pt}\hspace{16pt}{\cal F}_{ABC} {\cal F}_{A'B'C'}\left( \frac{1}{4}\op \mathcal{S}^{AA'} \eta^{BB'} \eta^{CC'} 
  -\frac{1}{12}\op \mathcal{S}^{AA'} \mathcal{S}^{BB'} \mathcal{S}^{CC'} 
    -\frac{1}{6}\hspace{4pt} {\eta}^{AA'} \eta^{BB'} \eta^{CC'} \right)\\[4pt]
 \label{dftactionfluxb}
 &\hspace{32pt}+ {\cal F}_A \op {\cal F}_{A'} \op \Big( \eta^{AA'}-\mathcal{S}^{AA'} \Big)
 \hspace{10pt}\bigg]\, , 
\end{align}
where we used $d^{2D} X\equiv d^D X \wedge d^D\hat X$. The definition of $e^{-2d}$ contains the 
ordinary dilaton $\phi$ and the
determinant of the metric $g_{ij}$, and reads
$\exp(-2d)=\sqrt{|g|} \exp(-2\phi)$. 
The  objects ${\cal F}_A$ are expressed as
\eq{
  \label{res_012}
     {\cal F}_A=\Omega^B{}_{BA}+2 \op \ov{\mathcal E}_A{}^I \partial_I d \,,
}
with 
$\ov{\mathcal E}^I{}_A$ the inverse of the vielbeins $\mathcal E^A{}_I$,
$\partial_I$ denoting the derivative with respect to the doubled coordinates $X^I$,
 and $\Omega_{ABC}$ being
the generalised Weitzenb\"ock connection
\eq{
  \label{flux_002}
      \Omega_{ABC}=\ov{\mathcal E}_A{}^I \bigl(\op \partial_I\op \ov{\mathcal E}_B{}^J \op\bigr) {\mathcal E}_{JC}\, .
}
Note that the frame-index of $\mathcal E_I{}^A$ has been lowered using $\eta_{AB}$.
The three-index object $\mathcal F_{ABC}$ appearing in \eqref{dftactionfluxb} 
is the anti-symmetrisation of $\Omega_{ABC}$, that is
\eq{
  \label{flux_001}
  \mathcal F_{ABC} = 3\, \Omega_{[\ul A \ul B \ul C]} 
  \,.
}

%%%%%%%%%%%%%%%%%%%%%%%%%%%%%%%%%%%%%%%%%%%%%%%
%%%%%%%%%%%%%%%%%%%%%%%%%%%%%%%%%%%%%%%%%%%%%%%

\subsubsection*{(Non-)geometric fluxes}

Since the three-index fluxes \eqref{flux_001} have upper and lower indices, it is natural to identify 
(the vacuum expectation value of)
$\mathcal F_{ABC}$ with the $H$-flux, geometric flux, non-geometric $Q$- and non-geometric $R$-flux as
\eq{
  \label{def_fluxes}
     {\cal F}_{abc}=H_{abc}\, , \qquad 
     {\cal F}_{ab}{}^c=F_{ab}{}^c\, , \qquad 
     {\cal F}_a{}^{bc}=Q_a{}^{bc}\, , \qquad  
     {\cal F}^{abc}=R^{abc}\, .
}
Double-field theory therefore includes geometric and non-geometric fluxes on equal footing. 
Furthermore, the fluxes have to satisfy consistency conditions of the form \cite{Geissbuhler:2013uka}
\eq{
  &0 = \mathcal D_{[\ul A} \mathcal F_{\ul B \ul C\ul D]} - \frac{3}{4} \, \mathcal F_{[\ul A \ul B}{}^{M} 
  \mathcal F_{M|\ul C\ul D]} \,,
  \\[6pt]
  &0 = \mathcal D^M \mathcal F_{MAB} + 2\op \mathcal D_{[\ul A} \mathcal F_{\ul B]}
  - \mathcal F^M \mathcal F_{MAB} \,,
}
where $\mathcal D_A =  \ov{\mathcal E}_A{}^I \partial_I$. These expressions are similar in 
form to the Bianchi identities discussed around equation \eqref{sg_dft_bianchi}.

However, even though the DFT fluxes \eqref{flux_001}
are similar to the fluxes in generalised geometry (cf. for instance \eqref{wim}), 
there are important differences. In particular, 
in double field theory the generalised vielbeins $\ov{\mathcal E}$ can depend 
not only on the ordinary coordinates $X^i$ but also on the dual winding coordinates $\hat X_i$. 
Let us illustrate this for the following 
DFT vielbein 
\eq{
  \ov{\mathcal E}_A{}^I = 
  \left( \begin{array}{cc} \delta_a{}^i & -\delta_a{}^m b_{mi} \\[2pt] 0 & \delta^a{}_i \end{array}\right) ,
}
in which $b_{ij}\equiv b_{ij}(X,\hat X)$ can depend on both types of coordinates. The corresponding
non-vanishing fluxes (in the coordinates basis) are then determined as follows
\eq{
  \label{dft_fluxes_24874}
  \begin{array}{l@{\hspace{50pt}}l}
  \mbox{double field theory}
  &
  \displaystyle
  \arraycolsep2pt
  \begin{array}{lcl}
  \mathcal F_{ijk} &=& -3\op  \partial_{[\ul i} \op b_{\ul j\ul k]} + 3\op  b_{[\ul i m} \hat\partial^m \op b_{\ul j \ul k]} \,,
  \\[4pt]
  \mathcal F_{ij}{}^k &=& - \hat \partial^k \op b_{ij} \,,
  \end{array}
  \\[26pt]
  \mbox{generalised geometry}
  &
  \arraycolsep2pt
  \begin{array}{lcl}
  \mathcal F_{ijk} &=& -3\op  \partial_{[\ul i} \op b_{\ul j\ul k]} \,,
  \end{array}
  \end{array}
}
where $\hat\partial^i$ denotes the derivative with respect to $\hat X^i$,
and where included the ge\-nera\-lised-geometry result 
from equation \eqref{gv_h_flux_case_vb}.
We therefore see that in double field theory there is an additional contribution to the fluxes 
coming from the dual winding coordinates.

%%%%%%%%%%%%%%%%%%%%%%%%%%%%%%%%%%%%%%%%%%%%%%%
%%%%%%%%%%%%%%%%%%%%%%%%%%%%%%%%%%%%%%%%%%%%%%%

\subsubsection*{Remarks}

We close this section with the following remarks:
\begin{itemize}

\item Ignoring how the DFT fluxes are realised in terms of vielbein matrices
and  considering only the expectation values of $\mathcal F_{ABC}$,
it has been shown in \cite{Blumenhagen:2015lta} that  DFT 
compactified on Calabi-Yau three-folds 
leads to the scalar potential \eqref{res_022} of four-dimensional $\mathcal N=2$ gauged supergravity.
(Here the supersymmetric extension of bosonic DFT to type IIB has been considered.)
Similarly, for compactifications of DFT on toroidal backgrounds 
with fluxes  the scalar potential of four-dimensional $\mathcal N=4$ supergravity 
has been obtained \cite{Geissbuhler:2011mx}.
These results show that double field theory correctly reproduces the scalar potential expected from
flux compactifications.

\item Double field theory is subject to the strong constraint \eqref{strong_c}. This 
constraint can be solved for instance by setting to zero all winding-coordinate 
dependencies, which for \eqref{dft_fluxes_24874}
implies that the DFT expressions agree with the generalised-geometry fluxes.

However, the strong constraint can also be solved by eliminating a linear combination 
of coordinates $X^i$ and $\hat X_i$. In generalised geometry this corresponds to 
choosing a different anchor projection for the Courant algebroid.

\item Double field theory is a rich subject with many applications. It is beyond the scope of this work
to go into more detail, and we have therefore only illustrated the appearance 
of non-geometric fluxes. The point we want to emphasise is that from a DFT point of view, 
non-geometric fluxes are on equal footing with ordinary geometric fluxes.

\item The generalised metric $\mathcal H_{IJ}$ and the $O(D,D)$ invariant metric
$\eta_{IJ}$ shown in \eqref{gen_met_098aaa} satisfy the following two relations
\eq{
  \mathcal H^T = \mathcal H \,, \hspace{70pt}
  \mathcal H^T \eta^{-1}\op \mathcal H = \eta \,.
}
Taking these now as  defining relations for two arbitrary $2D\times 2D$ matrices, in 
\cite{Morand:2017fnv} the corresponding geometries have been classified. 
Such geometries are in general non-Riemannian and have been called 
doubled-yet-gauged space-times \cite{Park:2013mpa}.

\end{itemize}

%%%%%%%%%%%%%%%%%%%%%%%%%%%%%%%%%%%%%%%%%%%%%%%
%%%%%%%%%%%%%%%%%%%%%%%%%%%%%%%%%%%%%%%%%%%%%%%
%%%%%%%%%%%%%%%%%%%%%%%%%%%%%%%%%%%%%%%%%%%%%%%
%%%%%%%%%%%%%%%%%%%%%%%%%%%%%%%%%%%%%%%%%%%%%%%
%%%%%%%%%%%%%%%%%%%%%%%%%%%%%%%%%%%%%%%%%%%%%%%
%%%%%%%%%%%%%%%%%%%%%%%%%%%%%%%%%%%%%%%%%%%%%%%
%%%%%%%%%%%%%%%%%%%%%%%%%%%%%%%%%%%%%%%%%%%%%%%
%%%%%%%%%%%%%%%%%%%%%%%%%%%%%%%%%%%%%%%%%%%%%%%
%%%%%%%%%%%%%%%%%%%%%%%%%%%%%%%%%%%%%%%%%%%%%%%
%%%%%%%%%%%%%%%%%%%%%%%%%%%%%%%%%%%%%%%%%%%%%%%
%%%%%%%%%%%%%%%%%%%%%%%%%%%%%%%%%%%%%%%%%%%%%%%
%%%%%%%%%%%%%%%%%%%%%%%%%%%%%%%%%%%%%%%%%%%%%%%
%%%%%%%%%%%%%%%%%%%%%%%%%%%%%%%%%%%%%%%%%%%%%%%
%%%%%%%%%%%%%%%%%%%%%%%%%%%%%%%%%%%%%%%%%%%%%%%
%%%%%%%%%%%%%%%%%%%%%%%%%%%%%%%%%%%%%%%%%%%%%%%
%%%%%%%%%%%%%%%%%%%%%%%%%%%%%%%%%%%%%%%%%%%%%%%

\clearpage
\section[Non-commutative and non-associative structures]{Non-commutative and non-associative \\ structures}
\label{cha_nca}

We now  discuss non-commutative and non-associative structures in string
theory. We explain how a non-commutative or  non-associative behaviour of  
closed-string coordinates is related to non-geometric fluxes, 
how such backgrounds can be obtained by applying T-duality transformations,
and how asymmetric orbifolds provide concrete realisations thereof. 
A review of these topics with focus on the underlying mathematical structures 
can be found in \cite{Szabo:2018hhh}.

%%%%%%%%%%%%%%%%%%%%%%%%%%%%%%%%%%%%%%%%%%%%%%%
%%%%%%%%%%%%%%%%%%%%%%%%%%%%%%%%%%%%%%%%%%%%%%%
%%%%%%%%%%%%%%%%%%%%%%%%%%%%%%%%%%%%%%%%%%%%%%%
%%%%%%%%%%%%%%%%%%%%%%%%%%%%%%%%%%%%%%%%%%%%%%%

\subsection{Non-associativity for closed strings}
\label{sec_nca_closed}

In this section we start our discussion by recalling  non-commutativity for the open string, 
and then discuss a similar reasoning for the closed string. For latter we will see 
a non-associative structure, which we describe using a tri-product. 

%%%%%%%%%%%%%%%%%%%%%%%%%%%%%%%%%%%%%%%%%%%%%%%
%%%%%%%%%%%%%%%%%%%%%%%%%%%%%%%%%%%%%%%%%%%%%%%

\subsubsection*{Open strings}

Let us start with non-commutativity  in the open-string sector. 
The world-sheet action describing open strings has a form similar to \eqref{action_01}, with the difference that
the two-dimensional world-sheet $\Sigma$ has a non-trivial boundary $\partial \Sigma\neq\emptyset$. 
Using the same conventions as in section~\ref{cha_t_duality} we have
\eq{
\label{0}
\arraycolsep2pt
\begin{array}{lcll}
 \mathcal S&=& \displaystyle
 -\frac{1}{4\pi \alpha'} \int_{\Sigma} &
 \Bigl[ \,G_{\mu\nu} \, d X^{\mu}\wedge\star d X^{\nu} -  B_{\mu\nu} \,dX^{\mu}\wedge dX^{\nu}
   + \alpha' \op\mathsf R\, \phi \star 1\,  \Bigr] 
 \\[12pt]
 &&\displaystyle-\frac{1}{2\pi\alpha'}\int_{\partial\Sigma} &\Bigl[\,  a_{\mu} \op  dX^{\mu}+
 \alpha' \op k(s) \op
   \phi \, d s
 \,\Bigr] \,,
 \end{array}
}
where the open-string $U(1)$ gauge field $a= a_{\mu}\op  dX^{\mu}$ is understood to be restricted to the boundary $\partial\Sigma$.
Coordinates on the world-sheet are denoted by $\sigma^{\alpha} = \{\sigma^0,\sigma^1\}$,
and  $t^{\alpha}$ and $n^{\alpha}$ are unit vectors tangential and normal to the boundary, respectively. 
The extrinsic 
curvature of the boundary is expressed as $k=t^{\alpha}\op t^{ \beta}\op\nabla_{\alpha} n_{\beta}$,
the coordinate on the boundary $\partial\Sigma$ is denoted by $s$, 
and  we have $ dX^{\mu} \rvert_{\partial\Sigma}= t^{\alpha}\partial_{\alpha} X^{\mu} \op d s$.
The gauge-invariant open-string field strength 
is a combination of the field strength $F=da$ for $a$ and the Kalb-Ramond field $B$, and can 
be expressed as\footnote{In this section we choose a different normalisation of $F=da$ as compared to 
equation \eqref{open_string_fs}.} 
\eq{
   \mathcal F =  F - B \,.
}
As boundary condition we can impose 
Dirichlet boundary conditions of the form 
$\delta X^{\mu} \rvert_{\partial\Sigma}=0$
or Neumann boundary conditions. 
Denoting the tangential and normal part of $ dX^{\mu}$ on the boundary by
$( dX^{\mu})_{\rm tan}  \equiv
   t^{\alpha} \op\partial_{\alpha} X^{\mu} \, d s \rvert_{\partial\Sigma} $ 
and
$\bigl( dX^{\mu})_{\rm norm}  \equiv
   n^{\alpha} \op\partial_{\alpha} X^{\mu} \, d s \rvert_{\partial\Sigma} $
we can summarise these boundary conditions as
\eq{
  \label{boundary_cond}
  \arraycolsep2pt
  \begin{array}{@{}l@{\hspace{30pt}}l@{}}
  \mbox{Dirichlet} & \displaystyle 0 = \bigl( dX^{\mu}\bigr)_{\rm tan}  \,,
  \\[6pt]
  \mbox{Neumann} & \displaystyle 0 = 
  \bigl( dX^{\mu}\bigr)_{\rm norm} +  \mathcal F^{\mu}{}_{\nu} \bigl( dX^{\nu}\bigr)_{\rm tan}
  + \alpha'\op k(s) \op G^{\mu\nu} \partial_{\nu} \phi\,d s\Bigr\rvert_{\partial\Sigma} ,
  \end{array}
}
where the index of $\mathcal F$ has been raised with the inverse of $G_{\mu\nu}$. 
Finally, a D$p$-brane is characterised by Neumann boundary conditions along the target-space time 
direction $X^0$ and along $p$  spatial directions. The remaining target-space directions are  
of Dirichlet type.

%%%%%%%%%%%%%%%%%%%%%%%%%%%%%%%%%%%%%%%%%%%%%%%
%%%%%%%%%%%%%%%%%%%%%%%%%%%%%%%%%%%%%%%%%%%%%%%

\subsubsection*{Non-commutativity}

Restricting now to a flat target space with $G_{\mu\nu} = \eta_{\mu\nu}$, 
constant Kalb-Ramond field $B_{\mu\nu}$, constant dilaton $\phi$
and constant open-string field strength $F_{\mu\nu}$,
one can determine the mode expansion of the open-string fields $X^{\mu}(\sigma^{\alpha})$. 
In particular, for Neumann boundary conditions we have
\eq{
\label{nca_open_746982}
&X^{\mu}(\tau,\sigma)=x^{\mu}_0+ 
\frac{2\pi \alpha'}{\ell_{\rm s}}\op
\bigl(\op p^{\mu} \op\tau - {\cal F}^{\mu}{}_{\nu}\op p^{\nu}\op \sigma\op \bigr) 
\\
&\hspace{50pt}+ 
i\sqrt{2 \alpha'}
\sum_{n\ne 0} \frac{1}{n} \, e^{-i \frac{n\op\pi\op\tau}{\ell_{\rm s}} } \left[ \,
        \alpha^{\mu}_n \cos \left( \frac{n\op \pi \op\sigma}{\ell_{\rm s}} \right) -
         i \,
        {\cal F}^{\mu}{}_{\nu}\op \alpha^{\nu}_n
        \sin  \left( \frac{n\op \pi \op\sigma}{\ell_{\rm s}} \right)
        \right] ,
}
where we normalised the world-sheet direction normal to the boundary as $0\le \sigma\le \ell_{\rm s}$,
and where the world-sheet time coordinate  $\tau$ is tangential to the boundary. 
As carried out in \cite{Chu:1998qz}, 
the commutation relations for  the modes appearing in \eqref{nca_open_746982}
can be obtained via canonical quantisation.
Using these relations,  the equal-time
commutator on the D-brane is evaluated as
\begin{equation}
   \label{comm_open_res}
   \bigl[X^{\mu}(\tau, 0),X^{\nu}(\tau,0)\bigr]=-\bigl[X^{\mu}(\tau,\ell_{\rm s}),X^{\nu}(\tau,\ell_{\rm s})\bigr]=
    2\pi i\hspace{1pt}\alpha'  \bigl(M^{-1}{\cal F}\bigr)^{\mu\nu}\, ,
\end{equation}  
where $\sigma=0,\ell_{\rm s}$ corresponds to the two endpoints of the open string on the D-brane. The matrix $M$ is defined as $M_{\mu\nu}=\eta_{\mu\nu}-  {\cal F}_{\mu}{}^{\rho} {\cal F}_{\rho\nu}$.
The relations \eqref{comm_open_res} show that the endpoints of the open string do not commute, and hence 
we find a non-commutative structure in the open-string sector.
Their low-energy limit can be described via a non-commutative gauge theory \cite{Seiberg:1999vs}, 
and for a review of such theories see for instance \cite{Szabo:2001kg}.

%%%%%%%%%%%%%%%%%%%%%%%%%%%%%%%%%%%%%%%%%%%%%%%
%%%%%%%%%%%%%%%%%%%%%%%%%%%%%%%%%%%%%%%%%%%%%%%

\subsubsection*{The Moyal-Weyl product for  open strings}

Another way to detect the non-commutative nature of the above setting  is to
consider the two-point function of two fields $X^{\mu}(\tau,\sigma)$. 
Going to an Euclidean world-sheet via a Wick rotation $\tau\to i\,\tau$ 
and  introducing a complex coordinate $z = \exp(\tau + i \sigma)$,
the two-point function of two open-string coordinates $X^{\mu}(z)$ inserted on the boundary  
takes the form \cite{Fradkin:1985qd,Callan:1986bc,Abouelsaood:1986gd,Seiberg:1999vs}
\begin{equation}
  \label{twopointa}
   \bigl\langle X^{\mu} (\mathsf z_1)\, X^{\nu}(\mathsf z_2) \bigr\rangle =  
  - \alpha' \mathsf G^{\mu\nu} \log (\mathsf z_1-\mathsf z_2)^2 + 
   \frac{i}{2} \op \theta^{\mu\nu}\,  \epsilon(\mathsf z_1-\mathsf z_2)\,,
\end{equation}
where $\mathsf z=\mbox{Re}\op(z)$ denotes the real part of the complex world-sheet coordinate $z$. 
The matrix $\mathsf G^{\mu\nu}$ is symmetric and can be interpreted as the (inverse of the) effective metric seen by the open string, and  $\theta^{\mu\nu}$ is  anti-symmetric and proportional to the two-form flux $\mathcal F$.
They are determined as the symmetric and the anti-symmetric part of the inverse of $G-\mathcal F$ as
\eq{
  \Bigl[ ( G - \mathcal F )^{-1} \Bigr]^{\mu\nu} = \mathsf G^{\mu\nu} + \frac{1}{2\pi\alpha'} \op\theta^{\mu\nu} \,.
}
The function $\epsilon(\mathsf z)$ is defined as
\eq{
  \epsilon(\mathsf z) = \left\{ \begin{array}{c@{\hspace{20pt}}c}
  +1 & \mathsf z \geq 0 \;, \\
  -1 & \mathsf z <0 \;,
  \end{array} \right.
}  
and it is  the appearance of the jump given by $\epsilon(\mathsf z_1 - \mathsf z_2)$ in \eqref{twopointa} which leads to  non-commutativity of the open-string coordinates on the D-brane.
The latter has also been discussed for instance 
in  \cite{Douglas:1997fm,Ardalan:1998ks,Ardalan:1998ce,Schomerus:1999ug}.

Next, we recall the form of an open-string vertex operator inserted at the boundary of an 
open-string
disc diagram.  Employing the short-hand notation $p\cdot X = p_{\mu} X^{\mu}$ and denoting normal-ordered products by $:\!\ldots\!:$, such a tachyon vertex operator can be written as
(for an introduction to conformal-field-theory techniques see for instance \cite{Blumenhagen:2009zz})
\begin{equation}
  \label{vertexa}
  \mathcal V(\mathsf z;p)= \::\! \exp\bigl(\op i\op p \cdot X(\mathsf z)\op \bigr)\!: \;.
\end{equation}
A correlation function of two such vertex operators in a background with non-vanishing $\mathcal F$-flux 
is found to be
\eq{
  \label{correla}
  \bigl\langle \,\mathcal V_1\, \mathcal V_2 \,\bigr\rangle 
  =\exp\Bigl[  \,i \op
    (p_1)_{\mu} \,\theta^{\mu\nu}\,   (p_2)_{\nu}\,  \epsilon(\mathsf z_1-\mathsf z_2)\Bigr]\times
  \bigl\langle \,\mathcal V_1\, \mathcal V_2 \,\bigr\rangle_{\theta=0} \,,
}
where $\theta=0$ denotes the result of the two-point function for vanishing flux.
The effect of the flux $\mathcal F$ encoded in $\theta$ is a phase factor 
due to the non-commutative nature of the theory.
This behaviour gives rise to the Moyal-Weyl star-product between functions $f_1$ and $f_2$ 
\eq{
  \label{Nbracketcon}
  \bigl( f_1 \star  f_2\bigr)(x) := 
  \exp\Bigl[  \, i \op     \theta^{\mu\nu}\,
      \partial^{x_1}_{\mu}\,\partial^{x_2}_{\nu}  \Bigr]
      \, f_1(x_1)\, f_2(x_2)\Bigr|_{x_1=x_2=x} \,,
}
which correctly reproduces the phase appearing in \eqref{correla}. 
Therefore, by evaluating correlation functions of vertex operators
in open-string theory, it is possible to 
derive the Moyal-Weyl product.

%%%%%%%%%%%%%%%%%%%%%%%%%%%%%%%%%%%%%%%%%%%%%%%
%%%%%%%%%%%%%%%%%%%%%%%%%%%%%%%%%%%%%%%%%%%%%%%

\subsubsection*{Non-associativity}

We now want to perform a computation analogous to the open-string case 
for the closed string. 
We are guided by the following observation:
%%%%%%%%%%%%%%%%
%%%%%%%%%%%%%%%%
\begin{figure}[t]
\centering
\begin{subfigure}{0.4\textwidth}
\centering
\includegraphics[width=100pt]{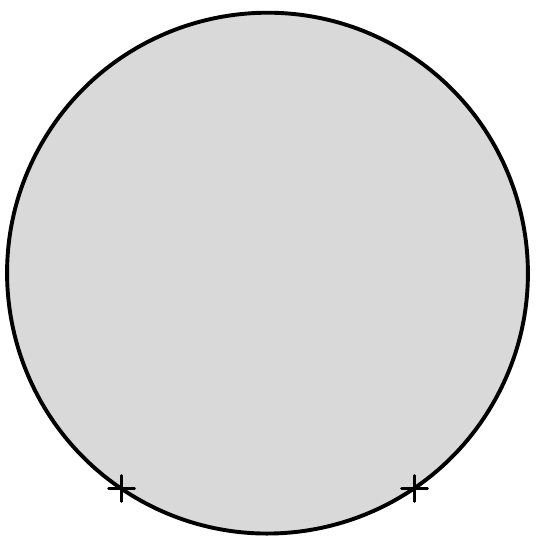}
\caption{World-sheet disc diagram.}
\begin{picture}(0,0)
\put(-41,48){\scriptsize$\mathcal V_1$}
\put(32,48){\scriptsize$\mathcal V_2$}
\end{picture}
\label{fig_disc_sphere_a}
\end{subfigure}
\hspace{40pt}
\begin{subfigure}{0.4\textwidth}
\centering
\includegraphics[width=100pt]{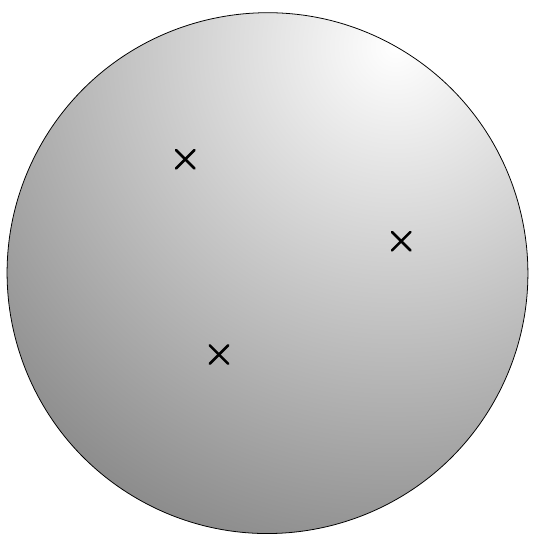}
\caption{World-sheet sphere diagram.}
\begin{picture}(0,0)
\put(-29,108){\scriptsize$\mathcal V_1$}
\put(-22,70){\scriptsize$\mathcal V_2$}
\put(28,90){\scriptsize$\mathcal V_3$}
\end{picture}
\label{fig_disc_sphere_b}
\end{subfigure}
\vspace*{-10pt}
\caption{Open-string disc diagram with the insertion of two open-string vertex operators 
on the boundary, and closed-string sphere diagram with the insertion of three 
closed-string vertex operators. \label{fig_disc_sphere} 
}
\end{figure}
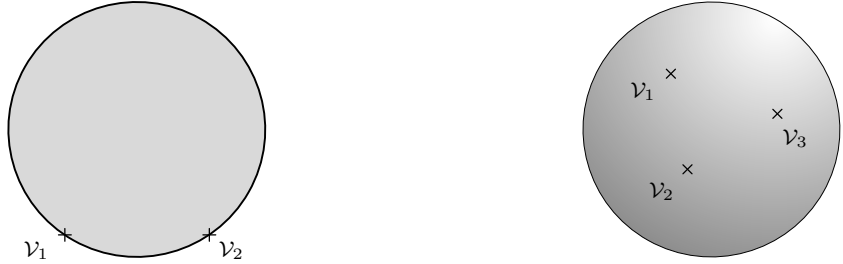
%%%%%%%%%%%%%%%%
%%%%%%%%%%%%%%%%
\begin{itemize}

\item For the open string, the non-commutativity arises because two operators 
inserted at the boundary of a world-sheet diagram (cf. figure~\ref{fig_disc_sphere_a}) do not commute. 
This non-commutativity is controlled by a two-form flux
$\mathcal F$.

\item For the closed string, operators are inserted on a  world-sheet without boundary (cf. figure~\ref{fig_disc_sphere_b}).
Here, one cannot define an ordering between two points -- however, 
for three-points  we can define an orientation. We therefore expect 
that for the closed string an object involving three closed-string fields is relevant,
which in turn should correspond to a three-form flux.

\end{itemize}
To make the latter point more precise, we consider the equal-time Jacobiator of 
three closed-string fields 
defined as 
\cite{Blumenhagen:2010hj}
\begin{equation}
  \label{nca_96924794}
  \bigl[X^\mu,X^\nu,X^\rho\bigr]:=
  \lim_{\sigma_i\to \sigma}\; \bigl[ [X^\mu (\tau,\sigma_1),
    X^\nu(\tau,\sigma_2)],X^\rho(\tau,\sigma_3)\bigr] +{\rm cyclic}\,. 
\end{equation}
Note that if this bracket is non-vanishing we have a non-associative structure, 
and we expect this three-bracket to be related to a three-form flux -- 
such as the $H$-flux, geometric flux, or the non-geometric $Q$- or $R$-fluxes.

This strategy has been followed in \cite{Blumenhagen:2010hj}, where the equal-time Jacobiator 
\eqref{nca_96924794}
 has been determined for the $SU(2)$ Wess-Zumino-Witten model (introduced around equation
 \eqref{tdual_su2_0t_99}).
This model is
described by an action of the form \eqref{action_01f}, 
and corresponds to a  string moving on a three-sphere $S^3$
with $H$-flux, where the radius of $S^3$ is related to the flux such that
the string equations of motion \eqref{eom_beta} are satisfied to all orders in sigma-model perturbation theory. 
For this background, the three-bracket \eqref{nca_96924794} was  found to be  \cite{Blumenhagen:2010hj}
\begin{equation}
  \label{nca_9864982984692386493864}
  \bigl[X^\mu,X^\nu,X^\rho\bigr]
   = \epsilon\, \theta^{\mu\nu\rho}\,,
\end{equation}
where $\theta^{\mu\nu\rho}$ is completely anti-symmetric in its indices and 
encodes the three-form flux. The parameter $\epsilon$ turns out to be 
$\epsilon=0$ for the $H$-flux background 
and $\epsilon=1$ for the background one
obtains after three T-dualities. 
Hence, for such backgrounds the closed string shows not only a non-commutative but 
a non-associative behaviour.

%%%%%%%%%%%%%%%%%%%%%%%%%%%%%%%%%%%%%%%%%%%%%%%
%%%%%%%%%%%%%%%%%%%%%%%%%%%%%%%%%%%%%%%%%%%%%%%

\subsubsection*{Tri-product}

Similarly to the open string, for the closed string the non-associativity can be encoded 
in a product of functions. 
This has been analysed in \cite{Blumenhagen:2011ph} using conformal field theory techniques, where  correlation functions of tachyon vertex operators 
have been computed for a background with $H$-flux (in a perturbative expansion).
Let us be more precise and state the definition of a closed-string 
tachyon vertex operator as
\eq{
  \label{def_vo7023}
  {\cal V}(z,\ov z;p) = \, :\!\exp \bigl( \op i\hspace{0.5pt} p_L\cdot  \mathcal X_L +
  i\hspace{0.5pt} p_R \cdot \mathcal X_R \op \bigr) \!: \,,
}
where $p_{L,R}^{\mu}$ are the left- and right-moving momenta. 
The coordinates  $\mathcal X^{\mu}_{L,R}$ are the left- and right-moving 
coordinates of the closed string, including linear corrections due to the $H$-flux.
The complex coordinate on the world-sheet is denoted by $z$.
For the three-tachyon amplitude one then finds \cite{Blumenhagen:2011ph}
\eq{
\label{nca_9864982984692386493864b}
  &\bigl\langle {\cal V}_1\,{\cal V}_2\,{\cal V}_3\op \bigr\rangle^{\pm} = 
   \exp \biggl[ \op -i  \op
   \theta^{\mu\nu\rho} (p_1)_{\mu} (p_2)_{\nu} (p_3)_{\rho}
   \Bigl[ {\cal L}\bigl({\textstyle \frac{z_{12}}{ z_{13}}}\bigr)
  \pm  {\cal L}\bigl({\textstyle\frac{\ov z_{12}}{ \ov z_{13}}}\bigr)
  \Bigr] \biggr] 
  \times  \bigl\langle {\cal V}_1\,{\cal V}_2\,{\cal V}_3\op \bigr\rangle_{\theta=0},
}
where we emphasise that  the exponential has been 
determined only up to linear order in $\theta$. 
The notation employed for the world-sheet coordinates is $z_{ij}=z_i-z_j$, where $z_i$ 
corresponds to the world-sheet coordinate of the closed-string vertex operator $\mathcal V_i(z_i,\ov z_i;p_i)$.
The function  $\mathcal L(z)$ is expressed in terms of the 
Rogers dilogarithm $L(z)$, which in turn is defined in terms of the usual dilogarithm ${\rm Li}_2(z)$
as follows
\eq{
   {\cal L}(z) &=L(z)+L\left({\textstyle 1-\frac{1}{z}}\right)
    + L\left({\textstyle \frac{1}{1-z}}\right)\,,
 \\
  L(z)& ={\rm Li}_2(z) + \frac{1}{ 2} \log (z) \log(1-z)\, .
}
The correlation function \eqref{nca_9864982984692386493864b} can be determined
reliably only for the $H$-flux background (corresponding to the $-$ sign) and
a background resulting after three T-dualities (corresponding to the $+$ sign). 
The latter is identified with an $R$-flux background.

Let us now study the behaviour of \eqref{nca_9864982984692386493864b}  under
permutations of the vertex operators ${\cal V}_i$. 
Using properties of the function $\mathcal L(z)$ and denoting by the $+$ sign again the 
$R$-flux background and by $-$ the $H$-flux background, 
one finds 
\begin{equation}
  \label{phasethreeperm}
  \bigl\langle \, {\cal V}_{\sigma(1)}   {\cal V}_{\sigma(2)}  {\cal V}_{\sigma(3)}  \bigr\rangle^{\pm}=
  \exp\Bigl[ \,i\left({\textstyle \frac{1\pm 1}{ 2}}\right)  \eta_\sigma\op  \pi^2\op   \theta^{\mu\nu\rho} (p_1)_{\mu} (p_2)_{\nu} (p_3)_{\rho}\Bigr]
  \bigl\langle {\cal V}_1\,  {\cal V}_2\,  {\cal V}_3  \bigr\rangle^{\pm} \,,
\end{equation}
where  $\eta_{\sigma}=1$ for an odd permutation and 
$\eta_{\sigma}=0$ for an even one. 
Thus,  for the $R$-flux background a non-trivial phase may appear
which in \cite{Blumenhagen:2011ph} has been established up to linear order in the flux. 
The phase in  \eqref{phasethreeperm}
suggests the definition of  a three-product of functions
$f(x)$ in the following way
\eq{
\label{threebracketcon}
   &\bigl( f_1\,\tri\, f_2\, \tri\, f_3\bigr) (x) \\
   &\hspace{60pt}
   := \exp\Bigl(
   {\textstyle \frac{\pi^2}{2}}\, \theta^{\mu\nu\rho}\,
      \partial^{x_1}_{\mu}\,\partial^{x_2}_{\nu}\,\partial^{x_3}_{\rho} \Bigr)\, f_1(x_1)\, f_2(x_2)\,
   f_3(x_3)\Bigr|_{x_1=x_2=x_3=x} \,.
}
The three-bracket \eqref{nca_9864982984692386493864} can then be re-derived  as
the completely anti-sym\-me\-trised sum of three-products (up to an overall constant) as
\eq{
\label{antisymtripcon}
    \bigl[X^\mu,X^\nu,X^\rho\bigr] =\sum_{\sigma\in P^3} {\rm sign}(\sigma) \;  
     x^{\sigma(\mu)}\, \tri\,  x^{\sigma(\nu)}\, \tri\,  x^{\sigma(\rho)} =
     3\pi^2\, \theta^{\mu\nu\rho}\; ,
}
where $P^3$ denotes the permutation group of three elements.
Let us recall that this three-bracket was defined
as the Jacobi-identity of the coordinates, which can
only be non-zero if the space is non-commutative and non-associative.

%%%%%%%%%%%%%%%%%%%%%%%%%%%%%%%%%%%%%%%%%%%%%%%
%%%%%%%%%%%%%%%%%%%%%%%%%%%%%%%%%%%%%%%%%%%%%%%

\subsubsection*{Remarks}

Let us close this section with remarks on the tachyon correlation function  and
on the tri-product \eqref{threebracketcon}:
\begin{itemize}

\item In CFT correlation functions operators
are understood to be radially ordered and so changing the order
of operators should not change the form of the amplitude.
This is known as crossing symmetry which is one of the defining
properties of a CFT.
In the case of the $R$-flux background, this
is reconciled with 
\eqref{phasethreeperm} 
by applying momentum conservation as
\eq{
   \theta^{\mu\nu\rho} (p_1)_{\mu} (p_2)_{\nu} (p_3)_{\rho} = 0
  \hspace{40pt}{\rm for}\hspace{40pt} p_1+p_2+p_3=0 \,.
}
Therefore, scattering amplitudes of three tachyons do not receive
any corrections at linear order in $\theta$  both for the $H$- and
 $R$-flux.  
(This is analogous to the situation in non-commutative open-string theory, 
where the two-point function \eqref{correla} does not receive any corrections.)
The non-associative behaviour for the closed string should therefore be understood
as an off-shell property of the theory (see also \cite{Aschieri:2015roa}).

\item In the above analysis the flux $\theta^{\mu\nu\rho}$ was assumed to be constant. For a
discussion with a non-constant flux in the context of double field theory
see \cite{Blumenhagen:2013zpa}.

\item Using Courant algebroids and regarding closed strings as boundary excitations of more 
fundamental membrane degrees of freedom, a non-associative star-product has been 
proposed in \cite{Mylonas:2012pg}.
This product can be related to the tri-product introduced in  \eqref{threebracketcon},
which has been established at linear order in the flux 
in \cite{Mylonas:2012pg}
and at all orders
in \cite{Aschieri:2015roa} (including extensions towards a non-associative differential geometry).
This star-product 
was also obtained via deformation quantisation of twisted Poisson manifolds in  \cite{Mylonas:2012pg}, 
but can also be found by integrating higher Lie-algebra structures \cite{Bakas:2013jwa}.
We also note that membrane sigma-models have  been used to study non-geometric fluxes
\cite{Chatzistavrakidis:2015vka} and properties of double field theory \cite{Chatzistavrakidis:2018ztm}.

\item In relation to the open string, we note that 
the result of a two-form flux inducing  non-commutativity of brane coordinates
is completely general, and has also been studied for codimension
one branes in the $SU(2)$ WZW model \cite{Alekseev:1999bs}. 
However, due to a background $H$-flux in this case,
 it turns out that the obtained structure is not only non-commutative but also
non-associative \cite{Alekseev:1999bs,Cornalba:2001sm,Herbst:2001ai}.

\item Using a non-associative star-product, one can try to construct a corresponding 
non-associative theory of differential geometry and a non-associative theory of 
gravity. This idea has been proposed in \cite{Blumenhagen:2010hj}, and 
further been developed in \cite{Barnes:2014ksa,Barnes:2015uxa,Blumenhagen:2016vpb,Aschieri:2017sug}.

\item Properties of non-associative star-products have been studied from a more mathematical
point of view also in \cite{Kupriyanov:2015dda,Kupriyanov:2016hsm}
and \cite{Bojowald:2016lnl}.

\end{itemize}

%%%%%%%%%%%%%%%%%%%%%%%%%%%%%%%%%%%%%%%%%%%%%%%
%%%%%%%%%%%%%%%%%%%%%%%%%%%%%%%%%%%%%%%%%%%%%%%
%%%%%%%%%%%%%%%%%%%%%%%%%%%%%%%%%%%%%%%%%%%%%%%
%%%%%%%%%%%%%%%%%%%%%%%%%%%%%%%%%%%%%%%%%%%%%%%

\subsection{Non-commutativity for closed strings}
\label{sec_nca_nc}

In this section, we discuss the non-commutative behaviour of the closed string in the context 
of torus fibrations. In particular, we consider parabolic and elliptic $\mathbb T^2$-fibrations 
over a circle and perform T-duality transformations. It turns out that the T-dual 
configurations can be interpreted as asymmetric orbifolds, for which the closed-string 
coordinates do not commute.

%%%%%%%%%%%%%%%%%%%%%%%%%%%%%%%%%%%%%%%%%%%%%%%
%%%%%%%%%%%%%%%%%%%%%%%%%%%%%%%%%%%%%%%%%%%%%%%

\subsubsection*{Main idea}

Let us consider a setting similar to the one in section~\ref{cha_torus_fib}, where we considered
$n$-dimensional torus fibrations over a base-manifold. 
Assuming that the non-triviality of the fibration is small in a certain parameter regime (for instance
a monodromy around a large cycle in the base), we can quantise the 
closed string on the fibre perturbatively. 
In this regime the equation of motion for the fibre-coordinates $X^{\mathsf a}(\tau,\sigma)$ reads 
(cf. equation \eqref{eom_closed_9494})
\eq{
 0=\partial_+\partial_- X^{\mathsf a}(\tau,\sigma) + \mathcal O(\Theta)\,,
 \hspace{60pt}\mathsf a = 1,\ldots, n\,,
}
where $\Theta\ll 1$ encodes the non-triviality of the fibration. 
For the geometric-flux background discussed in section~\ref{sec_twisted_torus_ex}, this parameter would be 
related to the flux-density shown in \eqref{ex1_2052}.
As discussed for instance in \cite{Blumenhagen:2011ph}, the solution to this equation of motion can then be split into left- and right-moving 
part similarly as in \eqref{cft_aa849}
\eq{
  X^{\mathsf a} (\tau,\sigma) = 
     X_L^{\mathsf a} (\tau+\sigma) + X_R^{\mathsf a} (\tau-\sigma) \,.
}
In general, the commutation relations of the left- and right-moving modes will take the following 
form \cite{Lust:2010iy}
\eq{
  \bigl[ X_L^{\mathsf a}, X_L^{\mathsf b} \op\bigr] = \tfrac{i}{2}\op\Theta_1^{\mathsf a\mathsf b} \,,
  \hspace{40pt}
  \bigl[ X_L^{\mathsf a}, X_R^{\mathsf b} \op\bigr] = 0\,,
  \hspace{40pt}
  \bigl[ X_R^{\mathsf a}, X_R^{\mathsf b} \op\bigr] =\tfrac{i}{2}\op \Theta_2^{\mathsf a\mathsf b} \,,  
}
where $\Theta_{1,2}^{\mathsf{ab}}$ are anti-symmetric in their indices and encode the non-triviality 
of the fibration. 
Now, for a purely geometric background -- such as the twisted torus with geometric flux --
one finds that $\Theta^{\mathsf{ab}}_1=-\Theta^{\mathsf{ab}}_2=\Theta^{\mathsf{ab}}$, and hence the fibre-coordinates commute
\eq{
  \label{nca_883582}
   \bigl[ X^{\mathsf a}, X^{\mathsf b} \op\bigr]
= 
   \bigl[ X_L^{\mathsf a} + X_R^{\mathsf a}, X_L^{\mathsf b} + X_R^{\mathsf b} \op\bigr]
   = \tfrac{i}{2}\op\bigl( \Theta^{\mathsf{ab}} - \Theta^{\mathsf{ab}} \bigr) = 0 \,.
}

Let us now perform T-duality transformations for the above situation fibre-wise. 
As discussed in the case of a single T-duality in section~\ref{sec_cft_circ}, for the
left- and right-moving coordinates this implies 
that the sign of the right-moving coordinate is changed
\eq{
  \label{nca_flip_24}
   \bigl(\,X^{\hat{\mathsf a}}_L\,,\,X^{\hat{\mathsf a}}_R\,\bigr) \quad \longrightarrow    
   \quad \bigl(\,+ X^{\hat{\mathsf a}}_L\,,\,- X^{\hat{\mathsf a}}_R\,\bigr) \,.
}
Here, $\hat{\mathsf a}$ denotes the direction along which a T-duality transformation 
has been performed, and in the following the dual coordinate
will be denoted by
$\tilde X^{\hat{\mathsf a}} = X^{\hat{\mathsf a}}_L - X^{\hat{\mathsf a}}_R$.
Coming back to the commutation relation \eqref{nca_883582}, 
we can now compute for instance \cite{Lust:2010iy}
\eq{
   \bigl[ \tilde X^{\hat{\mathsf a}}, X^{\mathsf b} \op\bigr]
= 
   \bigl[ X_L^{\hat{\mathsf a}} - X_R^{\hat{\mathsf a}}, X_L^{\mathsf b} + X_R^{\mathsf b} \op\bigr]
   = \tfrac{i}{2}\op\bigl( \Theta^{\hat{\mathsf a}\mathsf{b}} + \Theta^{\hat{\mathsf a}\mathsf{b}} \bigr) = i \op \Theta^{\hat{\mathsf a}\mathsf{b}} \,.
}
This means that after one T-duality for a purely geometric background the coordinates 
for the dual background do not need to commute anymore, and in general one 
obtains a non-commutative geometry.

%%%%%%%%%%%%%%%%%%%%%%%%%%%%%%%%%%%%%%%%%%%%%%%
%%%%%%%%%%%%%%%%%%%%%%%%%%%%%%%%%%%%%%%%%%%%%%%

\subsubsection*{Examples I -- parabolic  $\mathbb T^2$-fibrations over $S^1$}

The main idea outlined above has been proposed in \cite{Lust:2010iy} for $\mathbb T^2$-fibrations 
over a circle, and has been checked  more systematically in \cite{Andriot:2012vb}.
In order to discuss these results, let us denote coordinates in the fibre torus by $X^1$ and $X^2$, 
and the coordinate in the base will be denoted by $X^3$
\eq{
  \label{nca_ex_8244}
  \begin{array}{l@{\hspace{30pt}}l@{\hspace{6pt}}l}
  \mbox{fibre $\mathbb T^2$} & \biggl\{ & \begin{array}{@{}l@{}} X^1 \\[2pt] X^2 \end{array} 
  \\[16pt]
  \mbox{base $S^1$} && X^3
  \end{array}
}  
The metric and Kalb-Ramond $B$-field can be parametrised similarly as in \eqref{back_ex_939393},
and the fibre-components can  
be expressed using 
the complexified K\"ahler and complex-structure moduli $\rho$ and $\tau$ as  in \eqref{monodro_392}.
Let us then recall from section~\ref{sec_t2_fibr_general}
that the three-torus with $H$-flux, the twisted torus with geometric flux and the T-fold with 
non-geometric $Q$-flux can be characterised by the following monodromies 
\eq{
  \arraycolsep2pt
  \begin{array}{l@{\hspace{40pt}}lcl@{\hspace{20pt}}lcl}
  \mbox{$\mathbb T^3$ with $H$-flux } &  \tau & \to & \tau \,,
  & \rho & \to & \rho + h \,,
  \\[16pt]
  \mbox{twisted torus with $f$-flux } &  \tau & \to & \displaystyle 
   \frac{\tau}{-f\op\tau+1} \,,  & \rho & \to & \rho  \,,
  \\[16pt]
  \mbox{T-fold with $Q$-flux} & \tau & \to & 
   \tau \,,  & \rho & \to & \displaystyle  \frac{\rho}{-q\op \rho+1}  \,,
  \end{array}
}
when going around the base-circle as $X^3\to X^3+2\pi$. In this list, $h,f,q\in\mathbb Z$ denote the flux-quantum 
for the $H$-, geometric and $Q$-flux, and we note that these monodromies are all of parabolic type (cf. page~\pageref{page_monodro_class}).

In \cite{Lust:2010iy,Condeescu:2012sp,Andriot:2012vb} the commutation relations between the fibre-coordinates 
$X^1$ and $X^2$ have been determined for all three backgrounds mentioned above.
In particular, for a sector of the theory in which the base-coordinate $X^3$ has winding-number $N^3$ 
the equal-time commutator in the limit of coincident world-sheet space coordinates takes the form
\eq{
  \label{nca_nc_2846}
  \arraycolsep2pt
  \begin{array}{@{}l@{\hspace{30pt}}lcl@{}}
  \mbox{$\mathbb T^3$ with $H$-flux } &
  \displaystyle \bigl[ X^1(\tau,\sigma) , X^2(\tau,\sigma') \bigr] &\xrightarrow{\;\sigma'\to\sigma\;} & 
  \displaystyle0 \,,
  \\[12pt] 
  \mbox{twisted torus with $f$-flux } &
  \displaystyle \bigl[ X^1(\tau,\sigma) , X^2(\tau,\sigma') \bigr] &\xrightarrow{\;\sigma'\to\sigma\;} & 
  \displaystyle 0\,,
  \\[10pt] 
  \mbox{T-fold with $Q$-flux } &
  \displaystyle \bigl[ X^1(\tau,\sigma) , X^2(\tau,\sigma') \bigr] &\xrightarrow{\;\sigma'\to\sigma\;} & 
  \displaystyle - \frac{i}{2} \op \frac{\pi^2}{3}\op N^3\op q \,.
  \end{array}
}
Hence, in this limit, the commutator of two fibre-coordinates of the T-fold is non-vanishing, indicating 
a non-commutative behaviour. Furthermore, the right-hand side of this commutator is proportional 
to the winding number $N^3$ which can also be expressed as 
$N^3 = \oint dX^3$. Denoting then the constant $Q$-flux by $q = Q_3{}^{12}$ we can express the 
commutator for the T-fold as \cite{Andriot:2012an}
\eq{
\bigl[ X^1(\tau,\sigma) , X^2(\tau,\sigma') \bigr]_{\mbox{\scriptsize T-fold}} \xrightarrow{\;\sigma'\to\sigma\;}  
   - \frac{i\op\pi^2}{6} \op \oint Q_3{}^{12} dX^3 \,.
}
It has furthermore been proposed that this expression also holds for flux backgrounds in which the $Q$-flux depends
on the target-space coordinates.

%%%%%%%%%%%%%%%%%%%%%%%%%%%%%%%%%%%%%%%%%%%%%%%
%%%%%%%%%%%%%%%%%%%%%%%%%%%%%%%%%%%%%%%%%%%%%%%

\subsubsection*{Examples II -- elliptic  $\mathbb T^2$-fibrations over $S^1$}

Another class of examples are $\mathbb T^2$-fibrations with elliptic monodromies, as 
opposed to the parabolic ones of the previous paragraph. 
As mentioned in section~\ref{sec_schsch}, elliptic monodromies have fixed points 
which have an orbifold description (see footnote~\ref{foot_orbifold}).
Let us consider for instance 
a background with vanishing Kalb-Ramond field and elliptic monodromy
\eq{
  \label{nca_3427551}
  \tau \to -\frac{1}{\tau} \,.
}
If we denote two lattice vectors generating the two-torus by $\omega_1,\omega_2\in \mathbb C$,
the complex structure is given by $\tau = \omega_2/\omega_1$. 
The monodromy \eqref{nca_3427551} can then be represented by 
for instance $\omega_1\to\omega_2$ and $\omega_2\to - \omega_1$,
which for the  coordinates of the $\mathbb T^2$-fibre shown in \eqref{nca_ex_8244} implies
\eq{
  \binom{X^1}{X^2} \to \binom{+X^2}{-X^1} \,.
}
This $\mathbb Z_4$-action can be conveniently expressed in terms of a complex 
target-space coordinate $Z = \frac{1}{\sqrt{2}}(X^1 + i X^2)$, for which one finds
that 
\eq{
  Z \to e^{-\frac{2\pi i}{4}} Z \,.
}
Similarly as above, the mode expansion of the world-sheet field $Z(\tau,\sigma)$ can be determined
which respects this orbifold action. However, when taking into account the non-trivial 
fibration of the $\mathbb T^2$-fibre over the base-circle, such an expansion 
can only be written down at lowest order in the twisting of the fibre. 
Following \cite{Lust:2010iy,Condeescu:2012sp}, one finds
that 
\eq{
  \arraycolsep2pt
  \begin{array}{lcl}
  Z_L(\tau+\sigma) & = & \displaystyle i\op \sqrt{\frac{\alpha'}{2}} \sum_{n\in \mathbb Z} \frac{1}{n-\theta}\,
  \alpha_{n-\theta}\op e^{-\frac{2\pi i}{\ell_{\rm s}} (n-\theta)(\tau+\sigma)} \,,
  \\[4pt]
  Z_R(\tau-\sigma) & = & \displaystyle i\op \sqrt{\frac{\alpha'}{2}} \sum_{n\in \mathbb Z} \frac{1}{n+\theta}\,
  \tilde \alpha_{n+\theta}\op e^{-\frac{2\pi i}{\ell_{\rm s}} (n+\theta)(\tau-\sigma)} \,,
  \end{array}
}
where $\theta = -f \op N^3$ with $f\in \frac14 + \mathbb Z$ and $N^3$ labelling the winding-sector 
of the base-coordinate $X^3$. 
Similar expressions are obtained for the complex-conjugate coordinate $\ov Z(\tau,\sigma)$, 
and the commutation relations for the oscillator modes
are found to be $[\alpha_{m-\theta},\ov\alpha_{n-\theta}]=(m-\theta)\delta_{m,n}$. 
Using these expressions,  the following  equal-time commutators are computed
\eq{
  \arraycolsep2pt
  \begin{array}{c@{\hspace{0.5pt}}c@{\hspace{0.5pt}}cccc@{\hspace{0.5pt}}c@{\hspace{0.5pt}}ccl}
  \bigl[ & Z_L(\tau+\sigma),\ov Z_L(\tau+\sigma')& \bigr] &=& -&   
  \bigl[ &Z_R(\tau-\sigma),\ov Z_R(\tau-\sigma') &\bigr] &=& \tfrac{1}{2}\op \Theta(\sigma-\sigma';\theta) \,, 
  \\[8pt]
  \bigl[ & Z_L(\tau+\sigma),\ov Z_R(\tau-\sigma') &\bigr] &=&&   
  \bigl[& Z_R(\tau-\sigma),\ov Z_L(\tau+\sigma') &\bigr] &=& 0 \,.
  \end{array}
}
The function $\Theta(\sigma-\sigma';\theta)$ depends on the difference of the world-sheet space coordinates $\sigma$ and 
$\sigma'$, and its explicit form can be found \cite{Condeescu:2012sp}.
However, here we are only interested in the limit
\eq{
  \Theta(\theta) = \lim_{\sigma'\to\sigma} \Theta(\sigma-\sigma';\theta)
  = \left\{ \begin{array}{c@{\hspace{20pt}}l}
  -2\pi \cot(\pi \theta) & \mbox{for~}\theta\notin\mathbb Z\,,
  \\[6pt]
  0 & \mbox{for~}\theta\in\mathbb Z\,.
  \end{array}
  \right.
}  
From these expressions one finds that the target-space coordinates for an elliptic monodromy 
commute and that therefore the background is geometric
\eq{
    \bigl[ Z,\ov Z \bigr] = 0 \hspace{40pt} \Longleftrightarrow\hspace{40pt}
     \bigl[ X^1, X^2 \bigr] = 0 \,.
}

Let us next perform a T-duality transformation along say the direction $X^1$, which implies that 
similarly as in \eqref{nca_flip_24}
we change the sign of the right-moving modes. 
This means that $Z_R \to - \ov { Z}_R$, which then leads to the equal time commutator
$[ \op Z(\tau,\sigma),\ov { Z}(\tau,\sigma') \op] \xrightarrow{\;\sigma'\to\sigma\;}\; \Theta(\theta) $.
Using the real target-space fields, this corresponds to \cite{Lust:2010iy}
\eq{    
  \label{nca_nc_2846b}
     \bigl[ \tilde X^1(\tau,\sigma), X^2(\tau,\sigma') \bigr] \xrightarrow{\;\sigma'\to\sigma\;} i\op\Theta(\theta) \,.
}
Hence, after a T-duality transformation a background with non-commuting coordinates is obtained.
The commutation relations of the T-dual background are the same as for
an orbifold with the following orbifold action
\eq{
  \label{nca_nc_229487}
  \binom{X_L^1}{X_L^2} \to \binom{+X_L^2}{-X_L^1} \,,
  \hspace{50pt}
  \binom{X_R^1}{X_R^2} \to \binom{-X_R^2}{+X_R^1} \,,
}
which in terms of the complex coordinate $Z$ is expressed as
\eq{
    Z_L \to e^{-\frac{2\pi i}{4}} Z_L \,,
    \hspace{50pt}
    Z_R \to e^{+\frac{2\pi i}{4}} Z_R \,.
}
Note that the transformation behaviour is asymmetric between the left- and right-moving sectors, and 
hence it is called an asymmetric orbifold \cite{Narain:1986qm,Narain:1990mw}.
We furthermore observe that the mapping \eqref{nca_nc_229487} can be understood using 
$O(D,D)$-transformations. Recalling from page~\pageref{page_odd_trans_003} how the left- and right-moving 
coordinates behave under the duality group $O(D,D)$, we can infer that for the above background (with vanishing Kalb-Ramond
field) the corresponding $O(2,2,\mathbb Z)$ matrix can be chosen as
\eq{
  \arraycolsep2pt
  \renewcommand{\arraystretch}{1.2}
  \mathcal O = \left(
  \begin{array}{cccc}
  0 & 0 & 0 & +1 \\
  0 & 0 & -1 & 0 \\
  0 & +1 & 0 & 0 \\
  -1 & 0 & 0 & 0
  \end{array}\right) .
}
This transformation does not belong to the geometric subgroup of $O(2,2,\mathbb Z)$, 
and hence the dual background is non-geometric. 
We therefore see in this example that a T-duality transformation applied to a symmetric orbifold 
compactification leads to an asymmetric one, which can then be interpreted as a non-geometric 
background.

%%%%%%%%%%%%%%%%%%%%%%%%%%%%%%%%%%%%%%%%%%%%%%%
%%%%%%%%%%%%%%%%%%%%%%%%%%%%%%%%%%%%%%%%%%%%%%%

\subsubsection*{Remarks}

Let us close this section with the following remarks and comments:
\begin{itemize}

\item It is worth emphasising that similar to the discussion in section~\ref{sec_nca_closed},
the analysis is done at lowest order in the fluxes. In particular, the 
non-triviality of the fibration has been taken into account only at linear order, 
which made it possible to obtain explicit mode expansions and compute 
the commutators. 
This approach is justified since in the $\beta$-functionals \eqref{eom_beta} the 
flux only appears at quadratic order \cite{Blumenhagen:2011ph}.
At higher orders in the flux the above analysis becomes more involved.

\item The equal-time commutators of the target-space coordinates discussed in \eqref{nca_nc_2846}
and \eqref{nca_nc_2846b} depend in general on the world-sheet coordinate $\sigma$. 
Hence, from a target-space point of view such a commutator is not well-defined. However, 
in the limit $\sigma'\to\sigma$ this dependence vanishes, and a target-space 
interpretation is possible.

\item Additional examples for elliptic monodromies and their T-dual asymmetric orbifolds
can be found in \cite{Condeescu:2012sp}. In particular, $\mathbb T^5$-fibrations over 
a circle with freely-acting asymmetric orbifold actions are analysed.
In \cite{Condeescu:2013yma} this analysis has been extended to include more general cases,
and an explicit  relation between asymmetric orbifolds and the effective gauged supergravity theory has been given. 
Other examples for asymmetric orbifolds and their relation to non-geometric backgrounds 
can be found for instance in 
\cite{Dabholkar:2002sy,Hellerman:2002ax,Flournoy:2004vn,Flournoy:2005xe,Hellerman:2006tx}.

\item In an approach similar to  \cite{Lust:2010iy}, the non-commutative behaviour of the string 
coordinates has been derived by computing Dirac brackets for doubled-geometry world-sheet theory 
in \cite{Blair:2014kla}.

\end{itemize}

%%%%%%%%%%%%%%%%%%%%%%%%%%%%%%%%%%%%%%%%%%%%%%%
%%%%%%%%%%%%%%%%%%%%%%%%%%%%%%%%%%%%%%%%%%%%%%%
%%%%%%%%%%%%%%%%%%%%%%%%%%%%%%%%%%%%%%%%%%%%%%%
%%%%%%%%%%%%%%%%%%%%%%%%%%%%%%%%%%%%%%%%%%%%%%%

\subsection{Phase-space algebra}
\label{sec_nca_phase}

The above results for the commutation relations of closed-string coordinates 
can  be generalised. In this section we discuss such a generalisation 
as a twisted Poisson structure.

%%%%%%%%%%%%%%%%%%%%%%%%%%%%%%%%%%%%%%%%%%%%%%%
%%%%%%%%%%%%%%%%%%%%%%%%%%%%%%%%%%%%%%%%%%%%%%%

\subsubsection*{Point-particle in a magnetic field}

Before coming to the non-associativity corresponding to the $R$-flux, it is worth 
pointing out that a point-particle in the background of a magnetic field  also leads to 
a non-associative behaviour. 
Denoting by $x^i$ the coordinates and by $p_i$ the corresponding  momenta, the 
phase-space algebra takes the form
\eq{
  \label{bca_ps_al_24b}
  [x^i, x^j]= 0 \,,
  \hspace{40pt}
  [x^i, p_j] = i\op \delta^i{}_j \,,
  \hspace{40pt}
  [p_i,p_j] = i\op e \op\epsilon_{ijk} B^k\,,
}
where $\epsilon_{ijk}$ denotes the Levi-Civita tensor, $e$ is the electric charge of the point-particle and $B^k$ contains the magnetic field.
For this algebra the Jacobiator of the momenta is computed as \cite{Jackiw:1984rd,Jackiw:1985hq}
\eq{
   [\op p_i,p_j,p_k ] := \bigl[ \op[ p_i,p_j],p_k \bigr] + \mbox{cyclic} = - e\op \epsilon_{ijk} \nabla_m B^m \,,
}
which in general is non-vanishing and which shows the non-associative behaviour of the setting. 
Furthermore, from Maxwell's equations we know that for a magnetic monopole one has $\nabla_m B^m = 4\pi \op\rho$, where $\rho$ denotes the charge distribution of the monopole.

Considering now the finite translation operator $U(a)= \exp(i \op \vec a\cdot  \vec p)$ along a distance $\vec a$, we can determine 
the associator of three such operators as
\eq{ 
  \label{nca_82479247}
  \bigl( \op U( a) U(b) \op\bigr)\op U(c)  = \exp\Bigl[- i \op e \op\Phi(a,b,c) \Bigr]  U( a)\op\bigl(\op U(b)  U(c) \op\bigr)\,,
}
where $\Phi(a,b,c) = \frac16 \epsilon_{ijk}  a^i b^j c^k  \nabla_m B^m$ denotes the magnetic flux through the
tetrahedron spanned by the three vectors $\vec a$, $\vec b$ and $\vec c$.
Using Gauss law, this is the magnetic charge $4\pi \op m$ inside the tetrahedron. 
We also note that the non-associative phase in \eqref{nca_82479247}  vanishes if 
$e\op m \in \frac12 \mathbb Z$, which is Dirac's quantisation condition of the electric charge in the presence 
of a magnetic monopole. 
Hence, the non-associative behaviour is related to a violation of the Dirac quantisation condition.

%%%%%%%%%%%%%%%%%%%%%%%%%%%%%%%%%%%%%%%%%%%%%%%
%%%%%%%%%%%%%%%%%%%%%%%%%%%%%%%%%%%%%%%%%%%%%%%

\subsubsection*{$R$-flux algebra}

Let us now turn to the string-theory setting. 
We recall  the commutator of two target-space coordinates for the T-fold background
shown in \eqref{nca_nc_2846}, and note that the right-hand side is proportional to the winding number 
of the field $X^3$ parametrising the circle in the  base. 
Even though this direction is not an isometry, one can perform a T-duality transformation
along the base-circle in an abstract way. As mentioned for instance in equation \eqref{duality_001},
such a transformation is expected to interchange the winding with the momentum number
and hence on the right-hand side of the T-dual commutator the momentum appears. 
This now motivates the following $R$-flux phase-space algebra \cite{Lust:2010iy}
\eq{
  \label{bca_ps_al_24}
  [x^i, x^j]= i \op R^{ijk} p_k \,,
  \hspace{40pt}
  [x^i, p_j] = i\op \delta^i{}_j \,,
  \hspace{40pt}
  [\op p_i,p_j] = 0\,,
}
where $p_i$ are again the  momenta conjugate to the positions $x^i$, and where $R^{ijk}$ denotes 
the $R$-flux obtained after T-dualising the $Q$-flux. 
These relations arise from a twisted Poisson structure on the phase space \cite{Lust:2012fp,Mylonas:2012pg}, 
and one can determine the Jacobiator \eqref{nca_9864982984692386493864} as
\eq{
  \label{bca_ps_al_24c}
  [x^i,x^j,x^k ] := \bigl[ [x^i,x^j],x^k \bigr] + \mbox{cyclic} = 3\op R^{ijk} \,,
}
showing that the phase-space algebra \eqref{bca_ps_al_24} is non-associative. 
Furthermore, we note that in \cite{Bakas:2013jwa} the Jacobiator \eqref{bca_ps_al_24c} has been identified as a three-cocycle in Lie algebra
cohomology.

%%%%%%%%%%%%%%%%%%%%%%%%%%%%%%%%%%%%%%%%%%%%%%%
%%%%%%%%%%%%%%%%%%%%%%%%%%%%%%%%%%%%%%%%%%%%%%%

\subsubsection*{Remarks}

Let us close this section with the following  remarks:
\begin{itemize}

\item  In order to construct a Hilbert space for a quantised theory, one usually requires 
associativity. However, this requirement is not necessary and 
one can in fact construct a non-associative form of quantum mechanics
\cite{Mylonas:2013jha,Bojowald:2014oea,Bojowald:2018qqa}.
An investigation of the point-particle in a magnetic field from a quantum-mechanical 
point of view can be found in \cite{Bojowald:2015cha}, 
and for a more formal analysis see \cite{Bunk:2018qvk}.

\item The $R$-flux algebra \eqref{bca_ps_al_24}, and more generally non-commutative and non-associative 
structures originating from non-geometric fluxes, have been discussed in the context of 
matrix models in \cite{Chatzistavrakidis:2012qj}.

\item A realisation of the $R$-flux algebra in M-theory based on octonions 
has been first discussed in \cite{Gunaydin:2016axc}.
In \cite{Kupriyanov:2017oob} a higher three-algebra structure expected to govern the 
non-geometric M2-brane phase space has been proposed, which
has been embedded into a covariant description in \cite{Lust:2017bgx}
and related to exceptional field theory in
\cite{Lust:2017bwq}.

\end{itemize}

%%%%%%%%%%%%%%%%%%%%%%%%%%%%%%%%%%%%%%%%%%%%%%%
%%%%%%%%%%%%%%%%%%%%%%%%%%%%%%%%%%%%%%%%%%%%%%%
%%%%%%%%%%%%%%%%%%%%%%%%%%%%%%%%%%%%%%%%%%%%%%%
%%%%%%%%%%%%%%%%%%%%%%%%%%%%%%%%%%%%%%%%%%%%%%%

\subsection{Topological T-duality}
\label{sec_nca_top_t}

In this section we discuss how non-commutative and non-associative structures arise
when studying T-duality transformations from a mathematical point of view.

%%%%%%%%%%%%%%%%%%%%%%%%%%%%%%%%%%%%%%%%%%%%%%%
%%%%%%%%%%%%%%%%%%%%%%%%%%%%%%%%%%%%%%%%%%%%%%%

\subsubsection*{Topology change from T-duality}

We have already seen that under T-duality  the topology can change \cite{Giveon:1993ph,Bouwknegt:2003vb}. 
For instance, the three-torus with $H$-flux can be seen as a trivially-fibred circle over 
$\mathbb T^2$ while its T-dual -- the twisted torus -- can be seen as a non-trivially fibred 
circle over $\mathbb T^2$. 
In order to formalise this observation, let us consider 
a circle bundle over some base-manifold $\mathcal B$
\eq{
  S^1 \; \varlonghookrightarrow\; \begin{array}[t]{@{\op}c}\mathcal M \\[4pt]
  \big\downarrow \\[8pt]
  \mathcal B
  \end{array}
  \begin{picture}(0,0)
  \put(-4,-17.5){\footnotesize $\pi$}
  \end{picture}
}
where $\pi:\mathcal M\to \mathcal B$ denotes the projection from the fibre to a point in the base. 
The non-triviality of the fibration is encoded in the first Chern class $c_1(\mathcal M)\in H^2(\mathcal B, \mathbb Z)$, which 
is a two-form in the cohomology of the base-manifold. (For a textbook introduction to these
concepts see for instance \cite{Nakahara:2003nw}.)
Next, we denote the vector-field along the $S^1$-fibre by $k$, and we perform
a T-duality transformation along the fibre. One then finds that \cite{Bouwknegt:2003vb}
\eq{
  c_1(\mathcal M) \quad \xleftrightarrow{\hspace{8pt}\mbox{\scriptsize T-duality}\hspace{8pt}} \quad \iota_k H \,,
}
where $\iota_k H$ denotes the contraction of the $H$-flux $H\in H^3(\mathcal M,\mathbb Z)$  with the vector-field $k$. 
Hence, under T-duality the two-form corresponding to the first Chern class is exchanged with a two-form 
constructed from the $H$-flux. 
For the example of the three-sphere with $H$-flux we discussed this point in section~\ref{sec_wzw_action}.

In order to describe T-duality in this way, it is convenient to introduce the following structure. 
Denoting by a tilde the T-dual background and by $\times_{\mathcal B}$ the fibre-wise product over the
base-manifold $\mathcal B$, one has
\eq{
  \arraycolsep15pt
  \begin{picture}(0,0)
  \put(80,-12){\vector(-1,-1){30}}
  \put(106,-12){\vector(1,-1){30}}
  \put(50,-74){\vector(1,-1){30}}
  \put(136,-74){\vector(-1,-1){30}}
  \put(56,-98){\footnotesize $\pi$}  
  \put(125,-98){\footnotesize $\tilde\pi$}  
  \put(56,-22){\footnotesize $p$}  
  \put(125,-22){\footnotesize $\tilde p$}  
  \end{picture}
  \begin{array}[t]{@{\hspace{24pt}}ccc}
  &\mathcal M \times_{\mathcal B} \widetilde{\mathcal M} &
  \\[45pt]
  \mathcal M && \widetilde{\mathcal M}
  \\[40pt]
  &\mathcal B &
  \end{array}
}
where $\mathcal M \times_{\mathcal B} \widetilde{\mathcal M} $ is also called the correspondence space
and $\pi$, $\tilde \pi$, $p$ and $\tilde p$ denote various projections. 
For the bundle $\mathcal M$  one can define a twisted cohomology $H^{\bullet}(\mathcal M, H)$, 
where the nil-potent operator is given by $d+H\wedge$ similarly as in 
section~\ref{sec_cy_flux_echt}. As shown in \cite{Bouwknegt:2003vb}, T-duality for 
a circle bundle can now be seen as an isomorphism 
\eq{
  T_* : \; H^{\bullet}(\mathcal M, H) \to H^{\bullet+1}(\widetilde{\mathcal M}, \widetilde H)
}
where the superscript $+1$ denotes  a shift in the degree of the cohomology group. 
This isomorphism can be understood by first lifting $H^{\bullet}(\mathcal M, H)$
to the correspondence space using $p^*$, then performing a transformation on the correspondence space, and 
finally projecting down 
to $H^{\bullet}(\widetilde{\mathcal M}, \widetilde H)$ using $\tilde p_*$. 
We also note that in order to describe T-duality for the Ramond-Ramond sector one has 
to use twisted K-theory, for which one obtains a similar isomorphism.

%%%%%%%%%%%%%%%%%%%%%%%%%%%%%%%%%%%%%%%%%%%%%%%
%%%%%%%%%%%%%%%%%%%%%%%%%%%%%%%%%%%%%%%%%%%%%%%

\subsubsection*{Torus fibrations}

One can now generalise the above discussion to $n$-dimensional (principal)
torus fibrations $\mathcal M$ over a base-manifold $\mathcal B$.
To do so, let us first split the $H$-flux into fibre- and base-components in the following way
\eq{
  \label{nca_24996}
  H = H_0 + H_1 \wedge b_1 + H_2 \wedge b_2 + H_3\wedge b_3 \,,
}
where $H_p$ is a $(p-3)$-form along the torus-fibre and $b_q$ is a $q$-form on the base-manifold 
$\mathcal B$ (this notation is chosen for comparison with \cite{Bouwknegt:2004ap}). Furthermore, $H$ is required to be closed with respect to the exterior derivative.
We can now distinguish the following cases:
\begin{itemize}

\item For $H_0=0$ and $H_1=0$ we have a situation similar as above \cite{Bouwknegt:2003zg}. 
The first Chern class 
of the $\mathbb T^n$-fibre 
can be interpreted as a vector-valued two-form on the base-manifold $c_1^{\mathsf a}(\mathcal M)$,
where the index $\mathsf a=1,\ldots, n$ labels the circles in $\mathbb T^n = S^1 \times \ldots \times S^1$. 
For this situation there is a uniquely-determined T-dual background, where under T-duality along a fibre direction $\mathsf i$ the first Chern class 
$c_1^{\mathsf i}(\mathcal M)$ and the two-form $\iota_{\partial_{\mathsf i}}H$ 
are interchanged.

\item For $H_0=0$ and $H_1\neq0$ the situation is different
\cite{Mathai:2004qq,Mathai:2004qc}.
A T-duality transformation along two directions 
supported by the $H$-flux results in a field of non-commutative tori over the base-manifold  $\mathcal B$.
The T-dual is not unique, however, it does not affect its K-theory. 

\item Finally, for $H_0\neq0$ and if $H=H_0$ the T-dual is a field of
non-associative tori over the base  \cite{Bouwknegt:2004ap}. The T-dual is again not unique, but it does not affect its
K-theory. If furthermore also $H_1\neq0$ then one obtains a combination of non-commutative
and non-associative tori.

\end{itemize} 
Let us remark that a possible connection between the non-commutative torus and the T-fold background
has been investigated in \cite{Grange:2006es}, and that 
the non-associative torus has be studied using D-branes in \cite{Ellwood:2006my}.

%%%%%%%%%%%%%%%%%%%%%%%%%%%%%%%%%%%%%%%%%%%%%%%
%%%%%%%%%%%%%%%%%%%%%%%%%%%%%%%%%%%%%%%%%%%%%%%

\subsubsection*{Non-commutative torus}

Let us now provide some explanation for a non-commutative and non-associative torus. 
Roughly speaking, a topological space can be characterised in terms of the algebra of functions on this space
(cf. Gelfand-Naimark theorem).
A commutative algebra of functions corresponds to a commutative space -- 
while a non-commutative or 
non-associative algebra corresponds to a  non-commutative or non-associative space.

Let us now consider the non-commutative torus.
To do so, we start from
an ordinary two-torus $\mathbb T^2$ with plane-waves $U_1=e^{2\pi i x_1}$ and  
$U_2=e^{2\pi i x_2}$ where $x_{1,2}\in [0,1]$. Functions $f\in C^{\infty}(\mathbb T^2)$ on $\mathbb T^2$ can be expressed using a Fourier expansion as
\eq{
  \label{nca_22442}
  f = \sum_{(m,n)\in\mathbb Z^2} a_{m,n} \op U_1^m \op U_2^n \,,
}
where $a_{m,n}$ are complex-valued Schwartz-functions and $U_1^m$ denotes 
the $m$'th power of $U_1$. The algebra of functions on $\mathbb T^2$ is commutative. 
Next, we promote $x_{1,2}$ to operators $\hat x_{1,2}$ which satisfy
\eq{
  \label{nca_22442b}
  [\op \hat x_1, \hat x_2] = \frac{\theta}{2\pi i}\,, \hspace{40pt}\theta\in\mathbb R\,.
}
Using the Baker-Campbell-Hausdorff formula, this
 implies for the corresponding plane-waves $\hat U_{1,2}$ that 
\eq{
  \label{nca_22442c}
  \hat U_1\op \hat U_2 = e^{2\pi i \op \theta} \hat U_2 \op \hat U_1 \,,
}
which characterises the non-commutative torus. 
Functions $\hat f$ can now be expanded similarly as in \eqref{nca_22442}, which 
now satisfy a non-commutative algebra.
Coming now back to our discussion of T-duality, for a duality transformation along two 
directions the two-form $H_1$ in \eqref{nca_24996} corresponds 
to the parameter $\theta$ in \eqref{nca_22442b} and \eqref{nca_22442c} on the dual side
and controls the non-commutativity.

%%%%%%%%%%%%%%%%%%%%%%%%%%%%%%%%%%%%%%%%%%%%%%%
%%%%%%%%%%%%%%%%%%%%%%%%%%%%%%%%%%%%%%%%%%%%%%%

\subsubsection*{Remark}

We close this section with two remarks:
\begin{itemize}

\item The non-associative torus is similar to the non-commutative torus, where however
the algebra of functions is non-associative. The technical details of this construction 
can be found in \cite{Bouwknegt:2004ap}.

\item The description of T-duality as an isomorphism between $H$-twisted coho\-mol\-o\-gies
can also be formulated as an isomorphism of Courant algebroids. 
A discussion of this result can be found in \cite{Cavalcanti:2011wu}.

\end{itemize}

%%%%%%%%%%%%%%%%%%%%%%%%%%%%%%%%%%%%%%%%%%%%%%%
%%%%%%%%%%%%%%%%%%%%%%%%%%%%%%%%%%%%%%%%%%%%%%%
%%%%%%%%%%%%%%%%%%%%%%%%%%%%%%%%%%%%%%%%%%%%%%%
%%%%%%%%%%%%%%%%%%%%%%%%%%%%%%%%%%%%%%%%%%%%%%%
%%%%%%%%%%%%%%%%%%%%%%%%%%%%%%%%%%%%%%%%%%%%%%%
%%%%%%%%%%%%%%%%%%%%%%%%%%%%%%%%%%%%%%%%%%%%%%%
%%%%%%%%%%%%%%%%%%%%%%%%%%%%%%%%%%%%%%%%%%%%%%%
%%%%%%%%%%%%%%%%%%%%%%%%%%%%%%%%%%%%%%%%%%%%%%%
%%%%%%%%%%%%%%%%%%%%%%%%%%%%%%%%%%%%%%%%%%%%%%%
%%%%%%%%%%%%%%%%%%%%%%%%%%%%%%%%%%%%%%%%%%%%%%%
%%%%%%%%%%%%%%%%%%%%%%%%%%%%%%%%%%%%%%%%%%%%%%%
%%%%%%%%%%%%%%%%%%%%%%%%%%%%%%%%%%%%%%%%%%%%%%%
%%%%%%%%%%%%%%%%%%%%%%%%%%%%%%%%%%%%%%%%%%%%%%%
%%%%%%%%%%%%%%%%%%%%%%%%%%%%%%%%%%%%%%%%%%%%%%%
%%%%%%%%%%%%%%%%%%%%%%%%%%%%%%%%%%%%%%%%%%%%%%%
%%%%%%%%%%%%%%%%%%%%%%%%%%%%%%%%%%%%%%%%%%%%%%%

\clearpage
\section{Summary}
\label{cha_concl}

In this work we gave an overview of non-geometric backgrounds in string theory. 
These are spaces which cannot be  described in terms of  Riemannian 
geometry, but which are  well-defined in string theory. 
For instance, transition functions between local charts do not need to 
belong to symmetry transformations of the action such as diffeomorphisms or gauge transformations, 
but can contain T-duality transformations.

Non-geometric backgrounds are an integral part of string theory:
they can be characterised in terms of 
non-geometric fluxes which 
fit naturally into the framework of $SU(3)\times SU(3)$ structure compactifications
and which complete mirror symmetry for Calabi-Yau compactifications;
and non-geometric torus fibrations are needed for
heterotic--F-theory duality.
On the other hand, constructing explicit non-geometric solutions of string theory is difficult as
typically supergravity approximations break down.
However, solutions for certain non-geometric torus fibrations are provided by asymmetric orbifolds, for which a CFT 
description exists. 
Non-geometric backgrounds belong to the string-theory landscape and understanding them
is  crucial for understanding the space of string-theory solutions. 
But also at a more practical  level such backgrounds are important: 
they lead to scalar potentials in lower-dimensional theories and can therefore 
be used to construct models of particle physics and cosmology in string theory.
Furthermore, non-geometry can give rise to non-commutative and  non-associative
structures relevant for theories of quantum gravity.

The material discussed in this review can be organised into three (sometimes overlapping) topics: non-geometric spaces
and their explicit realisation, non-geo\-me\-tric fluxes and their effect on string-theory 
compactifications, and non-commut\-ative and non-associative structures.
We summarise them in some more detail in the following.

%%%%%%%%%%%%%%%%%%%%%%%%%%%%%%%%%%%%%%%%%%%%%%%
%%%%%%%%%%%%%%%%%%%%%%%%%%%%%%%%%%%%%%%%%%%%%%%

\subsubsection*{Non-geometric spaces}

In the introduction we have mentioned on page~\pageref{page_defs_nongeo} that non-geometric spaces 
are configurations which do not allow
for a geometric interpretation.
\begin{itemize}

\item Non-geometric spaces have been discussed in the context of torus fibrations
in section~\ref{cha_torus_fib}. As illustrated for instance in figure~\ref{fig_forus_fibration}, for these 
constructions the monodromy group acting on a $n$-dimensional torus fibre is given 
by $O(n,n,\mathbb Z)$, which is the T-duality group identified in section~\ref{cha_t_duality}.
This group contains symmetry transformations such as diffeomorphisms and gauge transformations as well as 
proper duality transformations.

\item The prime example for non-geometric spaces is the three-torus with $H$-flux
with its T-dual backgrounds, 
which have been studied in detail in section~\ref{sec_first_steps}. 
Other explicit examples are the compactified NS5-brane together with the 
Kaluza-Klein monopole and $5^2_2$-brane which have been considered in 
section~\ref{sec_tw_r2}, and the $\mathbb T^2$-fibrations discussed in sections~\ref{sec_t2_fibr_general}
and \ref{sec_t2_p1}.

\item A framework to discuss such non-geometric spaces is that of doubled geometry 
reviewed in section~\ref{cha_dg}. Here one doubles the dimensions of the fibre, which 
allows for a geometric description of non-geometric spaces.

\end{itemize}

%%%%%%%%%%%%%%%%%%%%%%%%%%%%%%%%%%%%%%%%%%%%%%%
%%%%%%%%%%%%%%%%%%%%%%%%%%%%%%%%%%%%%%%%%%%%%%%

\subsubsection*{Non-geometric fluxes}

To the backgrounds mentioned above one can often associate 
geometric as well as non-geometric fluxes. 
This has been made explicit for the example of the three-torus with $H$-flux discussed 
in section~\ref{sec_first_steps}.
\begin{itemize}

\item The $H$-flux encodes the non-triviality of the Kalb-Ramond $B$-field,
and the geometric flux (related to the first Chern class of torus fibrations)
encodes the non-triviality of the geometry. 
Both belong to the geometric fluxes. The non-geometric $Q$- and $R$-fluxes arise from
applying T-duality transformations.

\item For more general backgrounds, fluxes can be defined in the framework of 
generalised geometry discussed in section~\ref{sec_gen_geo}. In particular, 
the Courant bracket of generalised vielbeins determines these fluxes.

\item In section~\ref{cha_sugra} we have analysed how non-geometric fluxes
modify the effective four-dimensional theory corresponding to Calabi-Yau compactifications: 
fluxes lead to a gauging of $\mathcal N=2$ and $\mathcal N=1$ supergravity 
theories in four dimensions, which in turn induces a scalar potential. 
Explicit examples in the context of Scherk-Schwarz reductions have been 
discussed in section~\ref{sec_schsch}.
We have furthermore seen that non-geometric fluxes appear on equal footing 
with geometric fluxes and that they are needed for mirror symmetry.

\end{itemize}

%%%%%%%%%%%%%%%%%%%%%%%%%%%%%%%%%%%%%%%%%%%%%%%
%%%%%%%%%%%%%%%%%%%%%%%%%%%%%%%%%%%%%%%%%%%%%%%

\subsubsection*{Non-commutative and non-associative structures}

In section~\ref{cha_nca}  we have illustrated how non-geometric backgrounds
give rise to non-commutative and non-associative structures. 
\begin{itemize}

\item Non-associativity for the closed string due to non-geometric $R$-flux 
has been discussed in section~\ref{sec_nca_closed}.
Such a behaviour can be detected by computing correlation functions of vertex operators, 
which in turn leads to a non-associative tri-product.

\item In a slightly different approach, non-commutativity for closed strings has been 
studied in section~\ref{sec_nca_nc} by quantising the closed string for torus 
fibrations. After performing T-duality transformations this leads to non-vanishing 
commutators between target-space coordinates parametrised by the $Q$-flux. 
These constructions  correspond to asymmetric orbifolds.

\item In section~\ref{sec_nca_top_t} non-commutative and non-associative structures 
originating from topological T-duality have been discussed.

\end{itemize}

%%%%%%%%%%%%%%%%%%%%%%%%%%%%%%%%%%%%%%%%%%%%%%%
%%%%%%%%%%%%%%%%%%%%%%%%%%%%%%%%%%%%%%%%%%%%%%%

\subsubsection*{Topics not covered}

In this review we  gave a pedagogical introduction to non-geometric backgrounds 
in string theory. While providing an overview on many aspects of such spaces,
it was not possible to discuss each of those topics in detail and some topics had to be omitted:
\begin{itemize}

\item We have discussed  (collective) abelian T-duality transformations and  
we commented on non-abelian T-duality starting on page~\pageref{page_nonab}.
For a more detailed discussion of the latter we refer the reader for instance to 
\cite{delaOssa:1992vci,Giveon:1993ai,Alvarez:1993qi,Curtright:1994be,
Sfetsos:1994vz,Alvarez:1994np,Klimcik:1995ux,Lozano:1995jx,Curtright:1996ig},
and we mention that non-abelian T-duality has been employed as a solution-generating technique 
for instance in  \cite{Itsios:2012zv,Itsios:2012dc,Lozano:2012au}.

\item Certain (non-abelian) T-duality transformations can be related to 
integrable deformations of supergravity backgrounds. This has been 
investigated  in \cite{Klimcik:2002zj,Sfetsos:2013wia,Sfetsos:2014cea,Hoare:2015gda,
Borsato:2016zcf,Osten:2016dvf}.
Relations to non-commutative geometries and non-geo\-met\-ric backgrounds have been addressed for instance in 
\cite{Hoare:2016wca,Fernandez-Melgarejo:2017oyu,Lust:2018jsx}.

\item A world-sheet formulation which incorporates non-geometric fluxes
has been proposed in 
\cite{Halmagyi:2008dr,Halmagyi:2009te}. 
The Courant bracket (encoding geometric as well as non-geometric fluxes, cf.~section~\ref{sec_gen_geo})
has been derived from the usual sigma model in \cite{Alekseev:2004np},
and a description of $R$-fluxes from a membrane sigma-model point of view
can be found in \cite{Mylonas:2012pg}.

\item In section~\ref{sec_fluxes_app} we gave an overview on 
non-geometric string-theory solutions in view of moduli stabilisation, 
realising inflation and constructing de Sitter vacua. 
We did not construct explicit models in this review, but on page~\pageref{page_lit_sol} referred to 
the existing literature.

\item In section~\ref{sec_dft} we gave a brief introduction to double field theory
and explained its relevance in regard to non-geometric fluxes,
however, a more thorough discussion is beyond the scope of this work.
For more details  we refer to the existing review articles
\cite{Aldazabal:2013sca,Berman:2013eva,Hohm:2013bwa}.
The extension of the T-duality covariant formulation given by DFT
to a U-duality covariant framework is called exceptional field theory. 
Review articles for these formulations can be found in 
\cite{Hohm:2013vpa,Hohm:2013uia,Hohm:2014fxa}.

\item In this review we have focused on bosonic string theory and on type II superstring
theory. But non-geometric backgrounds also appear for
the heterotic string 
as studied in \cite{McOrist:2010jw,Malmendier:2014uka,Gu:2014ova,Lust:2015yia,Font:2016odl,Kimura:2018oze}.
In the context of M-theory non-geometric fluxes have been discussed for instance
in \cite{Blair:2014zba,Gunaydin:2016axc,Lust:2017bgx,Lust:2017bwq}.

\end{itemize}

%%%%%%%%%%%%%%%%%%%%%%%%%%%%%%%%%%%%%%%%%%%%%%%
%%%%%%%%%%%%%%%%%%%%%%%%%%%%%%%%%%%%%%%%%%%%%%%
%%%%%%%%%%%%%%%%%%%%%%%%%%%%%%%%%%%%%%%%%%%%%%%
%%%%%%%%%%%%%%%%%%%%%%%%%%%%%%%%%%%%%%%%%%%%%%%
%%%%%%%%%%%%%%%%%%%%%%%%%%%%%%%%%%%%%%%%%%%%%%%
%%%%%%%%%%%%%%%%%%%%%%%%%%%%%%%%%%%%%%%%%%%%%%%
%%%%%%%%%%%%%%%%%%%%%%%%%%%%%%%%%%%%%%%%%%%%%%%
%%%%%%%%%%%%%%%%%%%%%%%%%%%%%%%%%%%%%%%%%%%%%%%
%%%%%%%%%%%%%%%%%%%%%%%%%%%%%%%%%%%%%%%%%%%%%%%
%%%%%%%%%%%%%%%%%%%%%%%%%%%%%%%%%%%%%%%%%%%%%%%
%%%%%%%%%%%%%%%%%%%%%%%%%%%%%%%%%%%%%%%%%%%%%%%
%%%%%%%%%%%%%%%%%%%%%%%%%%%%%%%%%%%%%%%%%%%%%%%
%%%%%%%%%%%%%%%%%%%%%%%%%%%%%%%%%%%%%%%%%%%%%%%
%%%%%%%%%%%%%%%%%%%%%%%%%%%%%%%%%%%%%%%%%%%%%%%
%%%%%%%%%%%%%%%%%%%%%%%%%%%%%%%%%%%%%%%%%%%%%%%
%%%%%%%%%%%%%%%%%%%%%%%%%%%%%%%%%%%%%%%%%%%%%%%

\clearpage
\section*{Acknowledgements}

The author is grateful for  many helpful and interesting discussions 
during the preparation of this work with 
D.~Andriot,
D.~Berman,
R.~Blumenhagen,
I.~Brunner,
G.~Cavalcanti,
A.~Chatzistavrakidis,
G.~Dall'Agata,
L.~Ferro,
S.~Groot-Nibbelink,
M.~Haack,	
R.~Helling,
C.~Hull,
L.~Jonke,
D.~Junghans,
E.~Kiritsis,
S.~Krippendorf,
V.~Kupriyanov,
M.~Larfors,
D.~L\"ust,
S.~Massai,
E.~Malek,
C.~Mayrhofer,
J.~McOrist,
A.~Orta,
F.~Rudolph,
C.~Schmidt-Colinet,
R.~Szabo,
V.~Vall Camell,
T.~van~Riet and
M.~Zagermann.
For very useful comments on the manuscript he thanks
A.~Chatzistavrakidis,
F.~Cordonier-Tello,
H.~Erbin,
D.~Klaewer,
\mbox{D.~Osten} and
M.~Syv\"ari,
and he  thanks his collaborators 
with whom he has worked on many aspects of non-geometric backgrounds over the last years:
I.~Bakas,
\mbox{P.~Betzler,}
R.~Blumenhagen,
F.~Cordonier-Tello,
A.~Deser,
A.~Font,
M.~Fuchs,
D.~Herschmann,
D.~L\"ust,
F.~Rennecke,
C.~Schmid,
Y.~Sekiguchi, 
V.~Vall Camell and
F.~Wolf.

The author is partially supported through a Maria-Weber-Grant of the Hans B\"ockler 
Foundation, and he is grateful to the Mainz Institute for Theoretical Physics (MITP) for its hospitality 
and its partial support during the completion of this work. 
He  also thanks the Lorentz Institute for Theoretical Physics in Leiden for its hospitality.

%%%%%%%%%%%%%%%%%%%%%%%%%%%%%%%%%%%%%%%%%%%%%%%
%%%%%%%%%%%%%%%%%%%%%%%%%%%%%%%%%%%%%%%%%%%%%%%
%%%%%%%%%%%%%%%%%%%%%%%%%%%%%%%%%%%%%%%%%%%%%%%
%%%%%%%%%%%%%%%%%%%%%%%%%%%%%%%%%%%%%%%%%%%%%%%
%%%%%%%%%%%%%%%%%%%%%%%%%%%%%%%%%%%%%%%%%%%%%%%
%%%%%%%%%%%%%%%%%%%%%%%%%%%%%%%%%%%%%%%%%%%%%%%
%%%%%%%%%%%%%%%%%%%%%%%%%%%%%%%%%%%%%%%%%%%%%%%
%%%%%%%%%%%%%%%%%%%%%%%%%%%%%%%%%%%%%%%%%%%%%%%
%%%%%%%%%%%%%%%%%%%%%%%%%%%%%%%%%%%%%%%%%%%%%%%
%%%%%%%%%%%%%%%%%%%%%%%%%%%%%%%%%%%%%%%%%%%%%%%
%%%%%%%%%%%%%%%%%%%%%%%%%%%%%%%%%%%%%%%%%%%%%%%
%%%%%%%%%%%%%%%%%%%%%%%%%%%%%%%%%%%%%%%%%%%%%%%
%%%%%%%%%%%%%%%%%%%%%%%%%%%%%%%%%%%%%%%%%%%%%%%
%%%%%%%%%%%%%%%%%%%%%%%%%%%%%%%%%%%%%%%%%%%%%%%
%%%%%%%%%%%%%%%%%%%%%%%%%%%%%%%%%%%%%%%%%%%%%%%
%%%%%%%%%%%%%%%%%%%%%%%%%%%%%%%%%%%%%%%%%%%%%%%

\clearpage

\addcontentsline{toc}{section}{\protect Bibliography}

\bibliography{references}
\bibliographystyle{utphys}

%%%%%%%%%%%%%%%%%%%%%%%%%%%%%%%%%%%%%%%%%%%%%%%
%%%%%%%%%%%%%%%%%%%%%%%%%%%%%%%%%%%%%%%%%%%%%%%
%%%%%%%%%%%%%%%%%%%%%%%%%%%%%%%%%%%%%%%%%%%%%%%
%%%%%%%%%%%%%%%%%%%%%%%%%%%%%%%%%%%%%%%%%%%%%%%
%%%%%%%%%%%%%%%%%%%%%%%%%%%%%%%%%%%%%%%%%%%%%%%
%%%%%%%%%%%%%%%%%%%%%%%%%%%%%%%%%%%%%%%%%%%%%%%
%%%%%%%%%%%%%%%%%%%%%%%%%%%%%%%%%%%%%%%%%%%%%%%
%%%%%%%%%%%%%%%%%%%%%%%%%%%%%%%%%%%%%%%%%%%%%%%
%%%%%%%%%%%%%%%%%%%%%%%%%%%%%%%%%%%%%%%%%%%%%%%
%%%%%%%%%%%%%%%%%%%%%%%%%%%%%%%%%%%%%%%%%%%%%%%
%%%%%%%%%%%%%%%%%%%%%%%%%%%%%%%%%%%%%%%%%%%%%%%
%%%%%%%%%%%%%%%%%%%%%%%%%%%%%%%%%%%%%%%%%%%%%%%
%%%%%%%%%%%%%%%%%%%%%%%%%%%%%%%%%%%%%%%%%%%%%%%
%%%%%%%%%%%%%%%%%%%%%%%%%%%%%%%%%%%%%%%%%%%%%%%
%%%%%%%%%%%%%%%%%%%%%%%%%%%%%%%%%%%%%%%%%%%%%%%
%%%%%%%%%%%%%%%%%%%%%%%%%%%%%%%%%%%%%%%%%%%%%%%

\end{document}